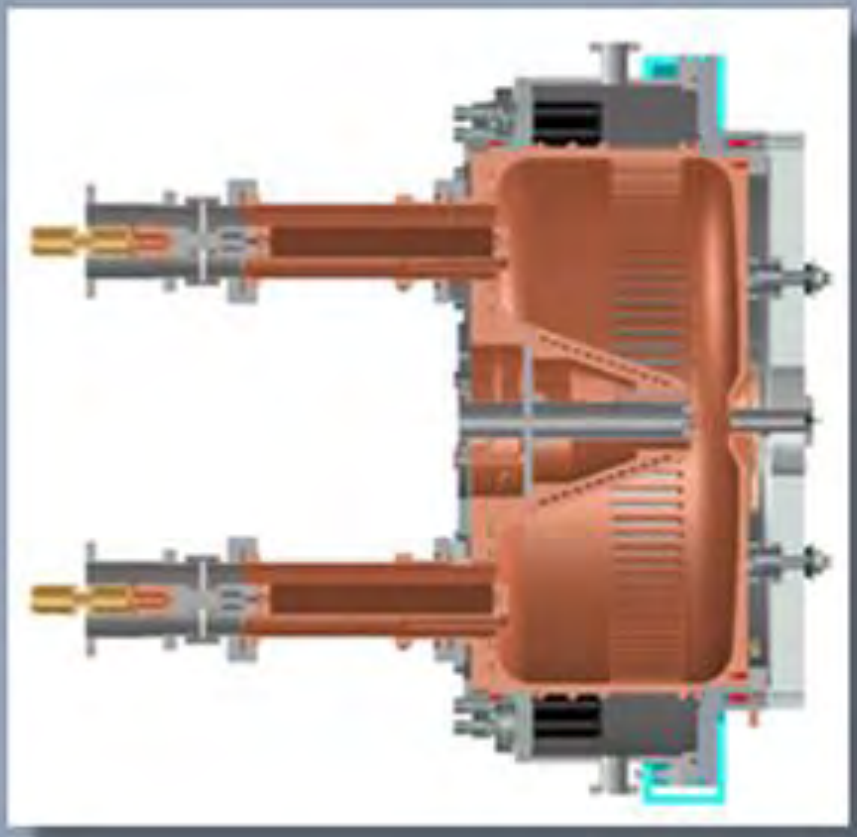
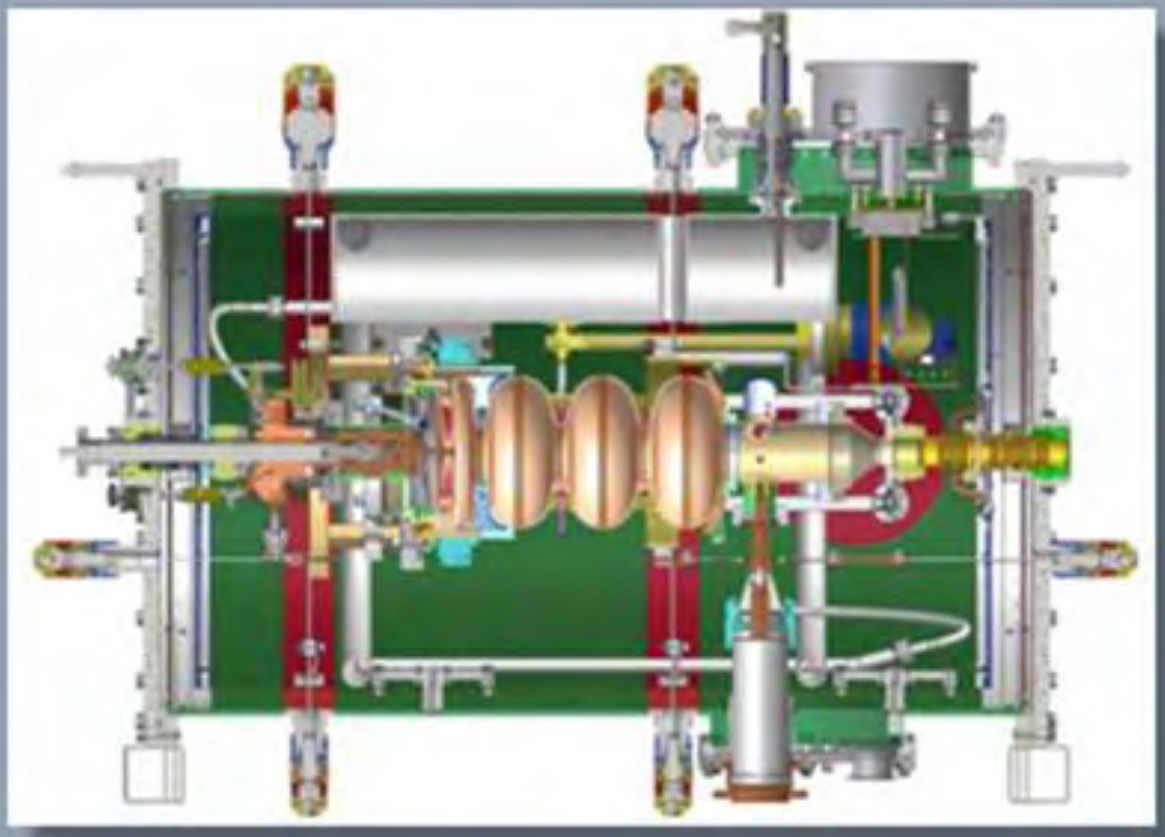

# AN ENGINEERING GUIDE TO PHOTOINJECTORS

Editors:

**Triveni Rao**

**David H. Dowell**

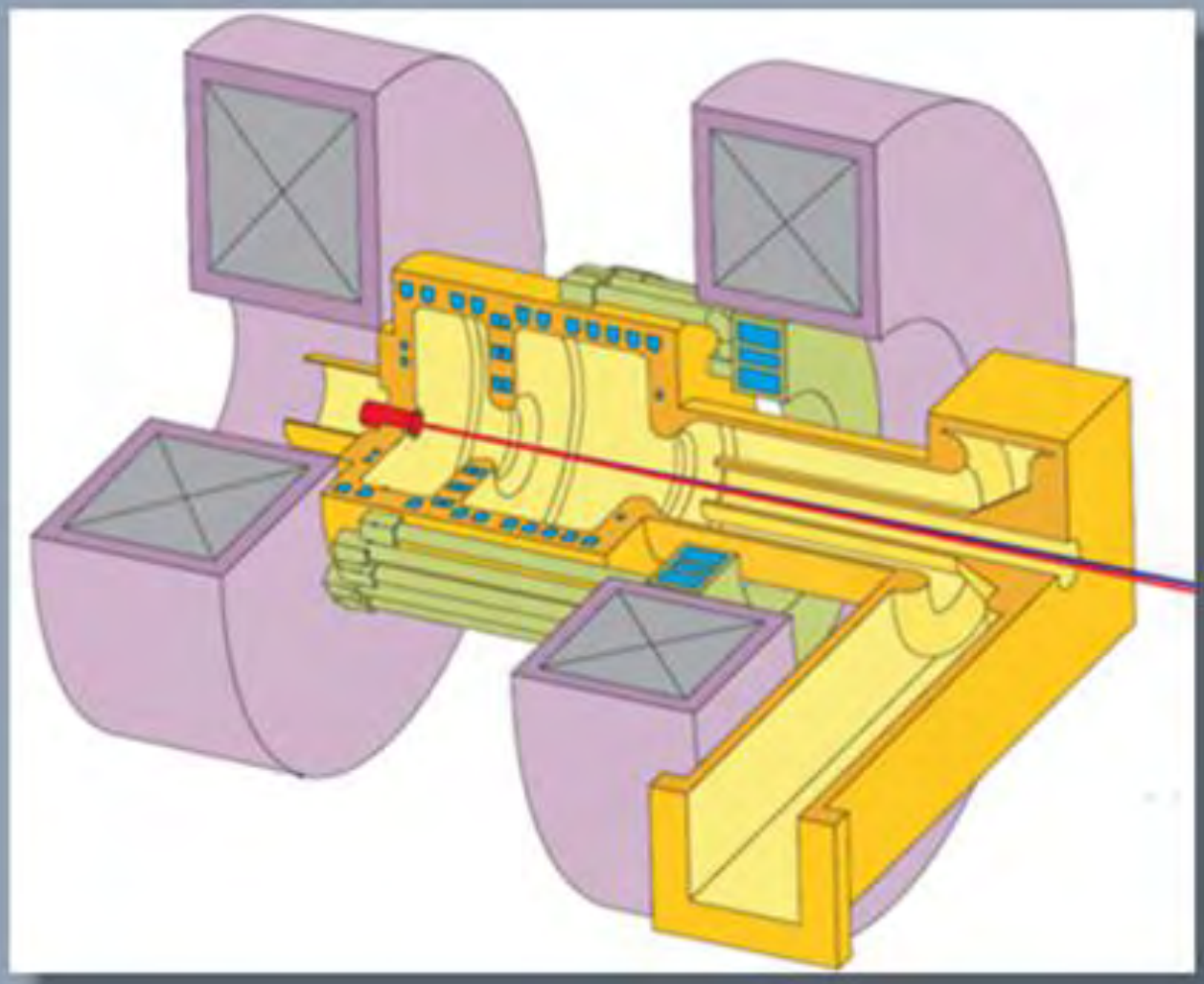
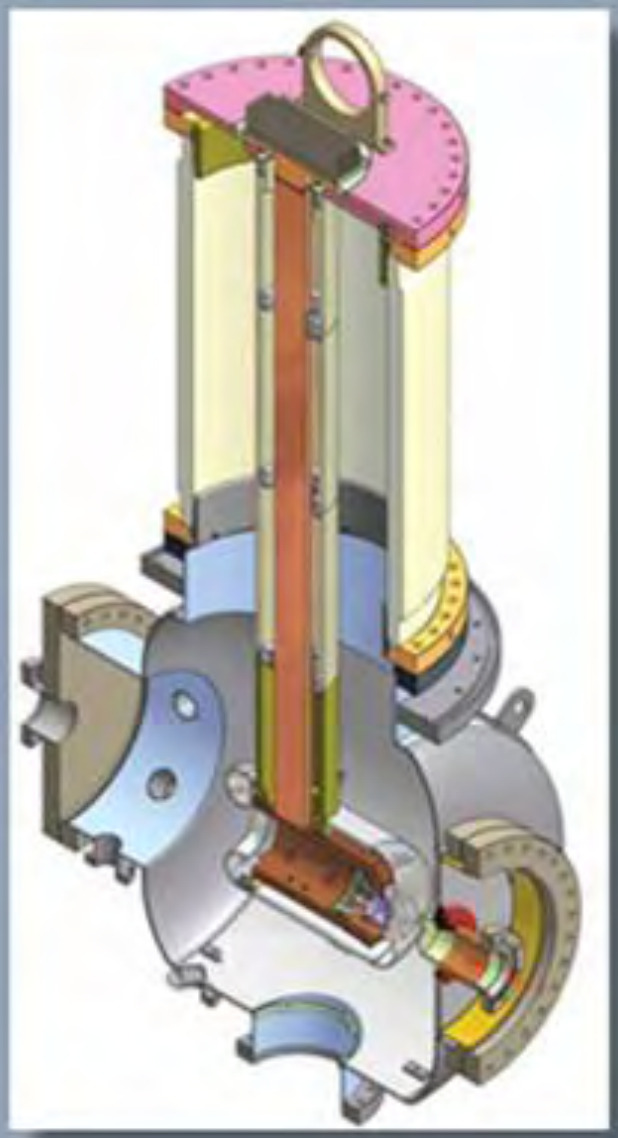

# AN ENGINEERING GUIDE TO PHOTOINJECTORS

Edited by
Triveni Rao
David H. Dowell

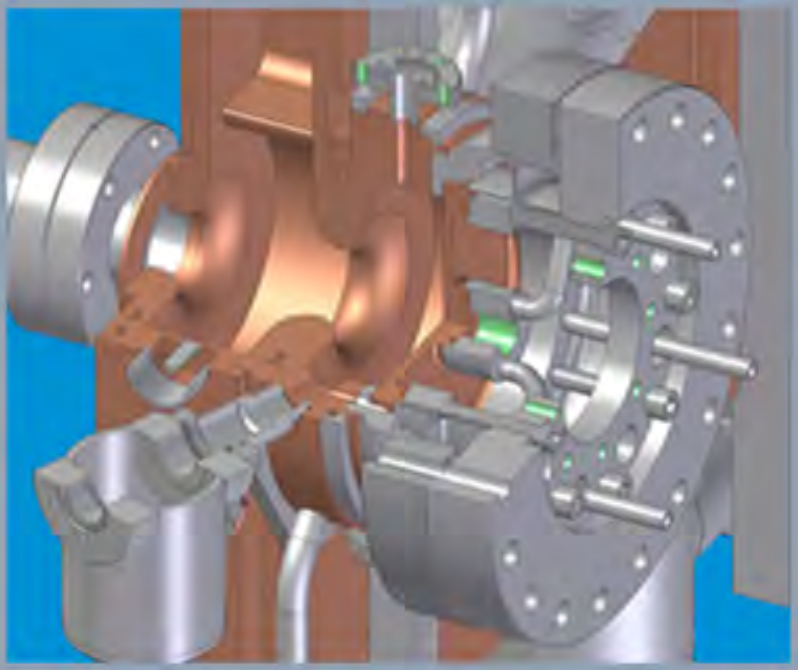

This book provides an introduction to the basic theory and engineering of advanced electron beam sources known as photoinjectors. Photoinjectors produce relativistic electrons for exciting new devices such as x-ray free electron lasers and the polarized beams for very high energy physics linear colliders. The chapters are written by renowned experts in the field who share their working knowledge of the technologies needed for designing and building photoinjectors.

Contents:

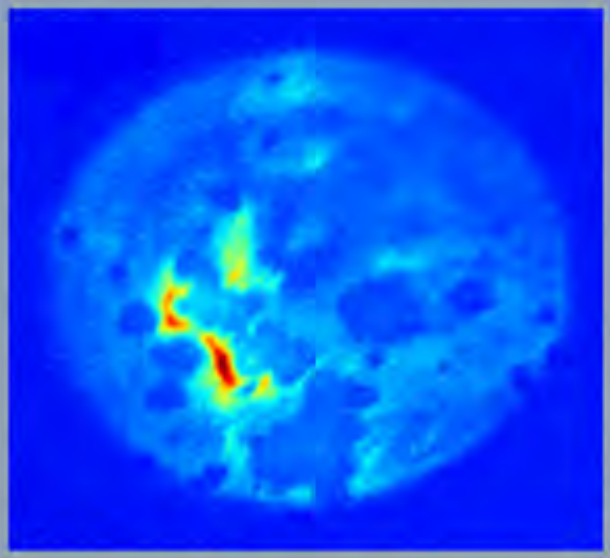

**The paperback printed version of this book can be found at:**
http://www.amazon.com/s/ref=ntt_athr_dp_sr_1?_encoding=UTF8&field-author=Dr.%20Triveni%20Rao&search-alias=books&sort=relevancerank

# DEDICATION

The editors dedicate this book to:

Rama Rao, Krishna Rao and Kavitha Rao for T. Rao

and

Alice Pitt and my parents, Harry and Betty Dowell for D. Dowell

# List of Contributors

**Wolfgang Anders**
Helmholtz-Zentrum Berlin for Materials and Energy
Albert-Einstein-Str. 15, D-12489
Berlin, Germany
Email: wolfgang.anders@helmholtz-berlin.de

**Ivan Bazarov**
Physics Department
373 Wilson Laboratory
Cornell University
Ithaca, NY 14853
Email: ib38@cornell.edu

**Luca Cultrera**
Cornell Laboratory of Accelerator-based Sciences and Education
Cornell University
Ithaca, NY 14853
Email: lc572@cornell.edu

**Dimitre Dimitrov**
TechX Corporation
5621 Arapahoe Ave. Suite A
Boulder, CO 80303
Email: dad@txcorp.com

**David H. Dowell**
SLAC National Accelerator Laboratory
Menlo Park, CA 94025-7015
Email: dowell@slac.stanford.edu

**Bruce Dunham**
Department of Physics and CLASSE, Wilson Laboratory
Cornell University
Ithaca, NY 14853
Email: bruce.dunham@cornell.edu

**John W. Lewellen**
Naval Post Graduate School
Monterey, CA 93943
Email: jwlewell@nps.navy.mil

**Axel Neumann**
SRF Science and Technology/BESSY II
Helmholtz-Zentrum Berlin for Materials and Energy
Albert-Einstein-Str. 15, D-12489
Berlin, Germany
Email: axel.neumann@helmholtz-berlin.de

**Matthew Poelker**
Thomas Jefferson National Accelerator Facility
M/S Bldg. 5A, Room 500-17
12050 Jefferson Ave.
Newport News, VA 23606
Email: poelker@jlab.org

**Triveni Rao**
Brookhaven National Laboratory
Upton, NY 11973
Email: triveni@bnl.gov

**Siegfried Schreiber**
Deutsches Elektronen-Synchrotron
Notkestraße 85, D-22603
Hamburg, Germany
Email: siegfried.schreiber@desy.de

**John Smedley**
Brookhaven National Laboratory
Upton, NY 11973
Email: smedley@bnl.gov

**Thomas Tsang**
Brookhaven National Laboratory
Upton, NY 11973
Email: tsang@bnl.gov

# PREFACE

Since the discovery of electrons more than a century ago, the generation, transport and characterization of electron beams have been an active field of research. Breakthroughs in this field have led to applications as far reaching as cancer treatment, investigation of nanoscale material and dark matter. In this book, we present a snapshot of the photoinjector technology which has contributed to these advances.

The brightness of a charged particle beam, defined as the number of electrons within the 6-D phase space volume, dictates its luminosity, a critical parameter in a linear collider. Likewise, since wavelength $\lambda$ of the free-electron laser (FEL) is governed by the inequality, $\varepsilon_n/\gamma < \lambda/4\pi$, the emittance $\varepsilon_n$ of the electron beam (6-D phase space volume) sets the lower limit on the wavelength deliverable by an FEL while the charge in this emittance envelope sets the gain of the FEL. As the wavelength of the FEL gets shorter, the required emittance gets lower and thereby brightness increases. Pushing the limits even further, recently, there has been greater emphasis on delivering not only low average current beams with high peak brightness, but also upon increasing the average current to a significant fraction of an ampere as well.

It has been shown consistently in the past two decades that a photoinjector can meet these needs set by the accelerator community for photon sources and high energy colliders. Over this period, the transverse emittance of the electron beam has decreased steadily to nearly the intrinsic limit of the electron source; its bunch length at the source has reduced to femtosecond and the charge per bunch greater than a nanocoulomb can be delivered routinely. Better understanding of the electron beam production and transport, resulting from rigorous bench marking of computer codes against experimental measurements, along with improved diagnostics have led to reliable performance in a user facility environment.

The rapidly advancing technology of high density, relativistic electron beams has made possible a wide range of new and exciting tools for research and industry. Some of the new applications include linear particle colliders, Compton scattering sources, electron cooling of protons and heavy ions stored in a ring, energy recovery linac (ERL) light sources, FELs, inverse FELs and ultrafast electron diffraction. In many cases these new devices would not be possible without the invention of the photocathode gun. This is because it alone has the capability to deliver the high quality beam necessary to drive an X-ray FEL, produce the low divergence beam required to observe an electron diffraction pattern or generate the high density polarized beams desired for the International Linear Collider (ILC).

This progress motivates the need for a comprehensive presentation of this critical photoelectron gun and injector technology. Therefore we have taken the audacious step toward writing the first such book on the technology of photoinjectors in collaboration with our distinguished colleagues. This is an ambitious undertaking since this broad topic covers subject matter ranging from the chemistry of photocathodes, to the lasers which drive them, to the HV and RF power systems which extract the electrons and accelerate them to high energy and myriad of other topics. Because of the wide breadth of this topic, there is no one person who can be an expert on all these subjects; and so we have collected contributions from renowned researchers who are builders of the instruments described. This collection of writings not only provides practical information, but also reflects each author's unique view of the technology to which they themselves have made significant contributions. We thank them for their efforts and hope this book becomes the resource accelerator engineers and scientists open first for information on cathodes, lasers, RF systems and other parts of the photoinjector.

This book primarily addresses the engineering aspects of a photoinjector. For the purpose of the book, the injector system is defined to consist of a cathode, a gun for rapidly accelerating the electrons from rest, followed by the booster accelerator that increases the electron energy to ~10 MeV. In some systems, it may also include low energy ballistic compression of the bunch.

The book is mainly directed to the practicing scientist or engineer interested in designing and building an injector. The comprehensive coverage makes it a useful tool supplementing practical considerations for theoretical understanding. Abstract mathematical equations are kept to a minimum in keeping with the objective of an engineering guide. For readers interested in delving deeper into the subject, sufficient references are provided for a more detailed account. Although the latest techniques and results are presented in the book, the reader should keep in mind that this field is still evolving.

The book is organized as follows: Chapter 1 gives a theoretical underpinning of an injector. Chapters 2, 3 and 4 describe the most common photoinjectors: i) Normal Conducting RF (NCRF) injector, ii) Superconducting RF (SCRF) injector, and iii) DC injector. Since the photocathode is a common element in all these injectors, Chapter 5 provides the fundamental physics of photoemission – namely the three-step model and its relevance to the typical cathodes used in operating systems. Chapters 6 and 7 describe the most common metal and semiconductor cathodes used to generate unpolarized electrons, their properties and preparation techniques, as well as how they are incorporated into the gun. Chapter 8 discusses the system to deliver polarized electron beams. Chapter 9 deals with the choices one has with the laser systems, the synchronization of the laser to the RF and longitudinal- and transverse-shaping of the laser profile for low emittance beams. Typically, the RF systems in the photoinjectors are commercial devices; hence, care must be taken in determining the system and its specifications that are appropriate for the application. This is discussed in Chapter 10, along with transmitting and coupling the high power to the cavity, controlling the amplitude and phase of the RF power delivered to the cavity, and design of higher-order mode dampers. All relevant diagnostics are dealt with in Chapter 11.

The editors wish to specifically acknowledge our co-authors and thank them for finding time in their busy day jobs to write, review and correct chapters for this book. It simply would not have been possible without the generous sharing of their knowledge, expertise and time. We are confident that students of the field will benefit from the unique insights provided by their intimate knowledge of the subject.

T. Rao would like to acknowledge the immense support provided by V. Radeka and Brookhaven National Laboratory. D. Dowell wishes to thank his many colleagues who have supported and inspired him over the years. And finally, all the authors are indebted to M. Rumore and A. Woodhead who have read the manuscript carefully, suggested changes, worked diligently to format the text, figures, references and equations and design of the final product. They vastly improved its readability and made it look great.

It has been a long process and we extend our heartfelt gratitude to everyone for their patience and persistence to make this book a reality.

Triveni Rao and David H. Dowell
Upton, NY and Seattle, WA
September 10, 2012

# Table of Contents

# CHAPTER 1: PHOTOINJECTOR THEORY

DAVID H. DOWELL

*SLAC National Accelerator Laboratory*
*Menlo Park, CA 94025-7015*

JOHN W. LEWELLEN

*Naval Post Graduate School*
*Monterey, CA 93943*

**Keywords**
RF Gun, RF Injector, Accelerator, Field Distribution, Beam Dynamics, Emittance, Brightness, Space Charge, Bunch Shape, Emittance Compensation, Chromatic Aberration, Geometric Aberration, Anomalous Field Aberration, Space Charge Shaping, Simulation Code, Injector Design Code, Particle-in-Cell Code, Envelope Code

**Abstract**
A comprehensive theory of photoinjectors involves a wide range of accelerator physics topics ranging from the material science of cathodes to the dynamics of electrons in magnetic, RF and DC fields as well as the strong effects the electrons have upon each other in their mutually repulsive fields, *i.e.* space charge fields. Whereas other chapters are concerned with subjects such as the physics of electron emission, this chapter concentrates upon electron beam dynamics from the cathode to the high energy accelerator after the gun. It briefly describes the history and components of the photoinjector as well as the basic beam parameters of emittance and brightness. The chapter then discusses beam dynamics without space charge (*i.e.* forces due to RF fields only), beam dynamics with space charge, the focusing and aberrations due to the magnetic solenoid lens and controlling beam quality with transverse shaping of the beam to eliminate non-linear space charge forces. The last section lists and describes the simulation codes available to the designers of photoinjectors. Two appendices giving tables of the chapter's formulae and mathematical symbols are included as a quick reference.

## 1.1 INTRODUCTION

Advances in the technology of high density, relativistic electron beams have made exciting new applications practical realities. A sampling of these new applications include Compton scattering sources, electron cooling of protons and heavy ions stored in a ring, energy recovery linac (ERL) light sources, free-electron lasers (FELs), inverse FELs and ultrafast electron diffraction. The first demonstration of a high average power FEL [1.1] and the operation of the first hard X-ray FEL as a $4^{th}$ generation light source [1.2] represent two challenging fronts on the frontier of high brightness beam applications. Both of these achievements have benefited from the invention and continued development of the photoinjector.

This book's goal is to describe the technological components of these new photoinjectors from an engineering perspective. These technologies involve fields as diverse as RF power, high voltage (HV) DC, lasers, chemistry, ultra-high vacuum (UHV), beam optics, and others which are ably discussed by the authors of the other chapters and detailed by their references. It is the goal of this chapter to collect the various aspects of the theory of photoinjectors.

The photoinjector consists of a laser generated electron source followed by an electron beam optical system which preserves and matches the beam into a high-energy accelerator, as shown in Figure **1.1**. Matching the electron bunch to the first high-energy accelerator is one of the photoinjector functions, since the proper phase space sizing of the beam into the first accelerator section is an essential element of emittance compensation. In the low-energy regime of the photoinjector, the beam is considered to be “space charge dominated.” In other words, its optical properties are strongly determined by the defocusing of space charge forces. Since the space charge forces are diminished as (*beam energy*)$^{-3}$, at relativistic energies, the electrons begin to follow ray optics and the beam is referred to as “emittance dominated.”

The generic configuration of the photoinjector is shown in Figure **1.1**. The photoinjector consists of a cathode fabrication and/or transport system and electron gun, powered by RF (Chapter 10) or biased at a high voltage, beam optics for transporting and matching the beam to the high-energy accelerator, and assorted diagnostics (Chapter 11) and controls. The photocathode can be either a metal (Chapter 6) or one of several semiconductor materials (Chapters 7 and Chapter 8). The gun can be a high voltage DC gun (Chapter 4), a normal conducting RF (NCRF) gun (Chapter 2) or a superconducting RF (SCRF) gun (Chapter 3). In addition, it is necessary to suit the drive laser to the type of cathode and the desired pulse format (Chapter 9). Although there is a wide range of options for the cathode, laser and gun, the underlying beam physics is quite similar, as shown in the following sections of this chapter.

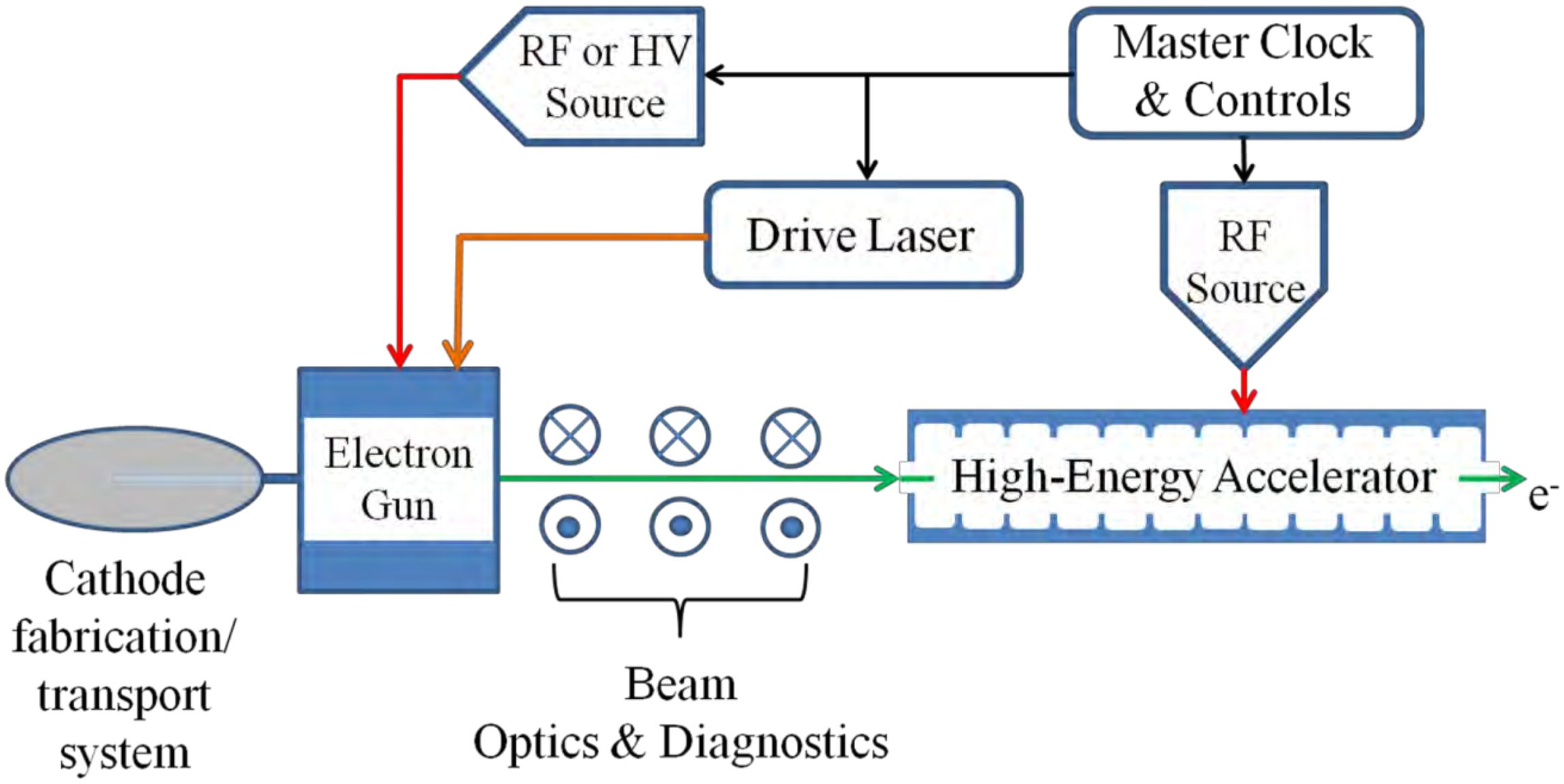

Figure 1.1. The parts of a photoinjector.

Electron beam quality is often specified in terms of three quantities: Emittance, peak current and brightness. The emittance is the area or volume of phase space the electrons occupy. In general, the phase space is the six dimensional space formed by an electron’s three positions and angles. In 6-dimensional (6-D) space, the phase space coordinates are $x$, $x'$, $y$, $y'$, $z$ and $z'$. Here, $x$, $y$ and $z$ are the electron position coordinates in the right-handed Cartesian coordinate system: the beam center is moving in the $z$-direction, the $y$-axis is vertical, and the $x$-axis pointing horizontal and to the left. The angles $x'$, $y'$ and $z'$ are given by the momentum along that axis divided by the total momentum. For example, $x' = p_x/p_{total}$, where $p_{total}^2 = p_x^2 + p_y^2 + p_z^2$. Strictly speaking, the 6-D emittance should be computed from the electron distribution in the 6-D phase space. While this may be possible theoretically, in experiments the complete distribution details are not known and one can only measure projections of the distribution to find the emittance and other beam qualities. Projections of the 6-D phase space onto the 2-D sub-spaces of $xx'$, $yy'$ and $zz'$ are called trace spaces.
The energy normalized trace space emittance is defined as

$$\varepsilon_n = \beta\gamma \sqrt{\langle x^2\rangle\langle x'^2\rangle - \langle xx'\rangle^2} \tag{1.1}$$

where $\beta = v\, c^{-1}$ is the electron velocity in units of the speed of light, $c$, and $\gamma = E_{total}/mc^2$ as the total beam energy, $E_{total}$, normalized to the electron mass, $mc^2$. Multiplying Equ. 1.1 by $\pi$ gives the area of trace space occupied by the beam distribution, so it's often stated that the $\pi$ is an "implied" factor in emittance unit. One can show that if there are no correlations between $x$, or $x'$, and the other four coordinates that the trace space emittance in $xx'$ trace space is conserved, that is $\langle xy\rangle = 0$, $\langle xy'\rangle = 0$ …, and similarly for the other trace spaces of $yy'$ and $zz'$. However, correlations between the separate trace spaces lead to an increase in the trace space emittance. An example of a $\langle xyx'y'\rangle$ correlation is given in Section 1.5.4. Since it includes the effects of the correlations mixing of the trace spaces, the 6-D phase space emittance, often called the canonical emittance, remains unchanged. For the rest of this chapter, the trace space emittance will be referred to as the normalized transverse emittance, or simply the emittance.

The peak current, $I_{peak}$, is the bunch charge divided by the bunch length $q_{bunch}$ and is usually calculated using the full-width at half-maximum (FWHM) bunch length. However, some authors compute it using the root mean squared (rms) or some other variant of the full width, or even reduce the charge to include only those electrons in the core of the bunch. In this book, we will use the total bunch charge and the FWHM bunch length. The peak current out of the gun should be as high as possible to avoid having to bunch the beam before acceleration in the high energy linac. The bunch length, and hence the peak current from the injector depends upon the RF frequency of the main accelerator since the bunch length should be a small fraction of an RF period. A sensible guideline is less than 10° RF[1] for the full bunch length. Hence, for a 3 GHz, linac the beam at the end of the injector should approximately have a maximum bunch length of 10° S or 10 ps. This then establishes a lower limit for the peak current at a given charge, ignoring bunch elongation. For example, at 1 nC the peak current from a 3 GHz gun would be 100 A. Most applications require $I_{peak}$ to be 10-100 A from the injector.

The beam brightness combines the emittance and the peak current into a single parameter measuring the electron volume density. There are various definitions for beam brightness, each having its own merits. The common practice is to define the transverse, normalized beam brightness, $B_n$, as given by Equ. 4.2 in Chapter 4

$$B_n = \frac{2I}{\pi^2 \varepsilon_{n,x}\varepsilon_{n,y}} \tag{1.2}$$

Here $\varepsilon_{n,x}$ is the normalized $xx'$ trace space emittance and $\varepsilon_{n,y}$ is the $yy'$ trace space emittance. The peak current is the bunch charge divided by the bunch FWHM.

While Equ. 1.2 is commonly used to define the beam brightness, a more accurate representation of beam brightness would include the bunch length and energy spread. Similar to the transverse emittance, these longitudinal characteristics can be represented by the longitudinal emittance, which in its simplest form (ignoring any correlations) is the product of the bunch length, $\Delta t$, and energy spread, $\Delta E$

$$\varepsilon_z = \sigma_z \, \sigma_{\Delta E/E} \tag{1.3}$$

[1] In this book, we will denote the RF phase in degrees as "° RF" and the temperature in degrees Celsius as "°C" to avoid the confusion of using "°" for both. When discussing a specific RF frequency band, such as the S-band, the RF phase in degrees at the S-band frequency is given as "° S" and similarly for the other frequency bands.

Here, $\sigma_z$ is the rms bunch length and $\sigma_{\Delta E/E}$ is the rms fractional energy spread. An alternative definition of the peak brightness would then be

$$B_{peak} = \frac{2q_{bunch}}{\pi^2 \varepsilon_{n,x} \varepsilon_{n,y} \Delta t \Delta E} = \frac{2q_{bunch}}{\pi^2 \varepsilon_{n,x} \varepsilon_{n,y} \sigma_t \sigma_E} \tag{1.4}$$

The factor of '2' results from the integration over the 4-D trace space enclosed by a hyperellipsoid. [1.3]

### 1.1.1 The First Photocathode RF Gun

In 1985, it was demonstrated that a photocathode could survive while delivering high current densities of over 200 A $cm^{-2}$. [1.4][1.3] This was rapidly followed by the first operation of a photocathode gun at Los Alamos National Laboratory as the electron source for an FEL experiment [1.5]–[1.7]. The gun was a single RF cell connected to a cross in which $Cs_3Sb$ cathodes were fabricated on the end of a long stick which could be inserted into the gun. The laser beam was reflected from an in-vacuum mirror through the solenoids and gun and onto the cathode. Downstream of the gun, the beam emittance was measured using the pepper-pot technique and a magnetic spectrometer measured the energy and energy spread. While this demonstration showed the many advantages of improved beams from photocathodes, it also illustrated the difficulties of working with a cathode material which is often hard to fabricate and always sensitive to its environment.

### 1.1.2 Summary of Advances

The invention of the photoinjector motivated a large increase in the number of laboratories studying electron guns. In 1983, there were only 1 or 2 gun projects, but by 1990 there were more than 25 laboratories along with a few companies actively building and testing guns [1.8]. The immediate improvement in electron beam quality with the advent of the photocathode gun was then followed by a methodical pace of small but steady steps toward the current state-of-the-art. Figure **1.2** illustrates the history of the 1 nC bunched beam emittance over the past 50 years.

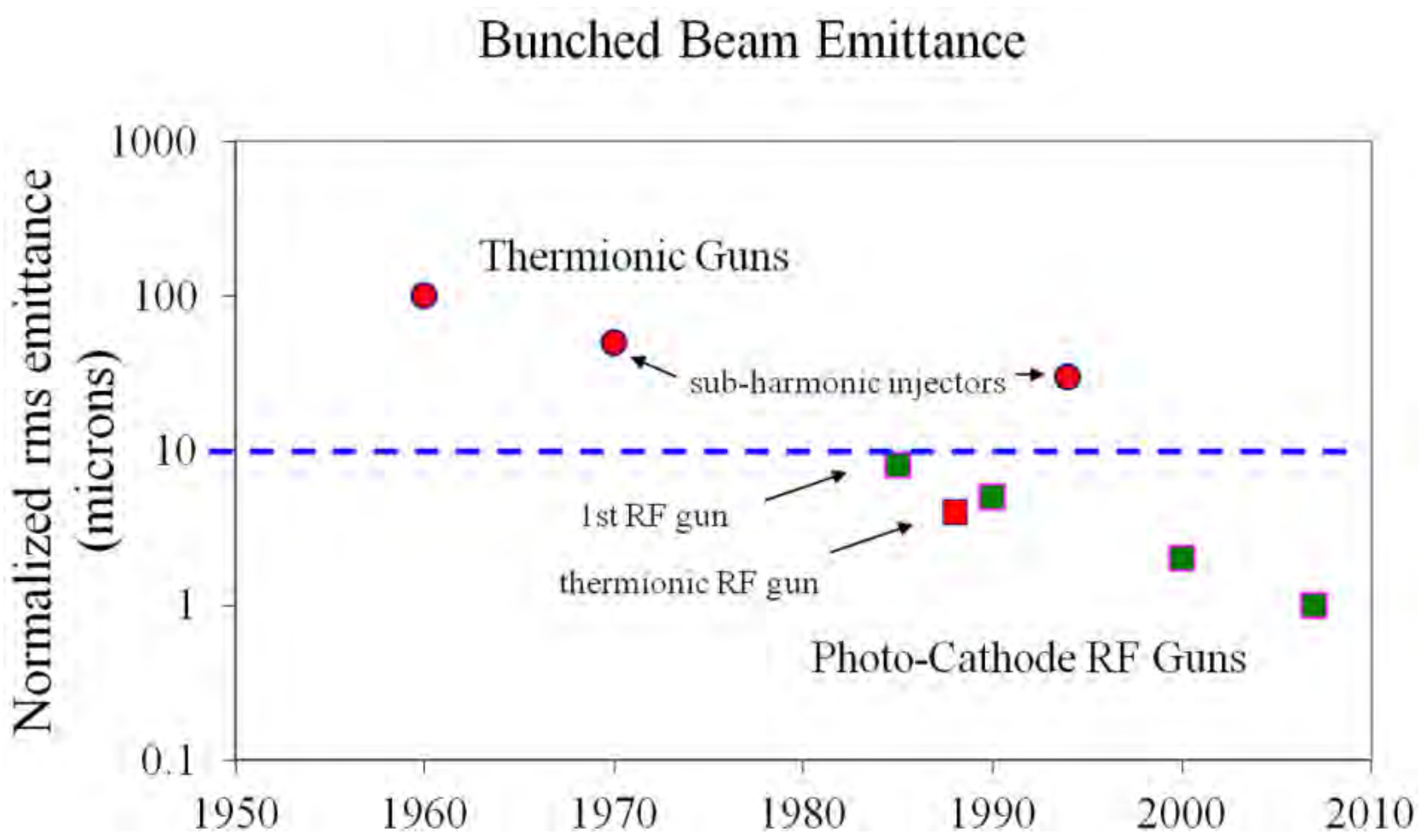

Figure 1.2. The demonstrated normalized emittance over the last 50 years for bunches compressed to 50 A to 100 A peak current.

The emittance is for a beam bunched from 50-100 A, since this is what's required for injection into a high energy linac. The projected emittance for 1 nC bunches is shown for thermionic and photocathode gun injector technologies in the figure. Thermionic guns were combined with sub-harmonic RF bunchers to achieve the 1 nC charge but could not achieve less than approximately 20 µm for the normalized emittance. In contrast even the first photocathode RF gun demonstrated better than 10 µm. This quick success was then followed by ~20 years of research to reach the goal of 1 micron at 1 nC.

In this book, we classify photoinjectors into three types: Normal Conducting RF (NCRF), Superconducting RF (SCRF) and high voltage DC (HV DC). There are one or more chapters devoted to each type. The proliferation and wide use of all three types of photoinjectors is illustrative of its success. Today there are more than a dozen facilities with NCRF guns, six operating or proposed SCRF guns and at least four using high brightness HV DC guns. A comparison of the beam emittance, charge, longitudinal phase space, repetition rate and other beam parameters for these three photoinjector types can be found in [1.9].

In addition, there are innovative injector designs which combine two or more of these three basic types of photoinjectors. A good example is the DC SCRF photoinjector at Peking University in Beijing. This system has a 90 kV HV DC gun placed close to the entrance of a 3½-cell SCRF accelerator and tries to combine the best features of the HV DC gun and the SCRF linac. [1.10]

The optimal RF frequency for the photocathode gun is often discussed, and it is often concluded that because higher frequencies produce higher electric fields that high frequencies should produce the best beam quality. There is no doubt that high electric field is important/essential for outrunning the space charge forces. However, there are other phenomena whose emittances scale upward with increasing RF field and frequency. For example, RF emittance is an unavoidable consequence of a gun with strong fields at a high RF frequency. As shown in Section 1.3, this emittance scales linearly with the peak electric field. Similarly, the emittance caused by the field enhancement due to rough cathode surfaces scales as the square root of the cathode field. Furthermore, for high average current injectors, the thermal management associated with high electric fields may even preclude operation in such a regime. Thus, high fields are not always advantageous.

Even with these advances, there remain opportunities for further improvements. Historically, the laser and the cathode have been the most problematic. However, recent developments in laser technology have resolved many of these issues. Diode-pumped solid state lasers now provide stable and reliable photons for photocathode guns. It is now possible to deliver the peak current required with a suitable choice of the cathode and the laser. Instead, further progress should concentrate on near the cathode. Improving the electron emission and mitigating the image and space charge effects will require a more complete understanding of the physical and chemical properties and reactions of the cathode material, both during fabrication and use in the gun. The goal is to develop cathodes with reliable and improved performance. The current cathode technology is discussed in Chapters 5 and Chapter 7.

Once the electrons are liberated from the cathode, they experience strong self-fields (space charge limited emission and emittance), fields in the accelerating cavities (RF emittance), and fields of the transport optics, all of which degrade the beam quality (chromatic and geometric aberrations). Space charge emittance is produced by the self-mutual repulsion of the electrons in the bunch, which is aggravated by the backward attraction of the image charge. The space charge limit occurs when the image charge electric field equals and cancels the applied electric field. Space charge emittance takes on various forms, but is always driven by non-uniformities in the charge density. The variation in the longitudinal charge density deforms the bunch with a variation in divergence, and hence emittance along its length. Emittance compensation in an RF gun refers to balancing the space charge divergence by focusing these divergence variations into alignment and compensate for the linear space charge force. This is possible if the space charge force is linear and the space charge force can be made linear if one uses special bunch shapes, for example the “beer-can.” Non-linear space charge forces increase the emittance due to a non-uniform charge distribution. The non-linear emittance usually cannot be corrected and instead is avoided by making the emission uniform.

The RF emittance results from the time-dependent focusing by the RF fields. The emittance of a thin time-slice of the bunch is unchanged and the slice merely obtains a change in divergence or an instantaneous kick in angle, as it would in a thin lens. However, the varying RF field gives a different divergence to each time-slice, which increases the projected emittance of the entire bunch. The RF emittance has a first- and second-order dependence upon the bunch length. This emittance is usually minimized by operating with short bunch lengths and optimal timing of when the laser produces the bunch with respect to the RF fields.

Since the gun itself acts as a strong defocusing lens, an equally strong focusing lens is needed to refocus the beam. This lens is usually a solenoid whose axial field focuses the electrons. Like all strong lenses this solenoid can produce a chromatic emittance due to the bunch energy spread and geometric emittance due to the transverse size.

This chapter attempts to discuss these effects with simple mathematical models. These models attempt to capture the underlying physics of the effects while providing useful formulas for estimating their contribution to the emittance. However, they are not meant to replace the need for numerical simulations using advanced multi-particle and mesh codes. The codes allow inclusion of the field shape and electron distribution minutia.

### 1.1.3 Organization of this Chapter

This chapter discusses the analytic theory of photocathode RF guns. It does this through the use of analytic models which accurately describe the physical phenomena for each portion of the photocathode gun system. The chapter begins with the definition of a photoinjector in Section 1.2. It is important to realize that producing a good beam from the cathode and gun is only the first part of emittance preservation of a larger system. There is also the matching of the beam into the first accelerator section and the damping of the emittance.

A discussion of beam dynamics without space charge forces is given in Section 1.3. The first- and second-order transverse RF emittances are derived and their relative sizes compared. The longitudinal emittance due to the RF is also computed.

Section 1.4 gives the space charge emittance for cylindrical (beer can) and Gaussian charge distributions. Analytic formulae are derived for both uniform and spatially varying transverse charge distributions. Space charge limited emission for a photocathode gun is shown to be different from the Child-Langmuir law. And a simple space charge model is given for non-uniform transverse emission. Emittance compensation is expressed in Section 1.4.4 in terms of a plasma oscillation being matched to the first linac section for minimum projected emittance. In a sense, this could be referred to as slice dynamics, given that it involves manipulating the slice parameters to make them the same.

Section 1.5 begins describing a linear optical model for the gun and solenoid. It then continues into an analysis of the solenoid's aberrations. The aberrations considered are chromatic, geometric and anomalous-quadrupole field. The anomalous-quadrupole field aberration results from a low strength quadrupole field which strongly couples the *x*- and *y*-emittances because of the large rotation angle in the solenoid. A means for "recovering" this emittance using normal and skew corrector quadrupoles is given.

The interesting and important topic of space charge shaping is presented in Section 1.6. Here, fundamental electromagnetic theory is used to compute ideal shapes with minimal non-linear space charge force. Often referred to as the "blow-out" regime, the linearization phenomenon is actually achieved by shaping the radial charge distribution so as to make the higher order space force terms zero.

The last section, Section 1.7, describes the capabilities of beam simulation codes. Sophisticated simulations allow high resolution computation of nearly all the details of the gun, solenoid and beamline components. And while the theory and models presented in the proceeding sections are important for understanding concepts and trends, all realistic gun designs require simulations with all the fine details.

# 1.2 THE RF PHOTOINJECTOR

## 1.2.1 The Photocathode RF Gun, Drive Laser and First Accelerator Section

A typical photocathode RF system shown in Figure **1.3** depicts a 1½-cell gun with a cathode in the ½ cavity being illuminated by a laser pulse train. At the exit of the gun is a solenoid which focuses the divergent beam from the gun and compensates for space charge emittance. The drive laser is mode-locked to the RF master oscillator which also provides the RF drive to the klystron. Other types of RF sources used to power RF guns are inductive output tube (IOT) and solid state RF amplifiers. (Not shown is the high voltage power supply powering the klystron.) How the RF power is coupled into the gun is an important technical aspect of the gun design. At high RF frequency, the coupling can be through the side-wall of one of the cavities with the cell-to-cell RF coupled through the irises between the cavities. Alternatively, the power can be coupled using coaxial coupler either at the beam exit or around the cathode [1.11]–[1.13]. At low frequencies, the RF can be launched using a coaxial cable [1.14]. These and other coupling schemes for NCRF guns are described in Chapters 2 and Chapter 10.

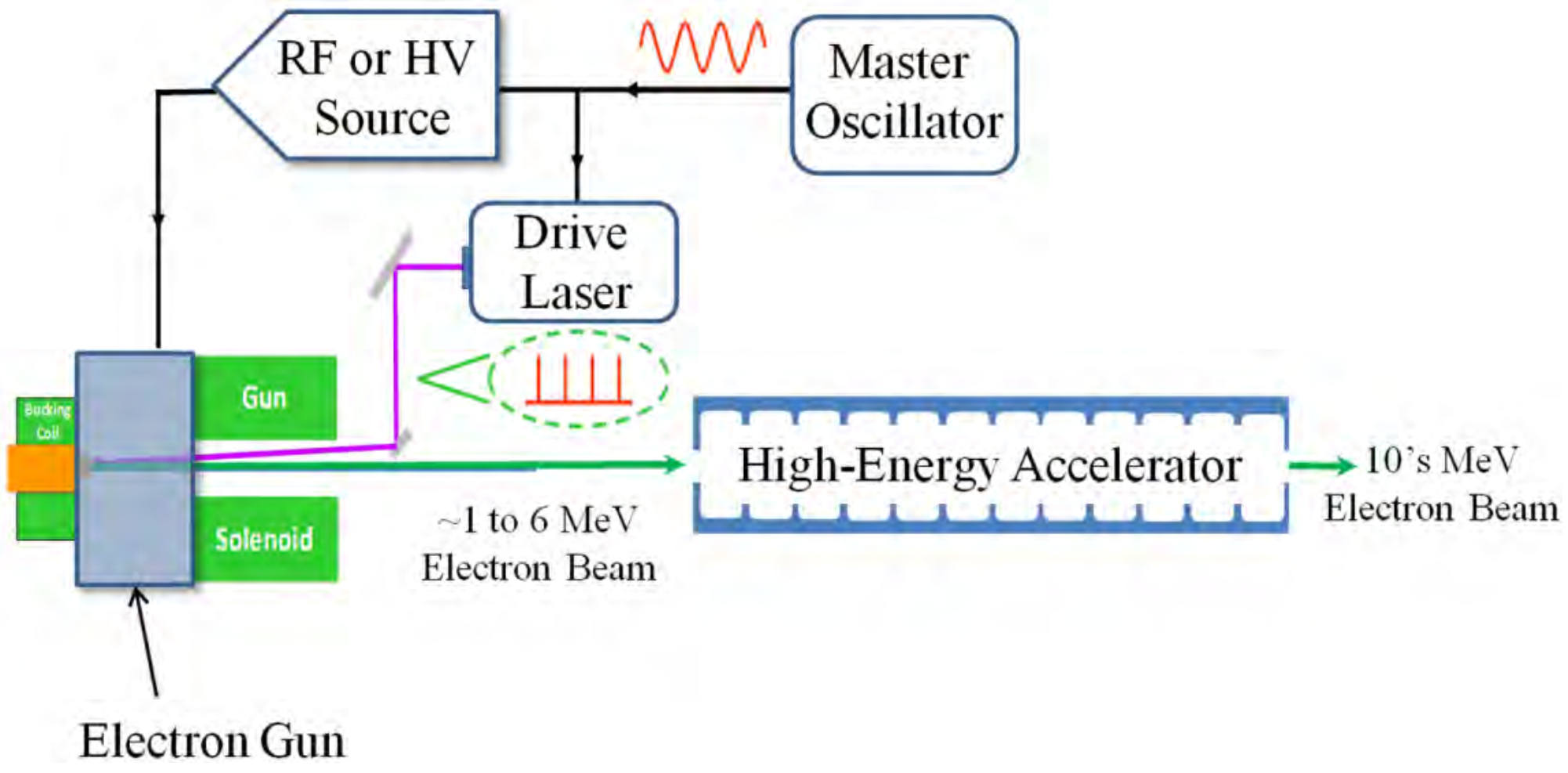

**Figure 1.3. Basic components of the photocathode RF gun injector.**

Equally important as the high field RF gun, cathode and laser is the optical matching of the beam size and divergence into the first linac section. The distance between the end of the gun and the entrance to the linac is determined by the bunch's plasma oscillation period. As discussed in Section 1.4.4, the bunch is matched when all the slices are aligned in transverse phase space, *i.e.*, have equal phase space parameters.

### 1.2.2 The ERL Injector System

Beam matching to the main linac is straightforward for a single pass accelerator. However, it becomes more problematic for circular machines, such as ERLs. In these accelerators, the spent, high-energy beam is decelerated in the same linac sections which accelerated them. Since ERLs are designed to operate at high average current, it is best to merge and un-merge the beam at beam energies which are below the neutron threshold of commonly used beam dump materials. Since neutron thresholds of most materials are from 10-15 MeV, 10 MeV is taken as the upper beam energy for the merger.

Due to the low beam energy and the relatively high peak current, there are significant space charge forces and other non-linear effects which can increase the emittance in the beam transport of the merger. Similar to emittance compensation technique, any emittance growth due to a correlation along the length of the bunch can be compensated for. The theory for the generalized dispersion produced by space charge dominated beams in bends has been developed in some detail for a merger with bi-lateral symmetry called the "zigzag" [1.15]. Here, this important topic is only briefly discussed for ERL mergers and the interested reader is directed to the references [1.16].

## 1.3 BEAM DYNAMICS WITHOUT SPACE CHARGE

### 1.3.1. RF Fields and Gun Geometries

The photocathode RF gun consists of a cathode in a half length cavity only, or the cathode ½-cell followed by one or more full length cavities. These cavities operate typically in a $TM_{011}$ transverse magnetic mode. Figure **1.4** shows the cell structure for a 2½-cavity gun. Full cells are added to raise the beam energy out of the gun.

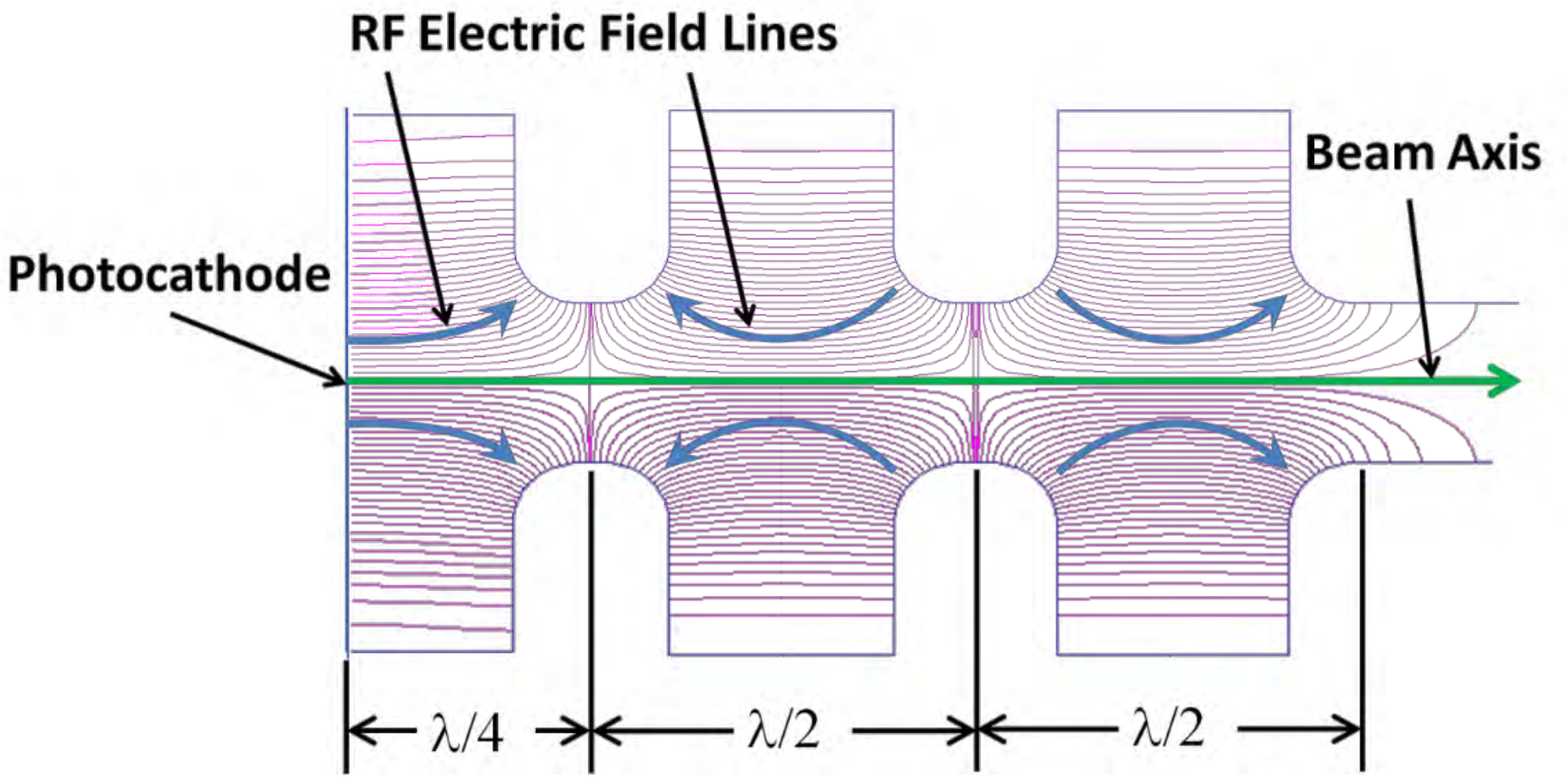

Figure 1.4. Field lines in a 2½-cell photocathode gun.

Figure **1.5** shows two commonly used geometries for the cell shape: Pillbox and re-entrant. The drawing on the left shows a pillbox cell, $\lambda/4$ long, creating the single cell pillbox RF gun. The re-entrant cavity shown on the right has higher shunt impedance than the pillbox and therefore has higher accelerating fields for the same RF power. However as a disadvantage, the reentrance shape has larger off-axis radial fields which can degrade the beam quality, but this effect is usually small. It is also more difficult to fabricate. On the other hand, the pillbox shape allows more freedom on the cathode design. This is advantageous in high voltage,

high frequency guns where the cell wall itself is used as for the cathode. In one common design, the BNL/SLAC/UCLA gun, the entire cavity wall with the cathode is replaceable. This is a useful feature since it allows placing the RF joint at the outer circumference of the cavity where the magnetic field or surface current is high, but the electric field is low. This reduces the likelihood of high voltage arcing, although it does require good electrical contact to avoid resistive heating. In the reentrant cavity, the cathode is usually a plug inserted into the nose cone as shown in the figure. This adds to the complication of the reentrant design. A more extensive discussion comparing these two cavity shapes is given in Chapter 2.

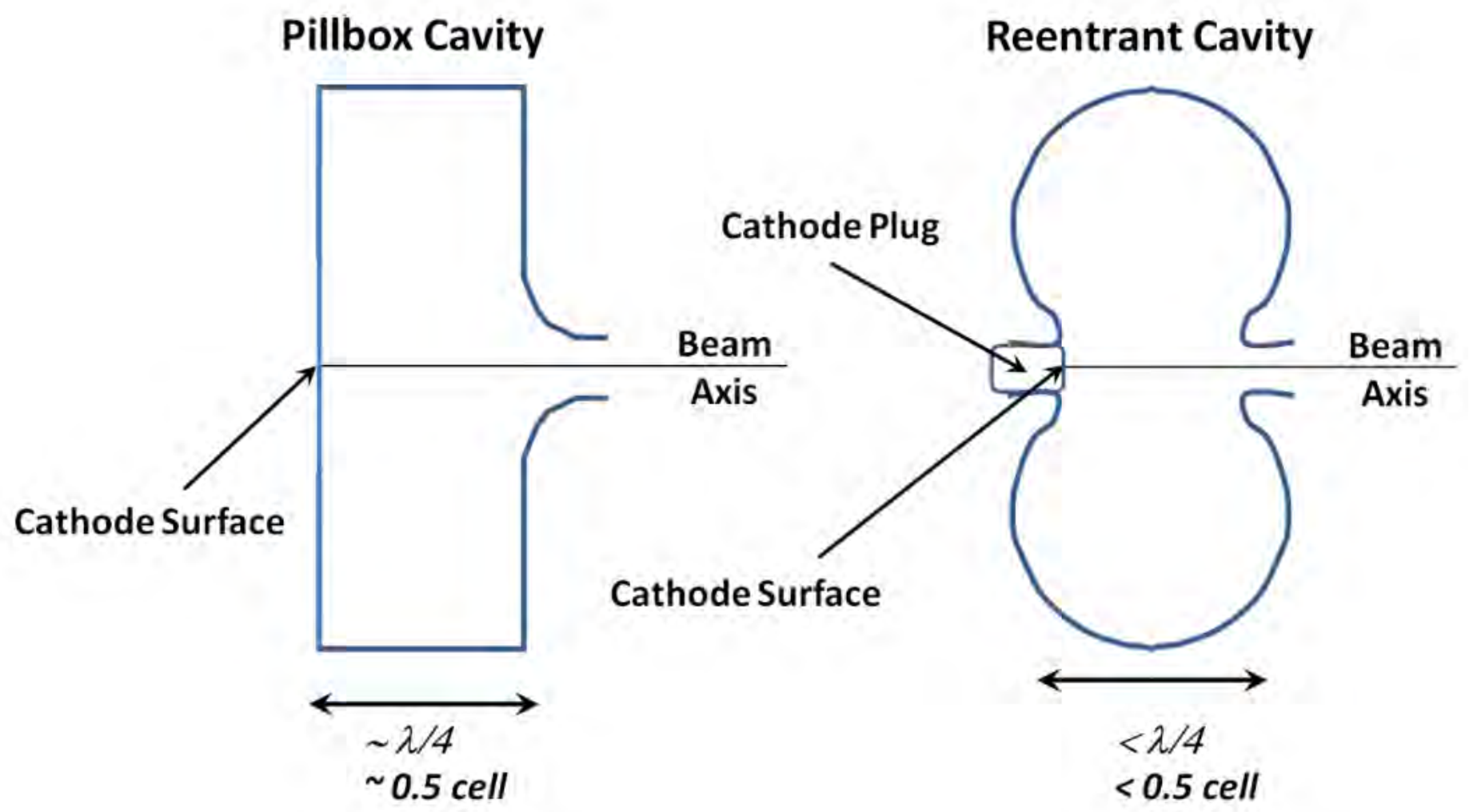

**Figure 1.5. The two common geometries for the RF gun half cavity: Pillbox (left) and re-entrant (right).**

Since we wish to accelerate electrons, the relevant modes are those with large longitudinal electric fields as shown in Figure **1.4**. These are the transverse magnetic (TM) modes. Writing out the electric and magnetic field components of the transverse magnetic modes, $TM_{mnp}$, of a pillbox cavity gives

$$E_z = E_0 J_m(k_{mn} r) \cos(m\theta) \cos(pk_z z) \exp[i(\omega t + \phi_0)] \tag{1.5}$$

$$E_r = -p \frac{k_z}{k_{mn}} E_0 J'_m(k_{mn} r) \cos(m\theta) \sin(pk_z z) \exp[i(\omega t + \phi_0)] \tag{1.6}$$

$$E_\theta = -mp \frac{k_z}{k_{mn}^2 r} E_0 J_m(k_{mn} r) \cos(m\theta) \sin(pk_z - z) \exp[i(\omega t + \phi_0)] \tag{1.7}$$

$$B_z = 0 \tag{1.8}$$

$$B_r = \frac{-i\omega m}{k_{mn}^2 c^2 r} E_0 J_m(k_{mn} r) \sin(m\theta) \cos(pk_z z) \exp[i(\omega t + \phi_0)] \tag{1.9}$$

$$B_\theta = \frac{-i\omega}{k_{mn} c^2} E_0 J'_m(k_{mn} r) \cos(m\theta) \cos(pk_z z) \exp[i(\omega t + \phi_0)] \tag{1.10}$$

These expressions assume the longitudinal origin at $z = 0$ is the cathode position. Here, $E_0$ is the field normalization, $J_m$ is the m$^{th}$-order Bessel function, $k_{mn}$ is the n$^{th}$ zero of the m$^{th}$-order Bessel function, $R_{cavity}$ is the internal radius of the cavity and $\omega$ is the RF angular frequency. $k_z$ is the longitudinal wave number, where $k_z = p\pi\, l^{-1}$, and $l$ is the cavity length. The dispersion relation relates the frequency to the radial and longitudinal wave numbers

$$\frac{\omega^2}{c^2} = k_{mn}^2 + k_z^2 \tag{1.11}$$

The $TM_{mnp}$ designation denotes the mode is transverse magnetic since $B_z = 0$. The $m$ mode number refers to the azimuth angle, $\theta$-dependence or rotational symmetry of the fields. Notice that the $m$ mode number also affects the radial dependence of the fields through the Bessel functions $J_m$ and their derivatives. Since we desire to produce a beam with rotational symmetry, $m = 0$ for all RF guns. The $n$ mode number has been given above, and with the cavity radius gives the position of the radial nodes. The mode shape along the $z$-axis of the cavity is given by $p$. For reasons of timing and efficient acceleration, the full cell length for most RF guns is $\lambda/2$ and $p = 1$. The above mode equations then give a π phase shift between cells. Since the cathode is at a high field position, its cavity length is half that of a full cell, or $\lambda/4$. Numerical studies show that more optimal performance is obtained if the cathode cavity is 0.6X the full cell length, rather than 0.5. And finally these spatial functions of the mode oscillate with the RF frequency, $\omega$, with a $\phi_0$ phase or time shift between the beam and the fields.

Thus most RF guns use the $TM_{011}$ mode whose non-zero field components are

$$E_z = E_0 J_0(k_{01} r) \cos(k_z z) \exp[i(\omega t + \phi_0)] \tag{1.12}$$

$$E_r = \frac{-k_z}{k_{01}} E_0 J_0'(k_{01} r) \sin(k_z z) \exp[i(\omega t + \phi_0)] \tag{1.13}$$

$$B_\theta = \frac{-ik_z}{k_{01} c} E_0 J_0'(k_{01} r) \cos(k_z z) \exp[i(\omega t + \phi_0)] \tag{1.14}$$

This mode is also called the π-mode, because the argument of the cosine function changes by π over a cell length. The fields for the π-mode oscillate at the same frequency, but with opposite sign. This cavity mode is often used for accelerators since the opposite field allows the electron bunch to catch the accelerating polarity of the field and remain synchronous with the accelerating field. Another mode is called the 0-mode has no phase change between the cells. For this mode, the fields oscillate in unison. Equ. 1.5 to Equ. 1.10 show there is no $z$-dependence with $p = 0$ for the $TM_{010}$ mode.

The π-mode electric field for a 1.6-cell S-band RF gun is given in Figure **1.6**. The top of the figure shows the field lines computed by the SUPERFISH code [1.17] which includes the details of the cavity shape shown. The lower portion of the figure compares the longitudinal and transverse fields computed with SUPERFISH and with those given by Equ. 1.12 and Equ. 1.13. The differences in the field shapes are due to the presence of the beam ports and details of the cavity shape. All these effects are ignored in the analytic expressions. However, the pillbox formulae do capture the main features of the fields, and later in the chapter they are used to compute the beam's dynamics.

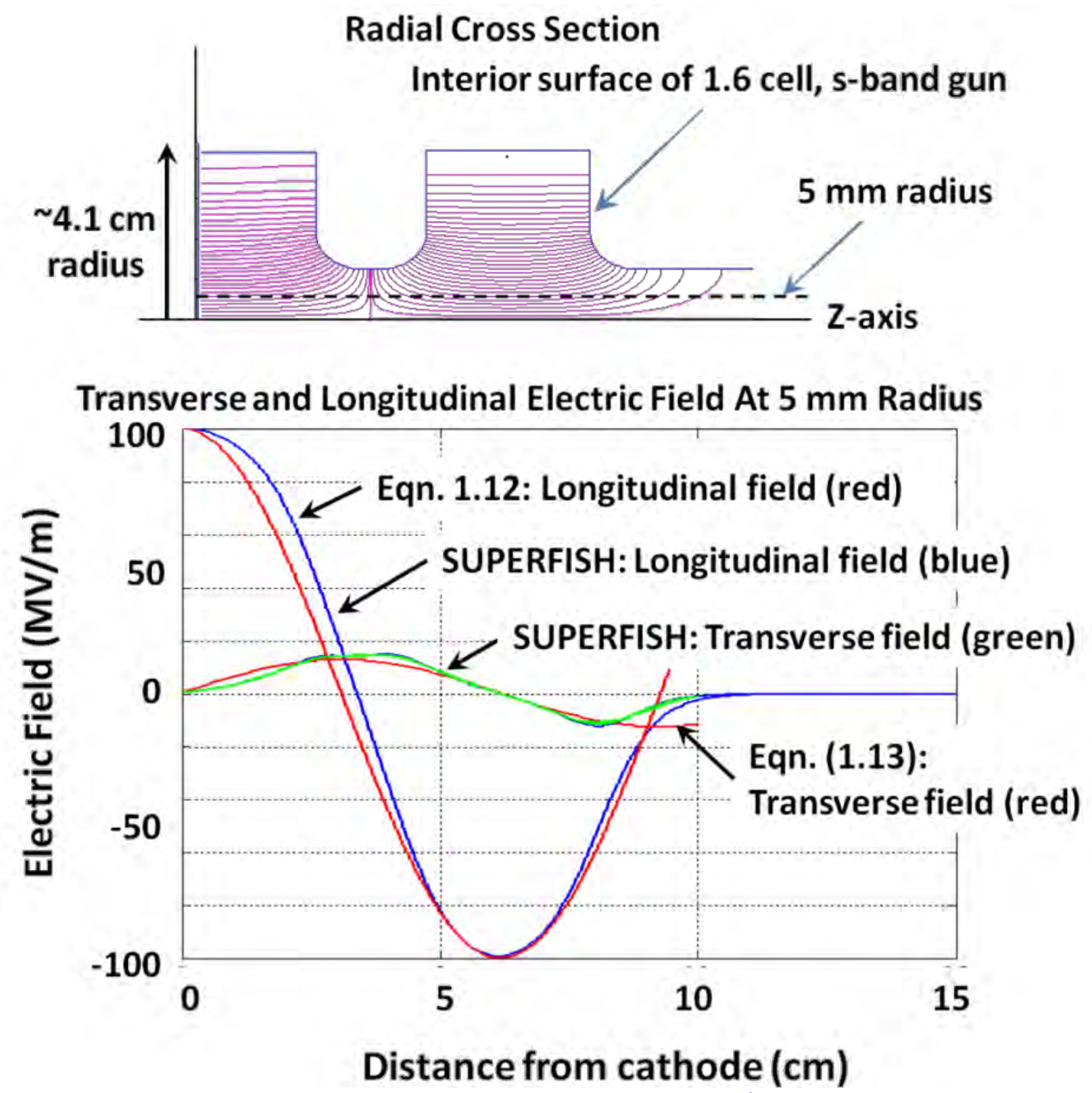

**Figure 1.6. RF fields for an S-band gun operating at a peak field of 100 MV m$^{-1}$ on the cathode. The upper portion of the figure is a cross section taken through the *r*-*Z* plane of a 1.6-cell, S-band gun showing the interior surface and the electric field lines. The field lines were computed by SUPERFISH [1.17] using the shown interior shape. The lower graph is a plot of the longitudinal and transverse electric fields at a radius of 5 mm as computed by SUPERFISH and given by Equ. 1.12 and Equ. 1.13.**

### 1.3.2. Transverse Beam Dynamics in the RF Field

The radial force is

$$F_r = e(E_r - \beta c B_\theta) \tag{1.15}$$

and inserting Equ. 1.9 and Equ. 1.10 for the fields gives

$$F_r = e\,\frac{k_z}{k_{01}} E_0[\beta \cos(k_z z)\sin(\omega t + \phi_0) - \sin(k_z z)\cos(\omega t + \phi_0)]\, J_0'(k_{01} r) \tag{1.16}$$

We consider the RF transverse force at the two locations where it is the largest: The center iris and the gun exit iris. There is little transverse force near the cathode since the low electron velocity makes the first term small and the second term as well since it is proportional to $\sin(k_z z)$ and $z \approx 0$ near the cathode. We argue that the transverse force at the center iris is also negligible since for the π-mode the field $E_z(z)$ is anti-symmetric about the iris. However, the transverse force at the exit iris is significant since $E_z(z)$ doesn't change sign across the exit iris. Thus, the total transverse force is an impulse given at the exit iris, $z = z_f$. In addition, for most RF guns $\beta \approx 1$ at the gun exit, and assuming a small beam size allows us to write the force in the region of the gun exit as

$$F_r = -eE_0 \frac{k_z r}{2} \sin(\omega t + \phi_0 - k_z z) \tag{1.17}$$

Next, we compute the change in radial momentum, $p_r$, using the equation of motion $\frac{\mathrm{d}p_r}{\mathrm{d}t} = \frac{F_r}{mc}$, where we're using Kim's definition of the dimensionless radial momentum $p_r = \frac{\gamma}{c}\frac{\mathrm{d}r}{\mathrm{d}t}$ [1.18]. The change in radial momentum is computed by integrating the force impulse over the position of the exit iris,

$$\Delta p_r = \frac{1}{mc^2}\int F_r dz = \frac{1}{mc^2}\int F_r(z)\delta\left(k_z\left(z - z_f\right)\right)dz \tag{1.18}$$

which gives

$$\Delta p_r = \frac{-eE_0}{mc^2} r\sin(\phi_e) \tag{1.19}$$

where $\phi_e = \omega t + \phi_0 - k_z z_f$. This is identical to Kim's expression although we began with the fields for the cylindrical pillbox cavity.

Following Kim, we convert from cylindrical to Cartesian coordinates to obtain the change in transverse momentum at the exit iris

$$\Delta p_x \equiv \beta\gamma x' = \frac{-eE_0}{2mc^2} x\sin(\phi_e) \tag{1.20}$$

Define the RF focal length in terms of the angular kick the beam gets at the exit of a cell,

$$x' = \frac{x}{f_{RF}} \tag{1.21}$$

which gives the focal length of the gun's RF lens as

$$f_{RF} = \frac{-2\beta\gamma mc^2}{eE_0 \sin(\phi_e)} \tag{1.22}$$

The RF focal strength can be quite strong for high field guns and for guns operating with high cathode field, but low exit energy. For example, for a gun operating at a peak field of 100 MV $m^{-1}$ and exiting the gun on crest with an energy of 6 MeV results in a defocusing focal length of 12 cm. Thus, the beam out of a RF gun requires a focusing lens which is usually a solenoid lens. Details of the solenoid are discussed later in this chapter; however, since the length of the solenoid is ~20 cm, it's nearly the same as its focusing focal length the solenoid adds an aberration to the beam.

The emittance is increased by the gun's RF defocusing due to different longitudinal sections or slices of the electron bunch arriving at the exit iris at different RF phases. If we consider thin slices longitudinally along the bunch, we can see that they will have received different angular kicks by the RF, and thereby increase the overall emittance of the bunch. The projected emittance is the term used to describe this overall emittance and the term slice emittance refers to the emittance of a short longitudinal sliver of the beam. The electron bunch is typically divided into 10-15 slices whose emittance can be determined experimentally using a transverse RF cavity or other techniques, such as chirping the beam energy and using a spectrometer to disperse the slices.

The increase in the transverse phase space area is shown in Figure **1.7**. The $xx'$ phase space is plotted for an exit phase of 0˚ are plotted as lines for the head (blue dash), tail (green dash) and the center (red solid) slices. The center slice (red solid) lies along the $x' = 0˚$ axis. Similar lines plotted along the diagonal line all lay on top of each other. The light shaded area illustrates the projected emittance for 0˚ which is much larger than the emittance for the 90˚ exit phase. This plot assumes the head-to-tail distance is 10˚.

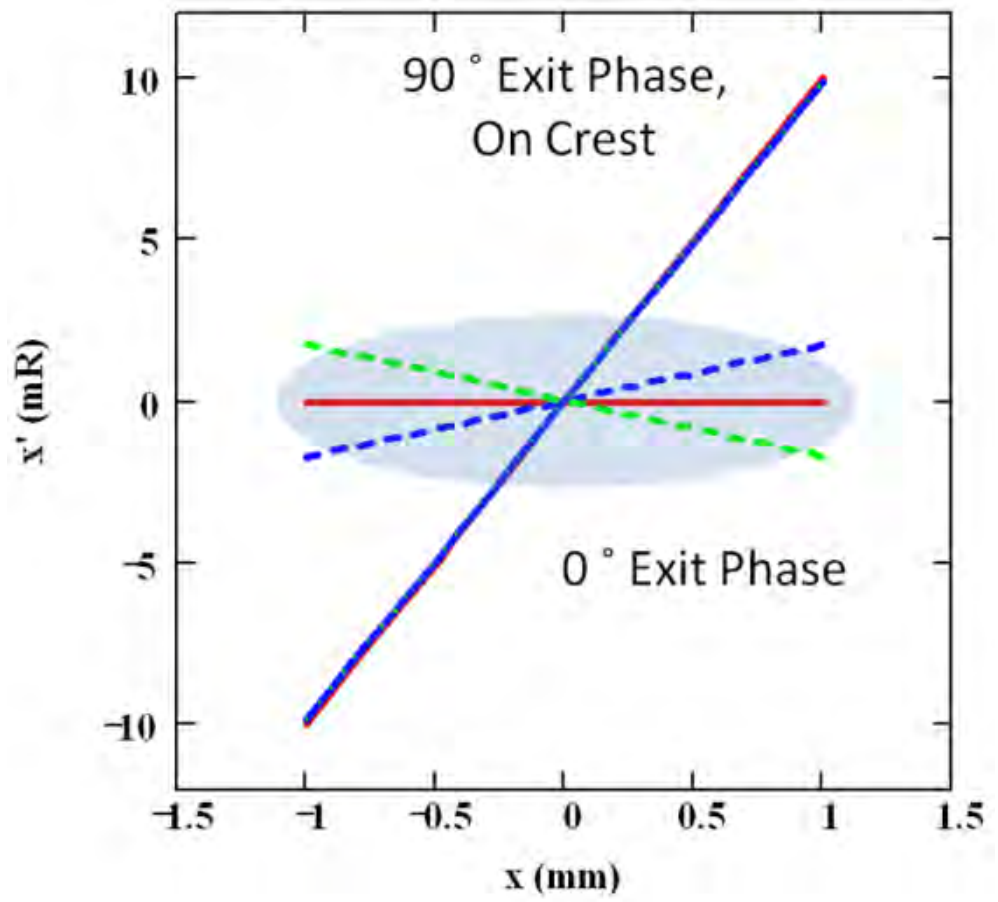

**Figure 1.7. The transverse phase spaces of head, center and tail slices for exit phases of 0˚ and 90˚. The three slice phase spaces for 0˚ exit phase are plotted for the center slice (red-solid), head slice (blue dash) and the tail slice (green dash). The same color scheme is used for the 90˚ phase spaces which all lie on the same diagonal line. The linear (first-order) emittance for an exit phase of 90˚ is zero, as shown by the diagonal line.**

The emittance can be computed beginning with the definition, $\varepsilon_n = \beta\gamma\sqrt{\langle x^2\rangle\langle x'^2\rangle - \langle xx'\rangle^2}$, which is normalized to the beam energy. For exit phases far from 90˚, the correlation term can be ignored

$$\varepsilon_n = \beta\gamma\sigma_x\sigma_{x'} \tag{1.23}$$

where $\sigma_x$ is the rms beam size at the exit of the gun and $\sigma_{x'}$ is the rms divergence. We can estimate the angular dispersion from the variation of the divergence with the exit phase

$$\Delta x' = -\frac{\mathrm{d}}{\mathrm{d}\phi_e}\frac{1}{f_{RF}}\Delta x\Delta\phi_e \tag{1.24}$$

Taking the derivative of the focal strength and converting to rms beam size and bunch length at the exit iris gives the rms divergence

$$\sigma_{x'} = \frac{eE_0\cos(\phi_e)}{2\gamma mc^2}\sigma_x\sigma_\phi \tag{1.25}$$

and the first-order RF emittance is

$$\varepsilon_{RF}^{(1)} = \frac{eE_0\cos(\phi_e)}{2mc^2}\sigma_x^2\sigma_\phi \tag{1.26}$$

Near 90˚, the phase space becomes highly correlated, as shown in Figure **1.7**, and the first-order emittance goes exactly to zero at 90˚. However, there remains a quadratic emittance due to the RF curvature which is a maximum on crest. For a Gaussian longitudinal bunch then the emittance due to the RF curvature is [1.18]

$$\varepsilon_{RF}^{(2)} = \frac{eE_0\,|\sin(\phi_e)|}{2\sqrt{2}mc^2}\sigma_x^2\sigma_\phi^2 \tag{1.27}$$

We can combine these expressions and write the total emittance for all values of $\phi_e$ as the quadratic sum of the first-order and second-order emittances

$$\varepsilon_{RF}^{total} = \frac{eE_0}{2mc^2}\sigma_x^2\sigma_\phi\sqrt{\cos^2(\phi_e) + \frac{\sigma_\phi^2}{2}\sin^2(\phi_e)} \tag{1.28}$$

Figure **1.8** graphs each of these emittances. The first-order RF emittance as a function of the exit phase becomes 0 on crest, or 90˚, where the second-order emittance is instead a maximum. The emittances shown have been calculated using a 4˚ rms (~10˚ FWHM) bunch length, a peak field of 100 MV m$^{-1}$ and rms beam size of 1 mm. The second-order emittance can be cancelled using a higher RF harmonic [1.19]. This is similar to the linearization of the longitudinal phase space is done for bunch compressors using a harmonic cavity [1.20].

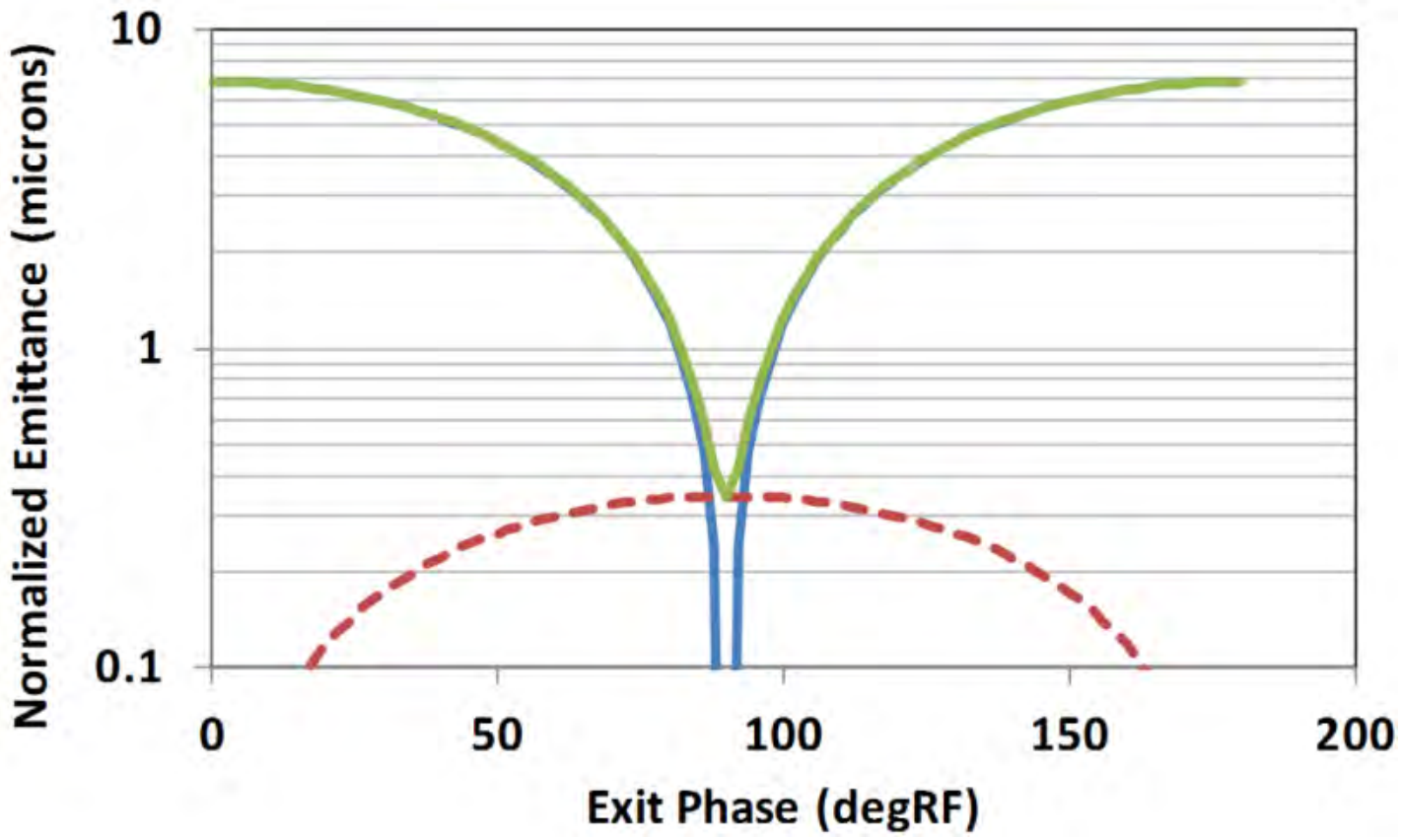

**Figure 1.8. The RF emittance as a function of the exit phase for a 100 MV m$^{-1}$ gun with a 1 mm rms size beam and a Gaussian longitudinal distribution of 4˚ rms at the exit iris. The total emittance (green solid) is the quadratic sum of the first-order (blue solid) and the second-order (red dash) emittances.**

### 1.3.3. Longitudinal Beam Dynamics in the RF Field

Beginning with the force equations written in cylindrical coordinates, one can write the longitudinal force on an electron in a rotationally symmetric π-mode as

$$\frac{d}{dt}(\gamma m\dot{z}) = -eE_0\left[J_0(k_{01}r)\cos(k_z z)\sin(\omega t + \phi_0) - \frac{\omega\dot{r}}{k_{01}c^2}J_0'(k_{01}r)\cos(k_z z)\sin(\omega t + \phi_0)\right] \tag{1.29}$$

Expanding the Bessel functions and keeping the linear terms gives

$$\frac{d}{dt}(\gamma m\dot{z}) = -eE_0\left[\cos(k_z z)\sin(\omega t + \phi_0) + \frac{k_z r\dot{r}}{2c}\cos(k_z z)\cos(\omega t + \phi_0)\right] \tag{1.30}$$

In most cases, $r/c$ is small and the second term can be ignored and we arrive at

$$\frac{\mathrm{d}}{\mathrm{d}t}(\gamma m\dot{z}) = -eE_0 \cos(k_z z) \sin(\omega t + \phi_0) \tag{1.31}$$

Kim emphasizes the importance of the backward-propagating wave in short linear accelerators such as RF guns. Since the field mode we're using is a standing wave mode, it naturally includes both forward and backward waves. It is easy to show that our equation for the longitudinal force agrees with Kim [1.18]. This is done by using a trigonometric identity to write Equ. 1.31 as

$$\frac{\mathrm{d}\gamma}{\mathrm{d}z} = -\alpha k_z\, [\sin(\phi)+\sin(\phi + 2k_z z)];\ \text{where}\ \alpha \equiv \frac{eE_0}{2mc^2 k_z} \tag{1.32}$$

Here, $\phi = \omega t - k_z z + \phi_0$ is the phase of the electrons with respect to the synchronous RF phase and we've defined the electric field parameter as $\alpha$. This expression is identical to Equation 4 in Kim's paper [1.18].

The phase slip is computed from the following integral

$$\phi - \phi_0 = \omega t - k_z z = k_z \int_0^z \left(\frac{1}{\beta} - 1\right) dz = k_z \int_0^z \left(\frac{\gamma}{\sqrt{\gamma^2 - 1}} - 1\right) dz_0 \tag{1.33}$$

Since the beam is rapidly accelerated, the bunch quickly becomes relativistic and becomes synchronous with the RF fields and the phase slip becomes constant. Thus, the integrand is significant only near the cathode and the phase slip, $\phi$ rapidly approaches the asymptotic phase, $\phi_\infty$, given by [1.18]

$$\phi_\infty = \frac{1}{2\alpha \sin(\phi_0)} + \phi_0 \tag{1.34}$$

Bunch compression can be found by taking the derivative of the asymptotic phase with respect to the initial phase

$$\Delta\phi_\infty = \left[1 - \frac{\cos(\phi_0)}{2\alpha \sin^2(\phi_0)}\right] \Delta\phi_0 \tag{1.35}$$

Equ. 1.34 and Equ. 1.35 should be used with some caution. When Equ. 1.35 is negative, the bunch head and tail are reversed however simulations show this does not happen. Thus these expressions appear to be valid only for initial phases giving a positive result for the bunch compression. However, experiments and simulations do show that there is significant bunch compression for initial phases near the phase corresponding to zero field at the cathode. For example, in an S-band gun operating at 115 MV $m^{-1}$, a bunch with an initial phase of 30˚ S with respect to the zero field phase has the same length as the laser pulse. Initial phases less than 30˚ S can compress the bunch a factor of five or more at high cathode fields with an asymmetric bunch shape [1.21].

The electric field parameter and the initial phase establish the RF gun's operating range with respect to the RF emittance. Curves of constant electric field parameter, $\alpha$, can be plotted in the plane formed by the asymptotic and initial phases as shown in Figure **1.9**. The figure shows the curves for electric field parameters, $\alpha$ = 0.5, 1.0.75, 1.0, 2.0 and 4.0. The asymptotic phase of 90˚ RF, where the RF emittance is a minimum, is shown by the horizontal dashed line. It can be seen that to reach the phase for minimum RF emittance requires $\alpha \approx 0.8$ or larger. In order for $\alpha > 1$ the product of the peak field and the RF wavelength needs to be greater than 6.4 MeV

$$eE_0\lambda_{RF} > 4\pi mc^2 \tag{1.36}$$

The peak field corresponding to $\alpha > 1$ is easily achieved for S-band (3 GHz, $\lambda_{RF}$ = 10 cm) guns where the requirement is 64 MV m$^{-1}$. Typical S-band guns commonly operate at 100 MV m$^{-1}$ and higher to give $\alpha > 1.5$. As the RF frequency increases, the field required for $\alpha = 1$ becomes more difficult. For example, the peak field needed at X-band (12 GHz) is 250 MV m$^{-1}$. This is possible, but on the high end of reliably achieved fields at this frequency. An X-band gun with a field of 200 MV m$^{-1}$ has an electric field parameter of 0.8, which is too low to minimize the RF emittance as shown in Figure **1.9** for a 250 pC bunch charge and a cathode radius of 1 mm.

Of course, this restriction only applies to guns consisting of a string of iris coupled cavities and not to single cell guns. In addition, if the cavities are independently powered, they can be timed to give the desired exit phase at the expense of producing a mismatch of the focusing strength at the irises between the cells.

For initial phase near the RF zero crossing the RF field becomes equal to the bunch electric field and the space charge limit is reach. As will be discussed in the next section, the space charge field is related to the surface charge density, $\sigma_{SCL}$, and can be written as

$$\sigma_{SCL} = \frac{q_{bunch}}{\pi R_c^2} = \varepsilon_0 E_0 \sin(\phi_{0,SCL}) \tag{1.37}$$

$\phi_{0,SCL}$ is the initial phase at which the RF field equals the space charge field produced by a bunch charge with $q_{bunch}$ and radius, $R_c$. Since $\alpha = \frac{eE_0}{2mc^2k_z}$, one can write

$$\alpha \sin(\phi_{0,SCL}) = \frac{eq_{bunch}}{2\pi\varepsilon_0 k_z R_c^2\, mc^2} \tag{1.38}$$

Thus bunches launched with an initial phase of $\phi_{0,SCL}$ and smaller are space charge limited and are not emitted. The space charge limited region is indicated by the shaded region in Figure **1.9** for a 250 pC bunch charge and a cathode radius of 1 mm.

The longitudinal emittance, $\varepsilon_z$, is defined in terms of the dimensionless longitudinal momentum, $p_z = \beta_z\gamma$, and the bunch length $\Delta z$

$$\varepsilon_z = \sqrt{\langle(\Delta p_z)^2\rangle\langle(\Delta z)^2\rangle - \langle\Delta p_z\rangle^2\langle\Delta z\rangle^2} = \sqrt{\langle[\Delta(\beta_z\gamma)]^2\rangle\langle(\Delta z)^2\rangle - \langle\Delta(\beta_z\gamma)\rangle^2\langle\Delta z\rangle^2} \tag{1.39}$$

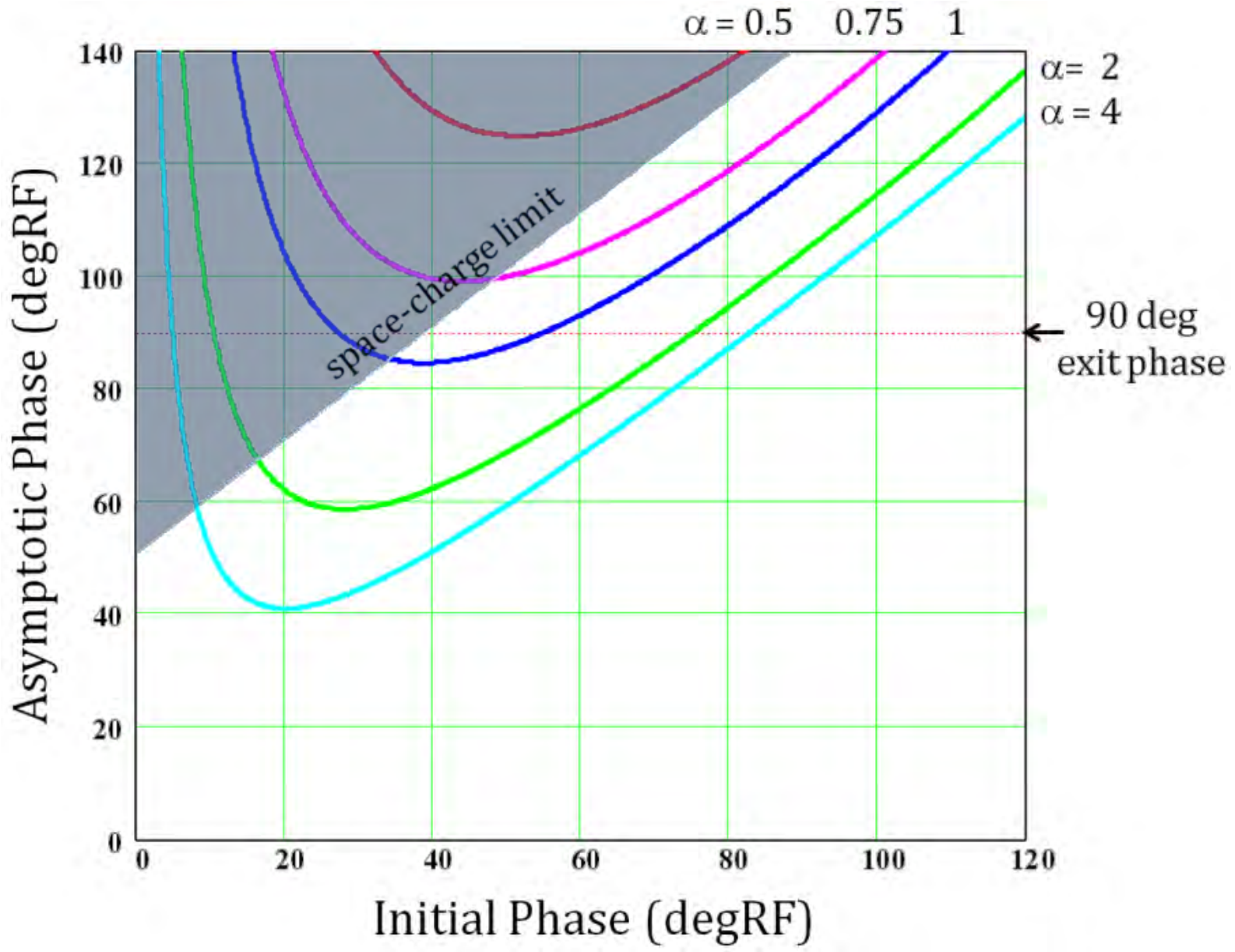

**Figure 1.9. The asymptotic phase for an S-band gun plotted as a function of the initial phase for $\alpha$ = 0.5, 0.75, 1, 2 and 4. The region for which the cathode field is below the space charge limit is indicated in the upper left region of the graph. The space charge limit is shown for a 250 pC bunch with a 1 mm cathode radius.**

The RF longitudinal emittance is derived from the beam energy for a small deviation $\Delta\phi$ from the mean phase, $\langle\phi\rangle$. For a Gaussian distribution and an on crest exit phase the longitudinal RF emittance is [1.18]

$$\varepsilon_z^{RF} = \sqrt{3}(\gamma_f - 1)k_z^2\,\sigma_z^3;\ \text{where } \langle\phi\rangle = 90^\circ\ \text{RF} \tag{1.40}$$

where $\gamma_f$ is the final normalized energy of the beam. The longitudinal RF emittance is seen to have a cubic dependence upon the bunch length at an exit phase of 90˚ where the 1st-order transverse emittance is zero.

## 1.4 BEAM DYNAMICS WITH SPACE CHARGE

The beam charge from a cathode is limited in two operating regimes. For thermionic cathodes, there are two limits: Temperature limited emission and space charge limited emission. In temperature limited emission, the current density is given by the Richardson equation; the space charge limit of the current density is given by the Child-Langmuir law. In the case of photoemission, the bunch charge can be photon limited or space charge limited. The photon limited emission is given by the quantum efficiency (QE) times the number of incident photons, and space charge limited emission is given by a sheet beam model.

### 1.4.1 Space Charge Limited Emission

Electron emission is strongly affected by the electric field produced by the electron bunch itself. Immediately at the cathode surface, the electrons experience their own image charge, which for metal cathodes, produces a field opposing the applied electric field. The magnitude of this field is easily estimated by considering the electron bunch as a very thin charge sheet very close to the cathode surface, as shown in Figure **1.10**. In this case, the space charge field is similar to that between the plates of a capacitor, $E_s$.

In this case, the space charge field, $E_{SCL}$, is similar to that between the plates of a capacitor

$$E_{SCL} = \frac{q}{A\varepsilon_0} = \frac{\sigma}{\varepsilon_0} \tag{1.41}$$

where $A$ is the cross-sectional area of the sheet beam, $q$ is the residual charge, and $\varepsilon_0$ is the permeability of free space. Electron emission saturates when $E_{SCL} = E_{applied}$, whether $E_{applied}$ is an RF or DC electric field

$$E_{SCL} = \frac{q}{A\varepsilon_0} = \frac{\sigma}{\varepsilon_0} = E_{applied} \tag{1.42}$$

Thus, for a RF photocathode gun, the bunch surface charge density is limited when

$$\sigma_{SCL} = \varepsilon_0 E_0 \sin(\phi_0) \tag{1.43}$$

Here, $E_0$ is the peak RF field on the cathode and $\phi_0$ is the laser launch phase.

At the space charge limit (SCL), the emitted charge saturates and the emission becomes constant. If the transverse distribution is non-uniform when the cathode is driven to the SCL, then different locations will saturate and other areas will not. In the RF gun, the signature observation of the SCL is the sub-linear dependence of the charge on the laser energy as shown in Figure **1.11**.

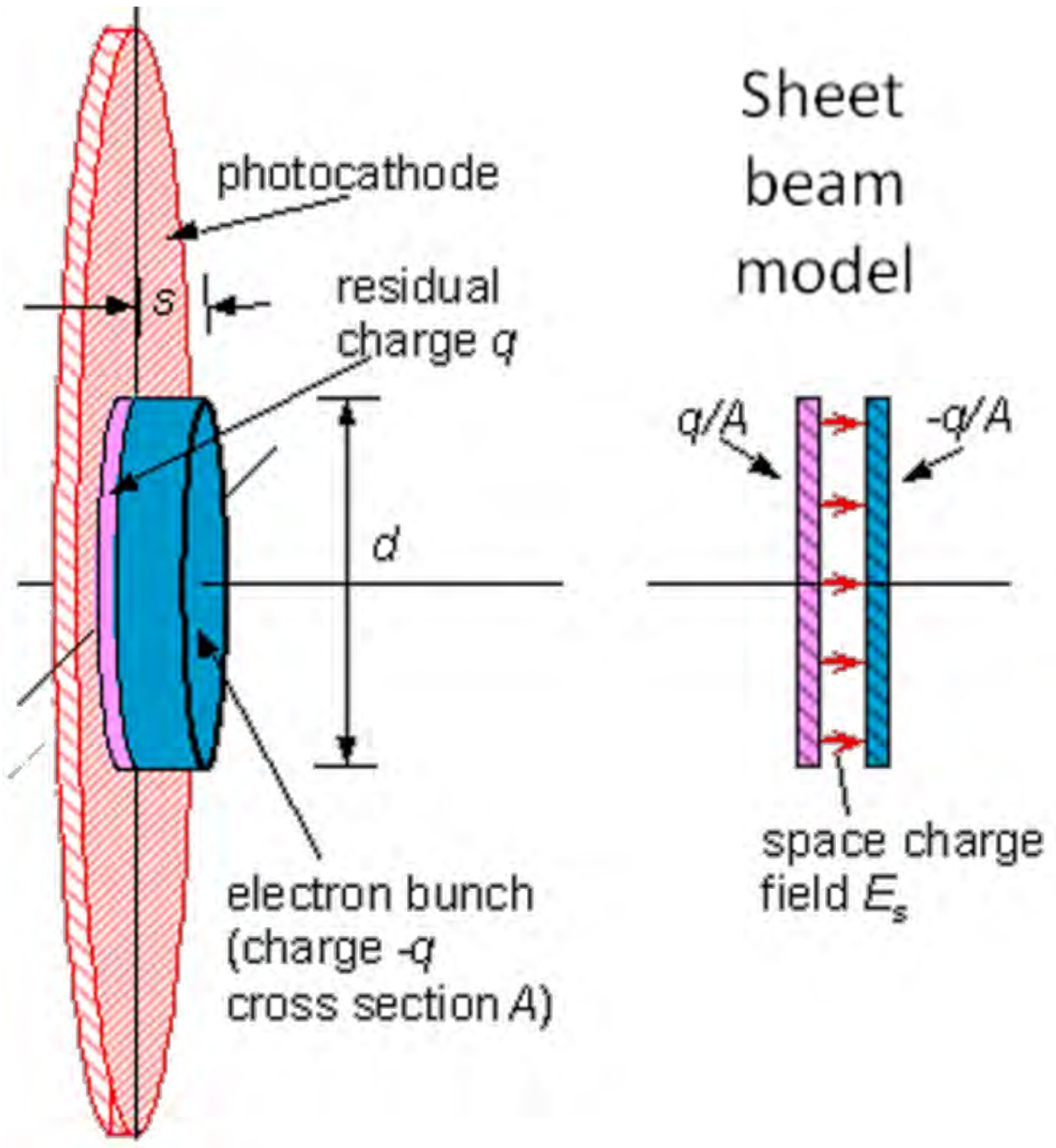

**Figure 1.10. Sheet beam model for short pulse photoemission. [Courtesy of A. Vetter]**

The space charge limit of the surface charge density in an RF gun is linear with the applied field. This differs from the 3/2-power for the voltage given by the Child-Langmuir law for surface current density, $J_{CL}$, thermionic emission [1.22]

$$J_{CL} = 1.67\times10^{-3}\sqrt{\frac{e}{mc^2}}\frac{V^{3/2}}{d^2} \tag{1.44}$$

$J_{CL}$ is the space charge limited current surface density of beam filling a cathode-anode gap as shown in Figure **1.12**. The voltage across the gap is $V$ and the gap length is $d$.

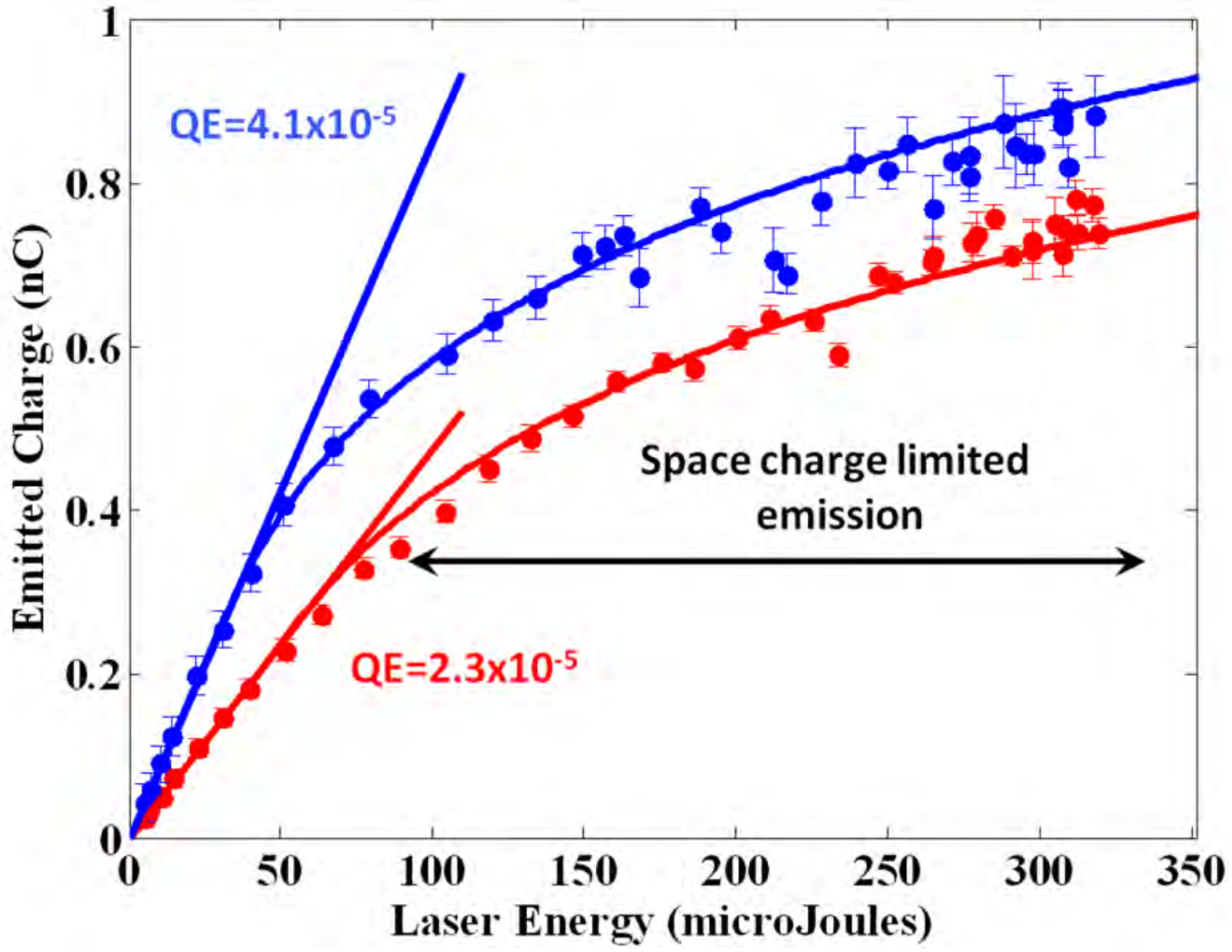

**Figure 1.11. The measured bunch charge vs. laser energy fit with an analysis for the QE and the SCL.**

Equ. 1.43 and Equ. 1.44 show the space charge limits are different for short and long electron bunches. Equ. 1.43 is derived assuming planar cathode and anode electrodes with a potential difference $V$ and a gap separation $d$ with a continuous stream of electrons between the electrodes. Thus, the SCL is linear with the electric field for a sheet-like beam, while the surface current density has a 3/2-power voltage dependence for a continuous beam in a planar diode.

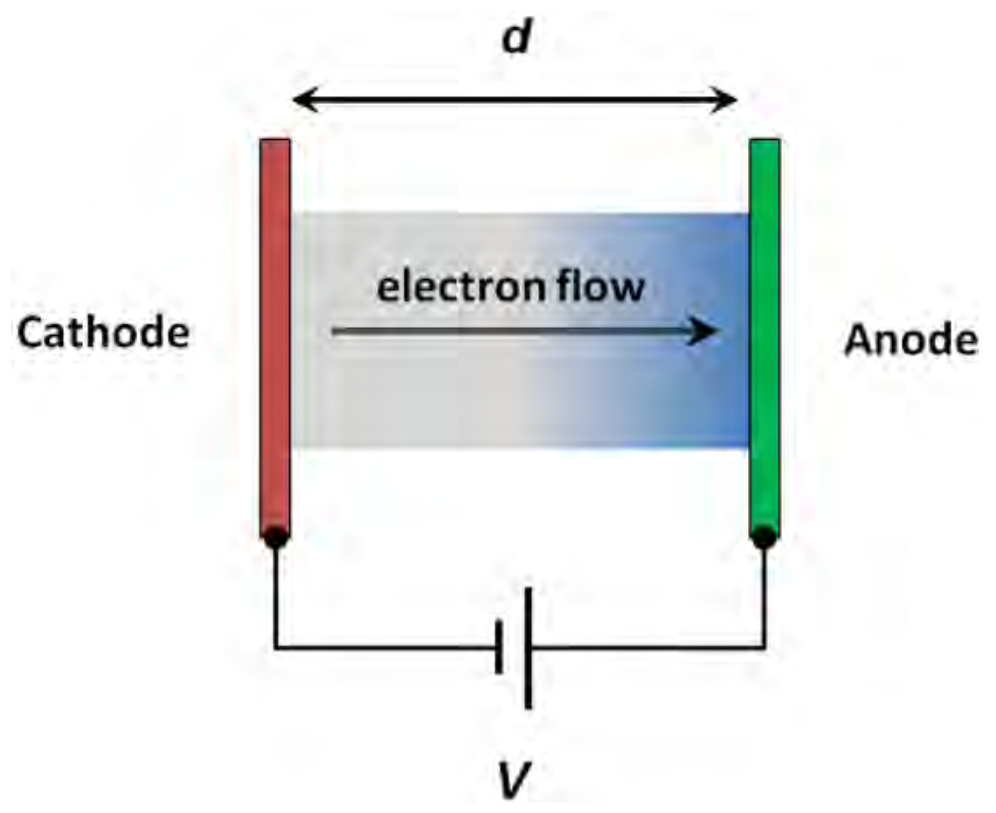

**Figure 1.12. Electron current in the diode region of a thermionic or long pulse gun.**

One computes the emitted charge as a function of the laser pulse energy by applying the simple assumption that emission saturates at the SCL. The emitted bunch charge as a function of the incident laser pulse energy is separated into two regions. For low laser energies below the SCL, the curve is linear with a slope related to the quantum efficiency, QE,

$$q_{bunch} = \frac{eE_{laser}}{\hbar\omega}\,\mathrm{QE} \tag{1.45}$$

where $q_{bunch}$ is the emitted bunch charge, $E_{laser}$ is the laser pulse energy and $\hbar\omega$ is the laser photon energy. Let us assume a Gaussian for the transverse distribution of the laser with $\sigma_r$ for the rms width of the Gaussian. Then when the laser pulse energy is high enough for the peak of the Gaussian to produce a surface charge density equal to the applied field and the SCL is reached. At the SCL, the bunch charge saturates or no longer increases with increasing laser pulse energy.

The situation is shown in Figure **1.13**, where the full Gaussian distribution (red line) is truncated to the SCL in its core (green dark). The total charge then, $q_{emitted}$, consists of the core, $q_{core}$, charge plus the charge emitted in the unsaturated tails, $q_{tail}$ [1.23]

$$q_{emitted} = q_{core} + q_{tail} \tag{1.46}$$

Radial integration of the core and tail regions of the transverse distribution gives the space charge limited bunch charge as

$$q_{emitted} = \pi r_m^2 \varepsilon_0 E_0 \sin(\phi_0) + \mathrm{QE}\,\frac{eE_{laser}}{\hbar\omega}\exp\left(\frac{-r_m^2}{2\sigma_r^2}\right) \tag{1.47}$$

with the radius of the saturated core, $r_m$, given by

$$r_m = \sigma_r\sqrt{2\ln\left(\frac{eE_{laser}\mathrm{QE}}{2\pi\varepsilon_0\sigma_r^2\hbar\omega E_0\sin(\phi_0)}\right)} \tag{1.48}$$

These expressions can be used to fit the charge vs. laser energy data such as shown in Figure **1.13**, to obtain the QE from the linear portion, as well as the rms radius of the equivalent Gaussian distribution. It also can crudely verify the strength of the electric field on the cathode when a direct measurement of the beam energy isn't possible.

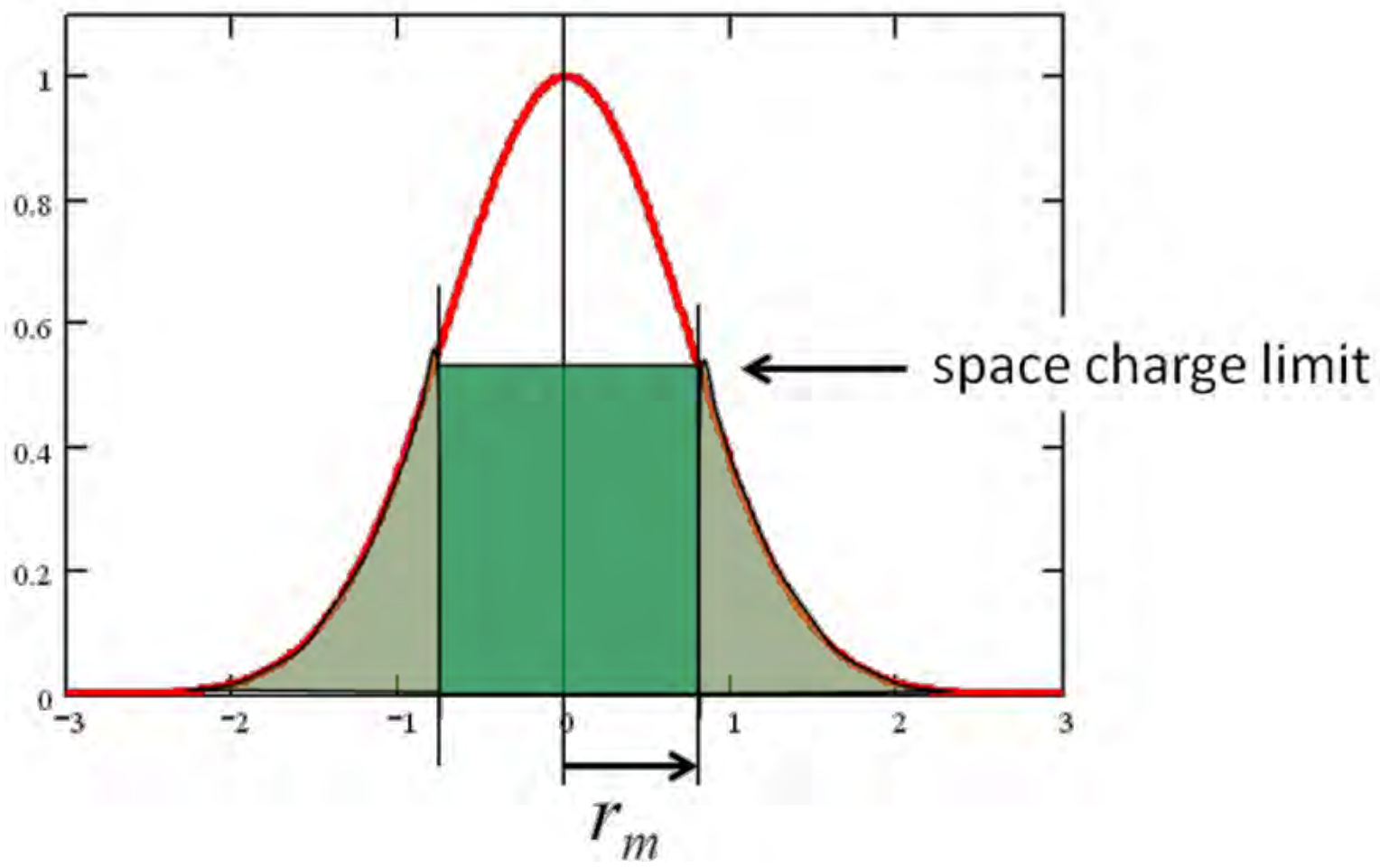

**Figure 1.13. The radial Gaussian distribution (red solid) showing the space charge limited core (green dark) and emission from the tails (green light).**

### 1.4.2 Space Charge Emittance due to the Bunch Shape

Using two models, we discuss the emittance growth due to the transverse space charge forces which can be separated into two spatial scales. The first, on the scale of the beam size, is due to the overall radial expansion of the bunch shape. It is dependent upon the transverse- to longitudinal-aspect ratio and the functional form of the charge distribution, for example, whether the bunch distribution is Gaussian or cylindrical. The second space charge emittance source occurs on a much shorter length scale during the homogenizing or smoothing of the transverse charge density during initial acceleration near the cathode.

In Kim's theory, both the transverse and longitudinal the space charge emittance is given as [1.18]

$$\varepsilon_i^{SC} = \frac{\pi}{4}\frac{1}{\alpha k_z \sin(\phi_0)}\frac{I}{I_0}\mu_i(A);\ \text{where } i = x \text{ or } z \tag{1.49}$$

where $I$ is the peak current at the bunch center and $I_0$ is the characteristic current of 17 kA. (Note: The critical current or Alfven current is energy dependent, $I_A = \beta\gamma I_0$.) The transverse- and longitudinal-space charge factors, $\mu_i(A)$, are the square root of the variance of the normalized transverse and longitudinal fields, respectively, in terms of the bunch aspect ratio, $A$.

For a rotationally symmetric, transverse and longitudinal Gaussian distribution with respective rms sizes of $\sigma_x$ and $\sigma_z$, the aspect ratio is

$$A_{gaussian} = \frac{\sigma_x}{\sigma_z} \tag{1.50}$$

To an excellent approximation, the space charge factors can be parameterized as

$$\mu_x^{gaussian}(A) = \frac{1}{3A + 5} \tag{1.51}$$

and

$$\mu_z^{gaussian}(A) = \frac{1.1}{1 + 4.5A + 2.9A^2} \tag{1.52}$$

In the case of a uniformly charged cylindrical volume of radius $a$ and length $L$, the aspect ratio is

$$A_{cylinder} = \frac{a}{L} \tag{1.53}$$

A cylinder's transverse factor is accurately described by

$$\mu_x^{cylinder}(A) = \frac{1}{35\sqrt{A}} \tag{1.54}$$

Returning to Equ. 1.49, the emittance equation, it can be seen that for a constant peak current, the bunch shape's effect on the space charge emittance is entirely contained in the space charge factor. In Figure **1.14**, we plot the Gaussian and cylindrical transverse space charge factors as a function of the aspect ratio and observe that the space charge factor of a uniformly charged cylinder is four or more times lower than for a Gaussian. The longitudinal space charge factor for a cylinder is approximately an order-of-magnitude smaller than the Gaussian longitudinal factor. This is the theoretical motivation behind needing a uniform laser pulse to produce low emittance beams.

The decline of transverse factors with increasing aspect ratio is shown for both bunch shapes, showing that the "pancake-like," $A > 1$, bunch has much lower space charge emittance than do "cigar-like" shapes, $A < 1$. Thus, the conclusions are the shorter the bunch the better and the shape should be a uniformly charged cylinder.

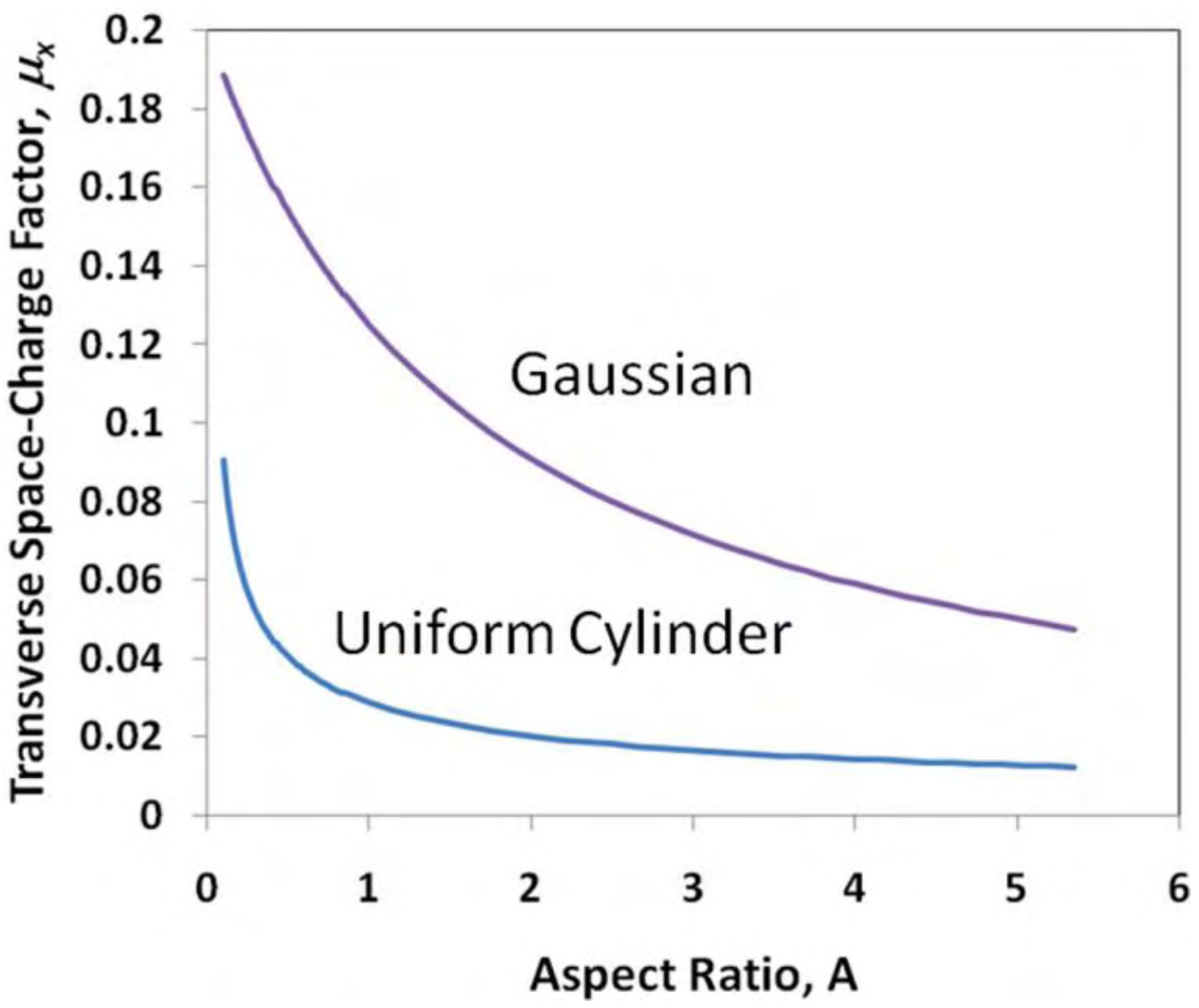

**Figure 1.14. The space charge factors for Gaussian and cylindrical bunch shapes.**

### 1.4.3 Space Charge Emittance due to Non-uniform Transverse Emission

A high spatial frequency intensity modulation across the emission area is a second general type of non-uniformity which is another important source of space charge emittance.

Experimental and theoretical studies have quantified the effect of non-uniform emission upon beam quality for space charge dominated beams [1.24], [1.25]. This work was done to establish the uniformity required to achieve low emittance beams for short wavelength FELs. Recent experiments performed at the SLAC LCLS FEL measured the effect emission uniformity has upon emittance and lasing at X-ray wavelengths. In this work, a space charge model has been developed which agrees well with emittance measurements for various mesh patterns of the drive laser projected onto the cathode [1.26]. Here, the model is formulated in terms of the spatial frequency of the non-uniform emission.

The beamlet space charge model is for a beam with overall radius $R$ composed of a large number of smaller beamlets arranged in a rectangular transverse pattern. Assume each beamlet has an initial radius $r_0$ and center to center spacing of $4r_0$ in a rectangular grid, as shown to the left in Figure **1.15**. Internal transverse space charge forces make each beamlet expand and merge with its neighboring beamlets, as illustrated in Figure **1.15** (right). This radial acceleration gives the beamlets transverse momentum leading to larger emittance for the total beam.

The radial expansion ends when the beamlets merge and form an approximately uniform distribution. At this point, the non-uniformity space charge emittance becomes constant. Simulations and analytic modeling of this geometry show the beamlets overlap within tens of picoseconds, therefore the non-uniformity emittance is generated very close to the cathode before the beam can become relativistic for even the very highest cathode RF fields. It is interesting to note that the electrons are still non-relativistic and the beamlets are merging at the head of each bunch, even while the tail electrons are just leaving the cathode.

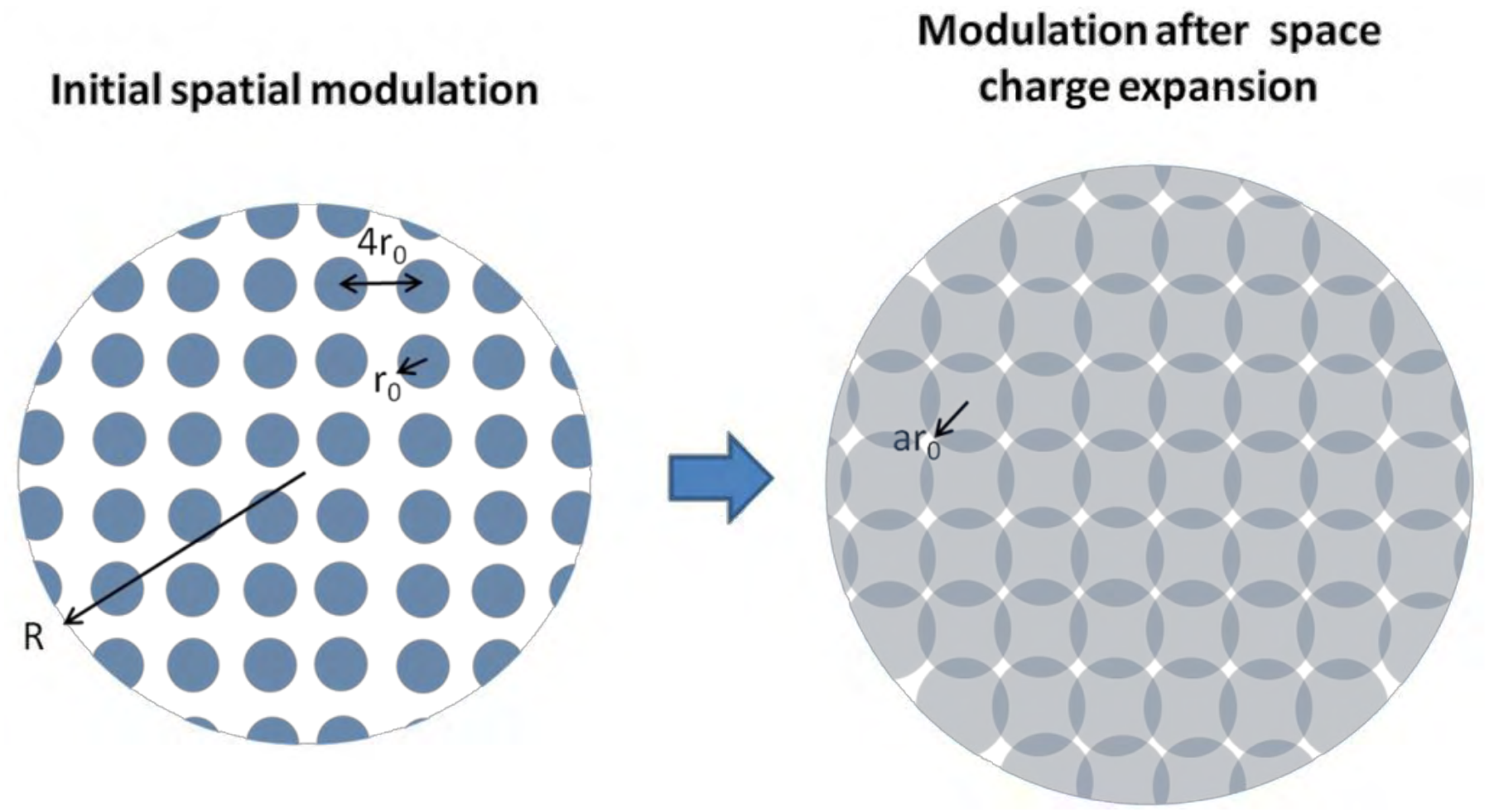

**Figure 1.15. Modulation patterns used to compute the space charge emittance. Left: The initial pattern on the cathode consisting of a rectangular array of circles with radius $r_0$ and a spacing of $4r_0$ within a full beam radius $R$. Right: Schematic view of the beamlet pattern after expansion due to transverse space charge forces. The integration of the transverse force ends when the beamlets with radius $ar_0$ begin to overlap and form a quasi-uniform distribution.**

The space charge emittance for the pattern shown in Figure **1.15** with a 100% depth of modulation in terms of the beam radius, $R$, the radius of a beamlet, $r_0$, and bunch peak current, $I$, is

$$\varepsilon_{n,sc} = \sigma_x \frac{4r_0}{\sqrt{\pi}R}\sqrt{\frac{I}{I_0}} \tag{1.55}$$

The characteristic current is $I_0 = ec\, r_e^{-1} \approx 17$ kA, where $r_e$ is the classical radius of the electron. The $\sqrt{I}$ dependence is similar to that found previously by Wangler [1.27]. In his theory, the space charge emittance grows linearly with beam position $z$ until the beam has travelled a quarter of a plasma period. At this point, the emittance "saturates" to an approximately constant value. In the beamlet model, the emittance saturates when the beamlets overlap and there's no charge gradient driving the transverse acceleration of the electrons.

It is useful to write the emittance in terms of the beamlet spatial number, or the number of beamlets, $n_s$, across the beam diameter, $2R$. Since the beamlet spacing is $4r_0$, $n_s$ is

$$n_s = \frac{R}{2r_0} \tag{1.56}$$

Therefore, the space charge emittance due to this transverse expansion immediately after emission can then be written as

$$\varepsilon_{n,s}(n_s,I) = \sigma_x \frac{2}{\sqrt{\pi n_s}}\sqrt{\frac{I}{I_0}} \tag{1.57}$$

Thus, the emittance decreases as the distance between the beamlets decreases or their spatial number increases. This clearly is because the shorter the expansion distance over which the beamlets can expand and undergo transverse acceleration, the smaller the final transverse velocity, and hence the smaller the emittance. Figure **1.16** shows the normalized divergence as a function of the spatial number for peak currents of 100 A, 40 A and 10 A which for LCLS parameters approximately corresponds to 1 nC, 250 pC and 20 pC bunch charge. This model is in reasonably good agreement with experimental results [1.26].

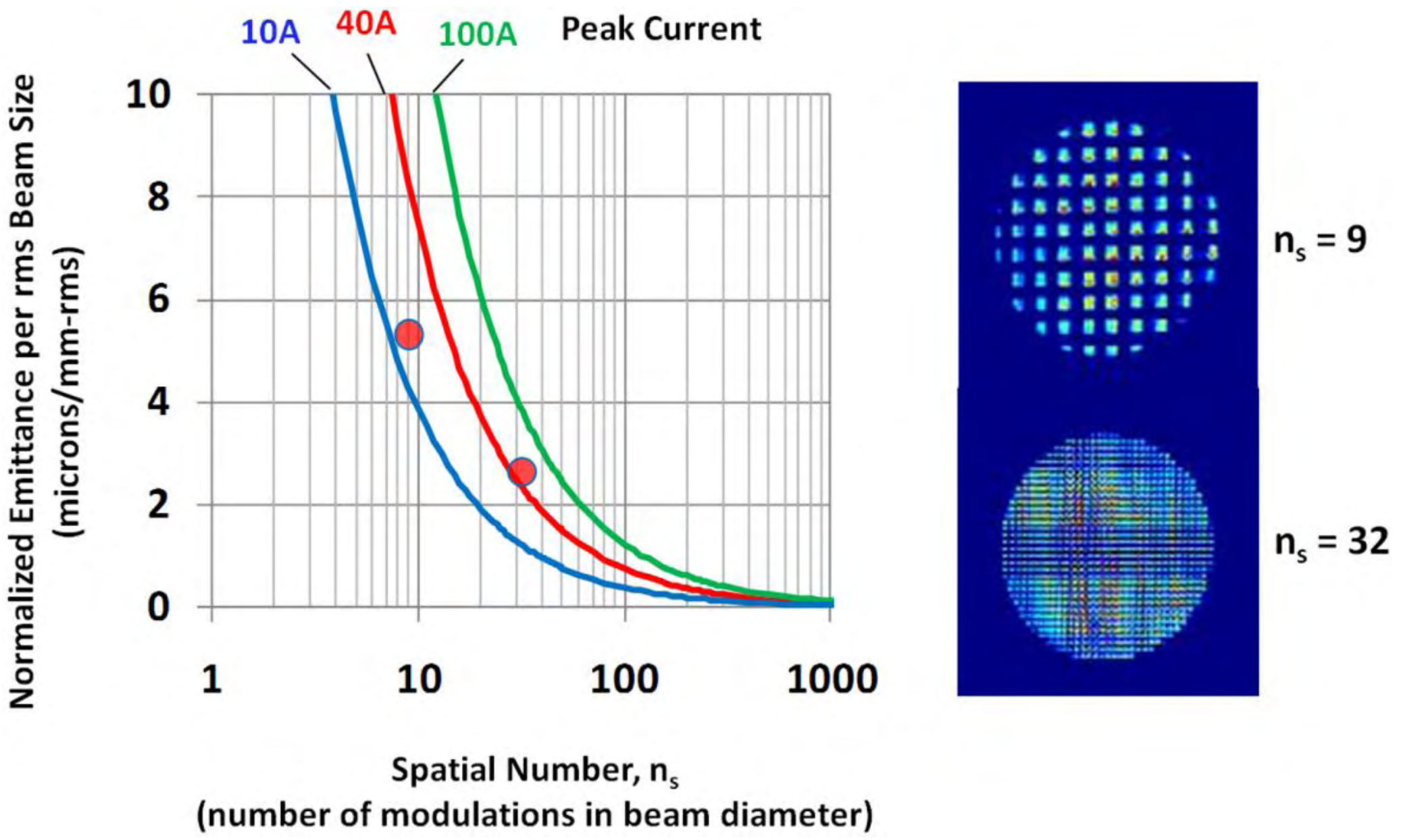

**Figure 1.16. The normalized divergence due to transverse space charge forces of a patterned emission distribution as a function of the number of modulations or beamlets across the beam diameter. Curves are shown for peak currents of 10 A (red), 40 A (blue) and 100 A (green). The red points are measured projected emittances, from which the uniform beam emittance has been subtracted in quadrature and then divided by the rms size of the laser spot to give the experimental normalized divergence. The right images are the laser patterns used to produce the experimental emittance shown by the respective points on left graph, modulation numbers 9 and 32. [[1.26]; Adapted under Creative Common Attribution 3.0 License (www.creativecommons.org/licenses/by/3.0/us/) at www.JACoW.org.]**

The thoughtful reader may notice that there is no energy dependence or cathode field strength in Equ. 1.57; this may seem wrong since it is generally assumed that the space charge force is mitigated by the rapid acceleration of the beam, and so any formulation of the space charge emittance should include the accelerating electric field or beam energy. However, this model assumes the emittance growth occurs before the beam can reach a relativistic energy. Therefore, there is no dependence upon the cathode field and this emittance is only reduced by making the emission uniform.

### 1.4.4 Emittance Compensation Theory

Emittance compensation in an RF gun was first explained by Carlsten [1.28] who used the concepts of slice and projected emittances to show how the projected emittance could be reduced by aligning the short longitudinal slices in transverse phase space. Emittance compensation theory divides the electron bunch into thin temporal slices and assumes they are not mutually interacting. Each slice's emittance is assumed to be the same, small (except in the presence of geometric aberrations), and nearly equal to the thermal emittance. Their relative orientation in transverse phase space differs from slice to slice, *i.e.*, the slices all have different Courant-Snyder parameters and betatron functions. Most analyses, such as the one presented here, assume the slice emittance and phase space parameters vary slowly along the bunch and that a constant peak current can be used for all bunches. An example of a configuration of slices with their phase spaces rotated along the bunch is shown in Figure **1.17**. While the emittance of each slice is quite small, the projected emittance of an ellipse enclosing all the slices is much larger. Aligning the slices in transverse phase space gives projected emittance which approaches the emittance of a single slice. However, it is typically larger depending upon the relative orientation of the slices in phase space.

The concept of emittance compensation is shown graphically in Figure **1.18**. At the cathode all the slices from the bunch head (green) to the bunch tail (red) are born with very low divergence over the size of the emission area (Figure **1.18(a)**). The projected emittance is nearly the same as the intrinsic emittance. As the beam expands from the cathode, the space charge forces give the head and tail different kicks in transverse angle depending upon the peak current of the bunch as shown in more detail below. This increases the projected emittance shown as the area enclosed by the ellipse as shown in Figure **1.18(b)**. Uncorrected, this emittance would persist and grow as the beam propagates down the beam line. However, a solenoid can recover the low emittance. Figure **1.18(c)** shows the angle kick given to the beam by the solenoid lens which gives the bunch head, middle and tail the same sign for the divergence. The beam now drifts a distance with the slices all converging at a beam waist (Figure **1.18(d)**).

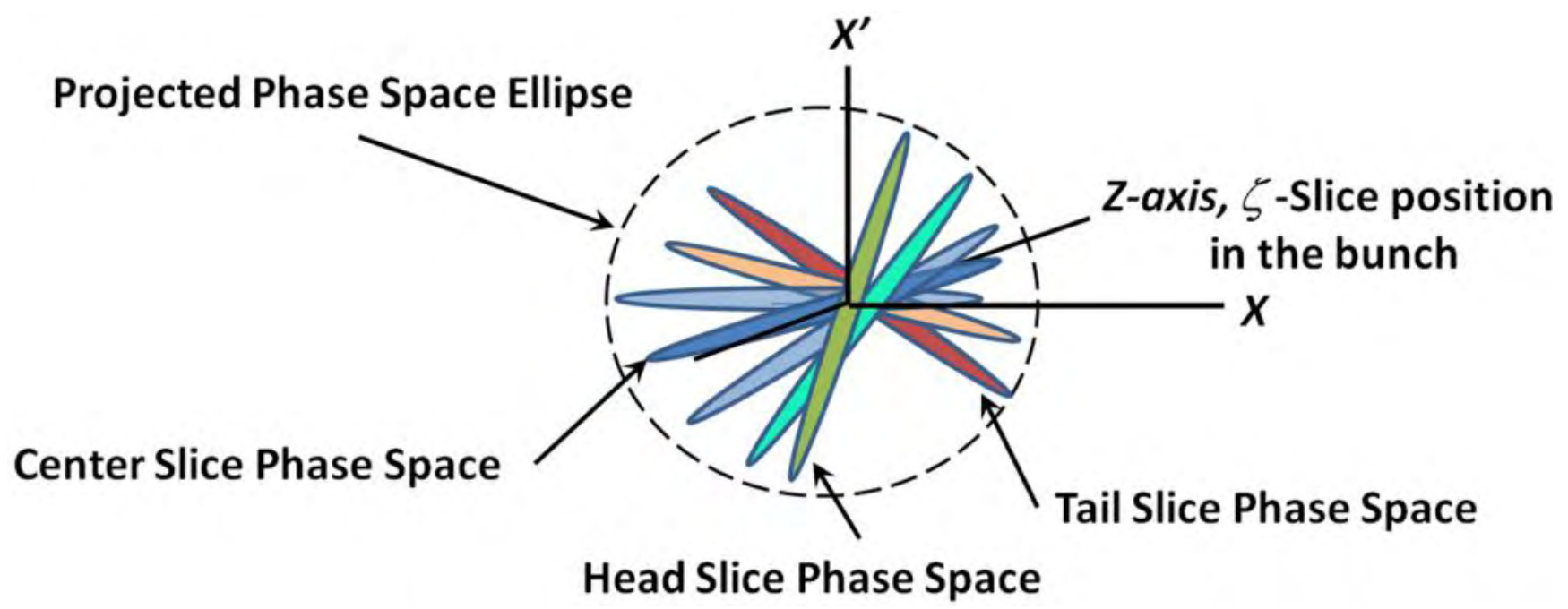

**Figure 1.17. Dividing the bunch longitudinally to form thin slices each with its own phase space distribution. The slice phase spaces are assumed to be independent and do not interact between themselves. The projected emittance is the phase space ellipse which encloses all the slice phase spaces.**

The beam waist is located at the entrance to the high-energy linac which rapidly accelerates the beam to relativistic energy from the space charge dominated to emittance dominated regimes. This beam optics is essentially producing a parallel-to-point image of the beam from the cathode to the linac. As long as the space charge force is linear, its kicks can occur anywhere along the beam line and still be corrected with a linear solenoid lens. The linearity of the space charge forces is a key ingredient in the emittance compensation technique along with proper matching of the bunch slices into the high energy accelerator to freeze the aligned slices.

Emittance compensation was first described by Carlsten [1.28] to explain simulations showing the projected emittance oscillating along the beamline. This theory was later expanded upon by Serafini and Rosenzweig who showed the oscillation period is related to the bunch plasma frequency, and that there is a preferred matching of the beam size and divergence to the beam line optics (called the invariant envelope) which minimizes the emittance [1.29]. Later work by Ferrario *et al.* showed there is an optimum beam size and divergence at the entrance to the first linac which minimizes the space charge emittance [1.30] and measured the emittance oscillations, thereby verifying the theory [1.31].

The Serafini and Rosenzweig formalism is based on balancing the space charge defocusing of the beam with an applied focusing force. The result is a beam whose radius and emittance oscillates with the bunch plasma frequency. In their analysis, the space charge kick shown in Figure **1.18(b)** is applied continuously along a channel of radial focusing fields. The outward radial space charge force is balanced by the focusing fields. If the space charge and focusing fields are linear, then the beam is transported without emittance growth. However, for most beam distributions, the space charge force is not linear leaving some emittance uncompensated for given the linear focusing fields.

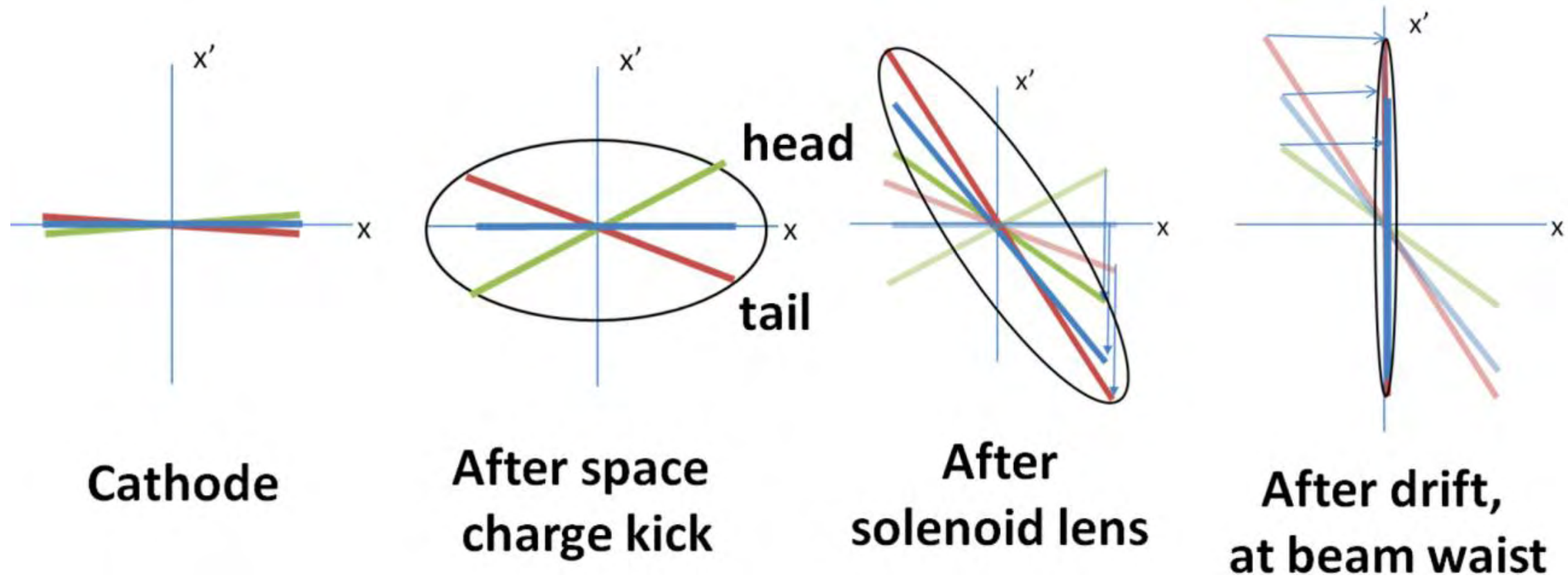

**Figure 1.18. Transverse phase space dynamics during emittance compensation. The transverse phase space is shown for different slices along the bunch. The bunch head slice is shown as a green line, the tail slice is red and the center slice is blue. An ellipse has been drawn around the three slices to indicate the projected phase space of the three slices.**

Compensation of the space charge force is best computed by beginning with the beam envelope equation for a slice with current $I(\zeta)$ in a uniform focusing channel

$$\sigma_r''(\zeta) + K_r\sigma_r(\zeta) = \frac{I(\zeta)}{2I_0(\beta\gamma)^3\sigma_r(\zeta)} + \frac{\varepsilon^2_{n,intrinsic}(\zeta)}{(\beta\gamma)^2\sigma_r^3(\zeta)} \tag{1.58}$$

where $\sigma_r(\zeta)$ is the rms radial beam size for a slice at position $\zeta$, $K_r$ is the channel focusing strength, $\varepsilon_{n,intrinsic}$ is normalized intrinsic emittance of the cathode, and $I_0$ is the characteristic current given by $I_0 = 4\pi\varepsilon_0 mc^3 e^{-1}$, which again is 17 kA for electrons. The intrinsic (a.k.a. thermal) emittance is also allowed to be different for different slices. The slice each is $\delta\zeta$ long at position $\zeta$ along the bunch as shown in the Figure **1.19(a)**. $\beta$ is the electron velocity in units of $c$ and $\gamma$ is the total energy normalized to the electron rest mass, $mc^2$.

The envelope equation given in Equ. 1.58 implies there is no transverse offset of the slice centroids with respect to each other. Including this in the calculation requires using the ray equation to take into account the relative displacement of position and angle between each slice. Although this aspect of the topic is not pursued any further in this chapter, it can be a significant effect in systems with unbalanced RF feeds, *etc.* [1.32].

Each slice can be characterized by its divergence and beam size in transverse phase space and current, as shown in Figure **1.19**. The slice current and slice radius determine the defocusing strength of the space charge field. The slices can have different sizes, divergences as well as different correlations between the size and divergence, as shown in Figure **1.19(b)**.

For a space charge dominated beam, the emittance term is small compared to the current term and the emittance dependence can be ignored. In balanced flow the space charge defocusing is exactly counteracted by the external focusing field with focusing strength $K_r$. In this case the beam drifts with laminar flow in which the electron trajectories do not cross. This balance between the radial space charge defocusing and external focusing is called Brillouin flow. [1.33] A beam in Brillouin flow has a static envelope equation for a given equilibrium beam size of a slice, $\sigma_{eq}$

$$\sigma'' = -K_r\sigma_{eq} + \frac{I}{2I_0(\beta\gamma)^3\sigma_{eq}} = 0 \tag{1.59}$$

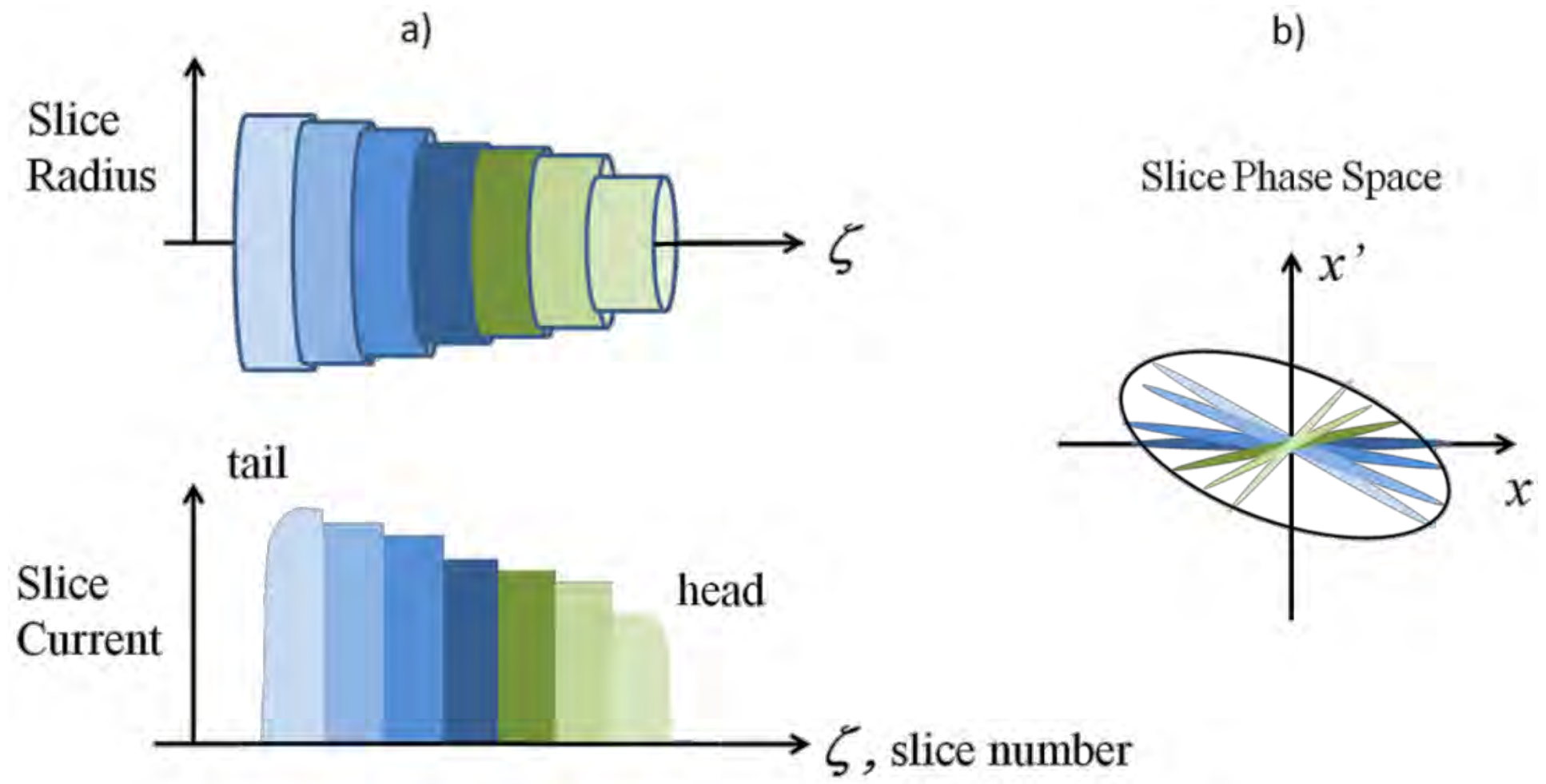

**Figure 1.19. a) The bunch is modeled by dividing it into thin sections or slices along the bunch ζ-axis. Each slice is δζ long. Simulations show that dividing the bunch into 10 or more slices accurately represents the beam dynamics. b) The areas and orientations of the slices in transverse phase space.**

Here, $K_r$ is the external radial focusing strength which balances the space charge force for an equilibrium rms beam size for each slice

$$\sigma_{eq}(\zeta) = \sqrt{\frac{I(\zeta)}{2I_0(\beta\gamma)^3 K_r}} \tag{1.60}$$

Since $K_r$ is the same for all slices, the only thing causing the equilibrium size to vary between slices is the slice current. Variations in the slice current correspond to different equilibrium sizes. This deviation of the actual size from the equilibrium size increases the projected emittance due to their different orientations of the slice in phase space as shown in Figure **1.19(b)**.

To derive the projected emittance consider a small change or perturbation in the slice radius away from the equilibrium radius

$$\sigma(\zeta) = \sigma_{eq}(\zeta) + \delta\sigma(\zeta) \tag{1.61}$$

Inserting this small deviation into the envelope equation for balanced flow and keeping the lowest order term in $\delta\sigma(\zeta)$ gives

$$\delta\sigma''(\zeta) + 2K_r\delta\sigma(\zeta) = 0 \tag{1.62}$$

The solution for this equation is the sum of sine and cosine functions times the initial (*i.e.*, at $z = 0$) slice deviation of size and divergence, respectively. The size oscillation of each slice as a function of distance along the channel is then

$$\delta\sigma(z,\zeta) = \frac{\delta\sigma_0'(\zeta)}{\sqrt{2K_r}}\sin\left(\sqrt{2K_r}\,z\right) + \delta\sigma_0(\zeta)\cos\left(\sqrt{2K_r}\,z\right) \tag{1.63}$$

As a reminder, $K_r$ is the external radial focusing strength and $z$ is the distance the bunch has traveled along the beamline.

To compute the emittance we assume that all the slices are identical except for their current. Specifically, all the slices initially have the same size, $\sigma_0$, and zero divergence variation at $z = 0$. Then, each slice's initial deviation from the equilibrium size is related to the slice current

$$\delta\sigma_0(\zeta) = \sigma_0 - \sigma_{eq}(\zeta) = \sigma_0 - \sqrt{\frac{I(\zeta)}{2I_0(\beta\gamma)^3K_r}} \tag{1.64}$$

With these initial conditions, each slice undergoes a small size oscillation about the equilibrium beam size, $\sigma_{eq}$, as it propagates along the focusing channel

$$\sigma(z,\zeta) = \sigma_{eq}(\zeta) + [\sigma_0 - \sigma_{eq}(\zeta)]\cos\left(\sqrt{2K_r}\,z\right) \tag{1.65}$$

The projected emittance is proportional to the rms of the slice currents normalized to the bunch peak current, $I_p$, and is given as

$$\varepsilon_{n,comp}(z) = \frac{1}{2}\sqrt{K_r}\sigma_0\sigma_{eq}(I_p)\frac{\delta I_{rms}}{I_p}\left|\sin\left(\sqrt{2K_r}\,z\right)\right| \tag{1.66}$$

Where $\delta I_{rms}$ is the rms current computed from the distribution of the slice currents. The projected emittance thus oscillates with the same wave number as does the beam size, but is shifted $\pi/2$ in phase [1.29].

It is useful to make some further simplifying assumptions to the compensation emittance. For instance, since the size deviation is small, the initial slice size is approximately equal to the equilibrium size. Therefore, Equ. 1.66 reduces to

$$\varepsilon_{n,comp}(z) = \frac{\delta I_{rms}}{\sqrt{2K_r}\,I_0\,(\beta\gamma)^3}\left|\sin\left(\sqrt{2K_r}\,z\right)\right| \tag{1.67}$$

It is interesting to consider the emittance as $\left(\sqrt{2K_r}z\right)$ becomes small and goes to zero. This is useful for estimating this projected emittance close to the cathode or in very weak focusing channels. As $\left(\sqrt{2K_r}z\right) \rightarrow 0$, the emittance further simplifies to

$$\varepsilon_{n,comp}(z) \propto \frac{\delta I_{rms}}{\sqrt{2K_r}I_0(\beta\gamma)^3}\,z;\ \text{where}\ \left(\sqrt{2K_r}z\right) \propto 1 \tag{1.68}$$

Therefore, solutions of the balanced envelope equation show an oscillating projected emittance as a function of $z$. The slice radii oscillate for small deviations from the balanced beam envelope. The solutions show all the slice envelopes oscillate about the balanced envelope with the same $z$-coordinate wave number, $\sqrt{2K_{r,eq}}$, and the amplitude results from the initial deviation in the beam size from equilibrium size. If the slices are aligned in transverse phase space at one $z$-position, then there are periodic locations where the slices re-align and the projected emittance is a local minimum. This occurs irrespective of the initial slice amplitude. Optimal compensation of the linear space charge emittance due to different slice currents occurs at these local minima.

Thus, in addition to compensating for the projected emittance due to misaligned slices, it is also necessary to match the beam into a high gradient booster accelerator and damp the envelope oscillations to a low emittance. The required matching condition is referred to as the Ferrario working point [1.30], formulated for the LCLS injector. In this scheme, the RF focusing of the linac is matched to the invariant envelope to damp the emittance to its final value at a relativistic energy. The working point matching condition requires the emittance to be a local maximum and the envelope to be at a waist at the entrance to the booster. The waist size is determined by the strength of the RF fields. RF focusing provides a weak focusing channel and acceleration damps the emittance oscillations.

Matching the beam to the first accelerator needs to be included as part of emittance compensation and should obey the following basic conditions at the entrance to the linac: the beam is at a waist: $\sigma' = 0$ and the waist size at injection is determined by a balancing of the RF transverse force with the space charge force. The second requirement establishes balanced flow similar to that used in emittance compensation.

As an example, assume the RF lens at the entrance to the booster is similar to the RF kick given at the gun exit (see Section 1.2). Assume the beam is injected into the linac on the crest of the RF waveform for maximum acceleration. In this case, the RF lens deflection is

$$\sigma'_{linac} = \sigma_{linac}\frac{eE_{linac}}{2\gamma mc^2} \tag{1.69}$$

where $\sigma_{linac}$ is the rms transverse beam size at the entrance of the linac, $E_{linac}$ is the linac peak accelerating field, and $\gamma mc^2$ is the total beam energy. Taking the derivative gives the RF force

$$\sigma'' = -\sigma_{linac}\frac{eE_{linac}}{2\gamma^2 mc^2}\gamma' = -\sigma_{linac}\frac{\gamma'^2}{2\gamma^2} \tag{1.70}$$

since $\gamma' = \frac{eE_{linac}}{mc^2}$. The envelope equation for the matched beam is then

$$\sigma'' = -\sigma_{match}\frac{\gamma'^2}{\gamma^2} + \frac{I}{2I_A\gamma^3\sigma_{match}} = 0 \tag{1.71}$$

and solving for the matched beam size gives

$$\sigma_{matched} = \frac{1}{\gamma'}\sqrt{\frac{I}{2I_A\gamma}} \tag{1.72}$$

as the waist size at injection into the linac. Once matched the beam emittance decreases along the accelerator due the initial focus at the entrance and Landau damping. This behavior has been verified using HOMDYN, an envelope code using slices, and the particle-pusher code, PARMELA. These codes are described in Section 1.7.

It is relevant to note that this technique of emittance compensation makes no assumption about the nature of the focusing channel. Therefore, this analysis applies equally well to both RF and DC guns. In fact, the same fundamental concept has been applied to the merger optics of a space charge dominated beams into energy recover linacs, as well as other beam transport systems.

## 1.5 GUN-SOLENOID OPTICS AND OPTICAL ABERRATIONS

The strong defocusing at the gun exit requires compensation by an equally strong focusing lens. The focusing is usually provided by a solenoid with a longitudinal magnetic field. It is relevant to comment on the dual role of the solenoid; it not only cancels the strong negative RF lens, but it also plays a crucial function of emittance compensation by aligning the slices transversely along the bunch to minimize the projected emittance.

This section begins with the description of a first-order model of the gun and solenoid system. This model is used to illustrate how the cathode uniformity of emission can be imaged when the solenoid is adjusted to image electrons from the cathode on a view screen. Section 1.5.2 presents a derivation of the first-order chromatic aberration of the solenoid. Section 1.5.3 shows simulation results for the geometric aberrations of the solenoid and the Section 1.5.4 discusses quadrupole field errors of the solenoid. It is shown that the emittance due to quadrupole field errors can be fully recovered with correction quadrupoles.

### 1.5.1 First-order Optics Model of the Gun

The RF gun can be assumed to be a series of thin lenses positioned at the entrance and exit of each cavity for an electron-RF phase which accelerates the beam. If the fields in each cavity are balanced, then the defocus at the exit of one cavity is cancelled by the focus of the next cavity. This is approximately true for all the internal cavities of the gun or any acceleration section. However, there is no cancellation of the defocus at the exit of the last cavity, which as shown in Section 1.3.2 results in linear and non-linear RF projected emittance. For a beam exit phase of 90˚ then the linear RF emittance is zero leaving only the second-order emittance (see Figure **1.8**). However, Figure **1.7** shows that while the emittance is minimal at 90˚, the RF strongly defocuses the beam, requiring an equally strong focusing provided by the gun solenoid. As noted in the discussion after Equ. 1.22, the focal length of the RF gun is 12 cm for an exit energy of 6 MeV and a peak field of 100 MV m$^{-1}$.

$$f_{RF} = \frac{-2\gamma mc^2}{eE_0 \sin(\phi_0)} \tag{1.73}$$

Equ. 1.73 is the same as Equ. 1.22, however, $\beta$ is considered to be unity at these energies. Due to the strong defocusing of the RF gun it is necessary to use a comparably strong focusing lens to collimate and match the beam into the high energy booster linac. If this focusing is done with a solenoid, then its focal strength, $K$, in the rotating frame of the electrons, $f_{sol}$, is [1.34]

$$\frac{1}{f_{sol}} = K \sin(KL_{sol}); \text{ where } K \equiv \frac{B(0)}{2(B\rho)_0} = \frac{eB(0)}{2p} \tag{1.74}$$

where $B(0)$ is the field of the solenoid, $L_{sol}$ is the solenoid effective length, $(B\rho)_0$ is the magnetic rigidity, and $p$ is the beam momentum with units of [GeV c$^{-1}$]. The rigidity can be expressed in the following useful units as

$$(B\rho)_0 = \frac{p}{e} = 33.356p \text{ [kG m]} \tag{1.75}$$

With the assumption that the focusing effects of adjacent cavities cancel, the first-order optics of the gun and solenoid can be modeled with a single thin defocusing lens for the RF and a thick solenoid. Besides using the solenoid to cancel the RF defocusing and for space charge emittance compensation, it is also useful for imaging the electron emission from the cathode with the configuration shown in Figure **1.20**.

The transformation of beam rays from the cathode to the view screen can be computed using linear matrix algebra of ray optics. With simple matrix multiplication, electrons emanating from the cathode with position and angle displacements relative to a central ray can be computed to the view screen.

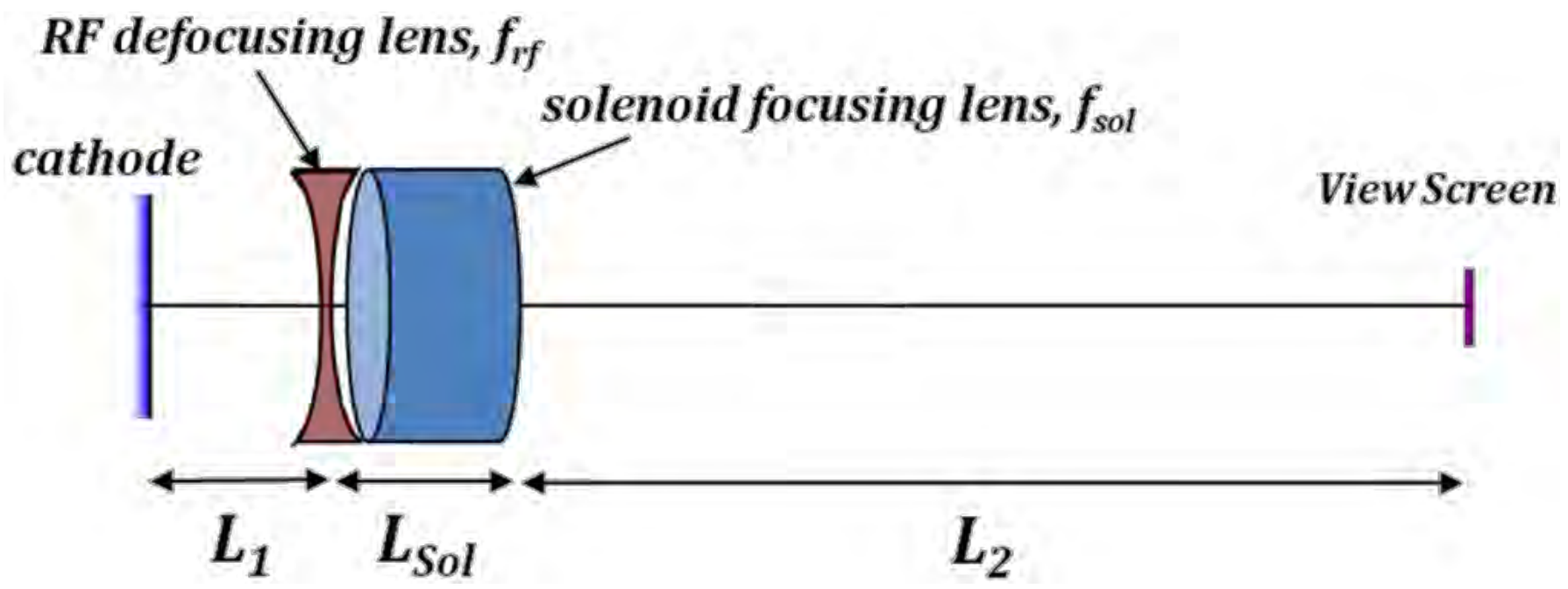

**Figure 1.20. Configuration of a first-order optical model showing principal optical elements of the gun, solenoid and drift distance to a view screen, which can be used to compute the cathode emission onto a view screen. For the LCLS S-band gun-to-linac region: $L_{sol}$ = 19.35 cm, $L_1$ = 12.3 cm and $L_2$ = 106.6 cm, which gives a magnification of approximately -3.6 for a point-to-point image.**

The calculation for the optical system when the solenoid is adjusted to form an image (point-to-point imaging) of the cathode emission on the view screen gives the magnification, $M$, as

$$M = \cos(KL_{sol}) - L_2 K \sin(KL_{sol}) - \frac{1}{f_{RF}}\left(\frac{\sin(KL_{sol})}{K} + L_2 \cos(KL_{sol})\right) \tag{1.76}$$

The magnification depends upon the solenoid and gun field parameters and not upon the distance to the cathode. The magnification is easily measured by inserting a target of known size into the laser beam optics at the object plane, which is then imaged onto the cathode. The size of this target is then measured on the view screen when an image is formed using the solenoid. A magnification of 3-4 is typical in S-band guns.

Electron beam images on a YAG view screen of a 6 MeV beam from an S-band gun with a peak cathode field of 115 MV m$^{-1}$ is shown in Figure **1.21**. The view screen images where taken using the second YAG screen shown in Figure **2.21**. The solenoid has been adjusted to produce an image of the emission pattern at the YAG's position. The electron magnification for the imaging from the cathode to the YAG screen is -3.6 in agreement with Equ. 1.76. The emission is the 2-D product of the QE and laser distributions. In these measurements, the laser distribution is known to have good uniformity and vary slowly over its diameter (low spatial frequencies); therefore, the observed images are good representations of the true QE map.

The two images in Figure **1.21** show very different emission patterns for the same cathode at different times in its two years of operation in the S-band gun. The image in Figure **1.21(a)** shows the QE map consists of small hot spots. These hot spots were observed for low QE, in the range of 10$^{-6}$. The emission image in Figure **1.21(b)** was measured after the same cathode was cleaned with the UV drive laser and low power RF and then continuous operated at high RF power for approximately 1½ years. The area illuminated by the 2 mm diameter laser beam is easily seen in the emission image. Its size gives the magnification of the electron optics between the cathode and the view screen. The dark, irregular regions approximately a few 100 μm in size are likely due to the different work functions of the grains of the poly crystalline composition of the copper cathode. The bright, red-yellow regions appear clustered around the edge of these dark grains and are likely due to enhanced photoemission by adsorbed molecules from the vacuum.

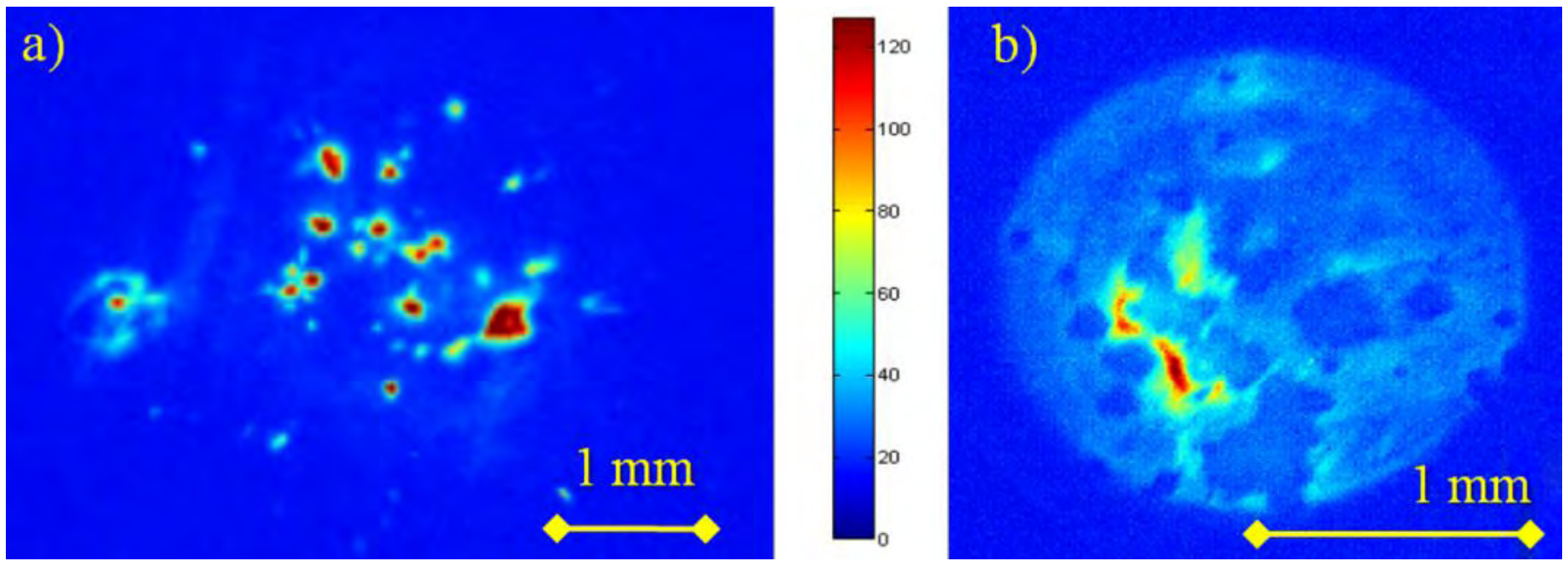

**Figure 1.21. Examples of electron beam images on a view screen with the solenoid adjusted to obtain an image of the electron image from the cathode when illuminated by a large laser spot. The bunch charge is 9 pC. The observed magnification was -3.6, the integrated field strength was 0.5165 kG m, $L_2$ = 1.066 m and the solenoid effective length is 0.1935 m. These parameters give a solenoid wave number of 6.75 per meter with the beam rotating 74.8° in the solenoid. The solenoid focal length is 15.3 cm.**

### 1.5.2 Chromatic Aberration of the Solenoid

The beam's energy spread can introduce additional emittance in the solenoid due to different electrons having different focal lengths. This emittance can be computed by starting with the symmetric transverse beam matrix, $\sigma_{beam}$, and the transformation for a thin lens. The beam matrix is defined as

$$\sigma \equiv \begin{pmatrix} \sigma_{11} & \sigma_{12} \\ \sigma_{12} & \sigma_{22} \end{pmatrix} \tag{1.77}$$

The transformation of the beam matrix though the thin lens is given by

$$\sigma(1) = R_{lens}\sigma(0)R_{lens}^{T} = \begin{pmatrix} 1 & 0 \\ -1/f & 1 \end{pmatrix}\begin{pmatrix} \sigma_{11} & \sigma_{12} \\ \sigma_{12} & \sigma_{22} \end{pmatrix}\begin{pmatrix} 1 & -1/f \\ 0 & 1 \end{pmatrix} \tag{1.78}$$

Performing the matrix multiplications gives

$$\sigma(1) = \begin{pmatrix} \sigma_{11} & \sigma_{12} - \dfrac{\sigma_{11}}{f} \\ \sigma_{12} - \dfrac{\sigma_{11}}{f} & \sigma_{22} + \dfrac{\sigma_{11}}{f^2} - \dfrac{2\sigma_{12}}{f} \end{pmatrix} \tag{1.79}$$

The change due to a variation in the beam momentum, $\Delta p$, can be obtained from the derivative of the beam matrix, $\Delta\sigma_{beam}$**(**1**)**, as

$$\Delta\sigma_{beam}(1) = \frac{d\sigma}{dp}\Delta p = \begin{pmatrix} 0 & -\sigma_{11}\dfrac{d}{dp}\left(\dfrac{1}{f}\right) \\ -\sigma_{11}\dfrac{d}{dp}\left(\dfrac{1}{f}\right) & \ldots \end{pmatrix}\Delta p \tag{1.80}$$

There is no need to compute the $\sigma_{22}$**(1)** matrix element because it gets multiplied by zero when the emittance is computed

$$\varepsilon_{n,chromatic} = \beta\gamma\sqrt{\det(\Delta\sigma_{beam}(\mathbf{1}))} = \beta\gamma\sigma_x^2 \left|\frac{\mathrm{d}}{\mathrm{d}p}\left(\frac{1}{f}\right)\right|\sigma_p \tag{1.81}$$

where $\sigma_{11} = \sigma_x^2$ has been used and $\beta$ is the beam velocity divided by the speed of light, $\gamma$ is the beam's Lorentz factor, $\sigma_{x,sol}$ is the transverse rms beam size at the entrance to the solenoid, and $\sigma_p$ is the rms momentum spread of the beam. This is a general expression for the chromatic emittance of a thin lens. For a solenoid lens in the rotating frame of the beam, the focal strength is given by

$$\frac{1}{f_{sol}} = K\sin(KL_{sol}) \tag{1.82}$$

Using this and other quantities for the solenoid in Equ. 1.81 results in the following expression for the normalized chromatic emittance of a solenoid

$$\varepsilon_{n,chromatic} = \beta\gamma\sigma_{x,sol}^2\, K\,|\sin(KL) + KL\cos(KL)|\,\frac{\sigma_p}{p} \tag{1.83}$$

Figure **1.22** is a plot the chromatic emittance as a function of the energy spread as given by Equ. 1.83 and by simulation [1.35]. There is excellent agreement between the analytic and numerical approaches. The simulation used an initial beam with zero emittance with zero divergence entering the solenoid. The plot assumes an rms beam size of 1 mm. The typical measured full bunch (projected) and time-sliced (slice) electron energy spreads are indicated showing the chromatic emittance to be ~0.3 μm for the projected emittance and 0.02-0.03 μm for the slice chromatic emittance. This should be compared with the measured LCLS projected emittance of 0.4-0.5 μm for 250 pC.

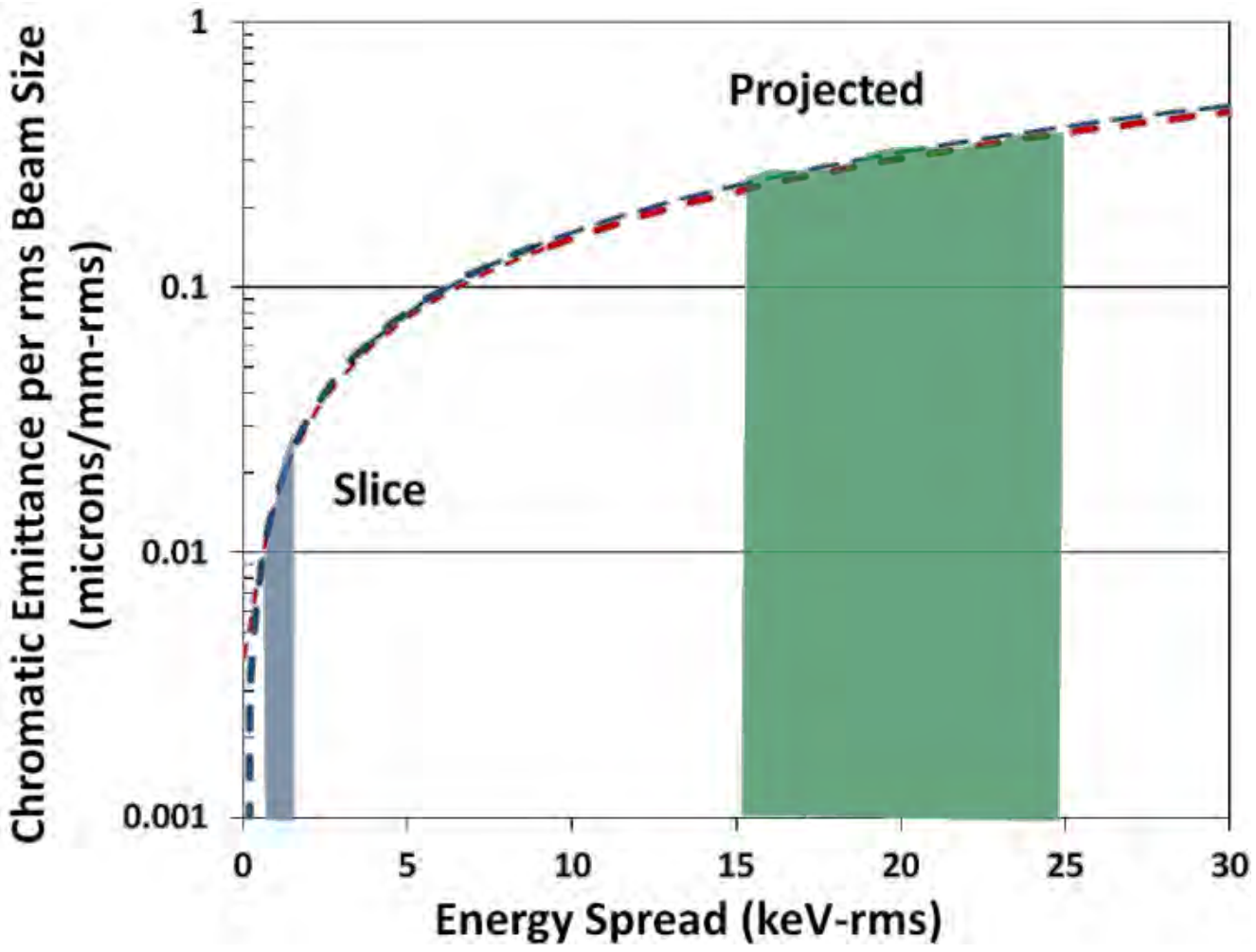

**Figure 1.22. Comparison of the chromatic emittance given by Equ. 1.83 (dashed red) and the emittance computed using the GPT particle pusher code (dashed blue) vs. rms energy spread. Both calculations assume the beam size at the solenoid is 1 mm-rms. The typical measured full bunch (projected) and time-slice (slice) electron energy spreads are indicated by the blue and green regions.**

While the solenoid's chromatic aberration can be a significant part of the projected emittance, its contribution is much less for the slice emittance. This is because the rms slice energy spread is small and thought to be 1 keV or less at 250 pC. Thus, the chromatic emittance for a slice is only ~0.02 µm (mm rms)$^{-1}$. It is important to note that because the beam size at the solenoid lens enters to the second power in Equ. 1.83, the chromaticity can introduce considerable emittance if the beam is large. Therefore, the beam size at the solenoid should be reduced in future gun designs.

### 1.5.3 Geometric Aberrations

It is known that all magnetic solenoids exhibit a 3$^{rd}$-order aberration, also known as the spherical aberration. This aberration is mostly located at the ends of the solenoid since it depends upon the second derivative of the axial field with respect to the beam direction [1.36]. In theory, the spherical aberration could be computed from the solenoid's magnetic field; in practice, this is difficult and doesn't take into account all the important details of the beam dynamics. Therefore, in order to numerically isolate the geometrical aberration from other effects, a simulation was performed with only the solenoid followed by a simple drift. Maxwell's equations were used to extrapolate the measured axial magnetic field, $B_z(z)$, and obtain the radial fields [1.35]. Following traditional optical analysis, an initial beam distribution of a square, 2 mm × 2 mm, was used assuming perfect collimation (zero divergence = zero emittance), zero energy spread and an energy of 6 MeV. The simulated transverse beam profiles given in Figure **1.23** show how an otherwise "perfect" solenoid has the characteristic "pincushion" distortion [1.37]. A 4 mm × 4 mm (edge-to-edge) object gives 0.01 µm emittance, while 2 mm × 2 mm square results in only 0.0025 µm.

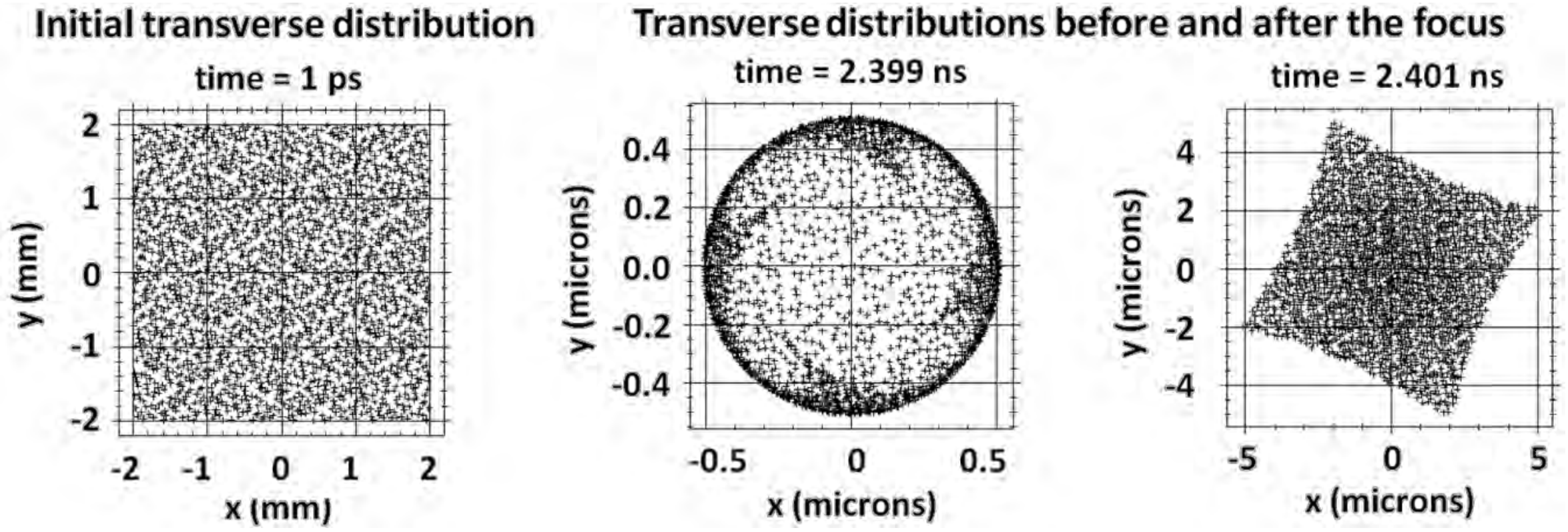

**Figure 1.23. Ray-tracing simulation of the transverse beam distribution due to the geometric aberration of a solenoid. Left: The initial transverse particle distribution before the solenoid with zero emittance and energy spread. Center: The transverse beam distribution occurring slightly before the beam focus after the solenoid illustrating the third-order distortion. Right: The beam distribution immediately after the beam focus showing the characteristic "pincushion" shape of the rotated geometric aberration.**

Figure **1.24** shows the simulation for a uniformly round beam with initially zero emittance as a function of rms beam size at the entrance of the solenoid. In addition to the simulation, the green curve gives a 4$^{th}$-order polynomial fit to the simulated emittance. It is still necessary to understand why the simulation indicates a 4$^{th}$-order dependence with beam size, rather than the expected 3$^{rd}$-order, spherical.

### 1.5.4 Aberrations due to Anomalous Quadrupole Fields and Emittance Recovery

Beam studies can show an astigmatic (unequal *x*- and *y*-plane focusing) beam from an RF gun due either to the single-side RF feed or to the magnetic field asymmetries of the gun solenoid. In order to understand and distinguish between these effects, the solenoid's multipole magnetic field was measured using a rotating coil. The magnetic measurements showed small quadrupole fields at the ends of the solenoid with equivalent focal lengths at 6 MeV of 20-30 m for the GTF solenoid. However, even though these fields were weak, it was decided to install normal and skew quadrupole correctors inside the bore of the solenoid to correct them. The details of how the correctors were incorporated into the gun are given in [1.38] and their use during operation is described in [1.39].

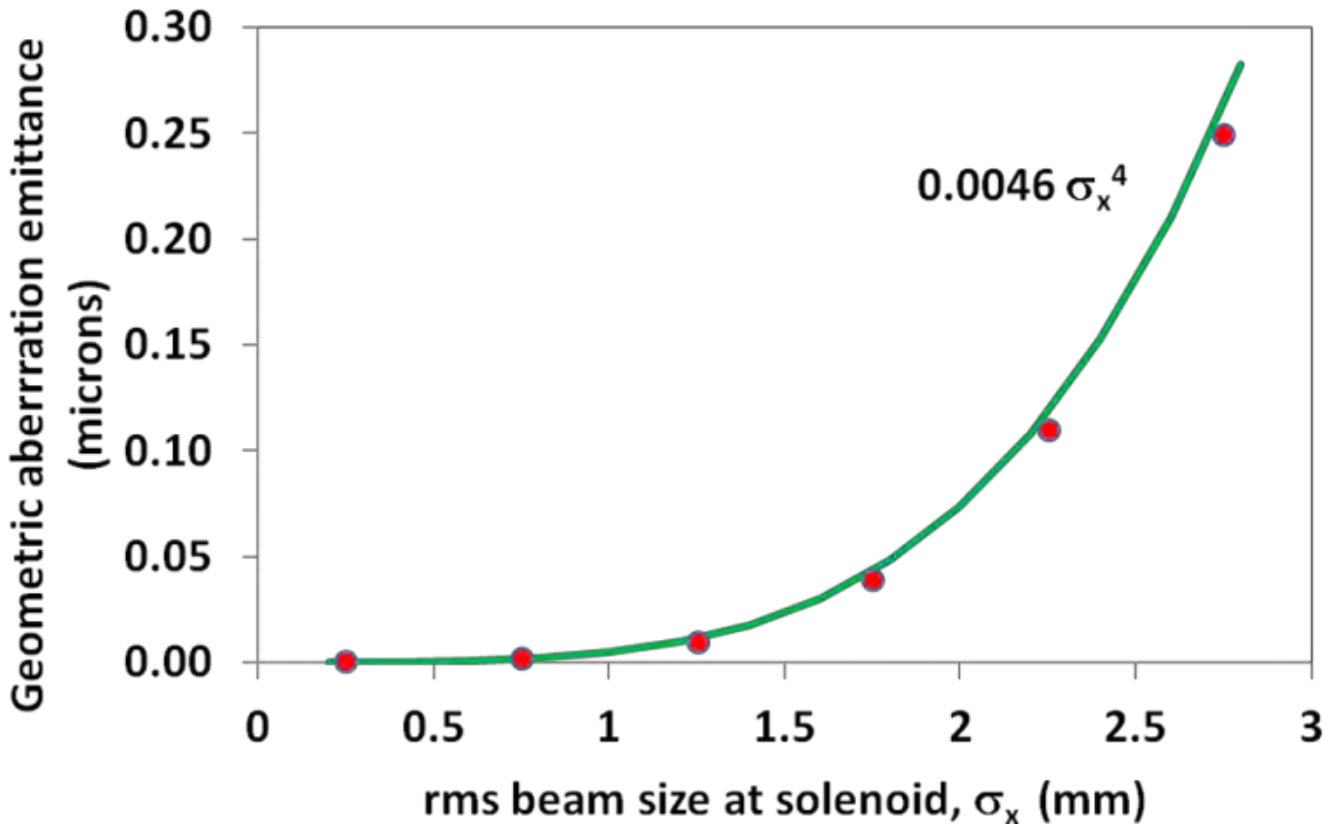

**Figure 1.24. The geometric aberration for the gun solenoid: emittance vs. the *x*-rms beam size at the lens. The emittance computed with GPT (points red) compared with a fourth order fit (solid green). The simulation used the axial magnetic field obtained from magnetic field measurements of a solenoid (Figure 1.25). The initial beam had zero emittance.**

Figure **1.25** shows the axial magnetic field and the quadrupole magnetic field and its rotation or phase angle along the beam axis of the LCLS solenoid. The quadrupole field was measured using a rotating coil with a 2.8 cm radius, which is the radius for which the quadrupole field is given. The quadrupole phase angle is the angular rotation of the poles relative to an aligned quadrupole. The phase angle is the angle of the north pole relative to the *y*-axis (left when travelling in the beam direction) for a beam-centric, right-handed coordinate system. In this coordinate system a normal quadrupole has a phase angle of 45˚. The difference in phase angle between the entrance ($z$ = -9.6 cm) quadrupole field and the exit ($z$ = +9.6 cm) field angle is close to 90˚ and both fields change sign when the solenoid's polarity is reversed. These are similar effects as measured previously for the GTF solenoid, although, the LCLS solenoid had weaker quadrupole fields with equivalent focal lengths of 50-70 m, instead of 20-30 m, as noted above for GTF.

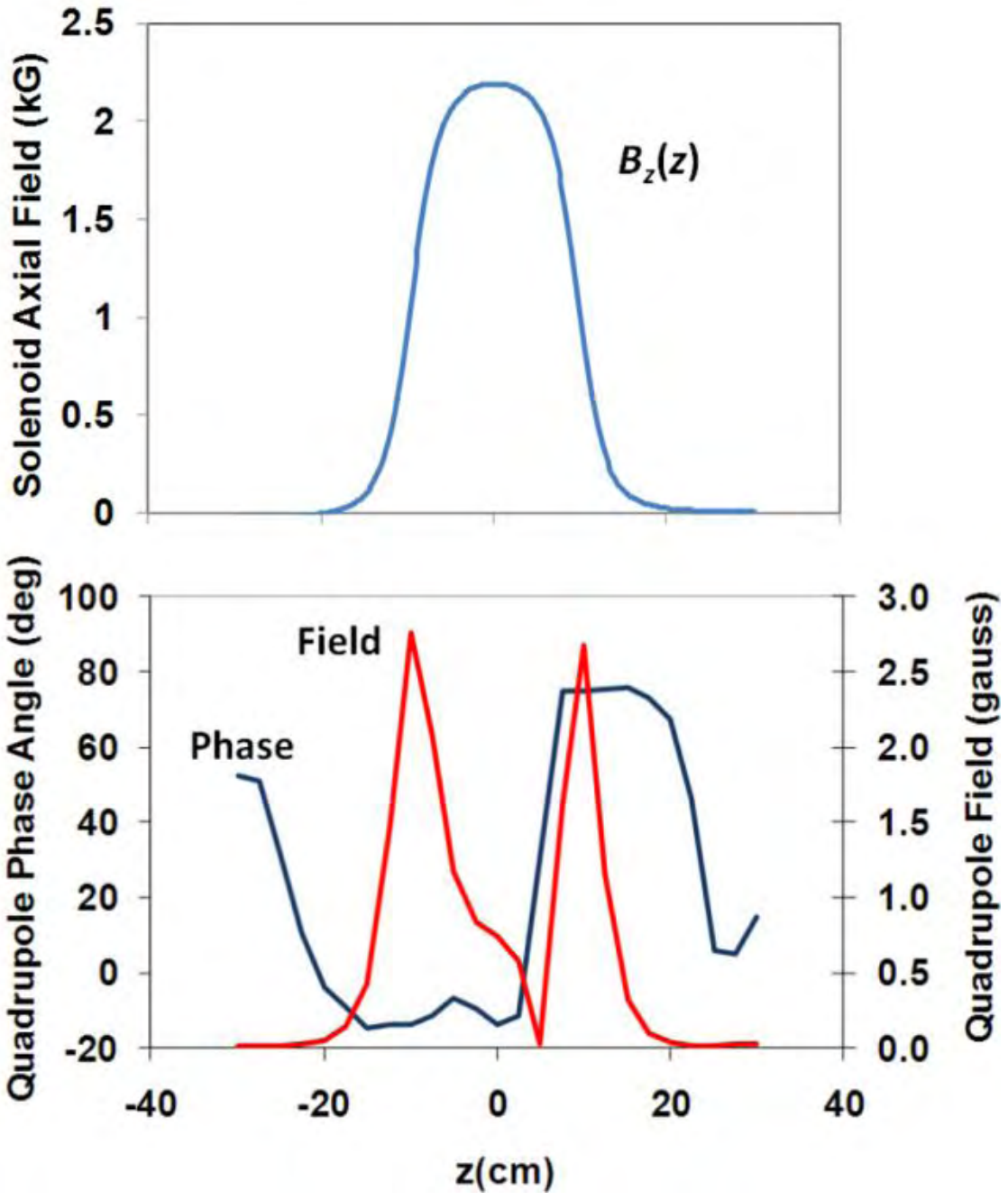

**Figure 1.25. Magnetic measurements of the LCLS gun solenoid. Top: Hall probe measurements of the solenoid axial field. The transverse location of the measurement axis (the *z*-axis) was determined by minimizing the dipole field. Bottom: Rotating coil measurements of the quadrupole field. The rotating coil dimensions were 2.5 cm long with a 2.8 cm radius. The measured quadrupole field is thus the average over these dimensions.**

As described earlier, the correction of these field errors was done by installing normal and skew quadrupoles inside the bore of the solenoid. The effect these correction quadrupoles have upon the emittance is quite profound, as can be seen in Figure **1.26** where the emittance for 1 nC and 250 pC are plotted vs. the normal corrector quadrupole strength.

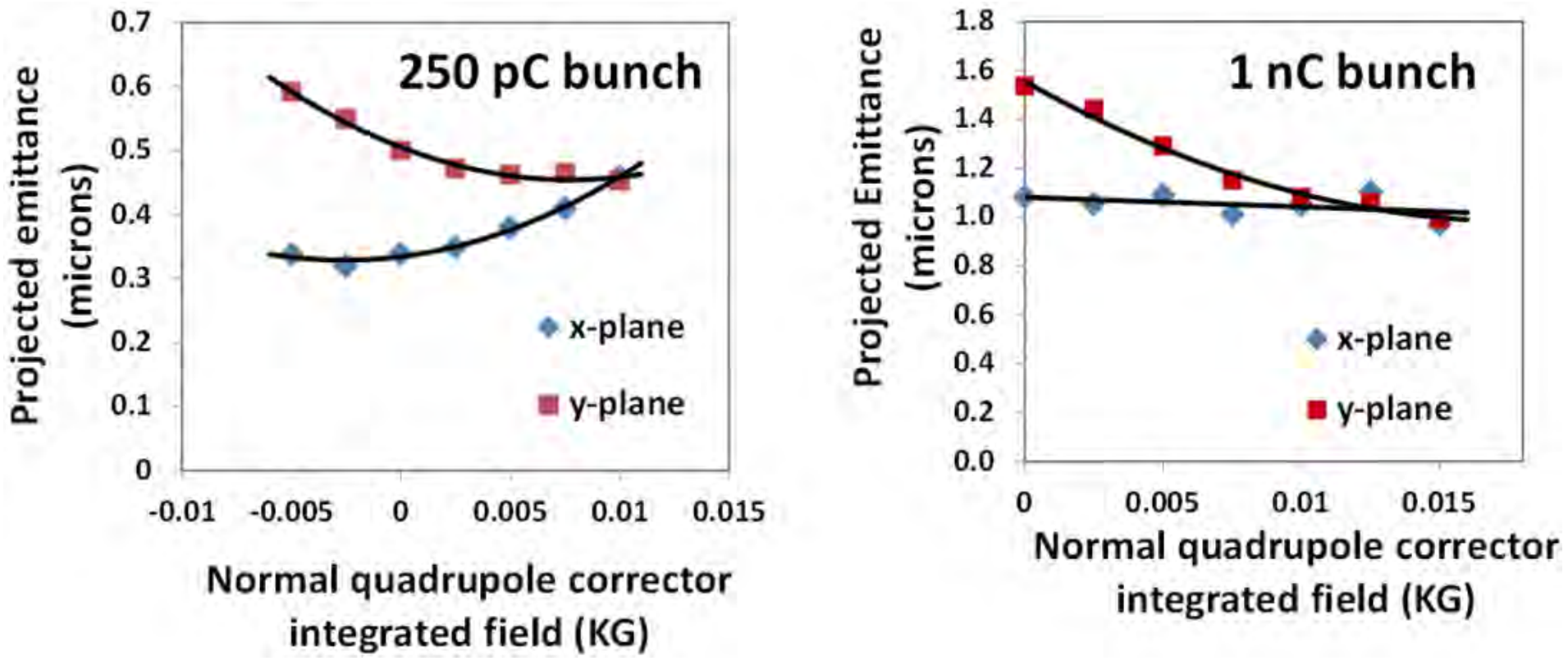

**Figure 1.26. a) Measured *x*-plane (blue) and *y*-plane (red) emittances vs. the normal corrector quadrupole strength for a 1 nC bunch charge. b) Behavior observed for 250 pC.**

The beam emittance due to these anomalous quadrupole fields can be computed both in simulation and analytically. The analysis begins by assuming a simple thin quadrupole lens followed by a solenoid with the 4×4 *x*-*y* beam coordinate transformation [1.34] given by

$$R_{sol}R_{quad} = \begin{pmatrix} \cos^2 KL & \frac{\sin KL \cos KL}{K} & \sin KL \cos KL & \frac{\sin^2 KL}{K} \\ -K \sin KL \cos KL & \cos^2 KL & -K \sin^2 KL & \sin KL \cos KL \\ -\sin KL \cos KL & -\frac{\sin^2 KL}{K} & \cos^2 KL & \frac{\sin KL \cos KL}{K} \\ K \sin^2 KL & -\sin KL \cos KL & -K \sin KL \cos KL & \cos^2 KL \end{pmatrix} \begin{pmatrix} 1 & 0 & 0 & 0 \\ -\frac{1}{f_q} & 1 & 0 & 0 \\ 0 & 0 & 1 & 0 \\ 0 & 0 & \frac{1}{f_q} & 1 \end{pmatrix} \tag{1.84}$$

As in the derivation of the chromatic emittance: $L$ is the effective length of the solenoid and

$$K = \frac{B_z(0)}{2(B\rho)_0} \tag{1.85}$$

where $B_z(0)$ is the interior axial magnetic field of the solenoid, $(B\rho)_0$ is the magnetic beam rigidity, and $f_q$ is the focal length of the anomalous quadrupole field. The beam rotates through the angle $KL$ in the solenoid.

The 4×4 beam matrix after the combined quadrupole and solenoid, is then

$$\sigma_{beam}(1) = (R_{sol}R_{quad})\ \sigma_{beam}(0)\ (R_{sol}R_{quad})^T \tag{1.86}$$

and the *x*-plane emittance after the quadrupole and solenoid is given by the determinate of the 2×2 sub-matrix,

$$\varepsilon_{x,qs} = \beta\gamma\sqrt{\det \sigma_x(1)} = \beta\gamma\sqrt{\det\begin{pmatrix} \sigma_{11}(1) & \sigma_{12}(1) \\ \sigma_{12}(1) & \sigma_{22}(1) \end{pmatrix}} \tag{1.87}$$

Finally, the normalize emittance is found to be

$$\varepsilon_{x,qs} = \beta\gamma\sigma_{x,sol}\sigma_{y,sol}\left|\frac{\sin(2KL)}{f_q}\right| \tag{1.88}$$

The *x*- and *y*-transverse rms beam sizes are the entrance to the solenoid are $\sigma_{x,sol}$ and $\sigma_{y,sol}$.

Figure **1.27** compares this simple formula (Equ. 1.88) with a particle tracking code [1.35] for the case of an initial beam with zero emittance, zero energy spread and assuming a round beam. The normalized emittance is plotted. For the comparison, assume the quadrupole focal length is 50 m, which is approximately the same as given by the magnetic measurements for the LCLS solenoid at 6 MeV. Both the analytic theory and the simulation assume a quadrupole field only at the solenoid's entrance. And of course, the simulation includes both this quadrupole effect and the geometric aberration described above. The good agreement verifies the model's basic assumptions and illustrates how a very weak quadrupole field can strongly affect the emittance when combined with the rotation in a solenoid field.

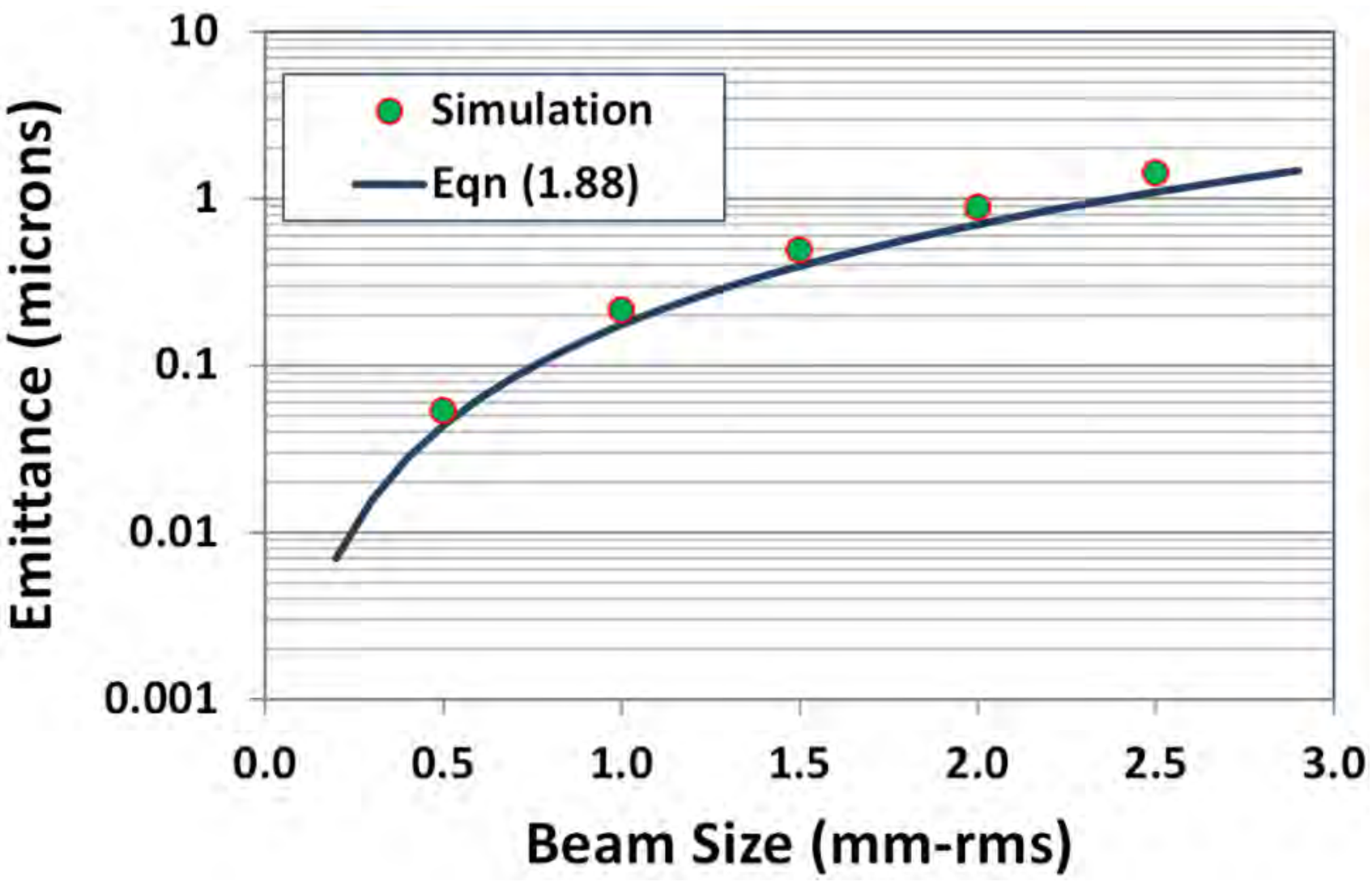

**Figure 1.27. Comparison of the emittance due to the quadrupole-solenoid coupling given by Equ. 1.88 with a particle tracking simulation for the case of the LCLS solenoid. For a beam energy of 6 MeV the quadrupole focal length was 50 m and the solenoid had an integrated field of 0.46 KG-m.**

The above expression is for the case of a quadrupole plus solenoid system where the quadrupole itself isn't rotated. When the quadrupole is rotated about the beam axis by angle $\alpha$ with respect to the normal quadrupole orientation, then total rotation angle becomes the sum of the quadrupole rotation plus the beam rotation in the solenoid. Then, the emittance becomes

$$\varepsilon_{x,qs} = \beta\gamma\sigma_{x,sol}\sigma_{y,sol}\left|\frac{\sin(2(KL+\alpha))}{f_q}\right| \tag{1.89}$$

Figure **1.28** compares Equ. 1.89 with a simulation for a 50 m focal length quadrupole followed by a strong solenoid (focal length of ~15 cm). Both show the emittance becoming zero whenever $KL + \alpha = n\pi$. The first

zero of the emittance occurring at negative quadrupole angle (not shown) is the beam rotation in the solenoid. The slight shift in angle between the theory and simulation results because the solenoid in the simulation has fringe fields which are ignored in the theory.

The final emittance is due to three effects: The skew angle and focal length of the entrance quadrupole ($\alpha_1$,$f_1$), the rotation in the solenoid ($KL$) and the skew angle and focal length of the exit quadrupole ($\alpha_2$,$f_2$). Combining the entrance quadrupole skew angle with the solenoid rotation, one obtains the emittance for a solenoid with quadrupole end fields,

$$\varepsilon_{x,total} = \beta\gamma\sigma_{x,sol}\sigma_{y,sol}\left|\frac{\sin(2(KL+\alpha_1))}{f_1} + \frac{\sin(2\alpha_2)}{f_2}\right| \tag{1.90}$$

It is relevant to point out some of the features of Equ. 1.90. First, consider the situation when both quadrupoles are perfectly aligned without any skew, *i.e.*, $\alpha_1 = \alpha_2 = 0$, then while there's no emittance contribution from the exit quadrupole, the entrance quadrupole still appears skewed by the beam's rotation in the solenoid and the emittance increases unless there is no entrance quadrupole field. For this case, the emittance does not depend upon the polarity of the solenoid field. However, this is not true for $\alpha_1$, $\alpha_2 \neq 0$. Equ. 1.90 also shows the emittance changes if the polarity of the solenoid field is reversed when there is a skewed quadrupole field: further details of this effect are discussed in the next section. Finally, the formula indicates that adding independently powered skew and normal quadrupoles after the solenoid can cancel this effect and recover the initial emittance or long wire skew and normal quadrupole correctors installed inside the solenoid can also be used for this cancellation, as was done in the LCLS solenoid [1.38].

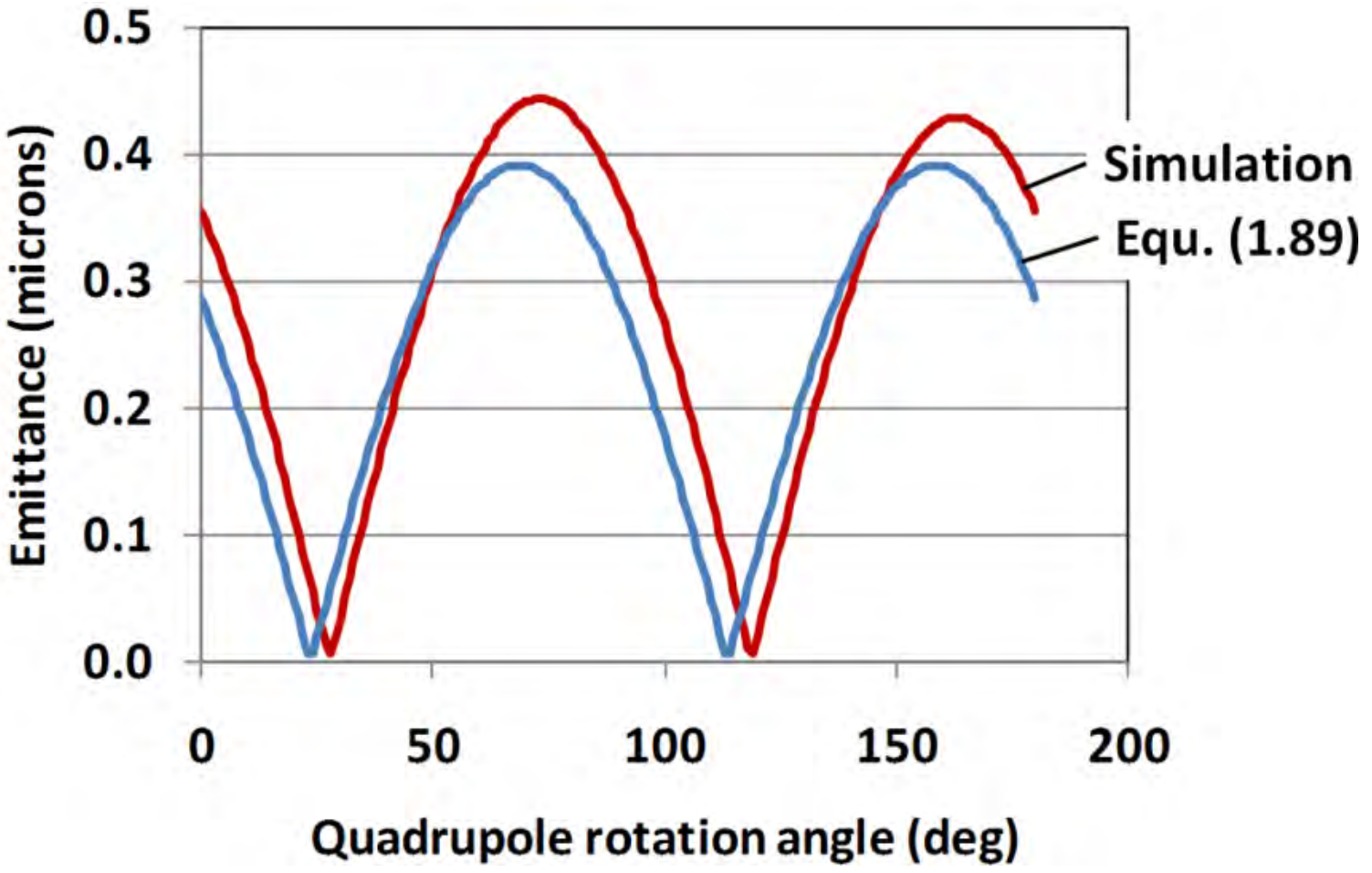

**Figure 1.28. The emittance for a quadrupole-solenoid system plotted as a function of the quadrupole rotation angle. The theory emittance (solid blue) is computed using Equ. 1.89 and the simulation (solid red) is done with the GPT code. The beam size at the solenoid is 1 mm rms for both the *x*- and *y*-planes.**

The emittance growth due to the solenoid's anomalous quadrupole fields can be compensated with the addition of skew and normal corrector quadrupoles, as shown by Equ. 1.90. Two quadrupoles, one normal and one skewed, are needed to produce the proper field strength and rotation angle. In the LCLS solenoid, these correctors consist of eight long wires inside the solenoid field, four in a normal quadrupole configuration and four arranged with a skewed quadrupole angle of 45˚. Thus, since corrector quadrupoles overlap the solenoid field, one would expect their skew angles should be added to $KL$, similar to the first term of Equ. 1.90. The emittance due to the composite system of a rotated quadrupole in front of the solenoid, the two corrector quadrupoles inside the solenoid, and the exit rotated quadrupole can be computed as the following sum

$$\varepsilon_{x,total} = \beta\gamma\sigma_{x,sol}\sigma_{y,sol}\left|\frac{\sin(2(KL+\alpha_1))}{f_1} + \frac{\sin(2KL)}{f_{normal}} + \frac{\sin(2(KL+\pi/4))}{f_{skew}} + \frac{\sin(2\alpha_2)}{f_2}\right| \quad (1.91)$$

The first and fourth terms inside the absolute value brackets are due to the entrance and exit quadrupoles with focal lengths $f_1$ and $f_2$ and skew angles of $\alpha_1$ and $\alpha_2$, respectively. The second and third terms are approximations for the normal and skew corrector quadrupoles with focal lengths $f_{normal}$ and $f_{skew}$, respectively, and of course the skew angles of the normal and skew corrector quadrupoles are 0 and π/4.

Figure **1.29** illustrates the emittance due to these effects as a function of the normal and skew corrector quadrupole focal lengths using Equ. 1.91. The entrance and exit anomalous quadrupole focal lengths are 50 m and their rotation angles as indicated by Figure **1.29** are -60˚ and 25˚, respectively. The red curves are for the normal corrector quadrupole only with the skew corrector quadrupole off, while the blue curves are given for the skew quadrupole only with the normal quadrupole off. The zero of emittance is shifted for the two correctors since the overall rotation necessary to correct the error fields is neither normal nor skewed, but something in between. Both solid curves asymptotically converge to the uncorrected emittance as the correctors are turned off (infinite focal length). The figure also shows the effect of reversing the polarity of the solenoid with corresponding emittances plot as dashed lines. In this case, the uncorrected emittance clearly approaches a much smaller emittance. As mentioned earlier, the skewed anomalous quadrupole fields make the resulting emittance growth and focusing of the solenoid dependent upon its polarity and provide an experimental signature that the fields are skewed. Therefore, if the anomalous fields are skewed, even with no quadrupole correction, one polarity of the solenoid results in a lower emittance than the other.

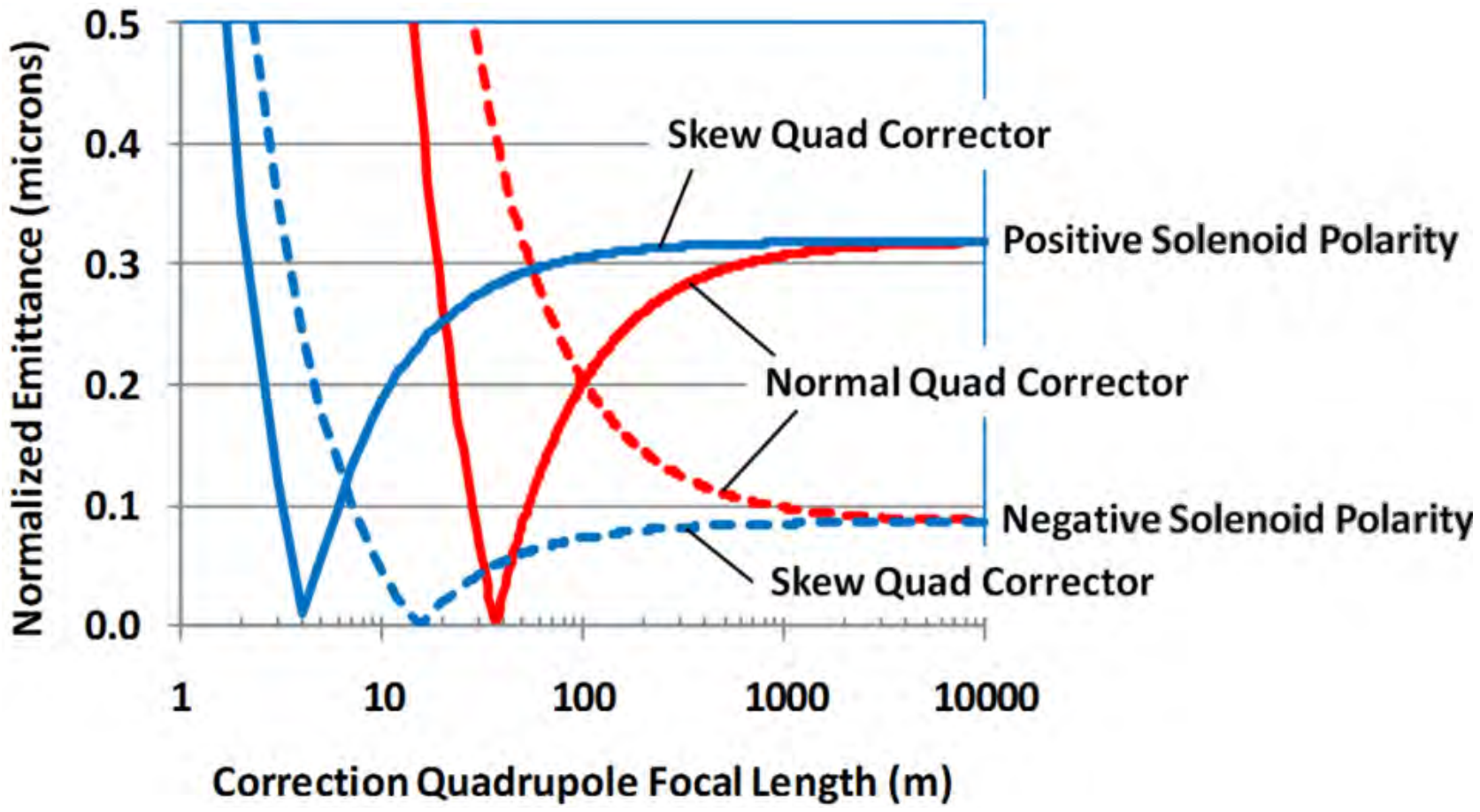

**Figure 1.29. The emittance as a function of the normal and skew quadrupole corrector focal lengths for positive and negative polarities of the solenoid. Anomalous quadrupole field errors with 50 m focal lengths are included at the ends of the solenoid with rotations of -60˚ and 25˚, respectively, as given in Figure 1.25. The *x*- and *y*-rms beam sizes at the solenoid entrance are assumed to be 1 mm.**

## 1.6 SPACE CHARGE SHAPING

The space charge force can defocus the beam, behaving similar to a negative focal length lens. While most emittance compensation techniques use external fields to cancel the linear space charge effects, it is the non-linear space charge forces which produce additional emittance. In this section, the radial and longitudinal electric fields inside a rotationally symmetric bunch are given in terms of a power series expansion about the bunch longitudinal axial. This derivation is for the steady-state case, without currents or time dependent fields. It gives the fields in the rest frame of the bunch, which for the example considered here, is thin and disk-like.

Here, we derive the radial force on an electron confined to a thin disk of charge. The surface charge density is assumed to be rotational symmetric with a radial quadratic dependence upon the surface charge density. The result is that the quadratic radial distribution can be adjusted to cancel the 3rd-order space charge defocus of the disk's distribution, Figure **1.30**. The technique is to first compute the electrical potential energy along the axis of rotation of the disk. Expanding this potential into a power series, we multiply each term by the appropriate order of Legendre polynomial to obtain the potential at any point on space. Finally, the divergence of the potential gives the radial electric field on an electron.

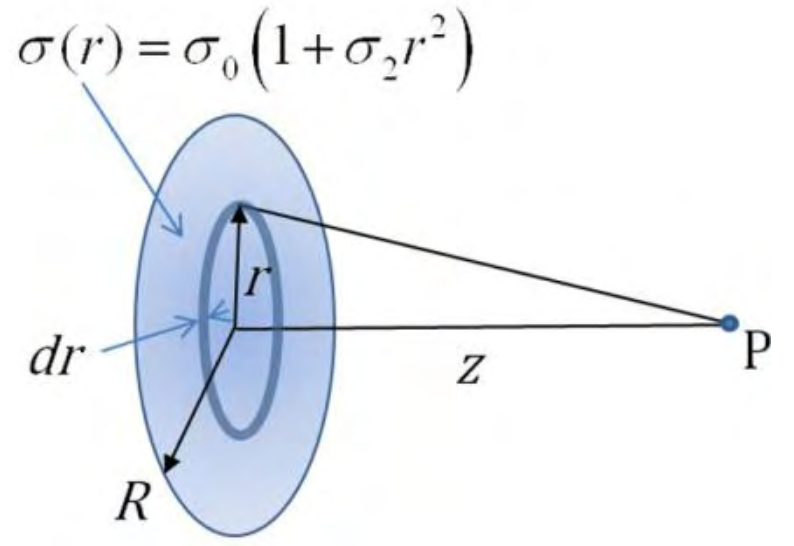

**Figure 1.30. The geometry for computing the on-axis electric field produced by a disk of charge.**

The electric potential along the disk's axis can be computed using the following integral

$$4\pi\varepsilon_0 V(z) = \iint \frac{\sigma}{\sqrt{z^2+r^2}}\,ds = 2\pi\int_0^R \frac{\sigma(r)r}{\sqrt{z^2+r^2}}\,dr \tag{1.92}$$

where surface charge density as a function of $r$ is $\sigma(r) = \sigma_0(1 + \sigma_2 r^2)$, then the integrand has two parts. One linear and the other third-order in $r$, as given by

$$V(z) = \frac{\sigma_0}{2\varepsilon_0}\left(\int_0^R \frac{r}{\sqrt{z^2+r^2}}\,dr + \sigma_2\int_0^R \frac{r^3}{\sqrt{z^2+r^2}}\,dr\right) \tag{1.93}$$

Performing the integration gives

$$V(z) = \frac{\sigma_0}{2\varepsilon_0}\left\{\sqrt{z^2+R^2} - z + \sigma_2\left[\frac{1}{3}(z^2+R^2)^{3/2} - z^2\sqrt{z^2+R^2} + \frac{2}{3}z^3\right]\right\} \tag{1.94}$$

The first two terms inside the outer brackets give the on-axis potential for a uniform disk. The $\sigma_2$ term is the potential coming from the parabolic radial part of the surface charge density. For a beam that is off-axis, we implement the following coordinate as seen in Figure **1.31**.

Consider the uniform part of the charge distribution first. Expanding the potential as a power series, grouping into terms with the same power and then multiplying each term with the Legendre polynomial of that power, gives the potential everywhere. The electric potential to 4th-order is then

$$V_0(\theta,r) = \frac{\sigma_0}{2\varepsilon_0}\left[R - rP_1(\cos(\theta)) + \frac{1}{2}\frac{r^2}{R}P_2(\cos(\theta)) - \frac{1}{8}\frac{r^4}{R^3}P_4(\cos(\theta)) + \ldots\right] \quad (1.95)$$

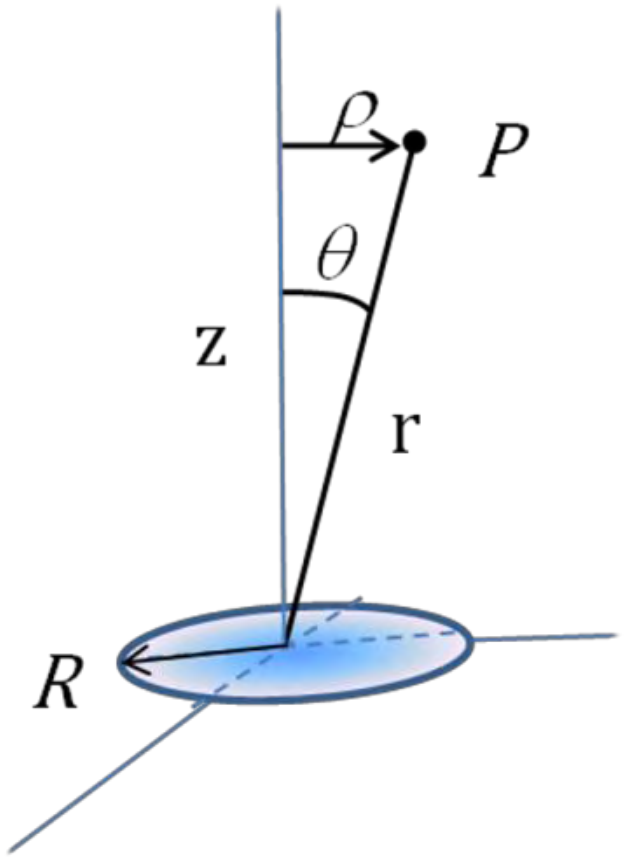

**Figure 1.31. Coordinates used to compute the off-axis electric field of a charged disk.**

Following the same procedure for the parabolic part of the charge density gives the electric potential due to the quadratic $\sigma_2$ in the plane for the disk as

$$V_2(\rho) = \frac{\sigma_0\sigma_2}{2\varepsilon_0}\left[\frac{1}{3}R + \frac{1}{4}R\rho^2 - \left(\frac{3}{8}\right)^2\frac{\rho^4}{R} + \ldots\right] \quad (1.97)$$

where the total potential is the sum, $V(\rho) = V_0(\rho) + V_2(\rho)$. The radial electric field is given by $E_\rho(\rho) = \frac{\partial V}{\partial \rho}$, or to third-order

$$E_\rho(\rho) = \frac{\sigma_0}{2\varepsilon_0}\left[\left(\frac{\sigma_2}{2}R - \frac{1}{2R}\right)\rho - \frac{3}{16}\left(3\sigma_2 + \frac{1}{R^2}\right)\frac{\rho^3}{R}\right] \quad (1.98)$$

Notice, that if $\sigma_2 = \frac{-1}{3R^2}$, then the third order term is zero when the radial charge density is parabolic

$$\sigma(\rho) = \sigma_0\left(1 - \frac{\rho^2}{3R^2}\right) \quad (1.99)$$

In this case and the radial space charge force becomes linear

$$E_\rho(\rho) = \frac{-\sigma_0}{3\varepsilon_0}\frac{\rho}{R} \quad (1.100)$$

and the beam expands linearly with little increase in the emittance. The parabolic radial distribution is plotted in Figure **1.32**.

The above description follows the seminal work of Serafini [1.40] who first proposed shaping in RF guns to reduce the non-linear space charge emittance. More recently, Luiten [1.41] has applied classical stellar dynamics to the problem and performed simulations showing that a hemispherical-shaped surface charge density at the cathode rapidly expands into a uniform 3-D ellipsoid distribution having linear space charge forces and no space charge emittance. Therefore, the precise radial shaping of the electron bunch using the drive laser should reduce the non-linear space charge emittance.

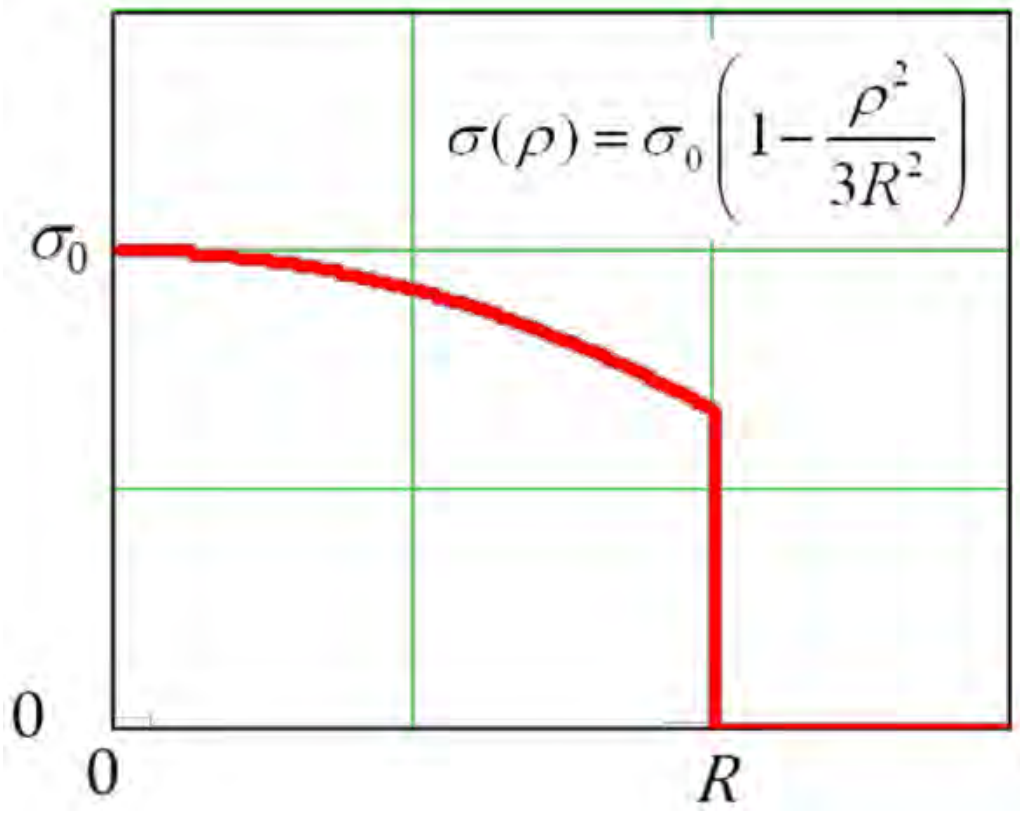

Figure 1.32. Plot of the radial charge distribution having no 3rd-order space charge force.

# 1.7 SIMULATION CODES

Simulation codes are critical components of the photoinjector design process and are an area of continual development. This section describes some of the more common codes used for photoinjector development, with an emphasis on codes that can be used effectively on a typical desktop computer.

## 1.7.1 General Comments on Simulation Fidelity

In the most general terms, operation of a photoinjector can be described by the following process:

1) A laser beam strikes the cathode.
2) Electrons are emitted from the cathode.
3) The emitted electrons interact, *via* electromagnetic fields, with
    a) the photoinjector,
    b) other components such as solenoids,
    c) each other, and
    d) the electron emission process.

Ideally, the process results in the production of a high quality electron beam. The simulation codes used in injector design to handle each of these steps with varying fidelity to the real world.

The typical goal of a beam dynamics simulation code is, broadly speaking, to provide the 6-D coordinates of particles within the beam at the end of the simulation. Every beam property of interest – emittance, energy spread, bunch duration, *etc.* – can be calculated from this distribution.

Most photoinjector simulations that work with distributions of particles are based upon stepping forward in time. At each time step, the motion of all particles and the forces acting upon them are calculated and updated. Depending on the simulation code used, time steps may be fixed or variable; typical time steps are on the order of 0.1-10 ps, with finer steps taken when the beam is being emitted from the cathode and larger steps when the beam is in regions of slowly varying external fields.

A "typical" bunch charge might be -1 nC, comprised of approximately $6.4\times10^9$ electrons. ~0.15 TB would be required to store every electron's position in 6-D phase space with quadruple-precision (32-bit) floating point numbers. Modern desktop computers do not typically have that much memory and only recently have cluster computers progressed to the point where each electron in a bunch could be independently tracked. Further, the time to execute the simulation generally scales at least as fast as the number of particles within the simulation. Thus, often the first reduction in fidelity that occurs is to use a "macroparticle" to represent a larger number of electrons in the bunch. In a desktop simulation, typically done as part of an initial design study, perhaps $10^4$-$10^5$ macroparticles might be used to provide a tractable number of particles from both computer memory and CPU time perspectives.

In most photoinjector simulations, all macroparticles are assigned the same charge, but with different charge density within the electron beam (say, from a non-uniform drive laser) reflected by different spatial density of macroparticles. This simplifies the macroparticle bookkeeping and various other computational tasks, but can yield "noisy" results and errors when the number density in the simulated beam is too low. An alternate approach is to keep the initial number density of macroparticles constant, but varying the charge per macroparticle to reflect density variations within the electron beam. The former approach is generally the one used in injector simulations, and the typical method of reducing noise is to increase the number of macroparticles or, *via* "quiet start," non-random distributions. The latter may be more suitable to the introduction of cathode physics into the beam dynamics codes.

Electron emission, encompassed by cathode theory and modeling, is a rich area of current development and is treated more fully in Chapter 5. Historically, most beam dynamics codes have not incorporated physics-based emission modeling, and so will not be discussed in great detail here. Rather, most beam dynamics codes allow the user to specify, for instance, the emission of macroparticles vs. time over a given area of the cathode; this may be done *via* supplying an external distribution or by specifying various parameters of the distribution.

Another commonly used approximation is the assumption of radial symmetry of RF and magnetic fields within the accelerating structures and beamlines, and of the beam's self-fields. For initial studies and for some photoinjector designs (such as RF photoinjectors with on-axis power couplers), this is not a bad approximation; however, it does represent an additional loss of fidelity with respect to the physical reality of the system being modeled, and of necessity excludes the impact of both TE modes and magnetic field aberrations such as those described in Section 1.5, as well as asymmetries arising in the accelerating fields from the presence of RF power couplers, field probes, viewports, *etc.* Radial symmetry of the beam's self-fields is clearly broken as soon as the beam passes through a quadrupole, or indeed any multipole beamline element.

The interaction of the beam with itself, a.k.a. "space charge effects," is central to the emittance compensation process. While in principle space charge forces can be calculated from every particle to every other particle, the time required to perform such a calculation scales as $N^2$, where $N$ is the number of macroparticles in the simulation. Most beam dynamics codes therefore use a variation of a particle-in-cell, or PIC, method to calculate space charge effects. The codes are grouped into one of two general categories. If the code ignores the interaction of the beam with the photoinjector structure, it is known as a particle-pusher (or sometimes pseudo-PIC) code. If such interactions are accounted for in a self-consistent fashion, the code is referred to as a particle-in-cell, or PIC, code.

Finally, electromagnetic (EM) design codes are used to simulate the physical structures, such as RF cavities, DC gaps, solenoids, *etc*., used in the photoinjector design process. As with beam dynamics codes, EM design codes differ widely in their capabilities, fidelity and ease of use. An important consideration is how readily information (primarily as field maps) can be transferred from EM design codes to beam dynamics codes.

Table **1.1** is a partial listing of simulation codes useful for photoinjector design. It is by no means a complete list and reflects the authors' experiences and predilections.

| **Name** | **Type** | **Notes** |
|---|---|---|
| POISSON / SUPERFISH | 2-D electro-, Magnetostatic and RF Code | Integrates well with PARMELA and GPT; TM RF Modes Only; Extensive Documentation. |
| CST Microwave Studio | 2-D and 3-D EM Modeling Code | General-purpose Very Powerful electromagnetic Modeling Code; Excellent Documentation; Some Beam Transport; Commercial Code |
| MAFIA | 2-D and 3-D EM Modeling Code | General-purpose Very Powerful Electromagnetic Modeling Code; Excellent Documentation; Some Beam Transport; Commercial Code |
| TRANSPORT | Envelope Code | The "Grandfather" Code; Manual is an Excellent Reference for 1$^{st}$-order Transport Matrices of Accelerator Components. |
| TRACE-3D | Envelope Code with Space Charge | Fast; Good Graphical Tools Available |
| HOMDYN | Envelope Code | |
| PARMELA | Particle Pusher | Includes many "Built-in" Accelerator Elements; Well-benchmarked; Good Documentation; Source Code not Available |
| T-Step | Particle Pusher | Upgraded Version of PARMELA; Commercial Code |
| ASTRA | Particle Pusher | Many Variations; Often the Code of Choice for Implementing Genetic Algorithm-based Optimization |
| GPT | Particle Pusher | Includes many "Built-in" Elements; New Elements can be Added by the User; Extensive Options for Importing Field Maps; Unusual, but Useful Coordinate Scheme; Commercial Code |
| IMPACT-T | Particle Pusher | Under Wide Development; Several Variants |
| SPIFFE | 2-D PIC Code | Basic Code; Fast; Good Learning Tool |
| VORPAL | 3-D PIC Code | Includes Updated Cathode Modeling; Commercial Code |
| MICHELLE | 3-D PIC Code | Focuses on Electron Gun Design; Commercial Code |
| MAD | High-energy | |
| ELEGANT | High-energy | Includes CSR and Longitudinal Space Charge Models; Used in LCLS Design |

**Table 1.1. List of simulation codes with some description.**

### 1.7.2 Particle Pushers

Particle pusher codes are beam dynamics codes which do not consider the interaction of the electron beam with the structure of the photoinjector. Injector beam dynamics codes must include space charge effects however, and this is often done *via* a PIC-like methodology.

In a typical pseudo-PIC calculation, a grid (2-D or 3-D) is overlaid upon the particle distribution. Each macroparticle's charge is assigned either to its nearest grid point, or split over the grid points of its encompassing cell according to its position within the cell. Maxwell's equations are then solved on the grid; the resulting fields are applied to the macroparticles and the next simulation time step is taken. Computation time scales approximately as $M * N$, where $M$ is the number of grid points and $N$ is, again, the number of macroparticles. A 2-D PIC implementation in essence treats every particle as a ring of charge; this approximation breaks down, to a greater or lesser degree, as soon as radially asymmetric charge distributions (from the cathode, passing through a quadrupole field, *etc.*) are encountered. In either a 2-D or a 3-D PIC code, there must be enough cells to sufficiently model local charge density variation of interest, but not so many that the statistics of assigning charge to the grid become poor. As importantly, there must be sufficient numbers of macroparticles to maintain good statistics for the space charge calculation.

Typical particle pusher codes, including PARMELA [1.42], T-Step [1.43], IMPACT-T [1.44], [1.45], GPT [1.35] and Astra [1.46] usually offer one or several PIC-like algorithm to calculate space charge effects. (PARMELA, for instance, can perform either a 2-D or a 3-D space charge calculation.) The "external" fields, such as those from DC gaps, RF cavities, solenoids and the like, are just that – typically provided by a field map generated by an external code, the fields from these elements are applied to the macroparticles, but the macroparticles cannot modify those fields. An example of such a calculated external field can be seen in the top of Figure **1.6**. The simulation neither knows nor cares where the physical boundaries of the photoinjector are, save perhaps specified radii beyond which macroparticles are assumed to have struck a wall and consequently be removed from the simulation. Likewise, there is no guarantee that the applied fields are consistent with the cavity geometry.

This pseudo-PIC approach has several advantages. Since the PIC mesh need to extend only over the electron beam, a high density of mesh cells can be used for modest memory expense; and if the mesh expands and contracts with the beam, the approximate macroparticle density within the mesh can be held steady, helping to preserve the statistics of the calculation.

To further save time, some pseudo-PIC codes perform a relativistic transformation to the average rest frame of the beam before applying the grid. The general assumption made is that in this frame the particles have negligible velocity, so only Poisson's equation need be solved and the fields are then transformed back to the laboratory frame. The main disadvantage of this approach is that it breaks down when beams have large velocity spreads, or large fractional spreads ($\Delta\gamma/\gamma$).

A general disadvantage of the particle pusher codes is they cannot self-consistently calculate the interaction of the beam with the structure of the injector. This can become very important when, for instance, attempting to simulate beam loading or wakefield effects. Also, as the electromagnetic fields are generally not self-consistent, incorporation of advanced electron emission models into these codes, particularly in the case of multi-bunch emission, can be problematic.

### 1.7.3 Particle-in-Cell Codes

Rather than applying a mesh over only the macroparticles, a true PIC code applies the mesh to the entire geometry, within which the beam can propagate, incorporating all boundary surfaces the beam can "see." The mesh is generally fixed in space rather than moving with the beam.

Depending on the PIC code, the fields used to accelerate and guide the beam can either be imported as with the particle pusher codes, calculated by the PIC code itself, or by some combination of the two methods.

PIC codes have several significant advantages. They are generally fully electromagnetically self-consistent, so beams with large energy or velocity spreads are handled properly. Boundary conditions are automatically incorporated as the injector geometry, so the PIC grids are properly terminated at their edges and impedance effects can be included, as well as particle beam wakefields. This also allows incorporation of cathode emission models that require accurate values for the electromagnetic fields at the cathode surface to calculate emission current density … whether or not the beam is within the vicinity of the cathode.

Finally, advanced 3-D PIC codes, such as VORPAL [1.47], can be used to perform most of the calculations required to model an injector, including RF power couplers, the buildup of accelerating fields, multipacting and the extraction of beam-induced higher order modes. This is perhaps the most self-consistent method of modeling an injector available. An interesting side-effect is that obtaining the actual cavity modes excited by wakefields can be challenging; a PIC code does not "know" about cavity modes, it simply knows the net charge and the electric and magnetic fields at each point of a grid at a given point in time: it will update those fields and particle positions and velocities self-consistently at each time step.

There are several significant disadvantages to PIC codes: first, they tend to be much more computationally intensive to operate, both from a memory and CPU time standpoint, than particle pusher codes of similar dimensionality (*i.e.*, 2-D or 3-D) because Maxwell's equations are being solved on every grid point in the model, whether or not there are macroparticles present.

Mesh generation is still in the realm of an art, and although much progress has been made with automatic mesh generation, it can still be challenging to generate suitable meshes. The difficulty lies in part with the ratio of the size of the beam to the size of the photoinjector, typically on the order of 100:1 in radius and 1000:1 or greater longitudinally. At the outer boundaries of the injector cavities, where no beam particles are liable to be present, the mesh can be relatively coarse; but, to resolve fine structures within the beam, the mesh density in the region of the beam must be relatively high. Thus, a uniform mesh will generally either be too coarse to properly resolve the electron beam, or too fine to allow the simulation to run in a reasonable amount of time and memory. Good progress is being made in non-uniform mesh generation; and concepts involving overlaid meshes are very interesting, but this is also an area of active development.

### 1.7.4 Other Types of Beam Dynamics Codes

Several other types of injector design codes should be mentioned. First, envelope codes, such as TRACE-3D [1.48] and HOMDYN [1.49], use externally generated, or analytic, fields and a simplified representation of the beam (an $M * N$ grid of charged rings for HOMDYN, or a uniformly filled ellipse for TRACE-3D) to model the injector. Historically these codes have been very valuable as "first step" modeling, however, with the increasing power of desktop computers, particle pusher codes running with small particle counts are nearly as fast in a practical sense and provide an easy method of model refinement by simply increasing the particle count.

Field mode codes represent a compromise between particle pusher and PIC codes. This type of code relies, as does particle pusher codes, upon local PIC grids and externally defined fields. However, the field code "knows" about cavity modes and can calculate the beam's contribution to, and influence from, multiple cavity modes at once. These codes are not currently in widespread use, however.

While many particle pusher codes theoretically incorporate enough accelerator component models to be useful for designing an entire accelerator, often they are used only for the photoinjector region, after which the beam is "handed off" to a high-energy accelerator design code.

High-energy accelerator design codes, such as TRANSPORT [1.34], MAD [1.50], [1.51] and ELEGANT [1.52], originated from the need to design complete accelerators comprising potentially thousands of elements. They typically propagate beam particles or rms envelopes using a 1$^{st}$-, 2$^{nd}$- or 3$^{rd}$-order matrix approach and often incorporate goal-seeking or parameter matching functionality. With certain limited exceptions, they generally do not perform space charge calculations and are therefore not suitable for injector design; however, they often include features critical for high performance accelerator design, such as coherent space charge radiation (CSR) modeling, which are missing from or have limited support in pseudo-PIC codes. Since these codes are often provided input from photoinjector design simulations, it is helpful to be at least somewhat familiar with their requirements and limitations.

### 1.7.5 Electromagnetic Design Codes and Accelerator Component Modeling

The electromagnetic fields used in particle beam simulations have to come from somewhere. In PIC codes, the fields in an RF cavity (for instance) can often be generated by the PIC code itself. In pseudo-PIC codes, the fields are generally either represented *via* analytic formulas, or are imported as external field maps.

In most photoinjector simulations, the RF fields used to accelerate the beam are provided *via* importing a field map generated by an external code. Critical magnetic elements near the photocathode, such as those generated by emittance compensation solenoid magnets, are also often generated externally and imported as maps. Other magnetic elements, such as dipoles and quadrupoles, often may (or must, depending on the code) be approximated by analytic expressions for “hard-edge” fields.

There are many EM codes available with varying degrees of fidelity, and the topic is well outside the scope of this book. When considering an EM design code, the photoinjector designer should consider both the fidelity with which an EM code will model elements of the injector, and also the ease with which the results of the calculations can be imported into the beam dynamics code. For instance, the POISSON/SUPERFISH codes [1.17] typically only calculate 2-D field maps (*e.g.*, *z-r* maps for RF cavities or solenoids). They are, however, tightly integrated with the POISSON beam dynamics code, and GPT has a number of useful tools to ease importation of field maps from POISSON/SUPERFISH. On the other hand, a “world’s most accurate” electromagnetic design code is of limited utility to the photoinjector designer if it cannot be used to generate the required field maps.

The line between EM codes and PIC codes for beam dynamics is not always well defined. As mentioned above, some beam dynamics codes, such as VORPAL, are capable of generating RF cavity fields. Some EM design codes, such as MAFIA [1.53], [1.1] and CST [1.55], can include electron emission modeling and transport with varying degrees of physical fidelity.

### 1.7.6 General Approach to Injector Modeling

Injector modeling can be approached in three phases: Conceptual development, tuning and final refinement.

Conceptual development can be performed with any type of injector design code, but the practice in common use as of this writing is to employ a code such as SUPERFISH, to generate radially symmetric acceleration and solenoid fields from simplified models of the injector geometry and use a particle pusher code to perform beam dynamics simulations. This can be used to quickly narrow down on a reasonable parameter space for more detailed exploration.

In tuning, or optimization, many simulations are run to identify optimal working points. Depending upon the sophistication of the optimization techniques used, this step can also adjust injector “physical” parameters, such as cavity length or cathode/anode geometry that are fixed once the injector is built, as well as parameters, such as accelerating gradients and solenoid field strengths, that can be altered without changing the injector's physical construction. Here again, the advantage to particle pusher codes is speed. The combination of fast particle pusher codes with concurrent computing and advanced optimization methods is extremely powerful, as demonstrated by the design of the Cornell ERL injector.

The final refinement of the design incorporates as many physical effects as reasonable, given the available resources and performance requirements. In the case of the LCLS injector, for instance, this stage included using 3-D electromagnetic field maps for both the $TM_{010}$ and $TM_{011}$ (accelerating) modes in the RF gun, quadrupole corrections to the emittance compensation solenoid, *etc.* This can be the most time consuming stage of the simulation process, but as the LCLS injector has demonstrated, the results are definitely worthwhile.

As a final note, it is well worth remembering that no simulation is a complete representation of physical reality. This is because neither the codes nor the researchers are perfect. That is to say, some things aren't in the model because the code doesn't support it; an example would be semiconductor cathodes in PARMELA. We know this and attempt to allow for it when interpreting our simulation results. Other things aren't in the codes because we do not implement them, although the code can support them. An example of this from the NCRF photoinjector design community is the influence of the $TM_{010}$ mode in the SLAC/BNL/UCLA-style RF guns. In practice, the $TM_{010}$ mode can have a noticeable impact on beam quality, but most early design studies did not include its effects in the simulations. As a result, obtaining 1 μm emittance beams at nanocoulomb bunch charges was considerably more challenging than was anticipated from the simulations.

Therefore, we close the section on simulation with two questions to keep in mind:

> "What does the simulation not include?" and
> "What am I not including that might matter?"

## 1.8 Conflict of Interest and acknowledgements

The authors confirm that this article content has no conflicts of interest. J. W. Lewellen acknowledges the support of the Office of Naval Research, the High-Energy Laser Joint Technology Office and the Naval Postgraduate School. D. H. Dowell thanks his understanding wife, Alice.

***References***

[1.1] G. Neal, C. L. Bohn, S. V. Benson *et al.*, "Sustained kilowatt lasing in a free-electron laser with same-cell energy recovery," *Phys. Rev. Lett.*, vol. 84, pp. 662-665, January 2000.

[1.2] P. Emma, R. Akre, J. Arthur *et al.*, "First lasing and operation of an ångstrom-wavelength free-electron laser," *Nature Photonics,* vol. 4, pp. 641-647, August 2010.

[1.3] M. Reiser, *Theory and Design of Charged Particle Beams,* Weinhiem: Wiley-VCH, 2007, pp. 16.

[1.4] C. H. Lee, P. E. Oettinger, R. Klinkowstein *et al*., "Electron emission of over 200 A/cm$^2$ from a pulsed-laser irradiated photocathode," *IEEE Trans. Nucl. Sci.*, vol. 32, pp. 3045-3047, October 1985.

[1.5] J. S. Fraser, R. L. Sheffield and E. R. Gray, "High-brightness photoemitter development for electron accelerator injectors," in *Proc. AIP Conf.,* vol. 130, 1985, p. 598-601.

[1.6] J. S. Fraser, R. L. Sheffield and E. R. Gray, "A new high-brightness electron injector for free electron laser," *Nucl. Instrum. Meth. A,* vol. 250, pp. 71-76, September 1986.

[1.7] R. L. Sheffield, "High brightness electron sources," in *Proc. 1995 Particle Accelerator Conf.,* 1995, pp. 882-886.

[1.8] C. Travier, "RF guns: bright injectors for FEL," *Nucl. Instrum. Meth. A*, vol. 304, pp. 285-296, July 1991.

[1.9] D. H. Dowell and J. G. Power, “High-average power facilities,” SLAC, Stanford, CA, Technical Report No. SLAC-PUB-15232, September 5, 2012.

[1.10] F. Zhu, S. W. Quan, J. K. Hao *et al.*, “Status of the DC-SRF photoinjector for PKU-SETF,” in *Proc. 2011 Superconducting RF*, 2011, pp. 973-976.

[1.11] D. Janssen, V. Volkov, H. P. Bluem and A. M. M. Todd, “Axial RF power input into photocathode electron guns,” in *Proc. 2005 Particle Accelerator Conf.*, 2005, pp. 743-745.

[1.12] R.A. Rimmer, “A high-gradient cw RF photo-cathode electron gun for high current injectors,” in *Proc. 2005 Particle Accelerator Conf.*, 2005, pp. 3049-3051.

[1.13] H. Bluem, “High power testing of a fully axisymmetric rf gun,” in *Proc. 2007 Particle Accelerator Conf.*, 2007, pp. 3142-3144.

[1.14] J. W. Staples, K. M. Baptiste, J. N. Corlett *et al.*, “Design of a vhf-band photoinjector with megahertz beam repetition rate,” in *Proc. 2007 Particle Accelerator Conf.*, 2007, pp. 2990-2992.

[1.15] D. Kayran and V. N. Litvinenko, “Novel method of emittance preservation in ERL merging system in presence of strong space charge forces,” in *Proc. 2005 Particle Accelerator Conf.*, 2005, pp. 2512-2514.

[1.16] D. Kayran and V. N. Litvinenko, “A method of emittance preservation in ERL merging system,” in Proc. 27th Int. Free Electron Laser Conf., 2005, pp. 644-647.

[1.17] J. H. Billen and L. M. Young, “Poisson SUPERFISH,” Los Alamos National Laboratory, Technical Report LA-UR-96-1834, updated 2003.

[1.18] K-J. Kim, “RF and space-charge effects in laser-driven rf electron guns,” *Nucl. Instrum. Meth. A*, vol. 275, pp. 201-218, February 1989.

[1.19] D. H. Dowell, M. Ferrario, T. Kimura *et al.*, “A two-frequency RF photocathode gun,” *Nucl. Instrum. Meth. A*, vol. 528, pp. 316-320, August 2004.

[1.20] D. H. Dowell, T. D. Hayward and A. M. Vetter, “Magnetic pulse compression using a third harmonic RF linearizer,” in *Proc. 1995 Particle Accelerator Conf.*, 1995, pp. 992-994.

[1.21] X. J. Wang, X. Qiu and I. Ben-Zvi, “Experimental observation of high-brightness microbunching in a photocathode rf electron gun,” *Physical Review E*, vol. 54, pp. 3121-3124, October 1996.

[1.22] M. Reiser, *Theory and Design of Charged Particle Beams*, Weinhiem: Wiley-VCH, 2007, pp. 10.

[1.23] J. Rosenzweig, N. Barov, S. Hartman *et al.*, “Initial measurements of the UCLA RF photoinjector,” *Nucl. Instrum. Meth. A*, vol. 341, pp. 379-385, March 1994.

[1.24] M. Reiser, *Theory and Design of Charged Particle Beams*, Weinhiem: Wiley-VCH, 2007, Section 6.2.1.

[1.25] F. Zhou, I. Ben-Zvi, M. Babzien *et al.*, “Experimental characterization of emittance growth induced by the nonuniform transverse laser distribution in a photoinjector,” *Phys. Rev. ST Accel. Beams*, vol. 5, pp. 094203-1–094203-6, September 2002.

[1.26] A. Brachmann, R. N. Coffee, D. H. Dowell *et al.*, “LCLS drive laser shaping experiments,” in *Proc. 2009 Free Electron Laser Conf.*, 2009, pp. 463-465.

[1.27] T. Wangler, *RF Linear Accelerators*, Weinhiem: Wiley-VCH, 1998, pp. 283-285.

[1.28] B. E. Carlsten, “New photoelectric injector design for the Los Alamos National Laboratory XUV FEL accelerator,” *Nucl. Instrum. Meth. A*, vol. 285, pp. 313-319, December 1989.

[1.29] L. Serafini and J. Rosenzweig, “Envelope analysis of intense relativistic quasilaminar beams in RF photoinjectors: a theory of emittance compensation,” *Phys. Rev. E*, vol. 55, pp. 7565-7590, June 1997.

[1.30] M. Ferrario, J. E. Clendenin, D. T. Palmer *et al.*, “HOMDYN study for the LCLS RF photo-injector,” SLAC, Stanford, CA, Technical Report No. SLAC-PUB-8400, March 2000.

[1.31] M. Ferrario, D. Alesini, M. Bellaveglia *et al.*, "Direct measurement of the double emittance minimum in the beam dynamics of the sparc high-brightness photoinjector," *Phys. Rev. Lett.*, vol. 99, pp. 234801-1–234801-5, December 2007.

[1.32] J. F. Schmerge, J. Castro, J. E. Clendenin *et al.*, "The s-Band 1.6 cell RF gun correlated energy spread dependence on π and 0 mode relative amplitude," in Proc. 46th Workshop INFN ELOISATRON Project, 2005, pp. 375-382.

[1.33] M. Reiser, *Theory and Design of Charged Particle Beams*, Weinhiem: Wiley-VCH, 2007, pp. 214.

[1.34] D. C. Cary, K. L. Brown and F. Rothhacker, "Third-order TRANSPORT with MAD input: a computer program for designing charged particle beam transport systems," SLAC, Stanford, CA, Technical Report No. SLAC-Report-530, April 1998, pp. 161.

[1.35] GPT: General Particle Tracer, ver. 2.82, Pulsar Physics.

[1.36] M. Reiser, *Theory and Design of Charged Particle Beams*, Weinhiem: Wiley-VCH, 2007, pp. 106-107.

[1.37] M. Born and E. Wolf, *Principles of Optics: Electromagnetic Theory of Propagation Interference and Diffraction of Light*, 6th Ed., pp. 217, Oxford: Pergamon Press, 1980.

[1.38] D. H. Dowell, E. Jongewaard, J. Lewandowski *et al.*, "The development of the linac coherent light source RF gun," *Beam Dynamics Newslett.*, No. 46, pp. 162-192, August 2008.

[1.39] R. Akre, D. H. Dowell, P. Emma *et al.*, "Commissioning the linac coherent light source injector," *Phys. Rev. ST Accel. Beam*, vol. 11, pp. 030703-1–030703-20, March 2008.

[1.40] L. Serafini, "Short bunch blow-out regime in RF photoinjectors," in *AIP Conf. Proc.*, vol. 413, 1997, pp. 321-334.

[1.41] O. J. Luiten, S. B. van der Geer, M. J. de Loos *et al.*, "How to realize uniform three-dimensional ellipsoidal electron bunches," *Phys. Rev. Lett.*, vol. 93, pp. 094802-1–094802-4, August 2004.

[1.42] L. M. Young, "PARMELA documentation," Los Alamos National Laboratory, Los Alamos, NM, Technical Report No. LA-UR-96-1835 (Revised June 8, 2004).

[1.43] L. M. Young, private communication.

[1.44] J. Qiang, I. V. Pogorelov and R. Ryne, "Recent improvements to the IMPACT-T parallel particle tracking code," in *Proc. 2006 Int. Computational Accelerator Physics Conf.*, 2006, pp. 185-187.

[1.45] J. Qiang, "IMPACT-T User Document Version 1.5," LBNL, Berkeley, CA, Technical Report No. 62326, 2007.

[1.46] K Flöettmann. (2011). ASTRA-Manual_V3 [Online]. Available: FTP: http://www.desy.de Directory: ~mpyflo/Astra_dokumentation/ File: ASTRA-Manual_V3.pdf.

[1.47] C. Nieter and J. R. Cary, "VORPAL – a multidimensional code for simulating advanced accelerator concepts," in *Proc. 2001 Particle Accelerator Conf.*, 2001, pp. 3105-3107.

[1.48] D. P. Rusthoi, W. P. Lysenko and K. R. Crandall, "Further improvements in TRACE 3-D," in *Proc. 1997 Particle Accelerator Conf.*, 1997, pp. 2574-2576.

[1.49] M. Ferrario. (1999). HOMDYN user guide [Online]. Available: FTP: http://nicadd.niu.edu Directory: fnpl/hondyn/ File: manual.pdf.

[1.50] F. C. Iselin, J. M. Jowett, J. Pancin *et al.*, "MAD version 9," in *Proc. 2000 European Particle Accelerator Conf.*, 2000, pp. 1042-1044.

[1.51] The Methodical Accelerator Design Group Home page. Available Online: http://mad.web.cern.ch/mad/.

[1.52] M. Borland, V. Sajaev, H. Shang *et al.*, "Recent progress and plans for the code elegant," in *Proc. 2009 Int. Computational Accelerator Physics Conf.*, 2009, pp. 111-116.

[1.53] F. Ebeling, R. Klatt, F. Karawzcyk *et al.*, "Status and future of the 3D MAFIA group of codes," in *Proc. 1988 European Particle Accelerator Conf.*, 1988, pp. 279-281.

[1.54] S. Setzer, R. Cee, M. Krassilnikov *et al.*, "FEL photoinjector simulation studies by combining MAFIA TS2 and ASTRA," in *Proc. 2002 European Particle Accelerator Conf.*, 2002, pp. 1664-1666.

[1.55] MicroWave Studio, ver. 4.2, available by Computer Simulation Technology GmbH, Bad Nauheimer Str. 19, 64289 Darmstadt, Germany. Available at www.cst.com/Content/Products/MWS/Overview.aspx.

Appendix 1.1: Useful Formulae

| Emittance type | Emittance formula |
|---|---|
| Intrinsic (a.k.a. thermal) | $\varepsilon_{n,\text{intrinsic}} = \beta\gamma\sqrt{\frac{\hbar\omega - \phi_{eff}}{3mc^2}}$ |
| Surface Roughness (High Field Enhancement) | $\varepsilon_{n,field} = \sigma_x \frac{\left\langle v_x^2 \right\rangle^{1/2}}{c} = \sigma_x \sqrt{\frac{\pi^2 a_n^2 eE_a}{2\lambda_n mc^2}}$ |
| Space Charge Emittance for Gaussian Distributions | $\varepsilon_{n,sc} = \frac{\pi}{4}\frac{1}{\alpha k_z \sin\phi_0}\frac{I}{I_0}\frac{1}{3A_{gaus}+5}$ <br> $\text{A}_{gaus} = \frac{\sigma_x}{\sigma_z}$ |
| Space Charge Emittance for Uniform Cylindrical Distribution | $\varepsilon_{n,sc} = \frac{\pi}{4}\frac{1}{\alpha k_z \sin\phi_0}\frac{I}{I_0}\frac{1}{35\sqrt{A_{cyl}}}$ <br> $\text{A}_{cyl} = \frac{a}{l}$ |
| Non-uniform Emission Space Charge Emittance | $\varepsilon_{n,sc}(\Delta I_{bunch}) = \sigma_x \frac{2}{n_s}\sqrt{\frac{2}{\pi}\frac{\Delta I_{bunch}}{I_0}}$ |
| 1st-order RF emittance | $\varepsilon_{n,1rf} = \frac{eE_0}{2mc^2}\left\|\cos\phi_e\right\|\sigma_{x,e}^2\sigma_\phi$ |
| 2nd-order RF emittance | $\varepsilon_{n,2rf} = \frac{eE_0}{2\sqrt{2}mc^2}\left\|\sin\phi_e\right\|\sigma_{x,e}^2\sigma_\phi^2$ |
| Geometric Emittance | $\varepsilon_{n,geo} = 0.0046\left(\mu m / mm^4\right)\sigma_{x,sol}^4$ |
| Chromatic Emittance | $\varepsilon_{n,chromatic}(\sigma_{x,sol}) = \beta\gamma\sigma_{x,sol}^2 K\left\|\sin KL + KL\cos KL\right\|\frac{\sigma_p}{p}$ |
| Anomalous Quadrupole Field Emittance | $\varepsilon_{n,quad-sol}(\sigma_{x,sol}) = \beta\gamma\sigma_{x,sol}^2\left\|\frac{\sin 2KL}{f_q}\right\|$ |

Appendix 1.2: Mathematical Symbols

| Symbol | Definition | Value |
|---|---|---|
| $\alpha$ | Electric Field Parameter, $\alpha \equiv \dfrac{eE_0}{2mc^2k_z}$ | - |
| $\beta$ | Velocity divided by Speed of Light, $\beta = \dfrac{v}{c}$ | - |
| $c$ | Speed of Light in Vacuum, $c = \dfrac{1}{\sqrt{\varepsilon_0\mu_0}}$ | $2.99\times10^8$ m/s |
| $\delta_{skin}$ | Skin Depth of the Transverse RF Field, $\delta_{skin} = \sqrt{\dfrac{2}{\sigma_{wall}\,\mu_0\,\omega_{RF}}}$ | - |
| $\varepsilon_0$ | Electric Permittivity of Vacuum | $8.85\times10^{-12}$ C/V-m, $5.526\times10^7$ e/V-m |
| $\mu_0$ | Magnetic Permeability of Vacuum | - |
| $\gamma$ | Total Energy Normalized to the Electron Rest Mass | - |
| $m$ | Rest Mass of the Electron | 0.511 MeV/$c^2$ |
| $\phi_{eff}$ | Effective Work Function for Photoemission, $\phi_{eff} = \phi_W - \phi_{schottky}$ | |
| $\phi_W$ | Material Work Function for Photoemission | ~4.6 eV for Cu |
| $\phi_{Schottky}$ | Schottky Work Function due to Image Charge and Field, $\phi_{schottky} = 3.7947\times10^{-5}\sqrt{E(V/m)}$ eV | - |
| $\varepsilon_n$ | Normalized Emittance | - |
| $\sigma_x$ | Transverse rms Beam Size | - |
| $\sigma_{wall}$ | Conductivity of Cavity Walls | - |
| $\upsilon_x$ | Velocity along the *x*-coordinate | - |
| $\sigma_z$ | Longitudinal rms Beam Size | - |
| $\sigma_{x,e}$ | Transverse rms Beam Size at Exit of RF Gun | - |
| $\sigma_{x,sol}$ | Transverse rms Beam Size in the Solenoid | - |
| $\sigma_p$ | Bunch rms Momentum Spread | - |
| $\sigma_\phi$ | rms Phase Length of Bunch | - |
| $\phi_0$ | Initial Launch Phase of the Electron relative to the RF Waveform | - |
| $\phi_e$ | Beam RF Phase when Bunch is at Exit of Gun | - |
| $\omega$, $\omega_{rf}$, $\omega_{RF}$ | RF Angular Frequency, $\omega = 2\pi f$ | - |
| $p$ | Bunch Average Momentum | - |
| $a_n$ | Amplitude of $n^{th}$ Spatial Frequency of the Surface Roughness | - |
| $e$ | Electron Charge | $1.6\times10^{-19}$ C |
| $E_a$ | Applied or External Electric Field, usually RF or HV DC. | - |
| $E_0$ | Peak Field at Cathode | - |
| $\lambda_n$ | Spatial Wavelength of Surface Modulation with Wave Number $k_n$ | - |

| **Symbol** | **Definition** | **Value** |
|---|---|---|
| $k_z$ | Longitudinal RF Wave Number, $k_z = \frac{p\pi}{l}$, where $l$ is the Cavity length; also, $k_z = \frac{2\pi}{\lambda_{rf}}; c = \lambda_{rf} f_{rf}$ for Standing Wave Cavity | - |
| $E_K$ | Kilpatrick Criterion Peak Field | - |
| $I$ | Beam Current, usually the Peak Current | - |
| $I_0$ | Characteristic Current for Electrons, $I_0 = ec\, r_e^{-1} \approx 17$ kA | - |
| $A_{gaus}$ | Aspect Ratio for Gaussian Bunch Shape, $\mathrm{A}_{gaus} = \frac{\sigma_x}{\sigma_z}$ | - |
| $A_{cyl}$ | Aspect Ratio for a Uniform Cylinder Bunch, $\mathrm{A}_{cyl} = \frac{a}{l}$ | - |
| $a$ | Radius of Cylindrical Bunch | - |
| $l$ | Length of Cylindrical Bunch | - |
| $\Delta I_{bunch}$ | difference of max and min local current of transverse spatial modulation | - |
| $n_s$ | Number of Spatial Modulations across the Beam Diameter | - |
| $K$ | Focal Strength of Solenoid, $K = \frac{eB_{sol}}{2p} = \frac{B_{sol}}{2(B\rho)_0}$ | - |
| $L$, $L_{sol}$ | Solenoid Magnetic Length | - |
| $f_q$ | Quadrupole Focal Length, usually for Solenoid Quadrupoles | - |
| $f_{RF}$, $f_{rf}$ | Radiofrequency (RF) in Hertz | - |
| $Q$ | Quality Factor of Resonant System, $Q = \frac{(Stored\ Energy)}{(Dissipated\ Energy)}$ | - |
| $r_{shunt}$ | Shunt Impedance for RF Power and Cavity Voltage, $r_{shunt} = \frac{V_0^2}{P}$ | - |
| $V_0$ | Cavity Voltage | - |

# CHAPTER 2: NORMAL CONDUCTING RF INJECTORS

DAVID H. DOWELL

*SLAC National Accelerator Laboratory*

*Menlo Park, CA 94025-7015*

**Keywords**

RF Cavity, Pillbox Cavity, Reentrant Cavity, Racetrack Cavity, Side Coupling, Thermal Management, Emittance Compensation, Wakefield, L-band Gun, S-band Gun, 144 MHz Gun, 1.3 GHz Gun, 433 MHz Gun, 700 MHz Gun, 187 MHz Gun

**Abstract**

This chapter provides engineering details for normal conducting RF (NCRF) photoinjectors. The basic gun design consists of a conducting cavity shaped to resonate at a desired RF frequency with a port for coupling RF power into the cavity. The two cavity shapes commonly used are pillbox and reentrant. And the two basic coupling schemes to pulse heating used are side-coupled and coaxial coupling. Various technical aspects of these four options in a gun design are discussed in some detail. The few sections describe the topics of the gun's thermodynamics related to pulse heating in a low duty factor gun and average power thermal management in a CW gun. The RF field mode purity and spatial linearity are discussed and the effect on beam quality is quantified. In addition to the gun, the emittance compensation solenoid, bucking coil and diagnostics after the gun are described. Engineering details of beamline components which minimize the emittance growth due to wakefields are presented. The chapter ends with descriptions of photoinjectors which have been built and successfully operated since 1985, when the NCRF gun was invented.

## 2.1 INTRODUCTION

The need for high brightness electron beams has driven the development of normal conducting RF (NCRF) gun theory and practice. The combination of high accelerating field and the fast switching of photoemission have made these high quality beams possible. In addition, emittance compensation has been shown to be a powerful technique for eliminating the emittance from linear space charge forces. This rapid progress in the electron source technology was driven by the invention of applications, such as the free electron laser, Compton sources and ultra fast electron diffraction. This chapter describes the current state-of-the-art technology for the two classes of NCRF guns: 1) High gradient guns operating at low duty factor, and 2) lower gradient, high duty factor guns. Here, high gradient implies 100 MV m$^{-1}$ or more cathode field and high duty fact means greater than ~10%.

NCRF guns operations span a broad range of RF frequencies ranging from 144 MHz to 30 GHz. Table **2.1** compares different NCRF guns and various properties each have, such as their operating RF frequency, peak cathode fields and duty factors. In general, the higher the duty factor, the lower the RF frequency. On the other hand, the higher frequency guns are capable of much higher peak cathode fields and hence brighter beams.

The peak field and average power of NCRF guns are commonly limited by field breakdown, pulsed heating and average power dissipation. Since the beam quality improves with higher cathode field, NCRF guns operate at the highest possible field strengths.

The RF frequency dependence of the limiting field is given by the Kilpatrick criterion [2.1], [2.2].

$$f_{RF}\,[\text{MHz}] = 1.64\, E_K^2 \exp\left(\frac{-8.5}{E_K}\right) \tag{2.1}$$

where $E_K$ is the Kilpatrick criterion peak field in MV m$^{-1}$. Figure **2.1** is a plot of the Kilpatrick criterion field versus the RF frequency. The peak cathode fields which guns have achieved are plotted versus their RF frequency in the same graph in Figure **2.1**. The operational gun fields are seen to be approximately a factor of two higher than the Kilpatrick limit above 1 GHz.

| **RF Frequency Cathode Cavity Shape [Reference]** | **Peak Cathode Field [MV/m]** | **Pulse Format** | | |
|---|---|---|---|---|
| | | **Macropulse Frequency** | **Micropulse Frequency** | **Macropulse Length** |
| 144 MHz Reentrant [2.3] | 25 | 10, 20 Hz | 14.4 MHz | 200 μs |
| 188 MHz Reentrant [2.4] | 20 | Continuous Wave | 1 MHz | (CW) |
| 433 MHz Reentrant [2.5] | 25 | 30 Hz | 27 MHz | 8300 μs |
| 800 MHz Pillbox [2.6] | 5 | CW | 800 MHz | CW |
| 1.3 GHz Pillbox [2.7], [2.8] | 40, 60 | 10 | 42 MHz | 100 μs |
| 2.856 GHz Pillbox [2.9] | 120 | 120 Hz | single bunch at 120 Hz | 2 μs |
| 12 GHz Pillbox [2.10] | 200 | 120 Hz | 120 Hz | 10 ns |
| 17 GHz Pillbox [2.11] | 150 | 10 Hz | 10 Hz | 100 ns |

**Table 2.1. Properties of NCRF guns operating from 144 MHz to 17 GHz.**

Dark current is another phenomenon which limits the field strength in guns and accelerators. It occurs at high surface field strength and electrons are produced by field emission from small imperfections in the surface or from particulates on the surface. Since the electrons are emitted at the high field portion of every RF period, this is a serious issue especially for long pulse, high duty factor guns. Besides fabricating a smooth surface, it is also important to clean all RF surfaces of small dust particles. This is an important issue in superconducting RF structures where the technique of high-pressure water rinsing is used. However, recent work shows that dry-ice cleaning is effective at removing hydrocarbons and particles without leaving the surface wet. [2.12] An impressive factor of 10 reduction in the dark current emission has been observed in the PITZ L-band NCRF gun. [2.13]

Since the beam needs to be rapidly accelerated from the cathode, it is standard for RF guns to use the standing wave $TM_{011}$ longitudinal electric mode as given by Equ. 1.8 through Equ. 1.10 in Chapter 1. The $TM_{011}$ mode is called the π-mode because of the 180˚ phase shift between adjacent cells. A gun has as many longitudinal modes as cells even if they are partial cells, for example, the 1.6-cell gun has 0- and π-modes. Because each mode will have a different spatial shape and RF frequency than the π-mode, these other modes can dynamically unbalance the gun fields and introduce additional emittance. This effect can be both detrimental and beneficial. For example, the introduction of a 3$^{rd}$-harmonic to the fundamental RF field can both eliminate the 2$^{nd}$-order RF emittance, as well as straighten the electron energy-time correlation in longitudinal phase space. The use of other harmonic TE modes may provide additional RF focusing, reducing the need for a magnetic solenoid. Details of these modes are discussed later in this chapter.

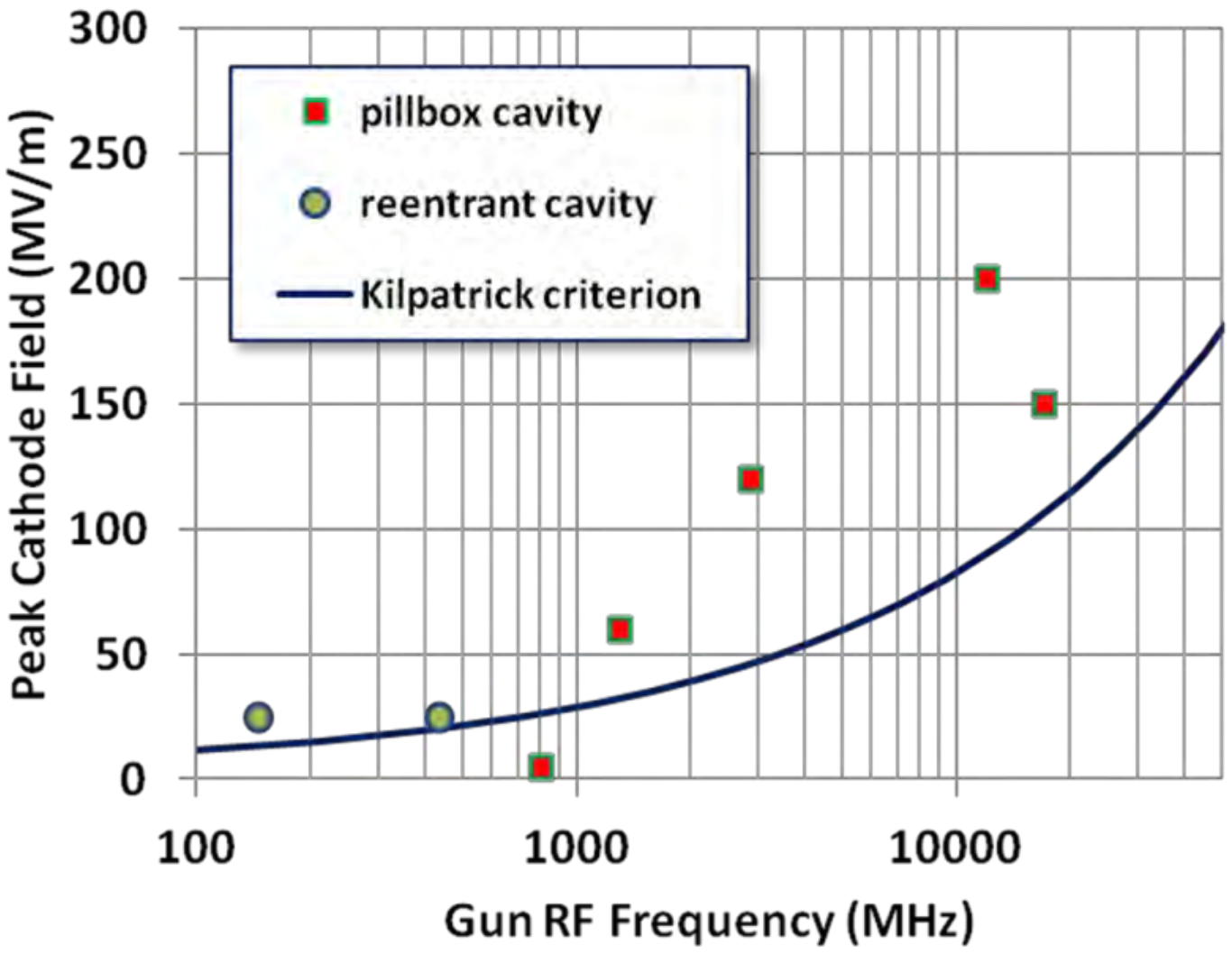

**Figure 2.1. Comparison of the frequency dependence of peak cathode fields achieved in operating NCRF guns with the Kilpatrick criterion.**

As discussed in Section 1.3.1, there are two basic interior shapes for RF guns: Pillbox and reentrant. These shapes are shown in Figure 1.5 of Chapter 1 and in Figure **2.2**. The pillbox cavity is relatively easy to fabricate, can operate at high electric fields for high RF frequencies and has less radial field distortion compared to the reentrant cavity. The reentrant shape has high shunt impedance, but larger non-linear radial fields. Generally, the pillbox is used in low duty factor guns which run at very high peak powers to produce high cathode field. The reentrant shape is used in high duty factor guns, since it produces higher fields per megawatt of RF power and has a large surface area for cooling.

This chapter ends with sections describing the two basic types of NCRF guns in terms of their duty factor, low or high. For the purposes of this discussion, we define high duty factor as a tens of percent to CW and low duty is everything lower. In general, high duty factor guns are used in Energy Recovered Linacs (ERLs) which operate in CW modes, while low duty factor guns are used for single pass linacs. The last sections of this chapter will describe the common features of these two operating regimes and provide examples of historical and current injector technology. The discussion on beam optics and emittance preservation is applicable to both gun types.

## 2.2 GUN CAVITY SHAPE AND RF COUPLING SCHEMES

### 2.2.1 Pillbox and Reentrant Cavity Shapes

As described in Chapter 1, guns are designed to have either a pillbox cavity shape or reentrant nose cones and elliptical side walls. These two shapes are shown in Figure **2.2**. Reentrant cavities tend have larger shunt impedances, and therefore require less RF power. The larger area of the reentrant nose cone improves the average power dissipation of the gun and is often used in high average power guns. The dissipated power of a cavity is usually the main factor limiting its duty factor. The disadvantages of the reentrant cavity include a more difficult shape to cool and fabricate, especially at higher RF frequencies. In addition, the reentrant shape produces greater non-linear radial fields compared to the pillbox. Generally, the long "tunnel" between cavity nose cones isolates the RF between cells which then requires a separate RF power coupler for each cell as illustrated in the right portion of Figure **2.2**. The power feed for each cell has its RF phase length adjusted to account for the electron's travel time in the relatively long drift length of the tunnel. For this reason, guns using a reentrant shape are configured as either an integrated pair of independently powered reentrant cells or a reentrant cathode cell followed by pillbox cells with a single RF coupling. The typical pillbox shaped cavity consists of closely spaced cells with intervening holes, or irises. These irises are designed to be large enough to allow the coupling of the RF power between cells. This greatly simplifies the waveguide design of the pillbox gun since it needs RF power fed to only one cell. A multi-cell gun using both cell types has been proposed; although in that design, the power is fed to each cell [2.14].

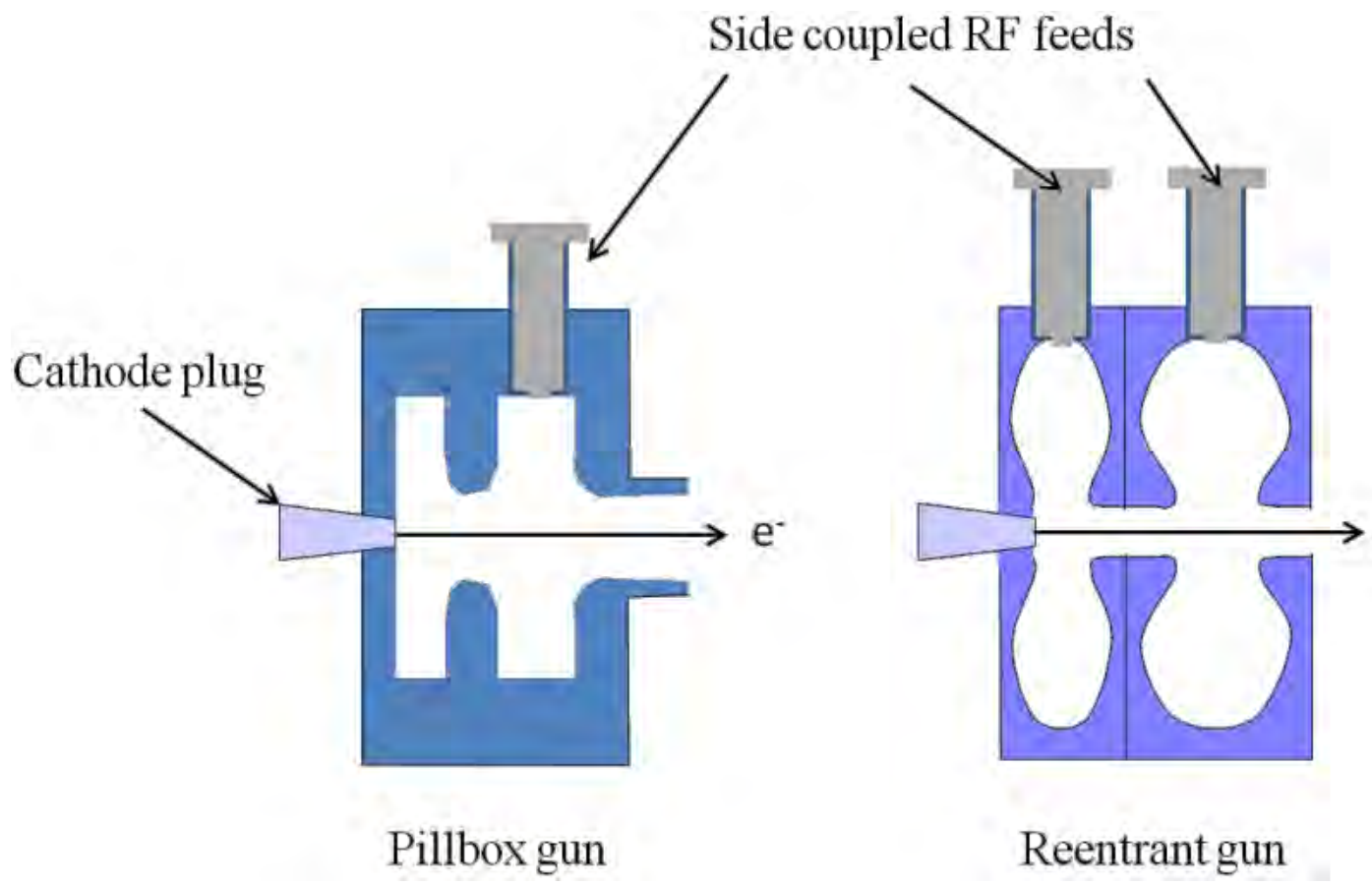

**Figure 2.2. Comparison of pillbox (left) and reentrant (right) cavity shapes for gun cavities. Because the pillbox cells are strongly coupled they can be powered by a RF feed on just one of the cells. For the reentrant shape the cell-to-cell coupling is small and an RF feed is needed on each cell.**

The RF power, $P$, it takes to accelerate electrons to an voltage gain of $V_0$ is

$$P = \frac{V_0^2}{r_{shunt}} \tag{2.2}$$

Here, the gun is considered to be a load with impedance $r_{shunt}$.

The RF quality factor, $Q$, is defined as the ratio of the stored RF energy to the averaged dissipated energy over one RF cycle. The $Q$ for cavities with the rotational symmetric $TM_{nm0}$ modes is [2.15]

$$Q = \frac{\mu_0}{\mu_c}\left(\frac{V}{S\delta_{skin}}\right) * \text{(geometric factor); where } \delta_{skin} = \sqrt{\frac{2}{\mu_0 \omega_{RF} \sigma_{wall}}} \tag{2.3}$$

where $V$ is the cavity volume, $S$ is its surface area, and $\delta_{skin}$ is the RF skin depth, where $\sigma_{wall}$ is the conductivity of the cavity walls and $\omega_{RF}$ is the RF frequency. The magnetic permeability of the vacuum and the cavity wall conductor are $\mu_0$ and $\mu_c$, respectively. The geometric factor is of order unity for a wide range of cavity aspect ratios and is approximately the same for pillbox and reentrant cavities excited by the same RF mode. The cavity $Q$ can be maximized by increasing the cavity volume while minimizing the surface area. This accounts for the nearly circular shape of reentrant cavities, giving them a larger ratio of volume to surface area than the pillbox. The peak surface field can be as much as 6X larger at the nosecone of a reentrant cavity producing a higher axial field than the pillbox for the same RF power.

Given these advantages, one may ask why all guns aren't built with reentrant shapes. Referring back to Figure **2.1**, the answer lies with the Kilpatrick criterion. Above ~1 GHz, the achieved peak cathode field is limited by RF breakdown of the cavity surfaces rather than by limitations of the power source. The observed fields are all approximately twice the Kilpatrick criterion for these few gigahertz RF frequencies. At these frequencies, the reentrant would breakdown at the same RF field as the pillbox cavity, although using a reentrant shape would require less RF power to attain the same accelerating field. However, there is a loss of energy gain per physical length since there is no acceleration in the long tunnels between reentrant cavities. Therefore, for the same gap fields, a pillbox gun will be shorter than a reentrant gun for the same acceleration. Since many guns in this frequency region are pulsed and operate at very low duty factor, power efficiency is not a consideration.

However, below ~1 GHz, the peak surface field achievable is no longer limited by breakdown, but rather by other practical issues, *e.g.*, excessive heating of the RF couplers or limited RF power. At these lower frequencies, the RF coupler can become problematic, thereby presenting a significant engineering challenge. In these cases, the reentrant shape improves the power efficiency while increasing the peak cathode and on axis field. As shown in Figure **2.1**, the peak cathode field of the pillbox gun at 800 MHz is below the breakdown criterion and the reentrant guns at 144 and 433 MHz have operated at or above the limit. Although it is generally more difficult to design and fabricate the reentrant cavity, the large cavity size at the low frequencies eases the engineering and fabrication tolerances as well as presenting different design options and requirements. For example, the large cavity size often requires placing the focusing solenoids inside the nose cones instead of around the outside of the gun. And since many low frequency guns are designed to operate at much higher duty factors, power efficiency is important.

### 2.2.2 Gun RF Theory

The typical layout of the gun RF power flow is shown in Figure **2.3**. The system consists of a RF source connected with a waveguide to a RF coupler. Depending upon the frequency and power levels the coupler can be an antenna probe, a coaxial line or, as shown in the figure, a side coupled waveguide. The RF source power generates the forward power, $P_{fwd}$, with some going into the gun, $P_{cav}$, and some being reflected as reverse power, $P_{rev}$. The total source power is the sum of the cavity and reverse powers, $P_{fwd} = P_{cav} + P_{rev}$.

The RF power enters the gun cavity through a coupling hole or port with a waveguide-cavity coupling parameter $\beta_{coupler}$ defined as

$$\beta_{coupler} \equiv \frac{P_{external}}{P_{cav}} \tag{2.4}$$

Here, $P_{external}$ is the rate the stored energy in the cavity radiates back into the waveguide when the RF source is turned off. $P_{cav}$ is the steady-state power flowing into the gun. Steady-state is defined as when the fields equilibrate and become time-independent. In steady-state, $P_{cav}$ equals the heat dissipation in the cavity walls and the RF power reflected back to the RF source, $P_{rev}$. The reverse (a.k.a. reflected) power is given by

$$P_{rev} = P_{fwd}\left(\frac{\beta_{coupler} - 1}{\beta_{coupler} + 1}\right)^2 \tag{2.5}$$

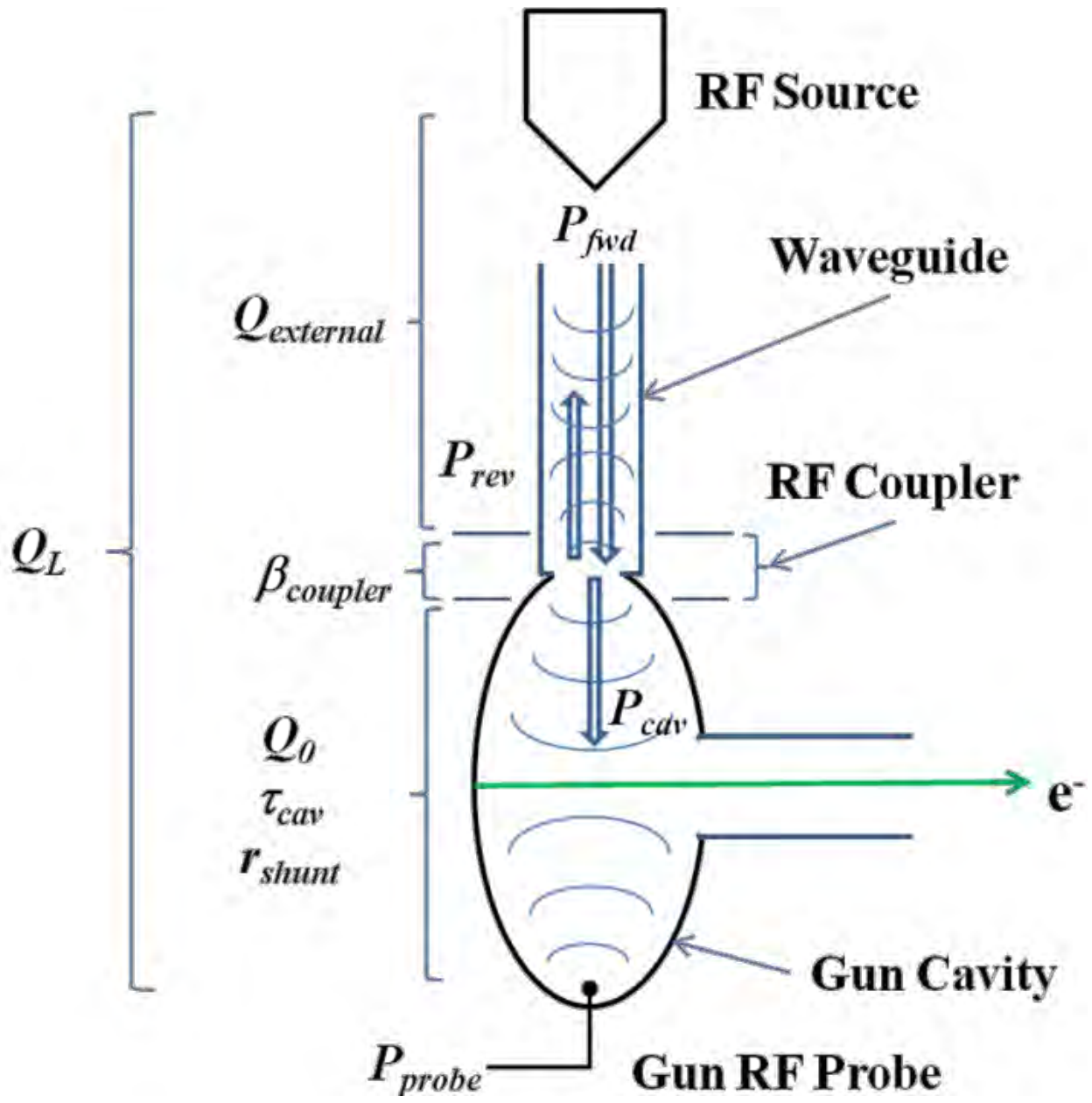

**Figure 2.3. Schematic layout of a gun showing the RF power flow and gun and waveguide parameters.**

The external and cavity RF powers depend upon the external and cavity quality factors, which are defined as $Q_{external}$ and $Q_0$ respectively. For a cavity with a stored energy of $U$ and a resonant frequency, $\omega_0$, the external power is given by

$$P_{external} = \frac{\omega_0 U}{Q_{external}} \tag{2.6}$$

and similarly for the steady-state power radiating into the cavity

$$P_{cav} = \frac{\omega_0 U}{Q_0} \tag{2.7}$$

With these definitions, the coupling parameter can be written in terms of the two quality factors

$$\beta_{coupler} = \frac{Q_0}{Q_{external}} \tag{2.8}$$

The quality factor of the entire system is called the loaded-$Q$, $Q_L$. In terms of the other quality factors the loaded-$Q$ is given by

$$\frac{1}{Q_L} = \frac{1}{Q_0} + \frac{1}{Q_{external}} \tag{2.9}$$

It can be shown using a resonant circuit model driven at its resonant frequency that the gun's response time to changes in the RF is the $1/e$ response time, $\tau_{cav}$, given as

$$\tau_{cav} = \frac{2}{\omega_0}\frac{Q_0}{1+\beta_{coupler}} \tag{2.10}$$

When the fields reach steady-state the cathode field can be related to the forward power using the shunt impedance, the gun effective length, $l_{gun}$, and the surface field ratio, $R_{field}$

$$E_{cathode} = R_{field}\left(\frac{\sqrt{P_{fwd}\, r_{shunt}}}{l_{gun}}\right) \tag{2.11}$$

The surface field ratio accounts for any enhancement of the cathode surface compared to the average accelerating field. This enhancement factor is determined by the cavity shape and as discussed earlier is larger for a reentrant cavity than it is for a pillbox. The voltage gain (see Equ. 2.2) is divided by the gun's effective length to give the cathode electric field. For example, the 1.6-cell, S-band gun the shunt impedance is 3.6 MΩ, the effective length is 8.68 cm and the surface field ratio is ~1.8. These parameters can be found for a specific gun design by using a field solving code such as SUPERFISH [2.16].

It is important to stress that the above relations apply only when the gun fields have reached their steady-state values, that is when $t >> \tau_{cav}$. When the RF power is initially turned on there are transients whose time-dependence is determined by the gun's response time often called the cavity fill time. Figure **2.4** shows measurements of the forward, reverse and gun probe powers for an S-band gun being driven by a 1.2 μs long RF pulse. This pulse is just long enough to fill the gun and launch a single electron bunch. This pulse structure minimizes the average power dissipation.

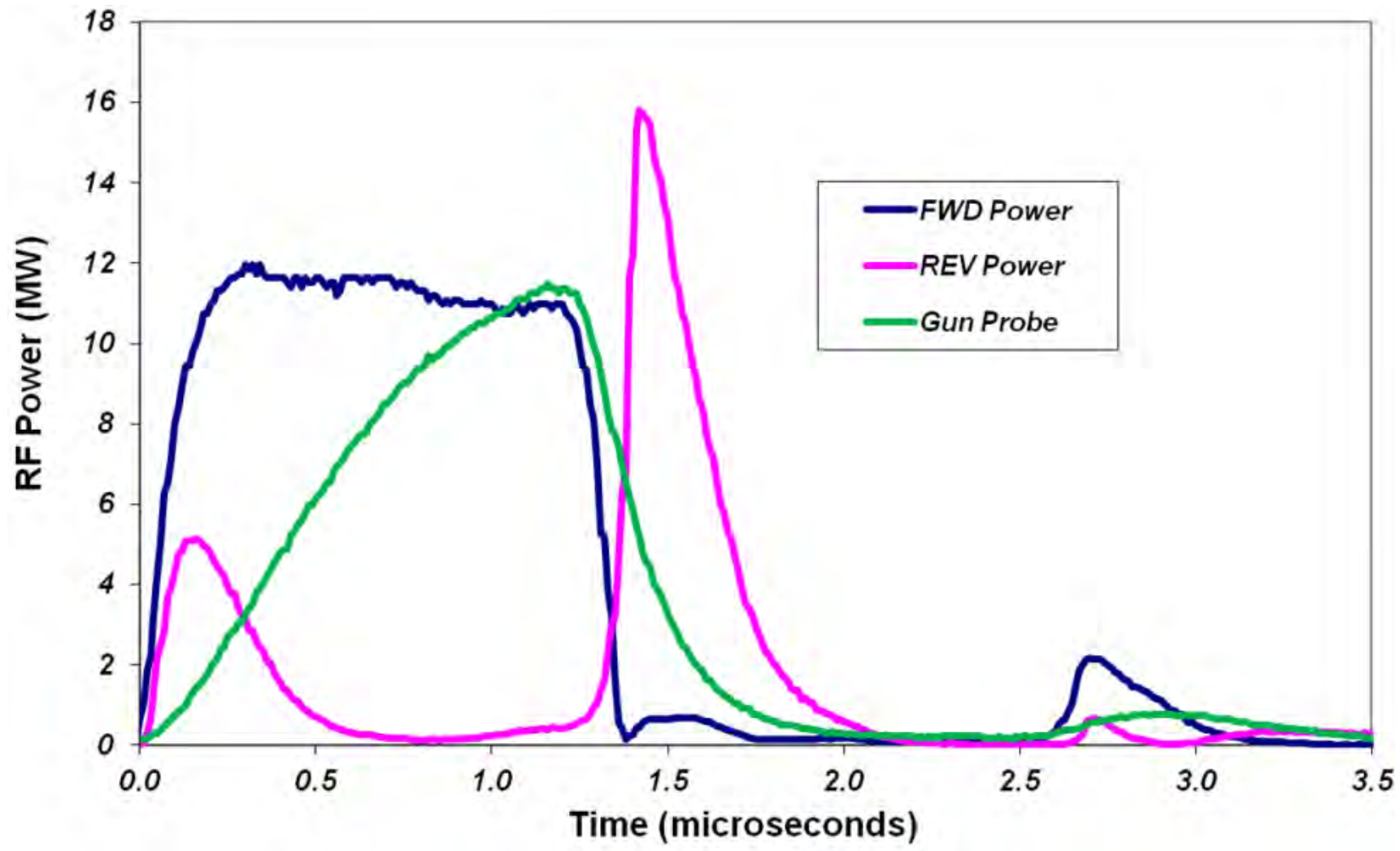

**Figure 2.4. RF waveforms measured during high power operation of an S-band gun. The small pulses near 2.7 μs are the reflections of the RF pulse from the klystron.**

The behavior of the reverse RF is determined by the voltage reflection coefficient, $\Gamma$, defined in terms of the forward and reverse voltages, respectively $V_{fwd}$ and $V_{rev}$ [2.17]

$$\Gamma(t) = \frac{V_{rev}(t)}{V_{fwd}} = (1 - e^{-t/\tau_{cav}})\,\frac{2\beta_{coupler}}{1+\beta_{coupler}} - 1 \tag{2.12}$$

This reflection coefficient produces a time-dependent reflection of power back toward the RF source given by the forward power times the voltage reflection coefficient squared,

$$P_{rev}(t) = P_{fwd}\,\Gamma(t)^2 \tag{2.13}$$

Consider Equ. 2.12 and Equ.2.13 at short and long times. At the start of the RF pulse, $t$ = 0, all the forward power is reflected independent of the coupling coefficient. For times long compared to $\tau_{cav}$ the system is in steady-state and the reflected power is given by Equ. 2.5 which has three regimes of interest. If $\beta_{coupler} < 1$, then the system is said to be undercoupled and there is always power reflected by the coupler. For $\beta_{coupler} = 1$, the system is critically coupled and the reflection becomes zero after several cavity times. If $\beta_{coupler} > 1$, then the system is overcoupled and after some time the reverse power goes first to zero and then at longer times increases again to the value given by Equ. 2.5. The measurements shown in Figure **2.4** are for an overcoupled gun where the reverse power becomes zero near 0.7 µs and then rises to its steady-state value.

At the end of the RF pulse, the forward power becomes zero and the reverse power immediately rises as the cavity's stored energy radiates back into the waveguide. The ratio of the radiated power in the reverse wave to the cavity power is

$$\frac{P_{rad}}{P_{cav}} = \left(\frac{2\beta_{coupler}}{1+\beta_{coupler}}\right)^2 (1 - e^{-t/\tau_{cav}})^2 \qquad (2.14)$$

For the S-band (2.856 GHz) gun data shown in Figure **2.4** the coupling coefficient is 2 and the $Q_0$ is 13 900, thus cavity time constant as 0.516 µs. Using these values in Equ. 2.14 gives $P_{rad}\, P_{cav}^{-1}$ = 1.45 when the pulse ends at $t$ = 1.2 µs. Since, at the end of the pulse $P_{fwd} \approx P_{cav}$, one sees in Figure **2.4** that the reverse power jumps to 1.4X $P_{fwd}$ in agreement with Equ. 2.14.

The reverse power reflects back and forth between the gun and the RF source producing distortions in $P_{external}$ and $P_{cav}$. These reflections can result in large variations in the phase and amplitude during the RF macropulse, and under some conditions constructively interfere with the forward RF wave to create locations of high field which electrically breakdown. In addition, a strong reflection back into the klystron can disrupt its output. For these reasons, it is usually necessary to protect the RF source from the reflected power. This can be done using a device called a RF isolator or a very long waveguide, in which the RF pulse ends before the reflection can travel back to the RF source. The topic of RF isolation is discussed again later in this chapter.

### 2.2.3 RF Coupling Schemes

There are three geometries which are commonly used for coupling the gun to the waveguide. These are: 1) Waveguide coupled to the cavity side wall; 2) coaxial RF feed using a door-knob mode transformer; and, 3) coaxial transmission cable and inductive loop coupling. The engineering details of RF fundamental power-couplers are discussed in Chapter 10, Section 10.4.

The most commonly used coupling is waveguide side coupling which has been used over the full frequency range RF guns have operated, *i.e.*, from 144 MHz to 17 GHz. A disadvantage of side coupling is that it induces a radial asymmetry in the axial RF fields which can increase the beam emittance. The RF field asymmetries can be minimized by using symmetric ports and a racetrack shape for the coupled cell as shown in Figure **2.5(a)**. The dipole and quadrupole field distortions on the beam axis are corrected by using dual feed ports and a racetrack cavity shape. The drawing shows the gun with both dual feeds and a horizontally wider coupler cell, a.k.a. a racetrack shape, to produce rotation symmetric RF fields [2.18].

The second coupler geometry is a coaxial coupler designed to maintain full rotational symmetry and has been successfully used in guns at 1.3 and 3 GHz. The design for an L-band gun is shown in Figure **2.5(b)**.

The coaxial coupler maintains rotation symmetry by having no asymmetric penetrations into the gun cavities. This design requires transforming the waveguide mode into a radial coaxial mode with a "door-knob" mode converter. Chapter 10, Section 10.4 provides more engineering details of the waveguide vacuum windows, low voltage breakdown (multipacting) and other practical considerations for the coaxial waveguide coupler design.

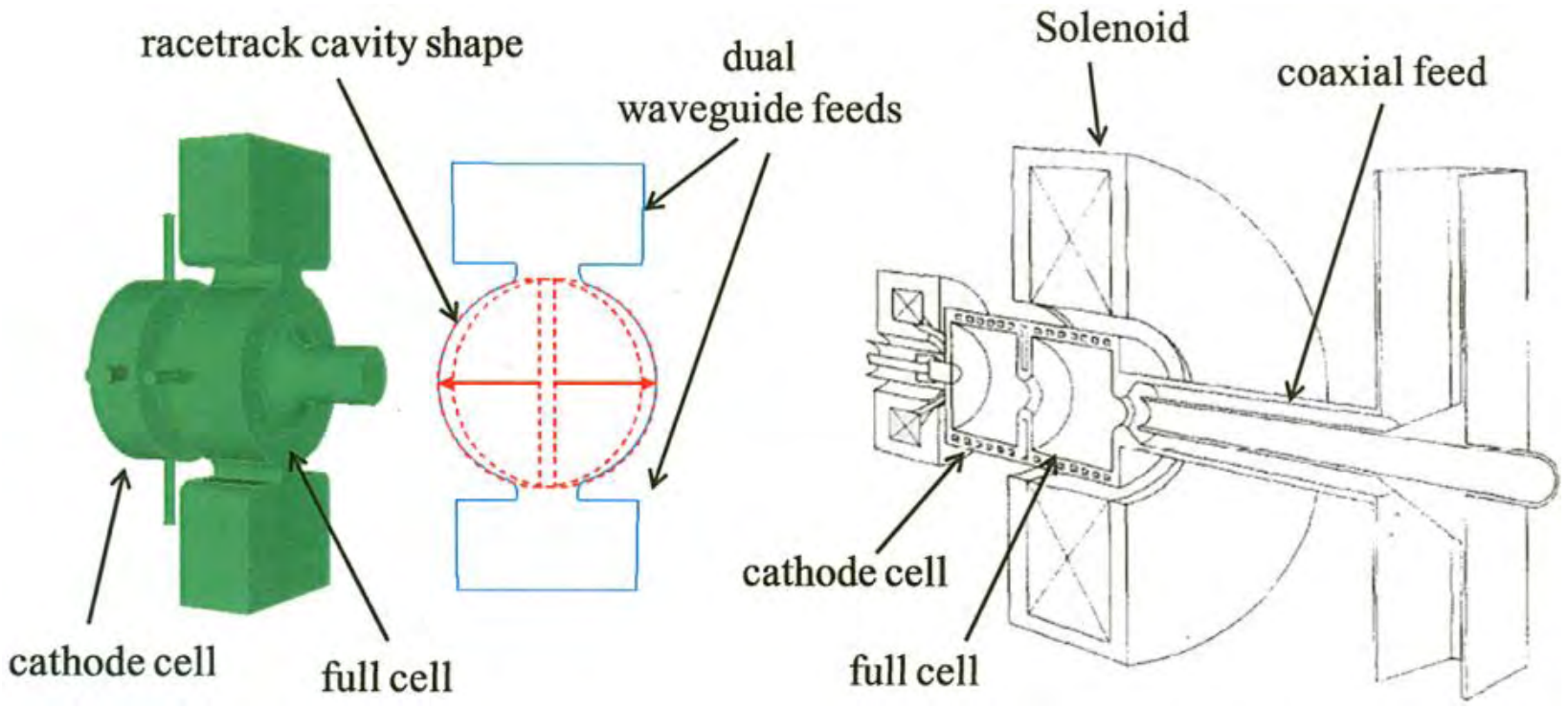

**Figure 2.5. The side-coupled (a) and coaxial (b) feed geometries for symmetrically coupling the RF power into the gun. The solenoids for the side-coupled gun are not shown. [2.19] [[2.9]; Adapted under Creative Common Attribution 3.0 License (www.creativecommons.org/licenses/by/3.0/us/) at www.JACoW.org.] [Adapted from [2.20], with permission from Elsevier.]**

The third coupler geometry consists of a coaxial cable connected to a loop antenna in the cavity. This approach is appropriate for low RF frequencies and is limited to lower levels of power than the other two methods. An example of coaxial lines feeding dual RF loop couplers in a 188 MHz gun is shown in Figure **2.6**. In this single cavity gun two couplers transmit RF into a cavity. Two loop antennae are used because the required power is more than a single loop can handle. Since the couplers are located far from the beam axis, they have negligible influence on the RF field symmetry.

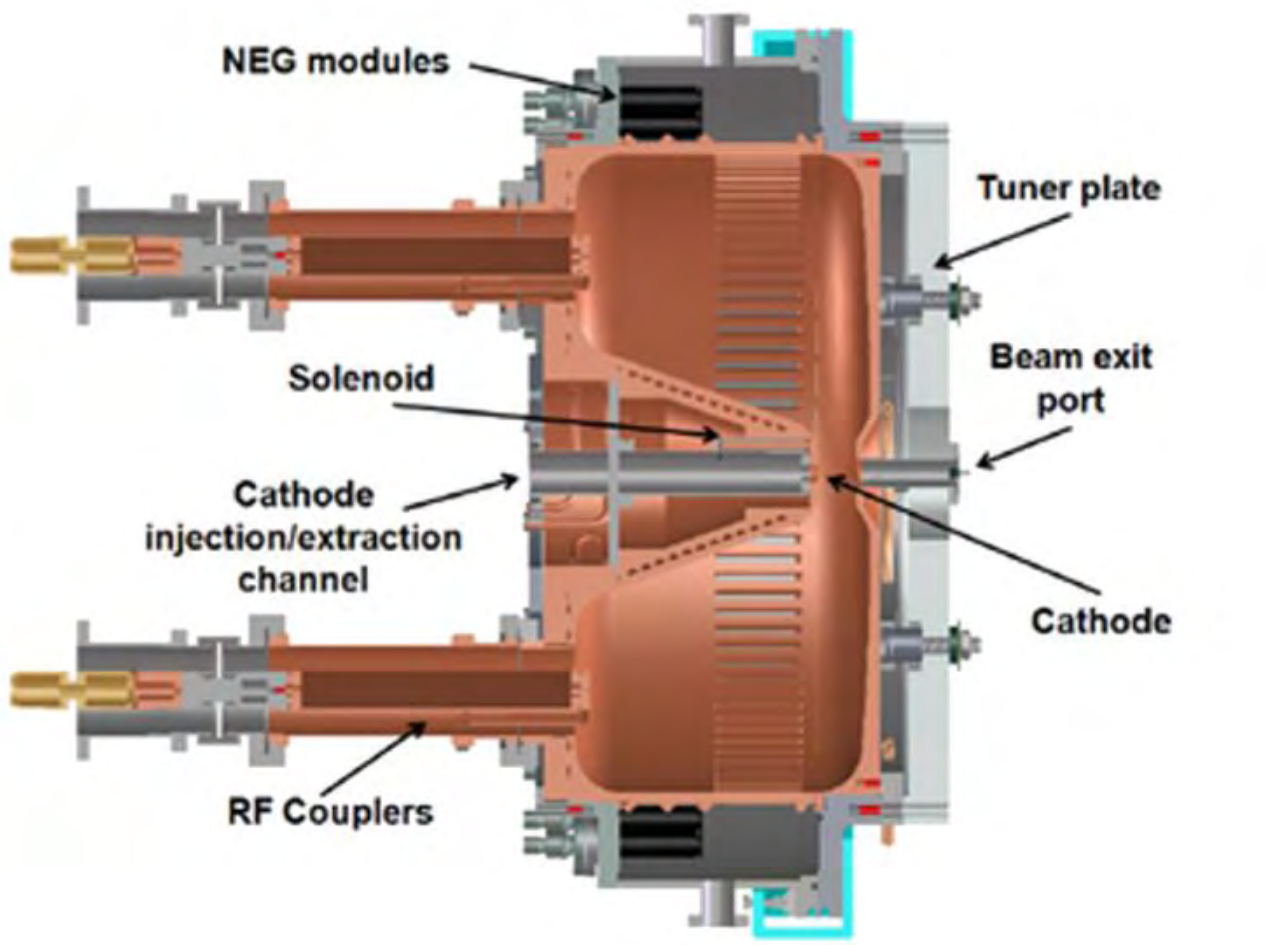

**Figure 2.6. Coaxial cable feed and loop coupling for the LBNL 188 MHz gun cavity. [[2.21]; Available under Creative Common Attribution 3.0 License (www.creativecommons.org/licenses/by/3.0/us/) at www.JACoW.org.]**

## 2.3 GUN THERMODYNAMICS AND RF POWER SYSTEM

While the gun resonant frequency is determined by the cavity shape, the stability of this frequency and the RF amplitude and phase is determined by the gun temperature. Since it is necessary to operate with a high cathode field, and in some cases high average power, there is significant heating of the cavity walls. This heating is especially problematic around the penetrations for the RF couplers, the RF probes and laser ports.

This heating manifests itself on two separate time scales, pulsed heating and average power heating. Generally, pulsed heating is confined to the cavity surfaces and occurs for RF pulses short compared to the mechanical response time of the cavity material. Average power heating affects the overall temperature of the gun. The RF power and cooling system are responsible for keeping the gun at resonance and stabilizing the phase and amplitude of the fields. This section discusses the gun's thermal design and RF control system.

### 2.3.1 Thermal Management of High Power RF

RF heating of the gun can be classified by the two time scales of pulsed and average power heating. Pulsed heating occurs during the RF macropulse that can range from sub-microseconds to milliseconds on a time scale which is short compared to the time it takes the material to mechanically expand. Between RF pulses, the heat diffuses away from the thin RF skin depth and the temperature falls until the next RF pulse. Thus, there are rapid cycles of expansion and contraction of the cavity surface. If the thermal stress induced by each RF pulses exceeds the elastic limit of the cavity material, then microcracks will form at the grain boundaries. With continued pulsing of the gun these microcracks can enlarge and propagate ultimately becoming sites for RF breakdown [2.22].

For guns operating in the long macropulse and/or high repetition rate regime, average heating becomes an issue. What is considered high average heating depends upon the RF frequency. The low RF frequency gun can dissipate higher average power than higher frequency guns. This is because the larger surface area reduces the surface power density making cooling easier. For NCRF guns the RF surface resistant will vary as $\sqrt{f_{RF}}$ and an acceptable surface heat density is about 20 W cm$^{-2}$. At very high average power the cavities will detune and reject the RF power, or if the gun is kept in tune the thermal stresses between different parts of the gun exceed the elastic limit and the mechanical interfaces can become distorted and fail.

RF heating at a penetration occurs where the wall current is deflected around holes and bumps in the cavity walls. These surface deformities locally increase the surface resistance to make hot spots. Since this heating is localized within the skin depth of the surface (a few micrometers), it is difficult to cool, especially for pulsed heating. The pulsed heating around penetrations can be mitigated by reducing the curvature of the surfaces and minimizing the number of penetrations.

The heat load on the gun walls will raise the gun body temperature, causing it to expand and shift the resonant frequency. If the gun is not designed and built with rotational symmetry, then it will flex asymmetrically at the higher operating temperatures and produce asymmetric RF fields which distort the beam. Thermo-mechanical analysis of the design using 3-D codes, such as ANSYS, and 4-D codes, like MAFIA, is an important aspect of most gun studies. Transient and steady-state simulations of the gun body temperature distribution and the mechanical stresses are typically performed.

As first example, we present thermodynamic calculations comparing two S-band (2.856 GHz) gun designs both operating at 4 kW of average power. This average power corresponds to a peak cathode field of 140 MV m$^{-1}$ and a 120 Hz pulse repetition rate with 3 μs long RF pulses. These are the requirements of the gun for an X-ray light source user facility.

We compare the steady-state surface temperature distributions of the BNL/SLAC/UCLA gun ( Figure **2.7(a)**) and the LCLS gun (Figure **2.7(b)**) operating at 4 kW average power dissipation. The temperature distribution for the BNL gun indicates the cathode is hot at 90 ˚C. Thus, there will be considerable expansion of the cathode plate relative to the gun body. The heating around the thin edges of the RF coupling port in the gun of Figure **2.7(a)** produces a peak von Mises stress of 14 kPSI [2.23]. This stress is 3-4X higher than the yield strength for brazed copper. In addition, a calculation of the pulsed heating around the coupler port gives a temperature rise of 137 ˚C per pulse which is 3X the acceptable limit established for X-band (11.8 GHz) structures [2.22]. The maximum heat flux of 70 W cm$^{-2}$ is more than 3X higher than the maximum average heat load of 20 W cm$^{-2}$. All these results indicate this gun design would have a limited lifetime operating continuous at the required 4 kW average power.

Figure **2.7(b)** shows an upgraded version of the gun which uses a long and wide coupling port for the RF and cooling has been added to the cathode. Temperatures and the mechanical stresses are significantly less in this design. Here, the maximum temperature at the coupler port is 36 ˚C for 4 kW of power dissipation compared to > 90 ˚F at the cathode in the previous design.

The pulse heating of the Z-coupling ports was optimized by increasing the radius of curvature of the edge of the port running the length of the cavity. Computer modeling gives a temperature rise of 42 ˚C at a 140 MV m$^{-1}$ cathode field. This temperature rise is safely below the recommended upper limit of 50 ˚C determined in the X-band cavity studies [2.22].

The results shown in Figure **2.7** are typical are for pillbox guns, but for a reentrant gun the distribution is distinctly different due to the nose cones. The higher cathode and on axis fields are achieved by increasing the current density with the nose cone shape. This higher current density also increases the surface heating in this region. The hotter area is a ring around the nosecone just behind the “nose.” A thermal model for a 1 300 MHz L-band gun operating at 5% duty factor gives a temperature rise of 60 ˚C just behind the bulge of the cathode nosecone. The same temperature rise is found around the RF coupler port. The analysis was done for 10 kHz repetition rate, a peak cathode field of 64 MV m$^{-1}$ and an average dissipated power of 29 kW. This design has areas with a peak power density greater than 110 W cm$^{-2}$ [2.6] which are over 5X the 20 W cm$^{-2}$ limit generally used for accelerator cavities.

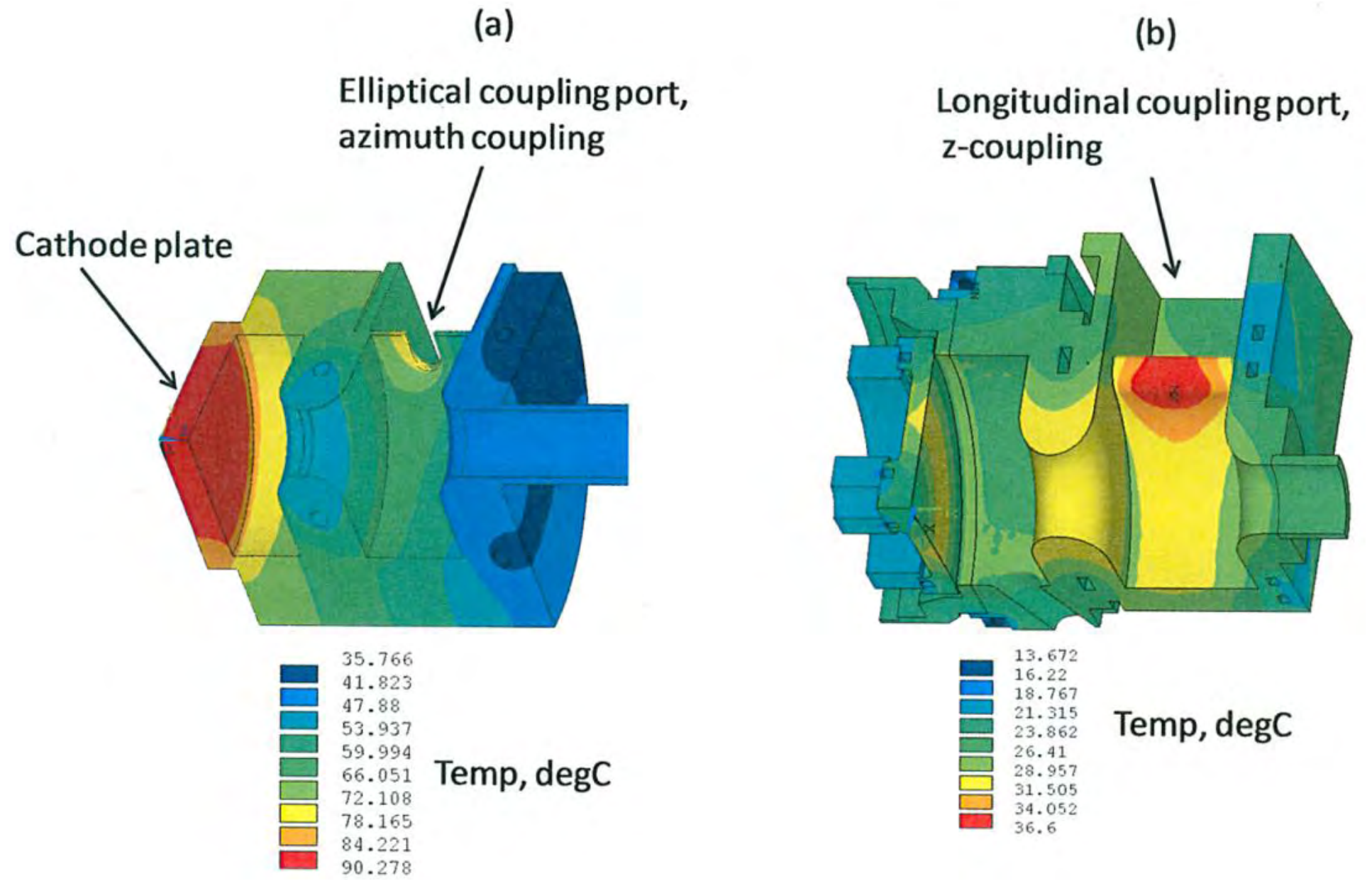

**Figure 2.7. a) A ¼ model of the body temperature for the BNL/SLAC/UCLA S-band gun operating at 3.7 kW of average power. [2.24] b) The ¼ model of the Z-coupled LCLS gun showing the steady state temperature distribution for 4 kW of average RF power dissipation. [[2.23]; Adapted under Creative Common Attribution 3.0 License (www.creativecommons.org/licenses/by/3.0/us/) at www.JACoW.org.] [2.19]**

The large power dissipation of high duty factor and CW NCRF guns requires careful thermal design of the gun cavities and the RF coupler. A good example of a CW coupler design is that used in the LANL/AES 2.5-cell gun operating at 700 MHz [2.25]. Dual RF feeds are used on the coupling cell to minimize any field asymmetry on the beam axis. The dual feed also allows splitting of the RF power, thereby sharing the heat load between two RF vacuum windows. The cavities are pillbox in shape and use a novel "dog-bone" shape for the RF coupling ports [2.26]. The waveguide transitions are tapered ridges to couple the RF power from the half height waveguide to the coupling ports. The ridge-loaded, tapered waveguide (RLWG) is carefully designed to minimize the reflected RF power as well as multipacting. Multipacting in the RLWG and excessive heating of the coupler holes was observed in CW operation of a Radio Frequency Quadrupole (RFQ) proton accelerator [2.27]. The electron gun was designed with 2-D and 3-D, frequency-domain and time-domain codes [2.28] to avoid these problems.

The dog-bone coupling port consists of two holes connected by a long, narrow slot through a sidewall of the cavity. The hole diameters are equal and adjusted to give the desired coupling coefficient while keeping their area small. The larger the area, the greater the surface current density increases around the hole. The narrow slot minimizes the axial field perturbation and allows cooling channels along its walls. The coupler consists of end holes 4.75 mm in radius at each end of a 5 cm long, 1.788 mm wide slot. The cavity wall thickness at the coupler is 12.7 mm. The coupling coefficient is 1.38 and the external $Q$ is 1 933[2.4]. The gun's $Q_0$ is 31 800 [2.25]. Enhanced cooling at the edges of the holes is necessary to avoid melting the opposing ends of the holes. [2.27]. Since the wall current dominantly flows parallel to the slot, this geometry only perturbs the wall current where it flows around the holes and thus is where much of the heating occurs. A thermal analysis for 461 kW of forward power shows a power density of 120 W cm$^{-2}$ at the outer edges of the holes.

While the pillbox cavity avoids the hot annulus behind the nose cone, there remains a large heat load of over 100 W cm$^{-2}$ covering large radial patterns upon the cavity walls. This is only slightly lower than the maximum power density of the coupler and produces stresses well above the tolerable yield strength of brazed copper. As a result, this and other high stress designs require the use of a higher yield strength copper-alumina sintered material called GlidCop® [2.29].

### 2.3.2 The Photoinjector Control System

All NCRF guns require active regulation and monitoring of a multitude of gun parameters, ranging from the RF power to the gun body temperature. A schematic layout of the gun control system is shown in Figure **2.8**. It's built from three basic parts: The sensors and actuators, the process variable database, and the high-level computer applications. The sensor & actuator level includes any feedback or feedforward controls which are either hardwired or embedded in local processors. An example of a hardware controller at the sensor & actuator level would be the fast feedback loop for maintaining the desired RF phase and amplitude during the RF macropulse. The field $\phi$-$A$ control box takes the forward- and reverse-RF power signals from the directional coupler along with the cavity RF probe signals and compares them with the desired RF phase and amplitude. All the information used by and varied by the RF field controller is available in the process variable database. High-level applications could change the controller's settings such as the desired RF power and the controller would regulate the RF to that value. In this architecture, the high-speed control is done by fast embedded hardware while the slower feedback and monitoring is done by the high-level applications.

The sensor and actuator signals are stored in a common database and are called "process variables," or PVs, for short. A PV can be a scalar or an array of values. Examples of scalar sensor PVs are the gun body water temperature and the peak forward klystron power. The digitized waveform of the RF macropulse shape would be an array PV. The PV information in the database is accessible by both the sensor/actuator hardware and the high-level applications. The local controllers continually update the sensor PVs on the database and read the values and status of the actuator PVs set by the high-level applications. The high-level applications monitor the sensor PVs and vary actuator PVs to modify and regulate the overall system.

The cooling system for high power guns can be fairly large and are usually located some large distance from the gun itself. This can make the circulation loop time fairly long compared to the RF power response time, leading to control loop delays and possibly slow and unstable temperature control. This problem is solved by installing a small temperature controlled chiller near the gun.

A small fraction of the water from the main loop is diverted through the chiller which the chiller heats or cools in order provide precise and fast control of the cavity temperature, and hence resonance frequency. The water system regulation needs to be stable and accurate enough to keep the resonance frequency within the control range of the LLRF system. For most cases the water system needs to regulate the gun body temperature with an rms temperature variation of 0.1 ˚C to be within the control range of the LLRF.

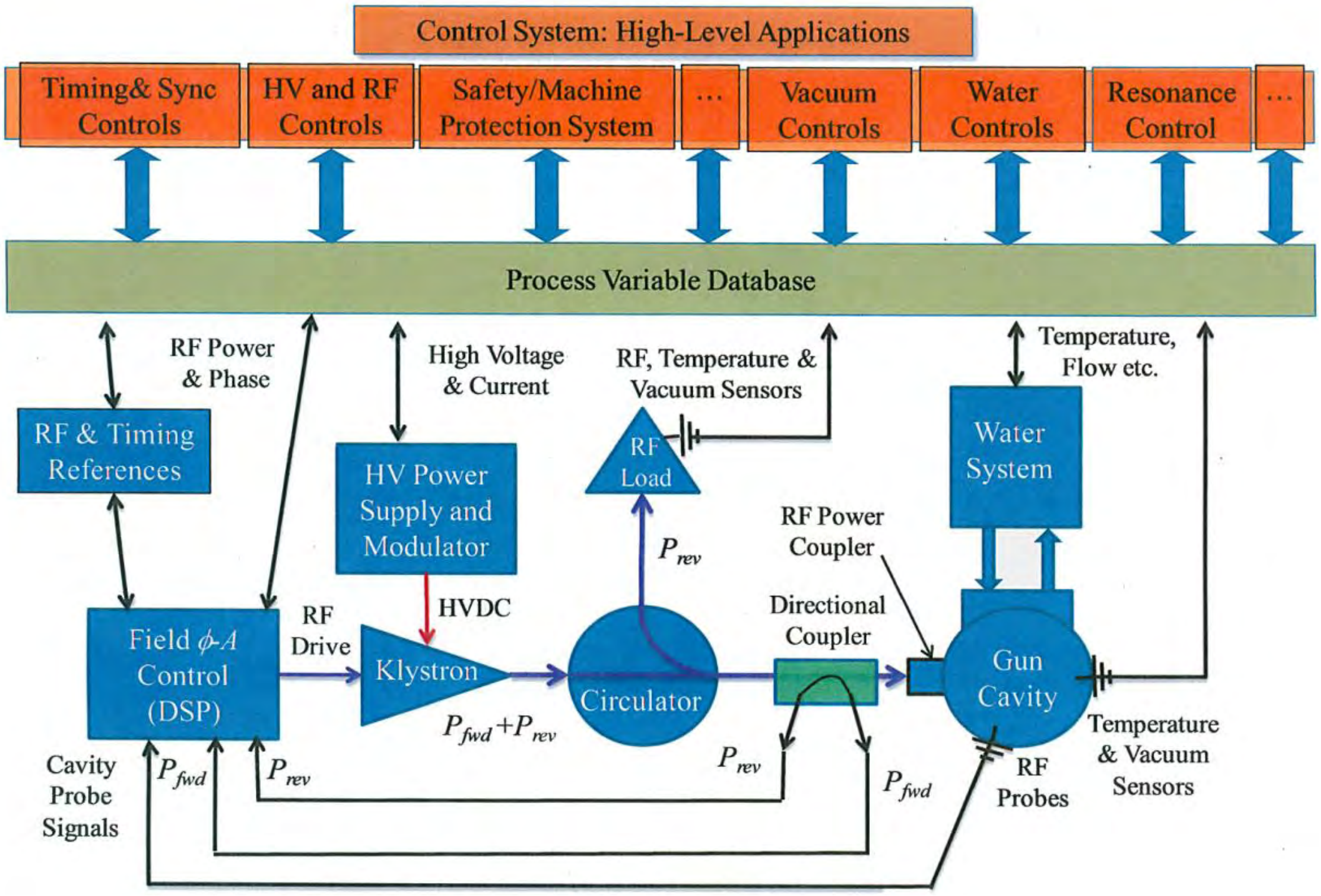

**Figure 2.8. Schematic drawing of gun RF and water control systems. The drawing shows the water and resonance controls being connected by an internal control application.**

The RF system includes a field phase and amplitude controller which compares signals from a gun cavity probe and a directional coupler with the RF reference to generate the low level drive for the RF amplifier (which is usually a klystron). A directional coupler near the gun coupling port samples a fraction of the forward- and reverse-RF power. The RF power source also requires a high voltage, high current power supply, and pulse modulator, which also requires a feedback circuit for stable operation. As discussed earlier, a large fraction of the RF power is reflected back from the gun at the start and end of the RF pulse and the RF source needs to be protected from these reflections. For short macropulses, the klystron can be isolated by time, that is, by delaying the reflection so it arrives back at the source after the RF macropulse ends. In practice the reflection is delayed by using a very long waveguide between the klystron and the gun which physically limits its application pulse formats with RF macropulses less than ~2 μs long. The isolation for long macropulses and CW operation uses a device called a three-port circulator which behaves like a RF diode. The circulator shown in Figure **2.8** passes the forward power into the gun, while directing the reflected, reverse power to an impedance matched RF load. RF isolation is discussed with more detail in Chapter 10, Section 10.3.

## 2.4 RF CAVITY FIELDS AND EMITTANCE

In this section, we discuss the effect the RF field has on the electron beam quality. In Chapter 1, the RF emittance was shown to be solely due to the time dependence of the fields; however, the RF field can also have a spatial variation which can distort the beam. The spatial distribution of the field is determined by the RF-mode being used, details of the cavity shape and the locations of penetrations, such as the RF coupler in the cavity walls. The emittance due to side-coupled RF ports is discussed along with a method for correcting the field asymmetry. It is similarly important to have symmetric RF fields in the first accelerator section after the gun and the racetrack design, for its field correction is also described. Later in this section, the effects of mode purity and the spatial field asymmetries are described.

### 2.4.1 RF Transverse Fields and Emittance

The RF coupling port can generate an asymmetric field distribution about the beam axis which can distort and reduce the beam quality. The dipole asymmetry can be removed by using a pair of phase and amplitude balanced opposing RF couplers, however, there remains the quadrupole and higher order fields. To aggravate the issue, the large, Z-coupled ports produce larger quadrupole fields than the smaller, oval ports of the BNL/SLAC/UCLA gun. The coupling coefficient for the Z-coupled is 2, and the coupling for the original BNL/SLAC/UCLA gun's oval ports is 1.3.

The RF quadrupole field can be canceled by making the coupler cavity slightly wider in the direction perpendicular to the coupler ports. This is called a racetrack cavity shape and was originally proposed for X-band travelling wave structures [2.30]. The RF quadrupole field introduced by the ports is compensated for by slightly elongating the cell in the direction perpendicular to the coupler ports. The cross section of the coupling cavity shape is shown in Figure **2.9**. In this design, the racetrack shape is formed by two circles whose centers are offset a distance $d_{sep}$.

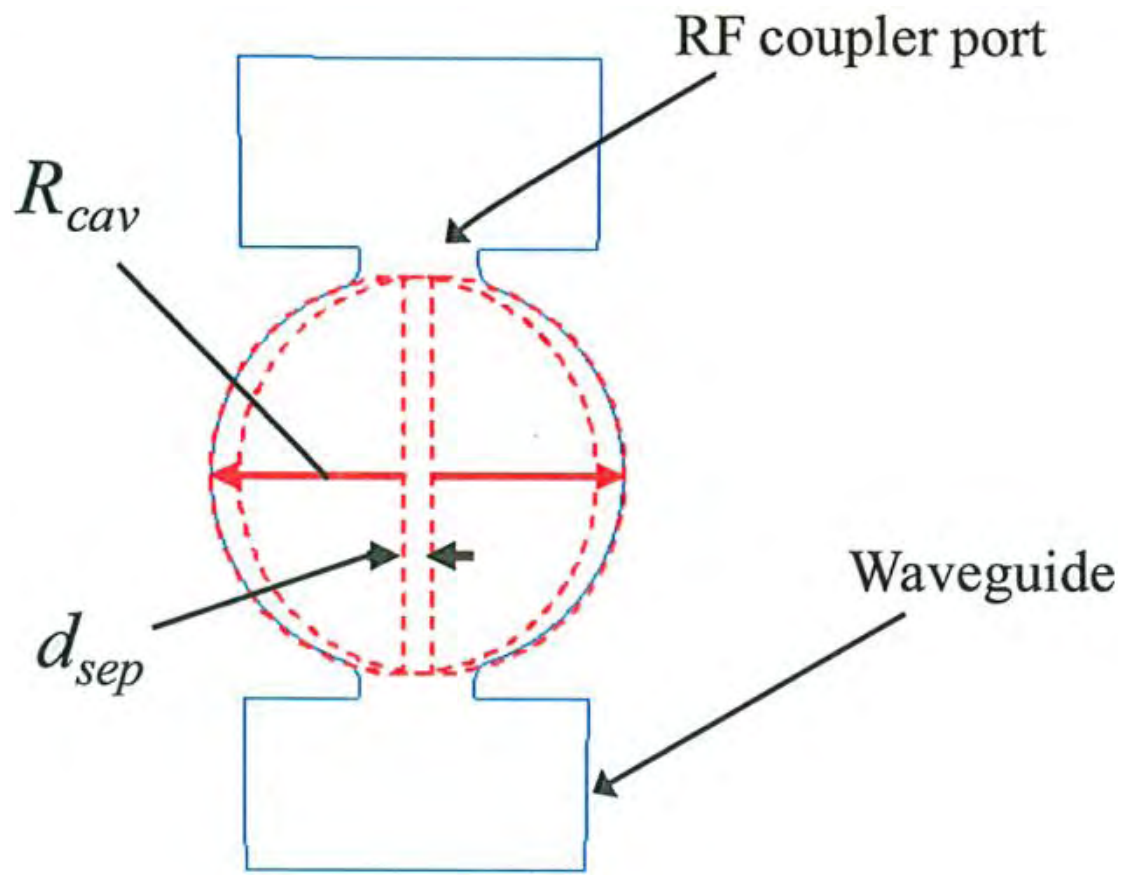

**Figure 2.9. Cross-section of the coupling cell for a side-coupled, dual RF feed gun.** [[2.9]; Adapted under Creative Common Attribution 3.0 License (www.creativecommons.org/licenses/by/3.0/us/) at www.JACoW.org.]

The radial kick from the quadrupole field is found by integrating electron trajectories through the time-dependent, 3-D cavity fields. The cavity fields are computed using details of the cavity surface shape and penetrations [2.29]. Analysis shows cavities with rotational symmetry behave as time-dependent electric lenses with focal lengths given in Chapter 1, Section 1. Adding opposing side ports for dual RF feeds is effective at eliminating the dipole RF kick, but it introduces a quadrupole field with a thin lens-like kick. Here, the quadrupole kick is defined as the focal length of a thin lens with the same angle change as computed from the 3-D field.

The quadrupole kick or lens strength, $1/f_{rfq}$, is given by

$$\frac{1}{f_{rfq}} = a_q \sin(\phi_{RF}) \tag{2.15}$$

The RF quadrupole focusing strength, $1/f_{rfq}$, has a unit of $m^{-1}$ (or equivalent units would be mR $mm^{-1}$). If the focal length is in meters, then the focal strength unit is the diopter. The amplitude of the sinusoidal focal strength is $a_q$ and $\phi_{RF}$ is the RF-bunch phase. Figure **2.10** shows the quadrupole focal strength computed from the 3-D fields for dual RF ports configured with ellipse-shaped, azimuth-coupled ports as shown in Figure **2.7(a)** [2.23]. The lens strength is plotted as a function of the RF phase for $d_{sep}$ = 0, 3.15, 3.56 and 3.40 mm. The plot for $d_{sep}$ = 0 is the case for RF ports (coupling factor of 2) in a round cavity. The quadrupole field is zeroed for $d_{sep}$ = 3.40 mm. Analysis shows the lens strength for the corresponding *Z*-coupled cavity is similar to that of azimuthally coupled cavities provided they have the same RF coupling factor. The peak quadrupole focal strength corresponds to a focal length of 222 mm which is approximately twice the radial RF focal length. See Chapter 1, Section 1.5.1.

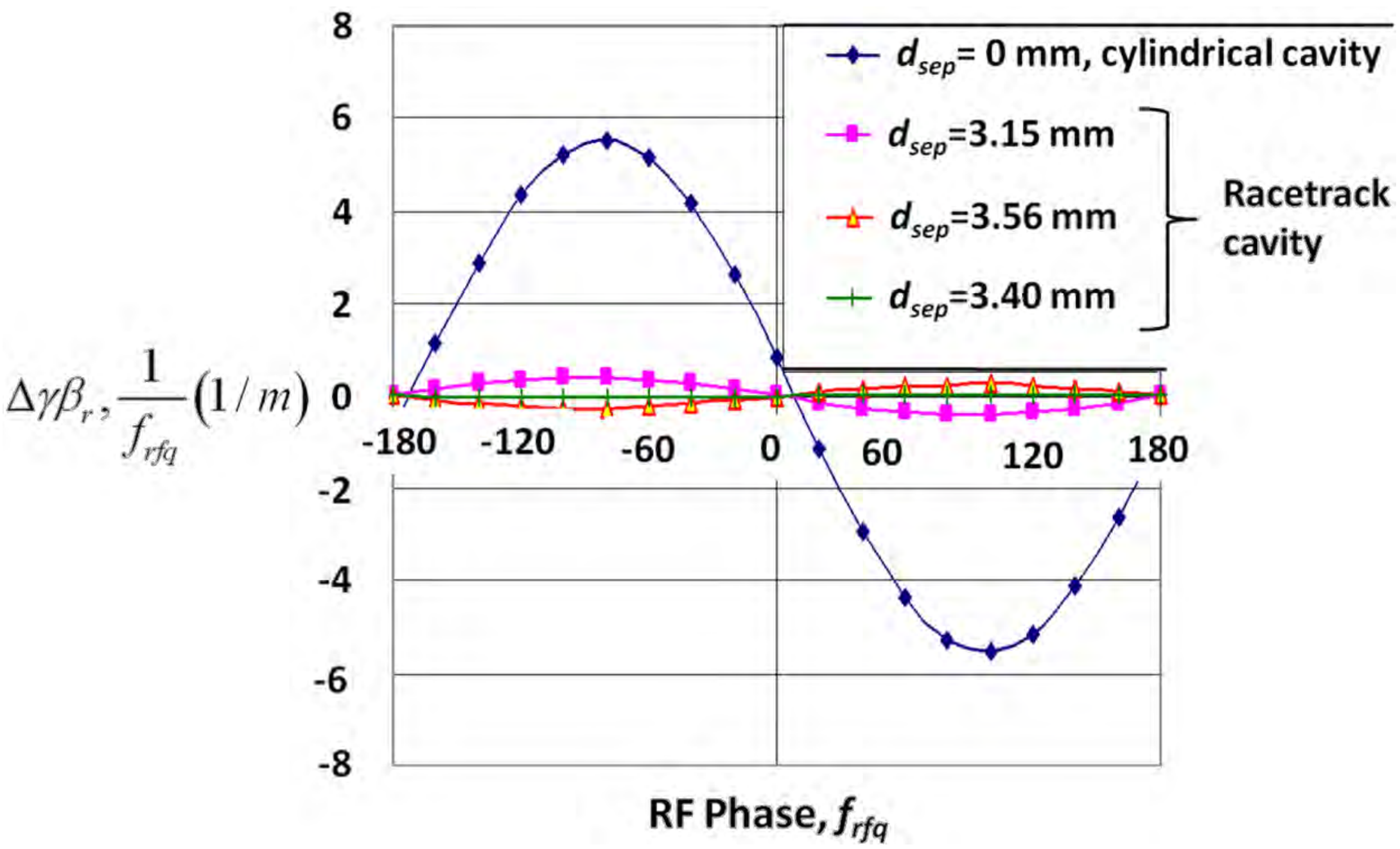

**Figure 2.10. The integrated transverse kick (1/m) due to the RF quadrupole field is plotted as a function of the RF phase. The quadrupole field is due to the RF field extending into the dual feed ports. The RF coupling factor is ~2 for both port configurations. RF crest is at zero degrees. The quadrupole lens strength for an elliptical port with its long axis in the azimuth direction as shown in Figure 2.7(a). [[2.23]; Adapted under Creative Common Attribution 3.0 License (www.creativecommons.org/licenses/by/3.0/us/) at www.JACoW.org.]**

As described in Chapter 1, a quadrupole field produces emittance effects somewhat different from those for rotational symmetric fields. Similar to the rotational symmetric fields, the time-dependence of the quadrupole field generates a first order projected emittance due to different focusing of each slice. The second order RF projected emittance due to the quadrupole field similarly results from the curvature of the RF waveform. However because the quadrupole field distribution does not have rotational symmetry, the beam's rotation in the solenoid field couples the *xx′* and *yy′* trace spaces. With this coupling the *xx′* phase space correlations with the *yy′* phase space appears to change the 2-D emittances, while the 4-D canonical emittance of would remain constant. The correlations affect both the slice and projected 2-D emittances. The correlated trace space emittance due to an anomalous quadrupole field is described in Chapter 1, Section 1.5.4. The emittance due to the RF quadrupole lens strengths shown in Figure **2.10** is discussed next.

Similar to the derivation for the RF emittance given in Chapter 1, Equ. 1.26, one can write the first-order RF quadrupole emittance as

$$\varepsilon_{rfq}^{(1)} = a_q\,\sigma_x^2 \left|\frac{\mathrm{d}}{\mathrm{d}\phi_{rf}}\left(\frac{1}{f_{rfq}}\right)\right| \sigma_\phi \tag{2.16}$$

The average transverse rms beam size in the quadrupole field region is $\sigma_x$ and the rms bunch phase length is $\sigma_\phi$ in radians. Inserting the relation for the quadrupole lens strength gives

$$\varepsilon_{rfq}^{(1)} = a_q \sigma_x^2 \sigma_\phi \left|\cos(\phi_{rf})\right| \tag{2.17}$$

Similarly the second-order emittance is found to be

$$\varepsilon_{rfq}^{(1)} = \frac{a_q}{\sqrt{2}}\,\sigma_x^2 \sigma_\phi^2 \left|\sin(\phi_{rf})\right| \tag{2.18}$$

Recall that $a_q$ is the amplitude of the sinusoidal quadrupole lens strength versus the RF phase. The graph of the cylindrical cavity quadrupole strength shown in Figure **2.10** indicates $a_q$ (cylindrical, azimuth) = 5.5 m$^{-1}$.

Adding the first and second order RF quadrupole emittances in quadrature gives the total emittance due to the quadrupole RF field,

$$\varepsilon_{rfq}^{(1+2)} = a_q \sigma_x^2 \sigma_\phi \sqrt{1 - \left(1 - \frac{\sigma_\phi^2}{2}\right)\sin^2(\phi_{rf})} \tag{2.19}$$

Since the asymptotic bunch-RF phase is always within ±20 ˚RF of the RF crest, the emittance in this region is ~0.3-0.4 μm for $\sigma_\phi$ = 0.0705 radians (4˚ RF-rms) and $a_q$ = 5.5 m$^{-1}$. This is all due to the first order quadrupole RF emittance.

Beyond emittance growth in the gun, it is also necessary to study the RF field effects in the first accelerator section (booster) after the gun. Field asymmetries can have a large effect, since the beam energy is still low. As an example, consider the SLAC S-band travelling wave linac after the LCLS gun. Applying the same field analysis as was done for the gun, one finds a similar head-tail, projected emittance. Figure **2.11** shows the transverse momentum change of the beam as it transits the coupler fields. The inserted drawing shows the dual feed, racetrack coupling cell and its location in front of the first cavities of the linac section. The radius of the port edge should be 1 mm or larger to minimize pulsed heating [2.31].

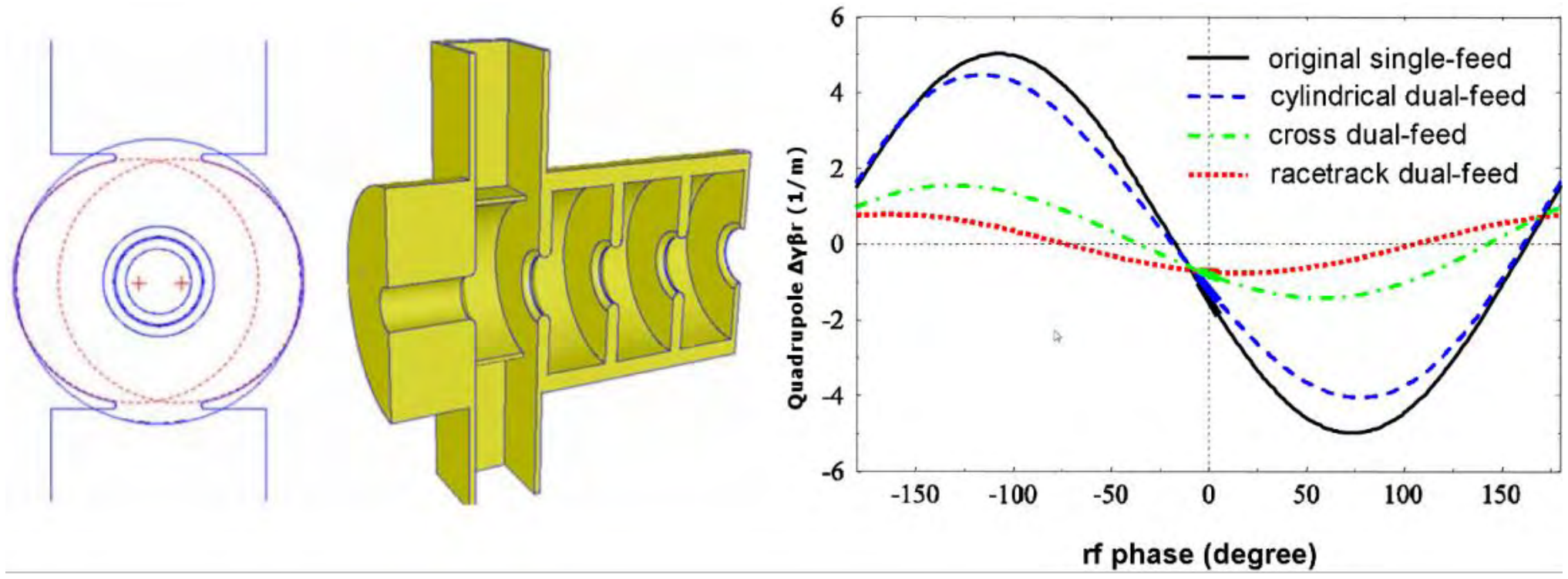

**Figure 2.11. Transverse momentum change of the beam after crossing the RF coupler quadrupole field. Beam energy crests at zero ° RF phase. (Left/Middle): Travelling wave S-band linac with a dual feed racetrack RF coupler. The coupling port runs the length of the coupling cell (Z-coupling) and the cavity inside shape is defined by two, offset circles to approximate a "racetrack" shape. [[2.23]; Available under Creative Common Attribution 3.0 License (www.creativecommons.org/licenses/by/3.0/us/) at www.JACoW.org.]**

### 2.4.2 RF Modes and the Emittance

In guns with n coupled cavities there are *n* longitudinal RF modes. Since it's important to excite only the π-mode, the other modes should be several MHz away in resonant frequency. The RF gun is said to have balanced fields if the peak field is the same in all cavities for the π-mode. The field balance is typically measured using Slater's perturbation technique, in which, a small dielectric or metal bead is inserted into the cavity and the resulting change in the RF resonant frequency is measured as a function of the bead's position in the cavity. Taking measurements for the 0- and π-mode resonances as a function of bead position gives the shape of each mode in the gun.

One set-up for performing these measurements is the bead drop technique as shown in Figure **2.12**. In the bead drop technique, the bead is held at the end of a thin surgical string thread and the resonant RF frequency measured for the π- and 0-modes as the bead is lowered to the cathode. In open ended RF structures, the bead is suspended in the middle of the string which is pulled through the structure; this is called the bead pull technique. Dropping the bead is preferred for guns since it measures the field very near the cathode wall without the fields being perturbed by a hole for the string.

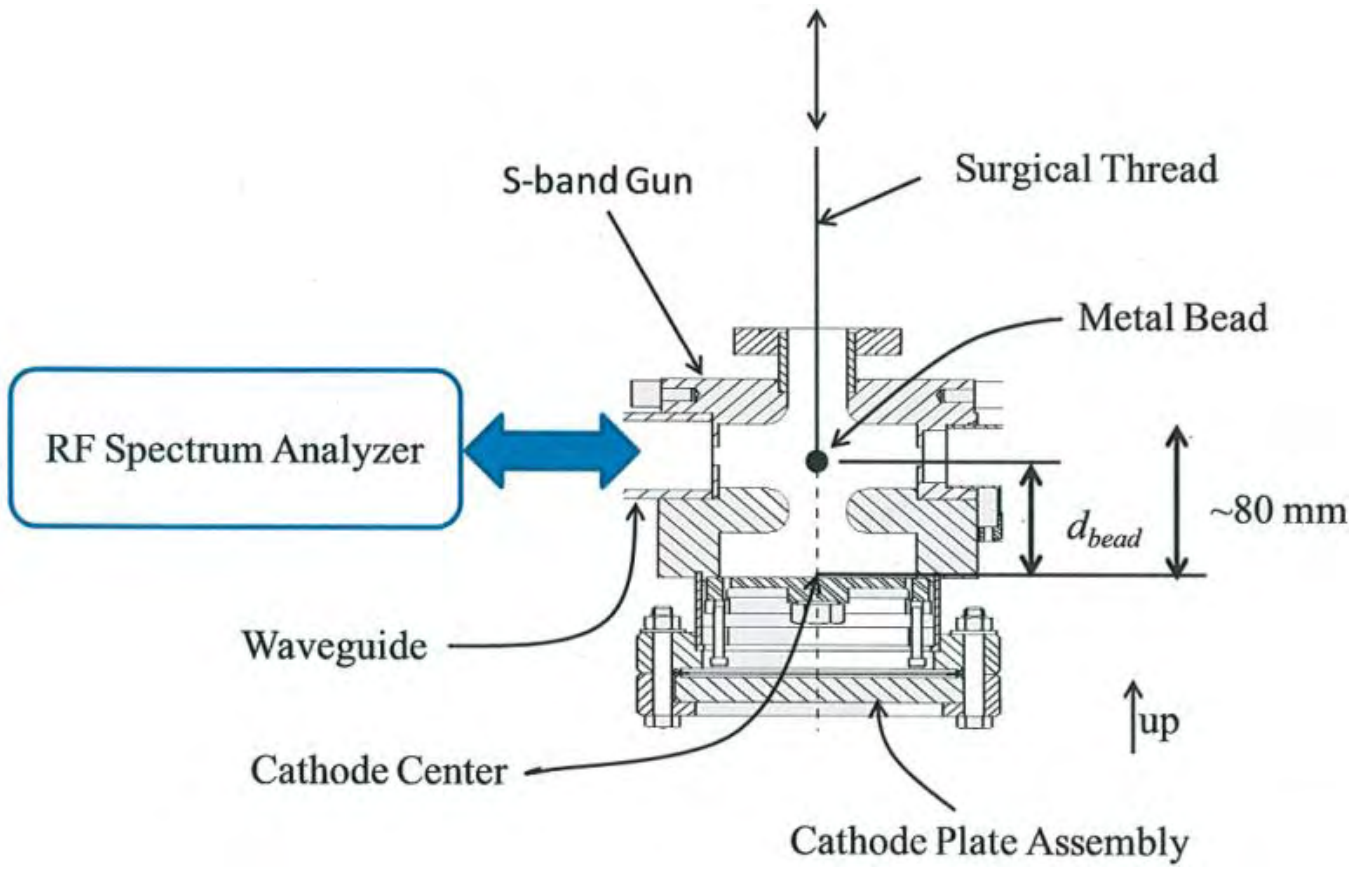

**Figure 2.12. Set-up for measuring the field shape using the Slater perturbation method.**

Bead-drop measurements for the π- and 0-modes of a 1.6-cell S-band gun are shown in Figure **2.13**. The π-mode, which is the desired accelerating mode of the gun, is seen to be balanced. That is, the electric field at the cathode (bead position = 0 mm) is equal to the peak of the electric field in the full cell. In contrast, the 0-mode shape is quite different and is not balanced between the two cavities.

The measurements shown in Figure **2.13** are performed using a narrow bandwidth (small frequency spread) CW RF source to selectively excite only the 0- or π-mode. However, if the gun is powered by a pulsed RF source both modes can be excited. Consider a 1.6-cell S-band gun with a 3.2 MHz mode separation being powered with a 2 μs long pulse with approximately 300 ns rise time. This fast pulse easily can excite the 0-mode since a ~300 ns rising edge has significant frequency components out to 3 MHz or more around the drive frequency. Measurements of an S-band gun in operation with a 2 μs RF pulse indicate the gun field is oscillating at frequency of the mode separation of 3 MHz and a peak-to-peak amplitude of 3-5%. [2.32]

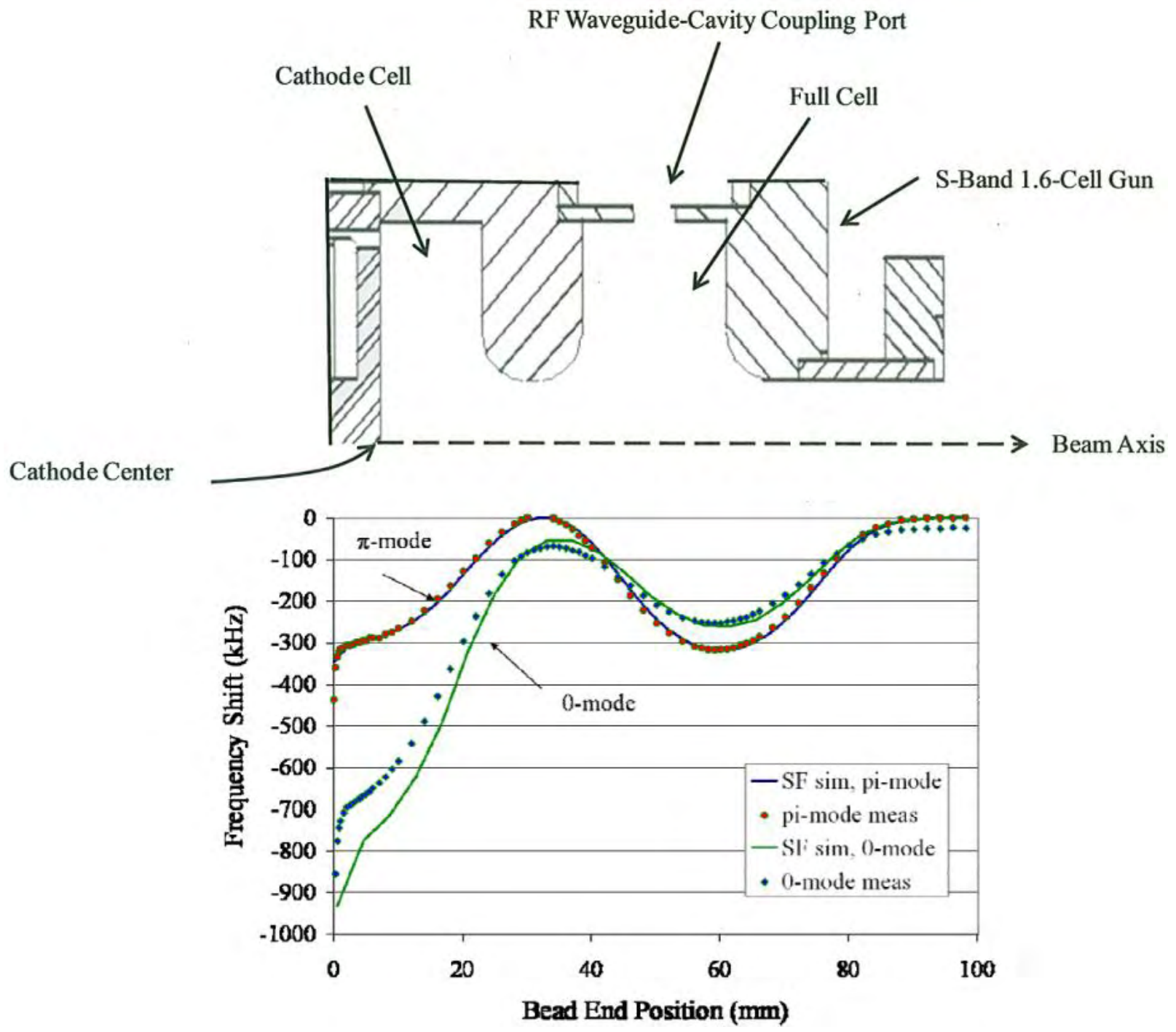

**Figure 2.13. Top: Upper half of cross section through a 1.6-cell S-band gun. Bottom: Bead drop measurements of the 0- and π-modes for a 1.6-cell S-band gun as a function of bead position from the cathode. The cathode is located at the 0 mm bead position. Points are the frequency difference between the resonance frequency measured at each bead position minus the resonance frequency when there is no bead in the cavity. The solid curves are a simulation using the SUPERFISH RF code [2.16]. [2.19]**

A consequence of exciting both RF modes can be seen in the energy spread and the emittance. Because the field balance requirement for the π-mode is 1-2%, a 3-5% contamination by the 0-mode unbalances the cavity fields and produces a correlated energy along the bunch. Since the energy spread is correlated it can be removed with the high energy linac, however at some energy cost of operating with the beam a few degrees away from the crest of maximum energy gain.

Ignoring longitudinal wakes, there will be a phase shift between maximum energy gain and minimum energy spread. This phase shift gives the correlated energy spread out of the gun. When compared with simulation, the correlation agrees qualitatively with the measured 0-mode. The simulation gives 11.8% for the 0-mode field relative to the π-mode; this is to be compared with the RF probe measurement of an S-band gun which indicated 3-5%.

Simulations were performed to determine how large a mode separation was needed were performed for an S-band gun. The studies used field maps for the 0- and π-modes generated by the SUPERFISH code [2.16] and the electron dynamics was simulated in PARMELA [2.33]. The case for 1 nC was optimized up to a beam energy of 135 MeV and the sensitivity to the difference between the 0- and π-modes determined for separation frequencies of 3.5 MHz and 15 MHz with 10% and 3% expected field strength compared to the π-mode field, respectively [2.34]. A reasonable metric is the tolerance of the emittance to the phase difference, therefore choose the range of 0-π phase, for which the normalized projected emittance is under 1 μm for a bunch charge of 1 nC. The simulations for 3.5 MHz and 10% field give ~10˚ S phase range. The 15 MHz and 3% 0-mode field has an emittance < 1 μm over the 70˚ S of phase studied. Interestingly enough, the optimum emittance for the 15 MHz cavity shape for the π-mode, but with no 0-mode field had lower emittance than did the 3.5 MHz cavity shape without a 0-mode. The difference in emittance is only 5% but suggests the beam dynamics appear to prefer the large mode separation.

## 2.5 SOLENOIDS, WAKEFIELD MITIGATION AND DIAGNOSTICS

This section discusses the gun solenoids used for focusing the beam out of the gun and cancelling any fringe fields there may be at the cathode. The section also will cover the major components in the beamline between the gun and the first accelerator. This beamline will be referred to as the Gun-To-Linac (GTL).

### 2.5.1 The Emittance Compensation Solenoid

As discussed in Chapter 1, the strong defocusing of the beam by the gun's RF requires compensating with an equally strong focusing lens. This focusing lens is usually an electromagnetic solenoid, which may or may not have iron to increase the axial field. Also discussed in Chapter 1 are the effects this solenoid can have on the beam emittance. The first and most important phenomenon is that of emittance compensation, in which the lens focuses the plasma oscillation of the slices to align them to minimize the projected emittance. Emittance compensation was described in Section 1.4.4. In addition, there are the aberration effects of the "perfect" solenoid. These included the chromatic and geometric aberrations, as well as the anomalous quadrupole emittance.

The discussion of the gun solenoid in Section 1.1.5 provides the general requirements for the engineering of the gun's solenoid. To facilitate the present discussion, the solenoid used in the LCLS gun will be used as an example. This solenoid is very similar to that used in the BNL/SLAC/UCLA S-band gun, with the addition of the quadrupole correctors described earlier in Section 1.5.4.

The LCLS gun solenoid consists of eight pancake coils wound from hollow, square copper tubing. In each pancake, the tubing is wound continuously from the outside to the inside to form two layers. Between the pancake coils are iron plates which concentrate and straighten the magnetic flux to give a more uniform magnetic field. There are also thin, iron mirror plates on the ends to contain the fringe fields. The potted coils and iron plates are stacked on a support tube and clamped together to form the solenoid assembly. Also, it is important that all the ends of the pancake coils exit along one side the solenoid and are kept close to each other to cancel fields due to the current in the ends of the conductors. An engineering drawing of the solenoid is shown in Figure **2.14**. Further design details can be found in [2.35].

An interesting feature of the LCLS solenoid design is the inclusion of quadrupole correctors. Inside the solenoid, running its full length, are the normal and skew quadrupole correctors. Each consists of a single run of 16 AWG enameled wire running down the solenoid, across a quarter the circumference of the bore, then back again to form the four poles of a quadrupole. The two correctors on their ~3.3" diameter support spool are shown in Figure **2.15**. This spool fits inside the solenoid's bore as indicated in Figure **2.15** and Figure **2.18**, where one can see the four electrical leads extending from the beam exit end of the solenoid. At 12 A, each corrector has a focal length of 20 m. During beam operation, the 12 A was to be increased to 15 A to minimize the emittance, thus future uses of this concept should roughly double this corrector integrated field strength of 15 G. Clearly, the power supplies for these correctors should be bi-polar.

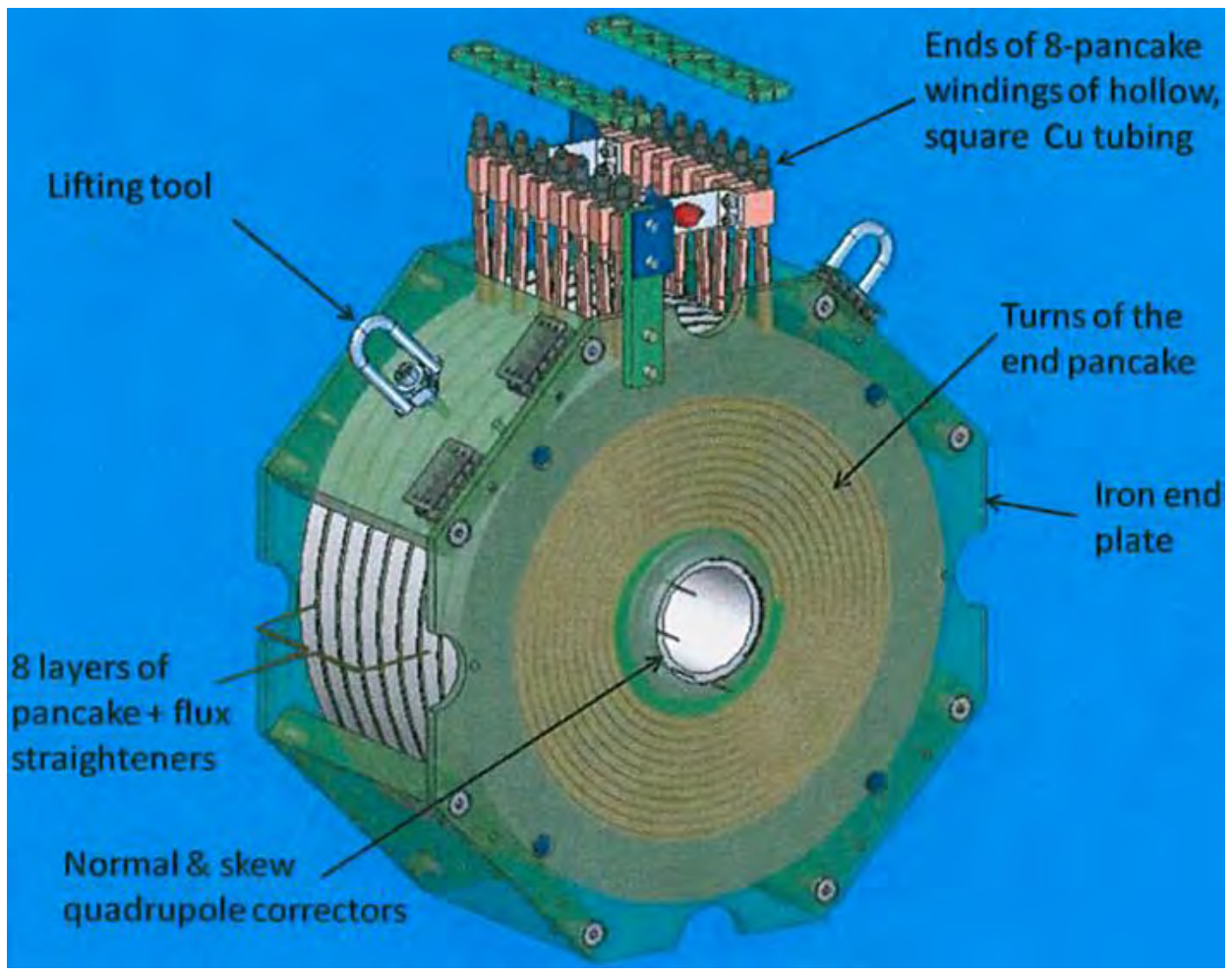

**Figure 2.14. Engineering drawing of the LCLS gun solenoid. [2.35]**

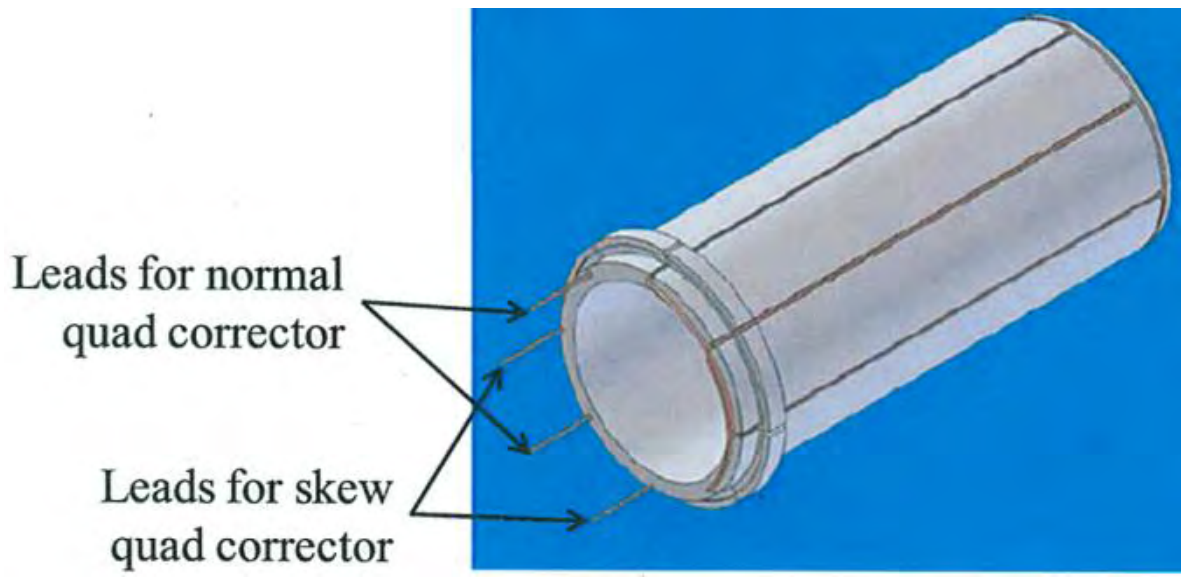

**Figure 2.15. Drawing of the 4-wire wire normal and skew quadrupole correctors for the LCLS gun solenoid. [2.35]**

In addition to the quadrupole correctors, there are horizontal and vertical steering coils placed on the beam tube inside the solenoid. These coils are capable of steering the 6 MeV beam 20 milliradians with a current of 10 A, and are shown in Figure **2.16**. While useful, it was found during beam commissioning that these steering coils produce oddly shaped beams due to the rotation in the solenoid, and in practice seldom used. Instead, the steering should be done with coils located immediately after the solenoid.

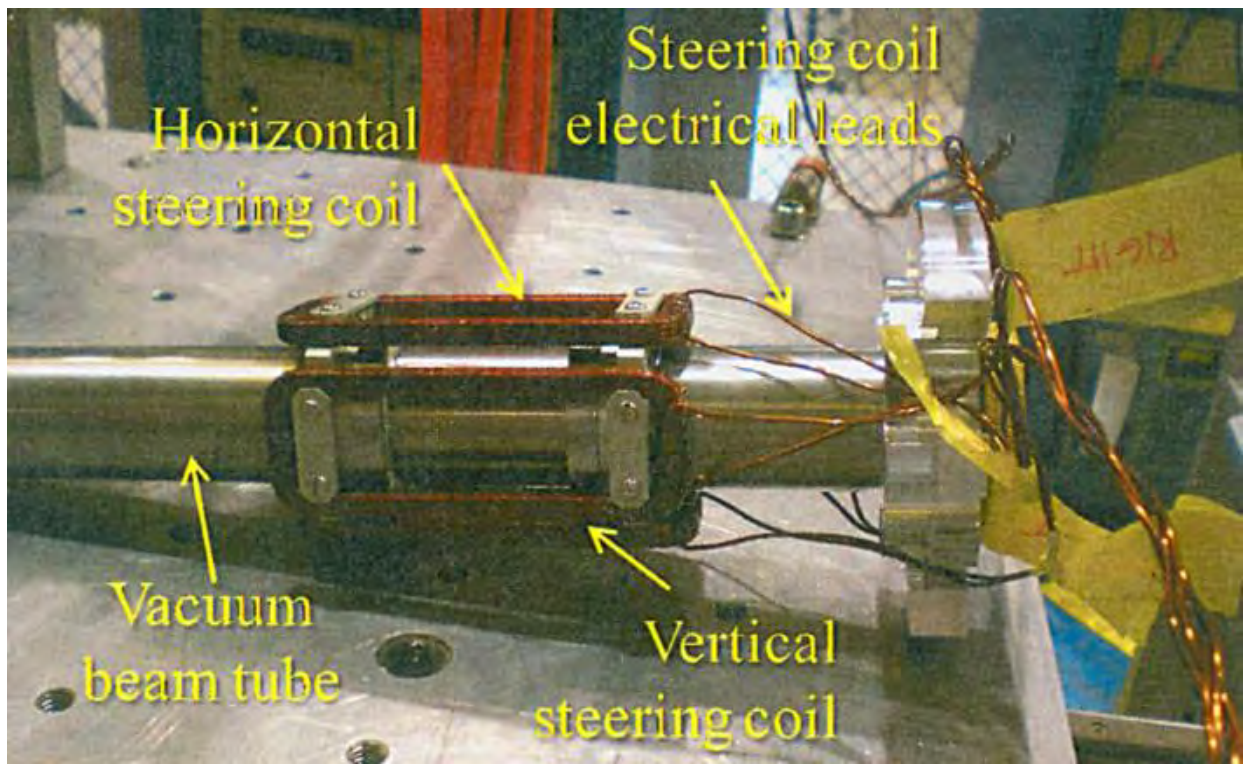

**Figure 2.16. Photograph of the dipole steering coils for the LCLS gun. These steering coils are attached to the beam tube which will be installed inside the bore of the gun solenoid.**

To achieve the best possible beam quality the gun system also requires a coil to cancel or “buck out” the fringe field from the gun solenoid at the cathodes. Any longitudinal magnetic field at the cathode will cause the electron beam to have an angular momentum about its longitudinal axis. This angular momentum is conserved as described by Busch’s theorem [2.36]. Such a magnetized beam has an emittance given by [2.37].

$$\varepsilon_N = \frac{eB_z r_0^2}{8mc} \tag{2.20}$$

This expression gives 0.07 μm for a magnetic field of 10 G and a beam radius of 1 mm. The black bucking coil is seen in Figure **2.17** during characterization in the SLAC magnetic measurements lab.

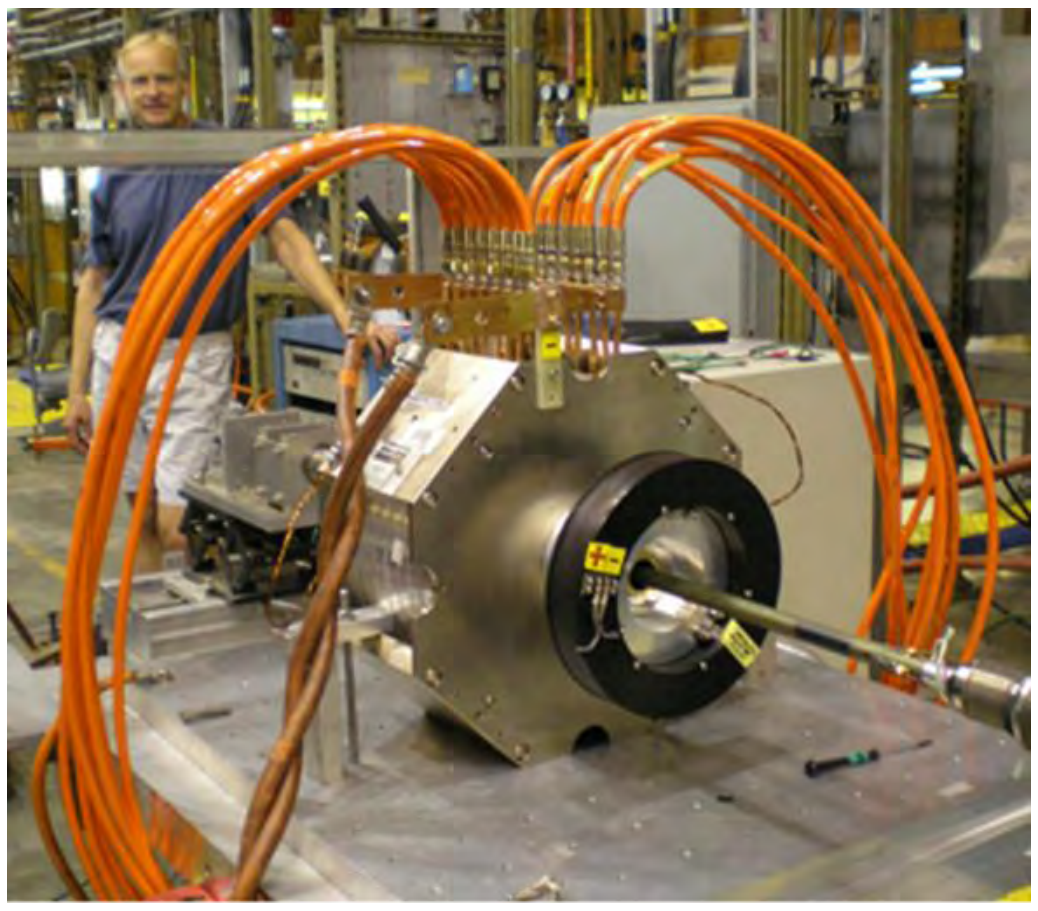

Figure 2.17. The emittance compensation solenoid and bucking coil in the SLAC magnetic measurements lab. [2.19]

Extensive magnetic measurements were made of the gun solenoid fields because of their critical role in compensation of the space charge emittance. Figure **2.18** shows these measurements for the LCLS gun solenoid. At the top of the figure is the Hall probe data for the axial field. This data gives the interior field calibration as a function of the current, shows any hysteresis effects, and provides the effective length of the solenoid. The measured field map is used in the beam simulations. The lower portion of Figure **2.18** gives the quadrupole field measured with a short (2.5 cm long) rotating coil. This type of coil gives the magnetic multipoles as a function of position along the solenoid. This device gives the field magnitude and the multipole phase at the radius of the coil. The phase angle is defined such that 0˚ corresponds to a skewed quadrupole and 90˚ corresponds to a normal quadrupole orientation.

The effects the solenoid quadrupole fields have on the beam have been discussed in some detail in Chapter 1, Section 1.5.4. As a reminder, they are the principle cause of the coupling between the *x*- and *y*-phase spaces, thus increasing trace space emittance. Since the canonical emittance is unaffected, the trace space emittance is fully recoverable using small normal and skewed corrector quadrupoles. The long corrector quadrupoles for the LCLS gun described above (see Figure **2.15**) were installed inside the solenoid’s bore as shown in the photographs of Figure **2.19**. For the section of the backup LCLS gun, there are short printed circuit board quadrupoles (PCquads) mounted inside and at each end of the solenoid’s bore. As shown in Figure **2.20**, these short quadrupoles are more effective at local cancelation of the anomalous quadrupole field than are the long, corrector quadrupoles. However, local cancellation is not necessary since the theory presented in Chapter 1, Section 1.5.4 shows that a pair of normal and skewed quadrupoles located anywhere near the solenoid can recover the trace space emittance.

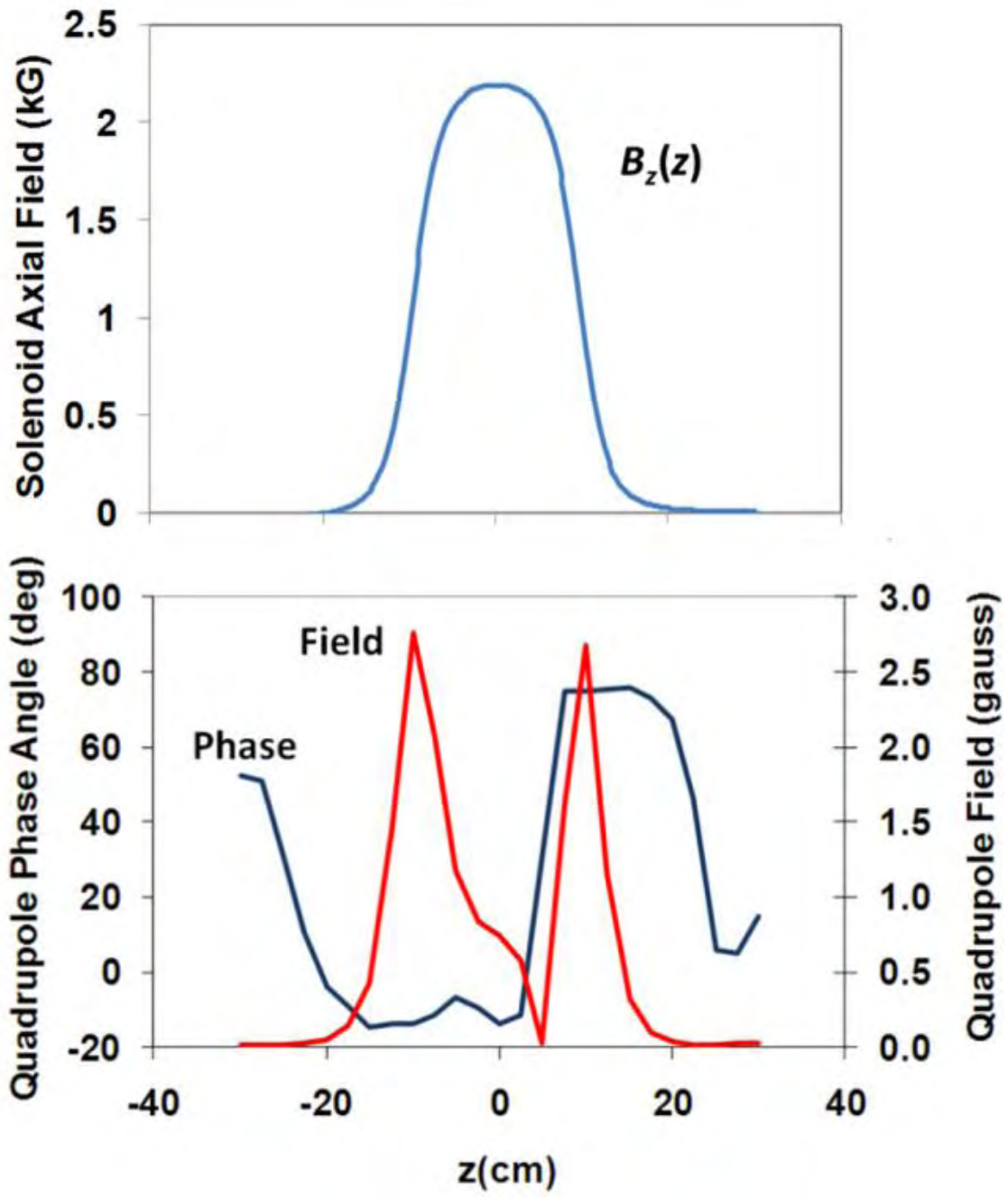

**Figure 2.18. Top: Hall probe field map of the LCS gun solenoid along the beam axis. Bottom: A short rotating coil measurement of the quadrupole field magnitude and phase angle. The short coil radius is 2.86 cm and has a length of 2.5 cm.**

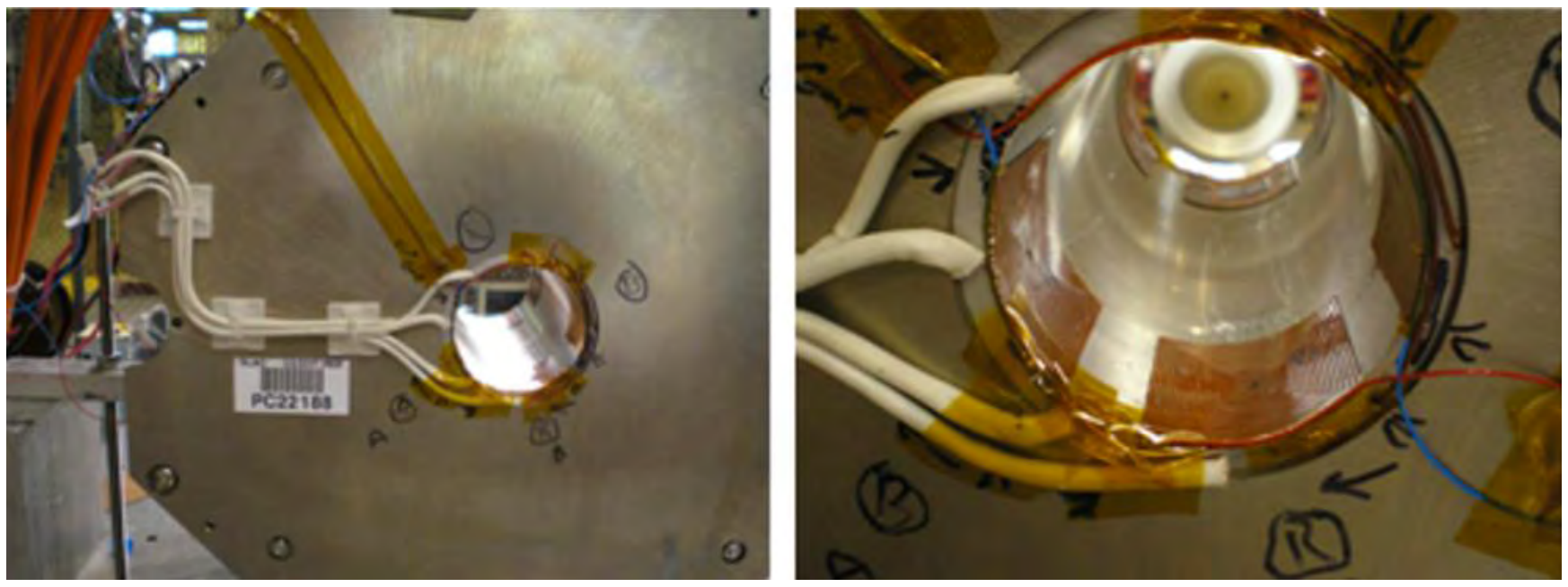

**Figure 2.19. Photographs showing the solenoid for Gun–2 showing the single wire and PC quadrupole correctors. The rotating coil used for measuring the field multipoles can be seen at the far end of the solenoid bore. [2.19]**

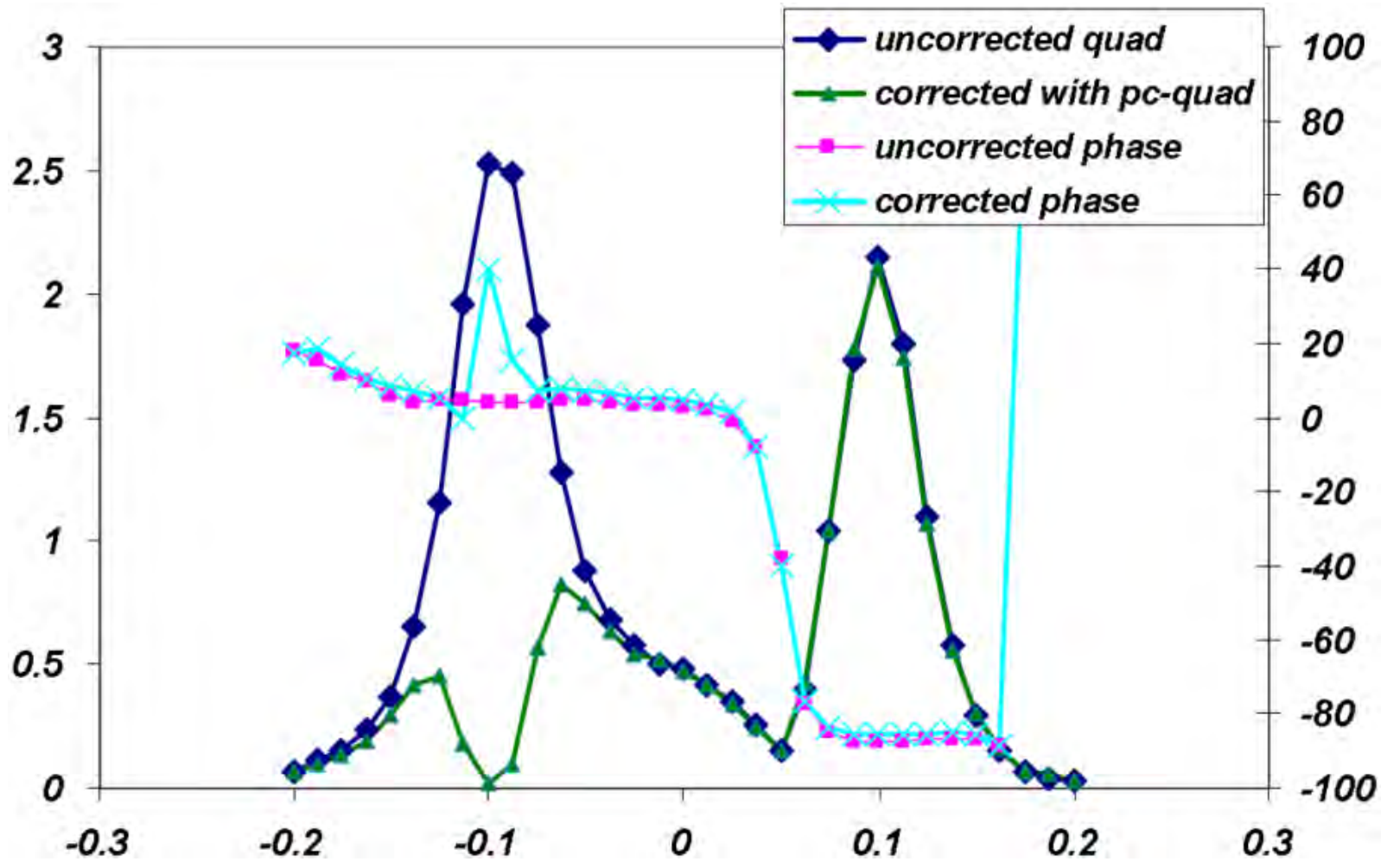

**Figure 2.20. Rotating coil measurements of the quadrupole field and phase along the solenoid's magnetic axis showing the uncorrected and corrected anomalous quadrupole field. The correction has been made only for the quadrupole field at the *z* = -0.1m end of the solenoid. The field at *z* = 0.1 m is uncorrected. [2.19]**

### 2.5.2 Wakefield Mitigation and Diagnostics in the Gun-to-Linac Beamline

The distance between the gun and the first linac section is determined by emittance compensation; it is ~1-2 m, depending upon the details of the gun voltage and solenoid configuration. A combination of diagnostics and the laser injection optics (for near-normal laser injection) need to fit into this region. For high bunch charge operation, it is important to design the beam pipe vacuum crosses and other vacuum chambers with wakefield liners allowing the beam a smooth and continuous boundary. The wakefield mitigation is even more critical for high duty guns, where the wake due to the preceding bunch can affect the subsequent bunches. As a design philosophy, one usually ignores longitudinal wakes as they can be compensated for by a small phase change of the beam relative to the linac section. On the other hand, it is more important to mitigate transverse wakes since they directly affect the emittance and are more difficult to correct later in the beam line.

Figure **2.21** shows a cutaway view of the GTL region built for the LCLS injector. This design includes dual laser in-vacuum metal mirrors, three Ce:YAG view screens, current toroid, three beam position monitors and an electron spectrometer. All these components have wakefield mitigation liners for the diagnostic and spectrometer vacuum chamber. A custom, all-metal valve was used after the solenoid to isolate the gun from the GTL. When this valve closes, it inserts a tube the same diameter as the vacuum pipe.

The dual laser mirrors on opposite sides of the beam balance the transverse wakes. These mirrors are all-metal to avoid electrostatic charging by low energy electrons and for high damage threshold. The mirror construction is Ni plating onto a BeCu substrate [2.38]. The surface flatness is $\lambda/10$ with a scratch and dig of 15-10. The surface roughness was less than 20 Å rms or $2\times10^{-9}$ m rms.

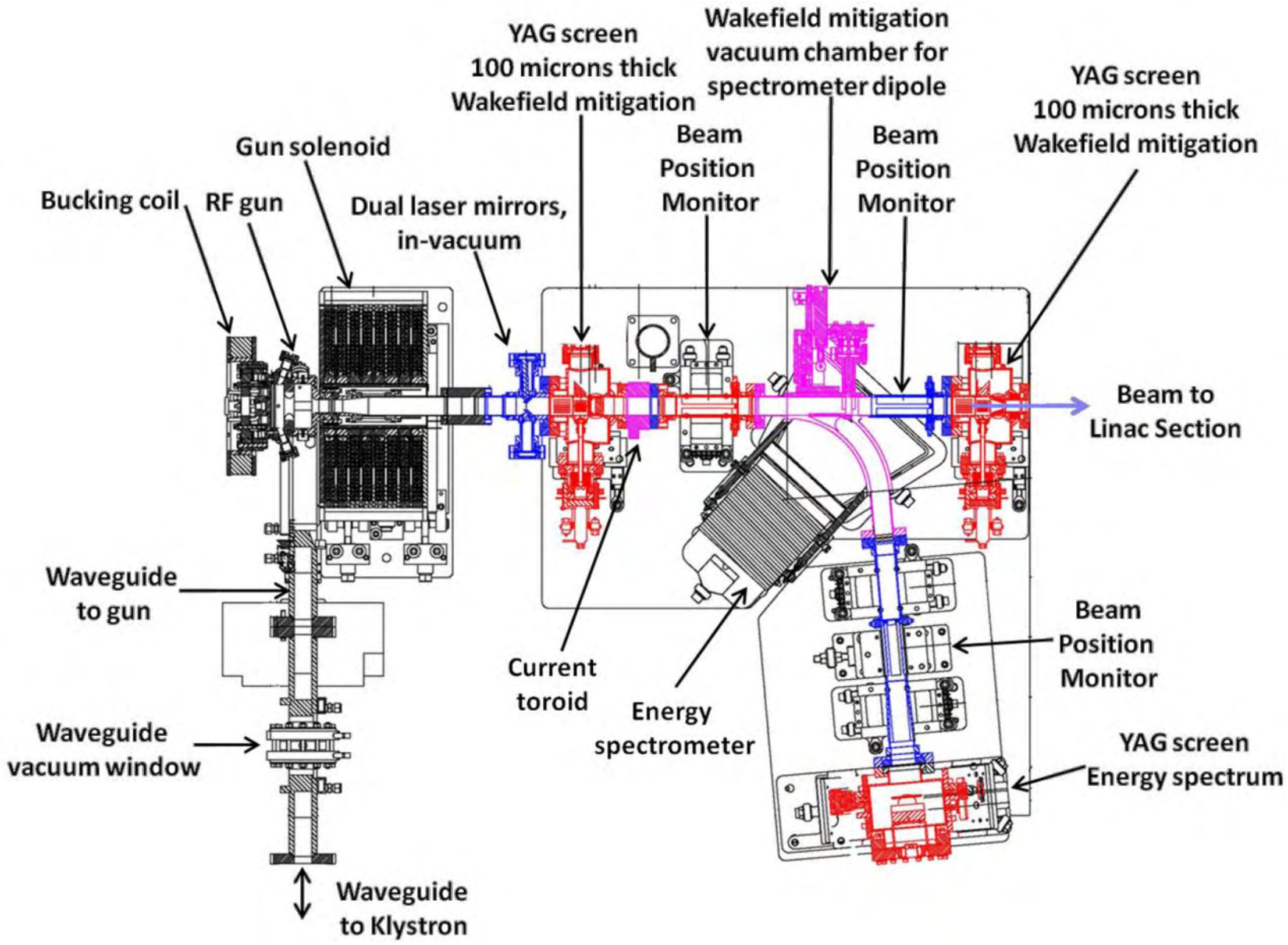

**Figure 2.21. Cutaway view of the LCLS Gun-to-Linac region. The overall length of the beam line including the gun is approximately 1.5 m. See [2.39] for more details.**

The Ce:YAG view screens were 100 μm thick and mounted normal to the beam. A 45˚ metal mirror behind reflects the fluorescent light from Ce:YAG through a quartz window to the digital camera. When these screens are retracted, the wakefield is mitigated by a cylindrical tube which gives a continuous wall, as seen by the beam. These tubes have longitudinal slots to increase the vacuum conductance to the arms of the cross. There is no electrical or RF connection between the wakefield plug and the beam tube. Instead, the length of the gap between the beam tube and plug was designed to be much less than a bunch length, which for 10 ps is 3 mm FWHM.

Although the space is limited, a dipole magnetic spectrometer was built for the LCLS GTL. This 85˚ dipole images the beam from the first YAG to the focal plane of the spectrometer in both the horizontal and vertical planes. The magnet requires a small trim coil for cancelling the remnant magnetic field. Equally important is determining a reproducible procedure for setting and cancelling this magnets field. The dipole magnet vacuum chamber also is wakefield mitigated for the straight through beam path. The mitigation is done with a plate which conforms to and covers the opening to the spectrometer.

# 2.6 EXAMPLES OF NCRF GUNS

## 2.6.1 Low Duty Factor Guns

### LANL 5½-Cell L-Band Gun

Based on the experience gained in building the first photocathode RF gun, LANL designed and operated a 1.3 GHz (L-band) gun with 5½-cells as shown in Figure **2.22** [2.7].

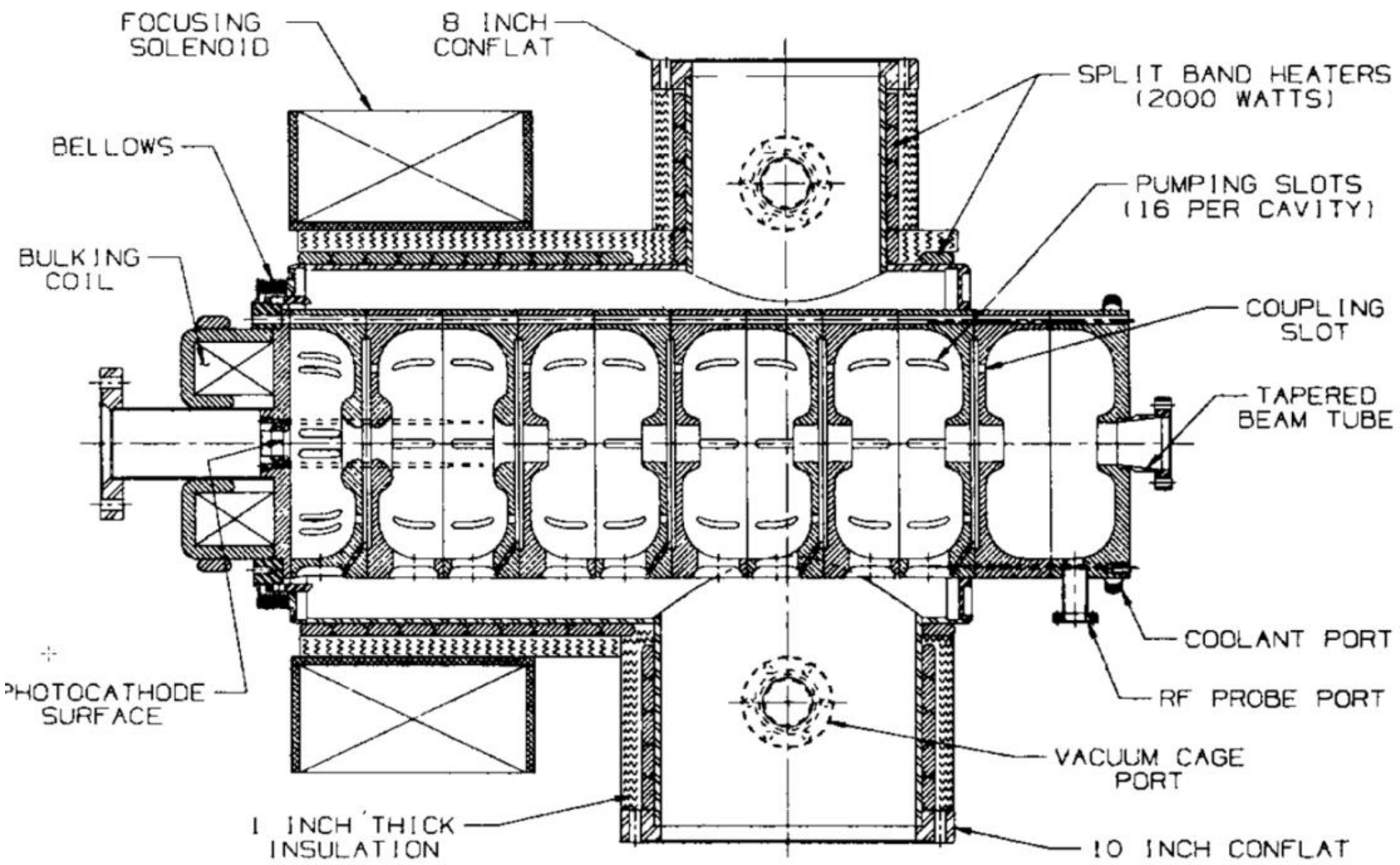

**Figure 2.22. The LANL 5½-cell RF gun with compensation solenoid and bucking coil. [Reprinted from [2.7], with permission from Elsevier.]**

This gun was built in the early days of emittance compensation, before it was realized that the beam quality is further improved if the first cell is extended from a 0.5 cell length to 0.6 long (0.6*$\lambda$/2). The gun $Q_0$ was 18 500 and the shunt impedance 53 MΩ m$^{-1}$. The gun operated at 0.1% duty factor requiring 2.4 MW of RF power to produce a 0.6 MW beam. The cell-to-cell RF fields are not balanced in this gun and average accelerating fields were 26 MV m$^{-1}$, 14 MV m$^{-1}$ and 10 MV m$^{-1}$ for cell 1 (cathode cell), 2 and 3, respectively. The gun was built with vacuum pumping preeminently in mind. The cavities were held inside a vacuum can. Slots in the cavities gave extra pumping of all the cavities into the common vacuum plenum.

The gun was used in several successful FEL experiments, most noteworthy being the regenerative amplifier FEL [2.40].

**CEA 144 MHz Gun**

At the low end of the RF frequency scale is the 144 MHz gun cell built and operated by the CEA facility at Breyères Le Châtel, France. A drawing of the gun and the test beam line is shown in Figure **2.23**. The RF structure used reentrant nose cones to boost the cathode field to 25 MV $m^{-1}$ and accelerate the beam in a single gap to ~1.5 MeV. A goal of this gun was to produce a long train of high charge bunches which are further accelerated and used to power an FEL. One operating mode was 2 nC bunches at 72 MHz giving an average current of 144 mA during a 200 μs macropulse. This remains a difficult goal to achieve, even given today's technology, with the cathode lifetime severely limiting the operating time. The first version of this gun had a very short cathode 1/*e* lifetime of only 15 min. Therefore, a second gun was built with a much better vacuum. In 2002, the new gun demonstrated 1 μm emittance for a 1 nC bunch charge, which is a difficult emittance to achieve at any frequency. The bunch length was 60 ps making the peak current ~17 A. [2.41]

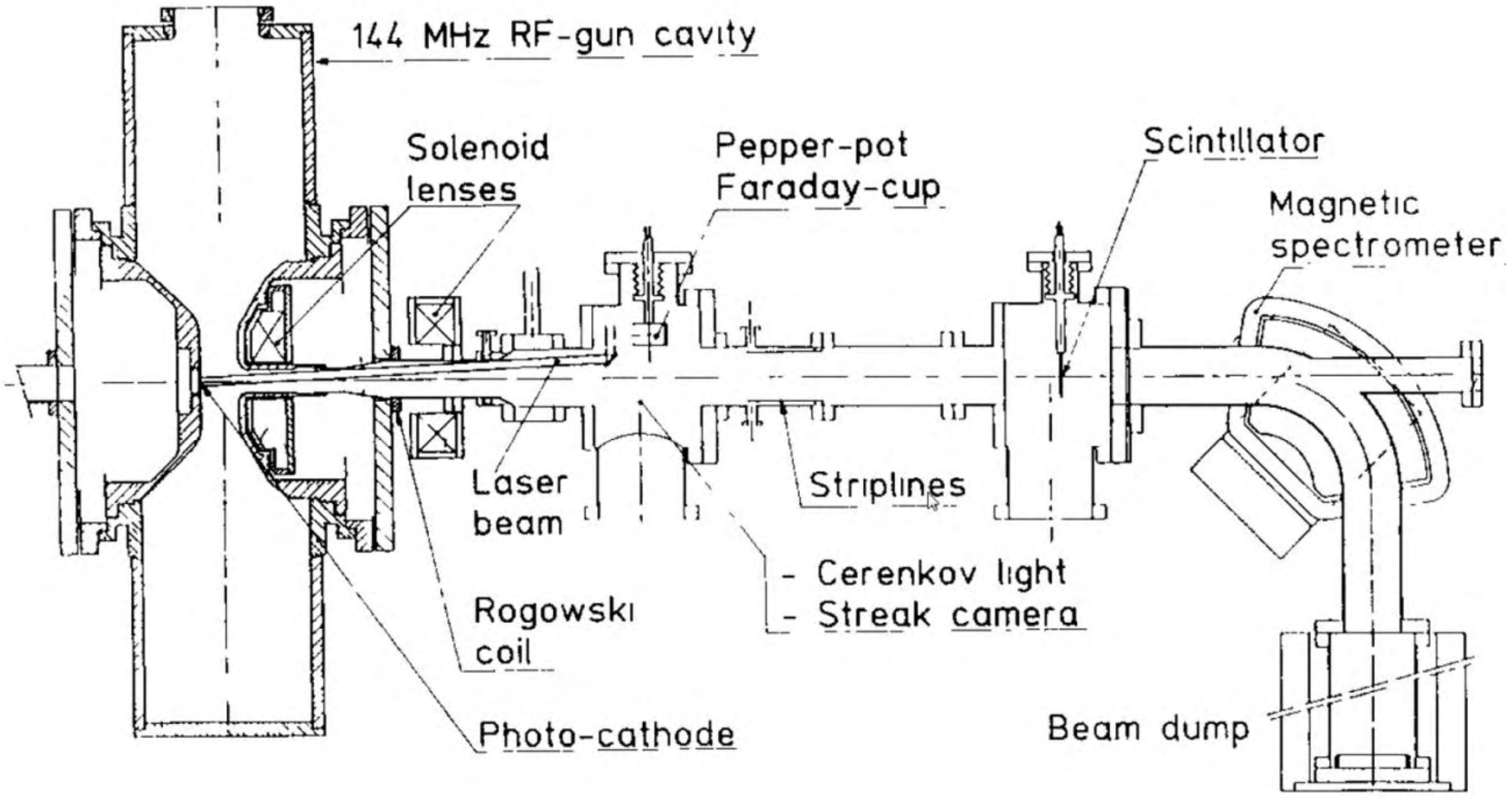

**Figure 2.23. CEA 144 MHz RF gun and beam line to first accelerator section [2.3]. [Reprinted from [2.42], with permission from Elsevier.]**

**The BNL, BNL/SLAC/UCLA, LCLS S-band guns**

In 1988, McDonald published a paper describing a pillbox S-band gun for the BNL Accelerator Test Facility (ATF) to provide electron beams for research on laser acceleration, inverse FEL's and nonlinear Compton scattering [2.43]. McDonald's paper is remarkable for all the things it reported on correctly, while missing other major factors. It described in detail the performance of a high field RF gun and was the first to quantify the RF and space charge emittances. This was done both numerically and analytically. Appendix A in [2.43] presents an interesting analysis of the radial RF fields and describes a cavity shape which minimizes the nonlinear components of these fields. But, what's missing is the gun solenoid: the paper never mentions the solenoid and its critical role in emittance compensation.

Four years later, in 1992, the Linac Coherent Light Source (LCLS) was proposed by Claudio Pellegrini and a collaboration of SLAC, Brookhaven and UCLA was formed to develop the gun. McDonald's gun became the obvious first choice for LCLS gun since it was at the same S-band frequency as the SLAC linac and simulations showed it could deliver the LCLS beam brightness. This led to the birth of the BNL/SLAC/UCLA S-band gun.

The BNL/SLAC/UCLA gun proved to be successful in the sense of being replicated in several labs and universities. In the end, it also was successful in its original goal of investigating how to make the bright beam required for LCLS and other FEL projects. Beam studies of the BNL/SLAC/UCLA gun showed what was limiting its performance and how to remove these limitations in the LCLS gun design.

**The DESY/PITZ Gun**

The DESY/PITZ gun is the first gun designed with symmetric RF fields and the first gun to use a coaxial feed for the RF power. This innovative pillbox L-band gun is illustrated in Figure **2.24**, which also shows the cathode transfer system and beamline downstream of the gun.

The coaxial feed gun has complete rotational symmetry, since there are no penetrations for laser ports or for RF probes. This simplifies the engineering design and avoids the problems of heating around the penetrations as discussed above. Another advantage of not having any ports or probes is this allows greater flexibility in placing the solenoid close to the cathode to maintain a small electron beam size. This advantage has been fully optimized in the S-band version.[2.44]

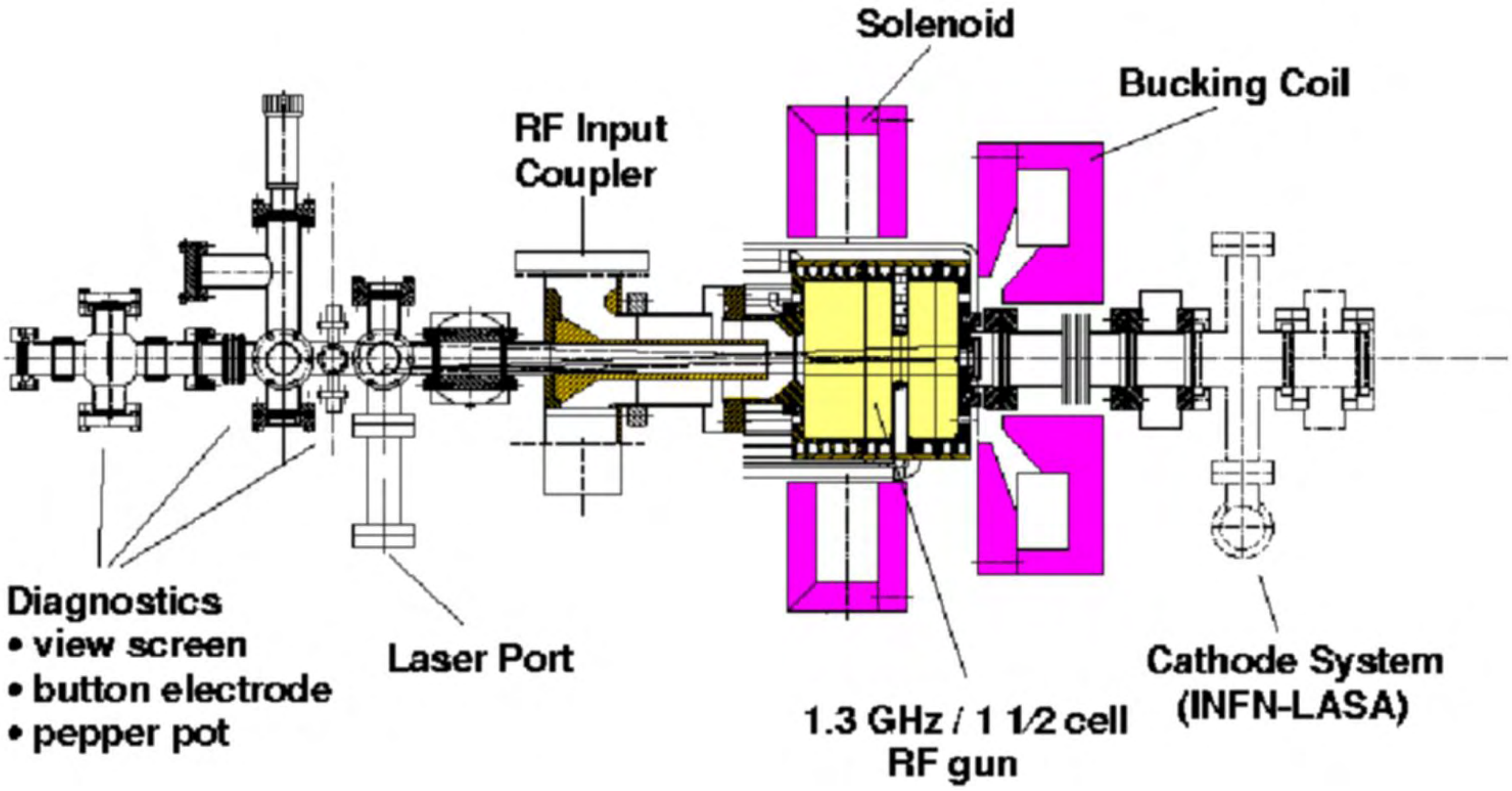

Figure 2.24. The DESY/PITZ coaxial feed gun. [Courtesy of F. Stephan (DESY/PITZ)]

However, there remain a few potential disadvantages of the coaxial feed. The first is the center conductor tube of the coax feed needs to be precisely positioned with respect to the gun cell. The coaxial is a low field region and low RF field regions are often susceptible to multipacting; this does not appear to be an issue since there is no evidence of multipacting during operation of the DESY/PITZ gun. And as an issue concerning RF controls, no penetrations means not having RF probe signals for the LLRF controls (see Section 2.3.2). Although the forward and reverse power can be used for RF controls, gun probes provide a direct measure of the cavity fields and are useful for at least the prototype version of the gun. Calculations and experience show that the field perturbation and heating around the probe port are small effects, but care is needed in locating the probe to avoid mechanical interference with the solenoid.

The DESY/PITZ gun was developed for the European X-ray Free Electron Laser and has successfully demonstrated sub-micron emittance at 1 nC which is required for that project [2.45], [2.46]. In addition this gun is capable of operating at 1% duty factor with 1 ms long RF macropulses at 10 Hz repetition rate. There can be up to 8 000 1 nC bunches per 1 ms long macropulse. The gun $Q_0$ is 21 500 and the peak cathode field is 40-60 MV m$^{-1}$.

### 2.6.2 High Duty Factor Guns

**The Boeing/LANL 433 MHz Gun**

Seven years after the first electrons were accelerated from a photocathode, a high duty factor gun demonstrated an average current of 32 mA in a 25% duty factor system. This giant leap of three orders of magnitude in duty factor was made possible due to aggressive funding by the Strategic Defense Initiative from 1983 to 1992. The electron gun was a 25% duty factor prototype of the CW electron source for a defensive free electron laser.

The gun consisted of two independently powered (and phased) reentrant cavities operating at 433 MHz. The waveguides for the two cavities dominate the photograph insert of Figure **2.25**. The accelerator and photocathode configuration used in the beam tests is shown in drawing. The beam energy out of the gun was 1.8 MeV with a peak cathode field of 26 MV m$^{-1}$. The beam was further accelerated to 5 MeV by two more independently powered 433 MHz cavities as shown in the figure. The 25% RF duty factor was produced with 8.3 ms long macropulses at a 30 Hz repetition rate and required 600 kW average RF power to produce the 26 MV m$^{-1}$ cathode field and 1.8 MeV beam. The bunch micropulse frequency was 27 MHz and with bunch charges from 1-7 nC [2.5]. $K_2CsSb$ cathodes were fabricated in a deposition chamber located behind the gun. QEs of 12% at 527 nm laser wavelength were obtained with this system. [2.47]

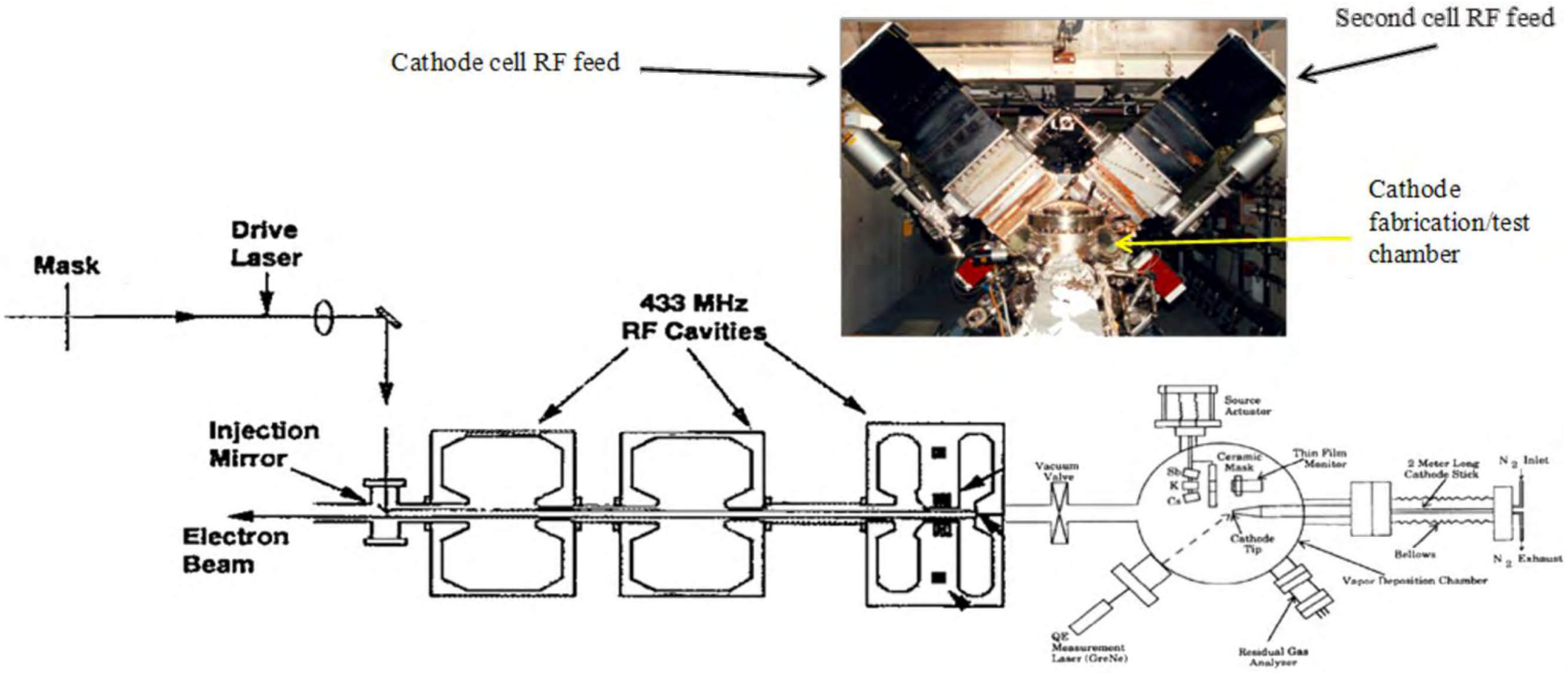

**Figure 2.25. Configuration of the 433 MHz high duty factor accelerator and cathode system. Insert: Photograph taken from the photocathode preparation chamber side (rear) of the gun. [Adapted from [2.48], with permission from Elsevier.] [Adapted with permission from [2.5]. Copyright 1993, American Institute of Physics.] [Adapted from [2.47], with permission from Elsevier.]**

**The LANL/AES 700 MHz Gun**

A truly CW NCRF gun is the 700 MHz gun designed by Los Alamos and built by Advanced Energy Systems, Inc. (AES). The unique RF coupler for this gun has been described earlier in this chapter and the full gun configuration is shown in Figure **2.26**. The gun has three cavities with accelerating fields of 7 MV m$^{-1}$, 7 MV m$^{-1}$ and 5 MV m$^{-1}$ for the cathode and second and third cavities, respectively. The low fields are required to manage the heat load and thermal stresses caused by the high dissipated power. Many of these issues were discussed earlier in the section on RF couplers. The gun has a dummy (unpowered) exit cavity functioning as a vacuum plenum with an array of eight vacuum pumps. The gun has been fabricated and is currently undergoing high power RF testing at LANL.

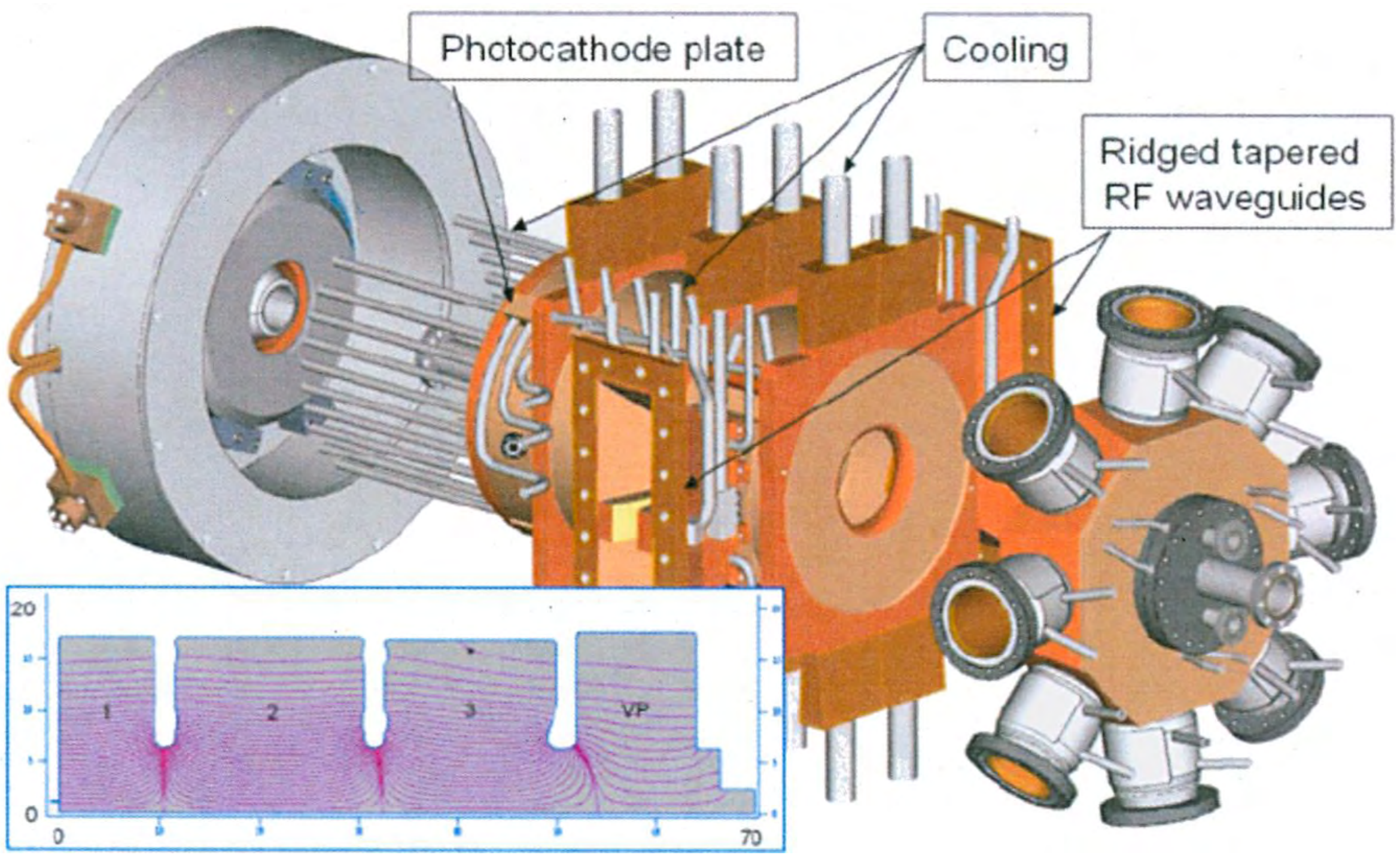

**Figure 2.26. The LANL/AES 700 MHz CW RF photocathode gun. [[2.6]; Available under Creative Common Attribution 3.0 License (www.creativecommons.org/licenses/by/3.0/us/) at www.JACoW.org.]**

**LBNL VHF Gun**

The LBNL VHF gun is at the low frequency of 187 MHz to allow CW operation. The gun was designed to have a very good vacuum. As shown earlier in Figure **2.6**, the 187 MHz gun has a row of 104 slots circling the circumference of the cavity. The outer plenum contains a battery of 24 NEG vacuum pumps with a CO pumping speed of 180 L s$^{-1}$ each, or 4 300 L s$^{-1}$ total. This immense pumping speed is necessary to reach the goal of $5\times10^{-11}$ Torr or $6.6\times10^{-9}$ Pa when operating a full CW power. This excellent vacuum is intended to give very long cathode lifetimes.

The gun is a reentrant single-cell cavity with a 40 mm accelerating gap. The cavity $Q_0$ is 30 877 and the shunt impedance is 6.5 MΩ. The cathode electric field is 19.5 MV m$^{-1}$ and produces a 0.75 MV gap voltage. At this low RF frequency, the electric field appears to be DC for a picosecond long electron bunch. Therefore, effects due to the time-dependence of the RF field, such as the RF emittance, will be small. The peak wall power density is a reasonable 25 W cm$^{-2}$ [2.21].

## 2.7 SUMMARY

This chapter reviewed the design options for RF guns which are seen to range over the entire RF spectrum. Generally, frequency and duty factor are related. Due to power dissipation limitations, the lower frequency guns are capable of higher duty factors than those of high frequencies. The maximum achievable field scales with the RF frequency, with the field limit given by the Kilpatrick criterion. Experience shows that operational fields with “bravery” factors of two and more times the Kilpatrick criterion is possible. The two cavity shape options are the pillbox and reentrant. The pillbox is easily fabricated and is generally used at high RF frequencies. The reentrant shape is more common at low frequencies and has higher on-axis field than the pillbox for the same RF power.

After reviewing the basics of RF gun theory, the two methods for coupling the RF power into the gun were discussed. These are side coupling and coaxial coupling. For the more commonly used side coupling, the RF power enters through a port hole located in the side of the cavity. For guns with multiple cavities, the power then flows into the other cavities through the irises between the cavities, or through slots in the walls between the cavities. The major disadvantages of side coupling are local heating by average and pulsed heating around the edge of the port, and the introduction of field asymmetries. The field asymmetries can be corrected using dual feeds and a racetrack shape of the coupling cavity. In addition, the waveguide’s location protruding from the side of the cavity limits the options for placement of the solenoid. In coaxial coupling, the RF power is brought into the gun on the cavity’s axis using a “door-knob” transition which converts the rectangular waveguide mode into an on-axis coaxial mode. The RF power fills the gun through the cavity irises. The advantages of coaxial coupling are the absence of penetrations in the cavity walls and their induced local heating and a symmetric field without the need for specialized cavity shapes. The solenoid can be placed much closer to the cathode, since the door-knob coupler is now downstream of the gun. A disadvantage is the possibility of multipacting in the low field regions of the center conductor.

Examples of two high performance guns are described in some detail: One operating at low duty factor and high field, and the other designed for high duty factor and low field. The low duty factor, high field gun was designed for an X-ray Free Electron Laser (FEL). The high field was essential to get the high brightness needed for the SASE FEL. The high duty factor, low field gun is intended for use in a CW FEL operating at IR wavelengths. In each case, the RF coupler issues dominate the design.

The RF power system is briefly described illustrating the need for control loops on both the low level RF and the cooling water to maintain the gun’s resonant frequency. This is followed by sections discussing the emittance due to asymmetric RF fields and solenoid effects in a working gun. The theory is given in Chapter 1.

Wake field mitigation and electron beam diagnostics in the gun-to-linac region is then briefly reviewed. The mitigation of the beam view screens is described in more detail in Chapter 10. Here, the added feature of shielding the spectrometer vacuum chamber is discussed.

The chapter finishes with a short descriptions and comments on examples of low- and high-duty factor RF guns built since the invention of the photoinjector. The low duty factor guns are represented by the LANL 5½-cell 1 300 MHz gun, the CEA 144 MHz gun, the BNL S-band gun and the DESY-PITZ L-band gun. The high-duty factor guns are the Boeing/LANL 433 MHz gun, the LANL 700 MHz gun and the LBNL VHF gun.

The intent of this chapter was to introduce the reader to the rich diversity of the NCRF guns which span nearly the entire range of RF frequency and operational duty factor. It is hoped that this short introduction educates new practitioners of this dynamic field and becomes a useful reference for the experts as they continue to advance the technology of NCRF photoinjectors and electron sources in general.

## 2.8 CONFLICT OF INTEREST AND ACKNOWLEDGEMENTS

The author confirms that this article content has no conflicts of interest.

***References***

[2.1] W. D. Kilpatrick, "Criterion for vacuum sparking designed to include both RF and DC," *Rev. Sci. Instrum.*, vol. 28, pp. 824-826, October 1957.

[2.2] T. Wangler, *RF Linear Accelerators*, Weinheim: Wiley-VCH Verlag, 199, pp. 1608.

[2.3] S. Joly, P. Balleyguier, Cl. Bonetti *et al.*, "Progress report on the BRC photo-injector," in *Proc. 1990 European Particle Accelerator Conf.*, 1990, pp. 140-142.

[2.4] J. W. Staples, K. M. Baptiste, J. N. Corlett *et al.*, "Design of a vhf-band rf photoinjector with megahertz beam repetition rate," in *Proc. 2007 Particle Accelerator Conf.*, 2007, pp. 2990-2992

[2.5] D. H. Dowell, K. J. Davis, K. D. Friddell *et al.*, "First operation of a photocathode radio frequency gun injector at high duty factor," *Appl. Physics Lett.*, vol. 63, pp. 2035-2037, October 1993.

[2.6] S. S. Kurennoy, D. C. Nguyen, D. L. Schrage *et al.*, "Normal-conducting high current RF photoinjector for high power cw FEL," in *Proc. 2005 Particle Accelerator Conf.*, 2005, pp. 2866-2868.

[2.7] L. M. Young, "Compact photoinjector accelerators for free electron lasers," *Nucl. Instrum. Meth. B*, vol. 56-57, pp. 978-981, May 1991.

[2.8] R. Bakker, M. v. Hartrott, E. Jaeschke *et al.*, "First measurements at the photo injector test facility at DESY ZEUTHEN," in *Proc. 2002 European Particle Accelerator Conf.*, 2002, pp. 1813-1815.

[2.9] L. Xiao, R. F. Boyce, D. H. Dowell *et al.*, "Dual feed RF gun design for the LCLS," in *Proc. 2005 Particle Accelerator Conf.*, 2005, pp. 3432-3434.

[2.10] A. E. Vlieks, V. Dolgashev, S. Tantawi *et al.*, "X-band RF gun development," in *Proc. 2010 Int. Particle Accelerator Conf.*, 2010, pp. 3816-3818.

[2.11] A. S. Kesar, J. Haimson, S. Korbly *et al.*, "Initial testing of a field symmetrized dual feed 2 MeV 17 GHz RF gun," in *Proc. 2003 Particle Accelerator Conf.*, 2003, pp. 2095-2097.

[2.12] A. Brinkmann, D. Reschke and J. Ziegler, "Various applications of dry-ice in the field of accelerator components at DESY," in *Proc. 2008 Linear Accelerator Conf.*, 2008, pp. 803-805.

[2.13] F. Stephan, J. W. Bähr, C. H. Boulware *et al.*, "New experimental results from PITZ," in *Proc. 2008 Linear Accelerator Conf.*, 2008, pp. 474-476.

[2.14] J. W. Staples, S. M. Lidia, S. P. Virostek *et al.*, "The LBNL femtosource (LUX) 10 KHz photoinjector," in *Proc. 2003 Particle Accelerator Conf.*, 2003, pp. 2092-2094.

[2.15] J. D. Jackson, *Classical Electrodynamics*, 3rd Ed., New York: Wiley, 1999, pp. 373.

[2.16] J. H. Billen and L. M. Young, "Poisson SUPERFISH," Los Alamos National Laboratory, Technical Report LA-UR-96-1834, updated 2003.

[2.17] T. Wangler, *RF Linear Accelerators*, Weinheim: Wiley-VCH Verlag, 1998, pp. 141-144.

[2.18] J. Haimson, B. Mecklenburg and E. L. Wright, "A racetrack geometry to avoid undesirable azimuthal variations of the electric field gradient in high power coupling cavities for TW structures," in *Proc.AIP Conf.*, vol. 398, pp. 898-911, March 1997.

[2.19] D. H. Dowell, E. Jongewaard, J. Lewandowski *et al.*, "The Development of the Linac Coherent Light Source RF Gun," LCLS, Stanford, California, Technical Report No. SLAC PUB-13401, September 2008.

[2.20] B. Dwersteg, K. Flöettmann, J. Sekutowicz *et al.*, "RF gun design for the TESLA VUV free electron laser," *Nucl. Instrum. Meth. A*, vol. 393, pp. 93-95, July 1997.

[2.21] K. Baptiste, J. Corlett, R. Kraft *et al.*, "Status and plans for the LBNL normal-conducting cw vhf photo-injector," in *Proc. 2009 Free Electron Laser*, 2009, pp. 470-472.

[2.22] P. Pritzkau and R. H. Siemann, "Experimental study of RF pulsed heating on oxygen free electronic copper," *Phys. Rev. ST Accel. Beams*, vol. 5, pp. 112002-1–112002-22, November 2002.

[2.23] D. H. Dowell, L. Bentson, R. F. Boyce *et al.*, "RF design for the linac coherent light source (LCLS) injector," in *Proc. 2004 Free Electron Laser Conf.*, 2004, pp. 538-541.

[2.24] R. Boyce, D. H. Dowell, J. Hodgson *et al.*, "Design considerations for the LCLS RF gun," LCLS, Stanford, California, Technical Report No. LCLS-TN-04-4, April 2004.

[2.25] S. S. Kurennoy, D. L. Schrage, R. L. Wood *et al.*, "Photoinjector RF cavity design for high power CW FEL," in *Proc. 2003 Particle Accelerator Conf.*, 2003, pp. 920-922.

[2.26] S. S. Kurennoy and L. M. Young, "RF coupler for high-power CW FEL photoinjector," in *Proc. 2003 Particle Accelerator Conf.*, 2003, pp. 3515-3517.

[2.27] L. M. Young, D. E. Rees, L. J. Rybarcyk *et al.*, "High power RF conditioning of the LEDA RFQ," in *Proc. 1999 Particle Accelerator Conf.*, 1999, pp. 881-883.

[2.28] MicroWave Studio, ver. 4.2, available by Computer Simulation Technology GmbH, Bad Nauheimer Str. 19, 64289 Darmstadt, Germany. Available at www.cst.com/Content/Products/MWS/Overview.aspx.

[2.29] SCM Metal Products, http://en.wikipedia.org/wiki/Glidcop.

[2.30] Z. Li, N. Folwell, L. Ge *et al.*, "High performance computing in accelerator structure design and analysis," in *Proc. 2004 Int. Computational Accelerator Physics*, 2004.

[2.31] Z. Li, J. Chan, L. D. Bentson *et al.*, "Coupler design for the LCLS injector s-band structures," in *Proc. 2005 Particle Accelerator Conf.*, 2005, pp. 2176-2178.

[2.32] J. F. Schmerge, J. Castro, J. E. Clendenin *et al.*, "The s-band 1.6 cell RF gun correlated energy spread dependence on π and 0 mode relative amplitude," in *Proc. 46$^{th}$ Physics Applications High Brightness Electron Beams Workshop*, 2005, pp. 375-382.

[2.33] L. M. Young, "PARMELA documentation," Los Alamos National Laboratory, Los Alamos, NM, Technical Report No. LA-UR-96-1835 (Revised June 8, 2004).

[2.34] C. Limborg, Z. Li, L. Xiao *et al.*, "RF Design of the LCLS Gun," LCLS, Stanford, California, Technical Report No. LCLS-TN-05-3, February 2005.

[2.35] J. Schmerge, "LCLS gun solenoid design considerations," LCLS, Stanford, California, Technical Report No. LCLS-TN-05-14, June 2005.

[2.36] M. Reiser, *Theory and Design of Charged Particle Beams*, 1$^{st}$ Ed., Weinheim: Wiley-VCH Verlag, 1994, pp. 313.

[2.37] S. M. Gierman, *Streak Camera Enhanced Quadrupole Scan Technique for Characterizing the Temporal Dependence of the Trace Space Distribution of a Photoinjector Electron Distribution*, Ph.D. Thesis, University of California, San Diego, CA, 1999, Appendix C.

[2.38] S. Gilevich, private communication.

[2.39] R. Akre, D. Dowell, P. Emma *et al.*, "Commissioning of the linac coherent light source," *Phys. Rev. ST Accel. Beams*, vol. 11, pp. 030703-1–030703-20, March 2008.

[2.40] D. C. Nguyen, R. L. Sheffield, C. M. Fortgang *et al.*, “A high-power compact regenerative amplifier FEL,” in *Proc. 1997 Particle Accelerator Conf.*, 1997, pp. 897-899.

[2.41] J.-G. Marmouget, A. Binet, Ph. Guimbal *et al.*, “Present performance of the low-emittance, high-bunch charge ELSA photo-injected linac,” in *Proc. 2002 European Particle Accelerator Conf.*, 2002, pp. 1795-1797.

[2.42] R. Dei-Cas, P. Balleyguier, J. Bardy *et al*., “Photo-injector, accelerator chain and wiggler development programs for a high peak power rf free electron laser,” *Nucl. Instrum. Meth. A*, vol. 285, pp. 320-326, December 1989.

[2.43] K. T. McDonald, “Design of the laser-driven RF electron gun for the BNL accelerator test facility,” *Trans. Electron Devices*, vol. 35, pp. 2052-2059, November 1988.

[2.44] J-H. Han, M. Cox, H. Huang *et al.*, “Design of a high repetition rate S-band photocathode gun,” *Nucl. Instrum. Meth. A*, vol. 647, pp. 17-24, August 2011.

[2.45] F. Stephan, C. H. Boulware, M. Krasilnikov *et al*., “Detailed characterization of electron sources yielding first demonstration of European X-ray free-electron laser beam quality,” *Phys. Rev. ST Accel. Beams*, vol. 13, pp. 020704-1–020704-33, February 2010.

[2.46] S. Rimjaemm, G. Asova, J. Bähr *et al.*, “Measurements of transverse projected emittance for different bunch charges at PITZ,” in *Proc. 2010 Free Electron Laser Conf.*, 2010, pp. 410-413.

[2.47] D. H. Dowell, S. Z. Bethel and K. D. Friddell, “Results from the average power laser experiment photocathode injector test,” *Nucl. Instrum. Meth. A*, vol. 356, pp. 167-176, March 1995.

[2.48] A. Todd, “State-of-the-art electron guns and injector designs for energy recovery linacs (ERLs),” *Nucl. Instrum. Meth. A*, vol. 557, pp. 36-44, February 2006.

# CHAPTER 3: SUPERCONDUCTING RF PHOTOINJECTORS

JOHN W. LEWELLEN

*Naval Post Graduate School*

*Monterey, CA 93943*

**Abstract**

High average current RF injectors typically operate at a higher duty factor, while maintaining as high an accelerating gradient as possible. This combination gives rise to increased wall current and RF loss leading to thermal management issues. One way is to address it is to operate it in superconducting RF (SCRF) mode. The development of SCRF injectors are still in its infancy. In this chapter, we establish the need for SCRF injectors and describe a number of systems that are currently in various stages of development. For those interested in building a SCRF injector, we list a number of questions to be answered prior to starting the design process and key issues that would impact the design, which is followed by the evolution of the design process and validation.

## 3.1 INTRODUCTION AND ASSUMPTIONS

This chapter centers on the beam physics design concerns for superconducting RF (SCRF) photoinjectors. An SCRF photoinjector is a specialized superconducting accelerator, and there are wealths of knowledge regarding various aspects of SCRF accelerator design and operation. Repeating that work in this chapter would not be fruitful in no small part because the author is a beam dynamicist and not an SCRF accelerator expert. Thus, many topics of high importance (such as multipacting and the resulting restrictions on cavity shape) to SCRF accelerator design and operation are only lightly touched upon here, simply to highlight issues of which the SCRF injector designer must be aware.

SCRF photoinjector design is still in its infancy. At least three different design paradigms are currently being developed; there are certainly other possibilities, as well. While the state of affairs results in an exciting area of research, unfortunately, it also limits the number of “guaranteed-to-work” suggestions one can offer. We do not know enough about how these devices will operate; it will be interesting finding out.

I start with some nomenclature and assumptions. A “cavity” refers to a single RF structure, and may consist of one or more “cells,” or regions in which the beam is accelerated. For instance, a standard 1.3 GHz TESLA structure is a 9-cell cavity. Generally, the phrase “cathode cell” refers to the first cell in a multi-cell photoinjector, within which the cathode is located. A “full cell” is taken to be a cavity of length ~$\lambda/2$, where $\lambda$ is the free-space RF wavelength corresponding to the injector’s operating frequency. The term “injector” refers here to SCRF photoinjectors unless stated otherwise. The “gun” is the portion of the injector containing the cathode cell. References to beam energy are to the beam’s kinetic energy, unless stated otherwise. For temperatures, “2K” refers to a 2.1 K superfluid liquid helium system. The term “4K” refers to either a 4.2 K system in which the liquid helium boils at standard atmospheric pressure, or to a 4.5 K system employing a supercritical refrigerator.

## 3.2 WHY SHOULD WE USE AN SCRF PHOTOINJECTOR?

A quick perusal of the preceding chapters illustrates that there is a large, thorough body of work, both theoretical and experimental, on developing and using normal conducting RF (NCRF) photoinjectors. Much of the initial theory rested upon analytical models of on- and near-axis RF fields with simple sinusoidal longitudinal components; aperture-coupled pillbox cavities present a convenient way to generate such fields. Typically, however, such cavity profiles are unsuitable for SCRF applications due to their susceptibility to multipacting. SCRF cell geometry in general, and cathode support structures in gun cells, in particular, can lead to electromagnetic field profiles considerably different from those assumed in classic emittance compensation theory. As a result, SCRF injector design places more emphasis on computer simulation rather than on theoretical modeling. There are fewer points of direct comparison with basic theory and correspondingly less guidance when moving into new operational regimes.

Peak electric and magnetic field strengths on the surface of the cavity are often of more immediate concern in an SCRF- than in an NCRF-cavity design. Too high of a peak magnetic field in an SCRF cavity can “quench” the cavity, or cause it to locally become non-superconducting. Multipacting can occur at modest surface electric fields, while high surface fields can lead to field emission. While field emission and multipacting can be problematic in any high gradient accelerator cavity, these effects can be especially pernicious within an SCRF accelerator for two reasons. First, even modest levels of field emission or multipacting can greatly increase the amount of RF power the cavity requires to attain a given gradient; second, the resulting heating from these phenomena increases the load on the cryogenics system and can quench the cavity. High multipacting currents can also pull the cavity resonant frequency off the design value.

Solenoidal magnetic fields are used in the immediate vicinity of essentially all NCRF photoinjectors as part of the emittance compensation process, as described in Chapter 1. Generally, these fields are applied as close to the cathode as possible. Typical installations also include employing a “bucking” solenoid sited on or near the back wall of the cathode cell to zero out the on-axis magnetic field at the cathode. In contrast, in designing SCRF accelerators, great efforts are made to keep magnetic fields away from the cavities, such as incorporating magnetic shielding into the accelerator’s support structures to reduce the earth’s magnetic field. This introduces challenges in designing both the accelerator structures and beam dynamics, as compromise is required.

Finally, an SCRF photoinjector must be operated at cryogenic temperatures; depending on the cavity resonant frequency and facility, this temperature might be 2K or 4K. However, in both cases, the injector cavity is surrounded successively by a liquid helium bath, a vacuum, a magnetic shield, a liquid nitrogen-cooled heat shield and another vacuum. The cryostat containing the gun will also incorporate layers of superinsulation, cavity tuning mechanisms and so forth. Both the design effort (for items such as cavity field probes, RF power couplers and laser injection ports) and routine maintenance (such as replacing the cathode) will be more complicated than for an NCRF injector.

None of these issues are insurmountable, but overall pose the question of whether it is worthwhile to choose an SCRF photoinjector since NCRF photoinjectors operate well. I believe the answer can be an emphatic “yes,” depending on the tasks the photoinjector is to perform. I consider several regimes in which an SCRF photoinjector is an eminently, or the only, practical choice.

### 3.2.1 High Average Current or High Duty Factor Operation

The amount of power required to generate a given beam current from an RF photoinjector is the sum of the RF power needed to generate the accelerating fields (which scales with the shunt impedance), the RF power delivered to the beam (beam loading) and the power required to extract waste heat from the cavity. For a superconducting accelerator, there is also a power cost associated with keeping the cavity cold, even when it is not operating.

Regardless of the required beam current, NCRF photoinjectors (and NCRF accelerating structures in general) typically require large amounts of RF power to maintain their accelerating fields at operating levels. For instance, a typical SLAC/BNL/UCLA "Gen-IV" S-band photoinjector has a shunt impedance of around 3 MΩ [3.1].[2] Therefore, to obtain a 5 MeV beam energy, approximately 8 MW of RF power must be delivered. This is not a problem if the photoinjector operates only for a few microseconds at a time. Most S-band NCRF linac installations use klystrons rated at 40-50 MW pulsed output, but with *average* power outputs on the order of several tens of kilowatts at most. 8 MW, continuous wave (CW) klystrons at any frequency are rare, which is one obstacle towards running such a photoinjector CW. Even assuming a 1 MeV beam energy, for example, was satisfactory, 350 kW would be required. This still is high power for a CW klystron.

Even positing the availability of a CW klystron (or other RF source) at the required frequency and power, the RF power lost into the walls of the photoinjector must be removed. At high duty factors, dissipating such heat is a non-trivial task; the Los Alamos CW NCRF photoinjector, operating at 700 MHz, requires more than 500 kW RF power simply to generate the desired accelerating fields of 5-7 MV $m^{-1}$ needed to obtain the design beam energy of 2.7 MeV [3.2]. The gun's structure arguably consists of more water passages than copper. At gradients and frequencies much above these levels, heat cannot be removed quickly enough from the cavity's inner surface to sustain CW operation.

In comparison, the ELBE SCRF photoinjector has a shunt impedance of around 3.4 TΩ when at its operating temperature of 2.1 K. Therefore, obtaining its nominal accelerating voltage of 9.6 MV requires a mere 26 W of RF power [3.3]. Since each watt removed from a 2K operating bath requires approximately 1 kW at room temperature [3.4], only ~26 kW of input power is required to dissipate the heat generated from running the photoinjector CW at high gradients.[3] While this is a non-trivial refrigeration system, it must be compared to the multi-megawatt CW RF system and water-cooling plant a conventional copper linac would need to operate CW at ~10 MeV.

The outcome is that for pulsed beams, up to 0.1-1 kHz, an NCRF photoinjector is an excellent choice. As the RF duty factor increases (either due to a demand for higher average beam currents, or for higher bunch repetition rates), the challenges associated with cooling an NCRF injector and providing the required RF power outweigh those associated with operating a superconducting injector. The crossover point in this decision will vary, depending upon particular requirements and system parameters[4] (and the opinions of those discussing the choices), but generally will lie between 1 μA and 1 mA average beam current. At 10 μA, a 10 MeV CW injector requires 100 W for the beam, in addition to that required to sustain the accelerating fields. For an SCRF photoinjector that value most likely already would exceed the cavity's RF power requirements, whereas for an NCRF photoinjector, it barely would be a noticeable additional load on the RF system.

[2] For shunt impedance, I use the definition $r_s = V_{acc}^2 P_c^{-1}$, where $V_{acc}$ is the accelerating voltage and $P_c$ is the RF power. A common alternate definition has an additional factor of two in the denominator.
[3] This assumes no problems with – or additional power consumption by – field emission or multipacting.
[4] Among other factors, the crossover point will depend upon the RF frequency, $Q$ of the cavity, the required electron beam bunch pattern, and average current.

Typical envisioned applications for high average current machines include Energy Recovery Linacs (ERLs) for light source applications and accelerators for producing radioisotopes.

### 3.2.2 Small Stand-Alone Installations

The potential applications discussed below are highly speculative and have not been explored with the thoroughness of light source-related applications. However, it is the author's belief that, there is great potential for small, stand-alone applications of superconducting injectors.

Electron microscopy is a very interesting "non-traditional" application for high-brightness photoinjectors. Intrinsically pulsed and intended to deliver high quality beams at megaelectron volt range energies, RF photoinjectors increasingly are viewed as candidates for next generation time-resolved electron microscope sources [3.5]. Initial experiments using S-band NCRF photoinjectors proved promising [3.6], [3.7]. SCRF injector-based electron beam sources are natural candidates for electron microscopy applications because they can operate in CW operation and offer a range of options for adjusting the beam's repetition rate, average current and bunch charge.

A related application would be a modest energy SCRF injector as an electron beam welder. Electron beam powers of 100-1 000 W (0.1-1 mA beam currents at 1 MeV), combined with very low emittance, would allow the precise application of extremely high power densities; the megavolt range beam energy would allow deeper penetration of the electron beam into the joint being welded. The net result could be more efficient, faster and produce cleaner welding. Since electron beam welding is a primary fabrication technique for superconducting cavities, there is a certain bootstrapping appeal to this application.

In any such application, the injector likely would be operated as a stand-alone apparatus in a small facility (relative to most accelerators). A suitable selection of operating frequency will allow an SCRF photoinjector to operate at 4K, simplifying the needed cryogenic system. CW operation would require only modest RF power to achieve high average beam power, as discussed above. The size of the overall installation needed would be comparable to that of a stand-alone pulsed NCRF photoinjector; the advantage of the SCRF photoinjector is that, in principle, it could be operated CW with relatively minor expense, while the NCRF photoinjector likely could not be. Whilst SCRF injectors, and particularly these types of applications, are in the very early stages of development, the ability to generate low emittance beams at useful currents and powers in small installations might lead to many applications beyond those discussed here.

## 3.3 PRESENT STATE-OF-THE-ART

Seemingly a greater variety of SCRF photoinjector types is under development than there are NCRF photoinjector variants under investigation. While refinements continue to be developed and implemented for the latter, their designs generally have converged toward a pillbox geometry, with a cathode cell typically slightly longer than $\lambda/4$ integrated with one to two full ($\lambda/2$) cells. (There are exceptions, such as the LBNL's VHF gun [3.8].) The NCRF gun typically is followed by a drift space and linac structures to freeze the emittance *via* additional acceleration. This basic design works very well in practice. In contrast, SCRF photoinjector development is comparable to the state of NCRF photoinjectors two decades ago; there is much to learn, and what works "best" has not been determined.

The two major SCRF injector design paradigms are single-cell and multi-cell guns, often, but not always, followed by booster linac sections before the beam's entry into the main linac (*via*, *e.g.*, an ERL merger). The designs of multi-cell SCRF guns are often similar in concept to those of multi-cell NCRF guns. There are also variations in the design of the individual SCRF cathode cells; the two main variants currently being developed are elliptical cells and quarter-wave cells.

This section exhaustively lists the current projects worldwide developing SCRF guns. Many laboratories, universities and companies are engaged in such research, and every conference seemingly brings new ideas and participants. Rather, this section surveys the current major design paradigms for SCRF injectors, in particular guns, illustrating them with existing projects.

### 3.3.1 Multi-Cell Gun Designs

**ELBE Injector**

The ELBE SCRF photoinjector (also termed the Rossendorf or Drossel SCRF injector) was the first SCRF photoinjector in the world to operate as an injector. It produced first beam in November 2007 [3.9] and in 2010, the ELBE accelerator started operating with this source [3.10]. A cross section view is shown in Figure **3.1(a)**. The ELBE injector is powered through an ELBE 10 kW RF power coupler [3.11] that is visible entering the beam pipe at the downstream (right) end of the injector. The three full cells are TESLA-type elliptical cells. The cathode cell is a foreshortened elliptical cavity.

The ELBE injector operating frequency is 1.3 GHz, the same as the ELBE linac. The design energy is 9.5 MeV.

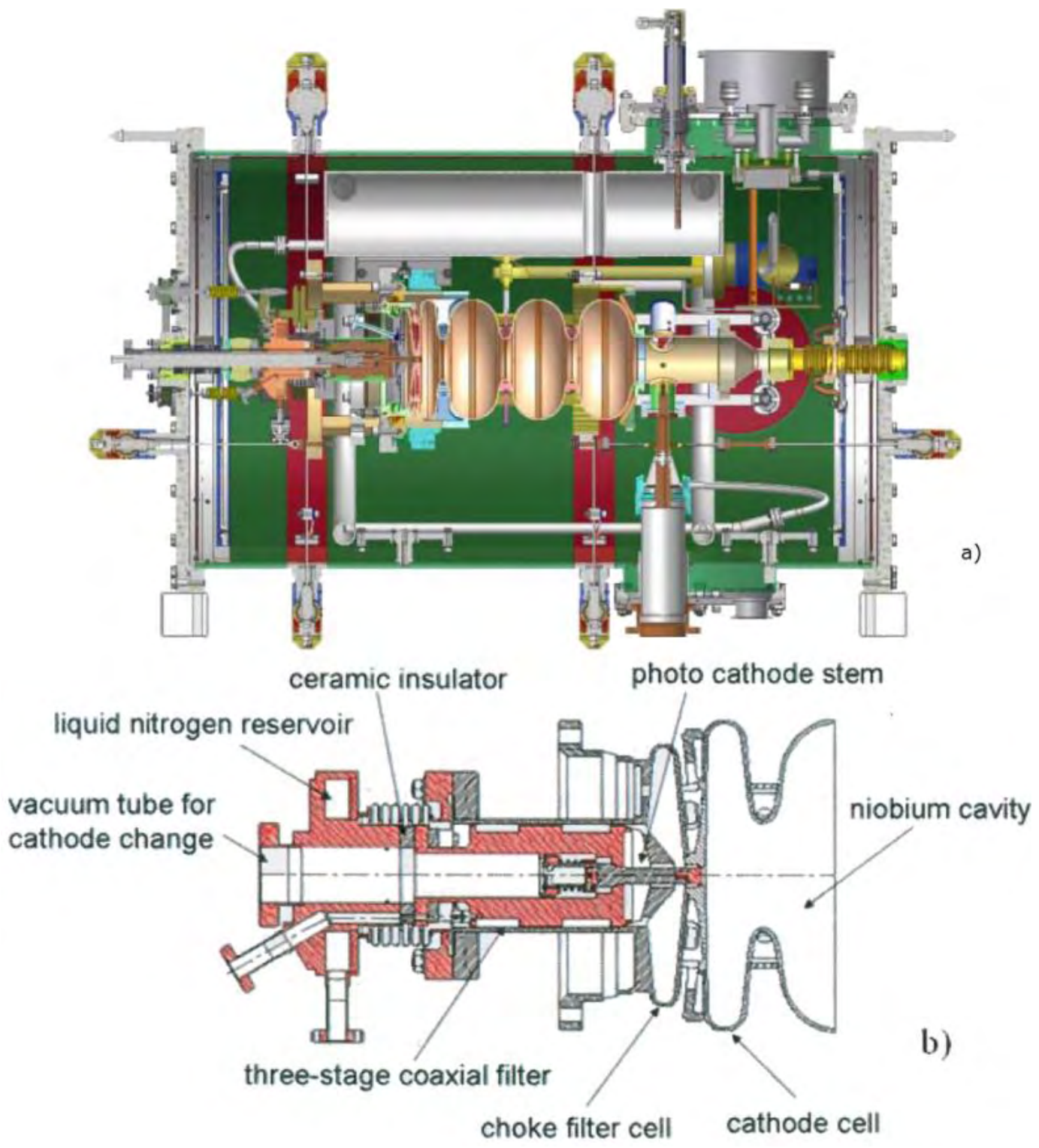

**Figure 3.1. (a) A cutaway view of the ELBE SCRF injector. The ELBE RF power coupler can be seen entering the beam pipe, to the right of the third full cell. (b) Design of the EBLE injector cathode choke joint and cathode insertion mechanism. [[3.12]; Available under Creative Common Attribution 3.0 License (www.creativecommons.org/licenses/by/3.0/us/) at www.JACoW.org.] [Courtesy of A. Arnold]**

The cathode in the ELBE injector does not touch the walls of the SCRF cavity; rather, an RF choke cavity (located to the left of the cathode cell in Figure **3.1**) provides RF isolation. The injector has operated with both Cu and $Cs_2Te$ cathodes; the latter is the cathode of choice for operation due to its higher quantum efficiency (QE). No degradation was observed in the cavity's performance due to contamination from the cathode after 500 hr of operation. An external liquid $N_2$ reservoir cools the cathode. The design of the choke joint and the cathode's insertion mechanism is shown in Figure **3.1(b)**.

Several early variations of this injector's design incorporated RF-based focusing schemes near the cathode [3.13]. In the ELBE injector as built, a solenoid is located in the beamline after the 3.5-cell gun. An interesting method of performing emittance compensation was proposed, using this injector as the design platform. Rather than employing an electromagnet to generate a solenoidal field, a TE (magnetic) mode is excited in one of the cells in the gun. As the beam traverses this cell, as well as being accelerated, it receives a focusing kick suitable for emittance compensation [3.14].

The beam from the 3.5-cell ELBE photoinjector is directed into the ELBE linac without further acceleration.

**BERLinPro Injector Test Gun**

The HZB BERLinPro injector is envisioned as operating with a 1.6-cell SCRF gun, followed by a higher energy booster linac. The injector for BERLinPro must deliver 100 mA average current, with 1 μm emittance and 77 pC bunch charge. To support high average beam currents, the BERLinPro injector's baseline cathode is $Cs_2KSb$ on a normal conducting insert. A staged approach towards attaining these parameters was adopted. In the first stage, the BERLinPro prototype gun is to be installed in the HoBiCaT cryovessel. The first prototype gun will have a portion of the back wall of the cathode cell coated with a Pb film to act as the cathode. RF power is provided *via* a coaxial coupler into the beam pipe.

The design frequency of the BERLinPro injector is 1.3 GHz for compatibility with the remainder of the linac. Its nominal beam energy is 1.5 MeV; the beam will be followed by a booster linac to raise the beam energy to 5-10 MeV [3.15]. The basic concept for the beam dynamics is similar to that of a typical 1.6-cell S-band NCRF injector.

As Figure **3.2** shows, a superconducting solenoid will be placed immediately downstream of the RF power coupler, providing the ability to perform emittance compensation. The solenoid is designed such that its fringe fields do not interfere when the cavity is transitioning to its superconducting state.

The next stage of the planned program of incremental improvements and upgrades includes implementing a gun with a $K_2CsB$ cathode to support operations at high average currents.

### 3.3.2 Single-Cell Gun Designs

Single-cell guns arguably are preferable when high average current beams are required because this minimizes the amount of power that a single RF coupler must supply. They also are often used when a project's goal is to test aspects of SCRF injector technology related to developing cathodes or addressing electron beam formation, as fabrication generally is simpler than for multi-cell structures.

The primary disadvantage to single-cell gun designs compared with multi-cell designs is the relatively low beam energy that is produced, due to limitations on achievable gradients. For high bunch charges (~1 nC) especially, this can lead to difficulties in transporting the beam from the gun to the remainder of the injector.

Currently (at least) two types of single-cell gun designs are in development: Cells based upon elliptical geometries and cells based upon quarter-wave resonator geometries.

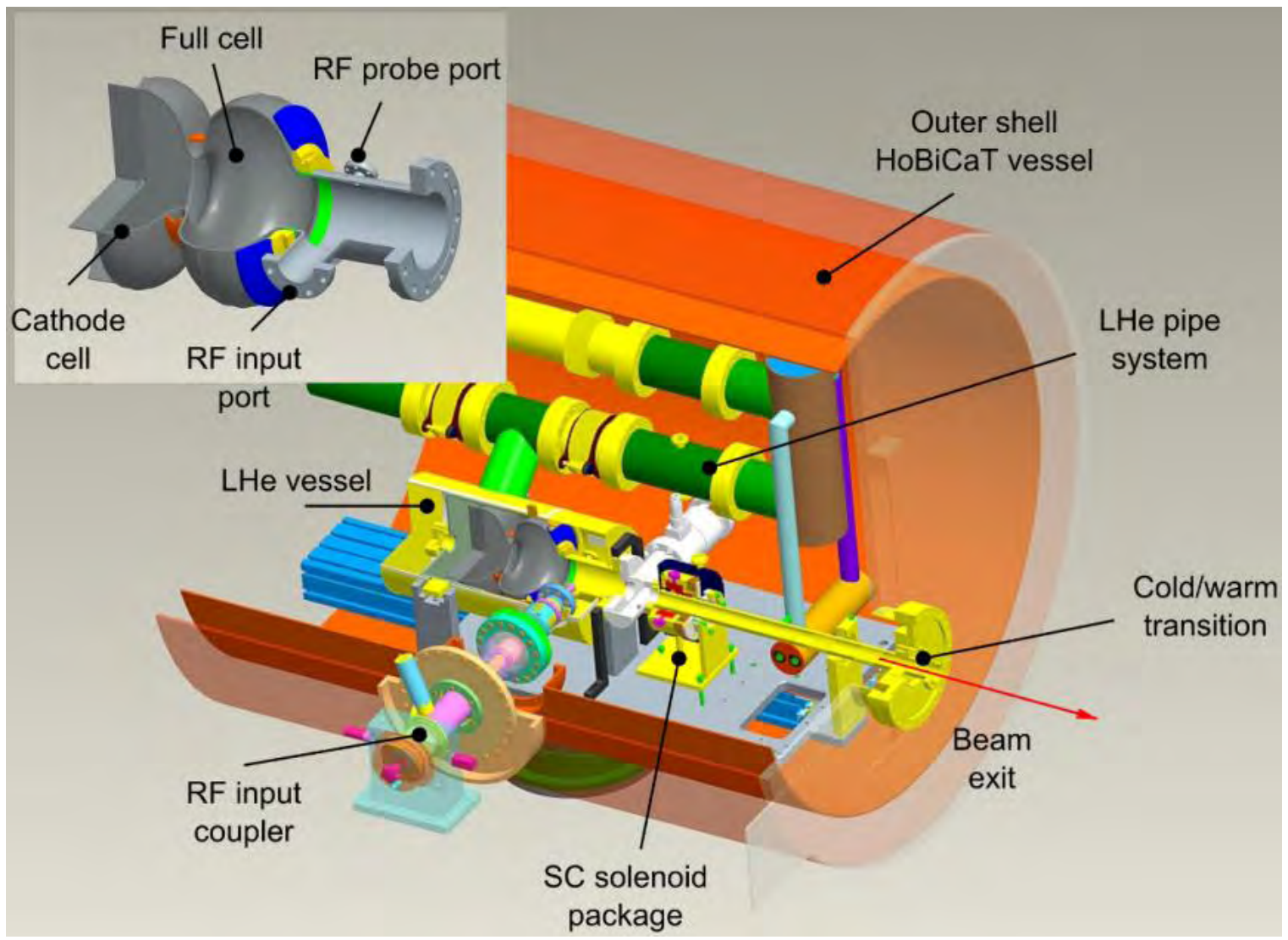

**Figure 3.2. Conceptual design of the HoBiCaT SCRF injector test installation. [3.16]Courtesy T. Kamps]**

#### *3.3.2.1 Elliptical Guns*

Elliptical guns, as the name implies, use elliptical-section cathode cells. The upstream and downstream ends of the cathode cell may or may not be identical in profile (*e.g.*, the wall angle may differ). As the beam starts out at rest from the cathode, the cavities typically have lengths of ~$\lambda$/4.

**BNL/AES and DESY 1.3 GHz Single-Cell Guns; JLab 1.5 GHz Single-Cell Guns**

The Brookhaven National Laboratory/Advanced Energy Systems (BNL/AES) single-cell 1.3 GHz SCRF guns, shown in Figure **3.3**, were not intended for use as part of stand-alone beam sources, or as parts of higher energy injectors. Rather, they were designed for research and development of SCRF gun technology. They have played, and are playing important roles in evolving the SCRF injector, with first beam being produced in 2005 [3.17].

Typical applications for these guns include testing cathode materials, such as gallium arsenide, in SCRF cavity environments [3.18], and cathode mounting techniques, such as quarter-wave chokes [3.19].

Similar guns were constructed at Thomas Jefferson National Accelerator Laboratory (JLab) and DESY [3.20], [3.21] as part of the path towards higher energy, multi-cell SCRF injectors.

A gun of this type, using a lead (Pb) cathode, will be installed in the HoBiCaT cryostat as the first stage of the BERLinPro Project.

**BNL/AES 703.75-MHz High Current Injector**

Unlike the 1.3-GHz test cells, the BNL/AES 703.75-MHz SCRF gun is intended to serve as a true stand-alone injector. The nominal application for this gun is to provide beam to a high current ERL (up to 500 mA). Hence, high bunch charges and high average currents are required. As envisioned, the gun sends its beam directly to an ERL merge and then into the main linac, at an energy of 2 MeV [3.22]. Figure **3.4** is a cross sectional view of the gun's cavity and its cryostat. The cavity fields and cathode choke joint are shown in Figure **3.5**, along with photographs of the two prototype cavities. One of the prototype cavities was fabricated from large grain niobium.

The 703.75 MHz injector employs two high power RF couplers, located in the beam pipe immediately after the gun cavity. With a nominal beam current of 0.5 A, each power coupler is required to deliver 500 kW.

The flared beam pipe shown in Figure **3.5** is intended to improve extraction of higher order mode RF power from the gun cavity.

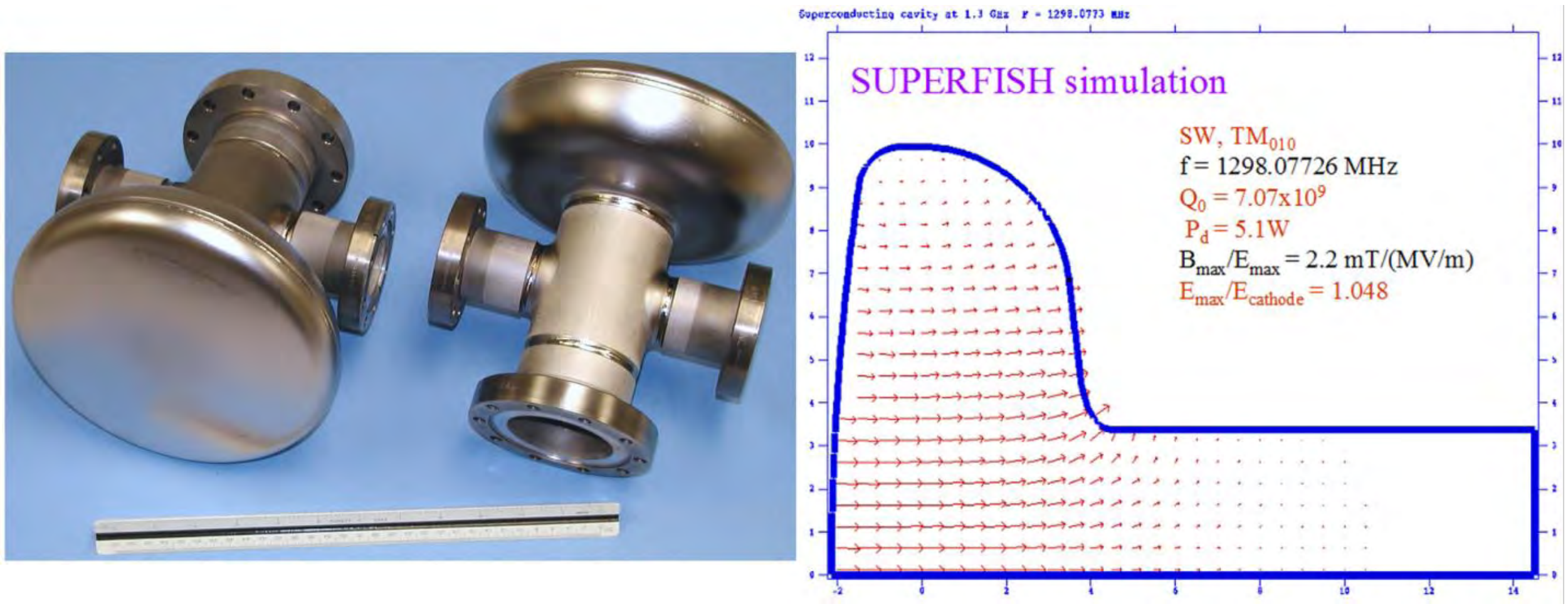

**Figure 3.3. Pictures of the BNL/AES 1.3-GHz SCRF guns (left) and as-designed field profile (right). The field profile plot, generated by SUPERFISH, shows the electric field *E*(*z*,*r*) vectors within the cavity. SUPERFISH assumes cylindrical symmetry about the *z* (horizontal) axis, so effects of the coupling ports are not included in the model. [Reprinted from [3.17], with permission from Elsevier.]**

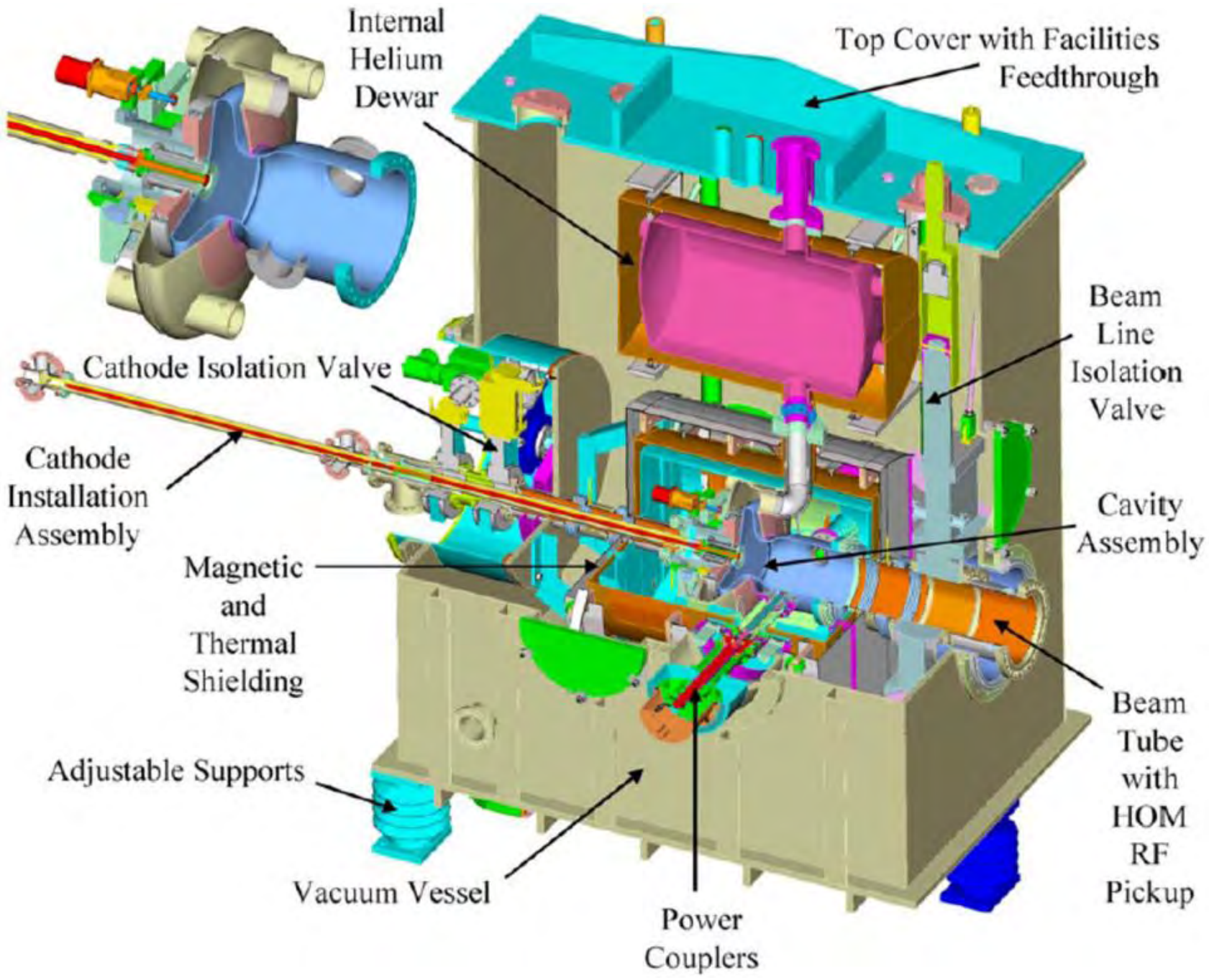

**Figure 3.4. BNL/AES 703.75 MHz SCRF gun design. The inset figure shows the gun cavity itself. The cryostat includes a built-in load-lock to facilitate cathode exchange. [Reprinted from [3.23], with permission from Elsevier]**

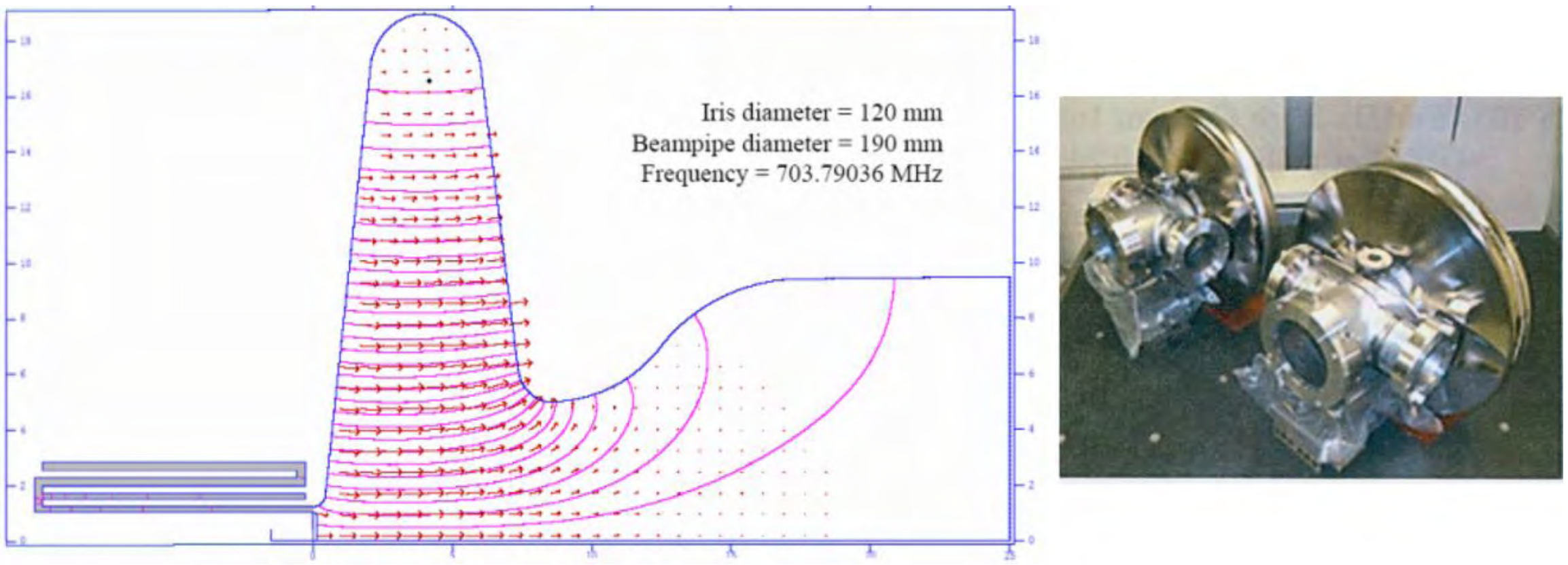

**Figure 3.5. SUPERFISH plot of cavity fields and quarter-wave choke joint of the BNL/AES 705.75 MHz gun (left), where the solid lines indicate constant $H_z$ and the red arrows indicate electric field strength and direction; photographs of the two prototype cavities (right); the rightmost prototype was fabricated from large grain Nb. [Adapted from [3.23], with permission from Elsevier] [Courtesy I Ben-Zvi]**

A novel cathode concept, the diamond amplifier cathode, was developed to be the beam source for this with this gun [3.24]–[3.26]; details are given in later chapters. As opposed to a choke cavity, as is used in the ELBE injector design, the 703.75 MHz gun uses a folded quarter-wave choke joint that offers a more compact installation, but does not as directly permit adjustment of the cathode's longitudinal position. This gun cavity has successfully completed its vertical dewar tests and first beam is anticipated in 2011 [3.27].

The ultra-high vacuum (UHV) load-lock system is an important part of any gun system employing high quantum efficiency cathodes; load-lock systems for superconducting gun present additional challenges due to the extreme sensitivity of the SCRF cavity to contamination. A load-lock system was developed for the BNL/AES 703.75 MHz SCRF gun that encompasses both a preparation chamber and a cathode transporter.

#### *3.3.2.2 Quarter-Wave Cavities*

Quarter-wave (QW) SCRF gun cavities are based upon designs originating in the heavy-ion accelerator community, wherein quarter-wavelength structures are used for low-β acceleration of ions. However, in an SCRF injector application, the direction of beam propagation is parallel to the axis of the QW cavity, rather than perpendicular to it.[5] Two such guns were fabricated, one of which has generated beam; at least two others are under development.

Compared to a cell with elliptical geometry of the same diameter, a QW cell can be designed to have a considerably lower operating frequency, permitting the use of a more compact cryomodule. Another potential advantage of the QW structure is the ability to easily fabricate cavities with small accelerating gaps, leading to high transit-time factors. (In contrast, for elliptical cells, the cell's length effectively is the gap length, so that attempting to reduce the accelerating gap tends to result in a "pancake" cavity shape.) This arrangement may allow beams to be launched from the cathode at larger RF phases (measured from the zero-crossing) and may be beneficial for alternate cathode types, such as field emitters.

**NPS / Niowave / Boeing (NNB) Mark I**

The Naval Postgraduate School Mark I SCRF injector, built by Niowave, Inc. and operated in collaboration with The Boeing Company, produced its first beam in June 2010. The Mark I was intended primarily as a research and development tool to explore issues in designing and operating SCRF guns in general, and QW SCRF guns in particular, although it can serve as an injector for the NPS Beam Physics Laboratory's linac. It uses an on-axis coaxial RF power coupler and has a resonant frequency of 500 MHz. This low frequency allows operation at 4K, thereby greatly simplifying the requirements of the cryogenic support system. Nominal beam energy is 1 MeV, with a maximum attained energy of 0.5 MeV to date. A cross sectional view and photograph of the NNB Mark I are shown, respectively, in Figure **3.6(a)** and Figure **3.6(b)**; the geometry of the cathode and power coupler geometry is depicted in Figure **3.7** [3.27].

[5] For researchers familiar with electron linac design, as employed as SCRF gun cavities, quarter-wave structures can be thought of as asymmetric highly reentrant cavities.

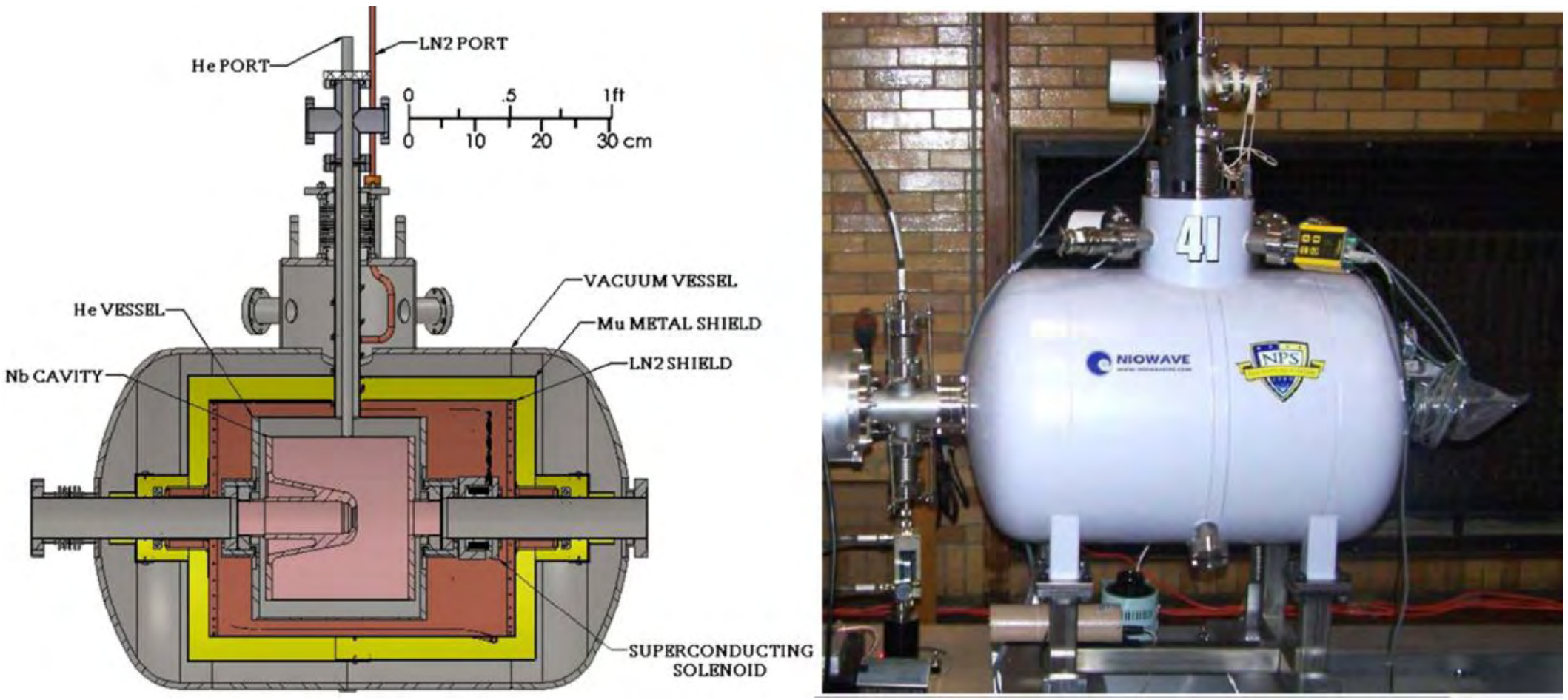

**Figure 3.6. NNB Mark I quarter-wave gun: (a) cross section of cryomodule and (b) assembled injector. [Reprinted figures with permission from [3.27]. Copyright 2011 by American Physical Society.]**

The accelerating gap in the Mark I is very short compared the wavelength; at ~$\lambda/7$, it is approximately half the length of a "normal" 500-MHz cathode cell, thereby resulting in a very high transit time factor compared to $\lambda/4$ cathode cells.

A significant advantage of the on-axis RF power coupler is that it allows placement of a superconducting solenoid close to the cavity, thereby aiding the emittance compensation process. The impact of the solenoid upon the cavity's performance was tested by repeatedly allowing the cavity to quench with the solenoid energized and measuring the cavity's $Q$ between quenches. A moderate decrease in cavity performance was noted, however, the original cavity $Q$ was recovered simply by degaussing the solenoid with the cavity quenched [3.29].

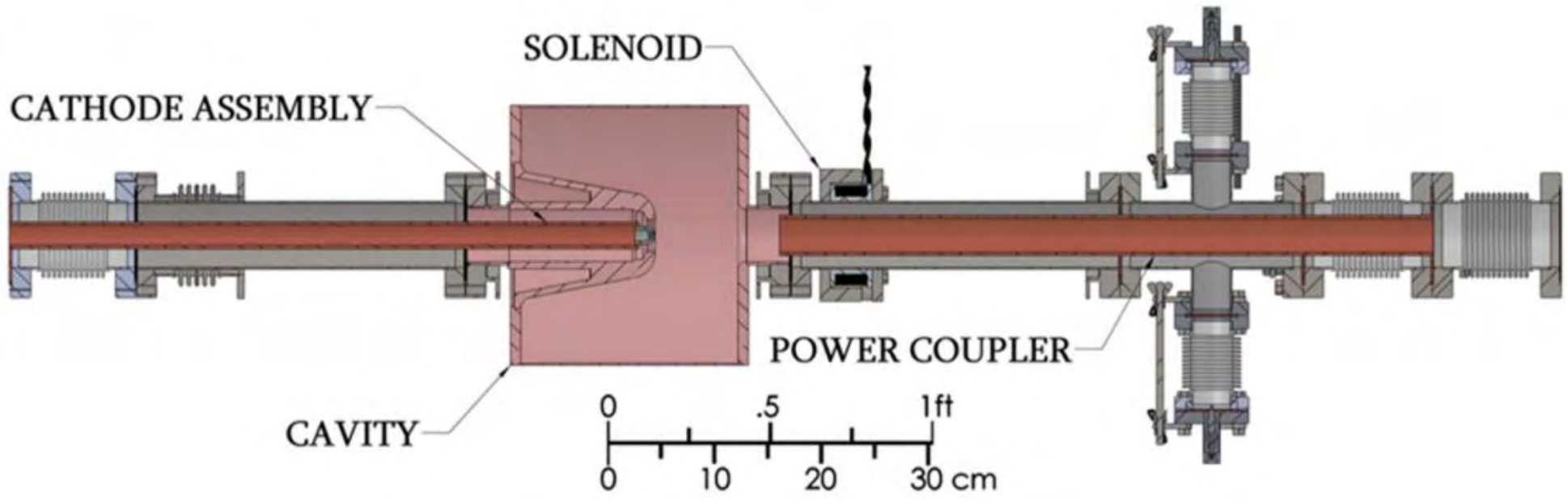

**Figure 3.7. NPS Mark I cathode and on-axis power coupler geometry. The cathode assembly can also serve as a cavity field probe, as it forms a coaxial transmission line. [Reprinted figure with permission from [3.27]. Copyright 2011 by American Physical Society.]**

The NPS Mark I does not use a sophisticated cathode joint arrangement. Rather, the cathode simply extends on a stalk of length $\lambda$. Since this configuration forms a coaxial transmission line, RF fields will be present all along the cathode stalk, resulting in up to several hundred watts of RF power dissipation on the normal-conducting cathode at the design cavity gradient. To reduce this RF power loss, the cathode's surface area was reduced and the radius of the cathode's stalk was stepped at several locations (not shown in Figure **3.7**).

First beam was produced with an Nb cathode; future plans include testing higher-QE metals, as well as semiconductor and field emission cathodes. Initial operation was limited to 0.5 MeV beam energy due to the available RF power, field emission and local radiation shielding, with detailed measurements performed at 0.3-0.5 MeV [3.27]. Continued testing is planned at higher energies.

**Other Quarter-Wave Guns Under Development**

Brookhaven National Laboratory and Niowave Inc. constructed a 112 MHz SCRF gun using a QW geometry, also intended as a candidate injector for an electron-cooling ERL. The nominal beam energy is 2.7 MeV. This gun is essentially completed and awaiting its initial cool-down tests. Its cross section is shown in Figure **3.8(a)**.

Figure **3.8(b)** shows the cross-section of the University of Wisconsin's (UW) SCRF gun, intended to serve as the beam source for the proposed WiFEL linac-based light source. The UW gun operates at 200 MHz and has a nominal beam energy of 4 MeV.

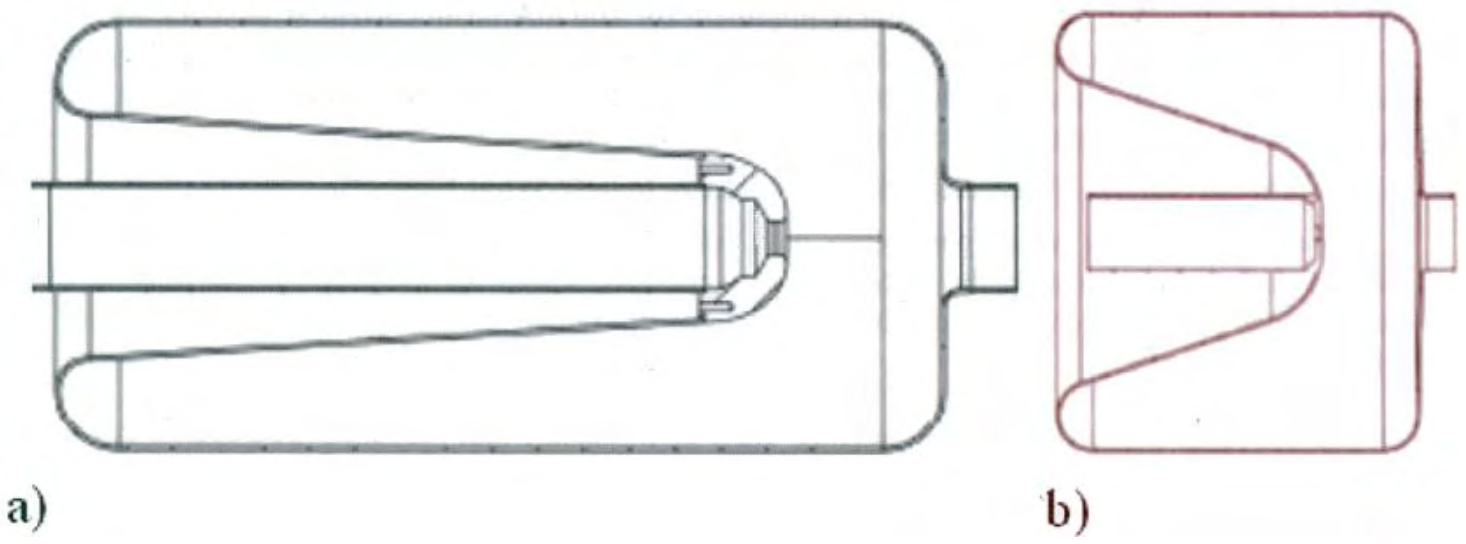

**Figure 3.8. Profiles of (a) the Brookhaven 112-MHz and (b) University of Wisconsin 200-MHz quarter-wave injectors. The shaded region in the center is represents the cathode. For scale, the accelerating gap in the 112-MHz Brookhaven cavity is approximately 16 cm. [Courtesy of Niowave Inc.]**

Finally, based on the rapid development cycle and initial success with the Mark I, NPS and Niowave have begun working on the design of a Mark II QW SCRF injector, to operate at 700 MHz. The Mark II resolves many of the issues encountered during the fabrication and initial operation of the Mark I version. The design retains the small accelerating gap at ~$\lambda$/6.

### 3.3.3 Hybrids

The hybrid DC/SCRF is an interesting concept that has seen substantial development. In this design, the cathode is replaced *in toto* by a miniature DC gun for the initial acceleration of the beam. The Peking University group reported developing such designs, most recently with a 3.5-cell SCRF cavity based on the design of the ELBE SCRF photoinjector [3.30].

## 3.4 INITIAL QUESTIONS TO CONSIDER

This section presents some questions that should be considered before the SCRF photoinjector design process begins. The answers are subject to change both because the project goals may change, and because as the design process evolves, better or alternate approaches may be found.

### 3.4.1 What are the Performance Goals for the Injector?

These are some of the most fundamental questions to ask at the start of the design process. Typical parameters include bunch charge, length and energy spread; transverse emittance; and, the required Twiss parameter values and tolerances at the entrance to the main accelerator.

Two parameters of particular note are the required average beam current and kinetic energy. Together, they establish the total RF power required to accelerate the beam. They also strongly influence the choice of whether to design the injector as a single multi-cell cavity (as in the ELBE design), as a series of single cells, each with its own RF power feed, or as a combined approach with (for instance) a single-cell "gun" containing the cathode followed by multi-cell structures.

There is insufficient experience as yet to suggest which approach might be "best," (furthermore, the answer most likely depends upon the application), but considering the state-of-the-art of RF power coupling suggests that requiring no more than 100 kW per power coupler should provide a good margin of safety.[6] Thus, having a requirement for 100 mA beam current at 10 MeV would suggest using a series of individual cells, while a 10 mA beam current at 10 MeV would suggest the feasibility of an integrated design. (A design based around individual cells might still be preferable for other reasons, depending upon the application.)

### 3.4.2 What System Characteristics are Already Fixed? What Might be Changed Later?

It is important to know which parameters have been previously fixed, and which are subject to change during the course of the injector design process. This does not refer to the injector's performance goals (which are also subject to change), but rather, to the physical conditions under which it will operate.

While it is quite likely that key parameters, such as frequency and operating temperatures, are already set, it is worth asking whether there are good reasons why the photoinjector could, or should, be different from those used by the rest of the accelerator.

#### *3.4.2.1 Is the Linac Frequency the Best Choice for the Injector's Frequency?*

The injector's operating frequency is seemingly a simple and obvious selection; make it the frequency of the remainder of the linac. However, using an integer subharmonic frequency to the main linac frequency might be considered if there are advantages in doing so for beam dynamics or other parameters.

For instance, the accelerator modules intended for use in the Naval Postgraduate School linac are Stanford/Rossendorf style cryomodules, with an operating frequency of 1.3 GHz. The NPS/Niowave Mark I SCRF QW injector, in contrast, was designed to operate at 500 MHz. The common frequency with the linac is 100 MHz, a suitable repetition rate for the intended experimental program.

The lower frequency of the Mark I photoinjector allows its operation at 4.2K (*e.g.*, at standard atmospheric pressure over the helium reservoir), so it can be tested and characterized well in advance of constructing the linac's cryoplant.

---

[6] This is another area of lively debate and anticipated experimental demonstration. Megawatt class couplers were built and tested on test stands, but to date no SCRF injector has been operated at even the level of 100 kW beam power.

#### *3.4.2.2 What is the Injector's Operating Temperature?*

Generally, lower frequency cavities can be operated at 4K, while higher frequency cavities must be operated at 2K to generate useful accelerating gradients. While there is no single cutoff point – and the answer will depend upon the frequency, gradient and duty factor needed – the general consensus is that below 500 MHz, 4K will suffice; above 1 GHz, 2K is required. The answer to temperatures in between depends on the system's particulars.

There are significant advantages to operating at 2K. Superfluid helium allows for excellent heat removal, and RF losses are lower at lower temperatures; a lower temperature allows higher operating gradients and peak surface magnetic fields. However, operating at 2K also has several disadvantages in terms of the required complexity, and corresponding cost, of the refrigeration system. A 4.2 K injector can be operated simply by supplying liquid He at room pressure into the injector's cryostat from a nearby dewar. At a minimum, an injector requiring 2K to operate requires a vacuum pump system to lower the gas pressure on the liquid helium bath.

If the photoinjector is to be part of a larger SCRF installation, then whatever temperature the remainder of the system uses probably is the best choice for the injector.

### 3.4.3 What Degree of Operational Flexibility is Desired?

Many injectors, if not most, are eventually operated under conditions different from their nominal design. Certainly this will be the case during initial commissioning of the injector, and is expected. However, in-service operation may be considerably different than originally envisioned because the application that the injector supports operates better under different conditions. A recent example is the Linac Coherent Light Source at the SLAC; by operating the injector at a reduced bunch charge from the nominal 1 nC case, wakefield effects in the main linac are reduced and a higher quality beam can be transported to the undulator.

Fortunately, most injectors tend to be flexible in terms of their operational regimes; SCRF photoinjectors may pose several extra challenges due to the constraints under which they are fabricated and operated. From the standpoint of beam dynamics, changes to the bunch charge density and average beam power are of particular concern.

Some SCRF photoinjector designs incorporate cathode-region focusing, *via* the RF field, to either augment or replace solenoid-based emittance compensation. As with the classic Pierce DC electrode geometry, the method attempts to balance the radial RF electric field against the radial space charge force within the electron bunch. If the electron beam's charge density is changed from its design value, the RF focusing strength will be incorrect. The ratio between longitudinal (accelerating) and radial (focusing) field strengths can be adjusted by moving the cathode longitudinally, but this action also impacts the cavity's tuning and, potentially, RF power dissipation on the cathode.

Similarly, operating the cathode cell at a different gradient from its design point will impact any RF-based emittance compensation scheme.

To first order, RF power coupling should not require modification so long as the average electron beam power (the product of the average beam current and voltage) remains constant. However, altering the bunch charge and repetition rate can change the generation and impact of wakefields, particularly for high average current injectors, even if the average current remains constant.

### 3.4.4 What Cathode(s) will be Used?

Photocathode selection is largely a function of the required average beam current. Traditional photoinjector cathodes include metals (usually copper or magnesium in NCRF injectors, but niobium and lead have also been used in SCRF photoinjectors); thin-film semiconductors ($Cs_2Te$, CsKBr); and crystalline structures (*e.g.*, Cs:GaAs). More complex cathodes are under development, such as the Brookhaven diamond amplifier, but have not been demonstrated in SCRF injectors. Cathodes are well treated in Chapters 5 through Chapter 8, so relative performance characteristics will not be reviewed in detail here.

However, there are two points to consider regarding the use of any type of cathode in an SCRF photoinjector.

First, the cathode will be operating within a cryogenic environment. If high average currents are required, high power photocathode drive lasers may also be required to obtain that current. At milliampere currents, several Watts of laser power may be required; at hundreds of milliamperes, a 100 W drive laser might be required. This laser power, deposited on the cathode, represents an additional heat load that must be removed from the cavity. A contact cathode joint will dump that heat, *via* conduction, into the cavity's liquid helium bath. A non-contact cathode may require external cooling, perhaps through a liquid nitrogen cooling circuit, to avoid overheating the cathode within the cavity. Most SCRF injectors intended to use removable cathodes have used non-contact cathode designs.

Second, SCRF cavities are sensitive to contamination from dust and from deposition of materials liable to increase field emission, such as cesium. Inserting and replacing cathodes introduces the possibility of generating dust. Concern has also been expressed regarding the use of cesiated cathodes, which includes essentially all non-metal cathodes, in SCRF cavities. The worry is that Cs emitted from the cathode will redeposit in high field regions. Results from the ELBE SCRF photoinjector indicates that operation with $Cs_2Te$ cathodes appears to viable; but again, experience with other cathodes in SCRF photoinjectors is very limited.

### 3.4.5 What is the Cathode's Position? Should it be Adjustable?

In most conventional NCRF photoinjectors, the cathode is usually flush, or nearly so, with the back wall of the cathode cell, providing an RF field commensurate with the basic emittance compensation theory. In particular, as shown in Figure **1.6** in Chapter 1, the radial RF fields near the cathode are very small.

By recessing a cathode somewhat behind the back wall of the gun cell, radial RF fields can provide initial focusing to the electron beam as it leaves the cathode, much in the way Pierce geometry functions in a DC electron gun. Early designs of the ELBE photoinjector considered implementing this scheme [3.13], and the initial simulations were promising. While some of these simulations assumed a curved cathode, the technique can operate well even with planar cathodes, as focusing is provided by the change in longitudinal gradient along the axis.

Moving the cathode longitudinally to adjust the RF focusing has two consequences. First, the cathode will act as a tuner, changing the cavity's resonant frequency; in a multi-cell photoinjector, compensating for this could be difficult as the RF field balance between the cells could also be affected. Second, moving the cathode changes the longitudinal field gradient on the cathode's surface. In addition to affecting initial acceleration, this can also strongly influence the RF power dissipated on the cathode in some designs and will change the space charge limited emission current density. Finally, altering the cathode's position can vary the strength of nonlinear radial components of the RF field.

If positioning the cathode at a particular location to obtain a given RF field configuration is an integral part of the beam physics design, early studies should include alternate charge scenarios to determine whether the cathode can remain in place under those conditions.

### 3.4.6 How will the Cathode be Mounted?

In several SCRF guns the cathodes are deposited directly upon the back wall of the cathode cell, as in the BNL/AES 1 300 MHz gun tests with Pb cathodes; however, this method does not lend itself to readily changing cathodes. The general assumption for high average current operation is that the cathode must be intermittently replaced or rejuvenated. In such cases, for instance in a user facility, a removable cathode is almost essential.

To date, three basic cathode mounting and isolation schemes have been used in SCRF injectors: Rossendorf-style choke cavities (as exemplified by the ELBE injector); quarter-wave folded chokes (as in the BNL/AES 703.75 MHz injector); and, cathode-on-a-stick (as in the NPS/Niowave/Boeing Mark I). None of these designs are contact cathodes, in the sense that there is no direct electrically conductive path between the cathode's outer rim and the body of the cathode cell. In all three schemes, the termination is far from the cathode's surface and the cathode forms a coaxial transmission line into and out of the cavity, along which RF induced currents will flow: the three schemes differ in the approach they take to dealing with, and terminating, that flow. All three schemes will impact the RF field near the cathode, to a greater or lesser extent, compared to a cathode flush with and in direct contact with the back wall of the gun cavity. At least an approximation of the selected cathode mounting scheme should therefore be included from the start of the design process.

A significant advantage of a non-contact cathode is that the cathode's temperature need not be the same as the cavity temperature. Indeed, the Mark I cathode is not cooled; the ELBE and BNL/AES injectors both incorporate provisions for liquid $N_2$ cooling. Particularly for operation at high average beam currents, when significant drive laser power may be directed onto the cathode, thermal isolation between the cathode and gun may be advantageous. A second, non-trivial advantage is that the cathode's contact point is far from the cavity. Barring a misalignment that scrapes the edge of the cathode during its insertion or removal, the non-contact cathodes minimize the potential for introducing dust or other contaminants into the cathode cell.

In principle, the cathode-on-a-stick approach allows ready change of the cathode's position along the axis of the cavity. This can aid in implementing RF focusing schemes, but the lack of a hard stop may make repeatable transverse and longitudinal positioning challenging. Further, in addition to changing the cavity's frequency, moving the cathode longitudinally changes the amount of RF power dissipated along the cathode stalk. Multipacting is also a serious concern with this design, although to date it has not been observed in the Mark I injector.

The choke cavity and folded choke designs both address the issue of RF power loss along the cathode stalk, and should provide improved repeatability and transverse alignment, but they are mechanically more complex. In operation, the Rossendorf-style choke cavity has proved successful; the folded choke design suffered from multipacting during initial testing, but the issue is being addressed.

### 3.4.7 Is a Single-Cell or Multi-Cell Gun Design Preferred?

The theoretical maximum accelerating gradient in a given SCRF cavity design depends upon the frequency, temperature, geometry and, in particular, the ratios between the peak on-axis and surface fields. The achievable beam energy gain will depend upon the gradient achieved in practice, the acceleration gap and the launch (or injection) phase of the electron beam.

Generally, the cathode cell in an SCRF photoinjector is expected to provide initial beam energies of 1-2 MeV, and each full cell ($\lambda/2$) is expected to deliver at most between 2-3 MeV additional energy gain. More conservative designs, multi-cell designs, or those which are expected to be heavily beam-loaded (such as the BERLinPro injector) often have lower energy gain design goals. Single-cell gun designs typically try to push the output energy as high as possible to help ameliorate space charge effects in the drift space to the first booster cavity.

In a multi-cell gun design, such as the ELBE 3.5-cell injector, RF power is supplied to the gun as a whole and cell-to-cell coupling provides RF power to each of the individual cells. This arrangement has the advantages of providing the highest possible real-estate gradient, and therefore the fastest acceleration to the injector's output energy. Minimizing the distance the beam spends at low energy helps to preserve beam quality, particularly for bunch charges in the nanocoulomb range. The gun may be followed by an energy booster (as in the BERLinPro design), or may serve as the entire injector (as in the ELBE design).

In many injector designs based on single-cell guns, such as the AES designs for high power ERLs, the injector consists of a cathode cell followed by a drift space and an energy booster. The booster consists of independently powered cavities, often (but not mandatorily) single cells; an example of such a booster is shown in Figure **3.9**.

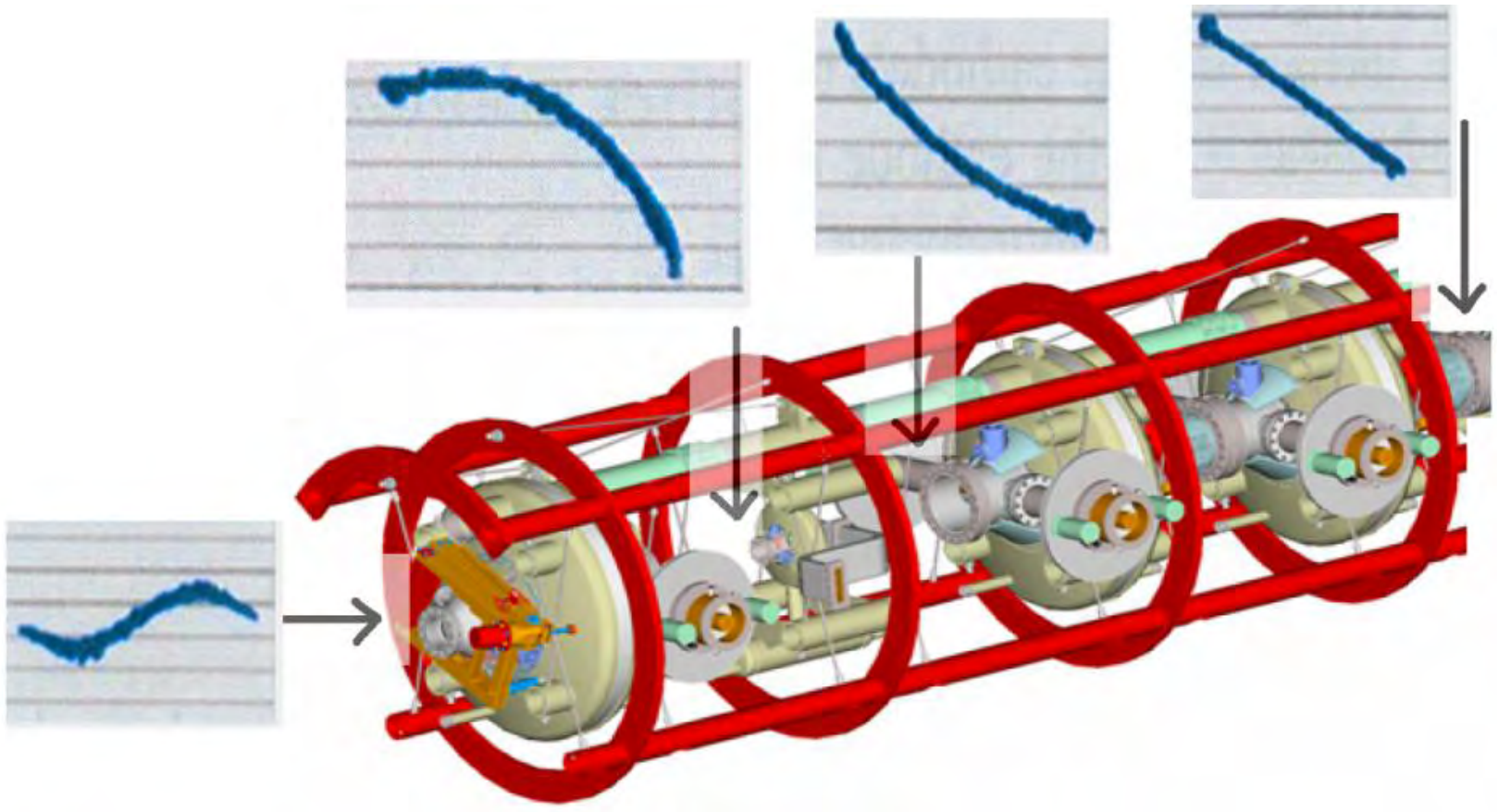

**Figure 3.9. An example of an energy booster, as designed by Advanced Energy Systems. In this particular design, the gun would sit in a separate cryomodule located to the left of the booster. This design incorporates three energy booster cavities at the frequency of the injector and a third-harmonic cavity used to assist with longitudinal phase-space control. [Adapted from [3.23], with permission from Elsevier]**

Independent control of RF power and phase to each cavity in the booster offers additional tuning "knobs" that can be used to exercise a high degree of control over the beam's energy spread and, to an extent, the bunch length at the exit of the injector. In Figure **3.9**, for instance, the booster cell gradients and phases have been selected so as to linearize the longitudinal phase space for the beam. There is also (theoretically) space between the booster cavities for additional diagnostics as well as RF power couplers, solenoids, *etc.* Thus, this approach may be very attractive for injectors for ERLs, as it permits greater control over the beam parameters at the end of the injector. However, greater control comes at the price of significantly increased complexity of the cryomodules and RF systems. Further, the increased distance between cells lowers the real-estate gradient, thus production of high charge, low emittance beams may be more difficult due to the increased drift distances at low energy.

## 3.5 CAVITY GEOMETRY CONSIDERATIONS

This section discusses several effects of particular concern to the design of the cavity itself, and is primarily written for the NCRF injector (and, in particular, gun) designer making forays into the realm of SCRF injector design.

As mentioned above, lack of space precludes a comprehensive discussion; the SCRF injector designer is strongly encouraged to obtain a copy of *RF Superconductivity for Accelerators***,** by Padamsee, Knoblock and Hays.

### 3.5.1 Field Emission and Multipacting

All RF cavities can experience field emission and multipacting, and are subject to performance limits based upon peak surface fields. However, there are significant differences in practice between SCRF- and NCRF-cavities.

For field emission, the electron current depends upon the gradient of the electric field applied to the surface. Field emission (FE) current generally follows the Fowler-Nordheim law, so small increases in gradients can lead to exponentially larger currents. FE current in an SCRF injector may exit the cavity into the beam pipe following the injector (in which case, it is also called dark current), or may impact on the wall of the cavity. FE current is therefore problematic for three reasons: First, it consumes RF power *via* beam loading; second, it deposits additional heat inside the SCRF cavity, potentially quenching it and definitely leading to higher liquid He consumption and cooling requirements; third, the impact of the field emission current inside the cavity can generate significant amounts of radiation, in turn driving radiation shielding requirements. The magnitude of field emission depends upon both the smoothness (in the sense of surface field enhancement) and cleanliness of the SCRF cavity's interior, as well as the operating gradient.

Multipacting, in contrast, is a resonance phenomenon, representing a buildup of electrons being emitted into the cavity and eventually returning to their starting point, with impact of each electron resulting, statistically, in the emission of more than one electron. It is encountered more often at low accelerating gradients, although any RF cavity will have low field regions, even when operating at its design gradient.

The probability of electron emission, and thus the buildup of multipacting current, depends on surface conditions and impact energy. As the multipacting current builds, RF power is absorbed and the cavity eventually will quench or move off resonance. The general solution is to design the cavity such that no simple resonant paths are available. This means, among other things, avoiding parallel walls and right angles. TESLA cavities, for instance, were designed with a primary goal of avoiding a multipacting-supporting geometry.

Both field emission and multipacting can sometimes be conditioned away by RF processing techniques, but this is not guaranteed. Multipacting can be addressed partly by choosing a cavity geometry that lowers the likelihood of a supporting a viable multipacting path. Field emission can be addressed, in part, by choosing a cavity geometry that is easy to clean and by using good cavity processing techniques.

### 3.5.2 Power Dissipation

Unlike NCRF cavities, typical power consumption by SCRF cavities is usually on the order of tens of Watts (barring field emission and multipacting). However, that heat must be removed if the inner cavity surface is to stay superconducting. This means choosing designs that minimize the distance between the interior surface of the cavity and the liquid He bath. In the NPS Mark I injector, for instance, the inner third of the nose cone is a concern as it is fabricated from relatively thick, solid niobium. While surface currents are relatively low there, the nose cone will still experience some heating, perhaps augmented by field emission as this is a high electric field region.

### 3.5.3 Emittance Compensation

Classic emittance compensation relies upon employing an externally applied radial focusing magnetic field, combined with evolution of the beam through a subsequent drift space, to “re-align” longitudinal slices of the electron beam’s transverse phase space. Ideally, this focusing is applied as soon as possible after the beam is generated and given its initial acceleration.

Unfortunately, this is problematic in SCRF injectors. Besides the potential for reducing the cavity’s $Q$ due to stray magnetic fields from the solenoid, the use of transverse coaxial power couplers further extends the distance between the cathode and the first available beamline location for a solenoid. Nevertheless, good beam dynamics solutions were found with the solenoid relatively far from the cathode, as in the Brookhaven 703.75 MHz design, while alternate power coupler geometries, such as the NPS Mark I on-axis coupler, allow closer placement of the solenoid.

One trick for modeling solenoids in the proximity of SCRF cavities may prove useful. Putting the SCRF cavity’s geometry (perhaps simplified) into the magnetostatic model of the solenoid with an infinitesimal magnetic permeability ($\mu$) will exclude the flux from the cavity walls. By re-running the model with the permeability set to unity, an easy analysis can be made of the potential for “flux leakage” inside the SCRF cavity if it were to quench with the solenoid turned on. This also determines perturbations to the solenoid field caused by the presence of a superconducting cavity, potentially important when the goal is to generate very low emittance beams.

Figure **3.10** shows a simple example, using the POISSON model of the NPS Mark I solenoid. The image on the left shows the field of the solenoid, distorted by a superconducting disk placed within the solenoid’s bore. The image on the right is the same model, but with the permeability of the disk set to unity.

The recent studies of emittance compensation using TE-mode magnetic fields are intriguing. To date, there is no published study of this technique applied to a single-cell SCRF gun, but, in principle, there is no reason why it could not work there as well as in the multi-cell ELBE SCRF injector cavity. Further study of this method is needed as it both avoids the issue of requiring DC magnetic fields near the cavity and could allow the application of magnetic focusing much closer to the cathode.

### 3.5.4 Power Coupler Options

Chapter 10 covers RF power coupling thoroughly: here, I briefly overview the advantages and disadvantages of two prevalent designs.

The TESLA-style transverse coaxial coupling scheme for SCRF cavities has been a popular choice for SCRF injector design. In the transverse coax coupling scheme, a coaxial line perpendicular to the axis of the cavity transfers power into the beam pipe immediately downstream from the gun cavity. (For an energy booster, couplers could be placed either upstream or downstream of the cavity.) It is well understood and successful in practice. Due to their location in the beam pipe, TESLA-style couplers also provide a relatively obvious location for RF field probes. The drawback of this scheme is that the couplers perturb the RF fields, breaking radial symmetry and potentially generating increased emittance. Their location generally restricts siting other components, such as solenoids, to locations further downstream from the cavity.

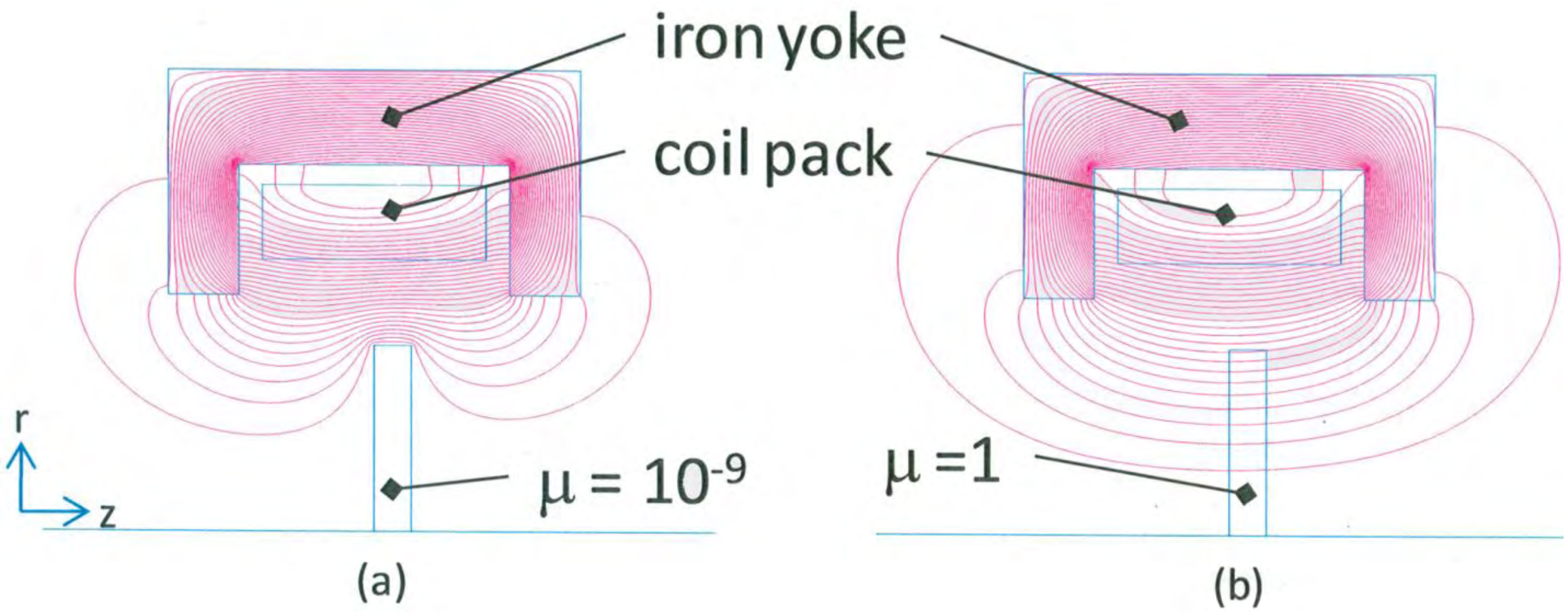

**Figure 3.10. POISSON model of the NPS Mark I solenoid, with a disk placed off-center along the axis. The disk is (a) superconducting and (b) non-superconducting and non-magnetic. Blue lines indicate geometry; pink lines are contours of constant $rA_\phi$ (radius times the azimuthal component of the magnetic vector potential).**

The on-axis coaxial coupler design (also used highly successfully in the TESLA Test Facility NCRF photoinjectors) uses a hollow coaxial line parallel to the beam axis to deliver RF power to the injector. RF power flows to the cavity between the outer wall and the center conductor; the beam travels through the center of the hollow center conductor. Under ideal conditions, this scheme maintains radial symmetry. Further, solenoids can be placed closer to the cavity if desired. Disadvantages include a smaller diameter passage through which the beam can travel, and the possibility of breaking radial symmetry anyway if the central conductor is misaligned.

Jefferson Lab-style (JLab) waveguide couplers have been conspicuous by their absence in SCRF photoinjector designs; the reason for their exclusion is not clear. Many SCRF photoinjector designs were undertaken at institutions that typically employ transverse coax power couplers for their other SCRF cavities rather than waveguide couplers, suggesting unfamiliarity with the latter might be a partial explanation.

### 3.5.5 Injector Cryomodule Diagnostics

Throughout the remainder of this chapter, "diagnostics" refers primarily to devices used to characterize the electron beam produced by the injector. Chapter 11 discusses such devices.

The cryomodule(s) containing the injector cavities will also require diagnostics to determine how well the injector is performing in terms of its cryogenics and RF characteristics. Cryogenic and RF diagnostics are important components of the SCRF photoinjector and should be incorporated into the design of the cryomodule(s) housing the photoinjector cavities from the start of the design process.

Typical cryogenic diagnostics include temperature sensors and helium level sensors. Temperature sensors should be placed not only on the injector cavities themselves, but also on critical components in close proximity to the injector cavities, such as superconducting solenoids, cathode support structures and electron beam diagnostics. Data from temperature sensors can be used to monitor conditions when the SCRF injector is operating normally, or to help resolve problems. For example, temperature sensors placed along the beam pipe could help diagnose problems with excessive FE current, or in a high current injector, problems with trapped modes or wakefields.

RF probes are very weakly coupled ports into the cavity radio frequency envelope. They can be placed in the beam pipe or directly embedded into the cavity wall to measure the evanescent field from the cavity. RF probes are potential sources of problems, in particular vacuum leaks, as they represent extra openings into the superconducting cavity's RF environment. However, RF probes, properly designed and placed, can provide important information about the electromagnetic fields within the SCRF injector cavities. As such, they are often of great use when attempting to diagnose why a cavity is not performing as anticipated.

## 3.6 DESIGN PROCESS

The notes and suggestions below are primarily from the perspective of a beam dynamicist, *i.e.* one whose area of expertise lies in the generation and transport of the electron beam itself, as opposed to the fabrication and operation of the structures generating the various required fields. In practice, the development effort must be a collaboration between beam dynamicists, SCRF scientists and the designers of the mechanical systems because a change in any part of the system can dramatically affect every part of the injector. Figure **3.1**, Figure **3.4**, Figure **3.6** and Figure **3.9** illustrate this point. Each injector development project will differ; this is not intended to be a prescription to follow as much as a highlight of some of the major stages involved.

### 3.6.1 Initial Modeling

Initial modeling encompasses the RF cavities used to generate the accelerating fields, as well as the initial layouts of the components within the injector's cryomodule(s).

Once the basic parameters for the injector are defined, initial cavity modeling is perhaps best undertaken with a fast 2-D code, such as SUPERFISH [3.31]. There are more advanced 3-D capable codes available, but SUPERFISH has three significant advantages: It runs quickly; it can be operated *via* command line (by other codes); and, its field maps can be easily imported into most of the commonly used particle-pusher beam simulation codes. SUPERFISH can also provide good initial guidance on the expected power dissipation within the cavity, which is important for beginning the cryogenic system's design. As shown on Figure **3.3** above, however, SUPERFISH does not incorporate features which break azimuthal symmetry. If maintaining a very low beam emittance is a primary goal, early incorporation of 3-D RF modeling codes into the design process is recommended. Experience with the LCLS S-band NCRF photoinjector, for instance, demonstrates the importance of fully understanding and controlling the fields experienced by the electron beam.

The initial mechanical design should focus on identifying and reserving beamline space for major items that are critical to operation, such as power couplers, but which are not yet included into the beam dynamics. This will avoid, for instance, finding out later that the optimized location for a superconducting solenoid is inside a power coupler.

In short, the goal for the initial cavity modeling should be to obtain a good, balanced design upon which to build further refinements. The cavity geometry should be reasonable from a fabrication and cleaning standpoint, incorporating known elements of good SCRF design. The required cavity field gradients should provide for a margin of safety relative to their theoretical maximum values; the electron beam parameters should approximate the targets established for the injector. Finally, there should be sufficient information to provide mechanical designers with an approximate component layout along the beamline.

### 3.6.2 Initial Beam Dynamics

Once the initial cavity model is defined from a "field map" perspective and constraints on component placement identified, beam dynamics calculations can begin. If the RF modeling and beam dynamics codes both allow command line operation, then both can be incorporated into a single optimization loop. This allows researchers to readily determine the effects of, say, cathode cell length or cathode-region focusing on the beam dynamics in a way that maintains consistency with the cavity model.

Some of the "knobs," or design parameters, can easily be adjusted during a photoinjector's operation, as well as in simulation; examples include RF power delivered to the cavity[7], or size of the laser spot on the cathode. Some knobs, such as the physical geometry of the cathode cell, can easily be adjusted in simulation, but are fixed in an operating photoinjector. Depending on details of the physical design, some knobs, such as the radius of the cathode, may be easy, difficult or impossible to adjust after the photoinjector is built, and over what range the adjustment may be made.

A major goal for the initial beam dynamics studies is to determine how meeting the design goals will affect the injector's physical design; this includes, ideally, how (or whether) new changes to the desired operating parameters can be accommodated once the injector is actually built.

For two examples, consider cathode-region focusing schemes and solenoid placement. If cathode-region focusing is to be used, this is the appropriate stage of the modeling process to set both a baseline position and nominal range of cathode locations, as this information is essential to designing the injector cavity tuner(s). The position of the cathode is an example of a knob that could be easy or difficult to adjust after the photoinjector is built. Simulating the injector's performance with the cathode at various locations provides important guidance to the engineering team as to how much effort to spend on making it easy. Arguably such studies are, or should be, made as a matter of course. While this is true, such studies are often aimed at meeting the initial design goals. Here, the emphasis is on thinking about what else the injector may be required to do.

If optimization includes altering the location of an emittance compensation solenoid, this will impact the placement of other components in the injector beamline (which is usually too short by about a factor of two for everything that needs to be placed in the beamline, but too long by about a factor of two for ideal beam size and emittance control). This is an example of a "knob" which is likely to be all but impossible to change once the photoinjector is built, so additional care should be taken to ensure that the solenoid placement doesn't unduly restrict the operation of the photoinjector. For instance, if it is very far from the gun, there may be a limit on achievable bunch charge due to space-charge-driven expansion of the beam. If it is very close to the gun, there may be a limit on the focusing field strength that can be applied due to quenching concerns.

[7] Whether the RF power is well-matched into the cavity, however, will depend on whether the RF engineers were told that variable power coupling would be required.

### 3.6.3 Refinement

Refinement should begin with examination of the preliminary beam dynamics in light of the original requirements, as well as the lay-out of the cryomodule and injector line.

Assuming the initial results are acceptable, refinement consists of successively improving the fidelity of modeling of the cavity fields and electron beam dynamics. Electromagnetic field simulations should be transitioned to 3-D and used to improve the fidelity of the beam dynamics simulations. Locations of RF power couplers, diagnostics ports and similar hardware should be finalized. More physical effects are incorporated at this stage, such as non-uniform current emission from the cathode. Particle-in-cell (PIC) codes are increasingly utilized, perhaps for wakefield analysis and multipacting studies, as well as for basic beam dynamics. The process is iterated, as needed, until a final design is obtained. If significant changes to the injector design are proposed, they should be evaluated quickly using the tools and techniques developed for the initial modeling process.

As the injector line will include diagnostics, it is well worth considering simulating the performance of the diagnostics. Beam dynamics codes usually can provide the full 6-D phase space of the beam at arbitrary locations along the beamline. However, simulating the particular technique to be used for, say, the emittance measurement (*e.g.* slits, quadrupole-, or solenoid-scan) helps to ensure the proposed diagnostics will properly measure the quantities in question at the desired resolution.

There is a phrase, “In a high power injector, any diagnostic can become non-intercepting.” This means that diagnostic devices which physically block the beam (such as viewscreens or emittance measurement slits) will have some physical limit to how much power they can absorb. Beyond that limit, they will be damaged. The damage can be done not only by the desired beam from the injector, but also from field emission current. Thus, this is also a good time to begin thinking about the interlocks that will need to be applied to the injector diagnostics, and how those interlocks will obtain the information they need to operate properly. Diagnostics which are able to characterize field emission current, at least to a basic level (for instance, position and average current), should be strongly considered for incorporation into the final design.

### 3.6.4 Validation

This is the stage at which the actual performance of the SCRF photoinjector is compared the models and simulation used to design it. (Construction is well beyond the scope of this chapter.) It encompasses both setting the physical parameters to match the assumptions in the simulation as closely as possible and repeating the simulations based on measured physical parameters of the as-built, as-operated system.

The validation process consists of many stages, and will range from testing of an individual superconducting cavity to measuring the output beam parameters of the injector as a whole. Details of the validation process will necessarily be different for every injector, but there are some common themes worth mentioning.

Thorough RF characterization of the SCRF cavities used in the injector is critical. Cavity performance has a critical impact upon the performance of the injector as a whole. The characterization should include how the cavity’s properties (quality factor, power dissipation, central frequency, *etc.*) change as parameters relevant to beam dynamics, such as cathode position, are adjusted, even if they are not initially envisioned to be altered significantly during normal operation. Particularly in user facilities, obtaining such information after normal operation commences can prove to be difficult due to time and scheduling constraints.

It is worthwhile to develop a comprehensive, but flexible, validation and commissioning plan for the SCRF photoinjector. Generating the plan helps to ensure that all of the desired measurements will in fact be made, in part because it provides a reasonable basis for estimating the required time to thoroughly commission the injector. The injector commissioning time requirements can then be incorporated directly into the overall project schedule.

## 3.7 CONCLUDING THOUGHTS

I have attempted to present an overview of issues specific to SCRF photoinjector design, along with a very brief review of designs presently in development or operation.

As high power accelerators enter into more widespread use and become more specialized in their designs, so too will their injectors. SCRF photoinjectors have crucial roles to play in the future of accelerator technology.

The SCRF photoinjector field is evolving very rapidly. Many of the current designs mentioned above borrow heavily from the experience and good results obtained *via* normal conducting photoinjector design. Given how well the NCRF injectors operate, this is an eminently reasonable starting point. However, there are fundamental differences – in terms of RF power consumption and interactions with external fields – between NCRF and SCRF accelerator technology. It is exciting to see how some of the new designs are beginning to explore and exploit those differences. An example is the use of a TE cavity mode to provide solenoid-like focusing, as proposed for the Rossendorf gun. (Of course, I am biased – I have enjoyed working with multi-frequency and multi-mode cavities in simulation and would like to see more of them implemented in practice.)

While this chapter focuses on beam dynamics, successful development of an SCRF photoinjector requires a collaborative effort. The ultimate goal of any injector design process is the same: provide a beam which meets the requirements of its associated accelerator. SCRF photoinjectors offer some unique capabilities that allow us to meet those needs, and it is my hope that this chapter provides some insights into the challenges that must be met to accomplish that goal.

## 3.8 CONFLICT OF INTEREST AND ACKNOWLEDGEMENT

The author confirms that this article content has no conflicts of interest and would like to acknowledge the support of the Office of Naval Research, the High-Energy Laser Joint Technology Office and the Naval Postgraduate School.

***References***

[3.1] A. Kumar, K. K. Pant and S. Krishnagopal, "Design and beam dynamics simulations of an S-band photocathode RF gun," *Phys. Rev. ST Accel. Beams*, vol. 5, pp. 103501-1–103501-6, October 2002.

[3.2] A. Todd, H. Bluem, V. Christina *et al.*, "High-power electron beam injectors for 100 kW free-electron lasers," in *Proc. 2003 Particle Accelerator Conf.*, 2003, pp. 977-979.

[3.3] A. Arnold, H. Buttig, D. Janssen *et al.*, "1$^{st}$ RF-measurements @ 3.5-cell SRF-photo-gun cavity in Rossendorf," in *Proc. 2006 Free Electron Laser Conf.*, 2006, 567-570.

[3.4] A. W. Chao and M. Tigner, Eds., *Handbook of Accelerator Physics and Engineering*, World Scientific: 1999, Section 5.9.1.

[3.5] M. L. Taheri, N. D. Browning and J. Lewellen, "Symposium on ultrafast electron microscopy and ultrafast science," *Microscopy and Microanalysis*, vol. 15, pp. 271, July 2009. [Online] Available: http://journals.cambridge.org. [Accessed April 4, 2011].

[3.6] W. E. King, G. H. Campbell, A. Frank *et al.*, "Ultrafast electron microscopy in materials science, biology, and chemistry," *J. Appl. Physics*, vol. 97, pp. 111101-1–111101-27, June 2005.

[3.7] J. F. Schmerge, private communication.

[3.8] K. Baptiste, J. Corlett, M. Huang *et al.*, "Status of the LBNL normal-conducting CW VHF photoinjector," in *Proc. 2009 Particle Accelerator Conf.*, 2009, pp. 551-553.

[3.9] J. Tiechert, A. Arnold, H. Buettig *et al.*, "Initial commissioning experience with the superconducting RF photoinjector at ELBE," in *Proc. 2008 Free Electron Laser Conf.*, 2008, pp. 467-472.

[3.10] R. Xiang, A. Arnold, H. Buettig *et al.*, "The ELBE accelerator facility starts operation with the superconducting RF Gun," in *Proc. 2010 Int. Particle Accelerator Conf.*, 2010, pp. 1710-1712.

[3.11] H. Büttig, A. Arnold, A. Büchner *et al.*, "Study of the ELBE RF-couplers with a new 1.3 GHz RF-coupler test bench driven by a resonant ring," *Nucl. Instr. Meth. A*, vol. 612, pp. 427-437, January 2010.

[3.12] J. Teichert, A. Arnold, H. Buettig *et al.*, "A superconducting RF photo-injector for operation at the ELBE linear accelerator," in *Proc. 2007 Free Electron Laser*, 2007, pp. 449-452.

[3.13] V. Volkov and D. Janssen, "RF focusing – an instrument for beam quality improvement in superconducting RF guns," in *Proc. 2000 European Particle Accelerator Conf.*, 2000, pp. 2055-2057.

[3.14] D. Janssen and V. Volkov, "Emittance compensation of a superconducting RF photoelectron gun by a magnetic RF field," in *Proc. 2004 European Particle Accelerator Conf.*, 2004, pp. 330-332.

[3.15] M. Abo-Bakr, W. Anders, T. Kamps *et al.*, "BERLinPro – a prototype ERL for future synchrotron light source," in *Proc. 2009 Superconducting RF Workshop*, 2009, pp. 223-227.

[3.16] T. Kamps, to be published.

[3.17] T. Rao, I. Ben-Zvi, A. Burrill *et al.*, "Design, construction and performance of all niobium superconducting radio frequency electron gun," *Nucl. Instr. Meth. A*, vol. 562, pp. 22-33, June 2006.

[3.18] J. Kewisch, I. Ben-Zvi, T. Rao *et al.*, "An experiment to test the viability of a gallium-arsenide cathode in an SRF electron gun," in *Proc. 2009 Particle Accelerator Conf.*, 2009, pp. 1-3.

[3.19] A. Burrill, I. Ben-Zvi, M. Cole *et al.*, "Multipacting Analysis of a Quarter Wave Choke Joint used for Insertion of a Demountable Cathode," in *Proc. 2007 Particle Accelerator Conf.*, 2007, pp. 3544-3546.

[3.20] J. Sekutowicz, J. Iversen, D. Klinke *et al.*, "Nb-Pb superconducting RF-gun," in *Proc. 2006 European Particle Accelerator Conf.*, 2006, pp. 3493-3495.

[3.21] J. Smedley, T. Rao, P. Kneisel *et al.*, "Photoemission tests of a Pb/Nb superconducting photoinjector," in *Proc. 2007 Particle Accelerator Conf.*, 2007, 1365-1367.

[3.22] R. Calaga, I. Ben-Zvi, X. Chang *et al.*, "High current superconducting gun at 703.75 MHz," *Physica C*, vol. 441, pp. 159-172, July 2006.

[3.23] A. Todd, "State-of-the-art electron guns and injector designs for energy recovery linacs (ERLs)," *Nucl. Instrum. Meth. A*, vol. 557, pp. 36-44, February 2006.

[3.24] I. Ben-Zvi, X. Chang, P. D. Johnson *et al.*, "Secondary emission enhanced photoinjector," Brookhaven National Laboratory, Upton, NY, Technical Report C-A/AP/#149, April 2004.

[3.25] X. Chang, "Studies in Laser Photo-cathode RF Guns," Ph.D. Thesis, Stony Brook University, Stony Brook, NY 11794, 2005.

[3.26] X. Chang, Q. Wu, I. Ben-Zvi *et al.*, "Electron Beam Emission from a Diamond-Amplifier Cathode". *Phys. Rev. Lett.*, vol. 105, pp. 164801-1–164801-4, October 2010.

[3.27] I. Ben-Zvi, private communication.

[3.28] J. R. Harris, K. L. Ferguson, J. W. Lewellen *et al.*, "Design and operation of a superconducting quarter-wave electron gun," *Phys. Rev. ST Accel. Beams*, vol. 14, pp. 053501-1–053501-25, May 2011.

[3.29] S. P. Niles, "Design and analysis of an electron gun/booster and free electron laser optical theory," Ph.D. Thesis, Naval Postgraduate School, Monterey, CA 93943, 2010.

[3.30] J. W. Lewellen, H. Bluem, A. Burrill *et al.*, "ERL 2009 WG1 summary paper: Drive lasers and RF gun operation and challenges," in *Proc. 2009 Energy Recovery Linac Workshop*, 2009.

[3.31] J. H. Billen and L. M. Young, "Poisson SUPERFISH," Los Alamos National Laboratory, Technical Report LA-UR-96-1834, updated 2003.

# CHAPTER 4: DC/RF INJECTORS

BRUCE M. DUNHAM

*Department of Physics and CLASSE*
*Wilson Laboratory*
*Cornell University*
*Ithaca, NY 14853*

**Keywords**
DC Electron Gun, High Voltage, Photocathode, Low Emittance, Photoemission, Superconducting RF, Laser, Laser Shaping, High Power, High Brightness

**Abstract**
A high voltage DC photoemission electron gun followed by an RF accelerating module is presently the best solution for generating high average power electron beams of moderate bunch charge, particularly for energy recovery linac based light sources and for free-electron lasers. Most modern injectors for accelerators use a photocathode illuminated by a pulsed laser which is synchronized to the RF system. This allows one to produce a variety of bunch trains that can be tailored to the needs of a particular machine by modulating the laser appropriately. High gun voltages, above 300 kV, are required in order to combat the emittance growth due to space charge in tightly bunched beams. The emittance growth varies greatly depending on the exact electron bunch dimensions, but DC guns are most often used for bunch charges up to a few hundred picocoulombs. High average beam powers are possible using a DC gun as most of the energy can be directly coupled into the beam, with very little lost in wall heating as is the case for normal conducting RF guns. The gun itself is followed by an emittance compensation section and a high average power RF accelerating section, often superconducting. This chapter will cover the history of DC gun based injectors, system level and sub-system level requirements and the practical details needed to design and construct a device.

## 4.1 INTRODUCTION

One of the first accelerator applications of DC photoemission guns was a GaAs polarized electron source at the SLAC [4.1]. This source used a relatively low voltage gun (~100 kV) followed by a standard RF bunching and accelerating section [4.2]. Chapter 8 details the design of a polarized source. Over time, these morphed into higher voltage, unpolarized DC photoemission sources that directly provided bunched beams, eliminating much of the complexity of RF manipulation associated with chopping a DC or quasi-DC beam to produce a bunch train. This chapter covers the design and implementation of an electron injector using such high voltage DC photoemission guns and an RF accelerator.

There are two main benefits of increasing the gun voltage well above 100 kV. For low bunch charges or quasi-DC beams 100 kV is adequate, but as the bunch charge increases, space charge forces become important and higher voltages are needed to reduce their promotion of emittance growth. The second benefit is that for a high enough voltage (say greater than 250 kV), a pre-acceleration cavity (often called a capture section) is not required, which shrinks the size and complexity of the injector. The RF guns discussed in other chapters also have these benefits.

The Lasertron was one of the first higher voltage DC photoemission guns [4.3]. This was a scaled-up version of a SLAC polarized source, operating at ~400 kV, and applicable as a driver for an RF power source. A similar device, designed and constructed as a high average power electron source for an FEL at Jefferson Lab [4.4], has operated at up to 9 mA and 350 kV for many years.

Problems with HV breakdown were common in early guns (and still are), as the cathodes (GaAs) were activated in the gun's vacuum chamber using cesium to generate high quantum efficiencies. Cesium contamination makes breakdown more likely, as it lowers the work function of the high voltage electrodes. To overcome this problem and make it easier to change cathodes without breaking vacuum, cathode load-lock systems (common in semiconductor industries) were introduced [4.5] so that the cathodes could be prepared in a separate vacuum chamber. The first systems were difficult to use as the load-lock hardware was attached to the gun's high voltage end, making changing and activating cathodes a slow process. In newer guns [4.6], the system's geometry was changed to allow the load-lock to be at ground potential, considerably simplifying the design and operation of that process. Today, one can swap cathodes in a matter of minutes, minimizing downtime and eliminating the HV problems caused by cesium.

Many currently operational DC photoemission guns use a scaled-up version of the original SLAC polarized source. They all continue to suffer from high voltage problems and several laboratories are devising new HV designs to overcome these deficiencies, which are described later in this chapter. Considerable work has been done on electron guns with pulsed high voltage [4.7], but much of this is not directly applicable to DC photoemission guns so will not be covered.

Over time, RF guns (see Chapters 1, 2 and 3) have become very popular due to their ability to produce high bunch charges and high beam energy with good emittance in a relatively small footprint. Recent detailed optimization studies [4.8] of DC photoemission guns followed by RF accelerating sections showing excellent results have renewed interest in this scheme, prompting many laboratories to pursue variants of the basic DC gun design. These studies revealed that it is possible to nearly re-capture the initial thermal emittance of the electron beam generated at the cathode after acceleration to many mega electron volts. DC guns also allow the use of GaAs-like cathodes that have the lowest thermal emittance of any current cathodes [4.9], thus providing a path for extremely bright electron-beam injectors. The other advantage of DC guns/RF injectors is that high average powers are relatively straightforward to obtain, with designs up to 100 mA of average current at 5-10 MeV being tested. In some instances, RF guns can produce high average powers [4.10].

In the next section, we will look at the parameters and requirements for electron injectors and demonstrate that the choice of injector depends on the desired operational goals.

## 4.2 SYSTEM-LEVEL REQUIREMENTS AND DEFINITIONS

The needs of the main accelerator, and ultimately the experiment's beam, will be used for determining the requirements for the electron injector. So, a common question is what type of electron injector is most appropriate to use from the types described in this book? Table **4.1** lists the important parameters.

For example, systems requiring high bunch charge (> 1 nC) and low average power will be best served by an RF gun. System needing a lower bunch charge (less than a few hundred picocoulombs) and high average current (> 1 mA) often will use a DC type gun, while superconducting (SC) RF guns fill the gap between NCRF- and DC-guns. If polarized beams are needed, then only DC guns are appropriate, which will limit some other choices of parameters.

For the rest of this chapter, I discuss only the DC gun/RF injector, within the parameter ranges in Table **4.2**. If polarized electrons are required, one must use GaAs-like photocathodes that have a very special list of additional requirements, covered in detail in Chapter 8. GaAs has several other special attributes that make it desirable even for non-polarized beams; in the rest of this chapter, I will assume semiconductor cathodes are used.

| **Parameter** | **Min** | **Max** |
|---|---|---|
| Polarized Beam? | | |
| Bunch Charge [pC] | $10^{-3}$ | $10^{4}$ |
| Average Current [mA] | $10^{-3}$ | $10^{2}$ |
| Duty Factor [%] | ~0 | 100 |
| Emittance [mm mrad rms, normalized] | 0.1 | > 10 |
| Final Beam Energy [MeV] | 0.5 | 20 |
| Bunch Length [Degree of RF phase] | 0.1 | 10 |
| RF Frequency [MHz] | 100 | 3 000 |

**Table 4.1. Important parameters for electron injectors, with typical minimum- and maximum-values of interest.**

| **Parameter** | **Range** |
|---|---|
| Polarized Beam? | No |
| Bunch Charge | 0-200 pC |
| Average Current | 0-100 mA |
| Duty Factor | 0-100% |
| Emittance | < 2 mm mrad rms normalized |
| Final Beam Energy | 5-15 MeV |
| Bunch Length | < 1 deg |
| Frequency | 1 300 MHz |

**Table 4.2. Range of parameters used for the discussions in this section.**

Before continuing, several terms and definitions need to be covered. An important concept is the difference between continuous wave (CW), DC and pulsed, when applied to high voltage and beams. DC means that the high voltage power supply is always on and connected across the gun's cathode-anode gap, or that the electron beam is always on. Pulsed signifies that the power supply is only on for a brief period, overlapping the time when the electron beam is extracted from the cathode. Many industrial devices are pulsed (*e.g.*, X-ray tubes, klystrons, modulators). For such systems, pulse durations are typically less than 1 μs, and pulsed power operators consider anything longer than that to be DC. For this chapter, DC means "always on". CW denotes that for each period of an RF cycle, a bunch of electrons is present. The DC/RF photoemission injectors discussed here are a hybrid system, since the gun voltage is always on (DC) and a laser generates electrons to fill each RF period (CW). The RF guns discussed in Chapter 3 typically are pulsed-mode devices, but CW versions can be built.

Other important concepts are the thermal emittance and beam brightness. Assuming that the emitted electrons follow a Maxwellian transverse velocity distribution, the transverse-momentum spread of the electrons is $\sigma_{p,rms} = \sqrt{mkT}$, where $kT$ is the electrons' effective transverse energy and $m$ is the mass of the electron. For a negative electron affinity cathode (NEA) like GaAs, illuminated with a wavelength near the band gap, $kT$ approaches the temperature of the crystal (at room temperature, $kT$ = 25 meV where $k$ is the Boltzmann constant and $T$ is the temperature). For non-NEA cathodes, $kT$ will be above the cathode temperature, and if GaAs is illuminated with a wavelength greater than the band gap, the effective temperature also will increase. Knowing $kT_{\perp}$, the normalized, transverse thermal-emittance of a beam from a photocathode, $\epsilon_{n,rms,th}$, is calculated from [4.11]

$$\epsilon_{n,rms,th} = \sigma_{laser,rms}\sqrt{\frac{kT_{\perp}}{mc^2}} \tag{4.1}$$

where $\sigma_{laser,rms}$ is the rms laser spot size at the cathode and $c$ is the speed of light in a vacuum. This holds only when the charge per bunch is low enough that space charge forces are not a concern. Thus, finding cathodes with the smallest $kT$ is crucial for low emittance guns.

For a given electric field at the cathode surface, $E_{cathode}$, Gauss' Law gives the maximum emission charge density, $\sigma$, that can be supported for a pancake beam, $\sigma = E_{cathode}\ \varepsilon_0$, where $\varepsilon_0$ is the permeability of free space. If a charge per bunch $q$ is required, then the area of the laser beam must be greater than or equal to $q\ \sigma^{-1}$ for a uniform laser distribution. This sets a minimum value on the thermal emittance, as the laser area (and spot size, $\sigma_{laser}$) is then determined from $E_{cathode}$ for the desired charge per bunch. This is a reason to push the electric field at the cathode as high as possible. Doubling the field allows the laser radius to be reduced by $\sqrt{2}$ for a constant charge, subsequently reducing the emittance by $\sqrt{2}$.

The brightness of the electron beam is another important term to understand, as the X-ray beam brightness produced by a light source (storage ring or Energy Recovery Linac) is related directly to it. The transverse, normalized beam brightness, $B_n$, is defined as

$$B_n = \frac{2I}{\pi^2 \varepsilon_{(n,x)}\varepsilon_{(n,y)}} \tag{4.2}$$

where $I$ is the beam current, and $\varepsilon_{(n,x)}$ and $\varepsilon_{(n,y)}$ are the normalized emittances in the transverse planes. The maximum transverse brightness for an electron bunch was recently shown to be given by [4.11]

$$\frac{B_n}{f} = \frac{mc^2\varepsilon_0 E_{cathode}}{2\pi kT} \tag{4.3}$$

where $f$ is the repetition frequency of the bunch train. The importance of this equation is that the maximum obtainable beam brightness depends not on the actual bunch charge, but solely on the transverse thermal energy of the emitted electrons and the electric field at the cathode surface at the time of emission. This again demonstrates the drive to increase the field at the cathode and to identify cathodes with the lowest possible $kT$.

Figure **4.1** is a block diagram of a DC/RF photoinjector. To control the space charge forces, the initial bunch length exiting the gun is set to ~30-40 ps (for 1 300 MHz) due to the relatively low initial beam energy (compared to an RF gun). The bunch is compressed further as it passes through an RF buncher cavity (normal conducting (NC) in this case) to 5-10 ps at the entrance to the accelerating cavities. The beam then

is compressed even more and accelerated as it passes through several SCRF cavities, to a final bunch length of 1-2 ps. The solenoidal focusing magnets control the beam size and compensate for emittance [4.12].

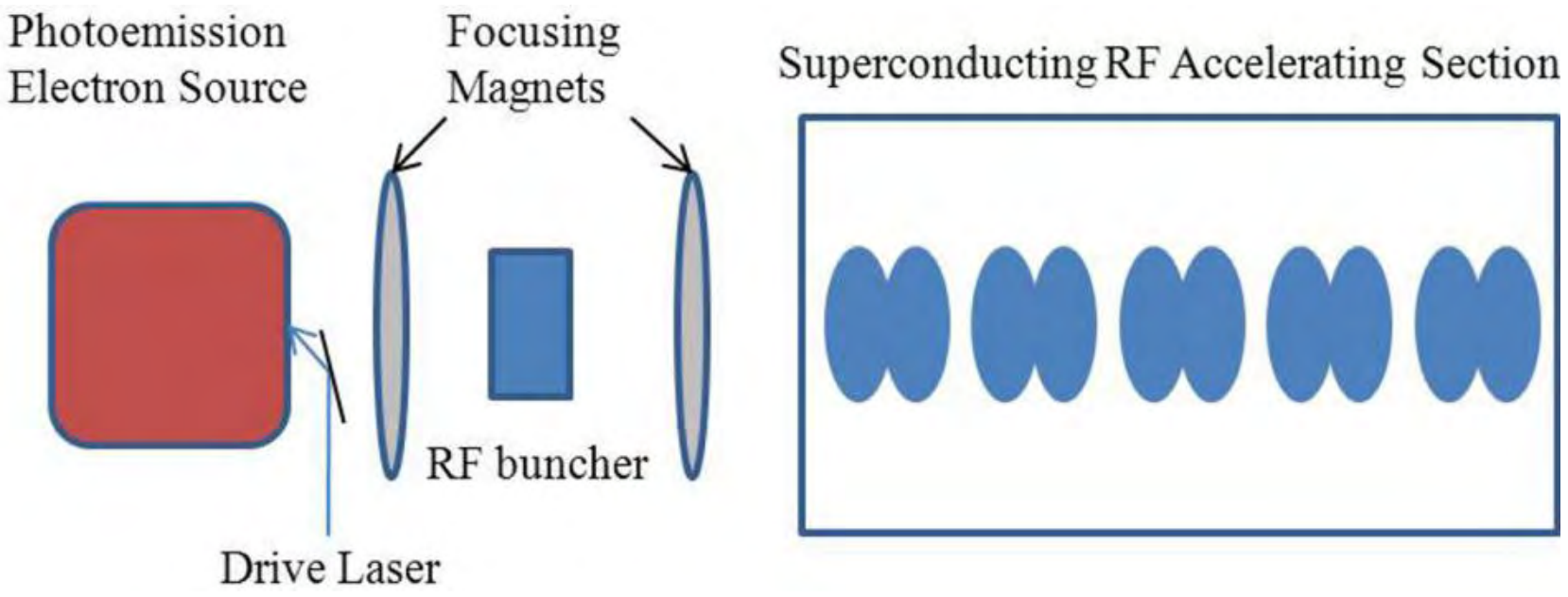

**Figure 4.1. Schematic of a typical electron injector using a DC photoemission gun and a superconducting RF accelerator section.**

For comparison, Figure **4.2** illustrates a thermionic electron gun-based injector [4.5]. In this arrangement the output beam is either DC or has a long pulse (for a gridded gun). To reduce the bunch length enough for injection into RF accelerating cavities, several extra steps are required. In some instances, a chopper system "chops" out a fraction of the beam in time, so that the bunches are at the proper repetition frequency (or sub-harmonic thereof). Then, one or more sub-harmonic- or harmonic-buncher cavities compress the bunches further until they are ready for acceleration. If the gun's voltage is too low, a pre-acceleration cavity may be necessary to boost the energy high enough for injection into the final set of accelerating cavities. Solenoid magnets along part or all of the beamline provide focusing. These systems have been used successfully for many years, but are more complex, less efficient and cannot provide the beam quality needed for some of today's demanding applications.

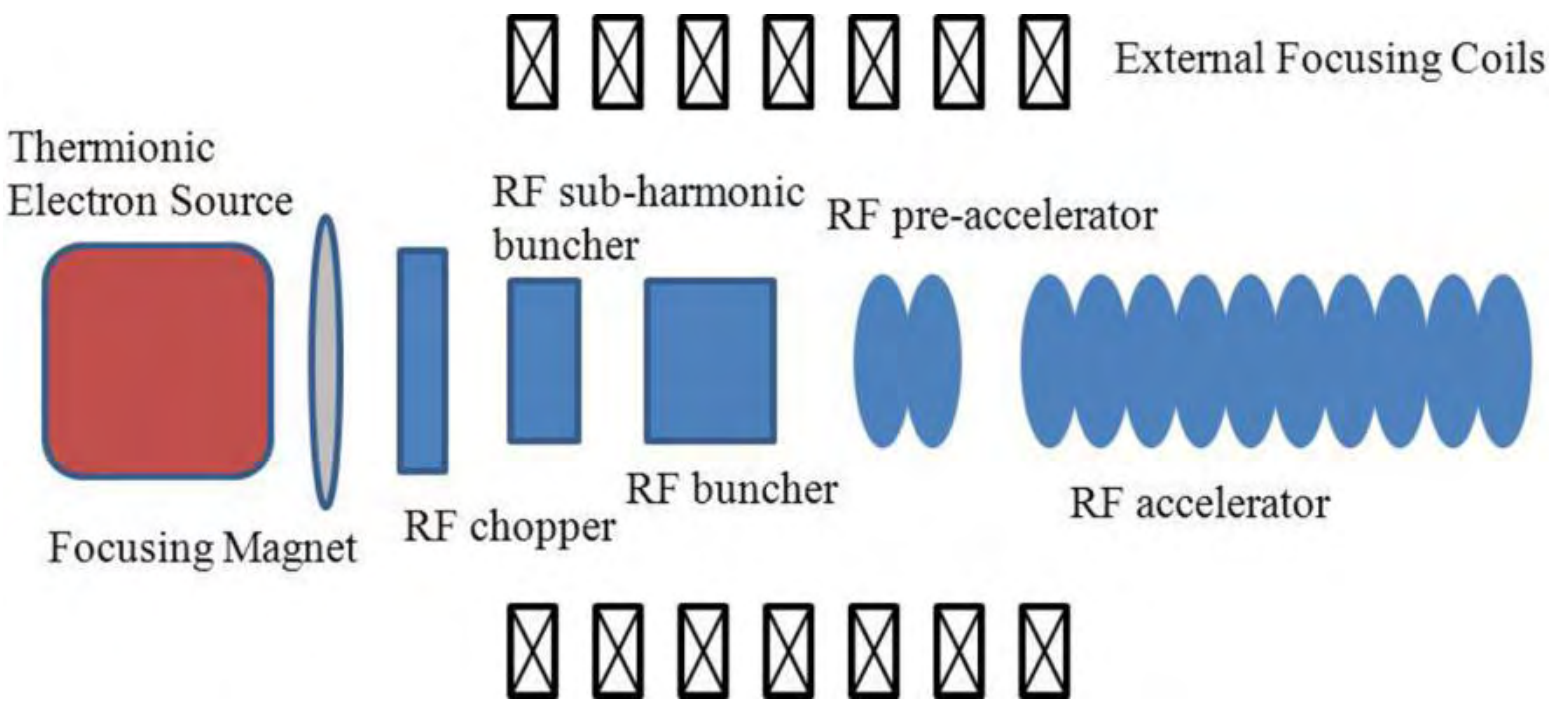

**Figure 4.2. A typical set-up for an electron injector using a thermionic gun and a complicated RF section for chopping, bunching and acceleration.**

# 4.3 SUB-SYSTEM DESCRIPTION

## 4.3.1 Overview

Designing and constructing the components of a DC photoemission electron-source-based, high average power injector is a difficult task: it still is evolving as more and more laboratories become involved. In this section, I describe the design and construction of the sub-systems of a DC photoemission gun/RF injector

## 4.3.2 High Voltage Gun

Simulations [4.8] show that, up to a certain point, a higher gun voltage is important for obtaining low emittance, after which the improvement is relatively small. Thus, a DC gun operating between 400-600 kV with the appropriate cathode should meet the emittance goals (Table **4.2**) for many high performance applications.

To operate in this range, the gun must be processed to 10-25% above the operating value to reduce field emission and arcing; consequently, a 600 kV operational voltage requires a gun designed to withstand 750 kV maximum voltage. Figure **4.3** is a schematic cutaway of a DC photoemission gun, designed to meet the requirements in Table **4.2**. This particular gun was operated for over a year using a test beamline to measure the performance of the gun and cathode before it was incorporated into the rest of the injector. Details of these measurements were published elsewhere [4.13]; the main result was that the emittance measurements at 77 pC per bunch and 250 kV beam energy match the simulations very closely, giving confidence that the latter are accurate.

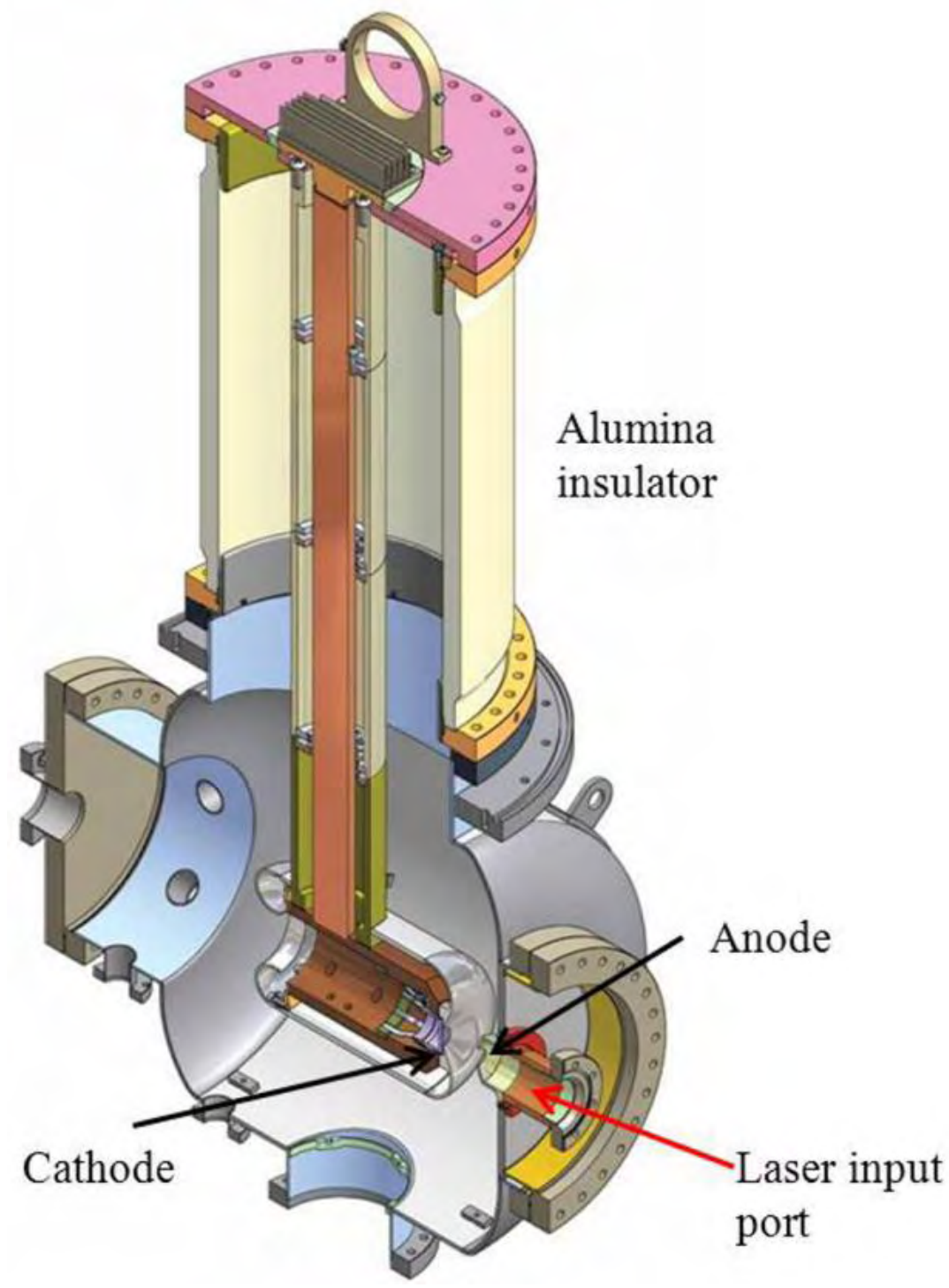

**Figure 4.3. A cutaway view of a DC photoemission gun. [Adapted with permission from [4.14]. Copyright 2008, American Institute of Physics]**

Sections 4.3.2.1 through 4.3.2.6 describe the gun's components.

#### *4.3.2.1 Photocathode Materials*

A perfect photocathode for an accelerator electron source would have high efficiency at a convenient laser wavelength, a fast response time, a long lifetime and a low thermal emittance. Unfortunately, no such cathode exists, although the search continues. Several different photocathodes meet some of these criteria, so tradeoffs must be made depending on a particular system's requirements. For an ERL application, obtaining 100 mA average current necessitates having high quantum efficiency (QE) photocathodes.

Currently, semiconductor photocathodes are the best choice for high QE and low emittance. Examples are GaAs, $Cs_2Te$, GaN and $K_2CsSb$. Both $Cs_2Te$ [4.15] and GaN [4.16] show promise, but require UV light; no laser systems yet can produce enough average power in the UV spectrum for 100 mA operation. Both GaAs and $K_2CsSb$ have 5-10% QE for ~520 nm light, where high average power lasers are readily available. In this chapter, only semiconductor cathodes are covered; Chapters 6, Chapter 7 and Chapter 8 discuss other types of cathodes for photoemission guns.

As mentioned earlier, nearly all the intrinsic thermal emittance from the cathode is recoverable using a carefully designed emittance compensation scheme. Of all the cathodes available, GaAs has the lowest thermal emittance [4.9], and is thus the one of choice for ultimate brightness. Unfortunately, the QE is near a minimum when the thermal emittance is smallest (close to the band gap), thus, it is unsuitable for high average current, so a compromise must be made between QE and emittance. Figure **4.4** demonstrates that the transverse electron energy at 520 nm is approximately 4X larger than at the band gap wavelength (850 nm). Quantum efficiencies of 5-20% can be reached at 520 nm, but fall quickly to less than 1% as the band gap wavelength is approached.

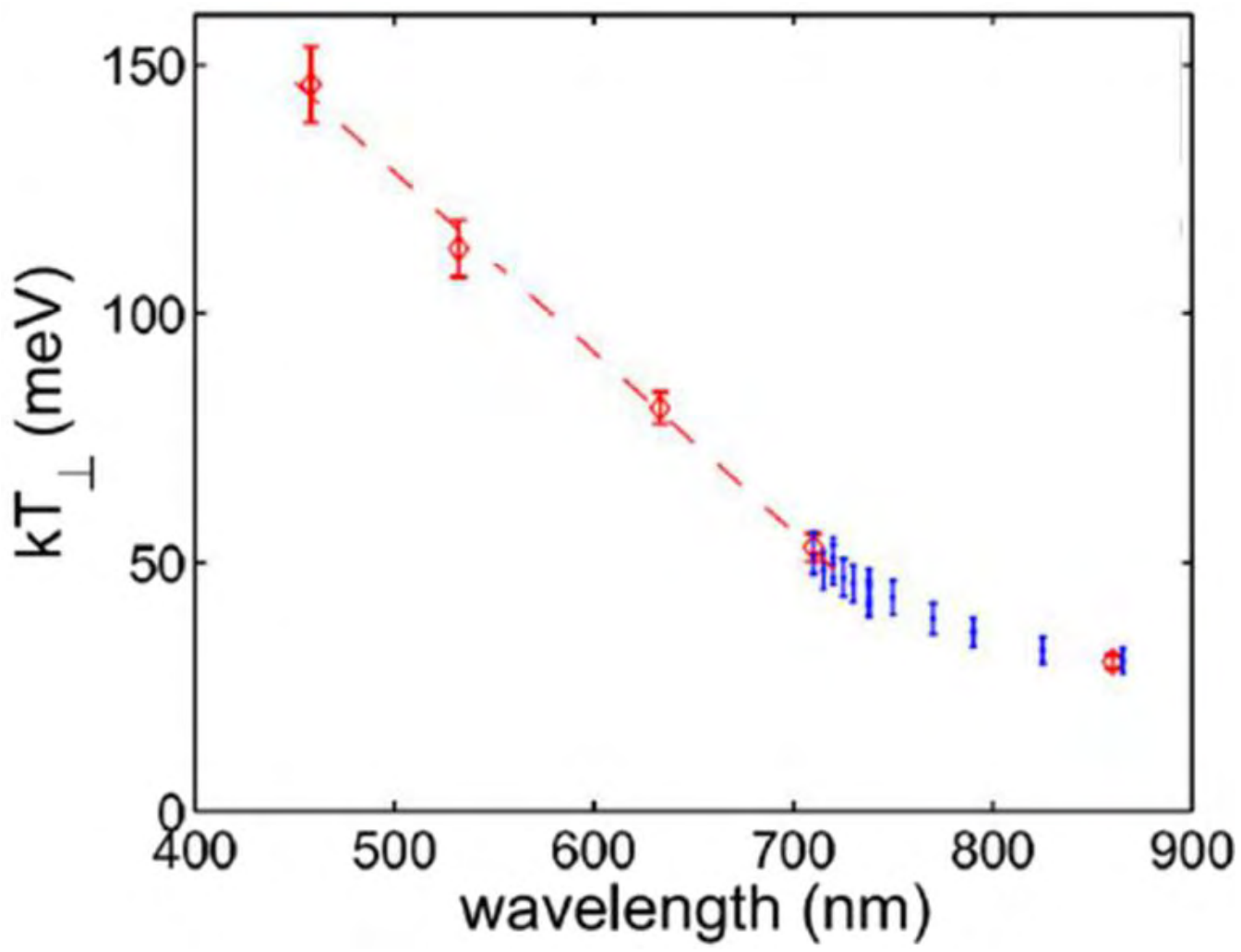

**Figure 4.4. Transverse thermal energy for the electrons photoemitted from GaAs as a function of wavelength [4.13]. The red points were measured using multiple laser spot sizes while the blue points were measured using a single spot size. The dashed line shows a fit over the linear region. $kT_\perp$ is the electron's effective transverse energy normal to the cathode's surface. [Reprinted with permission from [4.13]. Copyright 2008, American Institute of Physics.]**

In addition, it is known that at ~780 nm a fast laser pulse will generate an electron beam with a ~10 ps long tail [4.17], which is unacceptable for low emittance operation. For shorter wavelengths (~520 nm), recent measurements [4.13] show (Table **4.3**) that the response time is quite fast (~1 ps) provided that the QE is not too high (< 10%); this phenomenon is related to the absorption depth of light versus wavelength. Near the band gap, GaAs is nearly transparent, and thus, the electrons are produced deep in the material and take longer to reach the surface. The longer extraction time, though, gives the electrons more time to thermalize into the conduction band minimum, reducing their effective temperature. This example highlights some of the tradeoffs that must be made even when using GaAs; an operating wavelength of ~520 nm often is chosen as the best compromise between QE, thermal emittance, response time and available lasers. Some people have suggested using sufficiently thin layers of GaAs to overcome this problem, thin enough that the electron's transit time is shorter than that due to the absorption depth [4.18]. These layered cathodes, grown on glass and used in transmission mode, may simplify delivery of the laser to the cathode.

| **Wavelength [nm]** | **$\tau$ at 250 kV [ps]** |
|---|---|
| 860 | 69 ± 22 |
| 785 | 9.3 ± 1.1 |
| 710 | 5.2 ± 0.5 |
| 520 | < 1.0 |
| 460 | < 0.14 |

**Table 4.3. GaAs response time, τ, for various laser wavelengths [4.13].**

While GaAs offers the very high quantum efficiencies necessary for high power guns, it also is prone to generating a "halo" from unintended areas of the cathode. The cathode's surface is usually larger than the size of the electron beam (and the laser spot) to assure a uniform electric field across the emission area. Any stray or scattered light hitting outside the desired spot generates accidental electrons (Figure **4.5**). This drawback turned out to be the biggest factor in the poor cathode lifetime in the Jefferson Laboratory's polarized source [4.19]. Electrons generated near the edge of the electrode from stray light experienced non-uniform fields and eventually hit the vacuum chamber walls, producing secondaries, X-rays, UV light and increasing the vacuum, which all degrade the cathode's longevity.

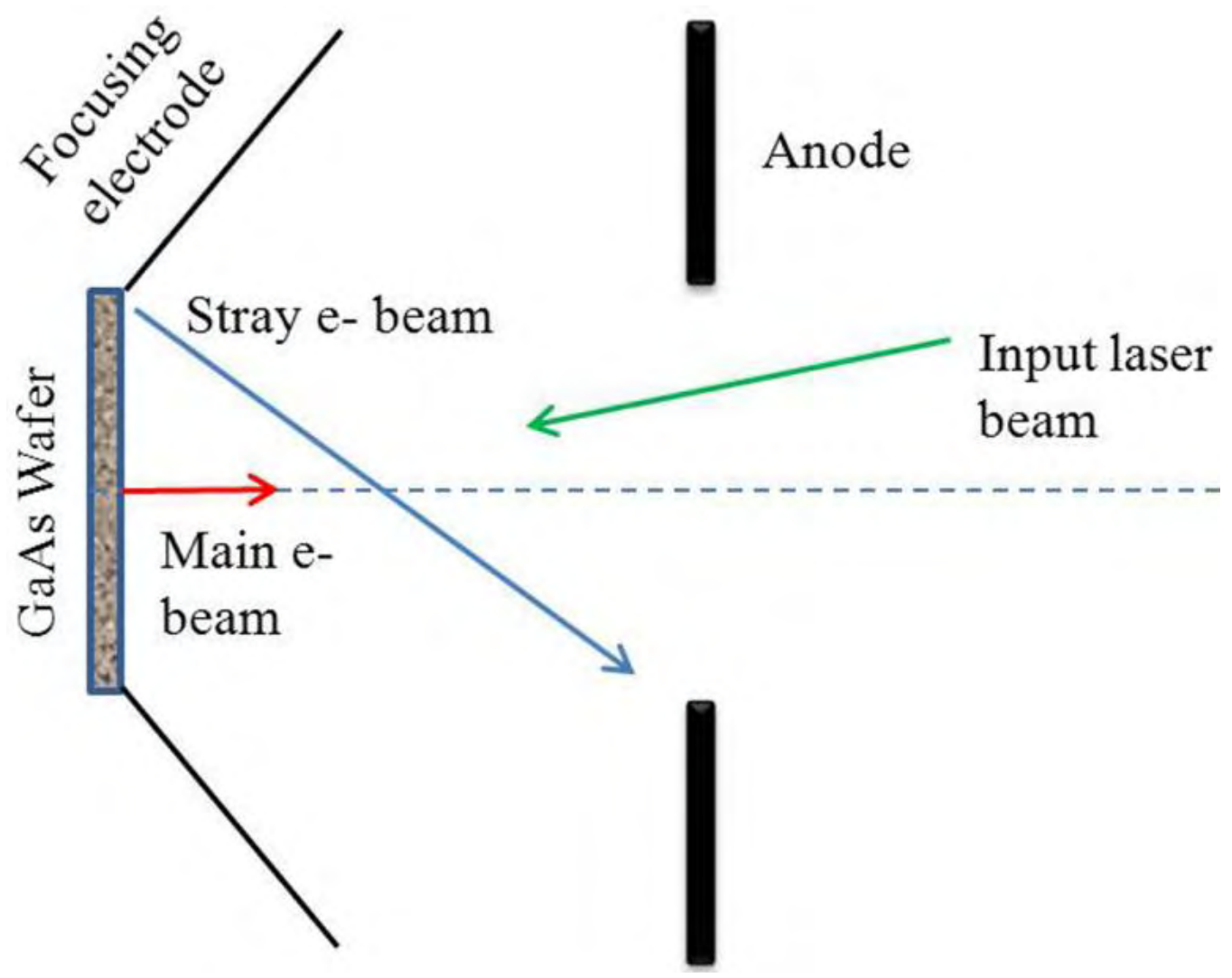

**Figure 4.5. Stray light can hit any part of the cathode's surface, thereby producing electrons. Electrons generated in non-uniform field regions (blue) can be deflected to hit the vacuum chamber walls.**

To overcome this problem, the outer area of the cathode must be inactive so it cannot emit electrons. One way to accomplish this is to cover the wafer with an oxide layer (deposited in an electrochemical cell using a weak solution of phosphoric acid) and then strip off the oxide at the center of the wafer (Figure **4.6**). Another way is to mask the wafer when depositing the cesium during activation.

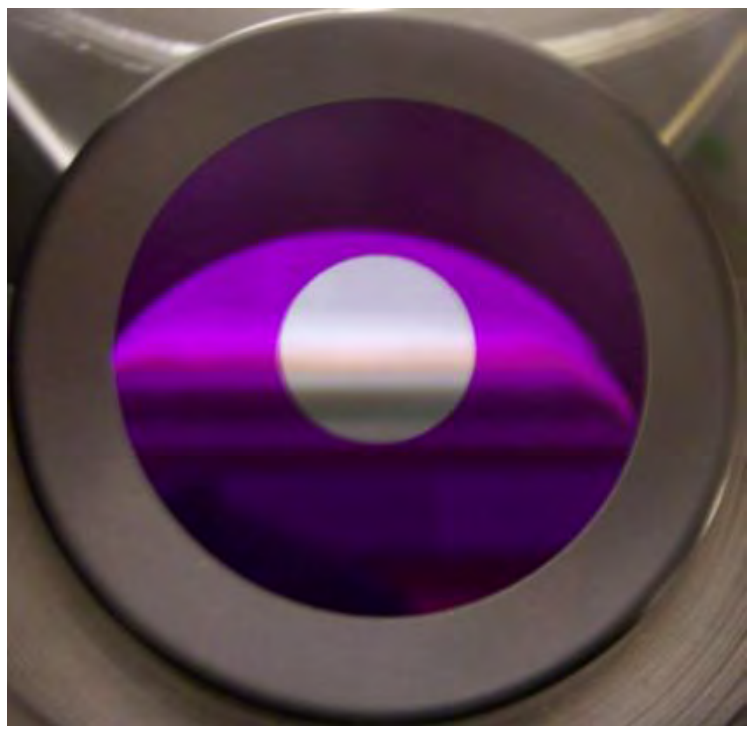

**Figure 4.6. The central silver circle is the active area of the GaAs cathode, while the purple area was anodized to inhibit electron emission. A tantalum retaining ring (gray) surrounds the wafer.**

The last important parameter is cathode lifetime. The cathodes described above are all somewhat sensitive to chemical poisoning, requiring an ultra-high vacuum (UHV) environment. GaAs is the most sensitive, unfortunately requiring vacuum levels < $1\times10^{-9}$ Pa to operate successfully. An additional lifetime limiter is ion back-bombardment that further lowers the QE. The electron beam can ionize residual gas molecules anywhere along their path, which then can be accelerated back towards the cathode's surface. Jefferson Laboratory's researchers [4.20] carried out extensive measurements at 10 mA average current and measured cathode lifetimes as high as $10^6$ C cm$^{-2}$ (the amount of charge extracted per square-centimeter when the QE has fallen by 1/*e*). We can use this data to estimate that a 10 W maximum power laser system should provide 100 mA over 100 hr with a 1.8 mm diameter laser spot [4.21]. Such performance in an operational environment has not been demonstrated; certainly, it is an optimistic estimate.

Figure **4.7** illustrates what happens during ion back-bombardment. An electron (red trajectory) is accelerated and can ionize any gas in its path. The ions (blue trajectories) are then accelerated back towards the cathode in a straight line due to their larger mass. The maximum in the ionization cross section is in the 50-100 V range, so many ions are created close to the cathode's surface, but also can be produced, with lower probability, anywhere between the cathode and anode, and beyond the anode. The right part of the figure shows an actual quantum efficiency map of a cathode after operating with the laser positioned off-center. The ion damage at the laser spot and a trough between it and the center of the cathode is clearly visible. The most straightforward method for reducing ion back-bombardment for GaAs is to lower the partial pressure of the background gas species as much as possible.

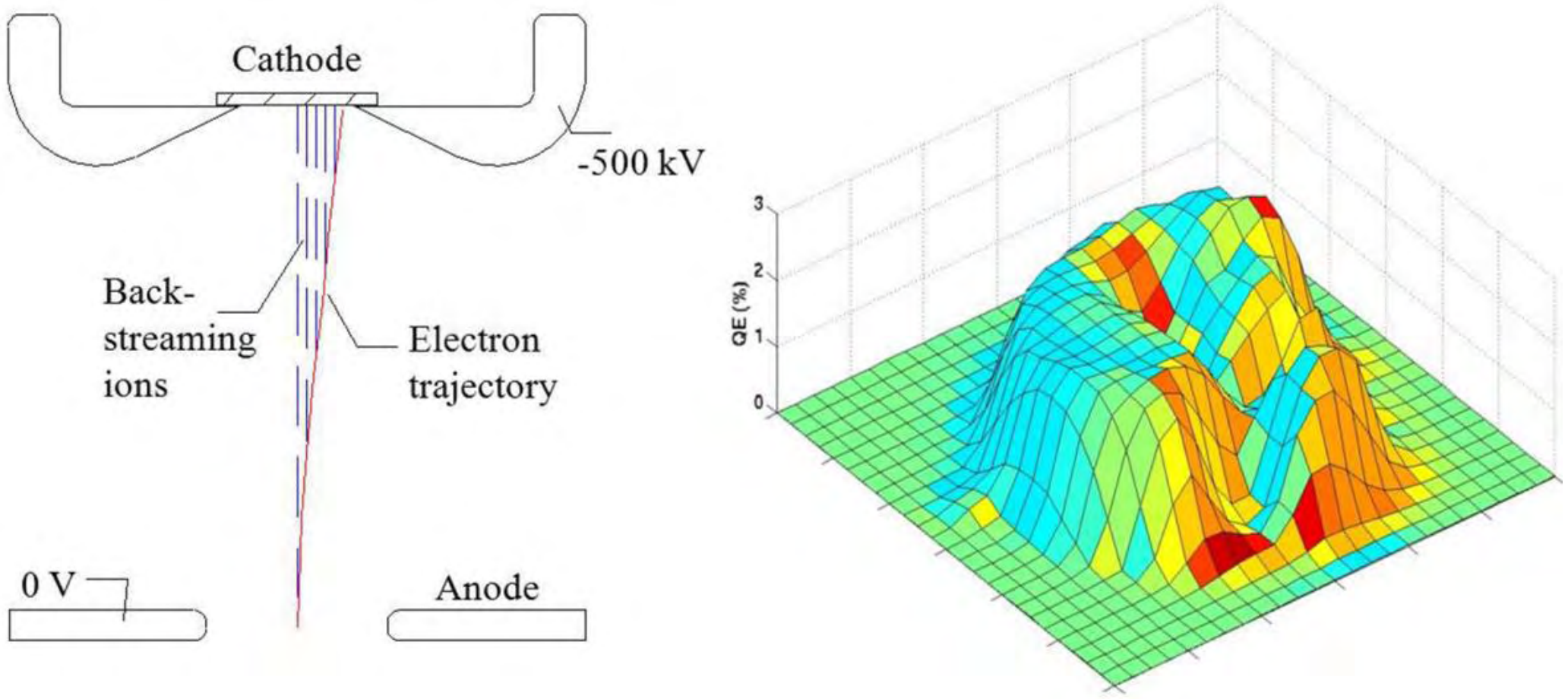

**Figure 4.7. Electrons leaving the cathode (red) ionize residual gases that then are accelerated back to the cathode (blue). The QE map on the right is from a cathode undergoing ion back-bombardment (arbitrary color map). The laser was positioned off-center, and the ions produced a channel from the laser's position to the electrical center of the cathode. [Adapted with permission from [4.22]. Copyright 1, American Institute of Physics]**

Many groups worldwide continue to investigate new cathode materials for electron sources and detectors. Reference [4.23] gives a good summary of recent work.

#### *4.3.2.2 Load-Lock System*

The semiconductor industry has used load-lock systems for many years to introduce wafers into the processing system without having to open the entire vacuum system. Similar reasons drove the development of load-lock systems for photoemission guns, and now they are a standard component for most new guns.

As part of GaAs activation, the cathode material is heat-cleaned to remove surface contaminants and then activated with cesium and an oxidant. For all the early photoguns, this was done inside the gun vacuum chamber [4.24]. However, this is detrimental to high voltage performance for two reasons. First, heating the cathode also heats high voltage surfaces that previously were conditioned; this can alter the electrode surfaces enough to degrade the voltage hold-off strength. Second, the cesium added to the cathode can end up on the high voltage surfaces, locally reducing their work function and increasing the probability of field emission. In addition, installing a new cathode meant breaking the vacuum and re-baking, a 1-2 week process, which is incompatible with accelerator operations.

While a number of designs were tried for keeping the cesium from reaching the electrodes, the only sure way is to separate the functions of preparing the cathode from the gun, and to use a load-lock system. The Stanford Linear Collider polarized source [4.25] tried several load-lock configurations to improve the gun's reliability. In one, the load-lock hardware was on the gun's high voltage end, requiring a large high voltage enclosure and limited access to when the gun was turned off. Now, several guns have altered the system geometry to allow the load-lock to be at ground potential, simplifying the design and operability (Figure **4.3**).

Below is a list of typical requirements for a photocathode load-lock system. Some of the functions may be combined in a single chamber, if desired. The functions may vary depending on the cathode type.

- separate chamber for loading new cathodes, including quick bake-out capabilities
- cathode cleaning chamber (heat cleaning and/or other cleaning methods)
- storage chamber for extra cathodes
- preparation chamber for activating cathode
- transfer mechanism between chambers and to the gun that must not generate any particles that can be transferred to the gun, nor ruin the vacuum in it

Figure **4.8** is a schematic of a simple but effective load-lock system that is mated to the gun shown in Figure **4.3**. Each chamber has its own ion pump, and the cathode-preparation system also has non-evaporable getter (NEG) pumps to keep the vacuum low during activation and storage.

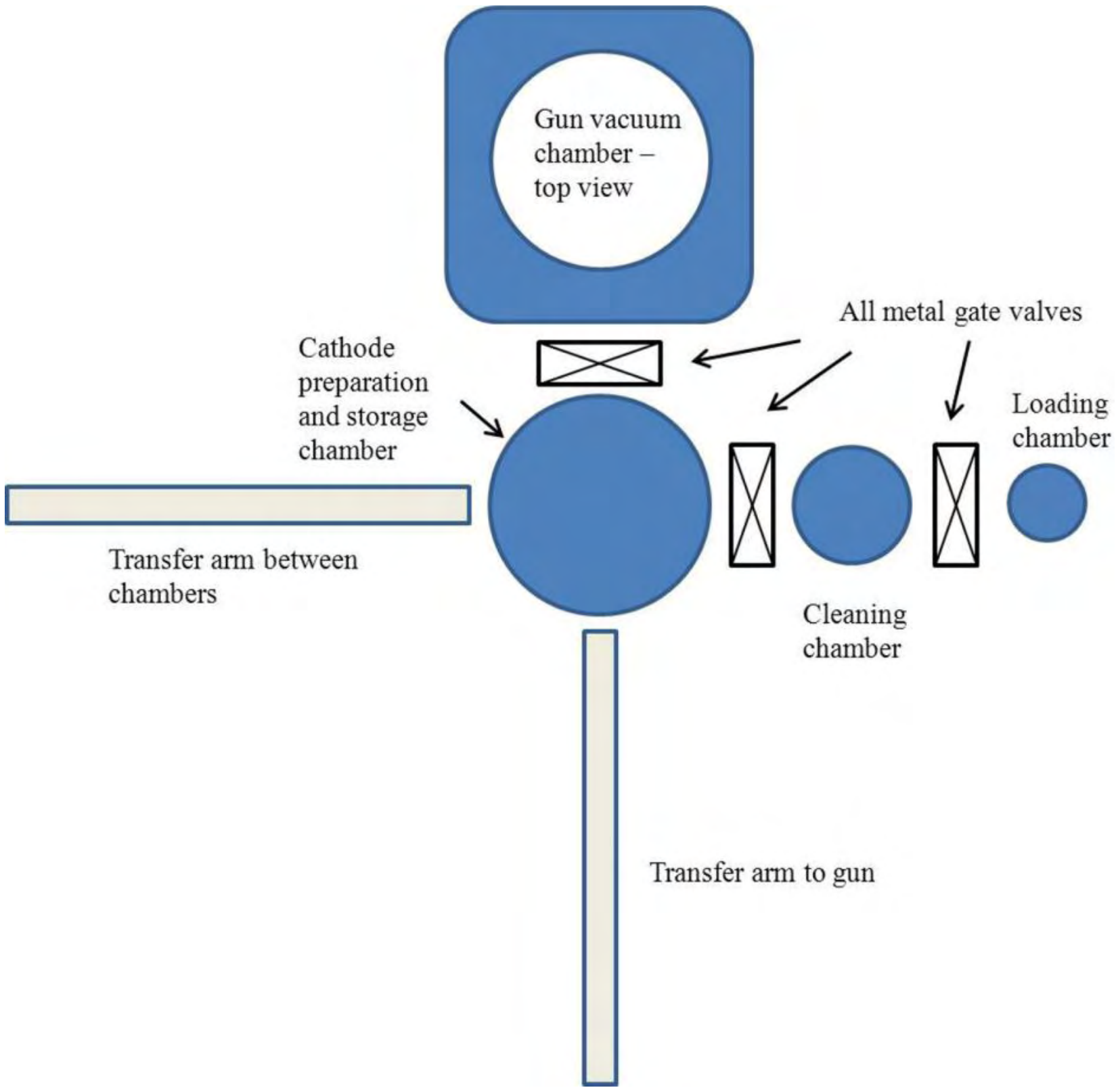

**Figure 4.8. An example of a cathode load-lock system.**

The load-lock chamber (Figure **4.9**) has an internal quartz-lamp heater for performing a quick bake-out, and the volume of the load-lock is minimized to facilitate rapid pump-down. Once the vacuum is low enough in the loading chamber (< $10^{-6}$ Pa), both gate valves are opened and the transfer arm is pushed through to the loading chamber to pick up the cathode holder and move it to the cleaning chamber. Thereafter, both valves are closed and the wafer is heated to 580-600 ˚C for 1-2 hr. Some laboratories also use atomic-hydrogen cleaning as the first step [4.26]. After the cathode has cooled to room temperature (using a cold finger to hasten the process), the transfer arm picks up the holder and moves it to the cathode preparation chamber for activation.

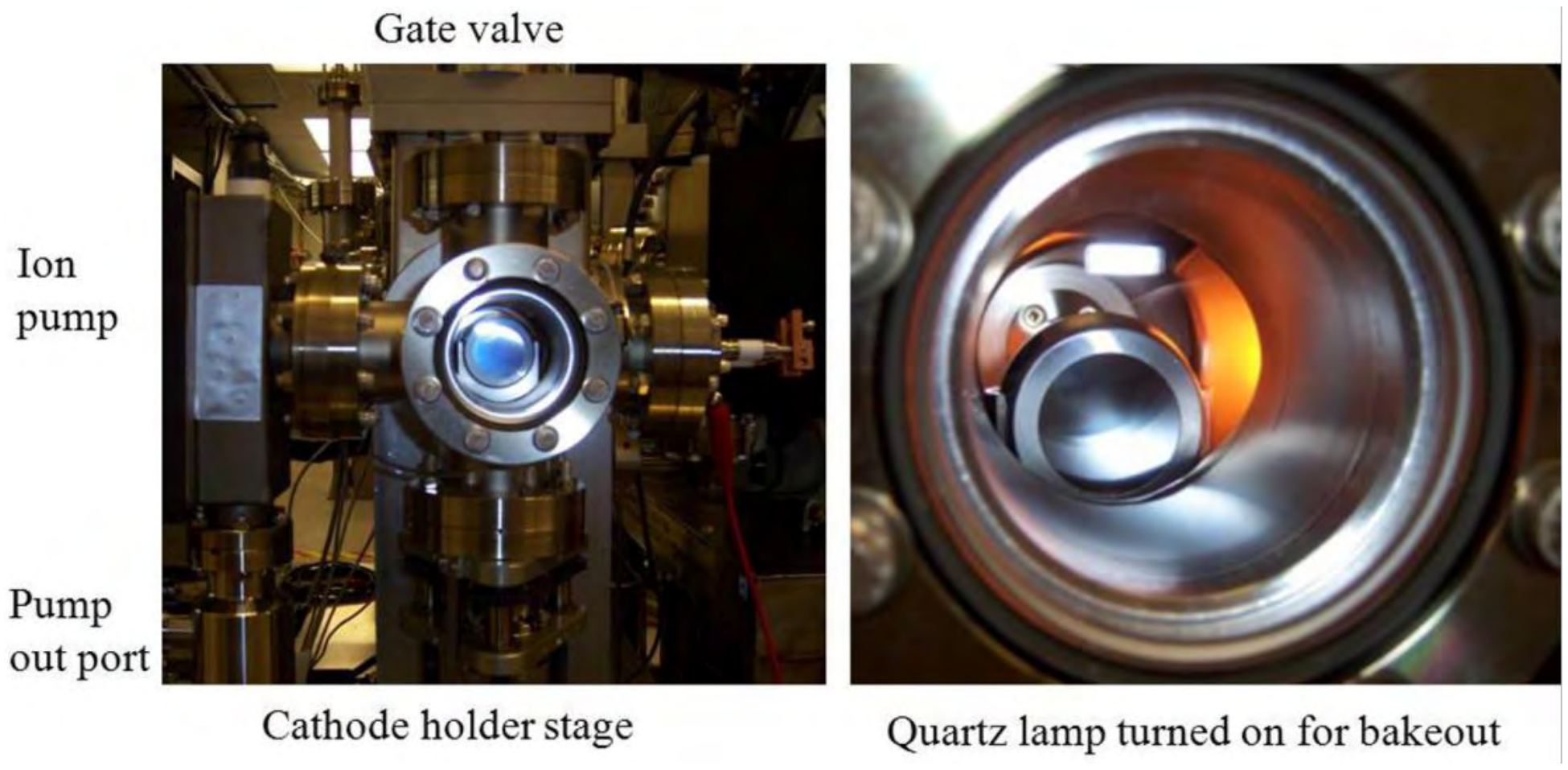

**Figure 4.9. The load-lock chamber is shown on the left and a close-up of the system during a bake-out is shown on the right.**

After activation, the valve between the gun and preparation chamber is opened and the transfer arm moves it into the gun, where it is registered and locked to the back of the electrode. There are numerous ways to secure the cathode, as long as the requirement of no particle generation is met. The materials in contact should be different to avoid cold welding. Molybdenum is used for the GaAs support and copper often is employed for the mating surface if heat must be removed from the cathode during operations. The arm is retracted, the valve closed, and then the gun is ready to go. Also, the cathode can be "topped up" if desired, by quickly retracting it, adding cesium, and then re-inserting it into the gun. During all of these procedures, the gate valve between the gun and the load-lock system should remain closed to reduce any chance of contamination, except when the cathode is removed from, or inserted into the gun.

At least two types of commercial transfer arms work well; one is a long, welded bellows, and the other a magnetically coupled rod. The former is somewhat more delicate of the two and requires care so as not to damage the thin-walled bellows. The magnetic type has internal bearings that potentially could produce particles, but the newest models generally are reliable in this respect.

#### *4.3.2.3 Vacuum Requirements*

The use of semiconductor photocathodes and their reliance on low work function surfaces impose strict requirements on the vacuum conditions in the gun during operation. For GaAs cathodes in particular, vacuum levels < $1\times10^{-9}$ Pa are necessary for good cathode lifetimes. One particularly damaging phenomenon, ion back-bombardment, as discussed in Section 4.3.2.1, is best combated by reducing the vacuum level as low as possible.

Several methods will assure extremely low vacuum levels. The base pressure in stainless steel vacuum systems is dominated by hydrogen outgassing from the system's thick metal parts [4.27]. It is preferable to use SUS316L stainless steel for the vacuum chamber and all the flanges (and, in some cases, SUS316LN, if the flange must be brazed) to avoid any issues with stray magnet fields. For critical high voltage components, vacuum remelted stainless steel should be considered as it has fewer inclusions and contaminants that can cause field emission.

The hydrogen outgassing rate for stainless steel can be minimized by following the procedure described by Park [4.27]. All the stainless steel components should be baked in air (a vacuum bake is acceptable also) at 400 ˚C for 100-200 hours, followed by a 150 ˚C 24-hour vacuum bake after final assembly. This procedure is most effective on thin walled parts (a few millimeters), and less effective on thick flanges. It can reduce the outgassing rate to as low as $2\times10^{-14}$ Torr L $s^{-1}$ $cm^{-2}$. Note that titanium also was used successfully to produce low hydrogen outgassing rate chambers [4.28]. These rates can be lowered further by cooling the vacuum chamber walls, since the rate drops exponentially with temperature.

Only selected materials may be used in a gun, chosen for various mechanical-, thermal- and electrical-needs, including stainless steel, molybdenum, copper, alumina, or other pure metals (*e.g.*, titanium, tantalum, niobium and tungsten), plus the cathode material. Procedures for cleaning various materials are given in the literature [4.29] and many laboratories have their own standard procedures. Unfamiliar materials and coatings must be avoided without extensive testing; it is best not to trust recommendations from others without first conducting one’s own tests.

The elimination of particles and dust is another critical process in preparing vacuum systems for high voltage guns. The SCRF community spent years perfecting particle-reduction techniques for superconducting cavities [4.30], many of which can be applied to guns. For example, the last step in cavity production is called high-pressure water rinsing (HPR). A moving jet of ultra-pure de-ionized water at $7\times10^{6}$ Pa is sprayed on the surface of the niobium cavity for several hours and then allowed to dry in a clean room. The physical process is very effective in removing dust and contaminants that cause field emission. This process is applicable to most parts of the gun vacuum system before assembly. Even ion-pump chambers should be taken apart and cleaned with HPR. Section 4.3.2.4 gives an example with high voltage electrodes.

Another method for removing particles and quantitatively determining their contamination level is depicted in Figure **4.10** [4.31]. High-pressure gas (nitrogen or dry air) at $5\times10^{5}$-$1\times10^{6}$ Pa is sent through a 0.003 µm particle filter and then through a clean stainless steel hose and nozzle. The cleanliness of the filter, hose and nozzle first is checked by directing the flow into a particle counter and waiting until the count drops to near zero. Then, the object under test is sprayed with the high-pressure gas stream with the particle counter downstream. This process is continued until the particle count drops to near zero, which may take anywhere from a few minutes to hours, depending on the object’s initial state of contamination. All of these procedures must take place in a cleanroom environment. This is an important final step before assembly, particularly for checking commercial parts, such as gate valves, that have many potential areas to trap particles and cannot easily be disassembled. Commercial $CO_2$ snow guns also can be used for this [4.32]; they utilize $CO_2$ as the gas and generate snow-like flakes that remove particles and degrease the surface.

Another technique to lower distributing particles from one place to another is to carry out the initial pump-down very slowly [4.33]. Evacuation rates of 5 000 Pa L $s^{-1}$ are slow enough to keep from stirring up particles and transporting them; this is accomplished using mass controllers and automated valves with inline 0.003 µm particle filters. Similar equipment is employed for venting the system.

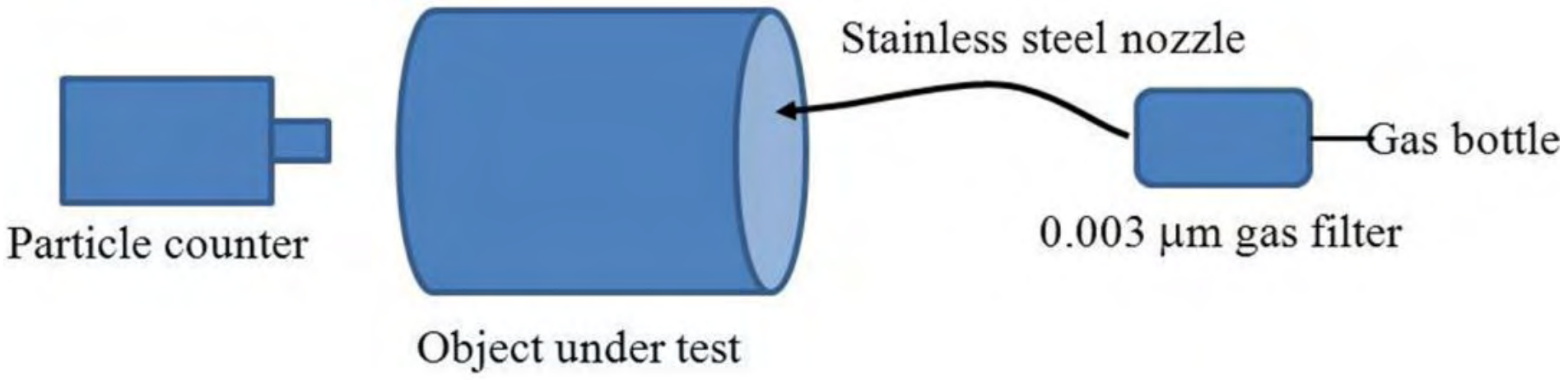

**Figure 4.10. Set-up for removing particles from an object. Filtered gas is directed at an object, and the downstream particle count is monitored until it drops to near zero.**

Even when using the procedures for hydrogen outgassing reduction, massive pumping still is required; often, an ion pump plus a large array of getter pump strips (such as SAES ST-707 [4.34], Figure **4.12**) are used. The getters must be arrayed around the beam area as close as allowed by the high voltage hold-off tolerances. At low pressures, the pumps will be effective in the beam area only if there is a direct line of sight to the getters. The ion pump is needed for bake-outs and to pump methane and argon, which the getters do not pump well. Before installation, the getter strips should be rinsed in methanol to ensure that they are particulate-free. The getters are activated by heating to 450 ˚C for 45 minutes after system bake-out. Often, the release of gas during the activation can overwhelm the ion pump, so an external, baked, turbo station should be prepared should this happen. After the final bake and getter activation, a pressure < $5\times10^{-10}$ Pa routinely is obtained, with a typical post-bakeout residual gas spectrum shown in Figure **4.11**.

Ion pumps are ineffective at very low pressures [4.35] and are difficult to restart at low vacuum levels if they shut off. Recently designed pumps have self-starters to keep them going, plus shrouds to block light and particles from escaping (Gamma Vacuum model 45S-IDIXTI). Turbo pumps and cryopumps typically are not used for photoemission guns since both types produce vibrations that can be transmitted to the cathode and cause beam motion. Also, mechanical devices require periodic maintenance, whilst ion pumps and getter pumps rarely do at the pressures involved. A cryopump recently developed by Oerlikon (COOLVAC BL-UHV) can be baked out completely; reportedly, it produces vacuum levels below $1\times10^{-10}$ Pa, and thus may warrant further investigation.

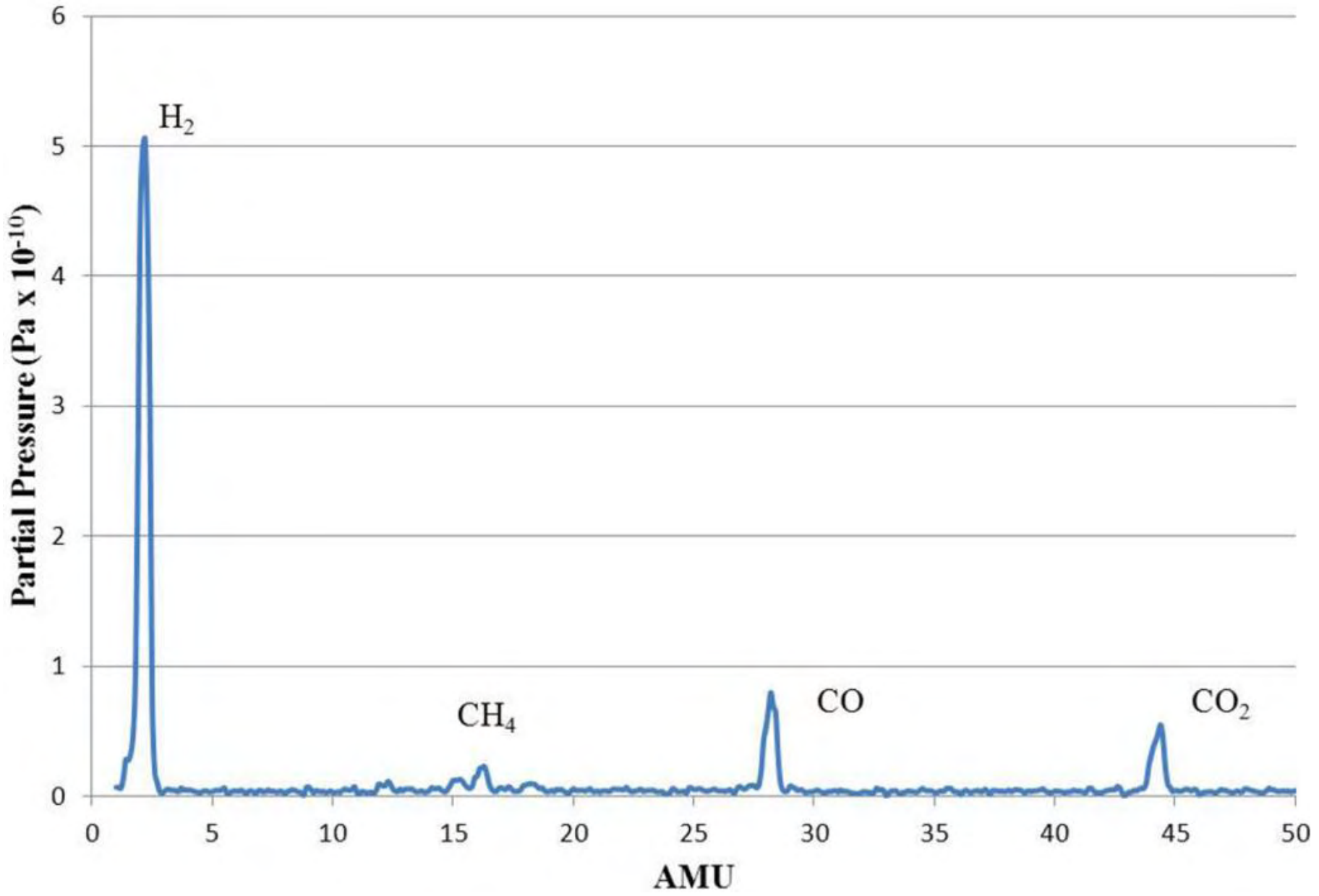

**Figure 4.11. Residual gas spectrum after bakeout and NEG pump activation, showing hydrogen, methane, carbon monoxide and carbon dioxide. The total pressure measured with an extractor gauge was $6\times10^{-10}$ Pa.**

Another important consideration is the quality of the vacuum downstream (beamline) and upstream (load-lock) of the gun. The latter often is of good quality due to the needs of the cathode, but careful attention must be paid to anything that generates particles or introduces any gasses or materials into the gun during cathode preparation (especially cesium). Downstream of the gun, many components may be needed for manipulating and measuring the beam, introducing the laser, RF cavities for bunching and so forth. Each of these must adhere to the same standards as the gun chamber, and there must be adequate pumping. Insertable devices, such as view screens and gate valves, often are the worst offenders, as particles can be generated during motion. For machines where wakefields are an issue, these devices need to have RF seals (usually BeCu spring fingers) that also may produce particles. Individual parts should be pre-tested to avoid such problems.

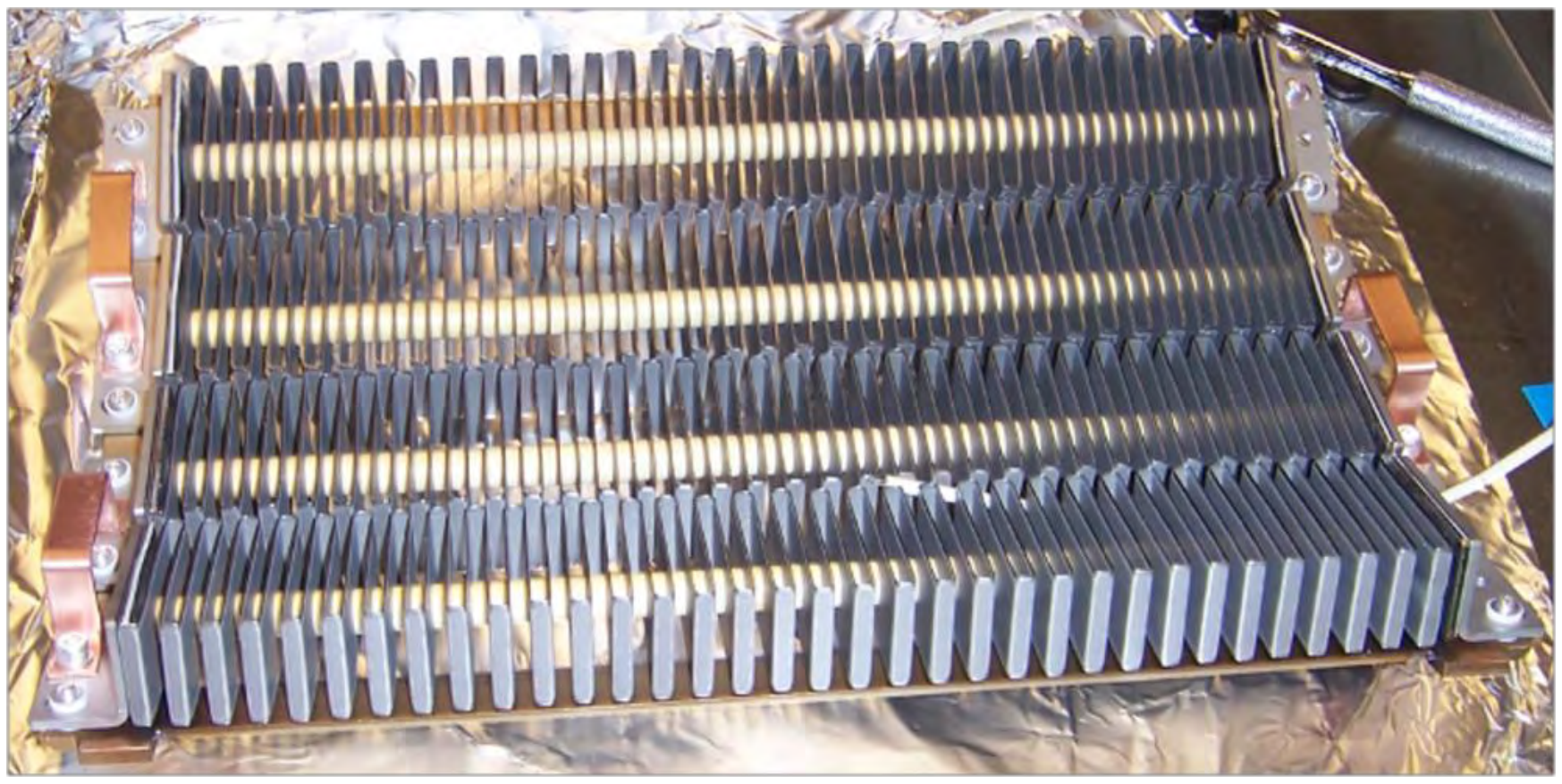

**Figure 4.12. An array of SAES ST-707 getter strips. They are connected in series to minimize the number of vacuum feedthroughs needed.**

List of General Vacuum Rules

1. Use high quality materials.
2. Follow standard practices for UHV and cleanroom.
3. Avoid coatings and plating without extensive testing. The coatings may come off, producing particles. Instead of using plated screws (silver), use screws of different materials to avoid binding.
4. Chemically or ultrasonically clean all parts to remove surface contaminants and particulates. The cleaning process is material-dependent.
5. Pre-bake parts if possible to outgas them.
6. Design parts to eliminate virtual leaks.
7. Clean, store and assemble all components in a clean room.
8. Use filtered gas to remove particles, with a particle counter for monitoring progress.

Procedure for Cleaning Stainless Steel, High Voltage Electrodes

1. Machine to a surface finish of 1 µm rms.
2. Use SiC paper to remove machining marks.
3. Rinse in methanol.
4. Ultrasonically clean in de-ionized water.
5. Store in water.
6. Electropolish, rinse and store in water.
7. Rinse with HPR for 4 hr to remove chemical residue then dry (in a clean room).
8. Air-bake at 400 ˚C for 100 hr.

9. Rinse in HPR for 4 hr then dry (in a clean room).
10. Store until installation.
11. Blow with high pressure $N_2$ until particle count is near zero.

A final source of vacuum degradation is the dark current that generates electrons at high voltage surfaces. These can hit other parts of the vacuum chamber, thereby degrading the vacuum by electron-stimulated desorption of ions, electrons, atoms and light. Section 4.3.2.4 discusses the high voltage aspects of reducing dark current. Dark current problems related to the vacuum are minimized by cleaning and assembling the gun in a clean room to eliminate particle contamination. As mentioned in Section 4.3.2.1, carefully designing the cathode and focusing optics will eliminate stray electrons from the cathode, which can have a similar effect as dark current.

Improving the vacuum level is an important topic for photoemission guns and research needs to be continued.

#### *4.3.2.4 High Voltage Electrodes*

As discussed earlier, the maximum charge that can be extracted from a cathode is determined by the electric field at the cathode and the laser spot size. Thus, much of the design work on various guns aims to make the electric field as high as possible (for both DC and RF guns). Also, if one can make electrodes that do not field emit, the demands on the insulator design are diminished – this is the Holy Grail for gun design (and for SCRF cavities). A few examples of DC gun electrodes are shown in Figure **4.13**.

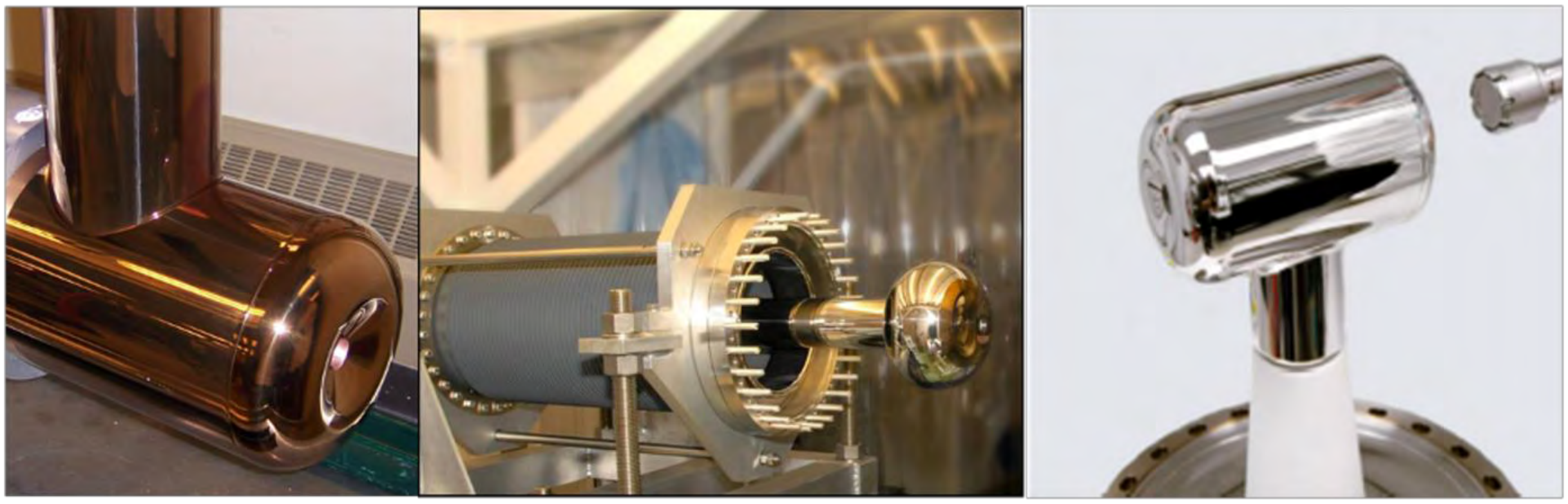

**Figure 4.13. The focusing electrode for three different DC photoemission guns: the left picture shows the Cornell University stainless steel electrode, with the copper color resulting from an air bake to remove hydrogen; the central picture shows the gun used at Daresbury Lab [4.36] [credit to ASTeC, STFC Daresbury]; and on the right is the electrode used in the Jefferson Lab inverted polarized source. [Reprinted figure with permission from [4.37]. Copyright 2010 by American Physical Society.]**

Volumes have been written about the best materials to use for electrodes, coatings to prevent field emission, cleaning procedures and geometries; unfortunately, most studies were for small anode-cathode gaps (millimeter range) and small-area electrodes. Incredible results were obtained that often led gun designers astray (myself included) when applying them to realistic geometries. The gun requirements covered earlier push toward the direction of very high voltages involving bigger gaps and large surface areas, which is not discussed often in the literature.

Figure **4.14** compiles data on HV breakdown for various size gaps [4.38]. For small gaps, the breakdown strength versus gap size is linear, but deviates from linear as the gap size increases. Hence, the experimental results from small gaps cannot be extrapolated to large ones. This phenomena often is called the "total voltage effect" [4.39], but the explanation is debatable. Additionally, this curve shows the breakdown

voltage for a gap. In a photoemission gun, the operation voltage must be well below the breakdown voltage to avoid field emission (which can cause electron stimulated desorption and vacuum increase, for example).

Figure **4.15** shows a test chamber for electrode studies used at Cornell University. Here, two large area (116 $cm^2$) electrodes with a Rogowski profile are held parallel to each other with adjustable gap spacing. The current hitting the anode is measured as a function of applied voltage (up to -125 kV), indicating the onset of field emission. A data set is displayed in Figure **4.16**. Such a test stand is ideal for testing various materials for cathodes and anodes and for exploring new cleaning techniques. In this particular experiment, a stainless steel cathode was cleaned by different methods, resulting in a field of ~30 MV $m^{-1}$ for a 4 mm gap at 125 kV.

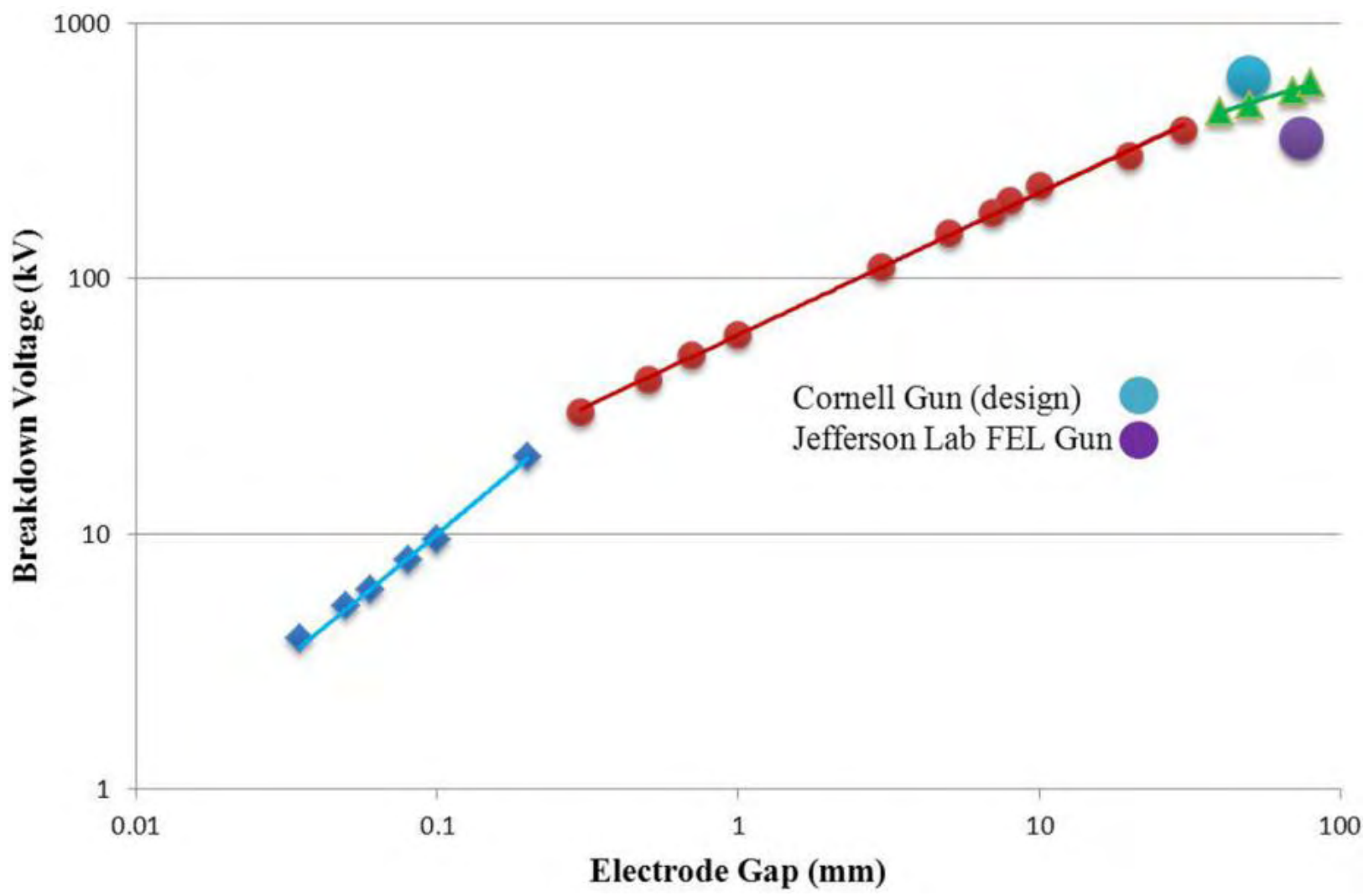

**Figure 4.14. A compilation of breakdown voltage (*V*) between electrodes for a given gap size (*d*) from various authors [4.38]. The different colors indicate regions where the data range follows the displayed transfer function. The points for the Jefferson Lab FEL gun and the Cornell ERL gun are shown.**

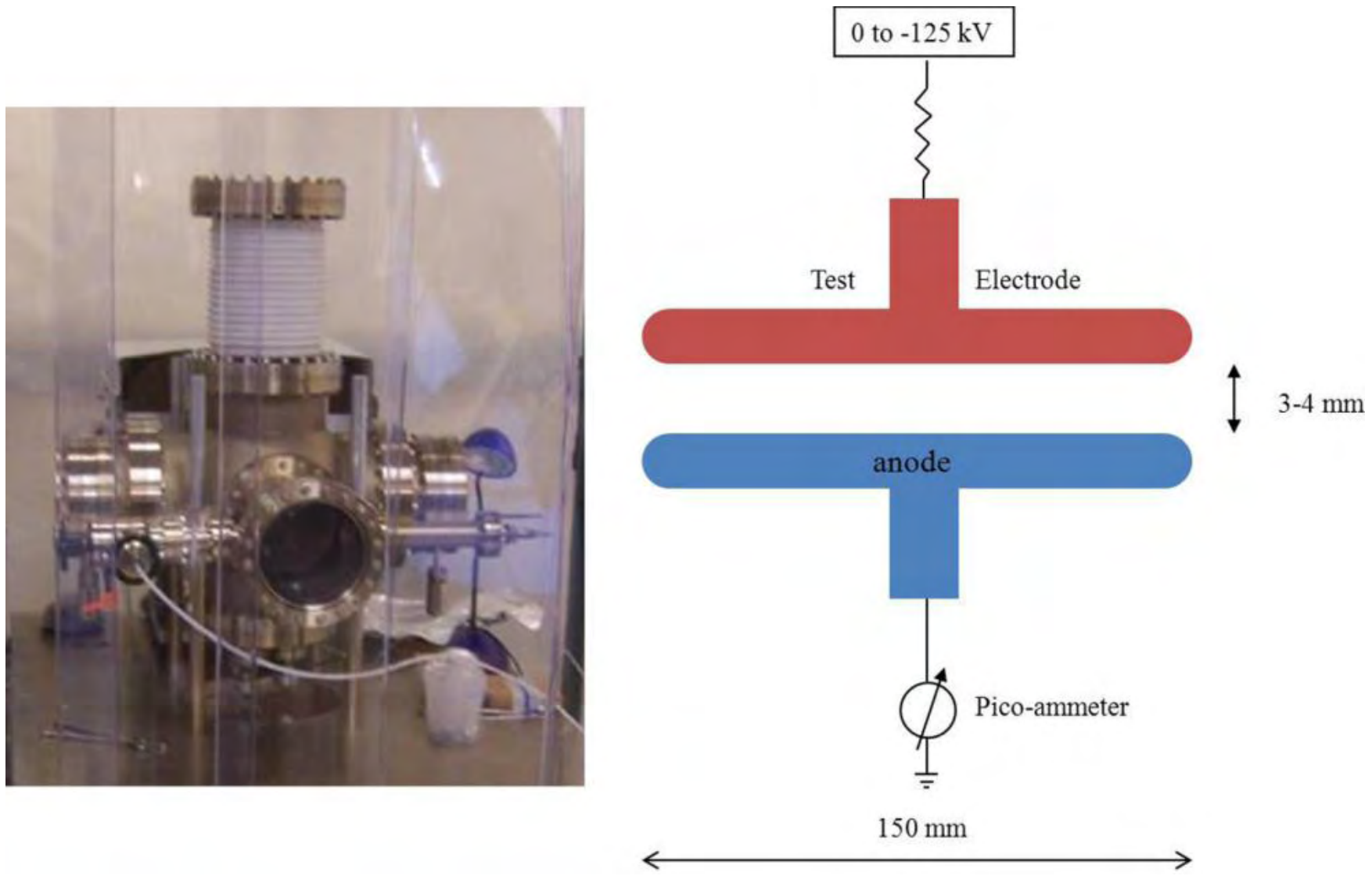

**Figure 4.15. Test chamber for measuring the field emission characteristics of large area electrodes.**

Extrapolating to 500 kV, a 16 mm gap for a large area electrode would seem achievable. Unfortunately, the best designs typically work at 5-10 MV $m^{-1}$ or less for 50 mm gaps, far from the small-gap results. Two important lessons emerge from this information: (1) While the findings from small area, small gaps can provide useful data, do not rely on them solely for realistic design parameters; and, (2) perform electrode tests on full-sized models.

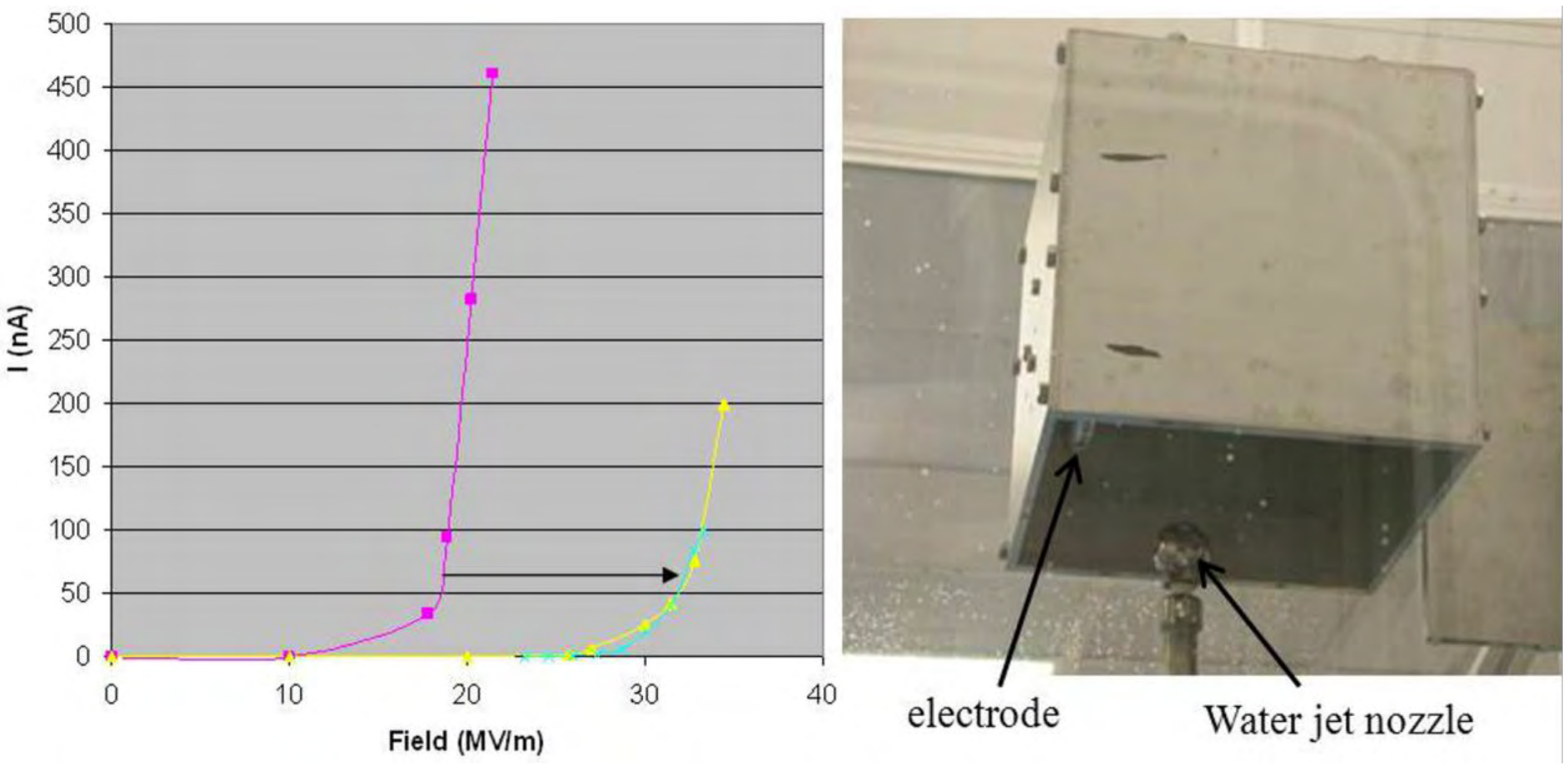

**Figure 4.16. Results from field emission tests (left). The pink curve shows the current hitting the anode as a function of electric field for a 4 mm gap, hand-polished stainless steel sample. The green and yellow curves show the improvement after high pressure rinsing the electrode (green was originally hand polished, yellow was originally electropolished, 4 mm gap). The HPR fixture is shown on the right.**

Figure **4.17** illustrates a chamber for testing large electrodes at large gaps that was constructed, in part, to study the total voltage effect [4.40]. The authors suggest that at higher voltages, the field-emitted electrons from the cathode can penetrate deeply into the anode, where eventually enough energy builds up to cause a catastrophic event. Such component-level test systems are crucial for gaining the necessary understanding for design very high voltage photoemission guns.

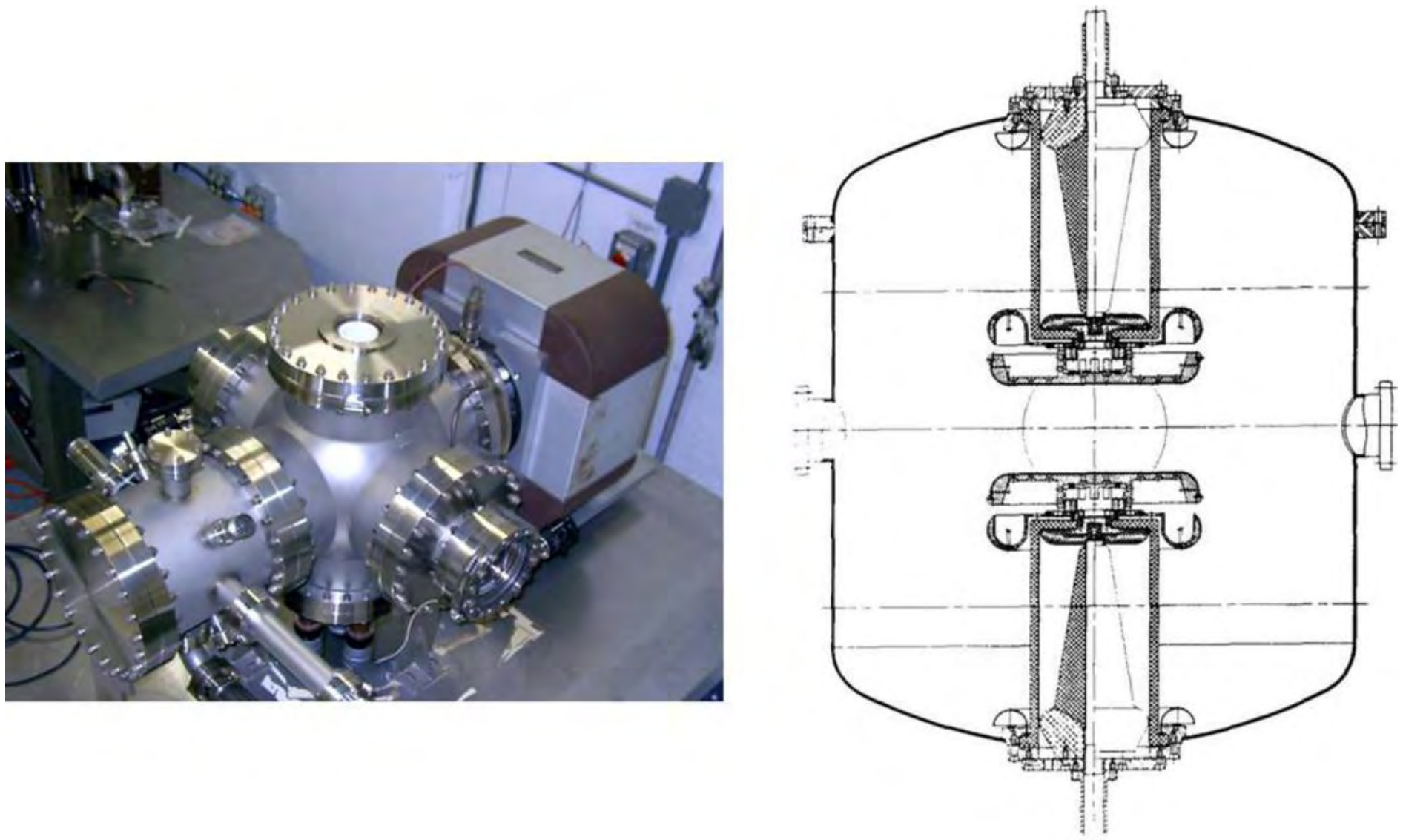

**Figure 4.17. Vacuum chambers for investigating the "total voltage effect" on large electrodes. The chamber on the left can reach -250 kV on the cathode (anode grounded), while the system at the right can reach 600 kV (cathode at -300 kV, anode at +300 kV). [[4.24]; Available under Creative Common Attribution 3.0 License (www.creativecommons.org/licenses/by/3.0/us/) at www.JACoW.org.] [[4.40] (© 1996 IEEE)]**

Several groups are investigating the best material to use: Niobium, stainless steel, or molybdenum for cathodes, and titanium or beryllium for anodes show promise [4.21], [4.41]. The gun in Figure **4.3** has a cathode of stainless steel with a beryllium anode. Using beryllium for the anode has two advantages: Excellent thermal conductivity to remove the heat produced during field emission events; and, low atomic number to reduce the production of X-rays when struck by field-emitted electrons from the negative electrodes [4.21]. Many authors have detailed research on field emission (*e.g.*, [4.39]); it is not further discussed here.

#### *4.3.2.5 Insulators*

Designing insulators for very high voltage DC photoemission guns is difficult. Many experts on high voltage engineering focus on pulsed power systems [4.42] that are not directly applicable to the DC case. Early work on large DC accelerating columns (Van de Graaffs) is not directly applicable, due to the special requirements for photoemission guns. Industrial devices, such as X-ray tubes for inspection work at 450 kV, operate in a bipolar geometry, with the anode at 225 kV and the cathode at -225 kV, and thus are not applicable.

In DC photoemission guns (Figure **4.3**), the electrons must be accelerated quickly to high energy after leaving the cathode to reduce the effects of space charge. Accordingly, the anode-cathode gap should be minimal, both to increase the electric field on the cathode surface and to accelerate the bunches in a short distance. These requirements eliminate many successful older HV accelerating tube designs (Figure **4.18**) because they lack a central support tube for holding the cathode close to the anode and their small diameter precludes inserting a tube while maintaining a reasonable field gradient. In these, the beam travels the length of the tube, gradually accelerating to the desired energy, and so is unsuitable for high bunch charges.

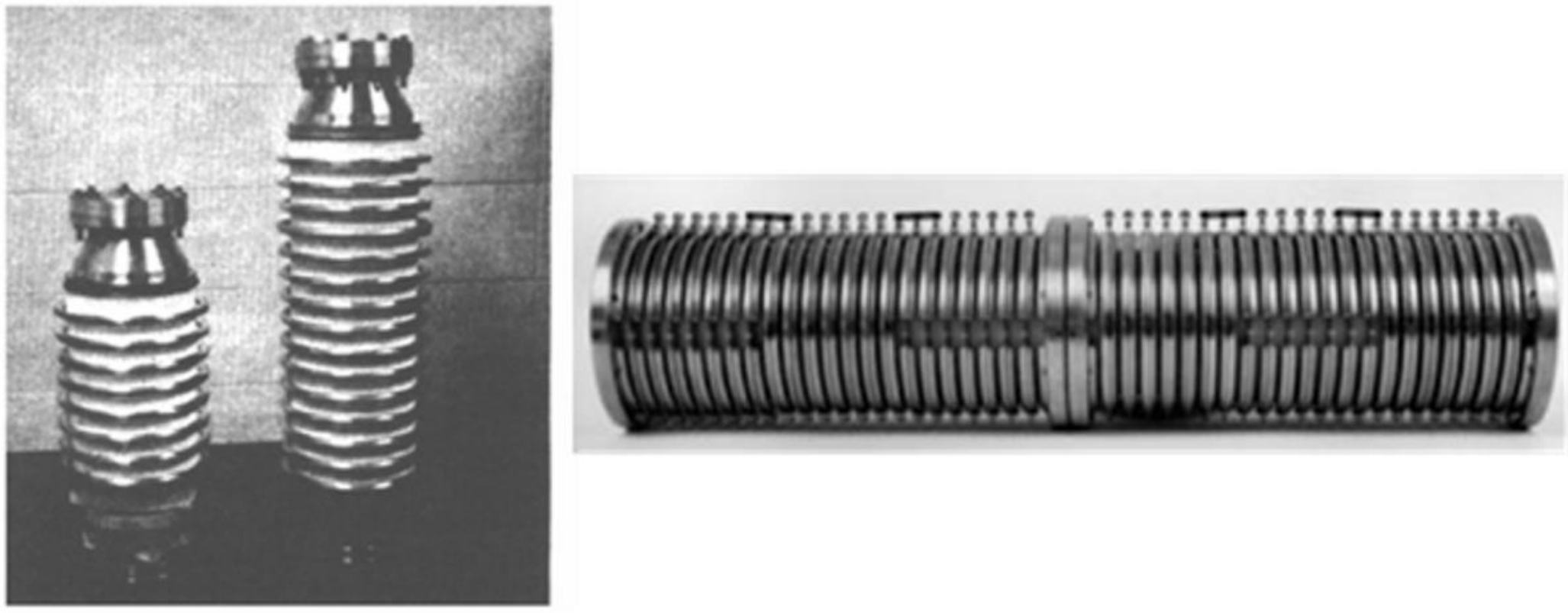

**Figure 4.18. 300 and 550 kV accelerating columns (left), and a 1 MV accelerating column (right) from NEC. [[4.43] (© 1975 IEEE)] [Courtesy of National Electrostatic Corp.].**

Since the insulator provides part of the vacuum envelope for the electron gun, the level of vacuum required (often set by the cathode type) greatly influences the insulator's materials and design. Here, I discuss the design covering the most difficult case of using GaAs-like photocathodes, which require a vacuum of $< 1\times10^{-9}$ Pa. It eliminates many materials and sealing designs; O-ring seals cannot be used, nor insulators made from plastics or epoxies. To obtain such a good vacuum means that the device must be bakeable at high temperatures. Alumina is the most common insulator material used today.

Alumina comes in several different types, with the alumina content varying from 92% up to 99.5% (sapphire is 100% alumina, for comparison). The material properties change slightly with alumina content, but for high voltage applications, values in this range are reasonable. For RF window applications, 99.5% alumina is often used due to the low RF loss factor. Table **4.4** lists the important properties. Several companies produce large alumina parts: CoorsTek; Kyocera; Friatec; SCT; and, Morgan Advanced Ceramics.

| **Property** | **92%** | **96%** | **99%** | **99.5%** |
|---|---|---|---|---|
| Compressive Strength [MPa] | 2 300 | | 2 160 | 2 350 |
| Tensile Strength [MPa] | 180 | 193 | 241 | |
| Young's Modulus | 280 | 320 | 360 | 370 |
| Thermal Expansion [40-800 ˚C, $\times 10^{-6}$ ˚$C^{-1}$] | 7.8 | 7.9 | 8.0 | 8.0 |
| Thermal Conductivity @ 20 ˚C [W $(m\ K)^{-1}$] | 18 | 24 | 29 | 32 |
| Dielectric Constant | 9.0 | 9.4 | 9.9 | 9.9 |
| RF Loss Factor ($\times 10^{-4}$) @ 1 MHz | 54 | 38 | 20 | 10 |
| Volume Resistance @ 20 ˚C [Ω cm] | $> 10^{14}$ | $> 10^{14}$ | $> 10^{14}$ | $> 10^{14}$ |
| Volume Resistance @ 300 ˚C [Ω cm] | $10^{12}$ | $10^{10}$ | $10^{10}$ | $10^{13}$ |

**Table 4.4. Properties of alumina with varying alumina content.**

To assure vacuum integrity, the insulator must be brazed to a metal ring that then either is brazed or welded to a vacuum vessel or vacuum flange. There are numerous examples of braze-joint geometries [4.29], brazing techniques and materials. Figure **4.19** shows a particularly good joint for the large insulators often used for photoemission guns. Many geometries that work fine for smaller diameters (200 mm and less) do not scale well as their dimensions increase with voltage due to the thermal expansion of various materials compared to alumina. Kovar offers the closest match for expansion versus temperature to alumina, and hence, is often used, however, it can become magnetized, so care is essential when low magnetic fields are required (electron microscopes, for example). Copper is often used even though the thermal expansion mismatch is much greater. The compressive- and tensile-strength of alumina is another important factor in designing a good braze joint. Alumina is ~6-8X stronger in compression than tension, so, after brazing, the joint should be in compression at its operating temperature. Figure **4.19** depicts a kovar braze-ring that is sandwiched between two cylindrical alumina rings, a geometry applicable to a single, large ceramic bushing, or a series of stacked ceramic rings. At the top and bottom, the kovar ring mates up with a vacuum chamber or flange, so that the two pieces can be welded to form the final vacuum seal. The bottom alumina ring helps equate the stress from brazing equally on both sides of the joint and affords a surface for registration with the vacuum flange. The stainless steel vacuum flanges should be treated to remove hydrogen before they are welded.

Finally, a vacuum bake (150-250 ˚C) and leak check should be undertaken. Surrounding the entire device with a large bag of helium will ensure the detection of the smallest leak. An RGA with an electron multiplier should show no helium signal above background after 1 hour.

An insulator must be able to hold off the maximum voltage applied without breakdown across its surface. Breakdowns occur on the vacuum side and the non-vacuum side, as discussed in [4.39]. The non-vacuum side can be air, $SF_6$, oil, or even a solid dielectric plastic/epoxy. The insulator's environment determines the size and geometry that will prevent breakdown. In Figure **4.27**, with $4\times 10^5$ Pa of $SF_6$, 30 cm affords a conservative buffer to hold off 750 kV between the gun and the grounded pressure vessel. In perfect conditions with uniform electric fields, $SF_6$ can hold-off $8.9\times 10^{-4}$ kV $cm^{-1}$ $Pa^{-1}$, but this is de-rated for practical situations. In air, the distance would be at least 2.5X larger. Oil is widely used in industrial devices, but is not recommended near UHV systems because of the potential for hydrocarbon contamination. Instead of placing the gun and power supply in a single tank, high voltage cables and connectors can be used with the gun and supply located in separate tanks; they are readily available only up to ~225 kV [4.44]. Cables and connectors for higher voltages can be obtained [4.45] or made in-house, but their design is difficult and can lead to reliability problems.

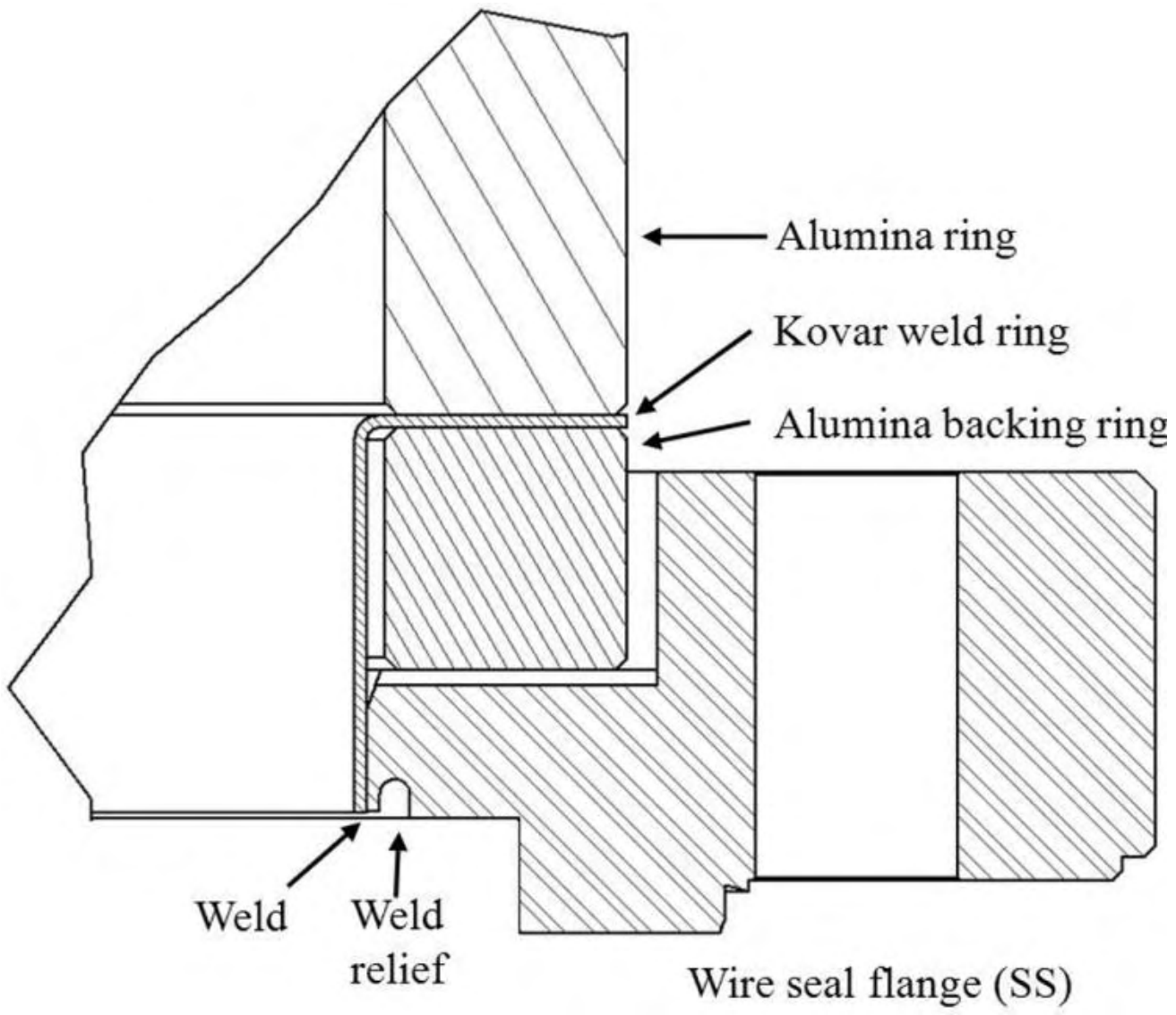

**Figure 4.19. Braze design for a large diameter insulator. A kovar ring is brazed between two alumina rings and ring is welded to the stainless steel flange.**

A rough estimate of the required insulator diameter can found by using the formula for a coaxial cylinder

$$E = \frac{\frac{V}{R_{inner}}}{\ln\left(\frac{R_{outer}}{R_{inner}}\right)} \tag{4.4}$$

where $V$ is the voltage on the central conductor, $R_{outer}$ and $R_{inner}$ are the respective radii of the outer- and inner-conductor, and $E$ is the electric field on the inner conductor. Since the outer conductor is part insulator and part conductor, this formula does not strictly hold and simulations are needed for the final design. A good upper limit for $E$ is 10 MV m$^{-1}$, with lower values being better. For 750 kV, values of $R_{inner}$ = 50 mm and $R_{outer}$ = 220 mm result in a field of 10.1 MV m$^{-1}$ on the inner conductor.

An insulator height of 250-300 mm is typically sufficient for 100 kV in air, or for 250 kV in $SF_6$. For 500 kV, 500-600 mm is adequate depending on the gas pressure. Again, simulations are needed to calculate the exact fields for the final geometry to ensure predetermined limits are not exceeded.

Important design considerations are the following:

1. Ensure ceramic parts are under compression.
2. Protect triple point junctions (limit the field to << 1.0 MV m$^{-1}$).
3. Make the diameter large enough to keep the central tube's field < 10 MV m$^{-1}$.
4. Use sandwich-type braze-joints for large diameters (> 200 mm).
5. Ascertain that no magnetic materials are near the beam.
6. Make braze joints accessible so any braze overflow can be cleaned up.

7. Minimize designs and materials that can generate and/or trap particles.
8. Consider the mechanical strength under vacuum and $SF_6$ loading and how it affects mechanical alignment and tolerance for placing electrodes.
9. Confirm resistance to flashover, field emission, UV, thermal stresses, secondary emission and corona, any of which can affect the vacuum and impact cathode lifetime.
10. Braze in a vacuum furnace, not a hydrogen furnace.

All guns of the type described incur the problem of controlling field-emitted electrons leaving the high voltage surfaces. These electrons can land on the insulator, and if the charge builds up there may be a punch-through, causing a vacuum leak (Figure **4.20**). Alternatively, or additionally, if heat builds up where the electrons strike the ceramic, the insulator can locally overheat and explode, also causing a leak [4.46]-[4.48].

A common design idea to bleed off these electrons is a cylindrical alumina bushing with an internal resistive coating. At high DC voltages, this was successful only up to 450 kV during processing, above which punch-throughs occurred [4.49]. The thin coating used in the reference exhibited a strong non-linear drop in resistance with voltage and the related increase in current passing through the ceramic can lead to thermal runaway. Further, the coating did not adhere well, leaving a layer of dust on the electrodes, certain demise for reaching 750 kV. Others adopted ion implantation to provide some conductivity [4.50]. For high voltages (> 300 kV), the electrons have enough energy to penetrate fractions of a millimeter into the material, beyond any thin surface layer. Thus, alumina with bulk conductivity is more desirable for higher voltages.

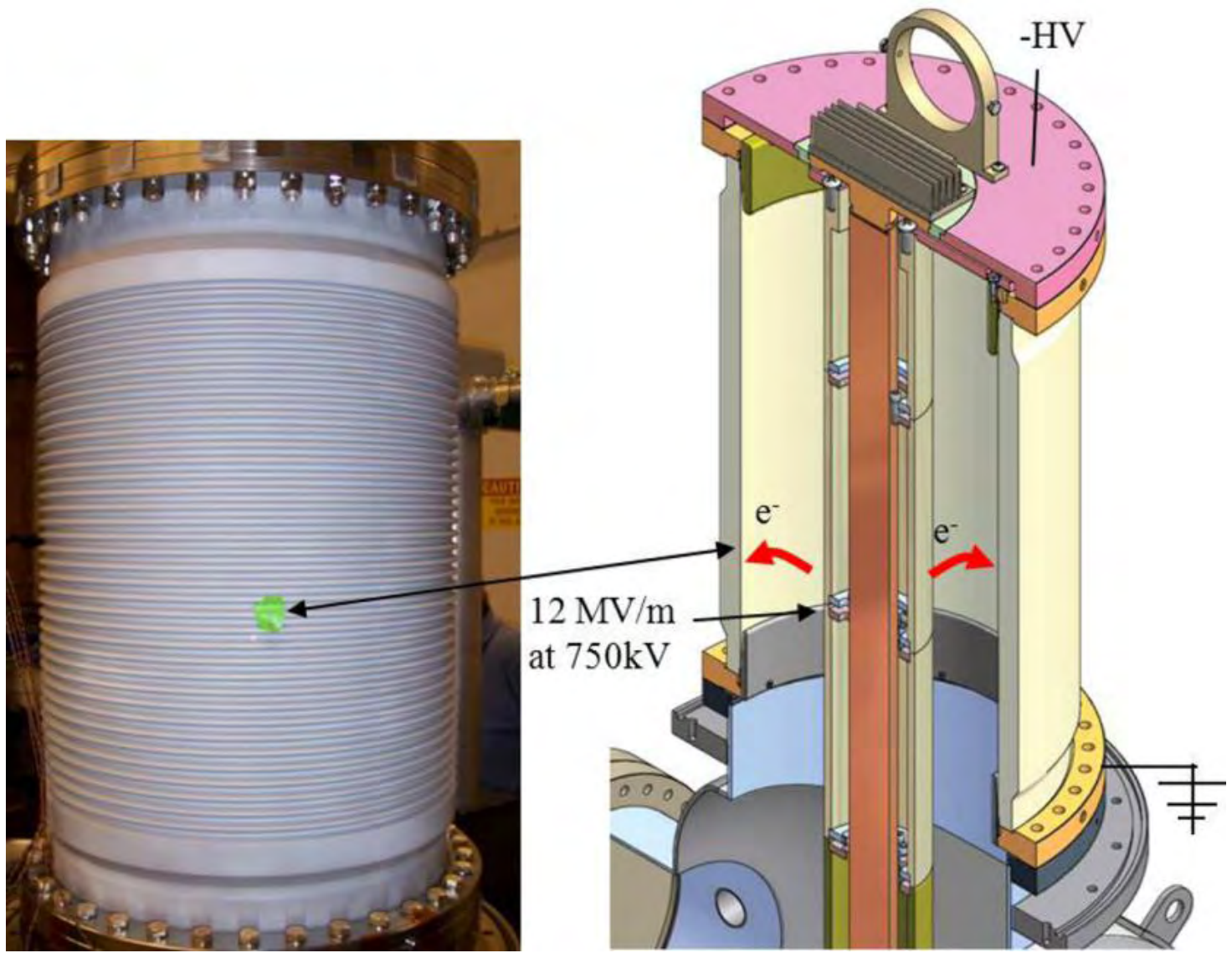

**Figure 4.20. Insulator damage due to field-emitted electrons from the HV center conductor.**

An insulator made from an alumina composite material from Morgan Advanced Ceramics (AL-970CD) was built at Daresbury Laboratory [4.36] that is resistant to field-emitted electrons up to 500 kV. It has a resistivity of 65 GΩ cm at room temperature that remains constant with changes in voltage. For the ~500 kV insulators discussed, this results in an overall resistance of 25-50 GΩ and draws 10-20 µA, limiting any potential thermal runaway. The same material was used at Cornell for a 750 kV gun, but it failed at 450 kV (Figure **4.21**). The importance of using a slightly conductive insulator is still an open question, as another failure mechanism for this system was discovered recently (section 4.3.2.7). If that turns out to be the root cause, this gun geometry may be able to reach higher voltages.

An alternative insulator, similar to an old accelerating tube, will completely block the line of sight between the electrodes and the insulator, thus preventing any field-emitted electrons from reaching it; such a device has been successfully tested to 550 kV [4.51]. Figure **4.22** and Figure **4.23** show a design for 750 kV.

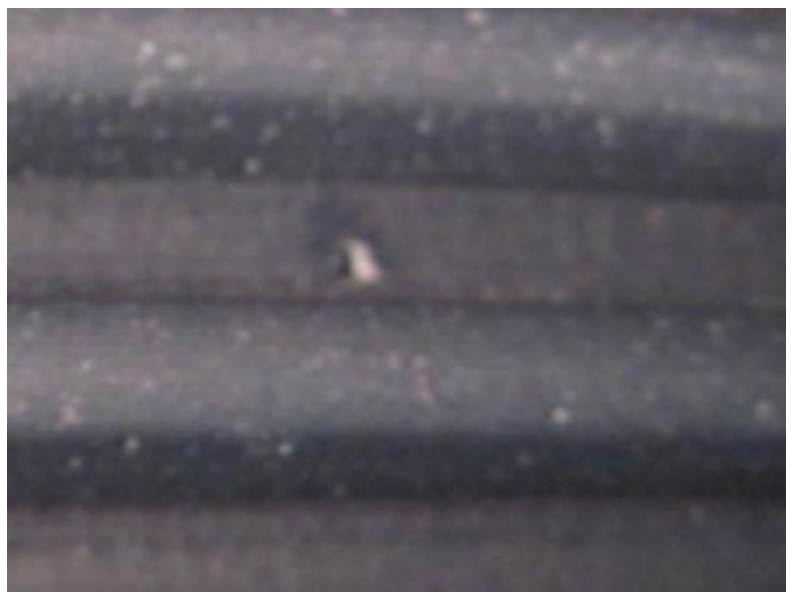

**Figure 4.21. A failure on a conductive insulator; the hole is ~1 mm$^3$ on the external surface.**

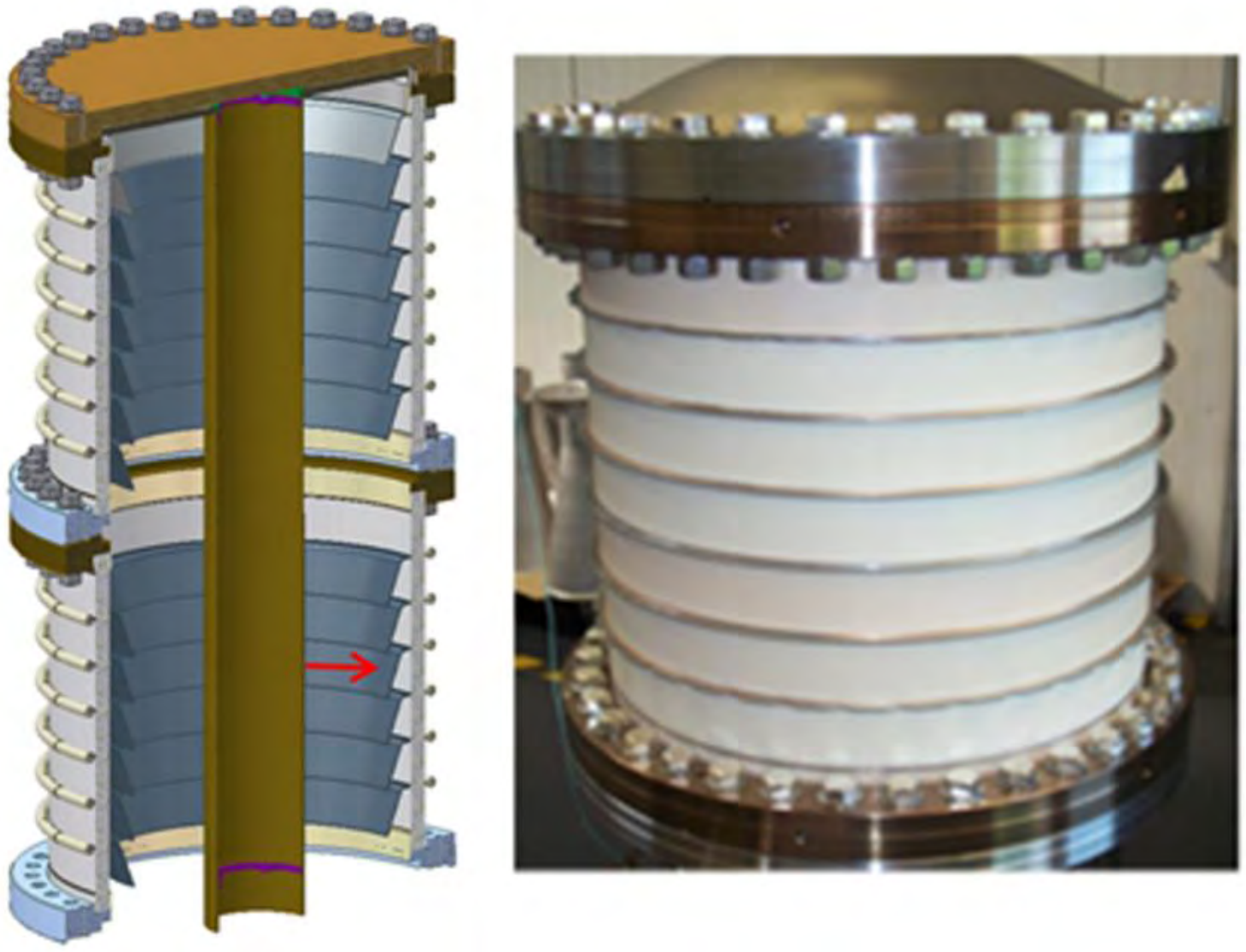

**Figure 4.22. A segmented insulator design. The wire seal flanges have a 560 mm outer diameter and the insulator rings are 50 mm tall and 20 mm thick. Any electrons field emitted from the central tube (red arrow) are intercepted by one of the metal rings. The photo on the right shows one of the insulator halves. [[4.49] (© 1996 IEEE)]**

The sizes were selected to keep the maximum field below 10 MV m$^{-1}$ and triple point junction fields below 1.0 MV m$^{-1}$ at the full 750 kV. The number of intermediate rings was chosen to limit the voltage between each ring to ~50 kV, where external resistors maintain a uniform voltage from the top to the bottom of the device. The central support tube was sized to minimize the gradient on the tube. If desired, the complete insulator can be made in two pieces to improve manufacturability and to provide a location for mounting an intermediate electrode. The angle and length of the intermediate rings were chosen by simulations so that any electrons emitted from the high voltage surface of the long central-support tube would strike the rings and never reaching the alumina (Figure **4.22**). There is also a likelihood that field emission between the rings will produce electrons that could land on the alumina, but they will be limited to 50 kV energy; the field gradients are low enough that field emission should be minimal. Kovar was used for the ceramic-to-metal joints for ease of brazing; any residual magnetic fields from the kovar will be far enough away not to affect the electron beam. The thermal conductivity of the material of the angled electrodes should be high enough to conduct away heat from field emission, such as copper. The tradeoffs of strength, vacuum properties, secondary emission yield and HV properties will determine the choice.

Although this design has the best chance for success at very high voltages with excellent reliability and low dark current, it has several disadvantages. Its large size (to keep the maximum field below 10 MV $m^{-1}$) necessitates very careful attention to cleanliness (particle reduction) and adequate pumping. Its large surface area at high voltage affords opportunities for field emission. A carefully designed external resistor chain is essential for maintaining the voltage at each ring, and for handling arcs during processing. Arc protection requires varistors or spark gaps between each ring.

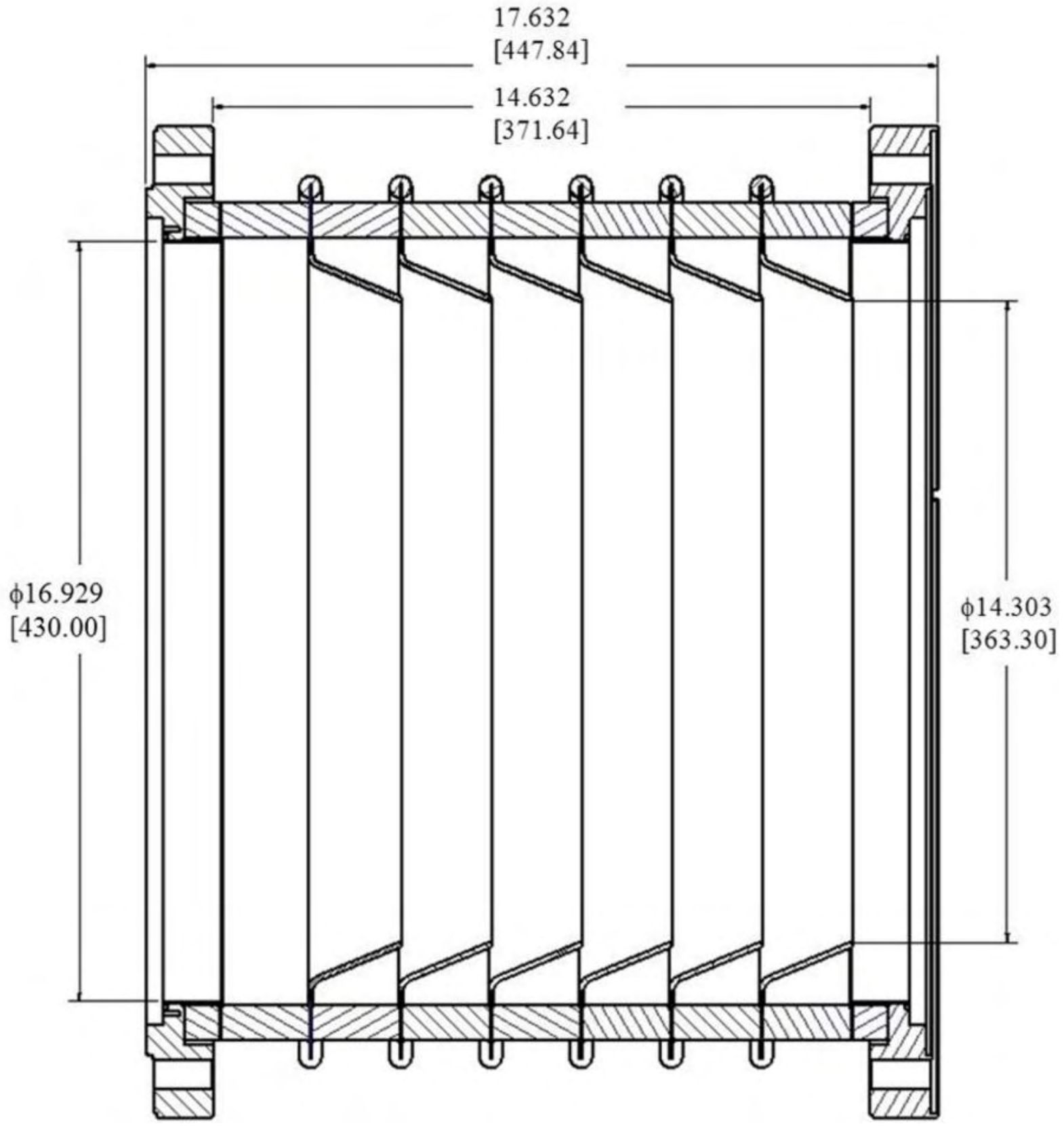

**Figure 4.23. A 2-D drawing of one section of the 750 kV segmented insulator. The dimensions are in inches and millimeters. The inner rings are at an angle of 23° to the insulator's surface.**

Another approach is adapting the design of an industrial X-ray tube insulator, rather than the bushing design. These are much smaller conical insulators with the cathode directly attached to the end of the cone, as Jefferson Laboratory demonstrated [4.37] (Figure **4.24**). The following are its main advantages: less surface area at high voltage (correspondingly, fewer field emission sites); commercial components that work up to 225 kV; low vacuum chamber volume; and availability of mating HV cables. Higher voltage inverted guns will require substantial redesign effort to assure reliability [4.52].

Procurement poses a final difficulty with HV DC insulators. Companies with the necessary expertise may be uninterested in making small quantities, and so the final cost can be high and delivery time long. Such constraints encumber building and testing multiple configurations. Prototypes using plastic or epoxy materials for the insulator might be possible. Rexolite is a common material in pulsed-power insulators. While vacuum constraints preclude them for the final design, they are adequate for quick prototyping.

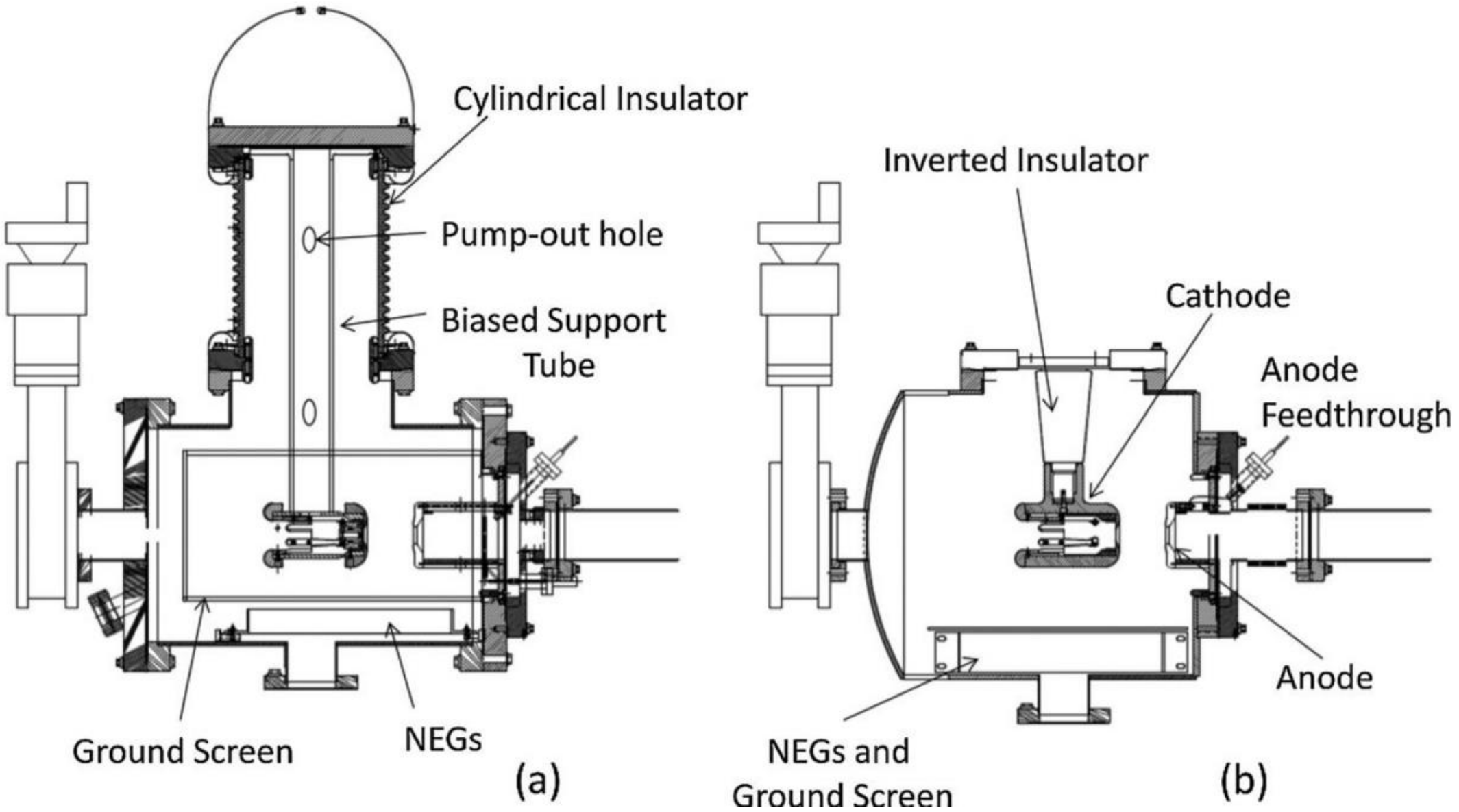

**Figure 4.24. (a) A polarized source using a bushing-type insulator design and (b) an inverted insulator design. [Reprinted figure with permission from [4.37]. Copyright 2010 by the American Physical Society.]**

### *4.3.2.6 DC HV Power Supplies*

The high voltage power supply for a DC photocathode gun is an important, but often overlooked component of the complete system. A sound set of requirements should be developed before considering what power supply to obtain. Table **4.5** lists many of the important requirements and representative values for a photoemission gun operating at 600 kV, 100 mA. Several of them are described later in the section.

The first item to determine is the highest voltage needed for routine operation and the overhead for conditioning. For stability, all HV electron tubes must be conditioned above the nominal voltage. Photocathode guns, particularly those using vacuum-sensitive cathodes, often require additional margin (up to 25%) for good cathode lifetime. Dark current (from field emission) in the nanoampere range is enough to produce noticeable local heating and light (X-rays and UV), contribute to secondaries and increases in pressure; a higher margin normally reduces the field emission at the operating voltage. For example, if 500 kV is the desired operating value, 600 kV would provide the minimum acceptable overhead.

For any gun that injects a beam into an RF accelerator, the control of the arrival time, or phase jitter, of the electron bunch is of critical importance. In terms of variation in gun voltage, the phase change $\varphi$ at a distance $L$ away from the gun is given by

$$\Delta\varphi = 2\pi f \frac{L}{c} \frac{\gamma - 1}{(\gamma\beta)^3} \frac{\Delta V_{gun}}{V_{gun}} \tag{4.5}$$

where $\varphi$ is in radians, $f$ is the RF frequency in Hertz, $L$ in meters, $c$ the speed of light, $\beta c$ is the electron velocity, $\gamma$ the Lorentz factor, $V_{gun}$ is the gun voltage in volts, and $\Delta V_{gun}$ is the magnitude of the voltage ripple. In terms of the RF phase, variations of the order $\pm\, 1^\circ$ are acceptable for low emittance beams. For example, at $f$ = 1.3 GHz, $\Delta\varphi = 1^\circ$ corresponds to a shift of ± 450 V, 1 m away from a 250 kV gun (0.18% variation). At 500 kV, the allowed variation increases to ± 1640 V (0.33%). The voltage variation, commonly called ripple, must be specified over the frequency ranges in the power supply, typically up to 60 kHz (or more) for modern switching supplies.

For a particular power supply, even monitoring ripple at the levels required, particularly higher frequencies, may be difficult. A time-of-flight detector (a beam-position monitor, for instance) downstream from the gun can be used to monitor the arrival time of the electron bunches. These devices easily measure the arrival time with sub-picosecond accuracy at many tens of kilohertz; the resulting signal is sent back to the power supply's feedback control loop. Both long term drifts (thermal) and short term variations (ripple) can be corrected to maintain a chosen arrival time at the detector. For known problem frequencies or instabilities in the HV power supply, feedforward methods might be useful.

| **Parameter** | **Specification** | **Notes** |
|---|---|---|
| Operating Voltage | 400-600 kV | |
| Conditioning Voltage | 750 kV | |
| Maximum Current | 100 mA | |
| Ripple | ± 0.2% at 500 kV | Value depends on the gun voltage |
| Voltage Reproducibility | 0.2% after 1 hr | |
| Ramp-up Time to Full Current | 50-100 ms | Depends on RF control system bandwidth and cavity fill time |
| Personnel Safety | At least 2 independent interlocks must be satisfied before turn on | |
| Machine Safety | Should have an external signal to shut down if an accelerator trip | |
| Arc Detection | A $dV/dt$ circuit to sense an arc and trip the supply | |
| Temperature Stability | 0.01% of volt per 1 ˚C | $SF_6$ temperature can be regulated, if needed |
| Environment | 100% $SF_6$ | Can use mixtures of $N_2$ and $SF_6$ |
| Current Measurement | Ability to measure with < 1 μA resolution for HV conditioning of electrodes | Need to separate the load current and the internal PS current draw |
| Calibration | 1% of full scale | Using external divider, difficult at this voltage and in $SF_6$ |
| Current Limit Trip | Exceeding preset limit trips the supply | |
| Line Regulation | $V_{out}$ ± 0.5% for a ± 10% line variation | |
| Voltage Stability | 0.25% from 10-100 mA | |
| | DC output ± 0.4% over an 8 hour period, including ripple | |
| Capacitance/Stored Energy | 100 pF, 10 J | Not a firm number, but lower is better |
| External Feedback Port | | Ability to apply an external signal to superimpose on control loop |

**Table 4.5. Specifications for high voltage power supply.**

Current stability and the response of the voltage to changes in the current are of utmost importance for photocathode guns. Drive lasers should be stable in power to < 1%, so the HV output should be insensitive to such changes in current level over a wide frequency range. As the cathode efficiency drops over time, laser power must be increased correspondingly to maintain output constant.

Another concern for photocathode guns is how to ramp up the current to reach the maximum operating value. Two strategies exist: 1) Start in CW mode at low current and ramp up the bunch charge; or, 2) start a full bunch charge in pulsed mode and increase the duty factor until the CW mode is reached. Both methods are problematic. For the first case, focusing changes as the bunch charge increases, necessitating either adjustments of the optics settings to compensate, or choosing a less-than-optimal setting that works for the full range of bunch charges. The second case requires a flexible laser-pulse generation system that can handle the full laser power without damage for duty factors from 0-100%. Existing systems can be turned on directly to a few milliamperes without tripping the RF systems. Beyond that, the HV power supply must be able to ramp-up the current quickly (50-100 ms is desirable) while maintaining a constant voltage at low ripple. The actual ramp-up time needed will depend on the accelerator's size, the RF control bandwidth and the cavity fill time.

A precision voltage divider can assure absolute calibration of the power supply. This often is difficult to realize because the supplies are normally held in a pressure vessel containing $SF_6$ or oil. The divider must be in the same vessel for the calibration. If absolute calibration is essential, the tank may need modification to leave the probe in place. For most photoemission guns, absolute calibration is not particularly important, but voltage repeatability and reproducibility are.

There are two types of DC supplies typically used in photoemission guns, and they cover the ranges of: lower voltage (< 225 kV) for a wide range of currents (0 to 100s of milliamperes); high voltage, low current (< 10 mA); and high voltage, high current (100 mA range). Staying at or below 225 kV is very convenient as it is the maximum voltage used in many commercial X-ray tubes. Above that range, the choices of commercial power supplies are very limited.

Cockroft-Walton voltage multipliers and variations thereof commonly have been used for many decades to produce voltages up to megavolts. The idea is simple [4.53], but implementation requires great skill to meet the strict requirements for photocathode guns (Figure **4.25** shows an example power supply).

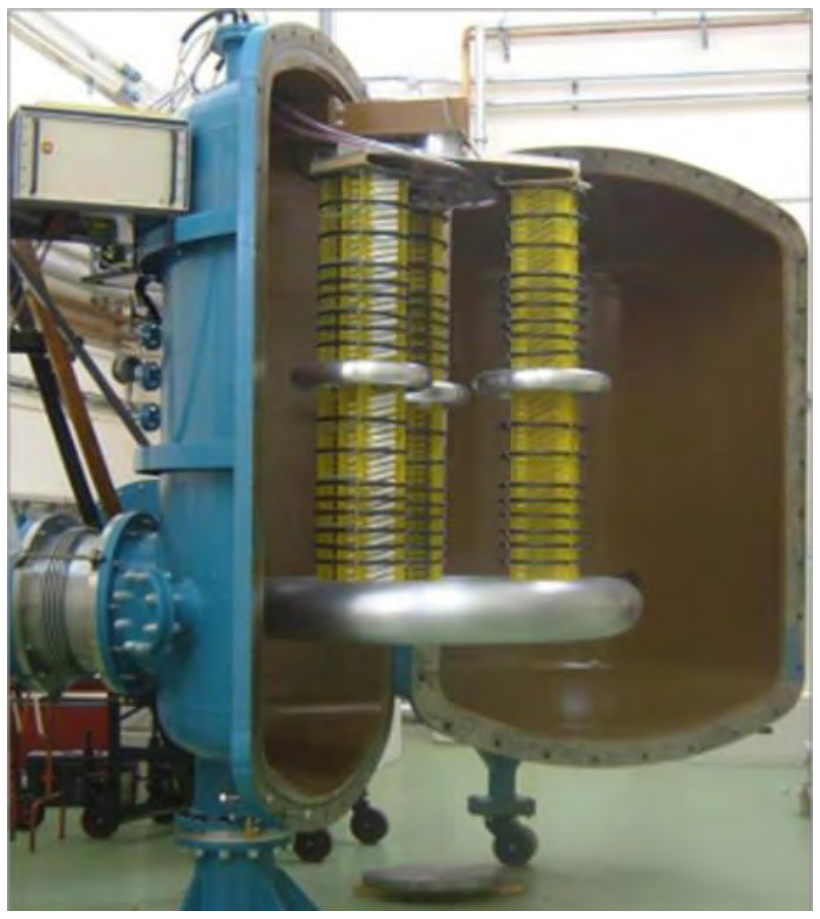

**Figure 4.25. 500 kV, 8 mA power supply (Glassman HV, Inc.) and $SF_6$ tank used at Daresbury Laboratory [4.54]. [Credit to ASTeC, STFC Daresbury]**

Industrial systems requiring high voltage and high current often use isolated core transformers (ICTs) [4.55]. Figure **4.26** is a schematic view of such a HV power supply. In contrast to the Cockroft-Walton type voltage multipliers, ICT supplies distribute line-frequency power to individual multiplier sections *via* a ferrite core transformer. A motor-driven autotransformer varies the primary voltage, thus regulation is slow (tens of Hertz). Flux leakage at the higher sections reduces the voltage per turn on the secondary, requiring changes to the sections to compensate for the losses.

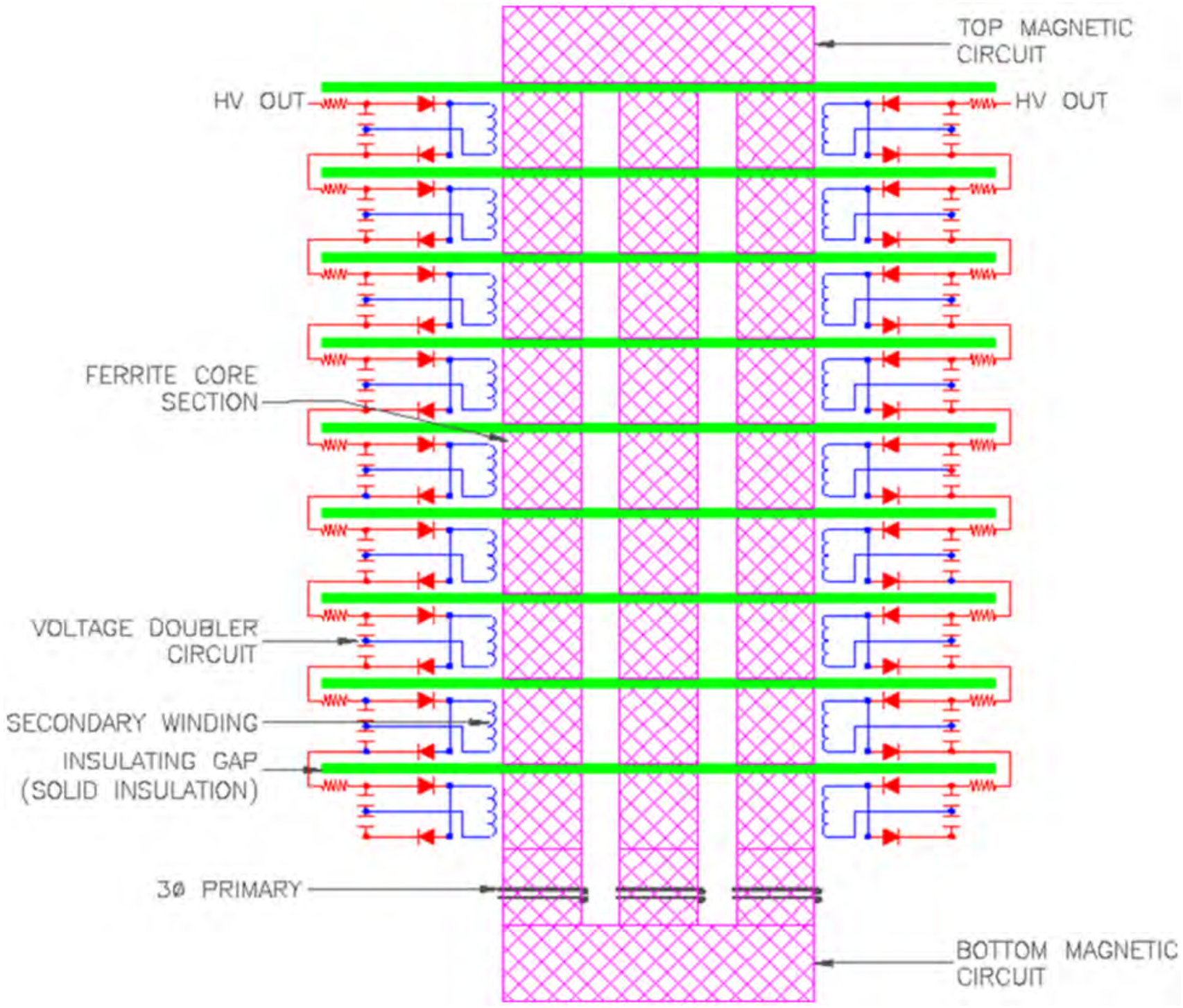

**Figure 4.26. A schematic view of an isolated core transformer. [Reprinted with permission from [4.56]. Copyright 2009, American Institute of Physics.]**

The Cross Transformer is a modern implementation of the ICT [4.56] that overcomes the limitations of an ICT, resulting in a very compact, convenient design for various accelerator systems. Its high-frequency PWM driver results in a much smaller footprint. A compensation capacitor corrects the flux-leakage problem so that each of the sections (circuit boards) is identical. The fault-tolerant individual multiplier sections are held at a low enough voltage to avoid corona, and the failure of one multiplier circuit on a board does not affect the entire board. The design has extremely low capacitance, resulting in low stored energy and a fast response time.

Figure **4.27** shows a -750 kV, 100 mA DC supply from Kaiser Systems, Inc., built using the Cross technology. The supply is very compact compared to other supplies operating at these voltages because all components are mounted on stacked-up printed circuit boards (Figure **4.28**) to reach the desired voltage. Each board generates up to 12.5 kV and 100 mA, with 60 boards needed for 750 kV. The system is mounted in an $SF_6$ tank (pressurized to $4\times10^5$-$5\times10^5$ Pa) next to the electron gun.

The same supply can also be used for electrode conditioning when a suitable current limiting resistor is placed between the gun and the power supply (Section 4.3.2.7). The low stored energy (< 10 J) is an advantage for conditioning, as it limits the amount of energy that can go into a field emission site, reducing potential damage to the insulator and electrodes.

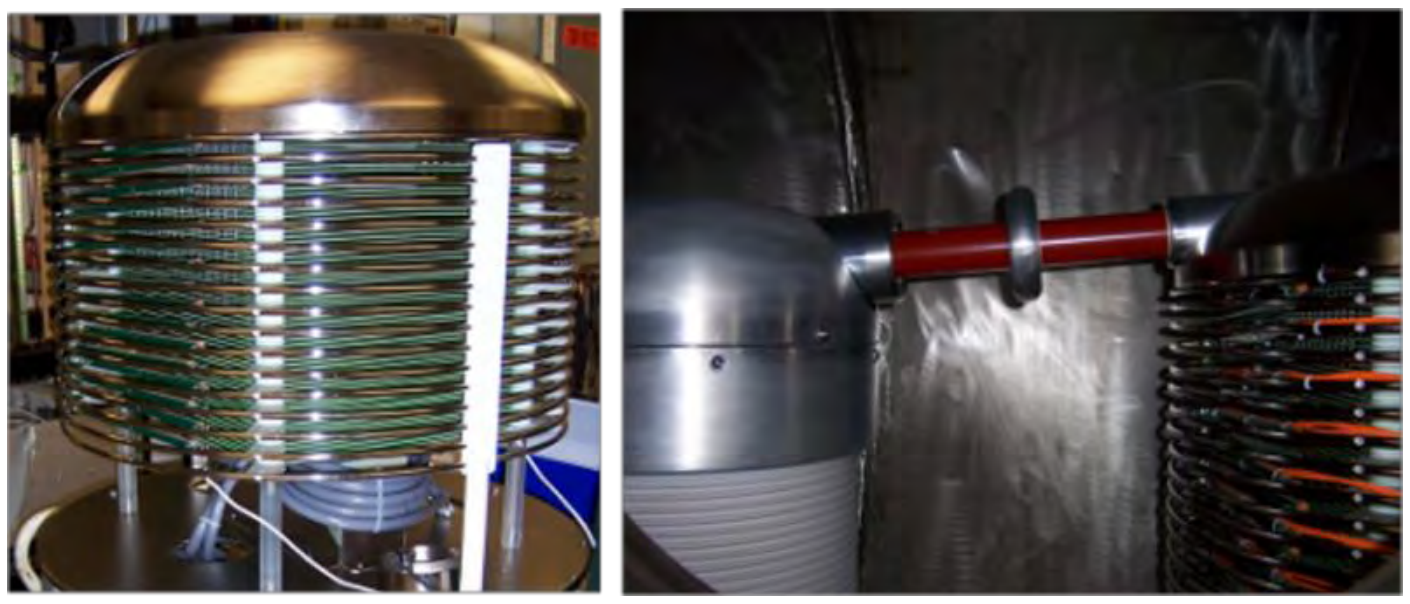

**Figure 4.27. Kaiser Systems, Inc. high voltage power supply, producing -750kV at 100 mA (left). On the right, the power supply is connected to the top of the electron gun insulator. The orange wires are fiber optic cables transmitting signals from a floating ammeter measuring the current entering the gun. [[4.49] (© 1996 IEEE)]**

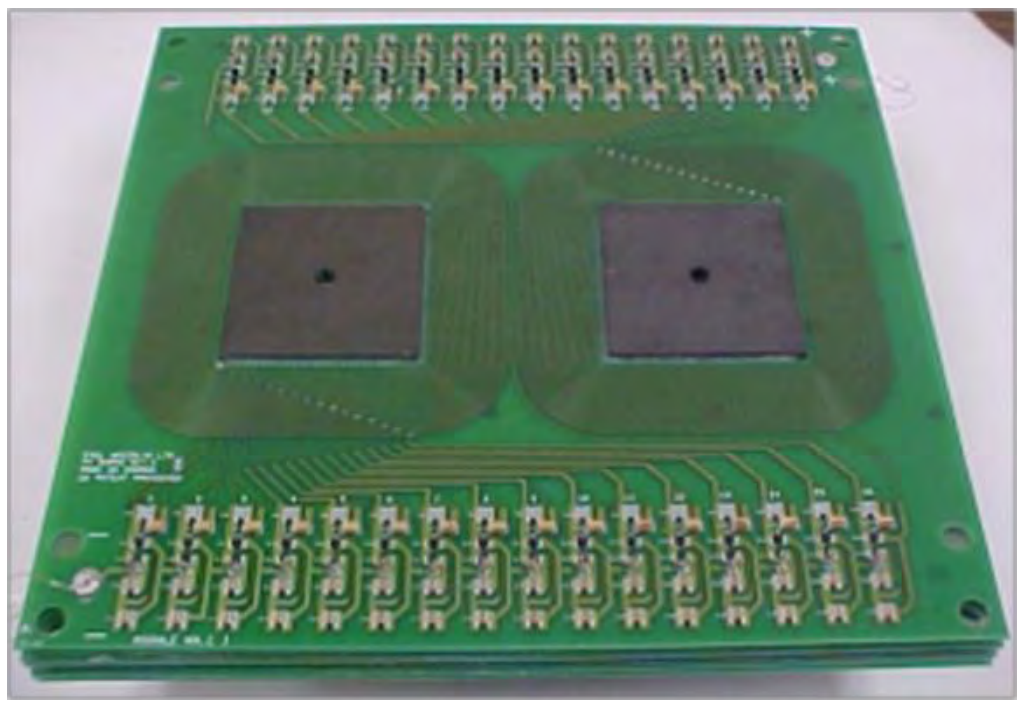

**Figure 4.28. A 40 cm × 40 cm circuit board from the Kaiser Systems -750 kV supply. Each board can generate up to 12.5 kV and 100 mA. [Reprinted with permission from [4.56]. Copyright 2009, American Institute of Physics.]**

#### *4.3.2.7 HV Conditioning*

All high voltage vacuum devices must undergo an initial conditioning so that the device is stable during normal operations. "Stable" can mean different things for different systems. For example, an X-ray tube used in a CT scanner can typically have a few high voltage arcs per exposure, as the image-processing software can correct for any artifacts due to the arcs. However, for a photoemission gun, the use of extremely sensitive electron-emitters like GaAs means that no arcs can be tolerated. In addition, even very low-level electron field emission from electrodes can impinge on the vacuum chamber's surface, generating X-rays, UV light and other particles that can increase the vacuum pressure and even damage the cathode's surface. Many studies [4.39] have shown that high voltage hold-off capability of electrodes and insulators is influenced by dielectric thin films (oxides), electrically stressed interfaces (triple points), metallic- and non-metallic-particulates, absorbed gases and dielectric inclusions. Their various influences were studied extensively, especially for niobium superconducting RF cavities [4.30]. Material properties and cleaning methods to minimize inclusions and particulates are critical to success.

HV systems are processed in several ways, including in-*situ* heating to remove absorbed gases, conditioning with multiple breakdowns, plasma processing, gas conditioning and pulsed processing. For pulsed power machines, reference [4.57] provides an excellent review of the processing. Typically several different methods are used for photoemission guns, partly determined by the extreme-vacuum requirements. Since, often only one or two units are built due to their cost and complexity, developing the best conditioning procedures for the guns is difficult; great care much be taken not to damage them. In contrast, multiple models of industrial devices can be built and tested to establish optimum procedures.

The goal for processing is to operate at a given voltage while drawing a minimum amount of current from field emission sources on the electrodes, preferably picoampere levels, but no more than nanoamperes. The lower the dark current, the more stable is the system, with fewer opportunities for vacuum bursts from electron-stimulated desorption. The processing voltage must be well above the operating point, typically 10-25%.

During processing, the gun vacuum is monitored with an extractor gauge (for total pressure), an RGA for measuring residual gas-content, arrays of photomultiplier tubes, Geiger counters for radiation, an accurate measure of the current from the power supply (with microamperes or better resolution), an arc counter, and perhaps an arc detector inside the $SF_6$ tank. The output of one or more of these monitors can be connected to interlocks for disabling the HV power supply. The power supply current limit should be set to ~100 µA. A low-capacitance HV power supply (described in Section 4.3.2.6) is beneficial in reducing stored energy, and thus the energy that can go into an arc (< 10 J is a good upper limit). A resistor between the gun and power supply limits the current to the arc with a value sufficient to always keep the current below a few milliamperes maximum. If the resistor is too large, processing will be very slow or non-existent; if too small, the electrodes and/or insulator may be damaged. Resistances of 50-200 MΩ often are used.

At the start of processing, the voltage is ramped up until there is a sign of activity, either an increase in vacuum, dark current, or radiation, which typically begins around 250 kV for the gun in Figure **4.3**, or when gradients of 3-4 MV $m^{-1}$ are exceeded somewhere in the gun. Then, the voltage is increased in small steps (say 100 V) at a rate of 1-10 kV per hour. The rate is determined by keeping the vacuum level below $\sim 1\times10^{-6}$ Pa, the dark current below a setpoint (< 100 µA) and the arc counter low enough to protect the resistors from overheating. The values determine how quickly the processing will proceed. Often, all the diagnostic signals will increase during an arc, but sometimes only one or two, depending on the source of the HV activity. When the signal(s) start going up, there are several methods for reacting. In the constant-current method, after the dark current jumps, the voltage is held steady until the current drops back to its baseline value before again stepping up the voltage. With spark processing, the voltage ramp is continued until either a discharge occurs, the dark current is knocked down, or one of the preset trip limits is reached. The procedure is very slow due to keeping limits on the allowed vacuum-pressure excursions.

Identifying a universally successful procedure is difficult because the type and location of field emission can vary depending on the cleaning methods, materials used, assembly procedures, *etc*. Even during processing, the system's response may change as the voltage rises, potentially requiring a different resistor value or new setpoints for the process variables. Field emission events can splatter material around the original emission site, generating multiple new sites requiring re-processing. Difficulties are compounded since only one or two guns are available and risks cannot be taken in assessing the best methods.

The vacuum level in the gun during processing warrants extra attention. Normally, the gun is vacuum baked after assembly and the NEG pumps activated, resulting in a vacuum level less than $10^{-9}$ Pa. Absorbed surface gas is removed from the electrodes during bake which should shorten HV processing time. If the pressure becomes too high during conditioning, the good pre-conditioned vacuum levels for GaAs-like cathode operation may not be recovered. Re-baking to improve the vacuum will change the surface condition of the electrodes, usually necessitating further processing. As higher and higher voltages are reached, recovery from field emission-induced vacuum events is prolonged, as there is more energy per event, which increases surface heating. The RGA spectrum should reveal only $H_2$, CO, $CO_2$ and $CH_4$, with $CH_4$ typically taking the longest to pump away, as NEGs do not pump methane.

Sometimes, processing reaches a point where there can be no further progress without endangering some system component. Several laboratories have experimented with processing using noble gases, such as helium and krypton [4.24], [4.58]. Introducing one of them at ~$10^{-2}$ Pa of pressure while the HV is on generates ions that either chemically react with, or damage field-emitter sites on the negative electrodes, quickly eliminating emission. A turbo pump, attached to the gun during this process, controls the pressure of the process gas. Since the getter pumps do not pump noble gasses, they do not become saturated. Although the ion pump is turned off to avoid overloading, it releases methane. Unfortunately, this process is “blind” to the residual gas content (other than the noble gas). It is easy to open up a vacuum leak without even knowing it. While this method is very efficient, more research is needed to verify its usage for the entire process. One problem is that a fix may not be permanent. If the gas absorbs or reacts with a particle on the HV surface, presumably raising its work function enough to kill the field emission, the gas eventually may desorb due to heat or other processing. Should the ion damage the emission site by melting, beneficial results will be maintained. After gas processing, regular processing can resume until another intractable spot occurs.

Guns of this type with cylindrical ceramic bushings are limited to well below the desired 500 kV operating points due to punch-throughs between 450-500 kV. Several insulator designs and coatings have been tried, and several news ones are ready. Another potential weak link in a HV system is the current limiting resistor used for processing.

The system (Figure **4.27**) failed at ~450 kV for several different insulators, no matter how carefully prepared, leading to the assumption that the insulator was the cause of the problems, not the effect. Recently, it was observed that each time the insulator experienced a punch-through, the current limiting resistor also failed. In fact, during an arc, the gun side of the resistor can drop to ground, resulting in the full voltage being dropped across the resistor, which was rated only for 40 kV continuous. After several such events, the resistor fails open, and seemingly, an arc occurs through the $SF_6$ gas across the resistor, dumping all of the power supply energy into the field emitter (tens to hundreds of Joules, depending on the power supply). The exact order of events is difficult to determine, but the resistor chain certainly was not designed properly.

The initial resistor used was a Cableform model CJE (40 kV, 35 W average power dissipation, Figure **4.29**); they normally withstand 2.5X the power rating for up to 5 s, with 1.5X over voltage. So, during 5 s, 438 J can be dissipated and if the power supply dumps 100 J per arc, then the resistor can handle up to 4 arcs per 5 seconds. With two resistors in series, the value is 8 arcs per 5 seconds. Thus, to assure survival (and the over-voltage condition), only 8 arcs per 5 seconds are allowed. Hence, an arc counter is needed ($dV/dt$ monitor); once the maximum arc rate is exceeded, the voltage must be reduced or turned off to allow the resistors to cool. The resistor chain also must be long enough to hold off the full voltage of the power supply during an arc. If the resistor is damaged (open), then a flashover could occur through the $SF_6$ if its length is too short. This particular design did not adhere to most of these conditions.

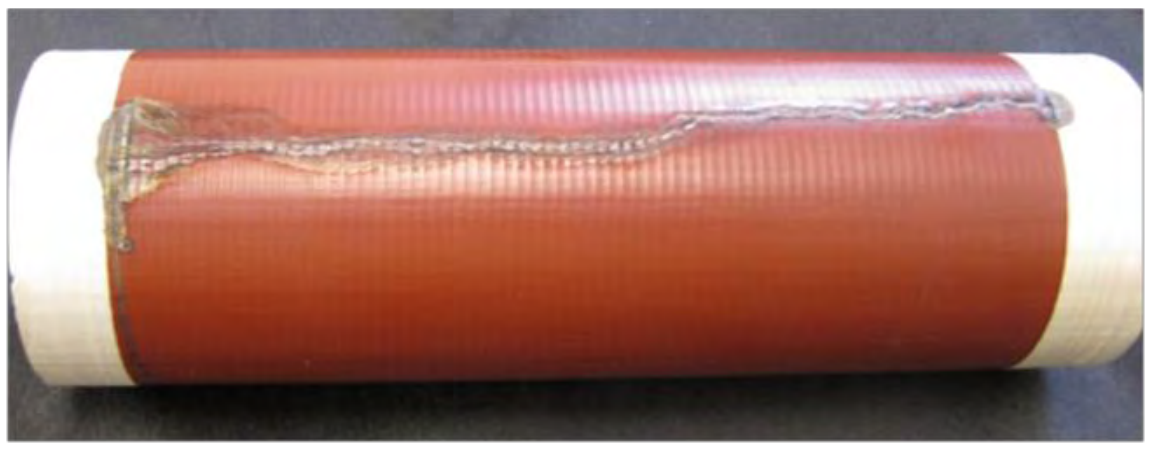

**Figure 4.29. A high voltage resistor damaged during processing.**

Multi-megaohm resistors that simultaneously withstand high voltages, high average power and multiple high-energy impulses are scarce. Nicrom Co. recently marketed very long thick film resistors capable of withstanding up to 400 kV in air continuously, with average power ratings up to 250 W, at the hundreds of megaohm range needed. The maximum energy a single resistor can withstand (assuming 2.5X surge power rating and 1.5X surge voltage rating for no more than 5 seconds), in 5 s is 3 125 J, or 31 arcs per 5-second period (100 J power supply). Adding several in parallel (Figure **4.30**) raises this number and provides redundancy. In addition, the 1 m length affords a 600 kV surge hold-off in air and substantially more in $SF_6$. For a conditioning supply with lower capacitance, the lower stored energy allows the resistors to withstand many more arcs per period.

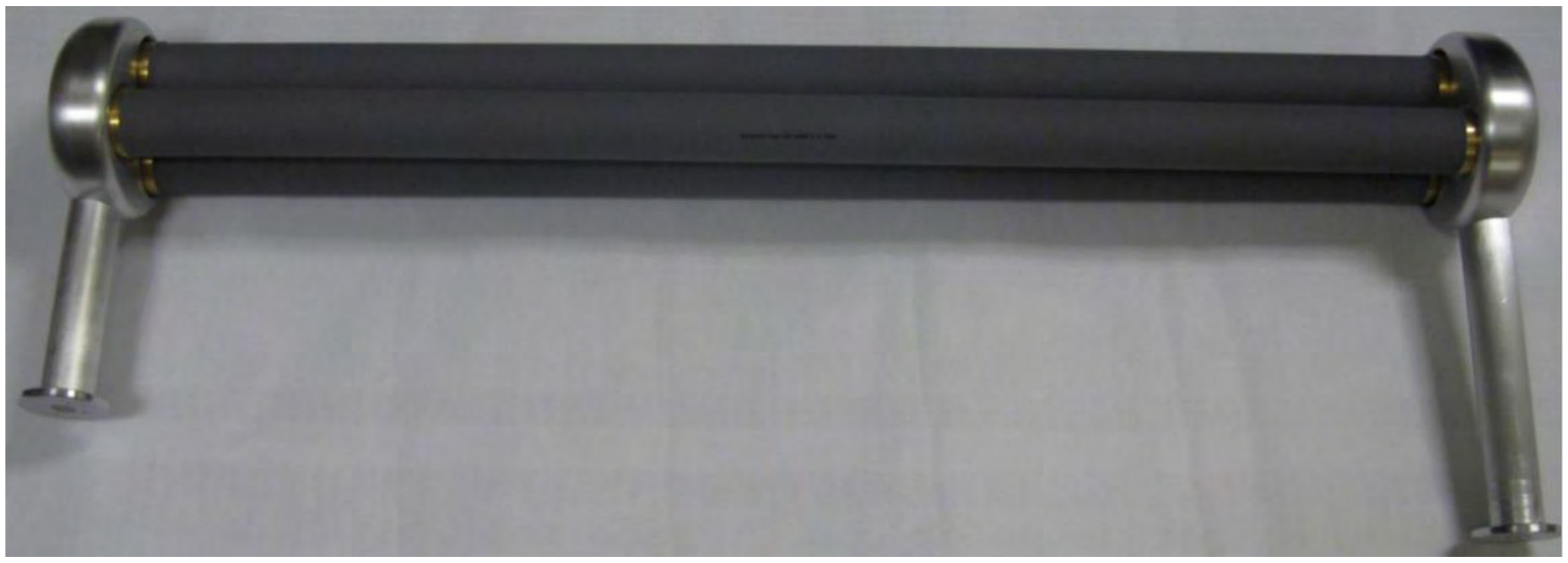

**Figure 4.30. Processing resistor schematic using three long resistors in parallel.**

Lacking such high power resistors, series chains of smaller resistors can be employed, similar to the construction of high voltage dividers (see Figure **4.31**). Caddock produces several reasonable high power, high voltage resistors, such as model MS-310 (1 MΩ, 10 W, 4500 V, 32 mm long body, 5X rated power with applied voltage not exceeding 1.5X rated for 5 s). Using 100 resistors in series gives 100 MΩ and 1 kW continuous power-dissipation (5 kW for 5 s, 450 kV $*$ 1.5 = 625 kV in air hold-off). Using 200-2 MΩ resistors in parallel offers redundancy.

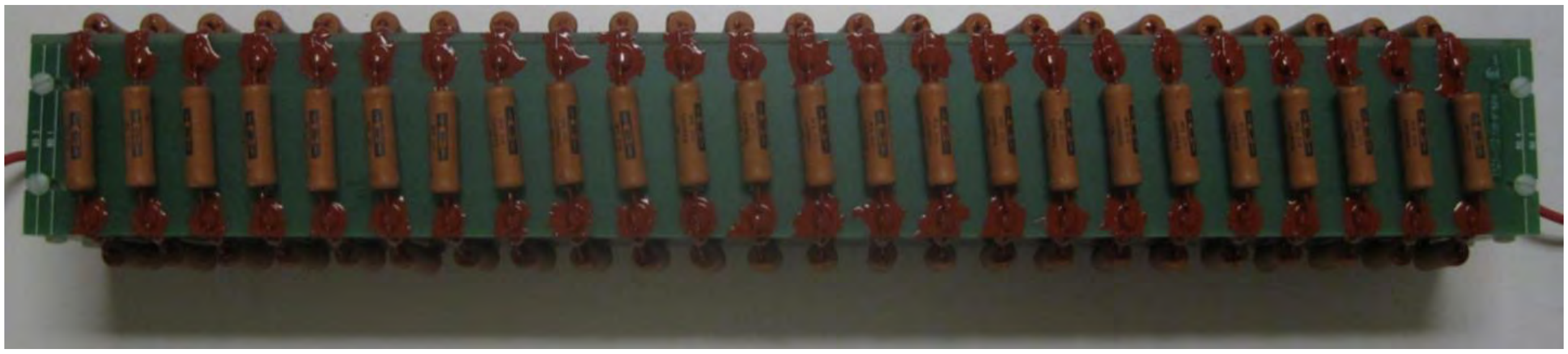

**Figure 4.31. A processing resistor constructed from 100 individual resistors**

While a poorly designed resistor is likely to be a major failure mechanism, other more robust insulator designs are being pursued presently.

### 4.3.3 Bunching and Focusing Section

The section after the gun has several functions: Focusing the beam to keep it from getting too large; focusing for emittance compensation; compressing bunches; diagnostics; and, focusing and steering into the first cavity of the cryomodule. The desire to shorten the distance from the gun to the cryomodule leads to a very congested beamline. One group [4.58] is planning considerable reduction by placing the gun directly inside the cryomodule (Figure **4.32**). The advantage of limited distance must be weighed against lack of access to the gun, fixed focusing distance and no diagnostics.

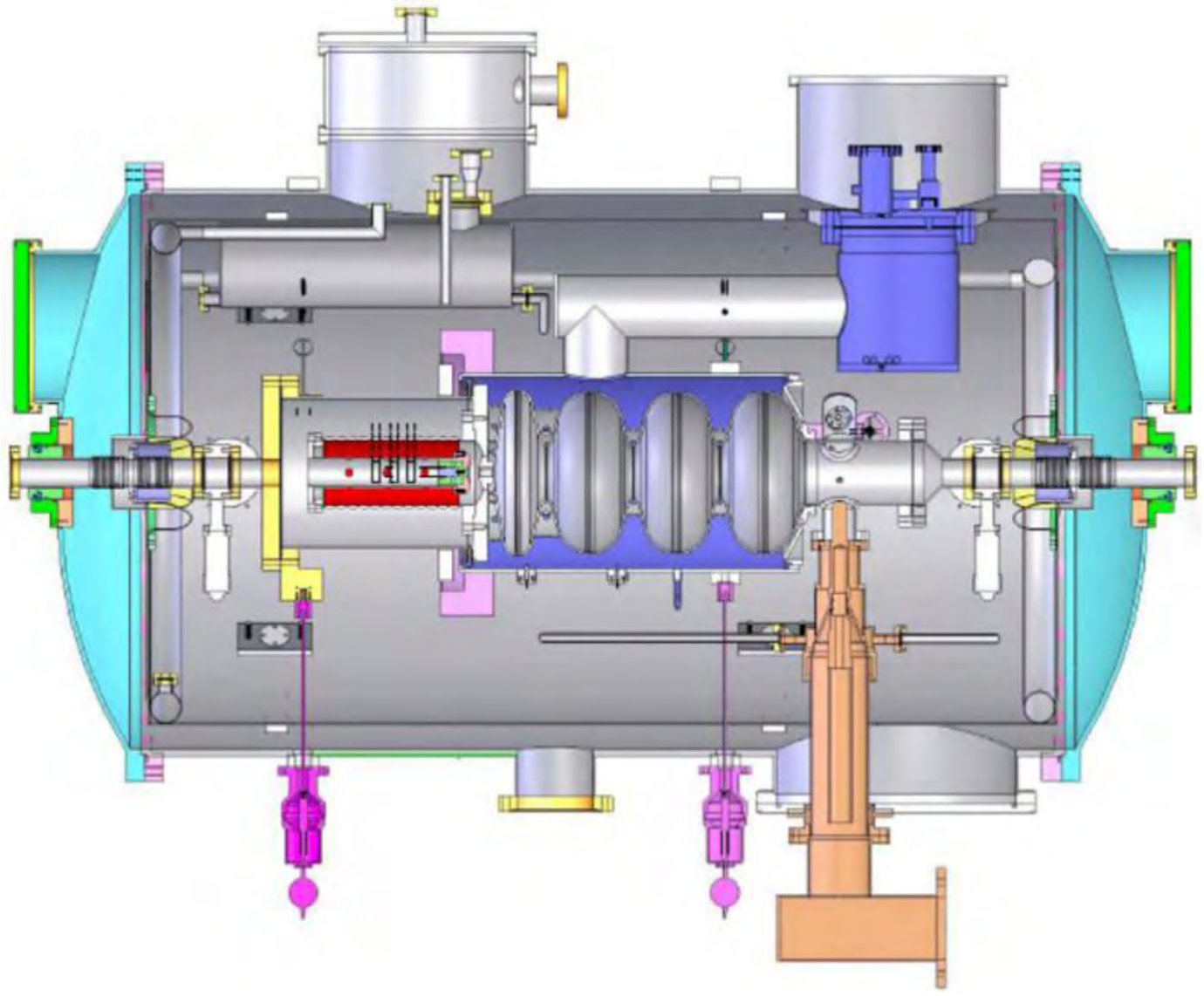

Figure 4.32. A combination DC gun/SCRF accelerator mounted inside a cryomodule. Such a design minimizes the possible drift distance after the anode. [[4.59]; Available under Creative Common Attribution 3.0 License (www.creativecommons.org/licenses/by/3.0/us/) at www.JACoW.org.]

Figure **4.33** shows an example of a more typical focusing and bunching beamline. A pair of corrector coils placed directly after the anode allows centering the beam on the axis of the first solenoid. Having enough degrees of freedom for steering is important; thus, two sets of correctors before the solenoid are preferable, to correct for angle- and offset-errors. In this illustration, a second set of correctors is located inside the solenoid to save space, along with a beam-position monitor. An RF-sealed gate valve protecting the gun is sited between the gun and first solenoid, but could be placed after the first solenoid to bring the focusing element closer. Simulations can determine the best location for all elements. A laser input box is next after the solenoid, followed by a normal conducting buncher cavity, a view screen, an RF-sealed sliding bellows joint and a final solenoid/corrector/bpm package.

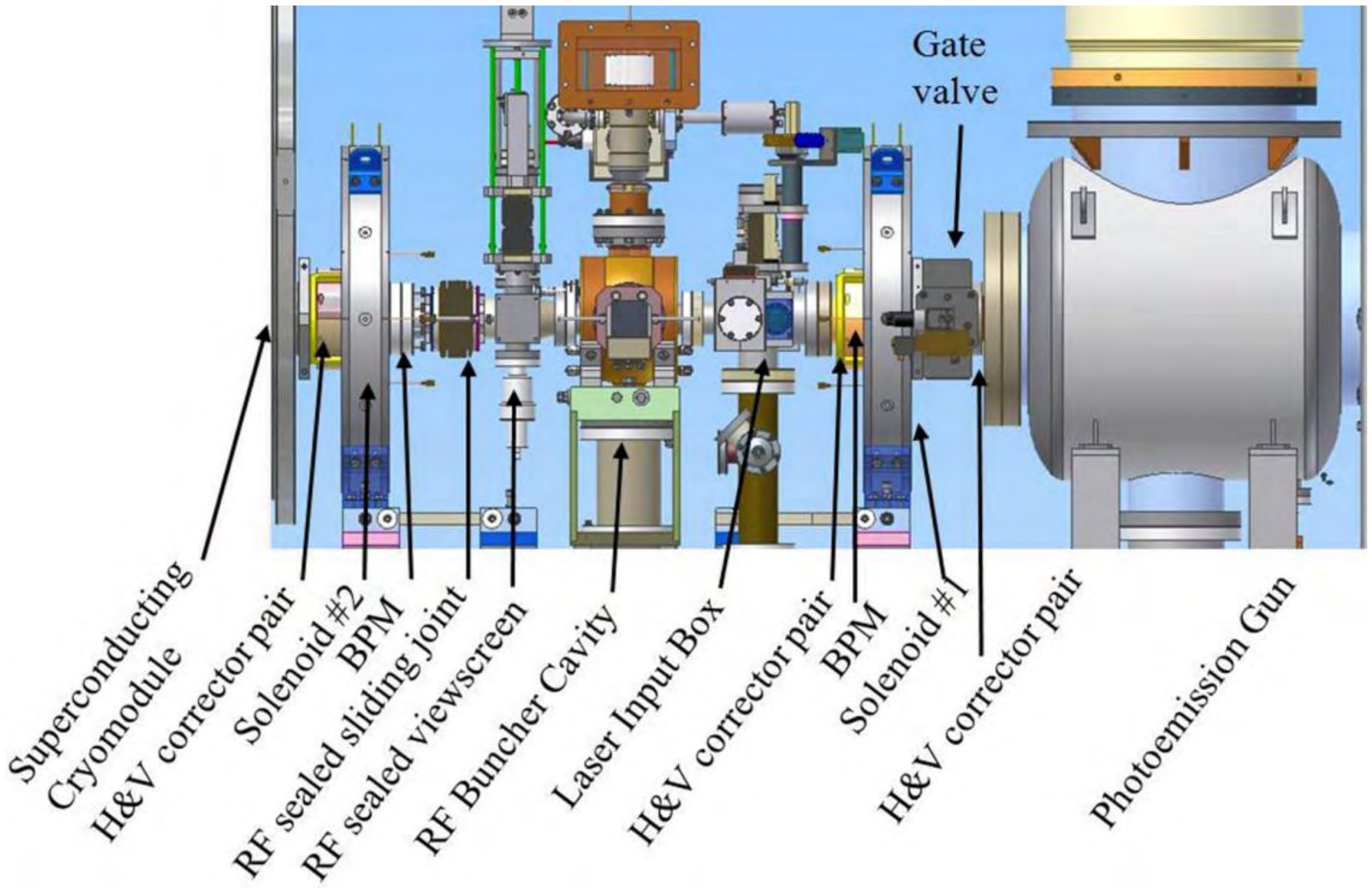

Figure 4.33. A bunching and focusing section between a DC gun (right) and a superconducting accelerating cavity (left).

The vacuum level in this section must be low enough that the gun vacuum is unaffected when the gate valve is opened, thus requiring strict adherence to the vacuum procedures for the gun. All of the moveable devices must be carefully designed and tested to minimize particle generation. For example, the tuners in the buncher cavity use BeCu spring fingers to maintain RF contact, but these scrape material from the tuner during motion. Coating the tuners with TiN eliminates this problem, and additionally, reduces RF multipacting.

To reach adequate vacuum levels, all parts should be heat-treated to remove hydrogen, and all designed to withstand bake-out temperatures of at least 200 ˚C. Similarly, the magnet coils must be potted with a high-temperature epoxy (such as Aremco Bond 526). Due to the limited conductance through the beam pipes, several smaller getter-pump cartridges are used, along with several ion pumps (the ion pumps must be magnetically shielded so they do not disturb the beam). Another option is to NEG-coat the interior of the beam pipes instead of using discrete getter pumps.

Solenoid alignment and the solenoid field quality are critically important for obtaining good beam emittance. Passing through an off-axis solenoid will cause asymmetric focusing across the beam and can affect the quality of the emittance compensation process. A remote control positioner can be used to put the solenoid on the same axis as the gun. One group reported how the quality of the solenoid field affects emittance [4.60]. They found the fringe-field region of the solenoid had a low-level quadrupole component, so they added correction coils to compensate for the irregularities, achieving a 10-30% drop in emittance.

The equations for ballistic bunching are well known for non-space charge dominated beams [4.61]. When space charge is important, the formulas provide only a rough estimate for the distance and power required to obtain the desired bunch length; simulations are needed to finalize the design. An example of a 1 300 MHz CW buncher (normal conducting) [4.62] is shown in Figure **4.34**.

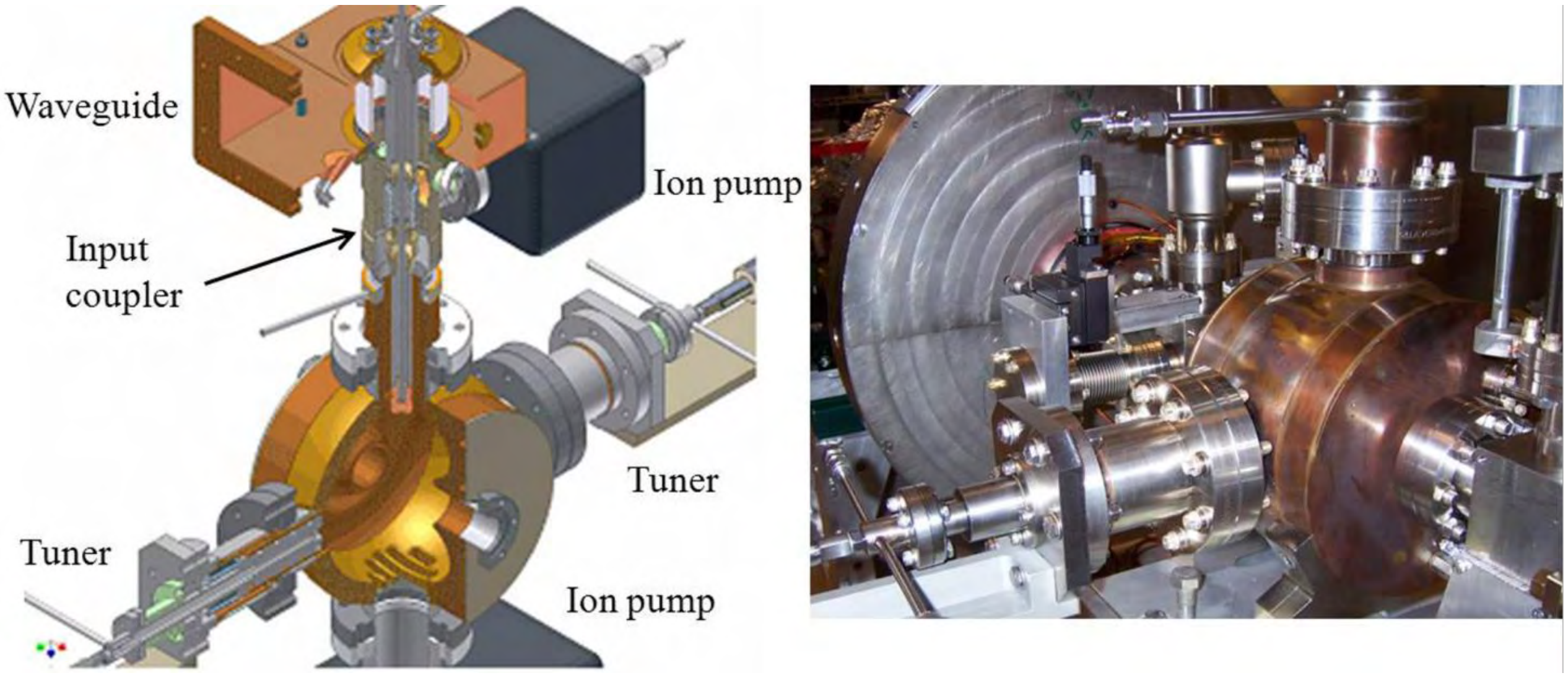

**Figure 4.34. A schematic cutaway view of a CW RF buncher (left). The actual buncher cavity is shown on the right, with the laser box and focusing solenoid in the background.**

Water-cooling is provided to the tuners, input coupler and the copper cavity. Compressed air blowing on the alumina RF window in the coupler keeps it cool. The water channels are designed such that the water is never in contact with a vacuum seal. Thus, if a braze joint or weld leaks to vacuum, the water cannot reach the vacuum space.

Two tuners provide better field symmetry, but only one is varied during operations, while the other is held fixed. Vertically, the input coupler (top) and the pump port (bottom) minimize asymmetries in that plane. The calculated angular kick to an off-center beam is $2.3\times10^{-4}$ $mm^{-1}$ (horizontal) and $3.5\times10^{-4}$ $mm^{-1}$ (vertical).

The last component is a view screen for observing the beam shape after it leaves the gun. Different materials can be used, including BeO, YAG, or CVD diamond. BeO and YAG are best used for low average currents (nanoampere levels), while diamond can withstand microampere average currents. For high bunch charge, short bunch-length beams, it is important to have a smooth surface along the inside of the beam pipe to minimize wakefield effects. Figure **4.35** illustrates how an RF trailer slides into place when the view screen is retracted. The trailer (aluminum), makes RF contact with the beam pipe using spring fingers.

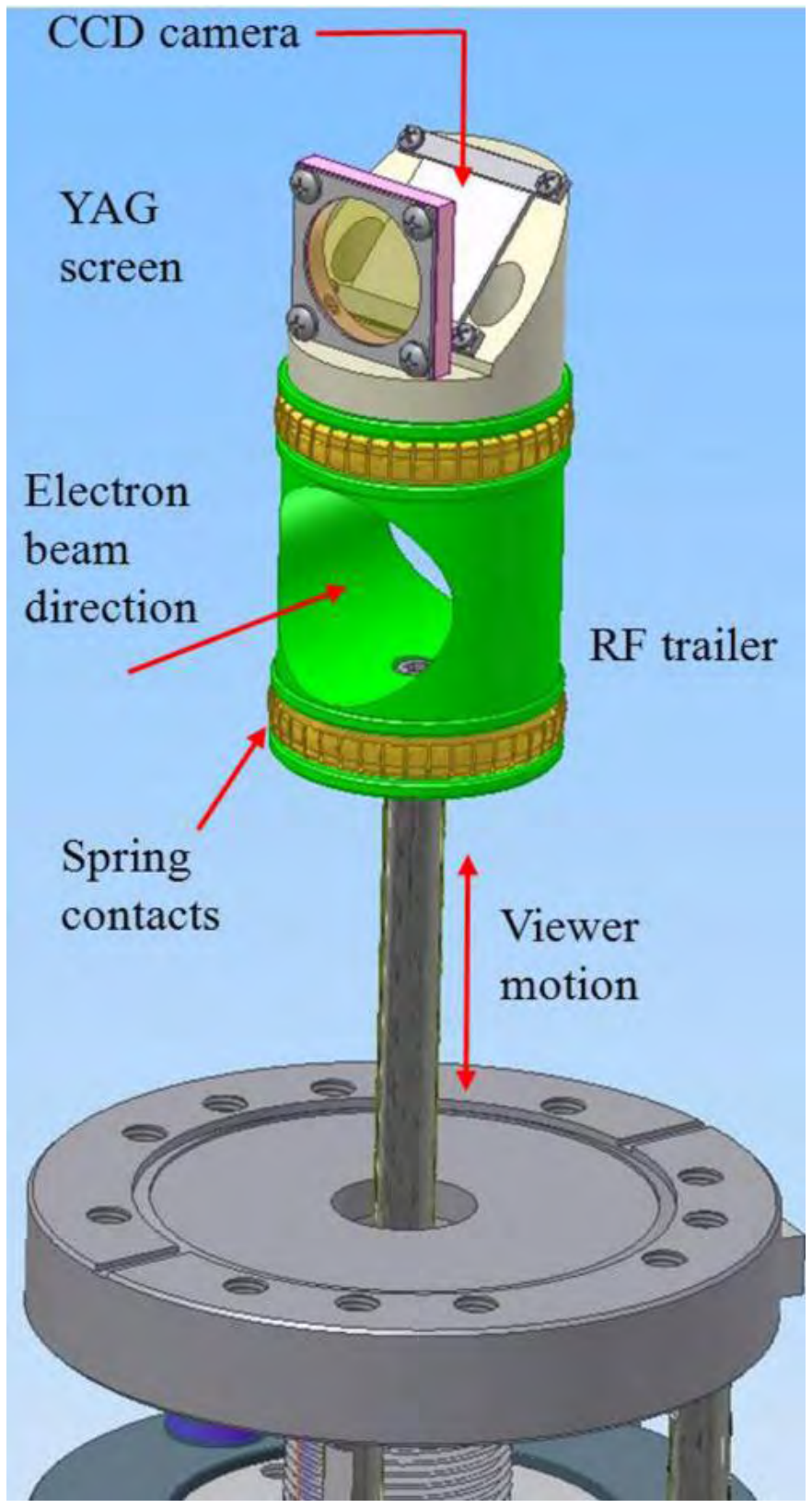

**Figure 4.35. A schematic of a retractable view screen mechanism. When the screen is retracted, an RF trailer slides into a cylindrical tube and provides a smooth surface along the beam pipe. The vacuum tube is not shown.**

The final subsystem of a DC/SCRF injector is the accelerating module, described in the next section.

### 4.3.4 Superconducting Acceleration Section

Many authors have detailed the design of superconducting cavities for accelerators (see [4.30] for examples). In this section, I discuss only those aspects related to high average power CW injectors, following the parameter set in Table **4.2**.

Accelerating a 100 mA average current CW bunch train to 5 MeV requires 500 kW of RF power. At 1 300 MHz, the highest average power input coupler available operates at 50 kW [4.63]; hence, at minimum, ten input couplers are required. At lower RF frequencies, higher powers can be used as the power per unit area scales as the square of the frequency. Similar to the buncher cavity, having opposing input feeds helps to symmetrize the fields in the cavity, meaning two couplers per cavity for a total of 5 RF cavities. Figure **4.36** shows an example of a 1 300 MHz input coupler for a superconducting cavity that can operate at 50 kW.

Each RF cavity can have one or more individual, coupled niobium cells. With 5 two-cell cavities, each cavity must provide 1 MV of energy gain, equivalent to a field gradient of 5 MV $m^{-1}$, well within today's capabilities. Figure **4.37** is an example of a two-cell cavity with two opposing ports for the input couplers [4.64], operating between 5 and 15 MV $m^{-1}$ with a quality factor ($Q_0$) of $1\times10^{10}$.

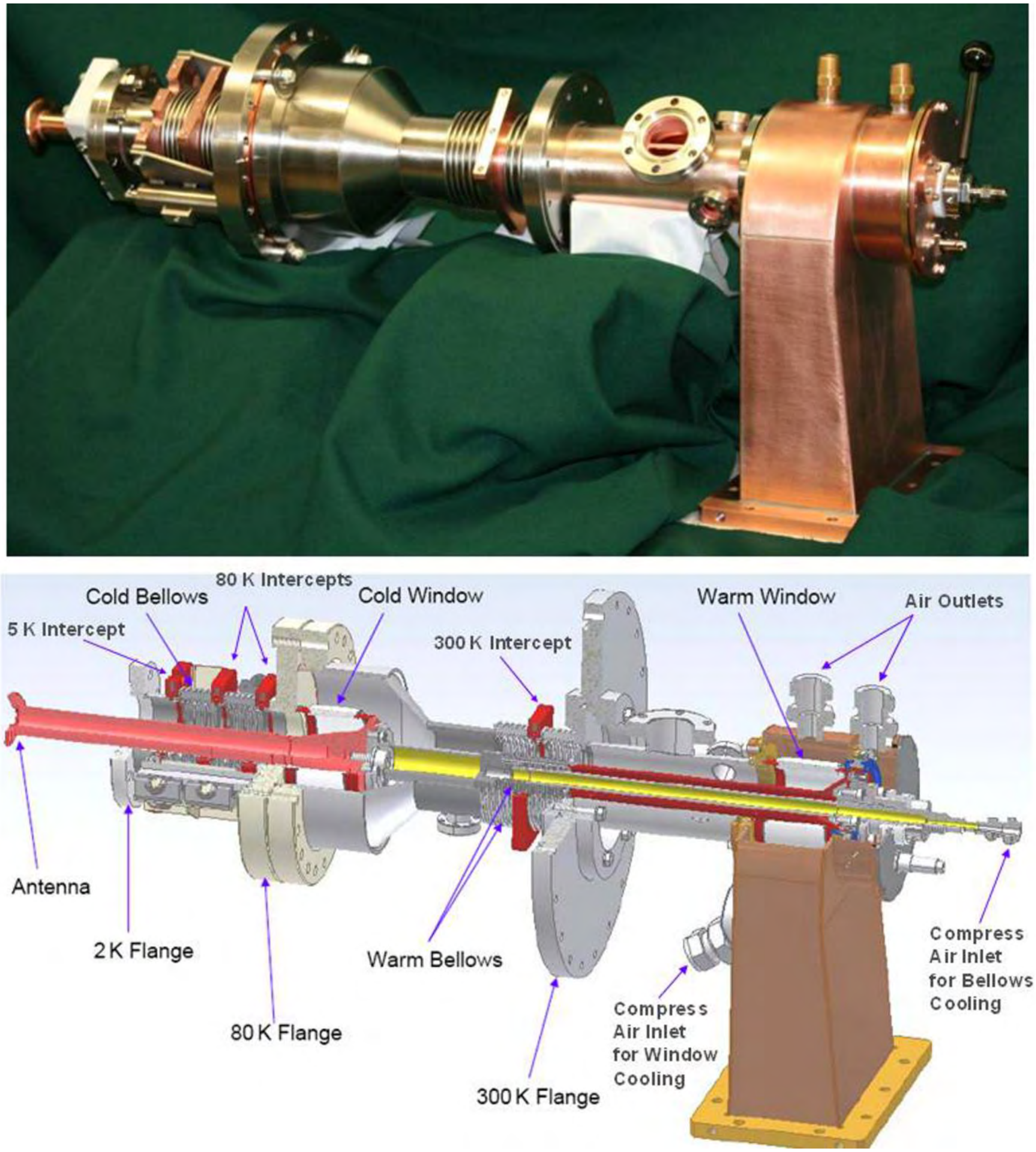

**Figure 4.36. A high average power 1 300 MHz input coupler. The top picture shows the actual device, while the bottom cutaway view indicates the inner details and temperature intercepts from warm to cold.**

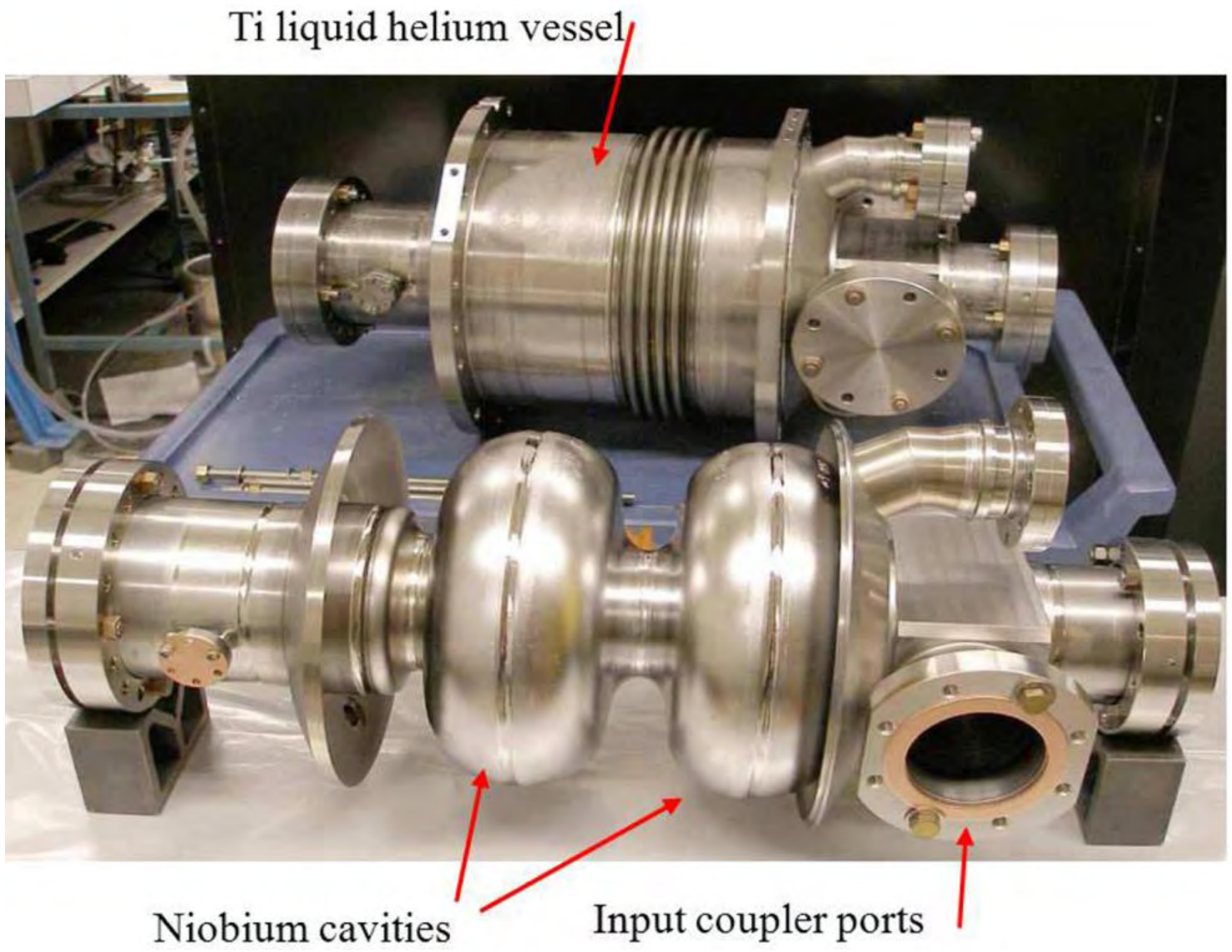

**Figure 4.37. The 2-cell niobium cavity with opposing input ports is shown at the bottom. Above, a titanium liquid-helium vessel is welded around the cavity. Its small size limits the amount of liquid helium required.**

For lower average power machines, the same cavity can be used for the injector and the main accelerator. At CEBAF, for example, cavities of 5 cells are used. An unintended benefit of having 5 two-cell cavities (opposed to 2 five-cell cavities) is that additional degrees of freedom are available for optimizing the beam properties, *viz.*, 5 gradients and 5 phases. With this configuration, the first and second cavities can be run off-crest to supplement the bunching process, providing an option of increasing the length of the initial bunch to reduce space charge forces. Further, this additional control supports a lower gun voltage, if desired. Beams as low as 200 kV can be straightforwardly captured from the gun by adjusting the phases and gradients of the first few cavities. With a five-cell cavity, the phase slippage due to the non-relativistic beam can decelerate the beam for low gun voltages.

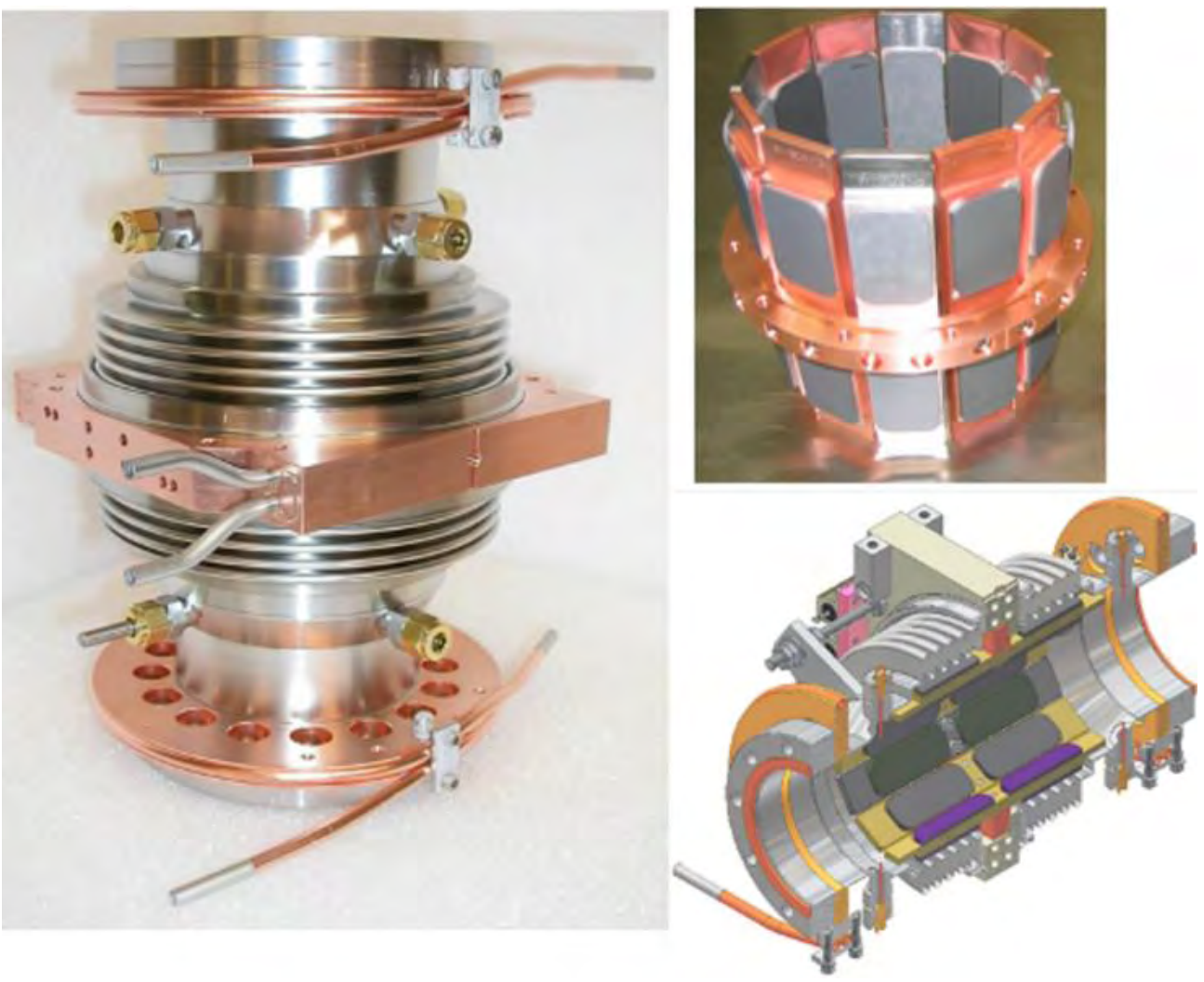

**Figure 4.38. An example of an RF HOM inline absorber. The picture at the upper right shows three types of RF absorber tiles mounted on copper heat sinks. The left shows the HOM vacuum vessel with plumbing for cooling gas to remove the heat from the tiles, while the lower right cutaway view shows how the device is assembled.**

The last critical component for an accelerating section is a way to absorb the power in the higher order modes (HOM). For the very short bunch lengths envisioned (< 1 mm), the beam can excite resonances at frequencies higher than the fundamental, which can increase beam emittance. There are several options for removing the power from HOMs, using either waveguide dampers, or inline absorbers. Figure **4.38** shows and example of the latter, using various types of absorbing tiles [4.64].

With all of the components in hand, the cavity string can be assembled (Figure **4.39**) in a cleanroom to minimize particle contamination, which can compromise the cavity's performance. The string is then inserted into the cryogenic vacuum vessel, where all the input couplers, helium plumbing and instrumentation are attached. Figure **4.40** shows the complete assembly ready to be shipped to its final destination.

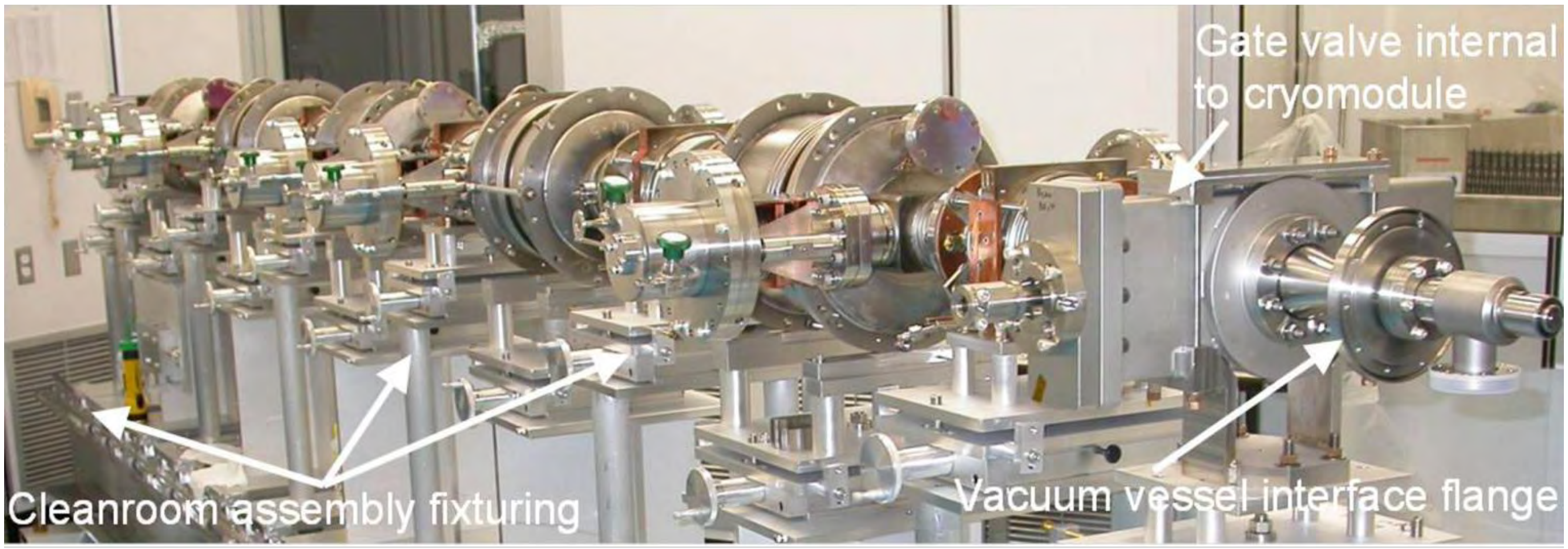

**Figure 4.39. An assembled cavity string, consisting of 2 gate valves, 6 HOM absorbers, 5 two-cell cavities. The input couplers are installed after the string is inserted into the cryomodule.**

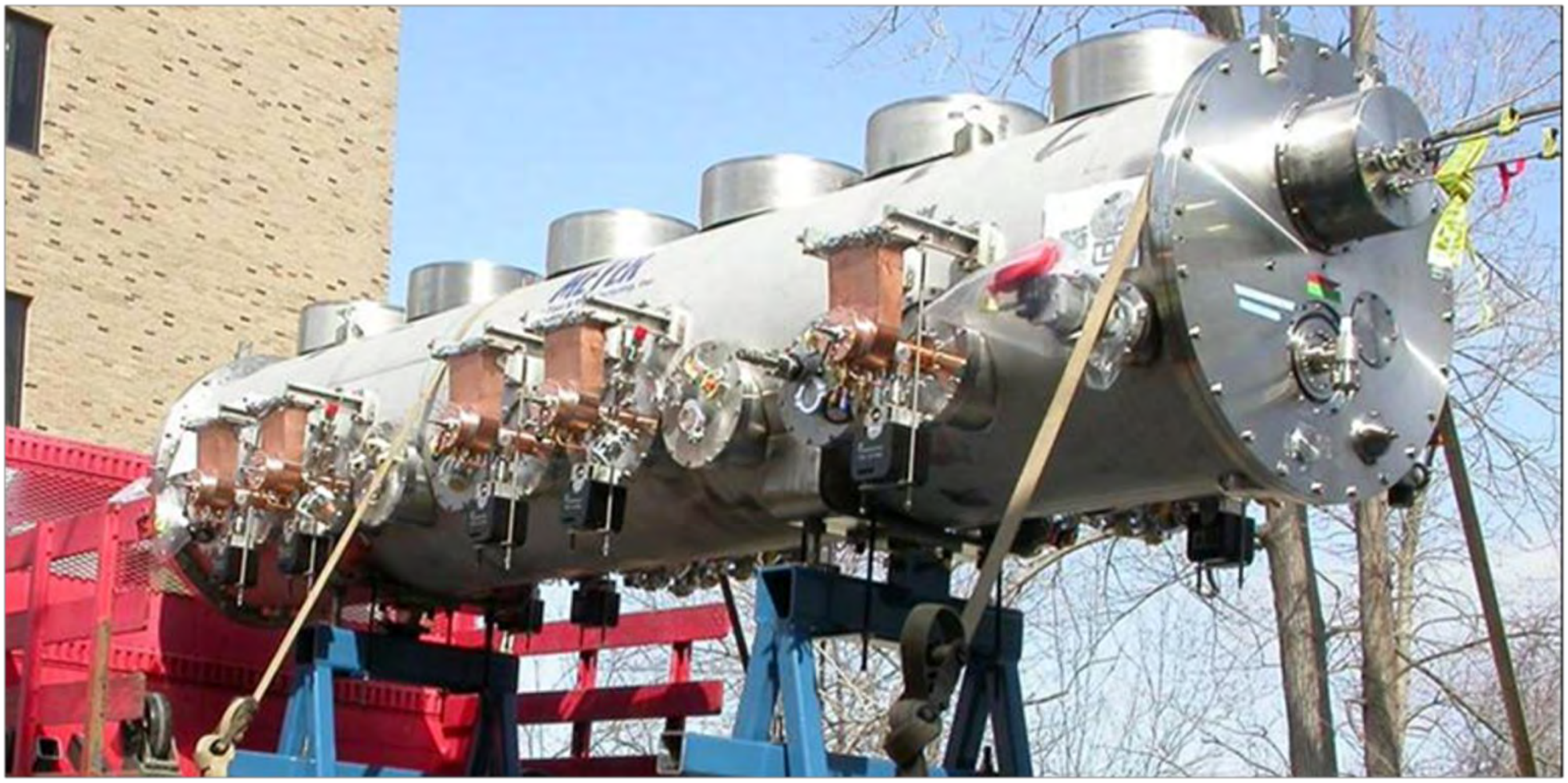

**Figure 4.40. The final fully assembled cryomodule ready to ship.**

There are other examples of RF designs for ~100 mA average current electron injectors. Figure **4.41** shows a design from Jefferson Laboratory [4.66]. It starts with a 500 kV DC photoemission gun, followed by a solenoid for emittance compensation. The accelerating cavities operate a 748.5 MHz, and at this lower frequency, three single-cell cavities can attain the 5-7 MeV desired beam energy. A third harmonic cavity located between cavities 1 and 2 linearizes energy variations along the bunch.

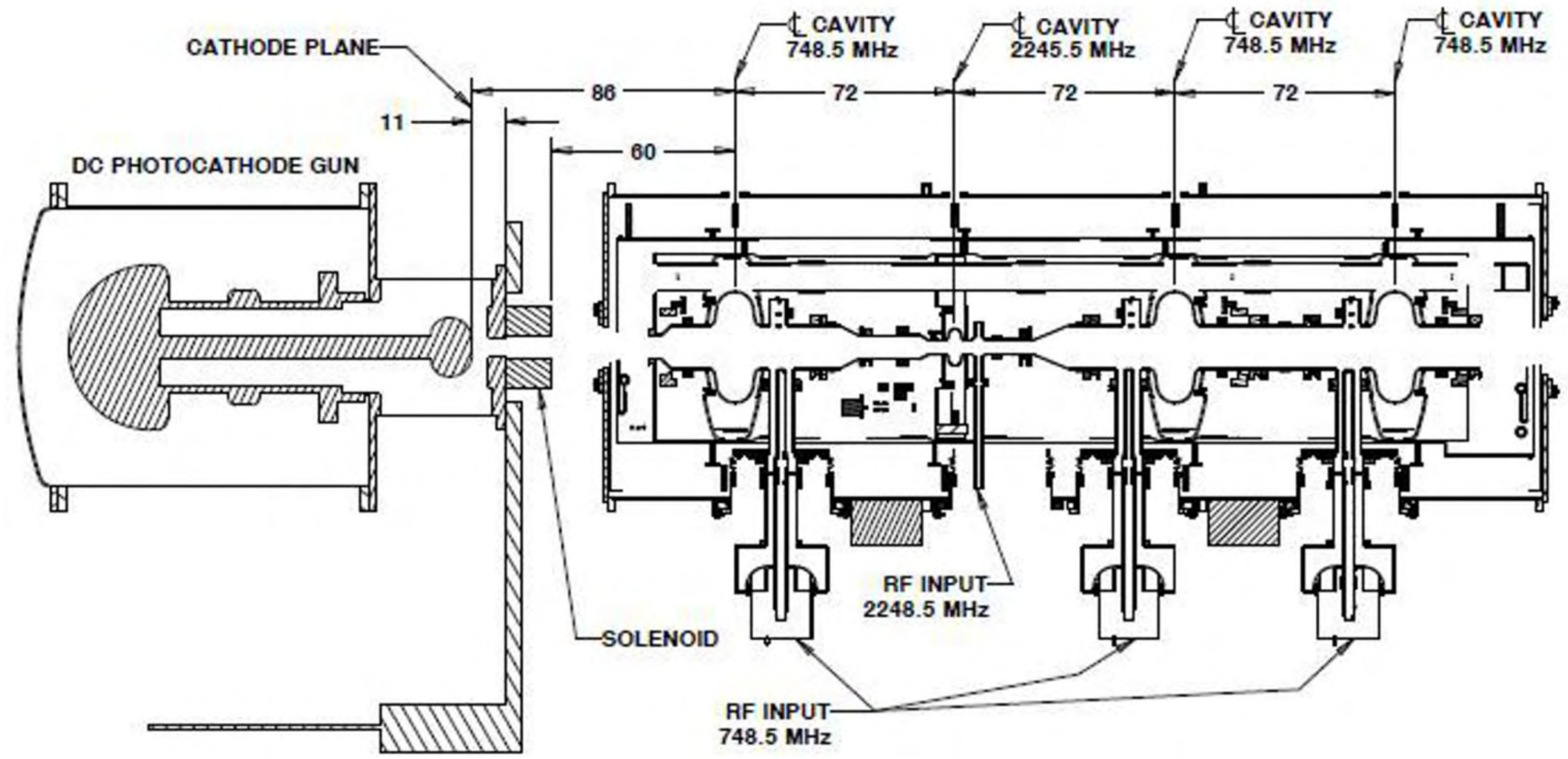

**Figure 4.41. Drawing of a Jefferson lab design for a 100 mA injector. [Courtesy of F. Hannon]**

## 4.4 SUMMARY

In this chapter, I covered the design and implementation of an electron injector using a high voltage DC photoemission gun and a superconducting RF accelerator. This type of injector is best suited for mid-range bunch charge, high average power applications that require high brightness beams. I described in detail the various sub-systems for a DC/RF injector; many of these parts are well defined. Obtaining ~500 kV operational gun voltages remains difficult, although rapid progress is being made. Cathodes exist with reasonable lifetime for several milliampere beams, but work continues to find low emittance, high efficiency cathodes with long operational lifetime at higher currents. For the RF- and SCRF-systems, tests are underway to prove that the designs are reliable between 10 and 100 mA.

Several groups are working on high average power machines to verify the technologies and to demonstrate the accuracy of the simulations. While DC gun/RF injectors have existed for many decades, the last ten years has seen many advances in gun-, laser-, cathode- and SCRF-technologies that promise to provide a path to reliable, high average power, high brightness electron injectors that can be used in many high performance accelerator applications.

## 4.5 CONFLICT OF INTEREST AND ACKNOWLEDGEMENT

I confirm that this article content has no conflicts of interest and would like to acknowledge the support by the National Science Foundation, grant number DMR-0807731.

***References***

[4.1] C. K. Sinclair, E. L. Garwin, R. H. Miller *et al.*, "A high intensity polarized electron soure for the Stanfor Linear Accelerator," in *Proc. AIP Conf.*, vol. 35, 1976, pp. 424-431.

[4.2] M. B. James and R. H. Miller, "A high current injector for the proposed SLAC linear collider," *IEEE Trans. Nucl. Sci.*, vol. 28, pp. 3461-3463, June 1981.

[4.3] C. K. Sinclair, "The SLAC lasertron project," in *Proc. AIP Conf.*, vol. 156, 1987, pp. 298-312.

[4.4] C. K. Sinclair, "A 500 kV photoemission electron gun for the CEBAF FEL," *Nucl. Instrum. Meth. A*, vol. 318, pp. 410-414, July 1992.

[4.5] R. Alley, H. Aoyagi, J. Clendenin *et al.*, "The Stanford linear accelerator polarized electron source," *Nucl. Instrum. Meth. A.*, vol. 365, pp. 1-27, November 1995.

[4.6] C. K. Sinclair, "DC photoemission electron guns as ERL sources," *Nucl. Instrum. Meth. A*, vol. 557, pp. 69-74, February 2006.

[4.7] J. Haimson, B. Mecklenburg, G. Stowell *et al.*, "A fully demountable 550 kV electron gun for low emittance beam experiments with a 17 GHz linac," in *Proc. 1997 Particle Accelerator Conf.*, 1997, pp. 2808-2810.

[4.8] I. V. Bazarov and C. K. Sinclair, "Multivariate optimization of a high brightness DC gun photoinjector," *Phys. Rev. ST Accel. Beams*, vol. 8, pp. 034202-1–034202-14, March 2005.

[4.9] B. M. Dunham, L. S. Cardman and C. K. Sinclair, "Emittance measurements for the Illinois/CEBAF polarized electron source," in *Proc. 1995 Particle Accelerator Conf.*, 1995, pp. 1030-1032.

[4.10] D. Dowell, S. Z. Bethel and K. D. Friddell, "Results from the average power laser experiment photocathode injector text," *Nucl. Instrum. Meth. A*, vol. 356, pp. 167-176, March 1995.

[4.11] I. V. Bazarov, B. M. Dunham and C. K. Sinclair, "Maximum achievable beam brightness from photoinjectors," *Phys. Rev. Lett.*, vol. 102, pp. 104801-1–104801-4, March 2009.

[4.12] B. E. Carlsten, "New photoelectric injector design for the Los Alamos National Laboratory XUV FEL accelerator," *Nucl. Instrum. Meth. A*, vol. 285, pp. 313-319, December 1989.

[4.13] I. V. Bazarov, B. M. Dunham, Y. Li *et al.*, "Thermal emittance and response time measurements of negative electron affinity photocathodes," *J. Appl. Physics*, vol. 103, pp. 054901-1–054901-8, March 2008.

[4.14] K. Smolenski, I. Bazarov, H. Li *et al.*, "Design and Performance of the Cornell ERL dc photoemission gun," in *Proc. AIP Conf.*, vol. 1149, 2008, pp. 1077-1083.

[4.15] S. H. Kong, D. C. Nguyen, R. L. Sheffield *et al.*, "Fabrication and characterization of cesium telluride photocathodes: A promising electron source for the Los Alamos Advanced FEL," *Nucl. Instrum. Meth. A*, vol. 358, pp. 276-279, April 1995.

[4.16] I. Bazarov, B. M. Dunham, X. Li *et al.*, "Thermal emittance and response time measurements of a GaN photocathode," *J. Appl. Physics*, vol. 105, pp. 083715-1–083715-4, April 2009.

[4.17] P. Hartmann, J. Bermuth, D. v. Harrach *et al.*, "A diffusion model for picosecond electron bunches from negative electron affinity GaAs photocathodes," *J. Appl. Physics*, vol. 86, pp. 2245-2249, August 1999.

[4.18] S. Pastuszka, D. Kratzmann, D. Schwalm *et al.*, "Transverse energy spread of photoelectrons emitted from GaAs photocathodes with negative electron affinity," *Appl. Physics Lett.*, vol. **71,** pp. 2967-2969, September 1997.

[4.19] C. K. Sinclair, P. A. Adderley, B. M. Dunham *et al.*, "Development of a high average current polarized electron source with long cathode operational lifetime," *Phys. Rev. ST Accel. Beams*, vol. 10, pp. 023501-1–023521-21, February 2007.

[4.20] J. Grames, M. Poelker, P. Adderley *et al.*, "Measurements of photocathode operational lifetime at beam currents up to 10 mA using an improved DC high voltage GaAs photogun," in *Proc. AIP Conf.*, vol. 915, 2007, pp. 1037-1044.

[4.21] C. K. Sinclair, "High voltage DC photoemission electron guns – current status and technical challenges," *Int. Committee Future Accel. Beam Dynamics Newslett.*, no. 46, August 2008, pp. 97-118.

[4.22] B. M. Dunham, P. Hartmann, R. Kazimi *et al.*, "Advances in DC photocathode electron guns," in *Proc. AIP Conf.*, vol. 472, 1998, pp. 813-822.

[4.23] D. H. Dowell, I. Bazarov, B. Dunham *et al.*, "Cathode R&D for future light sources," *Nucl. Instrum. Meth. A*, vol. 622, pp. 685-697, October 2010.

[4.24] N. Nishimori, I. Bazaraov, B. Dunham *et al.*, "DC gun technological challenges," in *Proc. Energy Recovery Linac Workshop*, 2009.

[4.25] R. E. Kirby, G. J. Collet and K. Skarpass, "An in-situ photocathode loading system for the SLC polarized electron gun," in *Proc. 1993 Particle Accelerator Conf.*, 1993, pp. 3030-3032.

[4.26] Z. Yu, S. L. Buczkowski, N. C. Giles *et al.*, "Defect reduction in ZnSe grown by molecular beam epitaxy on GaAs substrates cleaned using atomic hydrogen," *Appl. Physics Lett.*, vol. 69, pp. 82-84, April 1996.

[4.27] C. D. Park, S. M. Chung, X. Liu *et al.*, "Reduction in hydrogen outgassing from stainless steels by a medium-temperature heat treatment," *J. Vacuum Sci. Tech. A*, vol. 26, pp. 1166-1171, August 2008.

[4.28] H. Kurisu, G. Kimoto, H. Fujii *et al.*, "Outgassing properties of chemically polished titanium materials," *J. Vacuum Soc. Japan*, vol. 49, pp. 254-258, February 2006.

[4.29] F. Rosebury, *Handbook of Electron Tube and Vacuum Techniques*, Woodbury: American Institute of Physics, 1993.

[4.30] H. Padamsee, *RF Superconductivity*: *Science, Technology and Application*, Weinheim: Wiley-VCH Verlag, 2009.

[4.31] Zachary Conway, personal communication, January 2010.

[4.32] Applied Surface Technologies, Inc., manufactures $CO_2$ snow guns.

[4.33] M. Böhnert, D. Hoppe, L. Lilje *et al.*, "Particle free pump down and venting of UHV vacuum systems," in *Proc. 2009 Superconducting RF Conf.*, 2009, pp. 883-886.

[4.34] SAES Getters, Inc., Viale Italia, 77, 20020 Lainate MI, Italy. Online: http://www.saesgetters.com/.

[4.35] M. L. Stutzman, P. Adderley, J. Brittian *et al.*, "Characterization of the CEBAF 100 kV GaAs photoelectron gun vacuum system," *Nucl. Instrum. Meth. A*, vol. 574, pp. 213-220, January 2007.

[4.36] D. J. Holder, "First results from the ERL prototype (ALICE) at Daresbury," in *Proc. 2008 Linear Accelerator Conf.*, 2008, pp. 694-698.

[4.37] P. A. Adderley, J. Clark, J. Grames *et al.*, "Load-locked DC high voltage GaAs photogun with an inverted-geometry ceramic insulator," *Phys. Rev. ST Accel. Beams*, vol. 13, pp. 010101-1–010101-7, January 2010.

[4.38] P. G. Slade, *The Vacuum Interrupter: Theory, Design and Application*, Raton: CRC Press, Boca, 2008.

[4.39] R. Latham, Ed., *High Voltage Vacuum Insulation: Basic Concepts and Technological Practice*, London: Academic Press, 1995.

[4.40] P. Spolaore, G. Bisoffi, F. Cervellera *et al.*, "The large gap case for high voltage insulation in vacuum," in *17th Int. Symp. Discharges Elect. Insulation Vacuum*, 1996, pp. 523-526.

[4.41] F. Furuta, T. Nakanishi, S. Okumi *et al.*, "Reduction of field emission dark current for high-field gradient electron gun by using a molybdenum cathode and titanium anode," *Nucl. Instrum. Meth. A*, vol. 538, pp. 33-44, February 2005.

[4.42] I. Smith, "The early history of western pulsed power," *IEEE Trans. Plasma Sci.*, vol. 34, pp. 1585-1609, October 2006.

[4.43] J. Haimson, "Recent advances in high voltage electron beam injectors," *IEEE Trans. Nucl. Sci.*, Vol. 22, pp. 1354-1357, June 1975.

[4.44] Dielectric Science, Inc., 88 Turnpike Road, Chelmsford, MA, 01824. Online: http://www.dielectricsciences.com/.
[4.45] Claymount Corporation, Anholtseweg 44, De Rietstap, 7091 HB Dinxperlo, The Netherlands. Online: http://www.claymount.com
[4.46] N. H. Malik, A. A. Al-Arainy and M. I. Qureshi, *Elect. Insulation Power Syst.*, New York: Marcel Dekker, Inc, 1997.
[4.47] J. J. Shea, "Punchthrough of ceramic insulators," in *Proc. IEEE Conf. Electronics Insulation Dielectric Phenomena*, 1990, pp. 441-450.
[4.48] J. Vines and P. A. Einstein, "Heating effect of an electron beam impinging on a solid surface, allowing for penetration," in *Proc. IEE*, vol. 111, 1964, pp. 921-930.
[4.49] B. M. Dunham and K. Smolenski, in *Proc. IEEE Int. Power Modulator High Voltage Conf.*, 2010, pp. 92-101.
[4.50] F. Liu, I. Brown, L. Phillips *et al.*, "A method of producing very high resistivity surface conduction on ceramic accelerator components using metal ion implantation," in *Proc. 1997 Particle Accelerator Conf.*, 1997, pp. 3572-3574.
[4.51] R. Nagai, R. Hajima, N. Nishimora *et al.*, "High-voltage testing of a 500-kV DC photocathode electron gun," *Rev. Sci. Instrum.*, vol. 81, pp. 033304-1–033304-5, February 2010.
[4.52] S. V. Benson, G. Biallas, D. Bullard *et al.*, "An inverted ceramic DC electron gun for the Jefferson Laboratory FEL," in *Proc. Free Electron Laser Conf.*, 2009, pp. 383-385.
[4.53] J. D. Cockcroft and E. T. Walton, "Experiments with high velocity positive ions. II. The disintegration of elements by high velocity protons," in *Proc. Royal Society London A*, 1932, pp. 229-242.
[4.54] C. Hernandez-Garcia, T. Siggins, S. Benson *et al.*, "A high average current DC GAAS photocathode gun for ERLS and FELS," in *Proc. 2005 Particle Accelerator Conf.*, 2005, pp. 3117-3119.
[4.55] R. J. van de Graaff, "High voltage electromagnetic apparatus having an insulating magnetic core," U.S. Patent No. 3,187,208, June 1, 1965.
[4.56] U. Uhmeyer, "KSI's cross insulated core transformer technology," in *Proc. AIP Conf.*, vol. 1149, 2009, pp. 1099-1103.
[4.57] M. E. Cunco, "The effects of electrode cleaning and conditioning on the performance of high-energy, pulsed-power devices," in *18th Int. Symp. Discharges Elect. Insulation Vacuum*, 1998, pp. 721-730.
[4.58] S. Bajic and R. V. Latham, "A new perspective on the gas conditioning of high-voltage vacuum-insulated electrodes," *J. Physics D: Appl. Physics*, vol. 21, pp. 943-950, 1988.
[4.59] J. Hao, F. Zhu, S. Quan *et al.*, "3.5-cell superconducting cavity for DC-SRF photoinjector at Peaking University," in *Proc. 2009 Superconducting RF Workshop*, 2009, pp. 205-207.
[4.60] D. H. Dowell, E. Jongewaard, J. Lewandowski *et al.*, "The development of the linac coherent light source RF gun," *Int. Committee Future Accel. Beam Dynamics Newslett.*, no. 46, August 2008, pp. 162-192.
[4.61] A. Chao and M. Tigner, Eds., *Handbook of Accelerator Physics and Engineering*, Singapore: World Scientific Press, 1999, pp. 102.
[4.62] V. Veshcherevich and S. A. Belomestnykh, "Buncher cavity for ERL," in *Proc. 2003 Particle Accelerator Conf.*, 2003, pp. 1198-1200.
[4.63] V. Veshcherevich, I. Bazarov, S. Belomestnykh *et al.*, "Input coupler for ERL injector cavities," in *Proc. 2003 Particle Accelerator Conf.*, 2003, pp. 1201-1203.
[4.64] V. Shemelin, S. Belomestnykh, R. L. Geng *et al.*, "Dipole-mode-free and kick-free 2-cell cavity for the SC ERL injector," in *Proc. 2003 Particle Accelerator Conf.*, 2003, pp. 2059-2061.

[4.65] V. Shemelin, P. Barnes, B. Gillet *et al.*, “Status of HOM load for the Cornell ERL injector,” in *Proc. 2006 European Particle Accelerator Conf.*, 2006, pp. 478-480.
[4.66] F. E. Hannon and C. Hernandez-Garcia, “Simulation and optimization of a 100 mA DC photo-injector,” in *Proc. 2006 European Particle Accelerator Conf.*, 2006, pp. 3550-3552.

# CHAPTER 5: PHOTOCATHODE THEORY

JOHN SMEDLEY
*Brookhaven National Laboratory*
*Upton, NY 11973*

TRIVENI RAO
*Brookhaven National Laboratory*
*Upton, NY 11973*

DIMITRE DIMITROV
*TechX Corporation*
*5621 Arapahoe Ave., Suite A*
*Boulder, CO 80303*

**Keywords**
Photocathode, Theory of Photoemission, Quantum Efficiency, Three-Step Model, Semiconductor Cathode, Metal Cathode, Diamond Amplifier, Photon Absorption, Electron Excitation, Electron Transport, Electron Emission, Density of States, Transverse Emittance, Schottky Effect, Surface Roughness, Field Enhancement

**Abstract**
The cathode material used in a photoinjector is chosen to deliver the electron beam parameters for the desired application. Since these parameters at the cathode are determined primarily by the cathode material and the laser, *a priori* calculation of the quantum efficiency (QE) and the intrinsic emittance (IE) at the cathode can be used to compare the theoretical prediction to the observed performance and to improve it. In this chapter, we derive, based on the three step model, the expressions for the quantum efficiency and the intrinsic emittance of the electron beam from the cathode held in high accelerating field. We describe the specific modifications required for calculating these values for the metal, semiconductor and negative electron affinity cathodes. We discuss briefly the impact of surface roughness, impurity and crystal orientation to both the QE and IE. We also describe the theoretical formalism for calculating the electron yield secondary emitters such as diamond, which shows promise for delivering ampere level currents.

One of the primary components in designing photoinjectors is the photocathode to be used for the injector. Several factors dictate this choice: The quantum efficiency of the cathode at the laser's wavelength, the lifetime of the cathode, the total charge deliverable from the cathode in this life time, and its compatibility with the injector. This chapter is devoted to developing the theoretical underpinning of the electron yield from a material and its relevance to the injectors.

## 5.1 HISTORY

The first publication on photoemission from metals was a study of the photoelectric effect in 1887 [5.1]. Einstein offered the first theory describing the major aspects of this process in 1905 [5.2], work that was mentioned in the citation for his 1921 Nobel Prize. His theory demonstrated the quantization of the energy of light and proposed the existence of a material-dependent escape energy (today called the work function) to explain why UV radiation caused electron emission, but visible light did not. In the century since this groundbreaking work, many advances have been made in studying photoemission. In particular, the advent of lasers allowed the exploration of processes unattainable with less intense light sources, such as multiphoton emission [5.3]. Models of photoemission now describe not only the expected electron yield from a given metal and light source, but also provide predictions of the energy- and angular-distributions [5.4] of the emitted electrons.

The theory of photoemission relevant to the present work is named the "three-step model" [5.5]–[5.7]. Bergland & Spicer used it to explain emission from several materials [5.8], [5.9]. This model breaks down the process into three independent steps. The first step is the absorption of a photon and the resulting excitation of an electron. The second step is transportation of the excited (hot) electron to the surface; it takes into account the collisions that the electron may undergo during transit. The third step is the escape of the electron from the material after it reaches the surface. Since the three steps are considered to be independent, the probability of a single incident photon generating an emitted electron, defined as the quantum efficiency (QE), is the product of the probability of each of these steps. Although exact calculation of the QE is feasible, it is extremely tedious. Hence, we derive the equation for the QE, making some basic assumptions, and then discuss their validity for different types of cathodes in subsequent sections.

**Quantum Efficiency**

For these discussions, QE typically is defined as the ratio of the number of emitted electrons ($n_e$) to the number of incident photons ($n_p$). We note that this method differs slightly from other treatments [5.8] wherein efficiency is calculated with respect to the absorbed laser energy. For a cathode used in an injector, the incident laser energy generally is known, but the material's reflectivity may change with the surface preparation technique. Therefore, in these applications, it is standard practice to define the QE with respect to the incident beam's energy.

$$\mathrm{QE} = \frac{n_e}{n_p} = \frac{h\upsilon[eV]}{E_{laser}[J]}\, q\ [\mathrm{C}] \tag{5.1}$$

where $h\nu$ is the photon energy, $q$ is the charge released by the cathode and $E_{laser}$ is the laser energy.

## 5.2 THREE-STEP MODEL

### 5.2.1 Step 1 – Photon Absorption and Electron Excitation

This step entails calculating the probability of a photon being absorbed to excite an electron to a higher energy state. Two assumptions are made in this step. First, that the states lying below the Fermi energy, $E_f$, are filled and the states above are empty. This is equivalent to treating the material as a conductor at zero temperature. The second is that every absorbed photon excites an electron within the material, with an excitation probability dependent on only three parameters: The photon energy, the number of electrons in the occupied states of the material able to absorb photons (*i.e.* states below $E_f$), and the number of available states into which the hot electron is excited. The latter assumption, called the random-*k* approximation, implies that no selection rules are needed based on the electron's momentum. This simplified version deals only with indirect transitions wherein only the conservation of energy between the photon and the electron is a necessary condition, not the momentum. For polycrystalline metal surfaces typically used as accelerator photocathodes, this is generally a good approximation.

In the first step of the model, two processes occur. The first is the reflection of the incident laser beam from the surface. The probability of transmission into the material, $T(v)$, and absorption in an infinitely thick cathode, $A(v)$, is calculated from the reflectivity

$$T(v) = A(v) = 1 - R(v) \tag{5.2}$$

The reflectivity of the various materials can be measured or obtained from the literature [5.10]. The depth over which the photons are absorbed is determined by the material's absorption coefficient at $hv$.

The second process is the probability of exciting an electron to a given energy within the material. For this treatment, only the kinematic (energy) restrictions are considered. No selection rule is imposed on the wave vector, $k$, and the momentum matrix elements are assumed constant. This is equivalent to supposing that the electron's transverse momentum is not conserved. This approximation makes the probability of excitation from an initial state of energy $E_0$ to a final state of energy $(E_0 + hv)$ solely a function of the material's electronic density of states (EDOS). For convenience, we take the zero of energy to be the lowest in the material's valence band. The EDOS provide the number of available electron states as a function of energy, $n(E)$. The probability of excitation to a final state $E = (E_0 + hv)$, from an initial state $E_0$ clearly is proportional to the number of initial states, $N(E_0)$, and the number of final states, $N(E)$

$$P(E,hv) \propto N(E)\, N(E_0) = N(E)\, N(E - hv) \tag{5.3}$$

To obtain the fractional number of electrons excited to an energy $E$, we must divide by the total number of possible transitions

$$\int_{E_f}^{E_f + h\upsilon} dE'\, N(E')\, N(E' - h\upsilon) \tag{5.4}$$

The lower limit of integration arises from the exclusion principle. There are no empty states below the Fermi level; thus, all excitations must be to states above this level. The upper limit comes from the conservation of energy, as the highest energy initial state has $E = E_f$. Thus we have

$$P(E,hv) = \frac{N(E)\, N(E - h\upsilon)}{\int_{E_f}^{E_f + h\upsilon} dE'\, N(E')\, N(E' - h\upsilon)} \tag{5.5}$$

Step one is largely the same for both metallic and amorphous semi-conductor cathodes. The EDOS will be distinct; in particular, the semiconductor will have a band of disallowed states between the highest energy-filled state and the lowest energy-unfilled state. In the special case of spin-polarized photocathodes, the EDOS for each spin state must be considered and a different QE is calculated for each spin orientation.

**Cathodes for Polarized Electrons**

Currently, the polarized electrons usually are obtained by photoemission from GaAs-based semiconductors [5.11], [5.12]. The primary differences in the theoretical calculation in this case are: a) In addition to energy, the angular and crystal momenta are also conserved during the absorption of photon; and b) the initial and final states are near the absorption edge to reduce electron-phonon (e-p) collisions and preserve the polarization. Electrons from a specific spin state in the valence band is resonantly excited with circularly polarized light to the final state in the conduction band allowed by the energy and momentum conservation rules. After polarized electrons are generated in the conduction band, they are transported to the emission surface of the semiconductor's photoemitter. This surface is treated to create a negative electron affinity which enhances the emission of electrons.

However, since GaAs is a direct band gap semiconductor and has degenerate heavy-hole- and light-hole-valence bands at the Γ point (the maximum of the valence band), the electrons' spin polarization is limited to 50%. To increase it, degeneracy is removed by using either a strained GaAs layer, or a strained superlattice structure. This splits the sub-bands of the heavy-holes and the light-holes, with the former possessing greater energy than the latter. Electrons are then excited from the heavy-hole band with circularly polarized photons at the absorption edge, as detailed in Chapter 8. The band structure and the EDOS for superlattices usually are calculated using a multi-band Kane model (*e.g.* [5.13]) that supports the computation of polarization spectra that can be compared [5.11] with spectra from polarized electron emission experiments.

### 5.2.2 Step 2 – Photon Absorption Length and Electron Transport

The purpose of the second step is to calculate the probability that an excited electron reaches the surface of the cathode while retaining sufficient energy to escape into vacuum. Unlike the first step, this one differs dramatically between metallic- and semiconductor-cathodes.

In both cases, the photon's absorption length, $\lambda_{ph}$, in the material determines the depth at which the hot electron is excited. In metallic cathodes, electron-electron (e-e) scattering limits the range of the hot electrons in the material. This is represented by an e-e scattering length, $\lambda_e$. Since the theory herein is being developed for "near threshold" emission, a single electron-electron scattering event is likely to remove sufficient energy from the excited electron, such that neither of the electrons involved will retain enough energy to escape from the material. Thus, electrons that undergo a scattering event in the metal are assumed to be lost and not emitted. The e-p scattering in the metal does not change the energy distribution significantly. The momentum transfer due to this scattering is not relevant, since we assume isotropic velocity distribution for the electrons in the initial state; only energy conservation, not momentum conservation, has been taken into account in the transition.

For semiconductor cathodes, if the photon energy used is less than double the band gap energy (Spicer's "magic window"), e-e scattering is forbidden. In this case, other scattering mechanisms dominate (principally e-p). Because these processes entail the interaction of the hot electron with the ion lattice of the semiconductor, they have a small energy transfer, but a large momentum transfer. Therefore, an electron potentially can undergo many scattering events while retaining the energy sufficient to escape. Each event essentially will randomize the electron's direction of travel, leading to the diffusion of carriers to the surface under a random walk. This process typically is treated with a Monte Carlo algorithm with a characteristic scattering length, though other closed-form approaches exist [5.14]. Each scattering event has an associated energy loss and the calculation continues until each excited electron has been emitted or has lost energy such that it cannot overcome the work function. For negative electron affinity materials, charge trapping must also be considered.

#### *5.2.2.1 Detailed Metallic Case*

During their journey to the surface, the electrons may undergo a scattering event with an electron, phonon, or an impurity. Only electron-electron interaction is considered here, as its typical scattering length [5.8] dominates in the energy region of interest, *viz.*, 4-6 eV above the Fermi level. We make several simplifying assumptions herein. We ignore the specifics of the interaction between the excited electron and the "valence" electrons and assume that the probability of the interaction is solely a function of the number of available initial and final states in the material EDOS (the available phase-space for the interaction). Specifically, the probability, $S$, of an excited electron of energy $E > E_f$ interacting with a valence electron of energy $E_0 < E_f$ and imparting energy $\Delta E$ is proportional to:

1. The number of electrons, $N(E_0)$, with energy $E_0$.
2. The number of empty states, $N(E_0 + \Delta E)$, with energy $E_0 + \Delta E$.
3. The number of empty states, $N(E - \Delta E)$ with energy $E - \Delta E$.

$$S(E,E_0,\Delta E) \propto N(E_0)\, N(E_0 + \Delta E)\, N(E - \Delta E) \tag{5.6}$$

To obtain the total probability of scattering for an electron of energy $E$ by an electron of energy $E_0$, we must integrate Equ. 5.6 over all possible energy transfers, $\Delta E$.

$$S(E,E_0) \propto \int_{E_f-E_0}^{E-E_f} d(\Delta E)\; N(E_0)\, N(E - \Delta E)\, N(E_0 + \Delta E) \tag{5.7}$$

The lower limit of integration fulfills the requirement that enough energy must be imparted to the target electron to move it into an unfilled energy state (a state with $E > E_f$). The upper limit requires the initial excited electron retaining enough energy to enter an unfilled state.

To obtain the total scattering probability of an excited electron, we integrate Equ. 5.7 over all possible "valence" electron energies with a lower limit of integration representing the kinematic limitation that $E + E_0 \geq 2E_f$, thereby yielding

$$S(E) \propto \int_{2E_f-E}^{E_f} dE_0 \int_{E_f-E_0}^{E-E_f} d(\Delta E)\; N(E_0)\, N(E - \Delta E)\, N(E_0 + \Delta E) \tag{5.8}$$

The lifetime of the excited state (with respect to e-e scattering), $\tau(E)$, is inversely proportional to this scattering probability [5.15]:

$$\tau(E) \propto \frac{1}{S(E)} \tag{5.9}$$

The scattering length is related to the lifetime by the velocity. Here, we assume that the electron velocity is proportional to the square root of its kinetic energy (as it would be for a nonrelativistic free electron of energy $E$). The kinetic energy is taken to be the energy above the bottom of the band occupied by the electron for a semiconductor and at the Fermi energy for a metal

$$\lambda_e(E) = v(E)\,\tau(E) = \frac{\lambda_0\sqrt{E-E_f}}{\int_{2E_f-E}^{E_f} dE_0 \int_{E_f-E_0}^{E-E_f} d(\Delta E) N(E_0) N(E-\Delta E) N(E_0+\Delta E)} \tag{5.10}$$

Here, $\lambda_0$ is a constant of proportionality that is chosen so that the e-e scattering length (the length over which the intensity of unscattered electrons is 1/*e* of the initial intensity) matches a known value of the electron's mean free path at a single energy for a given material.

The depth at which an electron was excited determines the mean distance that it must travel to reach the surface of the metal. $\lambda_{ph}$ in the material determines this distance. The absorption length is calculated from the complex index of refraction, $\mathcal{N} = n + ik$, of the material for each incident wavelength $\lambda$ [5.10]:

$$\lambda_{ph} = \frac{\lambda}{4\pi k} \tag{5.11}$$

The probability that an electron created at a depth $d$ will escape is $e^{-d/\lambda_e}$, and the probability per unit length that a photon is absorbed at depth $d$ is $\lambda_{ph}^{-1}\, e^{-d/\lambda_{ph}}$. Integrating the product of these probabilities over all possible values of $d$, we obtain the fraction of electrons that reach the surface without scattering, $T(E,\nu)$,

$$T(E,\nu) = \frac{\lambda_e(E)/\lambda_{ph}(\nu)}{1+\left(\lambda_e(E)/\lambda_{ph}(\nu)\right)} \tag{5.12}$$

#### 5.2.2.2 Detailed Semiconductor Case

The number of scattering processes involved complicates the derivation of an analytical expression for electron transport in semiconductor cathodes. As mentioned, the Monte Carlo method often is applied to investigate charge transport in semiconductors. Overall, the method consists of simulating in two main steps, the trajectories of charge particles in an applied field. In the first, the free flight of carriers is calculated by generally solving Maxwell's equations; in the second step, a scattering event is identified (*e.g.* with phonons) and executed with specific probability. For a cathode with minimal field penetration, due to relatively high conductivity (as is typical for GaAs, the alkali antimonides and tellurides), the solution to Maxwell's equations in the bulk leads to (effectively) field-free drift between scattering events. When the energies of conduction electrons and valence holes are sufficiently close to the bottom of the conduction band and the top of the valence band, respectively, only one conduction band and one valence band are used. This Monte Carlo approach to charge transport in semiconductors is reviewed in Chapter 5 of [5.16]. Depending on the energy of absorbed radiation and the energy gap in a semiconductor photocathode, electrons and holes of created electron-hole pairs could have sufficiently large enough energies to produce additional secondary pairs due to impact ionization scattering. This does not occur in standard photocathodes irradiated with near-threshold radiation, but occurs in special cases such as diamond amplified photocathode, described below separately. These free charge carriers then move towards the emission surface *via* diffusion, and also drift when an applied electric field is present. In such a case, the full-band structure approach is used to generally model charge transport.

### 5.2.3 Step 3 – Escape

Our objective in this section is to calculate the probability that an electron, which has reached the surface, will be moving in the correct direction with sufficient energy to overcome the potential barrier and escape from the cathode's surface. This third step in the model accounts for the direction of travel of the electron as it approaches the surface. To escape, the electron's direction of motion must lie within a cone determined by the electron's energy and the material's work function.

The escape criteria are written [5.4] as

$$\frac{\hbar^2 (k_\perp)^2}{2m} \geq E_T \tag{5.13}$$

where $E_T$ is the energy required to escape ($E_T = \phi$ for a metal). The definition of $k_\perp$ we use here differs slightly from others in the literature [5.4]: for us, it represents the component of the excited electron's momentum directed perpendicular to the surface of the cathode material resulting in an escape cone with an opening angle described by [5.17]

$$\cos(\theta) = \frac{k_{\perp min}}{\left|\vec{k}\right|} = \sqrt{\frac{E_T}{E - E_f}} \tag{5.14}$$

Here, $k_{\perp min}$ refers to the value of the perpendicular component of the electron momentum for which Equ. 5.13 is an equality representing the minimum perpendicular momentum required to overcome the work function. Only those electrons whose trajectory falls within the cone described by Equ. 5.14 will escape the surface. The excitation and scattering processes are assumed to produce electrons with an isotropic angular distribution. Thus, the fraction of electrons that escape is given by the fraction of the total solid angle subtended by this cone

$$D(E) = \frac{1}{4\pi}\int_0^{\theta} \sin(\theta')d\theta' \int_0^{2\pi} d\varphi = \frac{1}{2}[1 - \cos(\theta)] = \frac{1}{2}\left(1 - \sqrt{\frac{E_T}{E - E_f}}\right) \tag{5.15}$$

The above expression holds as long as $(E - E_f) > E_T$. If $E < E_T$, the electron does not have enough energy to escape the surface, so $D(E) = 0$.

### 5.2.4 Yield and QE in the Three-Step Model

To calculate the total electron yield from a material for known photon flux $I$, and photon energy, $h\nu$, we integrate the product of the probabilities calculated in each step:

$$Y(\nu) = I(\nu)\, A(\nu) \int_{E_T}^{h\upsilon + E_f} dE\, P(E)\, T(E,\nu)\, D(E) \tag{5.16}$$

Taking the ratio of the electron yield to the incident photon flux, we obtain the QE

$$\mathrm{QE}(\nu) = A(\nu) \int_{E_T}^{h\upsilon + E_f} dE\, P(E)\, T(E,\nu)\, D(E) \tag{5.17}$$

For electrons with energy very close to the threshold for emission ($E - E_T << E_T$), the escape cone described in this section represents the dominant factor in determining the energy dependence of the QE [5.4]. In this case, the probability of an electron with energy, $E$, leaving the metal is proportional only to $D(E)$. The total QE then is proportional to the integral of $D(E)$ over all possible electron energies.

$$\text{QE}(\nu) \propto \int_{\phi+E_f}^{h\upsilon+E_f} D(E)dE \tag{5.18}$$

This integration yields

$$\text{QE}(\nu) \propto (h\nu - \phi) + 2\phi - 2\phi\sqrt{1 + \frac{h\nu - \phi}{\phi}} \tag{5.19}$$

Expanding for $(h\nu - \phi) << \phi$, and keeping the terms to second order

$$\text{QE}(\nu) \propto (h\nu - \phi) + 2\phi - 2\phi\left[1 + \frac{h\nu - \phi}{2\phi} - \frac{1}{8}\left(\frac{h\nu - \phi}{2\phi}\right)^2\right] \tag{5.20}$$

Collecting the terms, we obtain

$$\text{QE}(\nu) \propto (h\nu - \phi)^2 \tag{5.21}$$

Thus, for small values of electron energy in excess of the threshold, we expect the QE to show quadratic dependence on the excess energy.

## 5.3 THE TRANSVERSE EMITTANCE IN THE THREE-STEP MODEL

The rms transverse emittance in terms of the transverse momentum (referred to as $k$ in material and $p$ in vacuum after emission) and position distributions of the electrons can be expressed as

$$\varepsilon_x = \sqrt{\langle x^2\rangle\langle x'^2\rangle - \langle xx'\rangle^2}; \text{ where } x' = \frac{\mathrm{d}x}{\mathrm{d}z} = \frac{p_x}{p_z} \tag{5.22}$$

If the electrons are accelerated rapidly from rest, the longitudinal momentum $p_z$ reaches relativistic values quickly and can be written at $\beta\gamma mc$; then

$$\varepsilon_x \equiv \frac{1}{\beta\gamma mc}\sqrt{\langle x^2\rangle\langle p_x^2\rangle - \langle xp_x\rangle^2} \tag{5.23}$$

If there is no correlation between the position and momentum of the electron, then the cross correlation term vanishes. The normalized emittance can then be written as

$$\varepsilon_n \equiv \beta\gamma\varepsilon_x = \frac{1}{mc}\sqrt{\langle x^2\rangle}\sqrt{\langle p_x^2\rangle} = \sigma_x\sigma_{p_x} \tag{5.24}$$

where $\sigma_x$ is the rms spot size and $\sigma_{p_x}$ is the dimensionless rms transverse momentum.

The intrinsic transverse momentum distribution of the electrons in the cathode gives rise to an emittance that is characteristic of the cathode material, the photon energy and the external field imposed on the cathode. This can be calculated by determining the variance of the transverse momentum of the electrons exiting the surface.

The product of the probability densities from each of the three steps (prior to energy integration) provides the number of electrons emitted at a given energy from the material. This distribution is called the energy distribution curve (*EDC*).

$$EDC(E,v) = A(v)\, P(E)\, T(E,v)\, D(E) \tag{5.25}$$

As described in Section 5.2.3 and Equ. 5.14, the maximum angle of escape inside the metal, $\theta_{max}$, is given by

$$\cos(\theta_{max}) = \frac{p_z}{p_{total}} = \sqrt{\frac{\phi}{E - E_f}} \tag{5.26}$$

$$p_x = \sqrt{2m(E - E_f)}\, \sin(\theta) \tag{5.27}$$

where $p_x$ is conserved upon emission. The variance of the transverse momentum can then be written as

$$\sigma_{p_x}^2 = \frac{\int_{E_T}^{h\upsilon+E_f}\int_0^{\theta_{max}} A(\upsilon)P(E)T(E,\upsilon)D(\cos(\theta))\frac{p_x^2}{(mc)^2}\, d(\cos(\theta))dE}{\int A(\upsilon)P(E)T(E,\upsilon)D(E)} \dots \tag{5.28}$$

$$\dots = \frac{1}{mc^2}\frac{\int_{E_T}^{h\upsilon+E_f}\int_0^{\theta_{max}} A(\upsilon)P(E)T(E,\upsilon)E\sin^2(\theta)d(\cos(\theta))dE}{\int A(\upsilon)P(E)T(E,\upsilon)D(E)} \qquad \text{(5.28 cont.)}$$

where $E$ is the energy of the electron and $\theta$ is the angle between the surface normal and its trajectory inside the cathode. $\theta_{max}$ is the angle of the escape cone containing the trajectories of all the electrons and is defined by Equ. 5.14. One should keep in mind that for a NEA surface, for which the vacuum level is below the bottom of the conduction band, $\theta_{max} \approx 90^\circ$ and almost all the electrons reaching the surface escape. The three-step model can be applied to a full energy-momentum band structure calculation, which yields an angle and energy distribution curve. Such a calculation is beyond the scope of this book.

Equ. 5.28 can be simplified following certain assumptions that are valid for a number of cathode materials. If the photon energy is close to the effective work function and the electron density of states near the Fermi energy can be considered a constant, then the above equation can be analytically integrated to result in

$$\sigma_{p_x} = \sqrt{\frac{h\nu - \phi_{eff}}{3mc^2}} \tag{5.29}$$

$$\varepsilon_n = \sigma_x \sqrt{\frac{h\nu - \phi_{eff}}{3mc^2}} \tag{5.30}$$

Experimental results agree with Equ. 5.30, as shown in Chapter 6 and Chapter 7.

## 5.4 MODIFICATIONS TO THE THREE-STEP MODEL

The previous section addressed the process of photoemission from materials with known properties, *i.e.* EDOS and work function. This section discusses the needed modifications to this model to understand the behavior of the QE of a cathode in a photoinjector. Two major physical characteristics differentiate an electron gun from the "ideal" treatment presented above: The roughness of cathode surface on some scale, which would impact both the charge and the velocity distribution of the electrons and the large electric field (sometimes greater than 100 MV $m^{-1}$) accelerating the electrons within the gun. In this section, we consider the reduction of the material work function caused by applying an external field (Schottky effect) [5.18], [5.19].

### 5.4.1 Schottky Effect

This section determines the change in the work function caused by an external electric field. This derivation uses the classical electrodynamic model of work function: the required amount of work to separate a charge from its image within the metal. For our calculations, we treat the cathode surface as a perfect conductor.

In classical electromagnetism, the work function is viewed as arising from the attraction of an electron to its image charge within the metal. In this definition, the work function is the amount of energy required to move an electron from some minimum separation (minimum is undefined in classical theory) to a point at infinity [5.19]. In the absence of an applied field, the potential energy between an electron at a distance from the surface of a perfect conductor, $x$, and its image charge is

$$W_{e\text{-}image}(x) = e\Phi_{image}(x) = e\left(\frac{-e}{2x}\right) \tag{5.31}$$

To avoid unnecessary constants and remain true to the reference material, this derivation is performed in CGS units. However, the result is stated in MKSA units.

By symmetry, the potential energy between the electron and the surface is half of this value:

$$W_{surface}(x) = \left(\frac{-e^2}{4x}\right) \tag{5.32}$$

The work function then is

$$\phi_0 = W_{surface}(\infty) - W_{surface}(x_{min}) = \left(\frac{e^2}{4x_{min}}\right) \tag{5.33}$$

Now, let us add the effect of an applied field at the surface. The potential corresponding to a constant electric field normal to the surface is

$$W_{field}(x) = e\Phi_{field}(x) = e(-Ex) \tag{5.34}$$

Here, $E$ is the magnitude of the electric field that is normal to the surface of the conductor. The total potential energy then is

$$W_{total}(x) = \left(\frac{-e^2}{4x}\right) - eEx \tag{5.35}$$

This combined potential has a maximum below vacuum level, which occurs at $x_0$, such that

$$\frac{\mathrm{d}}{\mathrm{d}x}W_{total}(x)|_{x=x_0} = \frac{e^2}{4x_0^2} - eE = 0 \tag{5.36}$$

$$x_0 = \frac{1}{2}\sqrt{\frac{e}{E}} \tag{5.37}$$

$$W_{total}(x_0) = -e\sqrt{eE} \tag{5.38}$$

This maximum represents a change in the work function due to the applied field

$$\phi = W_{total}(x_0) - W_{total}(x_{min}) = \frac{e^2}{4x_{min}} - e\sqrt{eE} \tag{5.39}$$

Here, we assume that $x_{min}$ is so close to the surface that $W_{field}(x_{min}) = 0$. This change in the work function is called the Schottky effect. In MKSA units

$$\Delta\phi_{Schottky}\ [\mathrm{eV}] = \alpha\sqrt{E\left[\frac{\mathrm{V}}{\mathrm{m}}\right]} \tag{5.40}$$

$$\alpha = e\sqrt{\frac{e}{4\pi\varepsilon_0}} = 3.7947\times10^{-5}\ \left[e\sqrt{\mathrm{Vm}}\right]$$

The modified work function is then

$$\phi = \phi_0 - \alpha\sqrt{E} \tag{5.41}$$

In the presence of an external applied field that modifies the work function from $\phi_0$ to $\phi$, Equ. 5.20 can be written as

$$\mathrm{QE}(v) \propto (hv - \phi_0 + \alpha\sqrt{E})^2 \tag{5.42}$$

$$\sqrt{\mathrm{QE}(v)} = A(hv - \phi_0 + \alpha\sqrt{E}) \tag{5.43}$$

where $A$ is a constant of proportionality. This expression predicts a linear relationship between $\sqrt{\mathrm{QE}}$ and $\sqrt{E}$; this prediction will be important in analyzing the experimental data in the following chapters.

### 5.4.2 Effect of Surface Roughness

The preceding analysis assumes that the field at the surface of the conductor is uniform and equal to the field in the center of the space between the electrodes, as is true for flat, idealized surfaces. Studies of field emission [5.21], [5.22] demonstrated that micro-protrusions and surface contaminants can effectively enhance the field at the emitting surface. In analyzing field emission curves (charge versus field), it is customary to introduce a factor $\beta_{FE}$ (the field enhancement factor) that multiplies the applied field to obtain the surface field; enhancement factors of 20-100 for field emission are common in the literature [5.23]. We note that some authors [5.22], [5.24] attributed the large "enhancement" to factors other than the emitter's geometry, such as non-metallic inclusions in cracks or grain boundaries. These enhancements generally do not impact the calculation of quantum efficiency significantly for a planar geometry. This is due to Gauss's law – unlike field emission, photoemission occurs over the entire illuminated surface. Surface non-uniformity leads to field enhancement in some areas and field reductions in others. The net effect on the calculation of quantum efficiency is negligible; indeed when a field enhancement factor is introduced into $Q$ versus field plots for photoemission, $\beta$ is invariably found to be near one.

An exception to this occurs when a cathode is illuminated with a photon energy below (but close to) the work function. In this case, the Schottky effect is required for any emission at all and the cathode will emit only from areas where the field is sufficiently enhanced to reduce the local work function below the photon energy. In this case the field enhancement factor will be a function of the photon energy used. Another exception occurs in the case of needle cathodes, where the field is enhanced over the entire needle tip and only the tip is illuminated. In these cases, a field enhancement factor for photoemission, $\beta_{PE}$, is sometimes introduced in front of $E$ in Equ. 5.41 – however, it should be noted that this is an approximation, as the field enhancement strictly applies only to a microscopic treatment of the surface, and therefore the enhancement should be $\beta(x,y)$, where $\beta$ is allowed to freely vary from point to point along the surface.

More important than the impact on the QE is the impact of surface roughness on the intrinsic emittance. The modification of the applied field at the microscopic photoemission sites has a number of implications to the intrinsic emittance of the electrons from the cathode: initial energy, transverse velocity and charge density of the emitted electrons become spatially dependent.

### 5.4.3 Effect of Field Enhancement on Emittance

On a macroscopic scale, the applied field is normal to the cathode surface. However, within a few nanometers of the surface, the field lines are normal, not to the macroscopic surface, but the microscopic contour of the surface. As described earlier, this results in a spatially dependent change in the work function and a corresponding spatially dependent charge density. This is shown schematically in Figure **5.1**. This change in the work function also gives rise to a spatially dependent excess energy as the electrons exit the cathode. In addition, once emitted, these low energy electrons follow the local field lines, leading to a spatially dependent velocity distribution transverse to the beam axis.

As can be seen from Equ. 5.28 and Equ. 5.43, the normalized emittance varies as the effective work function and the field distribution change from the peak to the valley of the rough cathode surface. One should keep in mind the $\theta$ occurring in Equ 5.28 refers now to the microscopic surface. The net effect of this variation is to introduce a dependence of the intrinsic emittance on the amplitude ($n$) and spatial period ($d$) of the roughness, and on the electric field. This effect can be characterized by a "roughness" parameter $\xi = 2\pi n\, d^{-1}$; a parameter of unity results in a significant (~30%) increase in the expected intrinsic emittance for reasonable fields (24 MV m$^{-1}$). [5.25] [5.26]

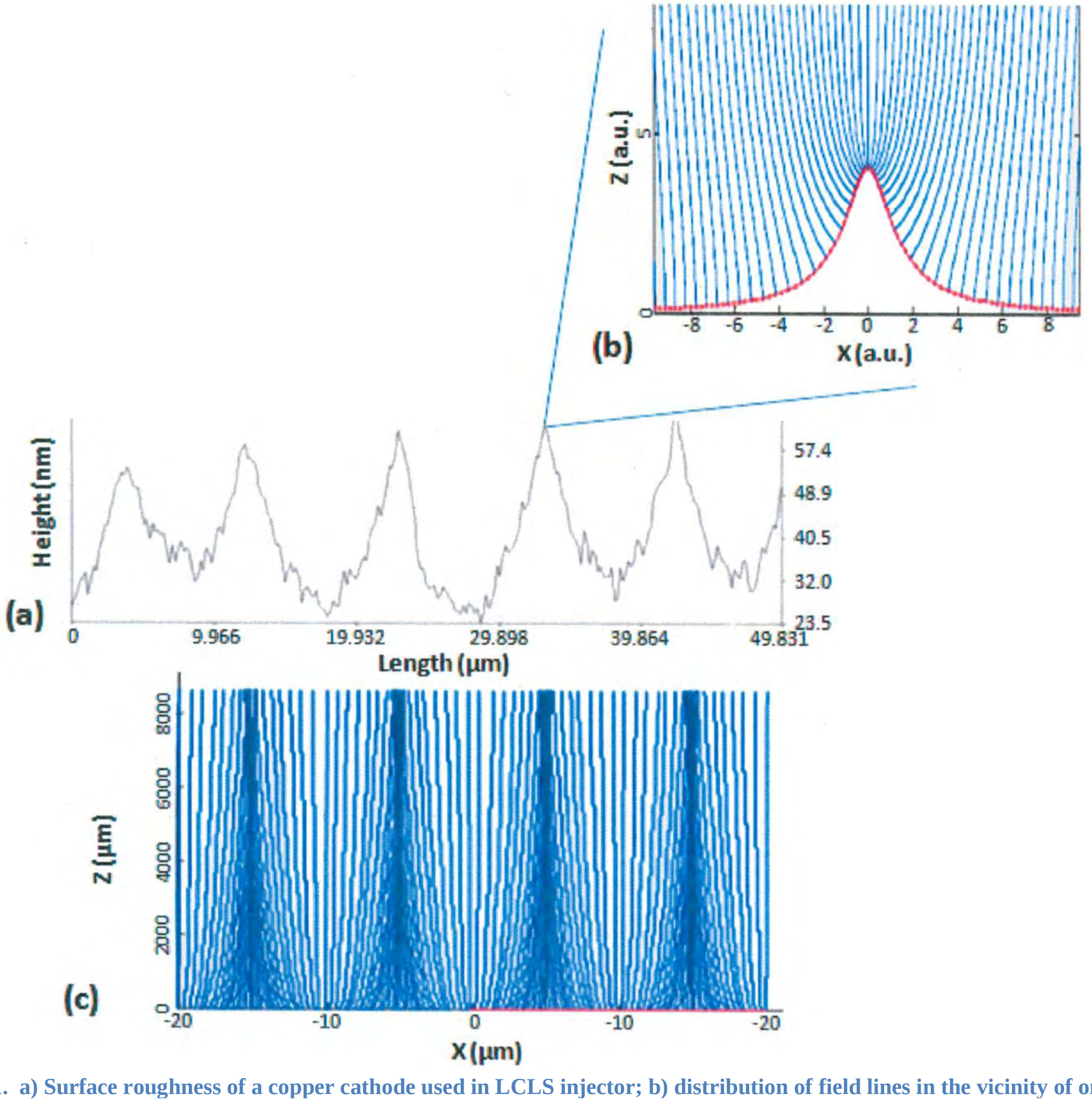

**Figure 5.1. a) Surface roughness of a copper cathode used in LCLS injector; b) distribution of field lines in the vicinity of one of the protrusions; c) distribution of the photoemitted charge in the vicinity and a few mm away from the cathode, indicating position dependent transverse velocity dependence in the vicinity of the cathode. [5.27]**

In addition to the microscopic protrusions, the roughness of the cathode could be caused by other effects such as random orientation of the crystalline cathode or impurities embedded in or on the cathode: all of these effects lead to non-uniform charge density at or near the cathode surface.

## 5.5 DIAMOND AMPLIFIER THEORY

A new concept (a diamond amplifier photocathode) to emit high average current, high brightness electron beams was proposed recently, and is currently being developed [5.28]. The underpinning of its operation is to first generate a primary beam of electrons (accelerated to ~10 keV) using a conventional photocathode and inject them into a diamond, with its entrance face coated with a metal. The energetic primary electrons scatter inelastically in the diamond, generating (through impact ionization) a number of secondary electrons and holes. These secondary electrons and holes initially relax their energies by producing more electron-hole pairs *via* impact ionization. When the energy of these free charge carriers closely approaches the energy gap value of diamond, scattering by phonons thereafter dominates the further relaxation of their energy. These free charge carriers then move towards the emission surface *via* diffusion and also drift when an applied electric field is present. In such a case, the full-band structure approach is used to generally model charge transport. Algorithms for full-band structure charge transport calculations are detailed in [5.16].

Investigating these phenomena requires modeling of the generation of secondary electrons, charge transport and electron emission.

Secondary electron generation (impact ionization) is treated using optical models (*e.g.* [5.29] and references therein). Where it is allowed, impact ionization continues until the energies of the free (conduction band) electrons and holes in the valence band fall below the energy gap. The scattering rate (or the mean-free-path) for impact ionization in semiconductors [5.29] of interest is such that the impact ionization is completed within several hundred femtoseconds. Modeling the impact ionization is important to understand when the free charge carriers are created (with the process initiated by primary electrons with energies of a few kiloelectron volts) near the metalized side of the amplifier [5.28] and then transported to its emission surface. Electrons recombine at the metal contact due to diffusive expansion, thereby reducing the total number of free electrons that reach the emission surface.

After the impact ionization phase is completed, (or electron-hole pairs created with energies below the energy gap,) the charge carriers drift under an applied electric field and scatter predominantly with lattice excitations (phonons). Scattering with charge impurities and trapping effects must also be accounted for when the densities of charge impurities and trapping centers are not negligible [5.30]. Trapping of charge carriers also leads to the modification of the effective field in diamond, which in turn, affects the transport of free electrons and holes. Both impact ionization and charge transport-related scattering processes are treated *via* Monte Carlo methods.

The emission of electrons from diamond requires modeling surface phenomena that include calculating the probability for emission, band bending, electron affinity, surface roughness and charge-trapping effects.

## 5.6 CONCLUSION

In this chapter, we have described the theoretical underpinning of the photoemission process, namely the three-step model, and developed the equations relevant to calculate the quantum efficiency of the cathode in a photoinjector. These can be used to evaluate the expected performance of the cathode in the injector, its optimization and even engineer the appropriate cathode using modern material science techniques. We also describe the modification to the model due to Schottky effect when operating the cathodes in the presence of high accelerating fields and its effect on the work function and quantum efficiency. We have derived the intrinsic emittance of the electrons arising from the transverse momentum distribution at birth, its relationship to the quantum efficiency and the impact of surface roughness on all these parameters.

## 5.7 CONFLICT OF INTEREST AND ACKNOWLEDGEMENT

We confirm that this article content has no conflicts of interest and would like to acknowledge the supported of US department of Energy under contract numbers DOE DE-SC0003965, DE-AC02-98CH10886, KC0407-ALSJNT-I0013 and DE-FG02-12ER41837.

***References***

[5.1] H. Hertz, "Ueber einen einfluss des ultravioletten lichtes auf die electrische entladung," *Annalen Physik*, vol. 267, pp. 983-1000, 1887.

[5.2] A. Einstein, "Über einen die erzeugung und verwandlung des lichtes betreffenden heuristischen gesichtspunkt," *Annalen Physick*, vol. 17, pp. 132-148, 1905.

[5.3] G. Farkas, A. Kohazi-Kis and C. Toth, "Above-threshold multiphoton photoelectric effect of a gold surface," *Optical Eng.*, vol. 32, pp. 2476-2480, October 1993.

[5.4] M. Cardona and L. Ley, Eds., *Photoemission in Solids: General Principles*, vol. 1, Berlin: Springer-Verlag, 1978, pp. 22-2, 237-264.

[5.5] H. Y. Fan, "Theory of photoelectric emissions from metals," *Phys. Rev.*, vol. 68, pp. 43-52, July 1945.

[5.6] H. Mayer and H. Thomas, "Zum äußeren lichtelektrischen effekt der alkalimetalle. II. Die spektrale verteilung der quantenausbeute bei kalium," *Zeitschrift Physik A*, vol. 147, pp. 419-441, August 1957.

[5.7] H. Puff, "Zur theorie der photoelektronenemission von metallen. II. Die verteilungen der photoelektronen," *Physica Status Solidi (b)*, vol. 1, pp. 704-715, 1961.

[5.8] W. F. Krolikowski, Ph.D. thesis, Stanford University, Stanford, CA, USA, 1967.

[5.9] C. N. Berglund and W. E. Spicer, "Photoemission studies of copper and silver: theory," *Phys. Rev. A*, vol. 136, pp. A1030-A1044.

[5.10] *CRC Handbook of Chemistry and Physics*, $65^{th}$ Ed., CRC Press., Boca Roton, Fl, 1984, pp. E-365.

[5.11] A. V. Subashiev, L. G. Gerchikov and A. N. Ipatov, "Optical spin orientation in strained superlattices," *J. Appl. Phys.*, vol. 96, pp. 1511-1520, August 2004.

[5.12] T. Nishitani, T. Nakanishi, M. Yamamoto *et al.*, "Highly polarized electrons from GaAs-GaAsP and InGaAs-AlGaAs strained-layer superlattice photocathodes," *J. Appl. Phys.*, vol. 97, pp. 094907-1–094907-6, April 2005.

[5.13] L.C.L.Y. Voonand and M. Willatzen, *The k·p Method: Electronic Properties of Semiconductors*, Springer-Verlag, 2009.

[5.14] K. L. Jensen, B. L. Jensen, E. J. Montgomery *et al.*, "Theory of photoemission from cesium antimonide using an alpha-semiconductor model," *J. Appl. Physics*, vol. 104, pp. 044907-1–044907-10, August 2008.

[5.15] D. Griffiths, *Introduction to Elementary Particles*, John Wiley & Sons, 1987, pp. 194-195.

[5.16] C. Jungemann and B. Meinerzhagen, *Hierarchical Device Simulation, The Monte-Carlo Perspective*, Chapter 5, New York: Springer-Verlag, 2003.

[5.17] N. V. Smith, "Photoemission properties of metals," *Crit. Rev. Solid State Material Sci.*, vol. 2, pp. 45-83, 1971.

[5.18] W. Schottky, "Über den austritt von elektronen aus glühdrähten bei verzögernden potentialen," *Annalen Physik*, vol. 44, pp. 1011-1032, 1914.

[5.19] C. Herring and M. H. Nichols, "Thermionic emission," *Rev. Modern Physics*, vol. 21, pp. 185-271.

[5.20] J. D. Jackson, *Classical Electrodynamics*, New York: John Wiley & Sons, 1962, pp. 45, 60.

[5.21] R. H. Fowler and L. Nordheim, "Electron emission in intense electric fields," in *Proc. Royal Society A*, vol. 119, May 1928, pp. 173-181.
[5.22] G. A. Mesyats and D. I. Proskurovsky, *Pulsed Electrical Discharge in Vacuum*, Heidelberg: Springer-Verlag, 1989.
[5.23] P. A. Chatterton, "A theoretical study of field emission initiated vacuum breakdown," in *Proc. Physical Society*, vol. 88, 1966, pp. 231-245.
[5.24] R. V. Latham, "The origin of prebreakdown electron emission from vacuum-iusulated high voltage electrodes," *Vacuum*, vol. 32, 1982, pp. 137-140.
[5.25] M. Krasilnikov, "Impact of the cathode roughness on the emittance of an electron beam," in *Proc. 2006 Free Electron Laser*, 2006, pp. 583-586.
[5.26] D. Xiang, W.-H. Huang, Y.-C. Du *et al.*, "First principle measurements of thermal emittance for copper and magnesium," in *Proc. 2007 Particle Accelerator Conf.*, 2007, pp. 1049-1051.
[5.27] D. H. Dowell, private communication.
[5.28] X. Chang, Q. Wu, I. Ben-Zvi *et al.*, "Electron beam emission from a diamond-amplifier cathode," *Phys. Rev. Lett.*, vol. 1054, pp. 164801-1–164801-4, October 2010.
[5.29] B. Ziaja, R. A. London and J. Hajdu, "Ionization by impact electrons in solids: Electron mean free path fitted over a wide energy range," *J. Appl. Phys.*, vol. 99, pp. 033514-1–033514-9, February 2006.
[5.30] M. Lundstrom, *Fundamentals of Carrier Transport*, 2nd ed., Cambridge: Cambridge University Press, 2000.

# CHAPTER 6: METAL CATHODES

TRIVENI RAO

*Brookhaven National Laboratory*

*Upton, NY 11973*

JOHN SMEDLEY

*Brookhaven National Laboratory*

*Upton, NY 11973*

**Keywords**

Photocathode, Quantum Efficiency, Metal Cathode, Copper Cathode, Magnesium Cathode, Lead Cathode, Niobium Cathode, Three-Step Model, Theoretical Calculation of Quantum Efficiency, Experimental Measurement of Quantum Efficiency, Intrinsic Emittance, Laser cleaning, Hydrogen Cleaning, Cathode Polishing, Arc Deposition, Friction Welding, Cathode Plug, Choke Joint

**Abstract**

A number of low to medium current, high brightness injectors use metal photocathodes for electron emission. The theory is developed in Chapter 5 and is applied in this chapter to calculate the QE of copper and lead, as examples. It has been shown experimentally that the QE of the as received cathode material can be improved significantly by appropriate processing. We describe in detail different cleaning techniques adopted in operational facilities, such as laser cleaning and ion cleaning. Incorporating the cathode in the injector is a non-trivial problem, especially if it is a dissimilar metal placed in a high field environment. We describe a number of techniques used including in-*situ* application, embedding the cathode in a plug, a flange, and a choke joint assembly and provide examples for each.

## 6.1 INTRODUCTION

Metal cathodes often are the default choice for low average current photoinjector applications, as they are easy to prepare and have a long operational lifetime with minimal vacuum requirements. They are prompt emitters, with an emission time of tens of femtoseconds; their high work functions lead to low dark currents even in high electric fields. Over the past two decades, copper and magnesium cathodes have been used [6.1]–[6.4] in high-brightness, low average current, low repetition rate normal conducting RF (NCRF) injectors. Table **6.1** lists some performance parameters using these photocathodes. [6.5]

Much of the recent interest [6.6]–[6.8] has centered on increasing both the repetition rate and the average current for several applications. Superconducting RF (SCRF) injectors were proposed for them so they may maintain a relatively high accelerating gradient and very low resistive loss of the RF power. Niobium, a natural choice for cavity material, has been tested [6.9]. The quantum efficiency (QE) measured [6.10] on Nb witness samples indicate average currents around 50 μA are possible with today's commercial lasers. Another metal cathode candidate is lead, a Type I superconductor, commonly used in ion accelerators. QE measurements with lead witness samples indicate that its QE can be as high as 0.5% for 6.4 eV photons. Hence, for attaining a higher average current, a lead cathode was tested both in a DC set-up with witness pieces [6.11] and in a 1.3 GHz SCRF gun [6.12].

One major drawback of the metal cathodes is their high work function, typically in the 4-5 eV regime that requires the operating wavelength of the driving laser to be in the UV range. Even at this large photon energy, the QE is only a fraction of a percent. The combination of these two factors and the power levels achievable with existing commercial lasers restrict their use primarily to intermediate average current applications. Furthermore, as explained in Chapter 5, of all the photo-excited electrons, only those that did not undergo inelastic electron-electron collision have enough energy to overcome the work function, and are emitted. Thus, assuming an isotropic velocity distribution for the electrons (for example, a polycrystalline cathode with no preferential emission directions), the maximum energy of the emitted electrons is ~$(h\nu - \phi)$, where $h\nu$ is the photon energy, and $\phi$ is the work function. As the QE scales roughly as $(h\nu - \phi)^2$ and intrinsic emittance as $(h\nu - \phi)^{1/2}$, there is a tradeoff between high QE and low emittance for these cathodes.

| Facility | Cathode Material | Preparation | Laser Wavele-ngth | Charge | Pulse Duration | Repetitio n Rate | Lifetime |
|---|---|---|---|---|---|---|---|
| LCLS | Copper | $H_2$ Ion Bombardment\ Laser Ablation | 253, 255 nm | Up to 1 nC | 0.7-3.7 ps | 120 Hz | 2 yr |
| BNL-ATF | Copper | Laser Ablation | 256 nm | Up to 1 nC | 7 ps | Up to 6 Hz | > 5 yrs |
| BNL-LEAF | Magnesium | Laser Ablation | 266 nm | 8 nC | 5 ps | 10 Hz | 18 months |
| UCLA | Copper with $MgF_2$ Coating | Laser Ablation | 266, 800 nm | 10 pc | < 1 ps | 5 Hz | > 1 yr |
| INFN: FERMI | Copper | Ozone Cleaning | 262 nm | 250 pC | 6 pc | 10 Hz | 3 months |

**Table 6.1. Facilities using metal photocathodes and their performance [6.5].**

In this chapter, we discuss the experimental measurements and theoretical calculations of the QE, its enhancement with surface treatment, the impact of the surface treatment on the emittance, and the incorporation of the cathodes in RF cavities.

## 6.2 EXPERIMENTAL MEASUREMENTS AND THEORETICAL CALCULATIONS OF THE QE

The QE of several metals have been measured in laboratory settings [6.13] and the QE of Mg, Cu, Nb, and Pb was measured in RF guns/injectors; the values are listed in Table **6.2**. We will describe here the theoretical calculations and the QE measurements for Pb and Cu.

| Metal @ Electric Field | Wavelength [nm] | QE [%] |
|---|---|---|
| Copper @ 100 MV m$^{-1}$ | 266 | 0.014 |
| Magnesium @ 100 MV m$^{-1}$ | 266 | 0.62 |
| Niobium @ 2 MV m$^{-1}$ | 266 | ~0.001 |
| Lead @ 2 MV m$^{-1}$ | 248 | 0.016 |

**Table 6.2. Quantum efficiency of selected metals measured in these respective electric fields in RF guns.**

As described in Chapter 5, three independent steps, developed by Spicer, describe the photoemission process to the first order [6.14]. The first step is the transmission of an incident photon of energy, $h\nu$, into the material, then the absorption of this photon by an electron and its excitation from an energy state ($E - hv$) to an energy state $E$. The second step is the transit of this electron to the surface; and the third is its escape from the cathode's surface. Explicit expressions for the probability of each of the steps were derived in Chapter 5. Here, we describe the methodology for calculating the QE of lead from these expressions. For this calculation, we made the following assumptions:

- The sample is sufficiently thick to absorb all the transmitted photons.
- The material is polycrystalline non-crystalline, and hence, only energy conservation is imposed in the excitation process.
- The excitation process is isotropic.
- Electron-phonon (e-p) collision is nearly elastic, and so can be ignored.
- Electron-electron scattering will result in both the electrons involved having insufficient energy to escape the material (so they are effectively lost).

### 6.2.1 Step 1

Two factors dictate the probability of the absorption of a photon of energy, $hv$: The transmission coefficient of the metal at $hv$, and the number of accessible initial- and final-states. The transmission coefficient is calculated from the optical constants of Pb. The accessible states are the product of the density of states (DOS) and the Fermi-Dirac distribution. To the first approximation, we can use the Fermi-Dirac distribution function at zero temperature, with all the states below Fermi energy filled, and all those above it empty. The simplified form of this probability is

$$P(E,hv) = (1 - R(v)) \frac{N(E)N(E-h\upsilon)}{\int_{E_f}^{E_f+h\upsilon} dE'\, N(E')N(E'-h\upsilon)} \tag{6.1}$$

where $R(v)$ is the reflectivity of the cathode at $hv$. This probability calculated for lead is shown in Figure **6.1**, and databases [6.15] such as the Naval Research Laboratory's (NRL) structures database are available to reference the DOS for a large variety of elements and compounds. The optical constants and work function are from [6.16], [6.17].

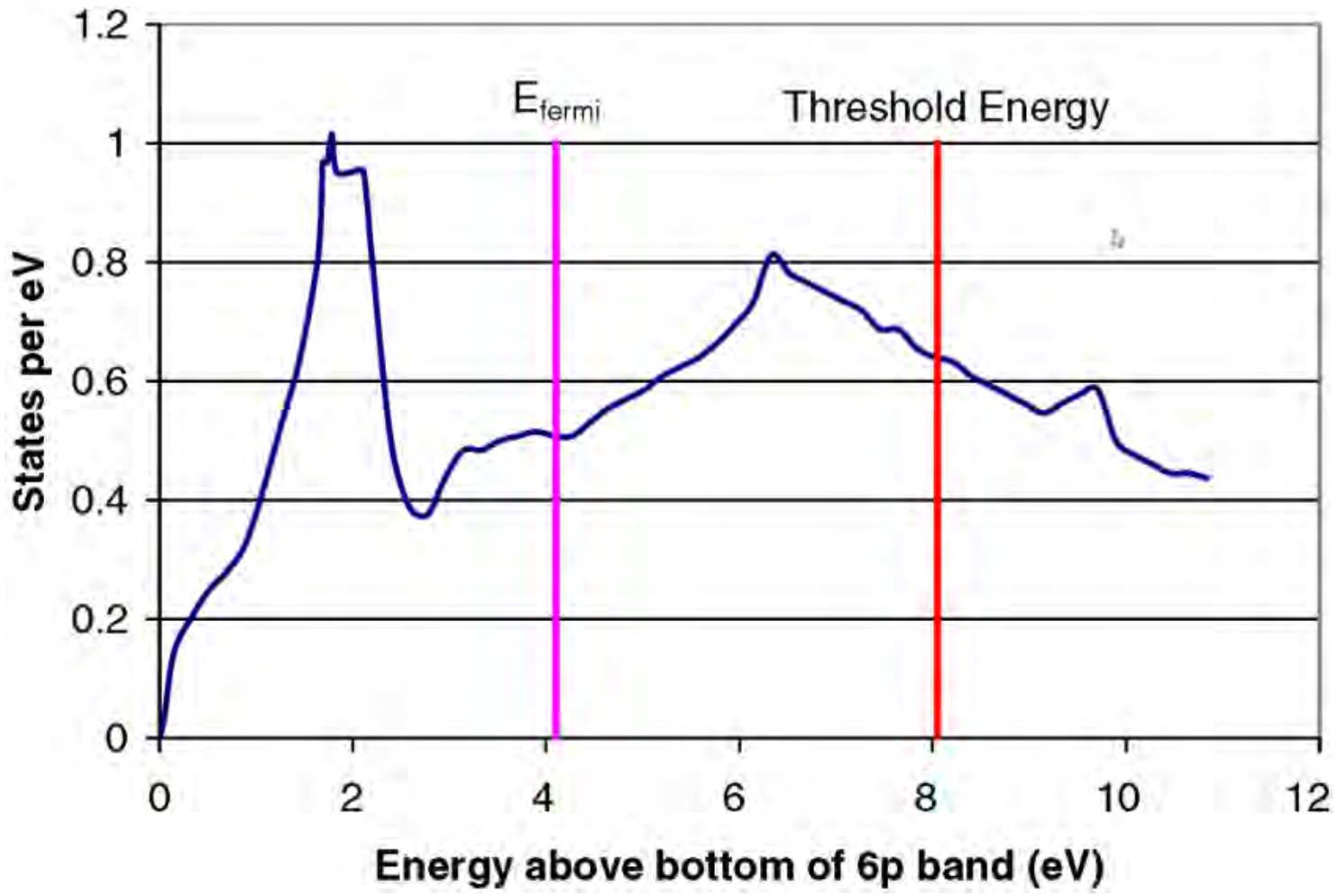

**Figure 6.1. Calculated Density of Electronic States for lead, with $E_{fermi}$ referenced to the bottom of the filled 6p band where the vertical axis is the number of states per electron volt within the material. [Reprinted figure with permission from [6.11]. Copyright 2008 by the American Physical Society]**

### 6.2.2 Step 2

The electrons with energy $E$ are generated in lead at depths dictated by the photon's absorption length. During their transit from the point of excitation to the surface, the electrons undergo e-e and e-p scattering; the latter primarily changes the electrons' trajectory. Since we assumed that the excitation was isotropic, we can neglect the impact of e-p collision. For the photon energies under consideration here, hot electrons undergoing inelastic e-e scattering will not be sufficiently energetic to overcome the work function. Hence, the fraction of the excited electrons that reach the surface is

$$F_{e\text{-}e}(E,h\nu) = C\,\frac{\lambda_e(E)/\lambda_{ph}(\nu)}{1 + \left(\lambda_e(E)/\lambda_{ph}(\nu)\right)} \tag{6.2}$$

where $\lambda_{opt} = \lambda\,(4\pi k)^{-1}$, $k$ is the imaginary part of the complex index of refraction, and $\lambda_{e\text{-}e}$ is the electron mean free path for e-e scattering. $C$ is the form factor that takes into account that the electron's trajectory may not be normal to the surface. For most of the near threshold cases, the escape cone is narrow enough that $C$ can be taken to be unity. The absorption depth is calculated using optical constants. The lifetime of the electrons is derived from first principles, assuming a free electron DOS and its mean free path are given by the product of the lifetime by the electron velocity.

### 6.2.3 Step 3

To escape the surface, the momentum of the electron normal to the surface, $k_\perp$, must satisfy the condition

$$\frac{\hbar k_\perp^2}{2m} > E_T \tag{6.3}$$

wherein $E_T = E_f + \phi$. In the presence of an accelerating field $E_{acc}$, the threshold energy, $E_{T,}$ is reduced by $\phi_{Schottky}$ [eV] $= 3.7947\times10^{-5}\sqrt{E_{acc}\,[\mathrm{V\ m^{-1}}]}$ due to the Schottky effect. The probability that an excited electron of energy $E$ escapes the surface is given by

$$D(E) = \frac{1}{2}\left(1 - \sqrt{\frac{E_T}{E - E_f}}\right) \tag{6.4}$$

The quantum efficiency is

$$\mathrm{QE}(h\nu) = \int_{E_f+\phi_{eff}}^{E_f+h\upsilon} dE\;\; P(E,h\nu)\, F_{e\text{-}e}(E,h\nu)\, D(E) \tag{6.5}$$

where $D(E)$ is the fraction of the electrons with sufficient longitudinal momentum to overcome the work function. Figure **6.2** shows the calculated quantum efficiency of lead as a function of photon energy using 3.95 eV for $\phi$.

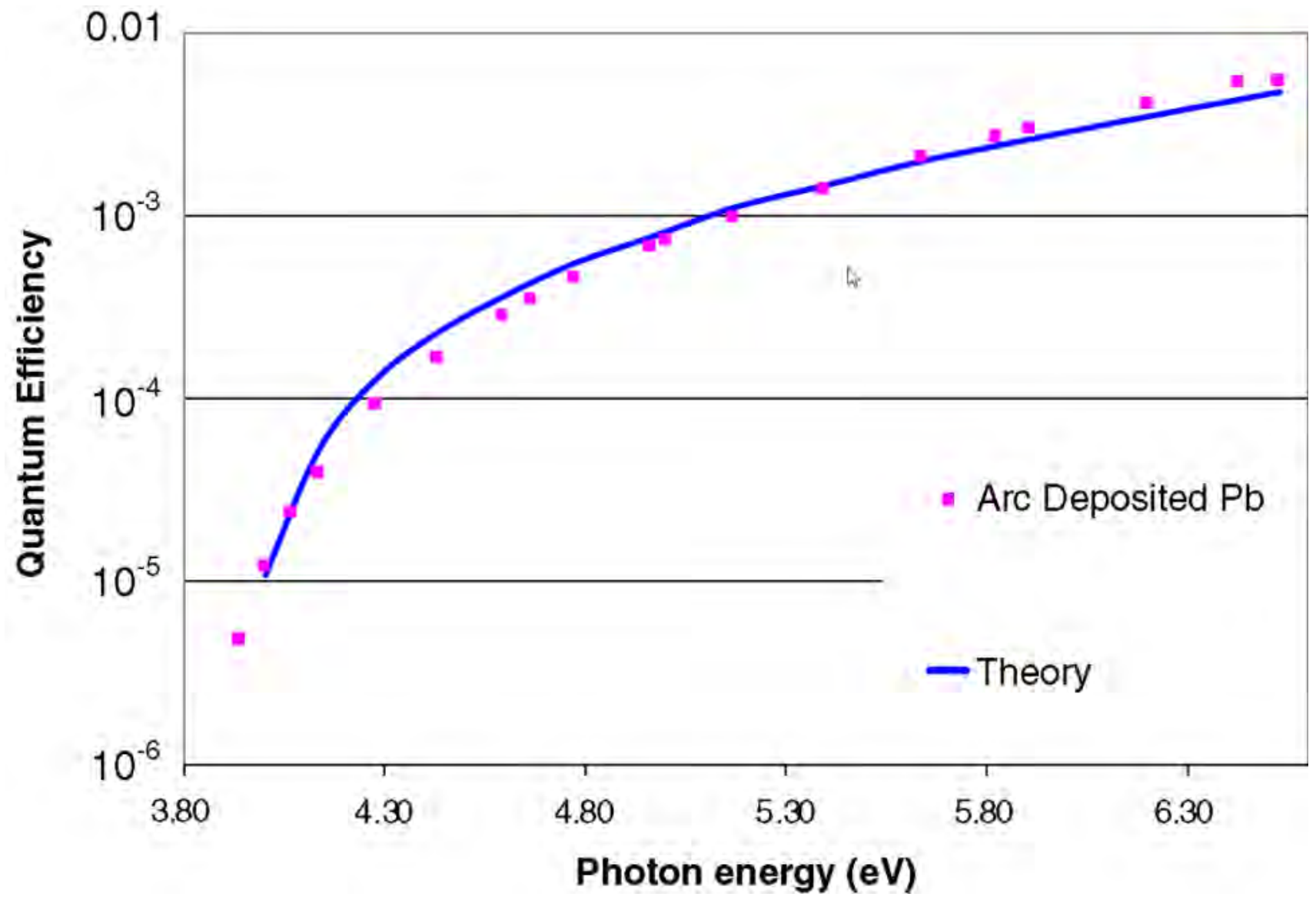

**Figure 6.2. Experimental- (■) and theoretical- (--) spectral dependence of QE on an arc-deposited lead witness sample. [Reprinted figure with permission from [6.11]. Copyright 2008 by the American Physical Society]**

QE is measured for an arc-deposited lead cathode in the laboratory with a mesh anode held in parallel to the cathode in an evacuated cube. The anode is held at a positive bias and a picoammeter measures the current leaving the cathode. A deuterium light source is fiber-coupled to a monochromator with a 300 µm exit slit. The desired wavelength $\lambda$ is selected on the monochromator. The output bandwidth is 2 nm, measured with an Ocean Optics HR2000 spectrometer. A fused silica lens focuses the light on the cathode through a vacuum window and the anode mesh. We measure, with a power meter, the output of the monochromator for each wavelength before and after each current measurement at a point after the lens, but before the vacuum window. The optical transmission of the vacuum window and the mesh are calibrated separately for each wavelength. Typical values for optical power $P$ (in a 2 nm band) are 10-100 nW, and those for current $I$ are 0.1-10 pA. For each wavelength, the QE is calculated *via*

$$\mathrm{QE} = \frac{Ih\nu}{P}; \text{ where } \nu = \frac{c}{\lambda} \tag{6.6}$$

Figure **6.2** compares the theoretical calculation and experimental measurements. Other researchers made similar calculations for copper [6.18]. The calculations for copper can be simplified greatly. For near-threshold emission, copper's DOS is nearly constant from 2 eV below the Fermi level up to the vacuum level; the Fermi-Dirac distribution can be assumed to be a step function. The probability expressions are simplified to generate an analytical expression for the energy distribution as well as the QE.

## 6.3 INTRINSIC EMITTANCE OF METAL CATHODES

The intrinsic emittance of the electrons at the surface of the cathode can be written as Equ. 5.30. As can be seen from the equation, the intrinsic emittance of the electrons from a cathode is dictated by the laser spot size, the photon energy and the applied field. The intrinsic emittance can be lowered by reducing the laser spot size and matching the laser photon energy closely to the effective work function of the cathode. Intrinsic emittance of 0.41 mm mrad mm$^{-1}$ and 0.68 mm mrad mm$^{-1}$ have been experimentally measured for hand polished polycrystalline copper irradiated by 282 nm and 262 nm laser, respectively [6.19]. However, one should keep in mind that matching the laser photon energy to the work function leads to a reduction in the quantum efficiency as well. The QE measured for this sample is $1\times10^{-5}$ at 262 nm and $5\times10^{-6}$ at 282 nm.

As mentioned in Chapter 5, the surface roughness can alter the intrinsic emittance as well, especially in the presence of an applied field. Dependence of QE and the thermal emittance of copper cathode in a RF gun has been measured by [6.20].

# 6.4 CATHODE PREPARATION TECHNIQUES

Since most metals, as received from a vendor, have surface contaminants, their QE is significantly lower than that predicted for a pure sample. The QE however can be improved by surface processing. A number of techniques have been tested and a few established ones are described below

## 6.4.1 Ex-*Situ* Preparation

### *6.4.1.1 Surface Preparation of Copper- and Magnesium-Cathodes*

The surface preparation described here, used in conjunction with laser cleaning, increases the electron yield by two to three orders-of-magnitude. Although we have used this procedure in a laboratory setting on several metal cathodes [6.13], it was tested in the RF injector only on copper- and magnesium-cathodes since these are the preferred ones for the applications. Comparison of the cathode's performance after using polishing products from different vendors revealed that the process is product-specific. We recommend using the ones specified in the process for optimum results.

1. Select high-purity bulk material (*e.g.*, high-purity, oxygen-free copper (OFC), high-purity Mg rod) as the source for the cathode.
2. After machining to the required dimensions, remove any fine scratches and ensure surface flatness using the following procedure:
   a. Mount three identical cathodes on a face plate (one intended cathode and two witness cathodes) in a circle at 120˚ apart. Dissimilar metals cross-contaminate the cathode's surface. Dissimilar sizes may degrade surface flatness.
   b. Adjust the three cathodes so that they are parallel to the face plate to within 0.025 mm.
   c. Remove surface scratches on the cathode by polishing its surface with Buehler Microcut 600 grit paper and then with Buehler Metadi fluid polishing extender. Using other polishing products did not reproducibly increase electron yield by two to three orders-of-magnitude.
3. The surface preparation is completed using a double platen, Buehler Ecomet 5 polishing grinder with Automet 2 power head, and a Buehler Mastertex polishing cloth. One of the platens and associated workstation is reserved for the final polishing with a 1-micron diamond polishing compound, and hence, is kept covered to avoid contamination, while the other station was used to polish the cathode successively with 9-micron- and 6-micron-polishing compounds. The polishing cloth is attached to the platen rotating at 120 RPM. The Buehler Metadi 9-micron suspension is sprayed on the polishing cloth, soaking it so that the liquid pools in the cloth. The Automet is set for a polishing time of 90 sec at a pressure of 3 lbs. The cathode in the chuck fixture is attached to the Automet and lowered on to the platens.

   If the cathode's surface is too large for faceplate to support, the positions of the cathode and the polishing cloth can be interchanged. If the cathode is integrated into a flange that prevents its being polished in the polishing machine, it can be mounted on a lathe and the cathode polished by holding the polishing cloth in front of it. In this case, the lathe should not rotate above 300 RPM.

4. Rinse the cathode and the fixture with hexane. Hexane was chosen as the cleanser because it does not contain oxygen and effectively removes the suspension residue from the cathode surface.
5. Repeat the polishing process with the 6-micron diamond suspension.
6. Switch over to the other head and workstation.

7. Repeat the polishing with the 1-micron diamond suspension, but reduce the polishing time to 60 sec; then inspect the surface and repeat if necessary. Over-polishing may introduce peeling, engendering a surface texture resembling an orange peel, observable even to the naked eye.
8. Rinse the cathode completely with hexane.
9. Remove the cathode quickly from the fixture and immerse it in a hexane bath in a beaker.
10. Place the beaker in an ultrasonic cleaner for 20 min to remove any embedded polishing material. Ensure that the temperature of the hexane does not rise above room temperature during this process.
11. Remove the cathode from hexane and dry the surface with high purity nitrogen at 60 psig.
12. Transfer the cathode to the vacuum system and start pumping down. The time lapse between drying with nitrogen and the pump down should be minimized (< 2 min).

The cathode is now ready for laser cleaning.

#### *6.4.1.2 Surface Preparation of Lead Cathodes*

Metals can also be coated on to the substrate *via* a variety of techniques, such as electroplating, evaporation, ion sputtering, or arc deposition. Although any metal can be deposited, lead coating has been tested for use in SCRF injectors (for example, at Helmholtz-Zentrum Berlin) and the process is presented here. Since arc deposition of lead was shown to deliver a higher QE than other techniques, it is being developed further. Figure **6.3** illustrates the arc-deposition system used for our measurements in three different configurations: Straight; 90˚ bend; and, 45˚ bend. We note that the left figure also shows the 1.3 GHz gun attached to the system in readiness for arc deposition. Figure **6.4** shows more details.

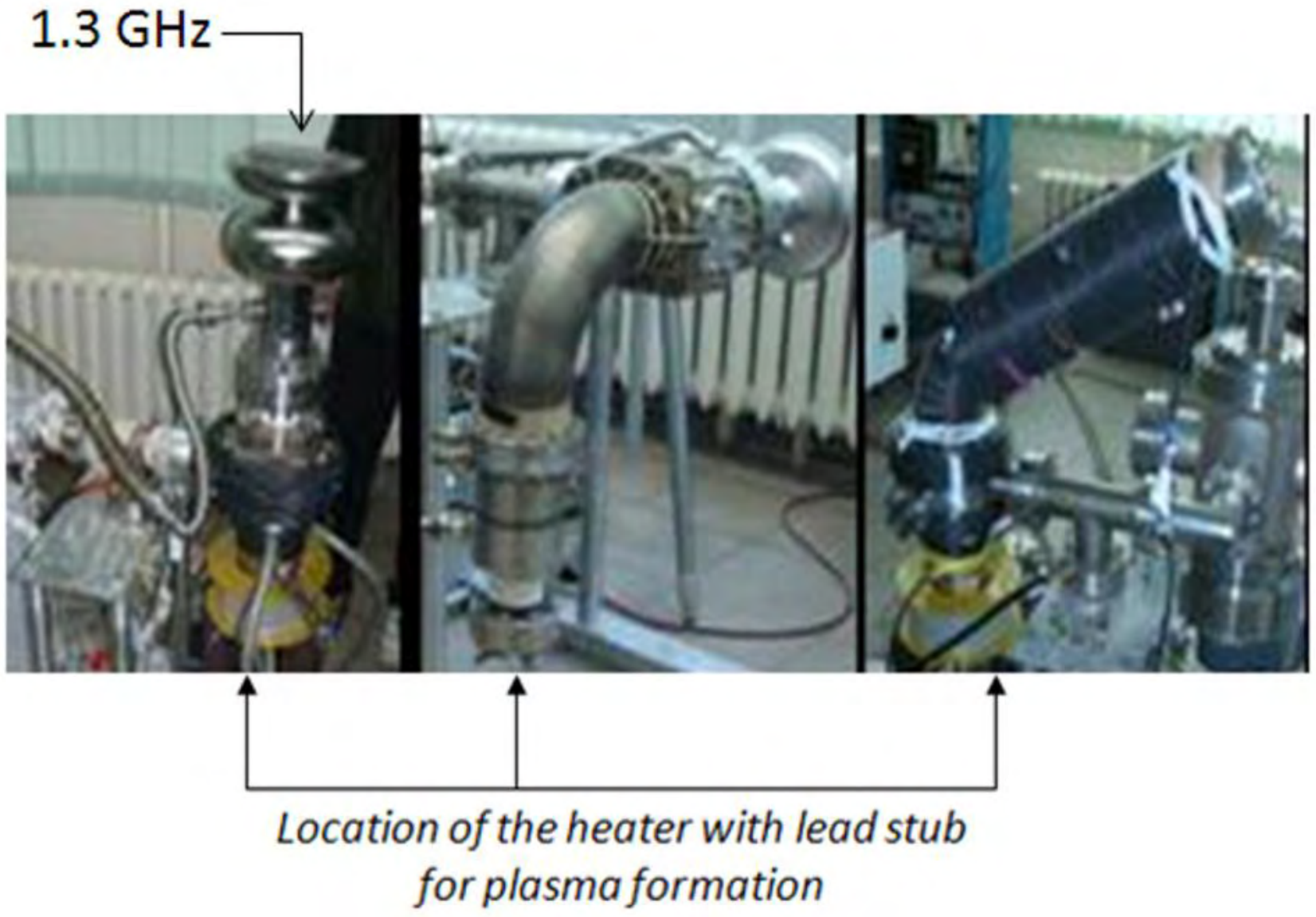

**Figure 6.3. Arrangements for plasma arc deposition: Straight (left), 90˚ bend (center), and 45˚ bend (right).**

In the arc-deposition system, a high-purity lead planar cathode is mounted on a water-cooled support inside a vacuum system capable of maintaining < $10^{-11}$ Torr of pressure. The cathodic arc of lead is guided through a magnetic filter that removes micro-droplets from the arc, and deflects the stream of plasma ions towards the gun. A specially designed shield assures that the central portion of the cavity's back surface is coated with lead, while protecting the rest of the cavity. The gun is electrically isolated from the walls of the vacuum chamber and is either DC- or pulse- (kilohertz range) biased to 100 V. The lowest possible arc current for stable operation in the DC mode is ~23 A, whilst the upper limit is set to 140 A by the anode's cooling system. The deposition rate depends on the distance between the cathode and the target, and is typically ~0.5 nm $s^{-1}$. Additional details of the system are given elsewhere [6.21].

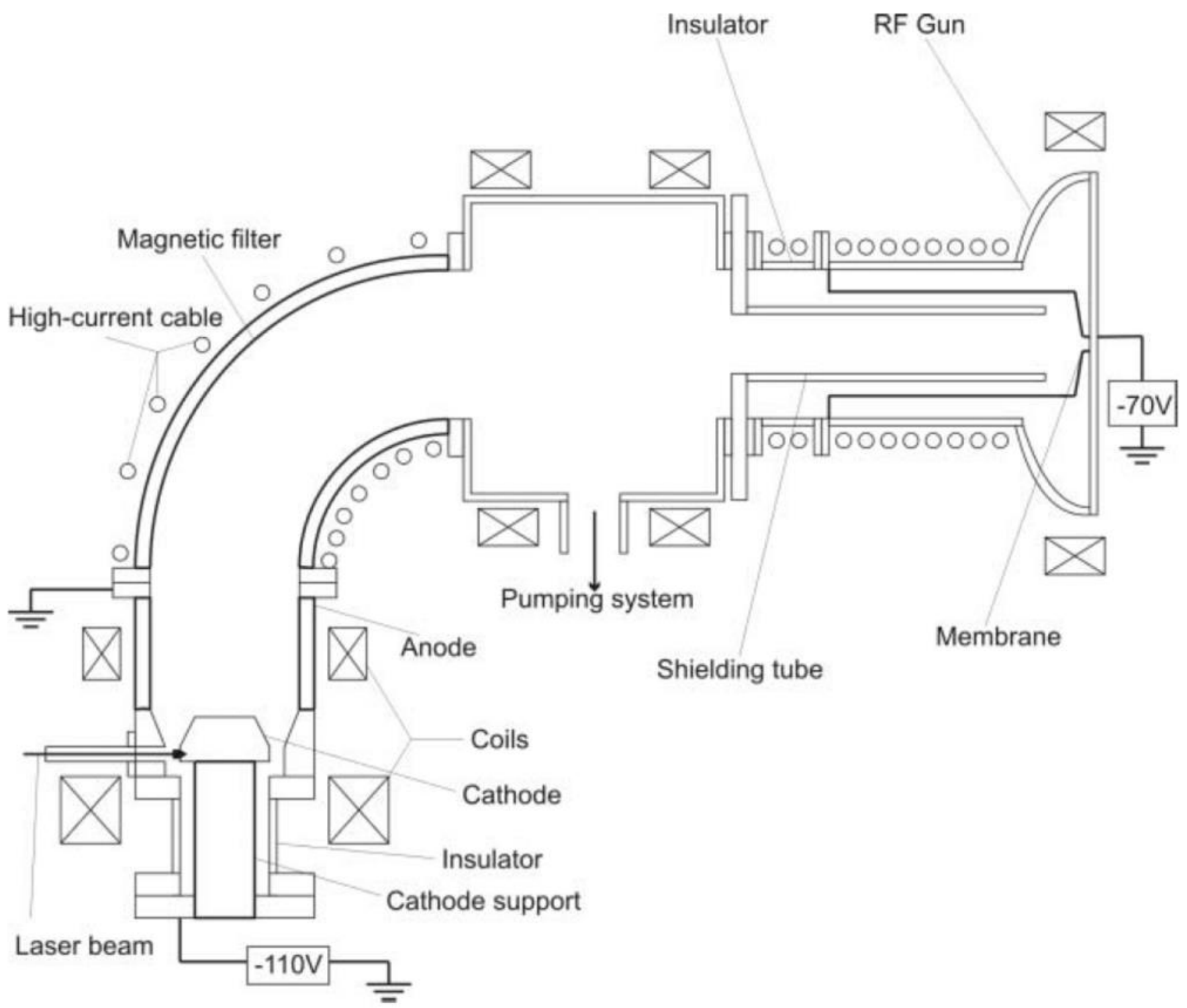

**Figure 6.4. Details of the arc deposition system, with 1.3 GHz gun attached, for depositing lead. [[6.21]; Available under Creative Common Attribution 3.0 License (www.creativecommons.org/licenses/by/3.0/us/) at www.JACoW.org.]**

In the straight set-up, there is no magnetic filter, the spacing between the cathode and the target is shorter, and the resulting coating thicker (approximately micrometers) than in the other configurations. However, the surface contains droplets of lead, making the coating's thickness slightly non-uniform. With the 90˚ bend, the droplets were eliminated, but the coating thickness was limited to 100 nm, based on SEM EDS measurements. This thickness is not adequate to ensure the efficient absorption of photons, as well as effective laser cleaning. Although the QE of the 100 nm thick sample improved significantly with laser cleaning, it did not reach the highest value (0.5% @ 193 nm) of the original arc-deposited witness sample. The deposition parameters must be optimized to obtain a micrometer thick coating without droplet formation and minimum contamination. However, lead cathode has been deposited in 1.3 GHz guns with this technique without significant degradation of the gun's RF properties; they were used to generate electron beams [6.12].

### 6.4.2 In-*Situ* Preparation

#### *6.4.2.1 Laser Cleaning*

The laser cleaning typically is done when the cathode is mounted in the RF cavity and the system has been pumped and baked to achieve a base pressure of $\leq 10^{-10}$ Torr. The laser is raster-scanned multiple times over the emitting surface with the scan step much smaller than the size of the laser spot, thereby smoothing out any spatial variation in the laser energy density to result in uniform emission. The density of the laser energy depends on the metal being cleaned. Table **6.3** lists the metals cleaned in the RF gun, as well as the minimum energy density required to clean the surface so that the QE is increased almost to its maximum value while the surface finish is preserved. Scanning with higher energy density may engender slightly higher QE, but will modify the surface morphology, which is undesirable in the injector. Laboratory measurements indicate that an excimer laser with a 248 nm wavelength and ~5 ns pulse duration is as effective as a 266 nm laser with a picosecond pulse duration. Since the injector facilities already are equipped with picosecond UV lasers, these are the lasers of choice for cleaning. However, the energy density required to clean the surface without altering the surface morphology for the sub-picosecond pulse duration must be reestablished for the cathode material.

| Metal | Energy Density of Cleaning Laser [mJ mm$^{-2}$] |
|---|---|
| Copper | 1 |
| Magnesium | 0.1 |
| Niobium | 0.6 |
| Lead | 0.2 |

**Table 6.3. Laser energy densities to clean the metal surface to improve the QE without altering the surface finish.**

In high average current injectors, the dark current from the cathode needs to be minimized since it poses a significant radiation risk. Experimental measurements with Pb coated Nb in all-superconducting cavity to generate 1 mA average current has shown [6.21] that laser cleaning reduces dark current and increases the onset field for field emission. This could be attributed to the removal of field emitter sites during laser cleaning.

#### *6.4.2.2 Ion Cleaning*

Hydrogen ion cleaning of copper has improved [6.18] the QE of OFC significantly, as shown in Figure **6.5**. However, cleaning increased sensitivity to surface contaminants, and hence, it is not being used for routine operation of the injector.

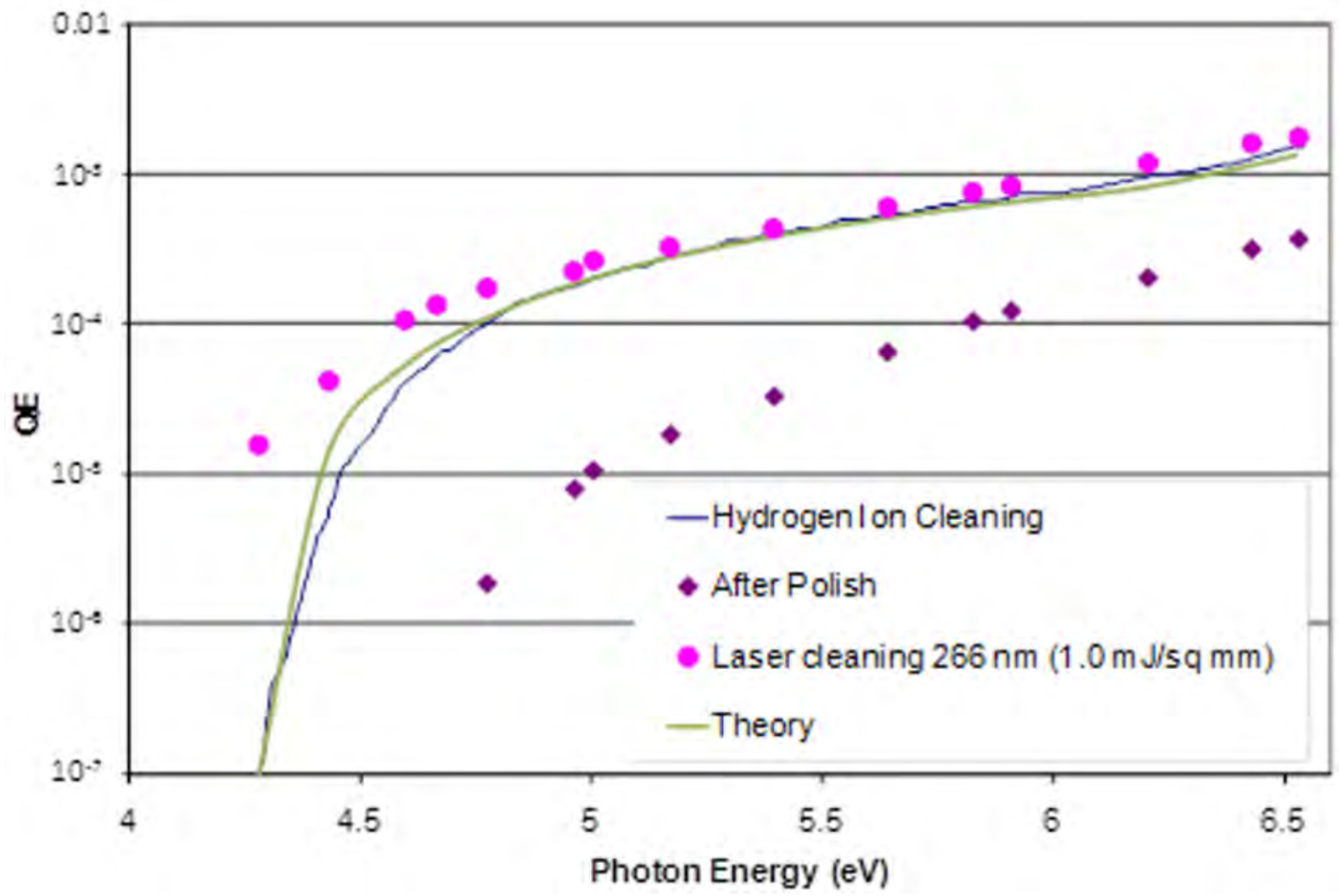

**Figure 6.5. Comparison of LCLS copper cathodes before cleaning, after H-ion cleaning, and after laser cleaning. Both modes of cleaning resulted in an ultimate QE consistent with theoretical expectations.**

## 6.5 IMPACT OF SURFACE FINISH ON EMITTANCE

The surface smoothness of the cathode may affect emittance in several ways. Any non-uniformity in the transverse profile of the cleaning laser can cause non-uniform electron emission, thereby degrading the emittance. Hence, it is imperative to check the emitting surface carefully after laser cleaning to assure the uniformity of emission. Any surface roughness could lead to field enhancement and a corresponding reduction in the work function due to the Schottky effect. This also would lead to a non-uniform emission and degraded emittance as discussed in Chapter 5. Furthermore, the field bending associated with the field enhancement could entail larger transverse energy and a larger emittance of electrons. For applications sensitive to the transverse emittance, care must be taken to minimize these effects.

## 6.6 INCORPORATING THE CATHODE IN A GUN

NCRF injectors are usually made of OFC making it relatively easy to incorporate the copper cathode. However, the cleaning procedures required to attain a high QE from copper requires surface preparation that is not fully compatible with good RF performance. The same is true for the niobium cathode in SCRF guns. Different designs have been used for a de-mountable cathode; the successful ones are described below. In all cases, the critical issues are to preserve the quality of the RF field, eliminate high voltage breakdown, and maintain ultra-high vacuum (UHV) conditions.

### 6.6.1 Plug

The plug arrangement used in the early operation of BNL-ATF's injector is shown in Figure **6.6**. A helical flex was used for the RF contact and metal gaskets to preserve the integrity of the vacuum. Figure **6.6(a)** shows the helical flex and the pitting developed over time, caused by electrical breakdown in this region. Spring fingers also will achieve good RF contact, but they were found to cause greater electrical breakdown and associated pitting. The square pattern in Figure **6.6(b)** depicts the result of laser cleaning. The entire plug can be made of the copper, incorporating the cathode, or a different material can be inserted at the center of the plug. An embedded magnesium cathode is illustrated in Figure **6.6(c)**. The four alignment fiduciaries in Figure **6.6(c)** help in centering the laser spot on the cathode.

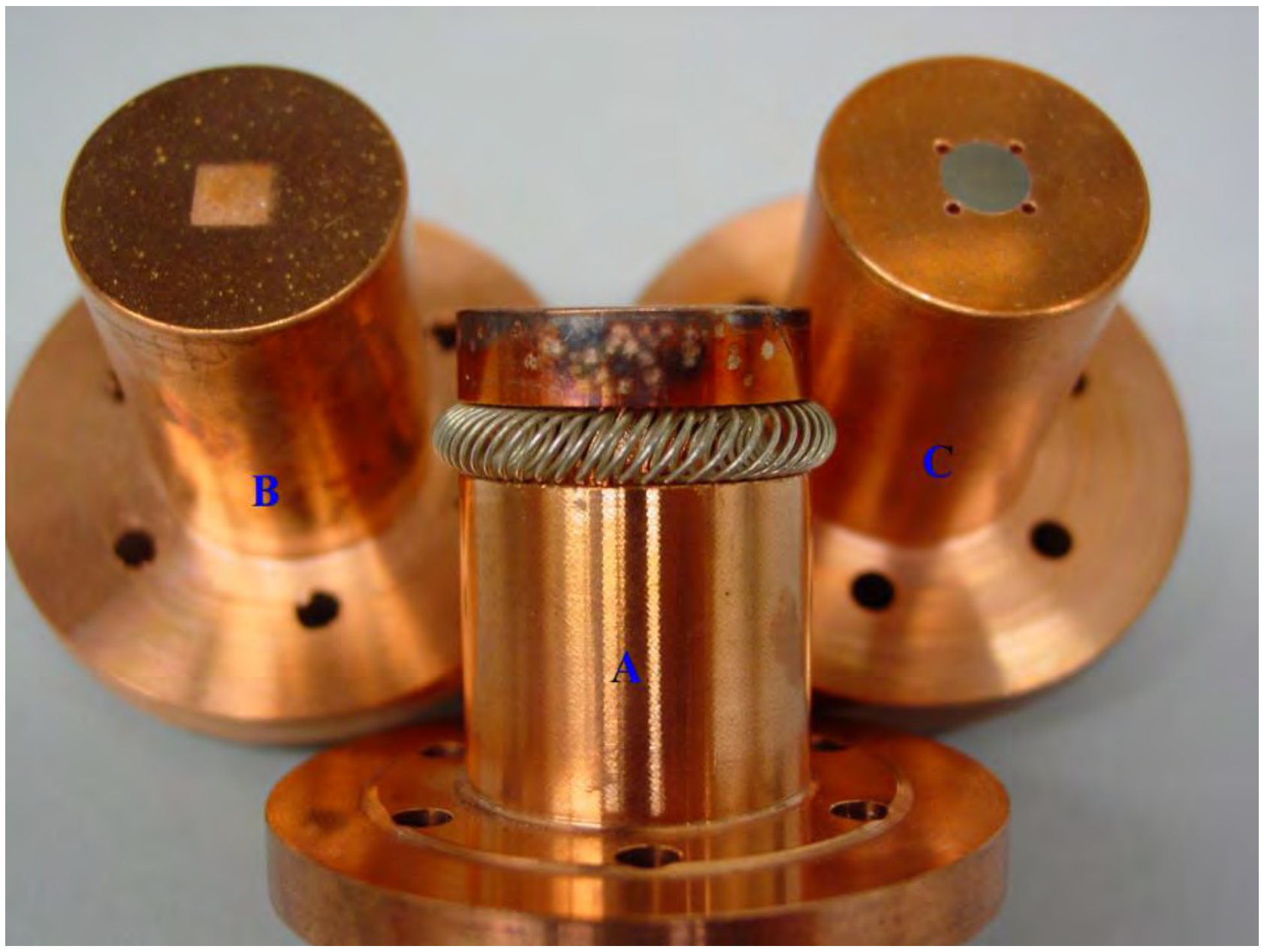

**Figure 6.6. Photograph of plug cathodes tested at BNL-ATF. (a) Copper cathode with helical flex; (b) Surface of the plug showing the laser-cleaned area; and, (c) Copper plug with magnesium cathode.**

Similar plug design also was used in the SCRF gun [6.23]. The niobium plug is pressure-fitted with an indium wire seal on the outside of the cavity. We also coated the Nb plug with lead and tested it in the same arrangement. Since the heat load and associated degradation of the cavity, *Q*, are of serious concern for SCRF injectors, its design must address the effective cooling of the plug.

### 6.6.2 Flange Cathode

To avoid the breakdown problems encountered in the plug design, in later designs of NCRF injectors, the entire back plane of the cavity was replaced by a flange containing the cathode. Figure **6.7** shows examples of this design.

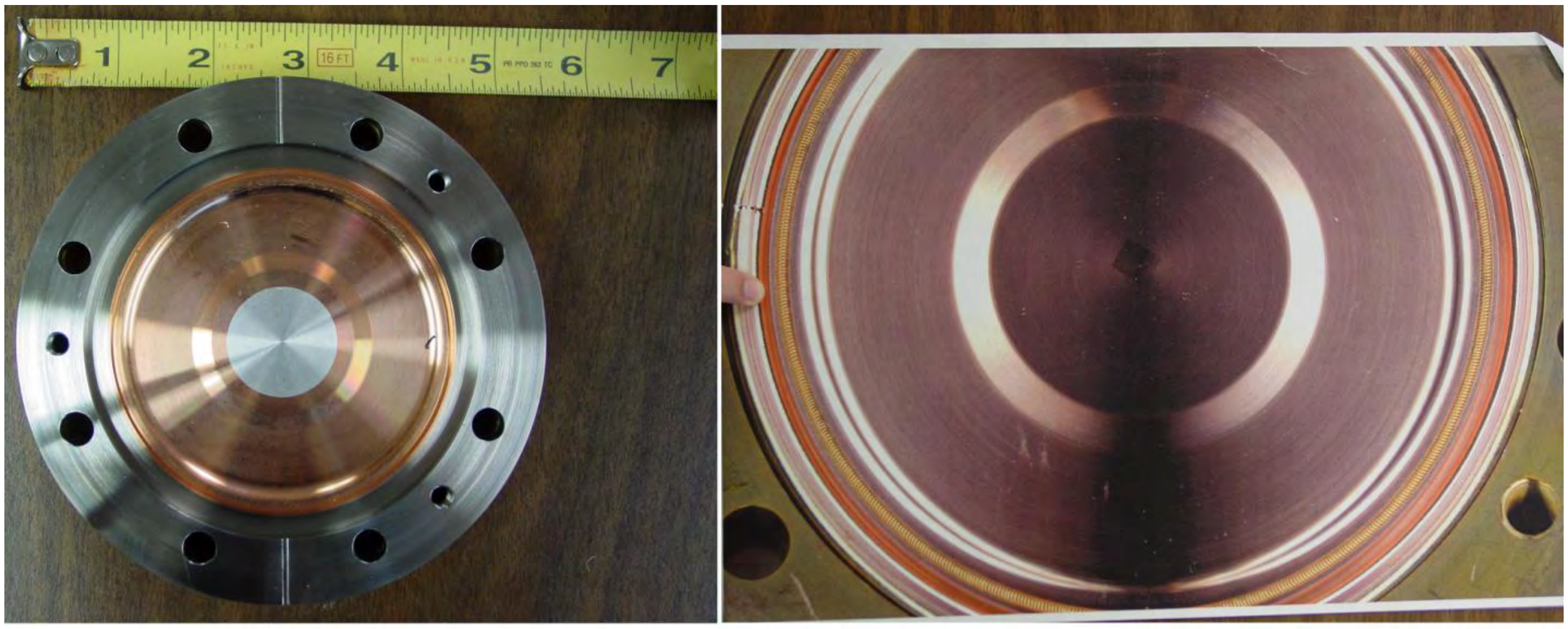

**Figure 6.7. Flange cathode. Left: Magnesium friction-welded to the copper flange. Right: Copper cathode. The square in the center is due to laser cleaning. In both cases, a helical flex is used for RF contact and gaskets to preserve the vacuum's integrity.**

The magnesium cathode, shown in left photograph of Figure **6.7**, is friction-welded to the flange. Figure **6.8** illustrates a cross-section of this friction-welded region.

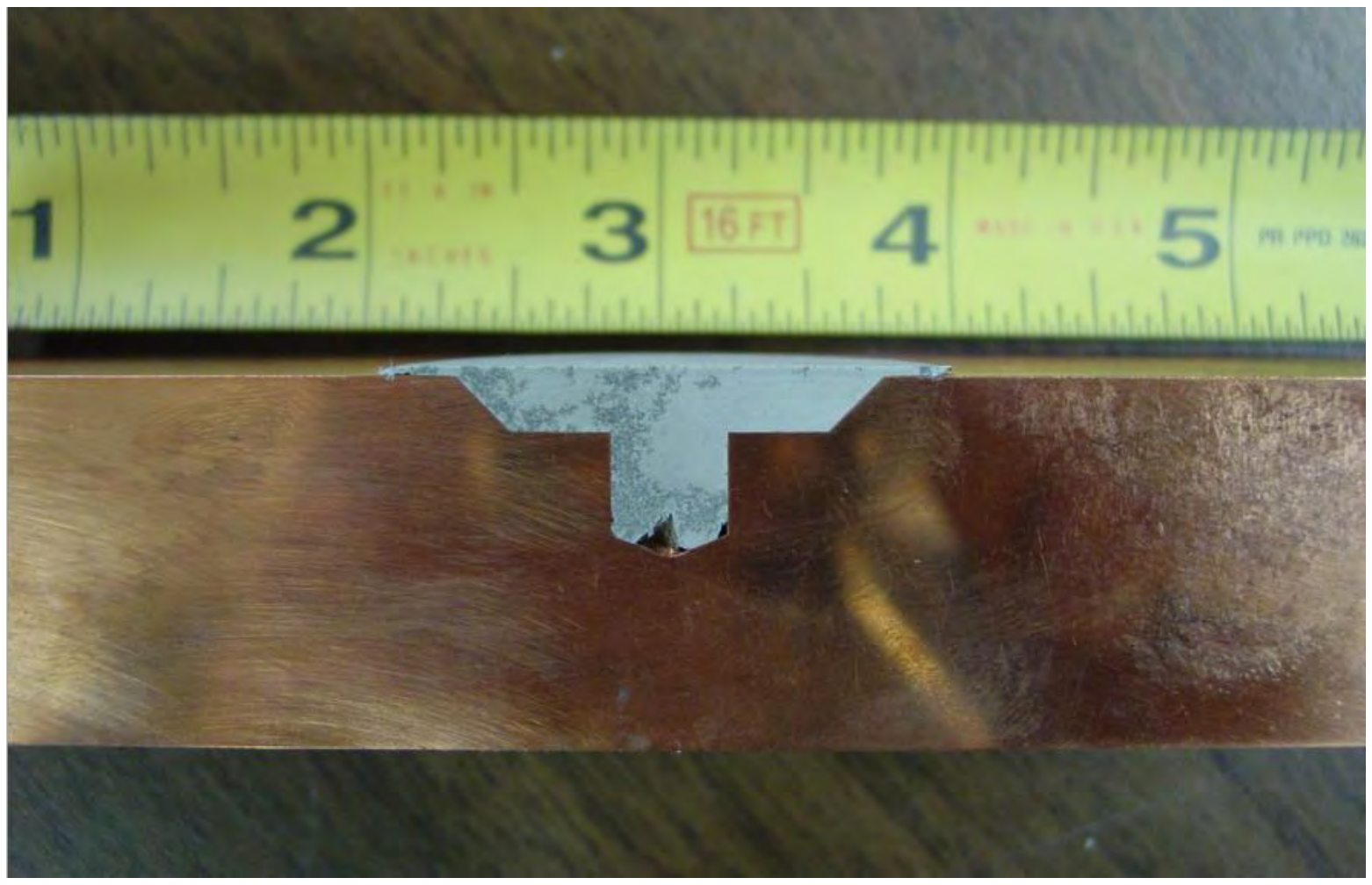

**Figure 6.8. Cross section of magnesium friction-welded flange.**

As evident in Figure **6.8**, the top surface of the magnesium is slightly higher than the back-plate, which was reduced and polished using the procedure described in Section 6.3.1.1. To avoid trapping gases, friction welding can be carried out under vacuum.

### 6.6.3 Choke Joint

Another method of integrating the cathode in a gun, especially in an SCRF gun, is using a choke joint. The cathode material is held in a separate support structure, thermally, and electrically isolated from the cavity; it can be kept at a temperature higher than the cavity and cooled by liquid nitrogen instead of helium. The coaxial line formed by the cathode's channel and its stalk would lead to RF power leakage, which is minimized by using a carefully designed filter. The $Cs_2Te$ cathode was successfully tested with such a gun at the Forschungszentrum Dresden-Rossendorf (FZD laboratory [6.23]–[6.26]. This scheme, as shown in Figure **6.9**, advantageously incorporates a demountable cathode, a requirement for guns using sensitive, short-lifetime cathodes while preserving the cryogenic temperature of the SCRF cavity.

Furthermore, this arrangement facilitates testing different types of cathodes in the same gun, supporting a one-to-one comparison. Power leakage also is ameliorated by incorporating a RF choke joint in the cavity. The design of the choke joint is non-trivial; multipacting could cause problems if the gun is operating at high accelerating fields [6.27]. Several approaches can resolve this issue, such as incorporating a grove to change the trajectory of the multipacting electrons, biasing the stalk to repel them, and coating the surface with a low-yield material.

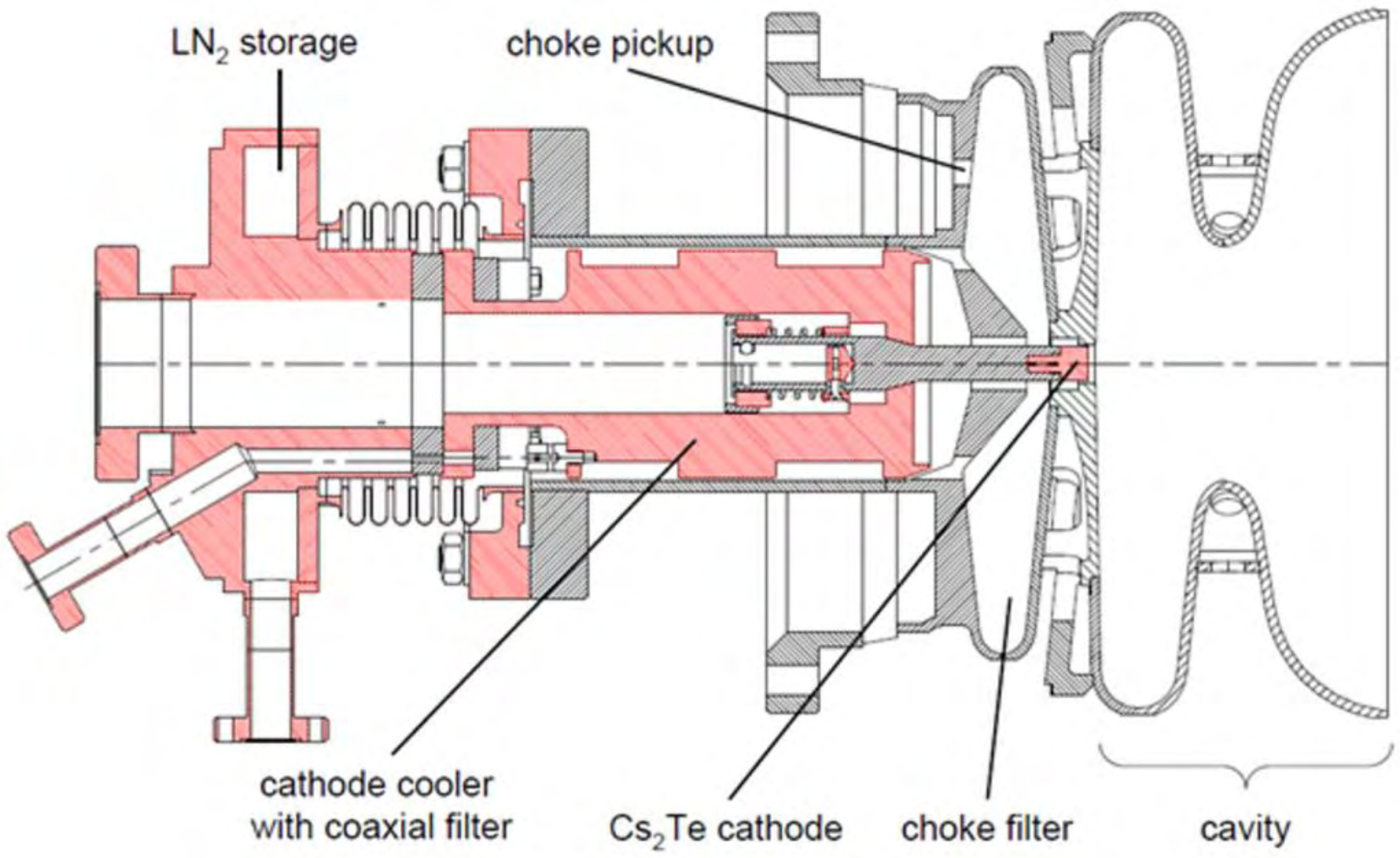

**Figure 6.9. Schematic of the choke joint used in SCRF gun. [6.28]; Courtesy of A. Arnold, Helmholtz-Zentrum Dresden Rossendorf, Institut für Strahlenphysik, Strahlungsquelle ELBE (FWKE)]**

## 6.7 CONCLUSION

We described in this chapter the relative merits and drawbacks of methods that currently are used to prepare metal photocathodes, to improve their electron yield, and to incorporate them into the injector. This field is still evolving, and the focus is on encompassing the physics behind the processes as well as the processes themselves. We anticipate that the next decade will bring better understanding and new and improved cathodes.

## 6.8 CONFLICT OF INTEREST AND ACKNOWLEDGEMENT

We confirm that this article content has no conflicts of interest and would like to acknowledge the support of the US Department of Energy under contract numbers DOE DE-SC0003965 and DE-AC02-98CH10886.

***References***

[6.1] K. Batchelor, I. Ben-Zvi, R. C. Fernow *et al.*, "Performance of the Brookhaven photocathode RF gun", *Nucl. Instrum. Meth. A*, vol. 318, pp. 372-376, July 1992.

[6.2] J. Rosenzweig, K. Bishofberger, X. Ding *et al.*, "The Neptune photoinjector", *Nucl. Instrum. Meth. A*, vol. 410, pp. 437-451, June 1998.

[6.3] X. J. Wang, T. Srinivasan, K. Batchelor *et al.*, "Photoelectron beam measurements for Mg cathode in a RF electron gun", in *16th Int. Free Electron Laser Conf.*, 1994, pp. 21-26.

[6.4] D. H. Dowell, E. Jongewaard, C. Limborg-Deprey *et al.*, "Results of the SLAC LCLS gun high-power RF tests", in *Proc. 2007 Particle Accelerator Conf.*, 2007, pp. 1296-1298.

[6.5] T. Rao, "Updates from facilities: ANL, BNL, Cornell, JLAB, LBNL, European Labs, Japanese Labs", presented at Workshop Photocathode Physics Photoinjectors, Brookhaven National Laboratory, October 2010.

[6.6] J. Sekutowicz, I. Ben-Zvi, J. Rose *et al.*, "Proposed continuous wave energy recovery operation of an X-ray free electron laser", *Phys. Rev. ST Accel. Beams*, vol. 8, 010701-1–010701-12, January 2005.

[6.7] I. Ben-Zvi, D. Barton, D. Beavis *et al.*, "Extremely high current, high-brightness energy recovery linac", in *Proc. 2005 Particle Accelerator Conf.*, 2005, pp. 1150-1152.

[6.8] I. Ben-Zvi, V. Litvinenko, D. Barton *et al.*, "Electron cooling of RHIC", in *Proc. 2005 Particle Accelerator Conf.*, 2005, pp. 2741-2743.

[6.9] T. Rao, I. Ben-Zvi, A. Burrill *et al.*, "Design, construction and performance of all niobium superconducting radio frequency electron injector", *Nucl. Instrum. Meth. A*, vol. 562, pp. 22-33, June 2006.

[6.10] J. Smedley, T. Rao and Q. Zhao, "Photoemission studies on niobium for superconducting photoinjectors" *J. Appl. Physics*, vol. 98, pp. 043111-1–043111-6, August 2005.

[6.11] J. Smedley, T. Rao and J. Sekutowicz, "Lead photocathodes", *Phys. Rev. ST Accel. Beams*, vol. 11, 013502-1–013502-9, January 2008.

[6.12] J. Smedley, Rao, T. Kneisel, P. *et al.*, "Photoemission tests of a Pb/Nb superconducting photoinjector", in *Proc. 2007 Particle Accelerator Conf.*, 2007, pp. 1365-1367.

[6.13] T. Srinivasan-Rao, J. Fischer and T. Tsang, "Photoemission studies on metals using picosecond ultraviolet laser pulses", *J. Appl. Physics*, vol. 69, pp. 3291-3296, March 1991.

[6.14] C. N. Berglund and W. E. Spicer, "Photoemission studies on copper and silver: theory", *Phys. Rev.*, vol. 136, pp. A1030-A1044, November 1964.

[6.15] T. Ogata, Computational Electronic Structure Database. Available Online: http://mits.nims.go.jp/matnavi/ [Accessed: January 30, 2012].

[6.16] C. Norris and L. Walldén, "Photoemissions from Pb", *J. Physics F: Metal Physics*, vol. 2, pp. 180-188, January 1972.

[6.17] H. G. Liljenvall, A. G. Mathewson and H. P. Myers, "The optical properties of lead in the energy range 0.6-6 eV", *Philosph. Mag.*, vol. 22, pp. 243, 1970.

[6.18] D. H. Dowell, F. K. King, R. E. Kirby *et al.*, "*In Situ* cleaning of metal cathodes using a hydrogen ion beam", *Phys. Rev. ST Accel. Beams*, vol. 9, pp. 063502-1–063502-8, June 2006.

[6.19] C. P. Hauri, R. Ganter, F. Le Pimpec *et al.*, "Intrinsic emittance reduction of an electron beam from metal photocathodes," *Phys. Rev. Lett.*, vol. 104, pp. 234802-1–234802-4, June 2010.

[6.20] H. J. Qian, C. Li, Y. C. Du *et al.*, "Experimental investigation of thermal emittance components of copper photocathode," *Phys. Rev. ST Accel. Beams*, vol. 15, pp. 04012-1–04012-8, April 2012.

[6.21] P. Strzyzewski, L. Langner, M. J. Sadowski *et al.*, "Deposition of lead thin films used as photocathodes by means of cathodic arc under UHV conditions", in *Proc. 2006 European Particle Accelerator Conf.*, 2006, pp. 3209-3211.

[6.22] A. Neumann, W. Anders, R. Barday *et al.*, "First characterization of a fully superconducting RF photoinjector cavity," in *Proc. 2011 Int. Particle Accelerator Conf.*, 2011, pp. 41-43.
[6.23] P. Kneisel, J. Sekutowicz, R. Lefferts *et al.*, "Preliminary results from a superconducting photocathode sample cavity", in *Proc. 2005 Particle Accelerator Conf.*, 2005, pp. 2956-2958.
[6.24] D. Janssen, H. Büttig, P. Evtushenko *et al.*, "First operation of a superconducting RF-gun", *Nucl. Instrum. Meth. A*, vol. 507, pp. 314-317, July 2003.
[6.25] A. Arnold, H. Büttig, D. Janssen *et al.*, "Development of a superconducting radio frequency photoelectron injector", *Nucl. Instrum. Meth. A*, vol. 577, pp. 440-454, July 2007.
[6.26] P. Murcek, A. Arnold, H. Buettig *et al.*, "Modified 3½-cell SC cavity made of large grain niobium for the FZD SRF injector", in *Proc. 2009 Superconducting RF Conf.*, 2009, pp. 585-588.
[6.27] A. Burrill, I. Ben-Zvi, M. Cole *et al.*, "Multipacting analysis of a quarter wave choke joint used for insertion of a demountable cathode into a SRF photoinjector", in *Proc. 2007 Particle Accelerator Conf.*, 2007, pp. 2544-2546.
[6.28] André Arnold, private communication, 2011.

# CHAPTER 7: SEMICONDUCTOR PHOTOCATHODES FOR UNPOLARIZED ELECTRON BEAMS

IVAN BAZAROV

*Physics Department*
*373 Wilson Laboratory*
*Cornell University*
*Ithaca, NY 14853*

LUCA CULTRERA

*Cornell Laboratory of Accelerator-based Sciences and Education*
*Cornell University*
*Ithaca, NY 14853*

TRIVENI RAO

*Brookhaven National Laboratory*
*Upton, NY 11973*

**Keywords**

Photocathode, Semiconductor, Antimonide, Telluride, Multi-alkali, Cesium, GaAs, GaN, Quantum Efficiency, Life-Time, Intrinsic Emittance, Thermal Emittance, Response Time, Electron Affinity, Negative Electron Affinity, Cathode Fabrication, $Cs_2Te$ Cathode, $K_2CsSb$ Cathode, $Cs_3Sb$ Cathode, Response Time, Activation, Surface Roughness, High-Current

**Abstract**

Semiconductor photocathodes offer great advantage for generating high average currents by the virtue of their high quantum efficiency in the UV or VIS spectral ranges. This chapter reviews basic properties of these photoemissive materials. Standard growth and preparation procedures for antimony and tellurium based alkali photocathodes as well as for III-V semiconductors activated to negative electron affinity are presented. Various technical requirements and instrumentation needed either for the synthesis or the activation of these materials are outlined.

## 7.1 PHOTOCATHODE TYPES

High quantum efficiency photocathodes are desirable for generating high average (greater than milliamperes) beam current. Essentially, all such photocathodes are semiconductors. Two general types of semiconductor photocathodes used in photoinjectors producing unpolarized electron beams are alkali-based ones (antimonides and tellurides), and bulk- or layered-III-V crystals (*e.g.*, GaAs or GaN) activated to the condition of negative electron affinity *via* the deposition of roughly a monolayer of cesium on the surface. In this chapter, we detail the important characteristics of these photocathodes, their growth procedures, and their performance in the accelerator environment.

## 7.2 DEFINITIONS AND PHOTOCATHODES' FIGURES OF MERIT

The intrinsic properties of the photocathode play a key role in establishing the quality of the electron beam (its emittance and time structure), the reliability of its operation, and the ease of maintenance.

The most relevant properties for a photocathode inside a photoinjector are its quantum efficiency and its spectral response, the photoemission response time, the thermal emittance (or intrinsic emittance), mean transverse energy, and the lifetime of the photocathode. Also of importance is its ability to withstand an adverse environment inside the photoemission gun, such as the back-streaming of ions, created *via* ionizing residual gas by the primary electron beam, towards the cathode. Its lifetime can also be negatively affected by residual gas inside the gun chamber. These effects determine the vacuum requirements for the photocathode (sensitivity to the partial pressure of various gasses encountered in ultra-high vacuum (UHV) systems) to maintain a high QE. Here, we summarize the main definitions used in this chapter.

Quantum efficiency (QE) is defined as the number of electrons per incident photon, given by the formula

$$\mathrm{QE} = \frac{I/e}{P/h\nu} \tag{7.1}$$

or, in practical units

$$\mathrm{QE}\ [\%] = 124\,\frac{I\ [\mathrm{A}]}{P\ [\mathrm{W}]\ \lambda\ [\mu\mathrm{m}]} \tag{7.2}$$

where $I$ is the current emitted by the cathode, $P$ is the laser power incident on the cathode, the electron charge $e = 1.602\times10^{-19}$ C, the Planck constant $h = 6.626\times10^{-34}$ Js, and, $\nu$ and $\lambda$ are the frequency and wavelength of light, respectively. Other related definitions commonly used are spectral responsivity $R$ (measured in units of ampere per Watt) and quantum yield (QY), defined as the number of electrons emitted per absorbed photon. For an optically opaque cathode, to convert QE to QY or to $R$, we use the following expressions

$$\mathrm{QY} = \mathrm{QE}\,\frac{1}{1-\varrho} \tag{7.3}$$

$$R\left[\frac{\mathrm{A}}{\mathrm{W}}\right] = \frac{\mathrm{QE}\ [\%]\ \lambda\ [\mu\mathrm{m}]}{124} \tag{7.4}$$

where $\varrho$ is the reflectivity of the photocathode (on the scale of 0 to 1).

Spectral response, not to be confused with spectral responsivity, is a measure of QE as a function of the incident photon’s wavelength. From the point of laser system and light-transport design, the preferred photocathodes have a good spectral response in the visible range.

Response time is defined as the time it takes for electrons excited by the incoming photons to escape from the cathode’s surface. (Techniques for measuring the electron bunch length are detailed in Chapter 11.) The photocathode’s response time is a critical parameter in producing very short electron pulses and for the temporal shaping of laser pulses required for a high brightness, low emittance beam.

### 7.2.1 Basic Band Gap Structure of Semiconductors

Semiconductors are characterized by having an energy structure wherein an energy gap separates the valence band (the highest occupied band) from the conduction band (the lowest empty band). Electrons are promoted from the valence to the conduction band by absorbing a photon of energy if $E_{ph} > E_{gap}$, where $E_{ph}$ is the photon energy and $E_{gap}$ is the gap energy. Electron affinity is defined as the difference between the vacuum energy level and the minimum of the conduction band. As depicted in Figure **7.1**, vacuum energy can lie either above the conduction band minimum (positive electron affinity), or below it (negative electron affinity). The three-step photoemission model, detailed in Chapter 5, [7.1] is also illustrated in the figure: 1) Photon absorption by an electron with its subsequent promotion from the valence band to the conduction band; 2) electron diffusion inside the crystal; and, 3) the electron's escape into the vacuum.

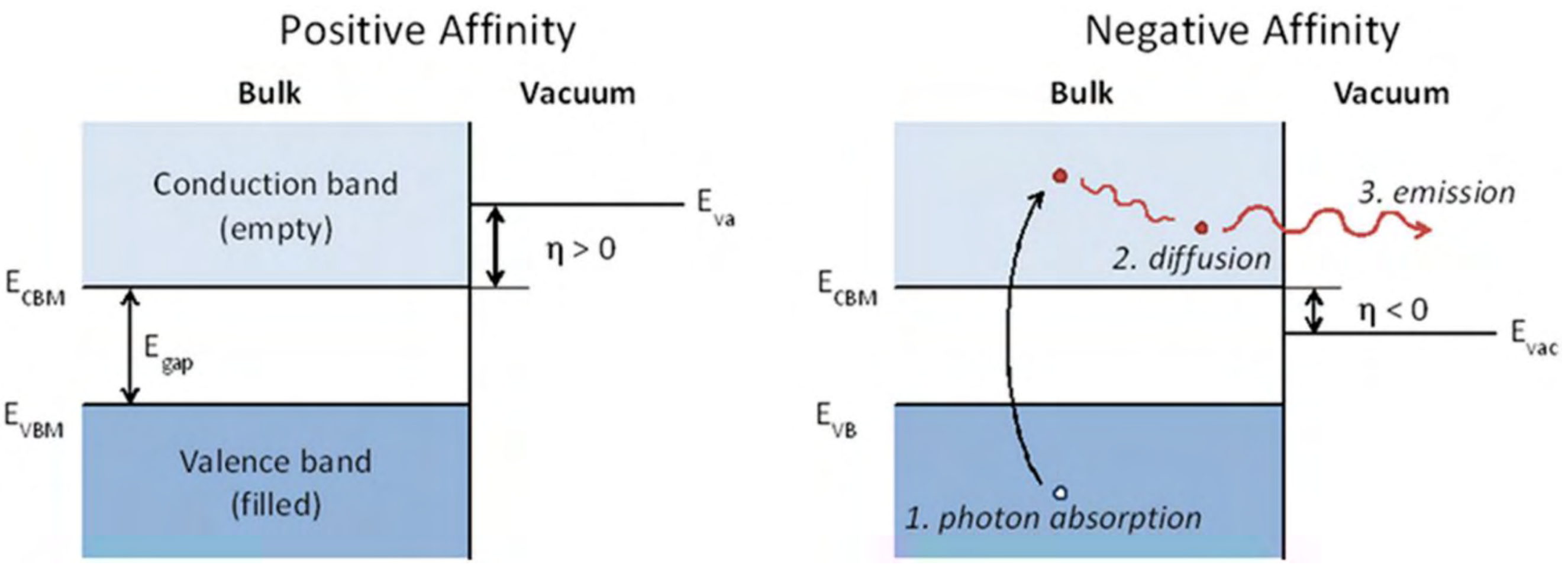

**Figure 7.1. Basic semiconductor's energy structure of positive electron affinity (left) photocathode, and negative affinity (right) photocathode. Energy gap is defined as the difference between the conduction band minimum energy and the valence band maximum, $E_{gap} = E_{CBM} - E_{VBM}$. Electron affinity is defined as the difference between vacuum level energy and conduction band minimum, $\eta = E_{vac} - E_{CBM}$. The three-step Spicer photoemission model is also shown (right).**

### 7.2.2 Transmission- and Reflection-Mode Photocathodes

Semiconductor photocathodes are realized in two basic configurations: Transmission- and reflection-mode. For the former, the laser light is coupled from the backplane of a thin film of photoemissive material deposited on a transparent substrate (typically glass, quartz, or sapphire). The film's thickness must be optimized to match the light's absorption length and the electrons' escape depth. The latter configuration can employ either a bulk crystal or a grown film. Most photoemission guns using high QE photocathodes have a translation mechanism that loads the photocathode from the back of the gun, and therefore, adopt the reflection-mode geometry of the photocathode.

Thermal emittance is defined as the photocathode's intrinsic emittance, *i.e.*, the emittance of the beam measured directly at the cathode in absence of any other degrading effects, such as aberrations from the electromagnetic fields of beam elements, or the space charge. The normalized root mean square (rms) thermal emittance of an electron beam in a direction transverse to the direction of propagation is defined as

$$\epsilon_{nx,th} = \sigma_x \sqrt{\frac{\text{MTE}}{3m_0c^2}} \tag{7.5}$$

or, in practical units

$$\epsilon_{nx,th}\ [\text{mm mrad}] = 1.40\sigma_x\ [\text{mm}] \sqrt{\text{MTE [eV]}} \tag{7.6}$$

Here $\sigma_x$ is the transverse rms spot size of the electron beam (same as the laser spot with a uniform QE distribution), MTE is the mean transverse energy, defined as MTE = $\langle \frac{1}{2} m_0 v_x^2 \rangle + \langle \frac{1}{2} m_0 v_y^2 \rangle$, with $x$ and $y$ denoting the directions perpendicular to the cathode's emission normal, and, $m_0c^2$ = 0.511 MeV is the electron's rest energy. The mean transverse energy, the average energy of the electrons in the transverse directions, is related to the mean transverse velocity spread of the beam. Unlike thermal emittance, which depends on the size of the laser spot, MTE is an intrinsic property of the photocathode and its condition at the surface. The significance of thermal emittance and the MTE is that they set an upper bound on the beam's brightness for photoinjectors delivering short bunches, that is, pancake-shaped ones near the photocathode with a diameter much larger than the longitudinal extent of the bunch after the illumination by an individual laser pulse is over [7.2].

### 7.2.3 Lifetime and Vacuum Requirements

The lifetime and vacuum requirements typically are defined by the time it takes for the QE of the cathode to drop to 1/*e* times its initial value. Additionally, one differentiates between the dark lifetime for the cathode storage, and the operational lifetime during the beam delivery. The cathode material and the conditions of the vacuum chamber housing the photocathode influence the dark lifetime. The photocathodes also are preferentially sensitive to certain residual gasses and their lifetime can be reduced greatly *via* chemical poisoning. Here, the most problematic species are $H_2O$, $O_2$, and $CO_2$. Carbon monoxide, CO, is also known to degrade the photocathodes albeit at a lesser rate than these others [7.3]. However, the noble gasses, $N_2$, and $H_2$ with $CH_4$ (commonly present in UHV systems) in general do not chemically poison the photocathodes. While nothing intrinsic in the photoemission process itself degrades the QE, the operational lifetime never is as good as the dark lifetime, provided all other conditions are identical. The mechanisms for QE degradation while the beam is running include: a) the back-streaming of ions created by the electron beam ionizing the residual gas (ion back-bombardment); b) increased temperature of the photocathode due to laser illumination; c) beam losses and scraping in the gun's vicinity and the corresponding vacuum-degradation; and, d) back-streaming of chemically active species towards the photocathode surface. Common remedies to improve operational lifetime include off-center illumination of the photocathode to avoid ion back-bombardment of the active cathode area, and minimizing the active area (by deliberately reducing the QE outside the active region), thereby minimizing the production of an unwanted beam halo by the scattered light from the laser.

### 7.2.4 Summary of Photocathode Parameters

In Table **7.1**, we summarize the various relevant parameters for different semiconductor photocathode materials described in this chapter (after [7.4]). $E_a$ is the electron affinity.

## 7.3 POSITIVE ELECTRON AFFINITY CATHODES: ALKALI-BASED PHOTOCATHODES

Some cathode materials actively pursued for photoinjectors are semiconductor cathodes with a positive electron affinity (PEA), such as cesium telluride and bi/multi-alkali antimonides. These cathodes have higher quantum efficiencies in the UV-Visible range of the radiation spectrum than metal cathodes (Figure **7.2**).

However, due to their sensitivity to contamination, they require UHV for fabrication, transfer, operation, and storage. We next describe the properties of these materials, their fabrication, transport into the photoemission gun, and performance.

| **Cathode Wavelength** | **$\lambda$ [nm] , $E_{ph}$ [eV]** | **QE [%]** | **$E_a + E_{gap}$ [eV]** | **Thermal emittance $\left[\frac{\text{mm mrad}}{\text{mm rms}}\right]$** | |
|---|---|---|---|---|---|
| | | | | **Theory (Equ. 7.6)** | **Experiment** |
| $Cs_2Te$ | 262, 4.73 | ~10 | 3.5 | 0.9 | 1.2 ± 0.1 |
| $Cs_3Sb$ | 532, 2.33<br>473, 2.62<br>405, 3.06 | ~4<br>~7<br>~9 | 1.6 + 0.45 | 0.42<br>0.62<br>0.82 | 0.56 ± 0.03<br>0.66 ± 0.03<br>0.80 ± 0.04 |
| $Na_2KSb$ | 330, 3.76 | ~10 | 1 + 1 | 1.07 | N/A |
| $Na_2KSb$:Cs | 390, 3.18 | ~20 | 1 + 0.55 | 1.03 | N/A |
| $K_2CsSb$ | 532, 2.33<br>473, 2.62<br>405, 3.06 | ~4<br>~11<br>~25 | 1 + 1.1 | 0.38<br>0.58<br>0.80 | 0.56 ± 0.03<br>0.69 ± 0.03<br>0.87 ± 0.04 |
| GaAs(Cs,F) | 532, 2.33 | ~10 | 1.4 ± 0.1 | 0.77 | 0.47 ± 0.03 |
| GaN(Cs) | 260, 4.77 | ~15 | 3.4 ± 0.1 | 0.94 | 1.35 ± 0.11 |

**Table 7.1. Commonly used high quantum efficiency photocathodes.**

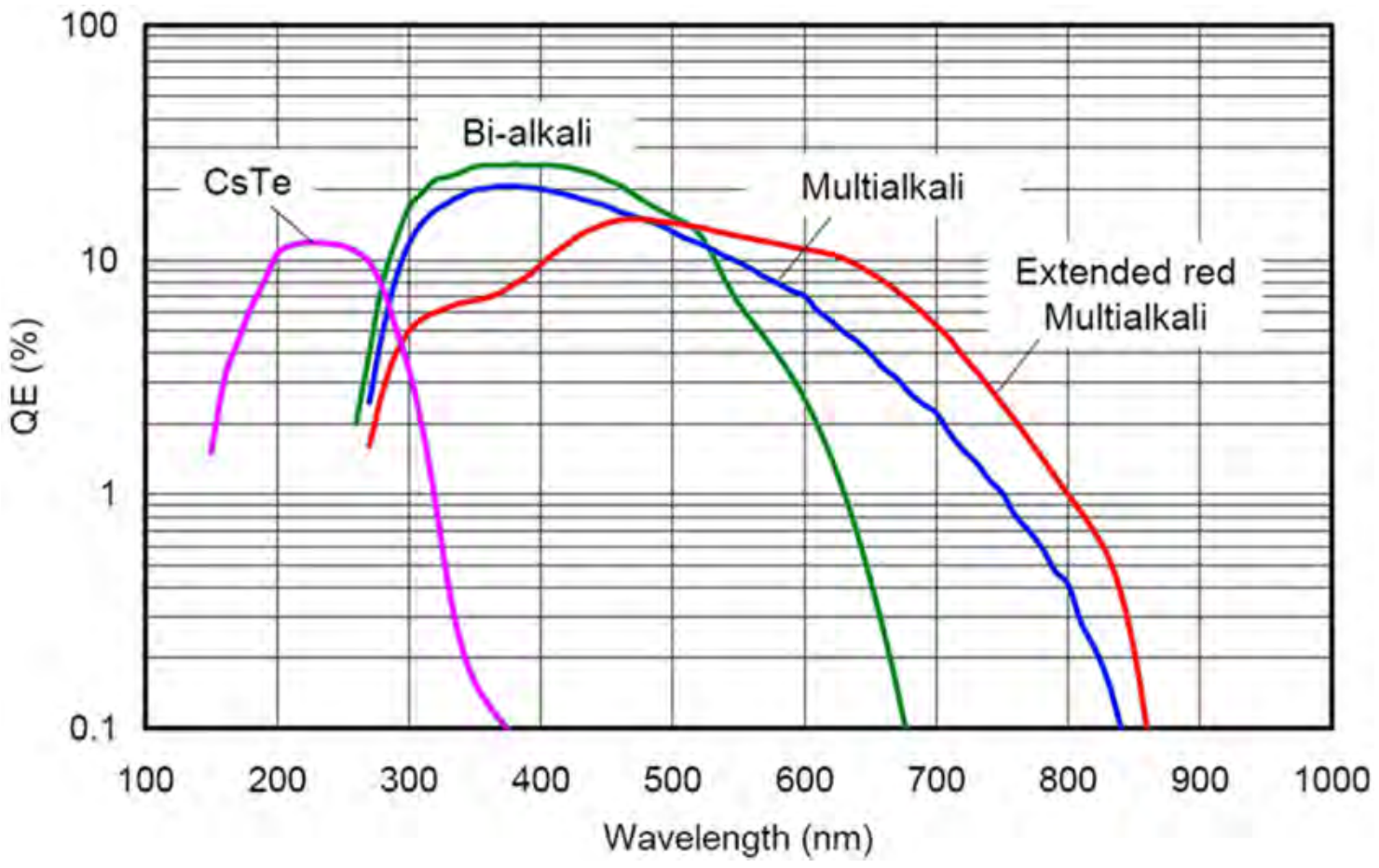

**Figure 7.2. Typical spectral response for different types of photocathodes' materials. [7.5]**

### 7.3.1 Materials Overview

Bi-alkali- and multi-alkali-photocathodes owe their development to the application of high sensitivity measurements of photons. Between the 1930s and '60s, considerable effort was made in developing and optimizing the performance of photosensitive materials, such as $Cs_3Sb$, $K_2CsSb$, $Na_2KSb$, and $Na_2KSb$:Cs. The fabrication process and the spectral response of these cathodes have been studied in depth [7.6]. However, due to the strong involvement of industry and the prospects of commercialization, much of the technical information is protected by intellectual property laws, and hence, details are not always readily available. Furthermore, since these cathodes were developed for detecting photons, *e.g.*, low current, low electric field applications, it was essential to reevaluate their performance in photoinjectors, wherein the peak current can be hundreds of amperes and the accelerating fields in excess of 100 MV $m^{-1}$.

The first successful use of these cathodes in an RF gun was in the APLE experiment in the early ’90s (the Boeing/Los Alamos Average Power Laser Experiment). By using a $K_2CsSb$ photocathode, researchers demonstrated a world record average current (32 mA) for a photoinjector with a 25% duty cycle [7.7].

$K_2CsSb$ is a stoichiometrically stable, p-type semiconductor with band gap energy of 1.0 eV [7.8], and an electron affinity of 1.1 eV. It has cubic crystal structure and high electron yield that is further increased by superficial oxidation. For these reasons, $K_2CsSb$ is considered as a suitable candidate for photoinjector applications.

As with other semiconductor photocathode materials, cesium telluride was studied extensively as a solar-blind, UV sensitive detector. The stoichiometry of the cathode, investigated using X-ray photoemission spectroscopy (XPS) [7.9] was assigned as $Cs_2Te$. The crystal structure is not well established, but could be considered cubic, with polycrystalline surface. Earlier work by Powell *et al.* [7.10] established that this cathode is a p-type semiconductor with $E_{gap}$ of 3.3 eV and $E_a$ of 0.2 eV. Cesium telluride received particular attention because of its demonstrated insensitivity to contamination from $O_2$ and $CO_2$ compared with other alkaline photocathodes. This insensitivity makes this material valuable for guns whose vacuum cannot be held below the $10^{-7}$ Pa range, for example, normal conducting RF (NCRF) guns with a higher thermal load on the accelerating cavities [7.11]. Considerable effort was devoted to their preparation and characterization over the last 20 years. Even with the drawback of requiring UV photons, this is one of the most reliable high QE photocathode materials available to date.

### 7.3.2 Photocathode Fabrications Recipes

The realization of a good antimonide photocathode, whether involving a single alkali species or multiple ones, always starts with the deposition of a high purity Sb film onto a suitable substrate. Due to their initial application in photomultiplier devices, *e.g.*, photomultiplier tubes (PMTs), much of the literature on photocathodes focuses on the synthesis of photosensitive materials related to the deposition of very thin, semitransparent films on glasses. For these applications, the substrate usually is a transparent glass and the Sb film’s optimal thickness is determined experimentally by measuring the decline in the transmitted intensity of a white light source to about 85% of its initial value. In photoinjectors, the cathodes often are frontally illuminated, and thus, preferentially, an opaque photocathode is deposited on a suitable conducting substrate. For this reflection-mode photocathode, the thickness of the Sb layer typically does not need to exceed 20 nm. During the deposition process it usually is measured in-*situ* using a calibrated quartz-crystal microbalance. Similar considerations apply for $Cs_2Te$ photocathodes.

#### *7.3.2.1 $Cs_3Sb$*

The synthesis of the cesium antimonide compound involves only a single source of alkaline metal and is rather simple. The differences in the recipes reported in literature are minimal. Once the substrate (the most common materials are molybdenum, stainless steel [7.12], and heavily doped silicon [7.13]) has been heated to about 160-150 ˚C, the Sb is evaporated until the film has the desired thickness. The substrate with the Sb film then is cooled down to ~130 ˚C, and exposed to Cs vapors. While Cs is reacting with Sb film to form the Cs3Sb compound, the photocurrent is measured. The Cs evaporation is stopped once the peak in quantum efficiency is observed (Figure **7.3**) [7.13].

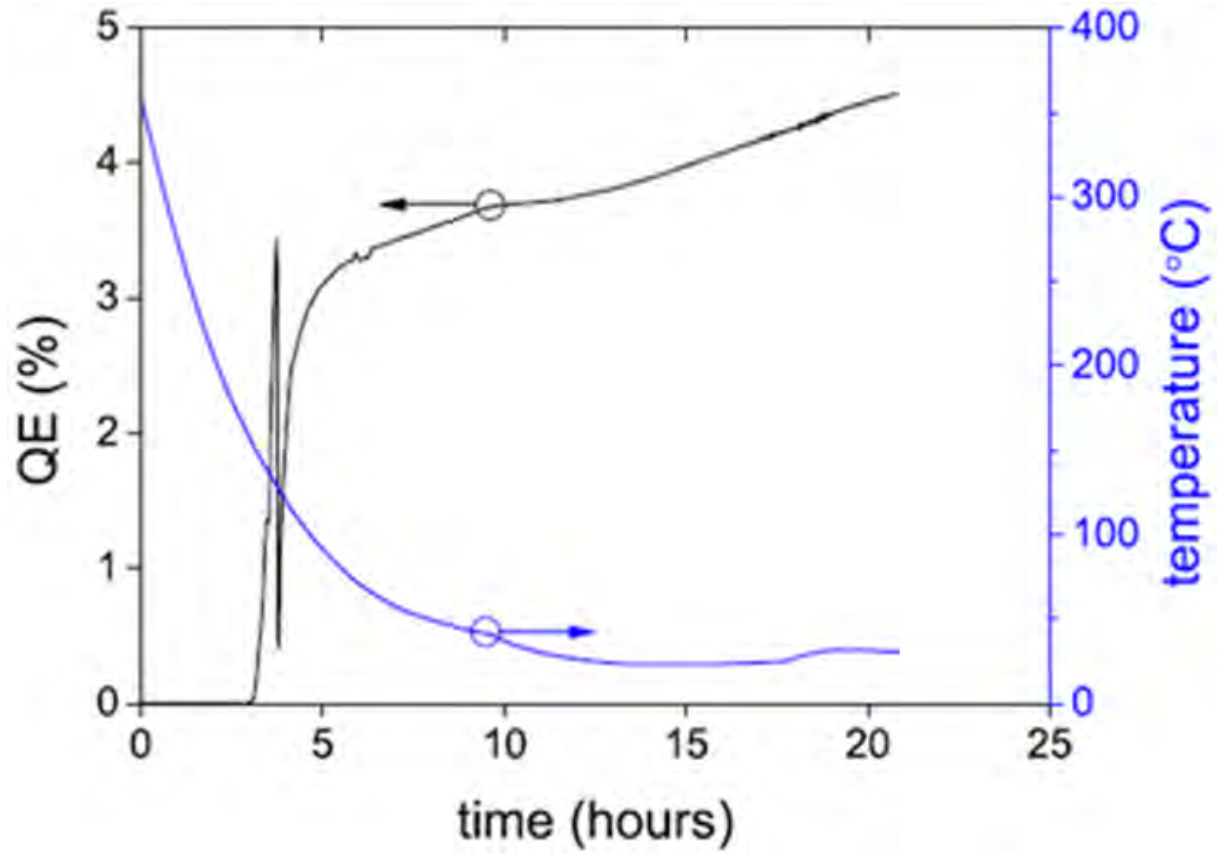

**Figure 7.3. Quantum efficiency during the synthesis of a cesium antimonide photocathode. The peak in QE is reached at 3.5% as monitored by a 532 nm laser. A sudden drop in QE is due to an over-cesiation of the surface, from which it recovers after several hours. Each plot's axis is designated by the color and direction of the arrow. [Reprinted with permission from [7.13]. Copyright 2011, American Institute of Physics.]**

The optimal thickness needed to maximize the light absorption, and hence the QE, from the photocathode layer at the desired wavelength is calculated using the complex index of refraction of $Cs_3Sb$ and of the substrate [7.14] and considering that the Sb film with the correct stoichiometry should have a final thickness about 8X larger than that of the original Sb film [7.15]. Exposing $Cs_3Sb$ to small amounts of oxygen (on the order of a fraction of a monolayer) can improve efficiency by an additional 50% [7.16].

#### *7.3.2.2 $Na_2KSb$*

While the literature contains only a few reports of $Na_2KSb$ being used in operating photoinjectors [7.17], its satisfactory performances in PMT applications, especially at elevated temperatures up to 150 ˚C [7.18], makes it an interesting candidate as a material of choice for accelerators.

In general, a bi-alkali compound typically is grown through both substitution- and diffusion-chemical reactions. In the former, a stable binary alkali-antimonide compound is formed first. Exposure to the second alkali metal then partially replaces the first alkali species inside the crystal structure. The diffusion reaction uses a succession of depositions with the inter-diffusion of the alkali species required to form the final photocathode [7.19].

The standard recipe [7.6] for fabricating this photocathode again starts with the evaporation of a thin Sb film (usually a few nanometers) on a suitable substrate heated to about 160 ˚C. The second step is the growth of the $K_3Sb$ photocathode (similar to the $Cs_3Sb$ cathode): The Sb film is exposed to hot K vapor at ~160 ˚C until a peak on the photocurrent is reached. Then, the substrate is heated up to 220 ˚C and the substitution reaction with Na vapors is carried out until another peak in the photocurrent is observed.

Another approach consists of alternating many cycles of Sb deposition while keeping the substrate with the photocathode in a low pressure environment alternately rich in the vapors of alkali metals (K first, and Na after). Sb vapor is then introduced to react with the alkalis, leading to the growth of the photosensitive film [7.20].

The co-evaporation of both alkali metals and antimony was also undertaken in a dedicated molecular beam epitaxy (MBE) chamber, demonstrating a product with a QE comparable to commercial ones [7.21]. In this case, the temperature of the substrate is kept at 115 ˚C (lower than described above), and a mass spectrometer is used to control the flux of the chemical species during evaporation. The ratios of the mass signal on a mass spectrometer are kept equal to 1:10:15, respectively for Sb:Na:K. Tailoring the ultimate thickness of the photoemissive layer can minimize the reflection of the incident light to further improve the QE [7.19].

#### *7.3.2.3 $Na_2KSb$: Cs*

The synthesis of this multi-alkali photocathode is complex. The starting point is the growth of a high QE, $Na_2KSb$ photocathode as described in the previous section. Thereafter, the cathode is left to cool down to about 160 ˚C and a very thin layer (a few nanometers) of Sb is grown on top. The Sb film is then allowed to react with Cs vapors to form a $Cs_3Sb$ top layer. Growth can also be achieved in successive steps by alternating the deposition of Sb and alkali, [7.6] or by evaporating Sb in the presence of vapors of alkali metals [7.20]. The presence of the thin $Cs_3Sb$ top layer is thought to lower the vacuum level of $Na_2KSb$ close to, or even below the minima of the conduction band [7.22], [7.23], thereby allowing the extraction of a larger number of electrons from the photocathode, and extending its spectral sensitivity to the longer wavelengths up to about 1 μm.

#### *7.3.2.4 $K_2CsSb$*

The recipes used by different photoinjector laboratories generally are similar [7.24]–[7.28]. QEs of several percent routinely are achieved with ~ 530 nm radiation. Growth, similar to that of the previously described antimonide photocathodes, begins with few nanometers of Sb at a rate of 0.1-0.2 nm $s^{-1}$ on a suitable substrate (molybdenum, stainless steel, and silicon are common choices) heated to temperatures between 100-180 ˚C. The Sb film then is left to cool to ~135 ˚C while being exposed to K vapors until the photocurrent reaches a maximum. Once the temperature drops below 130 ˚C, the exposure to Cs vapors begins. During this step, the cathode is left to cool down further, and the Cs flux is lowered to avoid its overexposure to Cs [7.29]. The exposure is stopped when the QE reaches the maximum value.

The temperature and the substrate material play a critical role in the formation of antimony layer. During evaporation at temperatures below 100 ˚C, the Sb layer on Mo and stainless-steel substrates appears patchy immediately after evaporation, but, with time, coalesces to form a more uniform, shinier surface. The rate of coalescence increases with temperature, taking up to a few minutes at room temperature to less than a few seconds at 100 ˚C. For substrates in this temperature range, the thickness of Sb measured using the crystal monitor and the SEM-EDX agree very well. Higher substrate temperatures result in lower sticking coefficient and correspondingly lower Sb thickness. For layer thickness of ~10 nm, SEM measurements show uniform film that conforms to the substrate, without any structure or crystal formation. The crystal structure of the initial Sb layer also seems to play a role in obtaining a high QE. Nakamura and collaborators demonstrated that the achievement of photocathode QEs as large as 40% in the visible range at 400 nm is related to the presence of well-defined Sb crystal peak on the X-ray diffraction spectra of the initial metal layer [7.30]. The role of the substrate is discussed in Section 7.3.3.2.

#### *7.3.2.5 $Cs_2Te$*

For a cesium telluride photocathode, the substrate (typically optically polished molybdenum) is held at 120 ˚C while several nanometers of Te (about 10 nm) are deposited at a rate of 1 nm per min. The film then is illuminated by UV light at 254 nm, and the Cs is evaporated at a similar rate while the photocurrent is monitored. When it reaches its maximum, the source and the substrate heater are turned off simultaneously [7.31]. Recent results [7.27] indicate that illumination with a longer wavelength (365 nm) supported the clearer identification of the QE peak (Figure **7.4**).

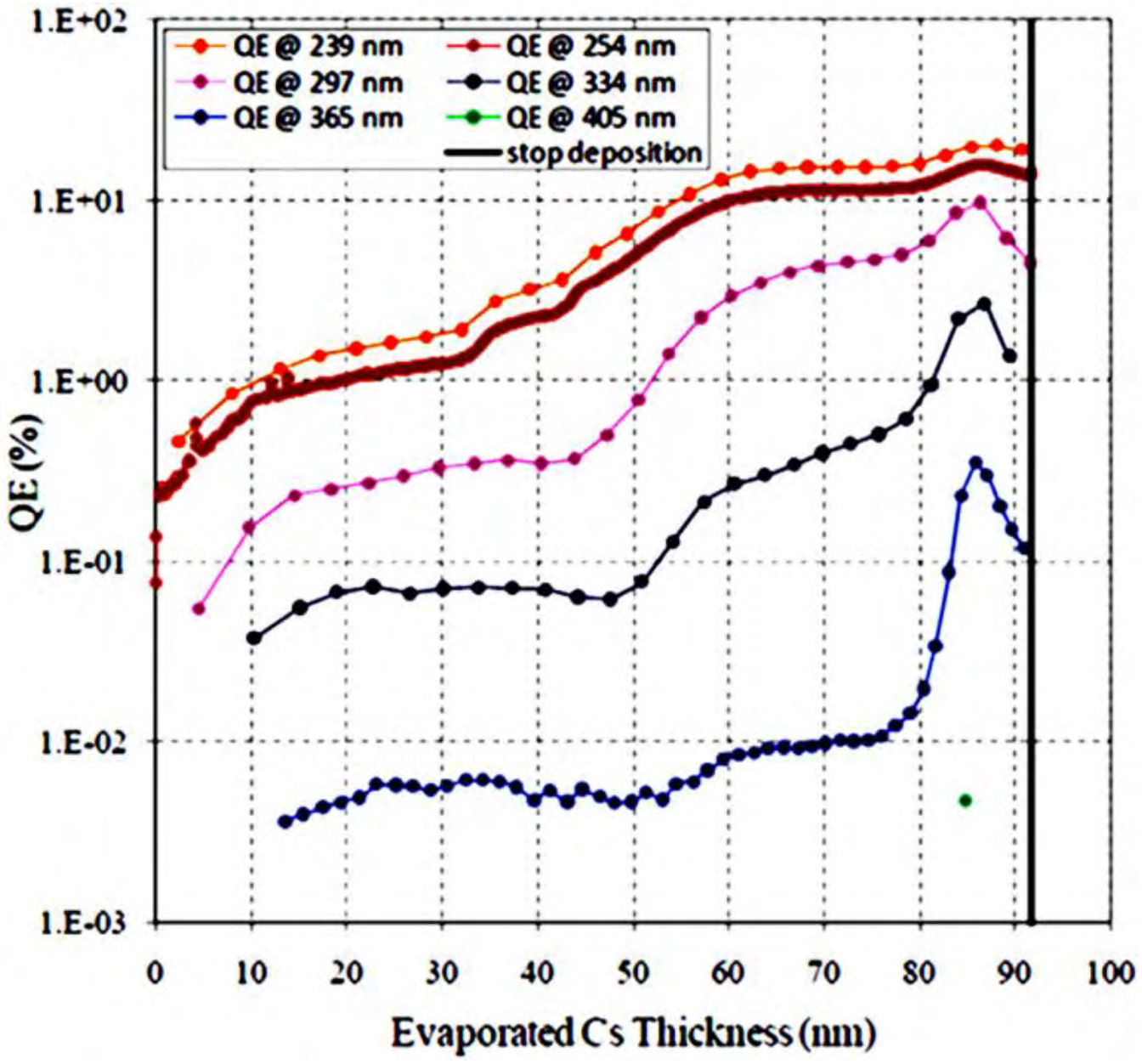

**Figure 7.4. Quantum efficiency of a cesium telluride photocathode during Cs deposition. The photocurrent is measured at different wavelengths. By using photons at 365 nm, the peak on the QE is more evident. [[7.32]; Available under Creative Common Attribution 3.0 License (www.creativecommons.org/licenses/by/3.0/us/) at www.JACoW.org.]**

Following this recipe typically ensures a QE of 10% at 264 nm [7.32]. The QE does not degrade if the cathode is stored in a UHV system with a base pressure of ~$10^{-8}$ Pa. Auger spectroscopy performed during the deposition showed that the stoichiometry of the photosensitive materials gradually evolves during the reaction with Cs, and the peak of the sensitivity is reached when the ratio between Te and Cs is about 1:2 [7.33].

### 7.3.3 General Considerations for Cathode Fabrication

**Requirements of the Deposition System**

There are several common aspects in designing a deposition chamber for either $Cs_2Te$- or alkali-antimonide-cathodes. The mechanical design is affected heavily by the size of the photocathode (which determines the distance required between the substrate and the sources to achieve good uniformity), and the load-lock mechanism for transferring the cathodes from fabrication to the photoemission gun while maintaining the UHV condition for the cathode.

#### *7.3.3.1 Metal Sources*

Due to the strong reactivity of the chemicals (especially the alkali metals) involved in the growth of photocathodes, the choice of the sources and the design of evaporators deserve special attention. Also, the purity of the evaporated species can strongly influence the photocathode's performance.

**Sb:** The source of high purity antimony either can be a metallic compound containing antimony, or high purity elemental antimony. Baumann *et al.* showed [7.34] that Sb from a metallic compound evaporates in small, possibly monatomic units of Sb, whereas that from elemental antimony appears as larger aggregates, such as $Sb_4$. For the simple thermal evaporation of commercial high purity (5N) Sb, the beads are placed in a tungsten- or tantalum-boat electrically heated up to a temperature where the equilibrium pressure of Sb under UHV is large enough to achieve a reasonable deposition rate. The latter depends on the size and distance of the substrate to the source. The temperature for evaporating Sb at a sufficient rate lies between 250-300 ˚C (the temperature for an equilibrium pressure of Sb at $10^{-6}$ Pa is 280 ˚C). Sb is evaporated mainly as tetramer ($Sb_4$) at these temperatures. If the photocathode recipe involves the co-deposition of Sb, it is convenient to use a cracker Knudsen cell that can break tetramers into dimers or even single atoms. Atomic Sb can improve reactivity with alkali metals.

Antimony also can be sputtered on to the substrate either in-*situ* or ex-*situ*. Substrates where antimony is layered ex-*situ*, the layer needs to be cleaned before evaporating the alkali metal, using techniques such as ion bombardment or heating. XPS measurements indicate that oxygen is the primary surface contaminant, most of which can be removed by heating the substrate to 400 ˚C. X-ray diffraction and SEM measurements of ex-*situ* sputtered Sb on Mo indicate that the Sb layer is amorphous for thicknesses up to ~ 7 nm, beyond which randomly oriented crystalline features were observed.

Tellurium also is evaporated from high purity beads (5N) but the needed temperature usually is lower than that required for Sb (the temperature for Te vapor pressure of $10^{-6}$ Pa is 155 ˚C). As with antimony, Te also sublimates as tetramer, so that using a cracker cell is recommended in MBE reactors. If the high vapor pressure of Te is a problem, an alternative approach is evaporating Te from a binary compound whose dissociation temperature is above the baking temperature. Indium telluride decomposes at 400 ˚C, above the bake temperature of most UHV systems, and its vapor pressure is under $10^{-6}$ Pa.

It is a good practice to degas the Sb/Te sources under vacuum before deposition to ensure the complete removal of surface contaminants that could be present either on the evaporation boat or the metallic beads.

**Alkali:** Different commercial sources are available for alkali metals: Alkali metal chromate powders, eutectic alloys, and pure metals.

Alkali-metal chromate powders mixed with a reducing agent are sold in small evaporators that must be resistively heated to a high temperature (can be more than 550 ˚C) before the chemical reaction yielding the metal vapor takes place. While the amount of alkali metals available in a single commercial dispenser is limited to a few milligrams each (usually suitable to generate a few photocathodes), they can be custom designed for larger capacity. These dispensers are not strongly reactive and can be exposed to air for assembling the source before placing them inside the vacuum chamber. For long term storage, the dispensers need to remain in a sealed, nitrogen filled container or be placed in a desiccator. Before starting the deposition, it is a good practice to degas these dispensers under vacuum by passing a current high enough to effectively heat them without starting the chemical reaction that generates alkali vapor. This current depends on the geometry and is usually specified by the manufacturer. During this process, the residual gas analyzer is used to monitor the partial pressure of the chemical species released (mainly CO, $CO_2$ and $H_2O$); degassing is complete once these are stabilized to partial pressures below $10^{-8}$ Pa. The presence of a getter (reducing agent Zr 84% Al 16%), however, usually precludes the complete release of the gasses from these dispensers. When designing the deposition chamber with short substrate-to-dispenser distances, the radiant heat load on the substrate during the high temperature heating of the dispenser should be taken into account. At source temperature in the range of 550 ˚C, the substrate temperature can exceed 100 ˚C inadvertently inducing unwanted changes in the specimen's surface during deposition.

Dispensers based on intermetallic bismuth alloy are an alternative to the metal-chromate sources. They offer the advantage of providing larger quantities (for example, 25 mg of K and 100 mg of Cs in a 3 mm diameter container are available from Alvatec; and, higher quantities are available in larger diameter containers) of evaporable material with a much lower gas load while releasing the alkali species. Their main disadvantage is that these alloys still are strongly reactive with oxygen-containing species, and hence, are sold in small tubes filled with argon and sealed with an indium cap. Once the seal is broken, the sources cannot be exposed to air without becoming completely contaminated. The seal is broken by passing a small current (~ 4 A for a 3 mm diameter source) through the dispenser in vacuum, sufficient to heat and melt the indium. This step can be done towards the end of the bake cycle, when pumping is primarily through the turbo pump that is able to handle argon, and the partial pressure of contaminants is low. These dispensers also need to be heated to temperatures > 400 ˚C for sublimation of the alkali metals, thus, an eventual heat load on the substrate surface should be taken in account.

For mass producing large quantities of photocathodes (tens of cathodes without replacing the sources), using pure alkali metal sources should be considered. Alkali metals, such as K, Na, and Cs, can be purchased in sealed glass-ampoules filled with argon. Typical purities of 3N and amounts of material from few grams to many grams are obtained readily allowing the deposition of a very large number of specimens. The hazards of handling these ampoules should be taken into account considering their high reactivity with oxygen and water.

Commercial and custom designed MBE Knudsen cells can be used for loading the alkali metals in the deposition chamber. Since the evaporation temperature of these species is very low (~30 ˚C for Cs, ~65 ˚C for K, and ~140 ˚C for Na), care should be taken during the baking procedure of the UHV deposition chamber. The high capacity of these sources (up to many grams) and the expected long term stability (several hours for a fixed temperature) of the fluxes coming from MBE sources in principle will allow tens of deposition cycles depending on the source capacity before needing to refurbish the alkali metals, thereby supporting the optimization of growth parameters for achieving the best QE photocathodes. Another valuable aspect of these sources is their ability to produce uniform fluxes with a larger evaporation cone, making the alignment of the sources with the substrate less critical compared to conventional dispensers.

Eutectic alloys of alkali metals (NaK or CsK) can be used as an alternative; however, there are no publications documenting the results of using these alloys as sources of alkali metals for synthesizing bi-alkali photocathodes.

#### 7.3.3.2 Substrate Requirements

The substrate for depositing a photocathode to be used in a photoinjector should incorporate the following abilities, *viz.*, to deliver high average replenishment currents (greater than milliamperes), and to conduct heat induced by the laser, which can be up to several tens of Watts. The glass substrates used in PMTs are unsuitable for this purpose due to their poor electrical- and thermal-conductivity. Stainless steel, molybdenum, or highly doped Si substrates, on the other hand, were shown to perform well.

The inertness of the substrate to the chemical species involved on the photocathode material growth should be considered to avoid diffusion at the boundary between the film and the substrate. Copper is a poor choice for the substrate, as it diffuses inside the photo-emitting layer [7.9], so explaining low efficiency of photocathodes grown on this material [7.35].

Assuring the cleanliness and flatness of the substrate is critical for photoinjectors. It is important to avoid unwanted contaminants on the substrate's surface that can diffuse inside the photosensitive layers or participate in the chemical reactions during growth. Oxygen-containing species, water vapors, CO, and $CO_2$ must be removed from the substrate surface before deposition. These molecules can be removed by thermal heating or ion sputtering, and their absence verified by surface analysis techniques, such as Auger spectroscopy. The most practical cleaning methods involve heating the substrate's surface above 400 ˚C [7.31]. For a Si substrate, the wafer is first rinsed in hydrofluoric acid to remove the native oxide layer, and then heated to 550 ˚C to remove the hydrogen passivation layer [7.36].

The flatness of the substrate plays an important role in determining one of the fundamental parameters of a photocathode, that is, its thermal emittance. Rough surfaces can cause considerably higher emittances of up to 60% [7.37]. Additionally, the chemical reactions leading to the formation of the photocathode seemingly are positively affected by good flatness in a substrate [7.35]. As the typical final thickness of the photoemitting layer is about 100 nm or less, the substrate's flatness is a key parameter, not only for low thermal emittance, but also to minimize the occurrence of field emission nucleated by surface defects.

The substrates for the $Cs_2Te$ cathodes used in the TTF, PITZ, and FLASH injectors were formed from sintered or arc-cast Mo, machined to design specifications [7.31]. The plug is then chemically cleaned by buffered chemical polishing to remove the contaminants from machining. The surface roughness, a cause of dark current in a high electric field, is reduced by optical finishing. Figure **7.5** illustrates the improvement in the dark current with a polished substrate compared to a rough surface. To obtain an optical finish, the surface is successively polished with a set of diamond polishing compounds of decreasing grain size, the smallest being 0.1 µm. The measured reflectivity at 543 nm is 56.3%. The polished surface is then rinsed with acetone and ethanol in an ultrasonic cleaner, and then heated at 500 ˚C in UHV for approximately 30 min to reduce surface contaminants before fabricating the cathode.

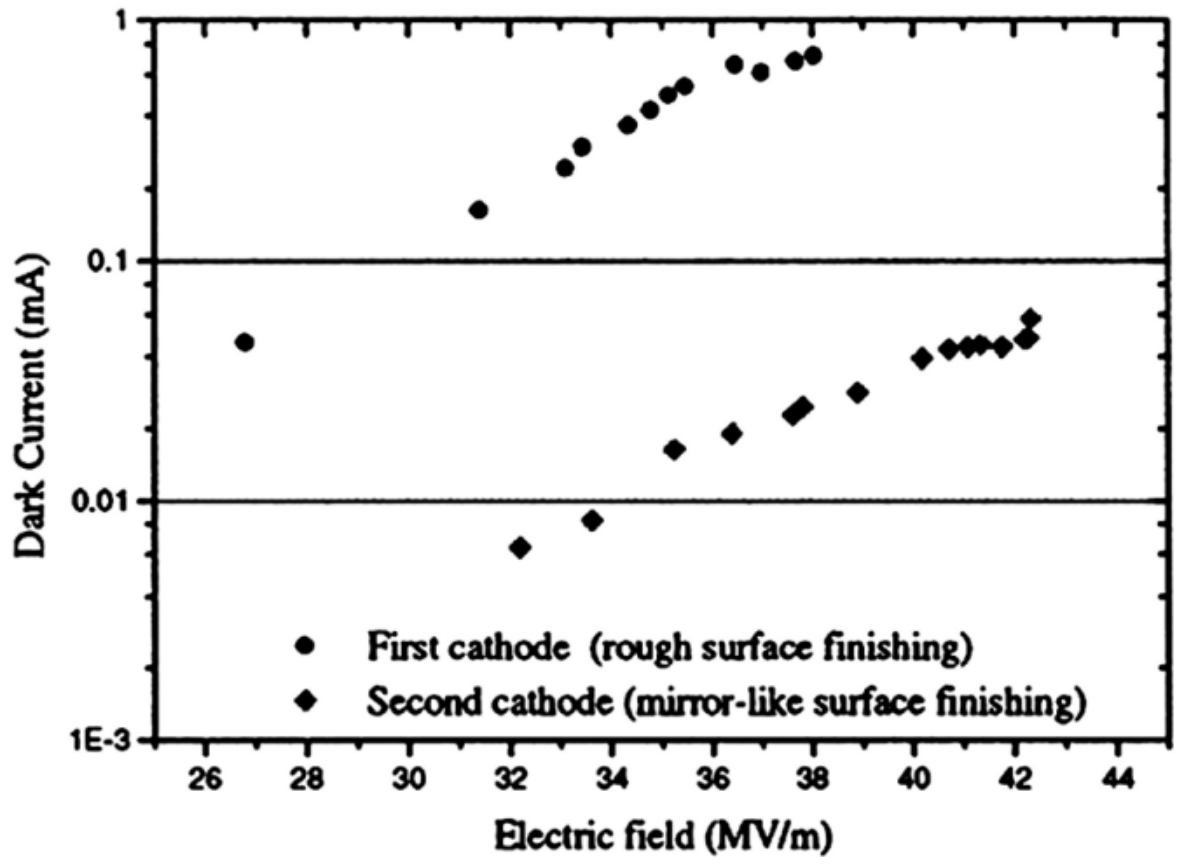

**Figure 7.5. Effect of surface finish of the substrate of a $Cs_2Te$ cathode on the dark current. [Reprinted from [7.38], with permission from Elsevier.]**

### *7.3.3.3 Vacuum Chamber Requirements*

The essential equipment for fabricating the cathode consists of a UHV chamber with high purity metals sources. The alkali sources and the cathodes themselves are very sensitive to contaminants (especially to water vapor, CO, and $CO_2$). The chamber should be able to reach UHV levels of $< 10^{-8}$ Pa. The system must be bakeable to ≥ 200 ˚C to completely remove water vapor. For vacuum isolation, all metal valves are preferable to reduce out-gassing, although metal bonnet valves are acceptable if there is a vacuum on both sides of the valve. Either magnetic- or bellow-translators are acceptable. The instantaneous increase in pressure during movement is higher for the latter than for the former. The system should have a sufficient number of ports to accommodate several viewports, electrical feedthroughs, and diagnostics. The viewports are used for observation and for irradiating the cathode to monitor the QE. The electrical feedthroughs are used for heating the sources and the substrate, connecting the thermocouple to measure relevant temperatures, and for measuring the current from the cathode to determine the QE. The electrical feedthroughs and connectors for heating should be chosen to minimize their heat load. The design and construction of the cathode preparation chamber should include the interface system for transporting the cathode from the preparation chamber to the gun. We discuss the details for the transport system later in this section.

#### *7.3.3.4 Required Basic Instrumentation*

The basic instrumentation needed to grow and perform initial characterization of the photocathode properties is the following:

- Metal Sources: Should provide uniform deposition profile with overlap from the other sources. The power supplies must be rated sufficiently high to heat the evaporators to the right temperature and ensure reliable evaporation. Equipping the sources with a thermocouple readout is optional, but useful for reproducing experimental results when the operational temperature of the sources is recorded. Again, it is preferable to isolate the sources to avoid their cross-contamination; their design should include a separate pumping system and isolation valve.
- Cathode (Substrate) Heater: The heater should be able to heat the cathode to higher temperatures (> 800 ˚C) when previous depositions must be removed from the substrate. A quartz heater is preferable since it reduces the accompanying gas load compared to tungsten filaments. A calibrated thermocouple to measure the temperature of the substrate is essential.
- Quartz Microbalance: This equipment is needed primarily to measure the thickness that is critical for forming stable stoichiometric compounds and growth rate of Sb or Te layers. The growth rate is measured either by placing the crystal monitor at the same location as the substrate and establishing the operating parameters to obtain this evaporation rate, or by positioning it at a different location and measuring the evaporation rate as the substrate is layered with Sb/Te. Since, in the latter approach, the monitor is not at the same place as the substrate, the thickness displayed by the microbalance must initially be calibrated against that on the substrate. Two different techniques can be used for calibration: 1) Measuring the optical transmission of a light beam of known wavelength through a transparent substrate while evaporating antimony; or, 2) taking the measurement externally. In the first technique, an optical beam of known power passes through the substrate before and after evaporation; thereafter, the thickness of Sb/Te is calculated *via* the formula $P_{out} = P_{in}e^{-\alpha t}$ ,where $P_{in}$ and $P_{out}$, respectively, are the incident- and transmitted-power, $\alpha$ is the absorption coefficient, and $t$ is the thickness of antimony/tellurium. Since the Sb aggregate from different antimony sources may differ, care must be taken to use the correct $\alpha$ in calculating the thickness. The sticking coefficient of the transparent substrate must be properly accounted for. In the second approach, the substrate with Sb/Te is removed from the vacuum chamber, and the signal (such as fluorescence, XRD) from elemental metal is measured against that of one with a calibrated thickness.
- Residual Gas Analyzer (RGA): The RGA is employed to monitor the composition of residual gas in the deposition chamber during all steps of the growth. It also is useful in indicating the onset of alkali-metal evaporation when the sources are heated. We recommend an RGA capable of detecting materials with a high atomic mass unit, such as Cs and Sb.

- Cold Finger: A cold finger aid in rapidly cooling the substrate and the photocathode after deposition, so that the alkali species do not evaporate from the finished photosensitive layers. Care must be taken to assure that it is not the coldest surface in the vacuum system because that would attract contaminants.
- Quartz Windows: These windows are used to illuminate the cathode's surface during the reaction with alkali metals to measure the photocurrent and QE while evaporating different alkali species. Quartz is essential when using the UV light for $Cs_2Te$ growth and for spectral response measurements. Mechanical shutters to these windows reduce any unwanted metal coating of their surfaces.
- Light Source and Lock-In Amplifier: A $D_2$- or Hg-lamp with a monochromator is useful for obtaining the cathode's spectral response. Ideally, the light source is equipped with a chopper so that a lock-in amplifier can be used to measure the photocurrent. The lock-in amplifier also removes the background DC current generated by stray photons of different wavelengths entering the UHV chamber.
- Vacuum Pumps and Gauges: These should be able to handle the range of pressure and the atomic species during evaporation. We note that for a given vacuum level, the base current of an ion pump may increase due to the metal coating of its electrodes, engendering a corresponding change in resistance. Hence, using ion pump current to extrapolate the vacuum level should accompany frequent calibration.

There are two possible source-substrate designs: All sources pointing to the substrate center, with substrate held in one position (Figure **7.6(b)**), or a substrate that can be translated in front of each of the sources in succession (Figure **7.6(a)**). Each design has its advantages and disadvantages. Figure **7.6(b)**, both the substrate and the crystal monitor are held in one position and the evaporation is achieved either by heating one source at a time or in parallel for co-deposition. The geometry of the sources and the substrate must be carefully designed to maximize the overlapping region of evaporation cones as shown in Figure **7.6(b)** and to minimize the evaporation time. The major disadvantage is non-uniformity of the cathode over the entire evaporated area. In the latter design, the substrate is positioned normal to each source, resulting in a better uniformity of the film. However, both the substrate and the crystal monitor should be translated when changing the sources, so complicating the design of the substrate holder. Both designs should allow the sources to be retracted away from the deposition chamber when not in use.

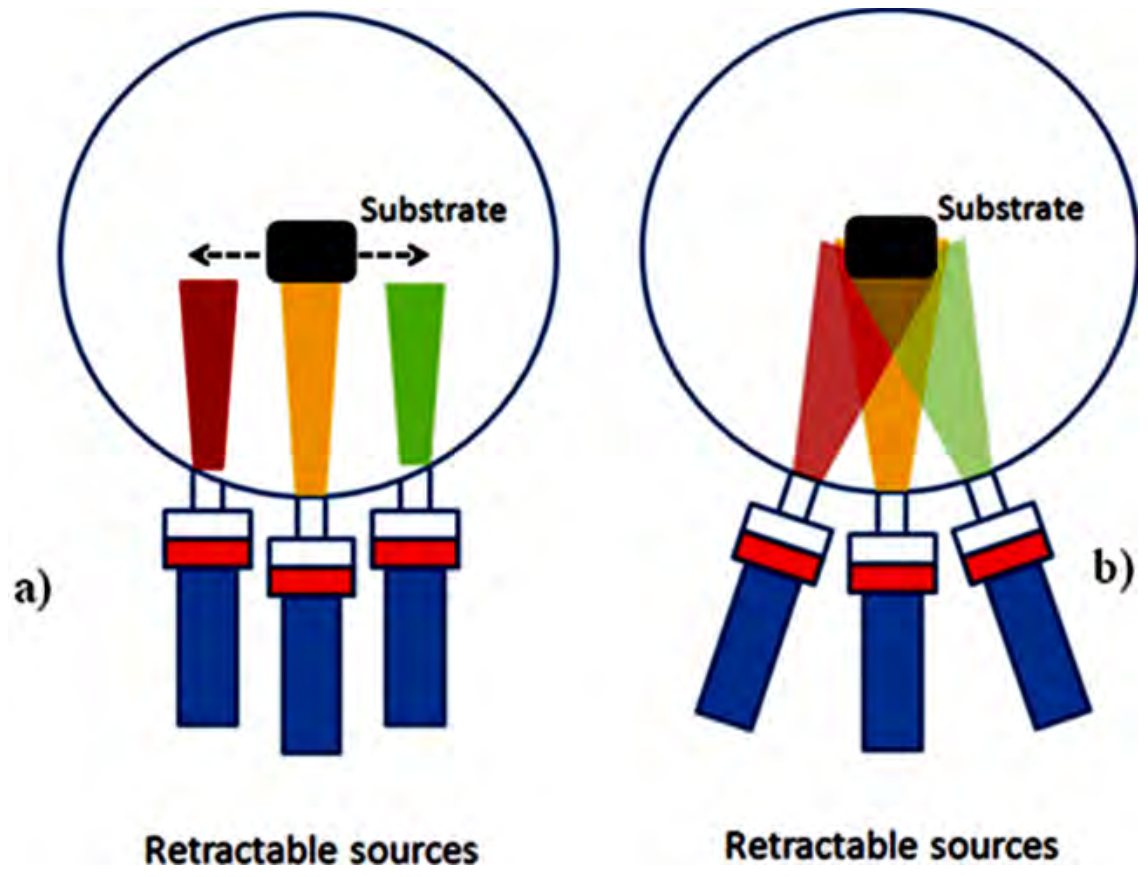

**Figure 7.6. Two possible configurations for growing alkali-based photocathodes. (a): The substrate is shifted frontally to each source for a sequential deposition of different elements. (b): The sources are pointing towards the center of the substrate with the evaporating fluxes overlapping over its surface.**

The current emitted from the cathode can be measured either from the cathode, or collected by a low voltage bias anode close to the substrate between the electrodes to overcome space charge effects. A QE map of the cathode area is obtained by irradiating light at a single wavelength and scanning a small laser spot on the cathode while measuring the QE. Figure **7.7** shows examples of such scans.

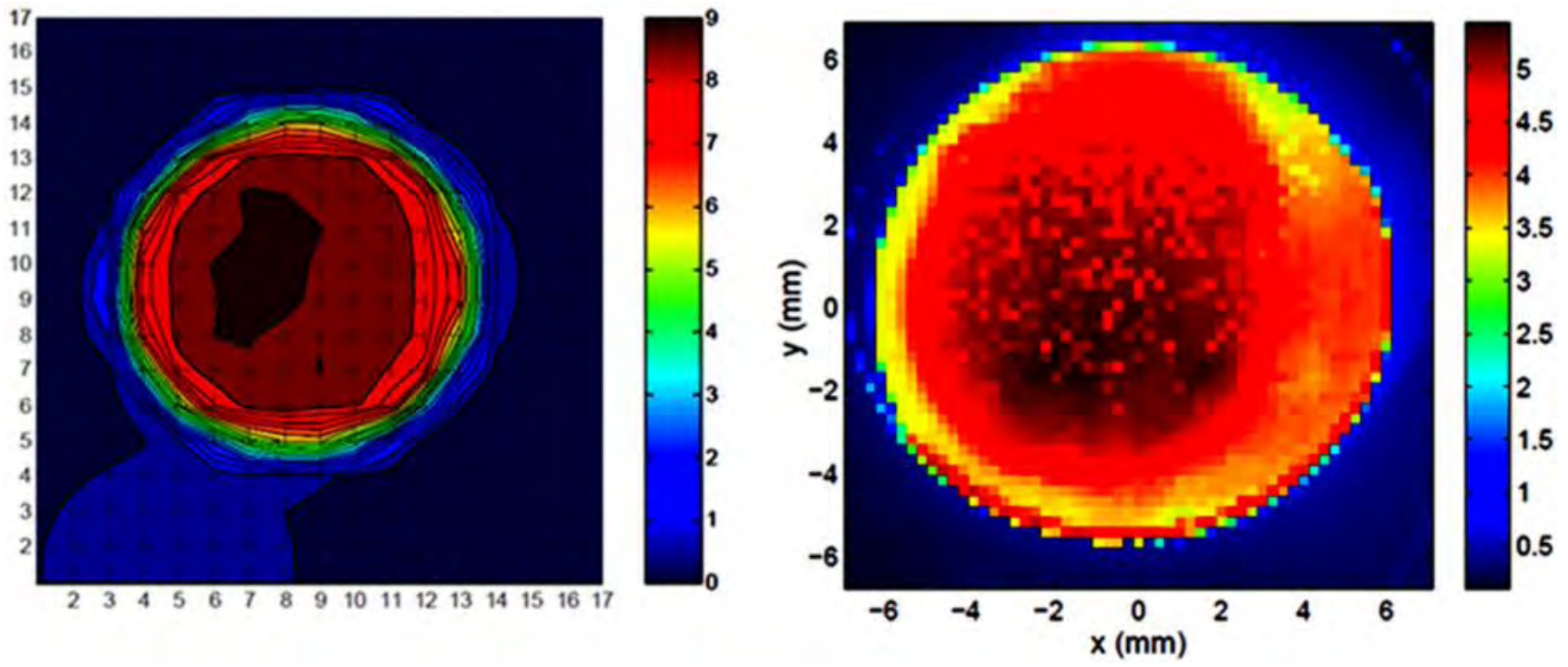

**Figure 7.7. Examples of 2-D scans to measure the QE uniformity: A $Cs_2Te$ photocathode at 254 nm (left), and a $Cs_3Sb$ photocathode at 532 nm. [Reproduced from [7.39]] [Reprinted with permission from [7.13]. Copyright 2011, American Institute of Physics.]**

Figure **7.8** is a schematic of an UHV growth chamber with the basic instruments needed to synthesize a semiconductor, high QE photocathodes. Therein, the main chamber is equipped with an RGA for monitoring the constituent gases, and gauges for measuring the overall vacuum level; it has two windows for light transmission (input and output), a quartz microbalance crystal to measure deposition rates and thicknesses of evaporated layers, and the substrate holder equipped with a heater and a cold finger. Several vacuum ports allow the connection of a pumping system (a combination of ion pump, titanium sublimation pump, and NEG modules) that can reach a pressure below $10^{-8}$ Pa. The metal sources facing the cathode surface can be retracted and isolated from the main chamber *via* vacuum gate valves to prevent cross-contamination. Another valve allows insertion and removal of the cathode substrate without breaking the vacuum of the main chamber.

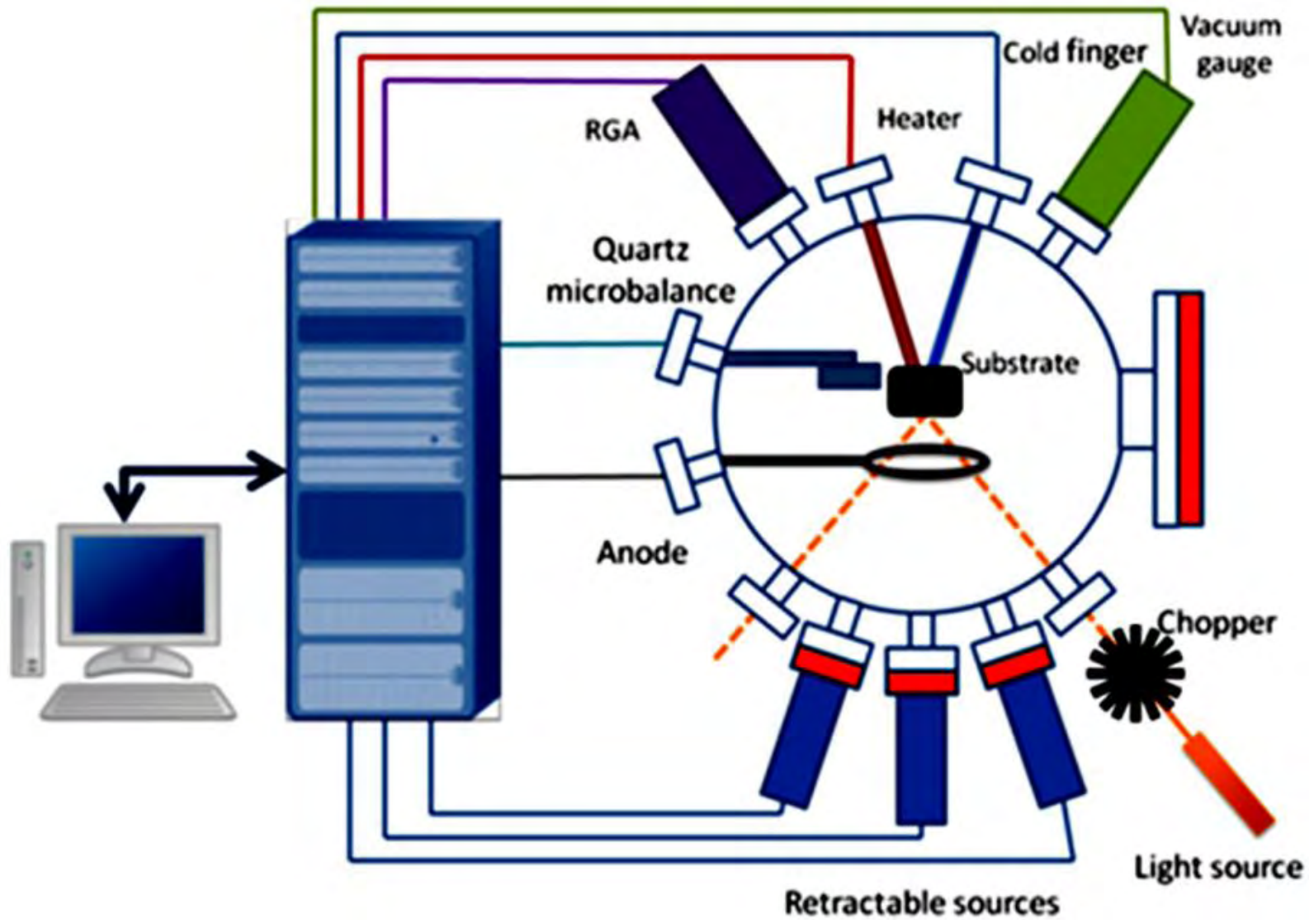

**Figure 7.8. Schematics showing a typical UHV chamber set-up for growing alkali-based photocathodes. All instrumentation is interfaced with a personal computer for setting the experimental parameters for the growth and recording the data for each deposition. Red rectangles indicate the gate valves used for sealing the metal sources to avoid cross-contamination and for loading the photocathode substrate. For simplicity, the pumping systems for both the UHV chamber and individual source assemblies are omitted.**

When the cathode in the gun must be replaced often, it is convenient to design a suitable vacuum suitcase and a photocathode storage chamber that can accommodate multiple substrates simultaneously. The suitcase is used to transfer the cathode between the storage or the deposition chamber and the photoinjector under UHV conditions. With such a set-up, $Cs_2Te$ cathodes are fabricated in LASA, Milan, Italy, and routinely transported to injectors in TTF, PITZ, and FLASH. Similarly, the $K_2CsSb$ cathode was fabricated at BNL and tested in a DC gun at JLab. It is desirable to have the ability to monitor the QE of the photocathodes during and after their transport to the photoinjector. We have used magnetically coupled translators successfully for moving the substrates back and forth from the vacuum suitcase and the cathode to and from the injector. The following are the essential components of the transfer system (Figure **7.9**):

- A UHV transfer chamber with translators to move the cathode from the fabrication chamber to the injector
- a pumping section to maintain vacuum during transport
- an intermediate section with two valves and pumping station connecting the fabrication chamber/injector. The temperature of the cathode must carefully be maintained close to room temperature when this intermediate section is baked and pumped before opening the valves to transfer the cathode.

### 7.3.4 Photocathode Performance

#### *7.3.4.1 Quantum Efficiency and High Current*

Several institutions have fabricated $Cs_2Te$ cathodes and transported them successfully the RF injector, for example, CLIC [7.40], TTF [7.41]**,** FLASH [7.42], PITZ [7.43], and Los Alamos [7.44]. In the injector, the cathode must meet additional requirements, such as delivering a large photocharge at low emittance, reliably and reproducibly with very low dark current. The publication describes how $Cs_2Te$ cathodes with an average QE of 9% at 264 nm are reproducibly fabricated at LASA, Milan. This average QE fell to ~2.5% upon transport to, and use in the gun at DESY. The initial dark current from these cathodes depends strongly on the cathode's substrate and the accelerating field in the cavity [7.44]. During operation in the gun, the dark current from the cathode also increases, probably due to cavity conditioning. Furthermore, the gun's vacuum also degrades during high current operation resulting in additional exposure of the cathode to gas and its subsequent degradation of QE and increased the dark current [7.42]. Single pulse charges of up to 50 nC without any saturation effects were delivered with this cathode material. Pulse train lengths of 30 μs with a repetition rate of 1 MHz and a charge per bunch of 1-8 nC were produced routinely with the $Cs_2Te$ cathode at FLASH; furthermore, the train length was extended to 300 μs. The operational life time of the cathode, defined as the time at which the QE drops to 0.5%, is between 100-200 days; hence, a vacuum suitcase with two or three cathodes can service the injector for over a year.

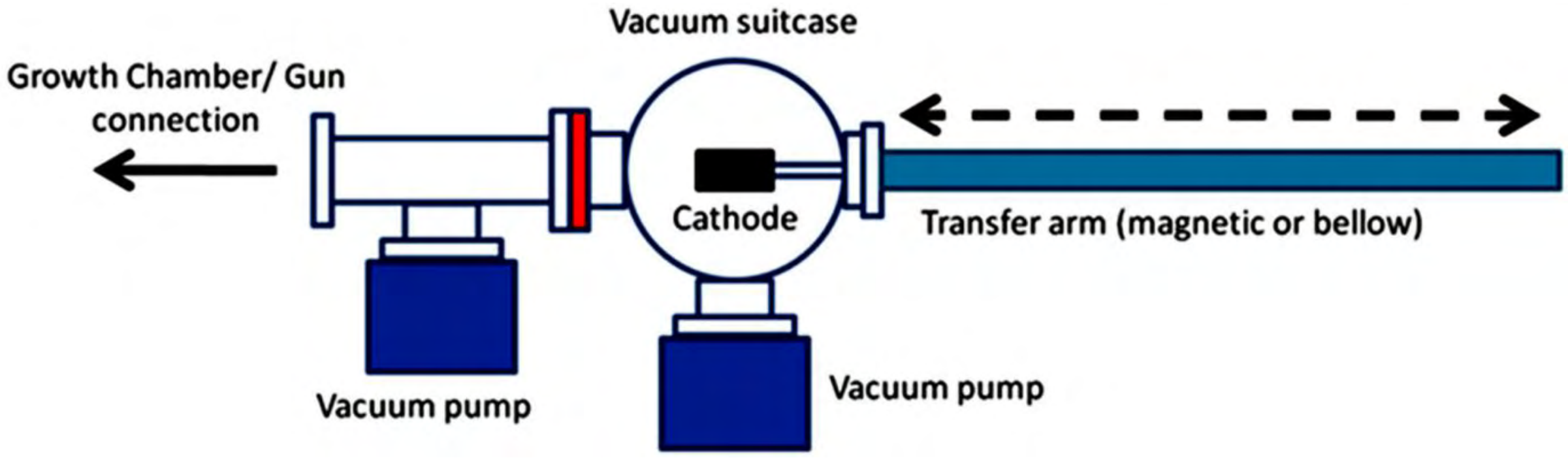

**Figure 7.9. Schematics of a vacuum suitcase for transferring photocathodes from the deposition chamber to the electron gun. The red rectangle indicates the gate valves used to seal and keep the photocathode under UHV during transport.**

Similarly, $K_2CsSb$ cathodes were tested in both NCRF [7.7], [7.24] and DC guns [7.45]. They achieved QEs in excess of 14% at 532 nm, average current of 32 mA [7.24], and current densities exceeding 160 mA $mm^{-2}$ [7.45].

Beam brightness is defined as a ratio between the beam's current and its emittance (either in 2-D-, 4-D-, or 6-D-phase space). Accordingly, the way to increase the beam's brightness is by raising its current, lowering the transverse emittance, and/or the pulse duration-energy spread. In addition to the QE, other fundamentals characteristics of a photocathode operating in a photoinjector are its thermal emittance, response time, and lifetime.

### *7.3.4.2 Thermal Emittance*

A simple relationship can be used to roughly estimate thermal emittance [7.4]

$$\epsilon_{nx,th} = \sigma_x \sqrt{\frac{\mathrm{MTE}}{3m_0c^2}}; \text{ where MTE} = E_{ph} - (E_{gap} + \eta) + \phi_{Schottky} \tag{7.7}$$

Here $\sigma_x$ is the laser rms spot size, MTE is the mean transverse energy of photoemitted electrons, $m_0c^2$ is the electron rest energy, $E_{gap} + \eta$ is photoemission threshold energy, given by the sum of $E_{gap}$ and electron affinity, $\eta$. Thus, $E_{ph} - (E_{gap} + \eta)$ represents the maximum excess kinetic energy imparted to electrons by the photons. $\phi_{Schottky}$ is the effective reduction in work function when strong external electric field is present.

The Schottky lowering of the work function is determined by the effective electric field, $E$, at the photocathode's surface

$$\phi_{Schottky} = \sqrt{\frac{e^2E}{4\pi\varepsilon_0}} \tag{7.8}$$

or, in practical units

$$\phi_{Schottky}\,[\mathrm{eV}] = 0.0379\sqrt{E\left[\frac{\mathrm{MV}}{\mathrm{m}}\right]} \tag{7.9}$$

As the physics of photoemission processes is complex, the Equ. 7.7 can only be used as a rough guide. Even though the photocathode materials based on alkali-antimonides and tellurides have been known for decades, only recently have staff at different laboratories determined MTE values experimentally. Some of these results are shown in Figure **7.10** and Figure **7.11**.

The former depicts the very good agreement with Equ. 7.7 for $CsK_2Sb$. The latter illustrates the dependence of the thermal emittance on the intensity of the extracting field: at the higher fields, for a given photon energy, thermal emittance increases because of the reduction in work function due to the Schottky effect, Equ. 7.8. Table **7.1** in Section 7.2.4 compares the data with Equ. 7.7, indicating a fair agreement between theory and experiment for alkali-based materials. Normalized emittance of 0.37 µm $\mathrm{mm\text{-}rms^{-1}}$ has been measured from a $K_2CsSb$ cathode held at ~2 MV $m^{-1}$ and irradiated by 534 nm optical beam by [7.26]. Similar results have been obtained for $Cs_2Te$ cathodes as well. [7.44]

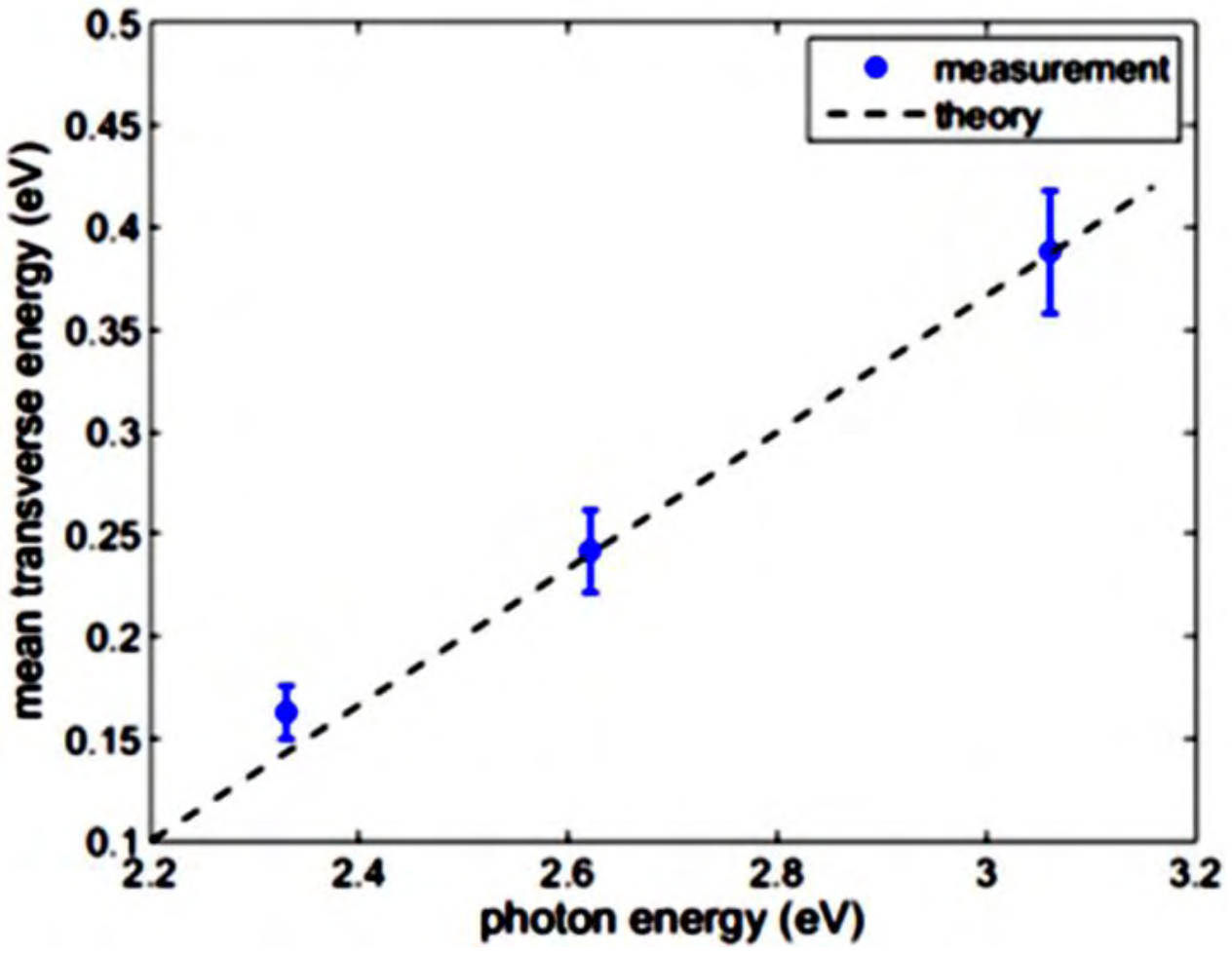

Figure 7.10. Comparison between the experimental values obtained for the MTE of electrons emitted by a $K_2CsSb$ photocathode at different wavelengths and the model prediction obtained using Equ. 7.7. [Reprinted with permission from [7.25]. Copyright 2011, American Institute of Physics.]

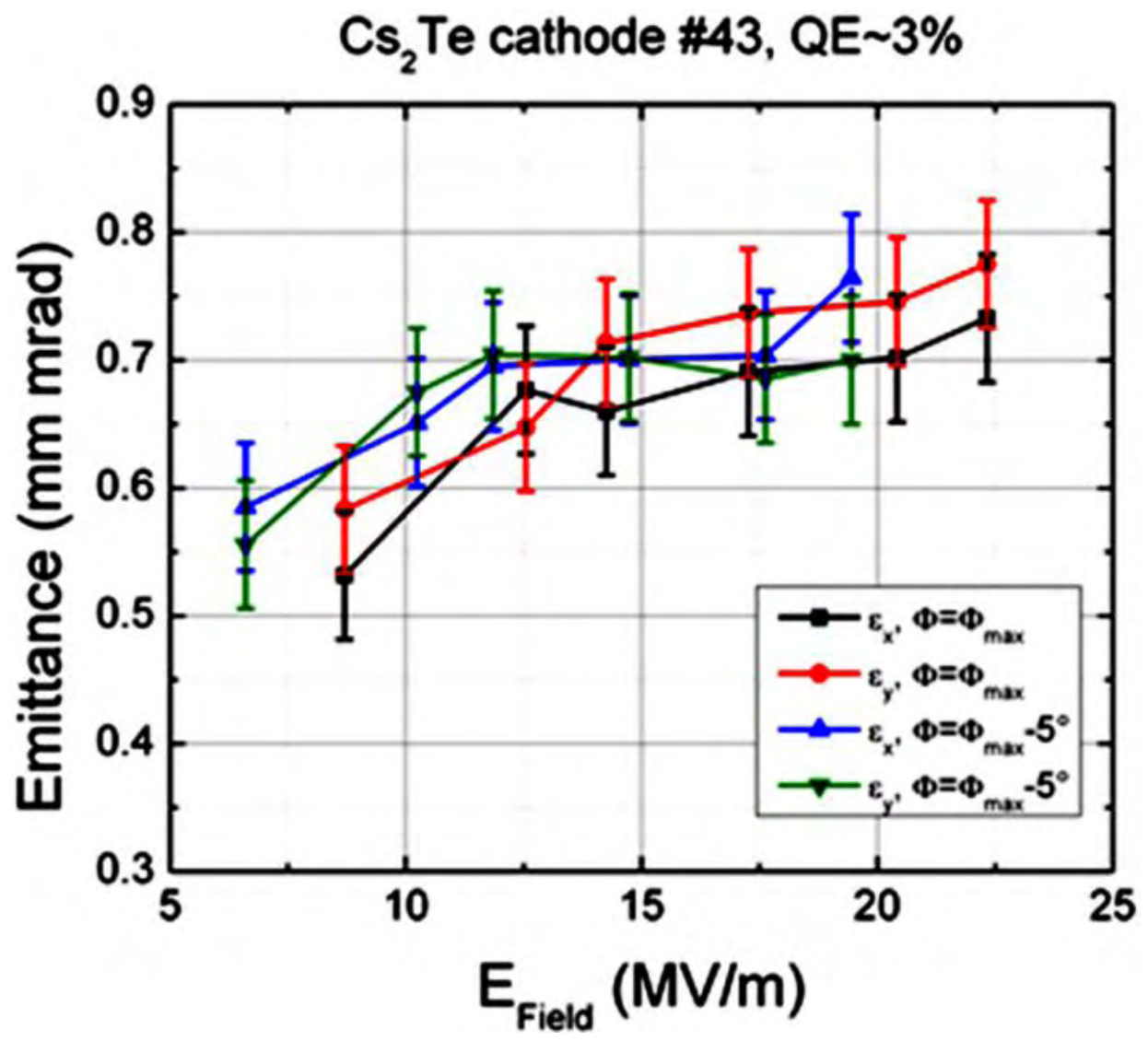

Figure 7.11. Thermal emittance measurement from a $Cs_2Te$ photocathode for different values of the extraction electric field. Thermal emittance increases at higher electric fields due to the Schottky effect. [[7.44]; Adapted under Creative Common Attribution 3.0 License (www.creativecommons.org/licenses/by/3.0/us/) at www.JACoW.org.]

#### *7.3.4.3 Response Time*

A photocathode response time of < 1 ps is desirable for most photoinjector applications so that the electron beam's temporal profile reflects the laser's temporal shape. A much longer response time (tens of picoseconds) usually is unacceptable unless a complex RF bunching and chopping system is being employed. Alkali-based photocathodes in the form of thin films are used in modern streak camera devices, demonstrating a very fast response time (a time resolution of 200 fs is commercially available) [7.46], [7.47]. Recent measurements carried out in photoinjectors by using an RF deflecting cavity confirmed that the response time from $Cs_2Te$ and $Cs_3Sb$ (Figure **7.12**) is on the picoseconds scale, or shorter [7.13], [7.48].

#### *7.3.4.4 Lifetime*

Lifetime is a critical parameter for a photocathode used for generating the beam in modern photoinjectors. For alkali-antimonide materials, the factor limiting the photocathode's dark lifetime essentially is poisoning, caused by the oxidizing species in the residual gases inside the gun chamber. Indeed, poor vacuum conditions during the operation of the Boeing RF gun restricted the operation of a $CsK_2Sb$ photocathode to about several hours due to reactions with water vapors inadvertently present in that RF gun's vacuum (Figure **7.13**) [7.24].

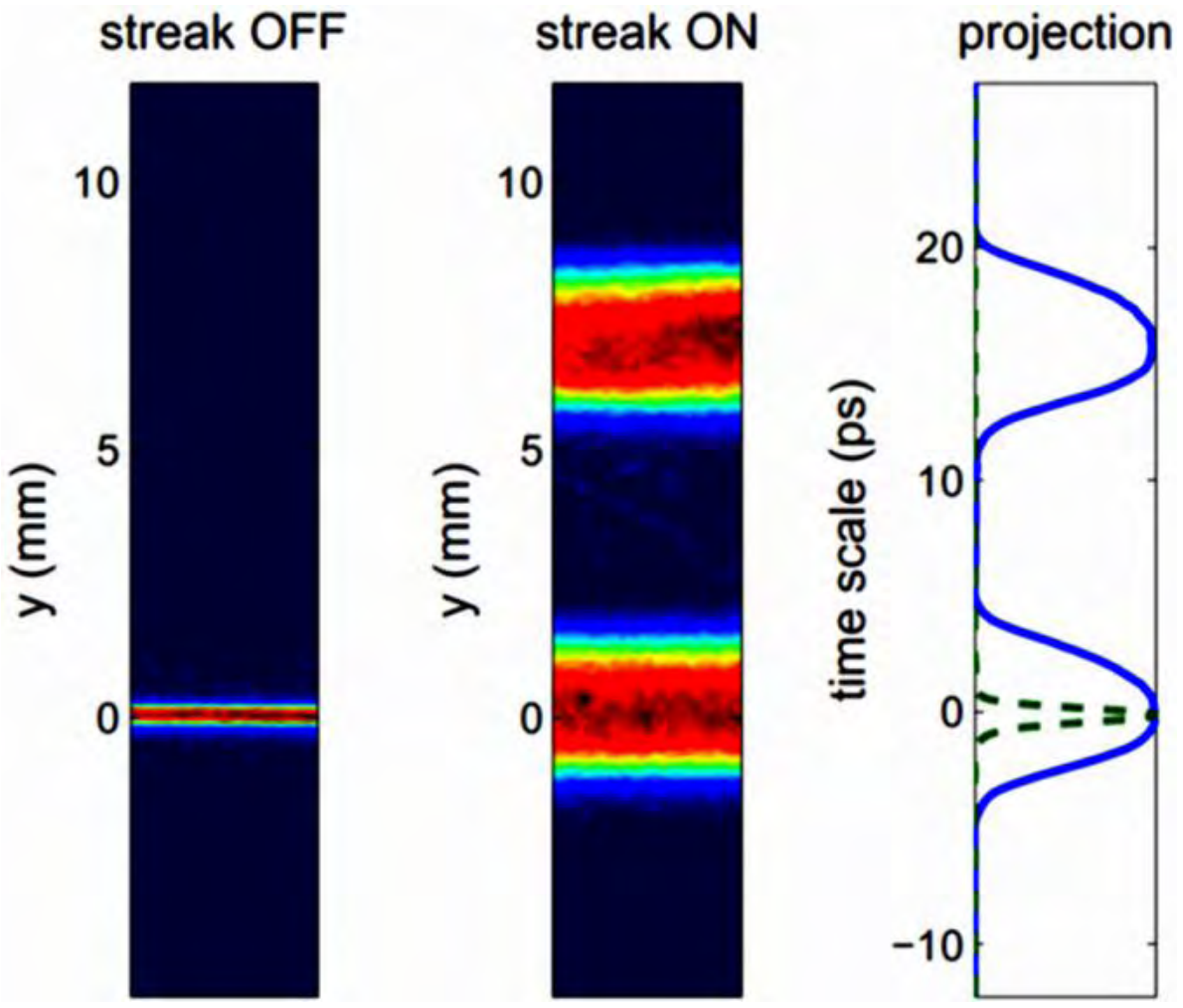

**Figure 7.12. Example of a measurement to determine the response time using an RF deflecting cavity. Electron beam generated by two laser pulses with a duration of about 1 ps rms and separated by 16 ps (green dashed line on the right plot) are vertically deflected by the RF cavity. The temporal profile is obtained using a view screen, and the upper estimate of the response time is obtained (additional causes contributing to the width are the RF laser synchronization, or fluctuations in arrival time). [Reprinted with permission from [7.13]. Copyright 2011, American Institute of Physics.]**

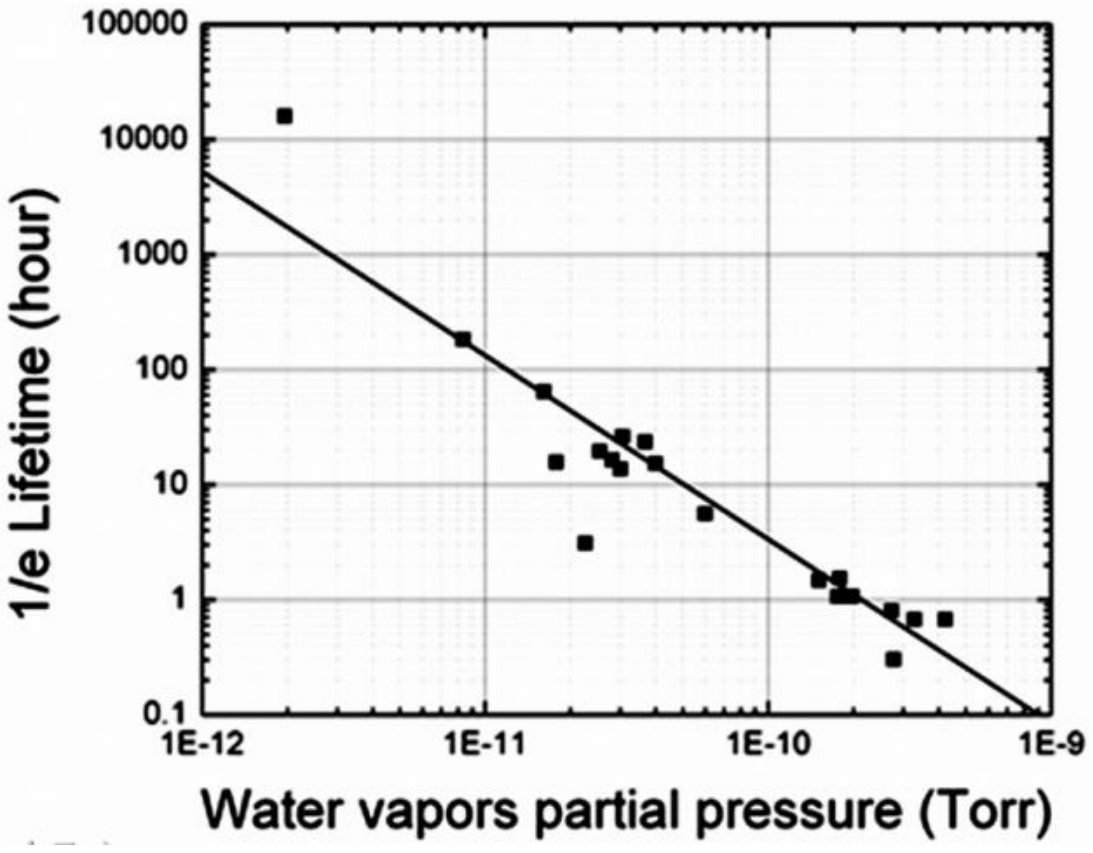

**Figure 7.13. Plot of experimental data of 1/*e* lifetime of $K_2CsSb$ photocathode operating at different levels of water vapor partial pressures (reproduced from). Recent results indicate possibly longer lifetimes [7.26]. [Adapted from [7.24], with permission from Elsevier]**

Recent studies indicate that the QE decays from 6% to ~1.5% over 20 hr in $10^{-10}$ Pa partial pressure of water when illuminated with 532 nm light [7.25].

The same poisoning effect due to poor vacuum might have contributed to the degradation of the QE of a $Na_2KSb$ photocathode operated at $10^{-6}$ Pa inside a RF gun at Tokyo University, even though the authors attributed the decay to the presence of high intensity electric fields, up to 100 MV $m^{-1}$ [7.49]. In a more favorable vacuum environment, $K_2CsSb$ cathodes are very stable and have demonstrated the ability to deliver high average beam currents for long periods. For example, the DC gun of the ERL photoinjector prototype operated at Cornell University with a base vacuum level of ~$10^{-10}$ Pa used a $K_2CsSb$ photocathode to deliver an average current of 20 mA and current density of 10 mA $mm^{-2}$ for 8 hr, with only a slight decrease in the QE (Figure **7.14**) [7.36].

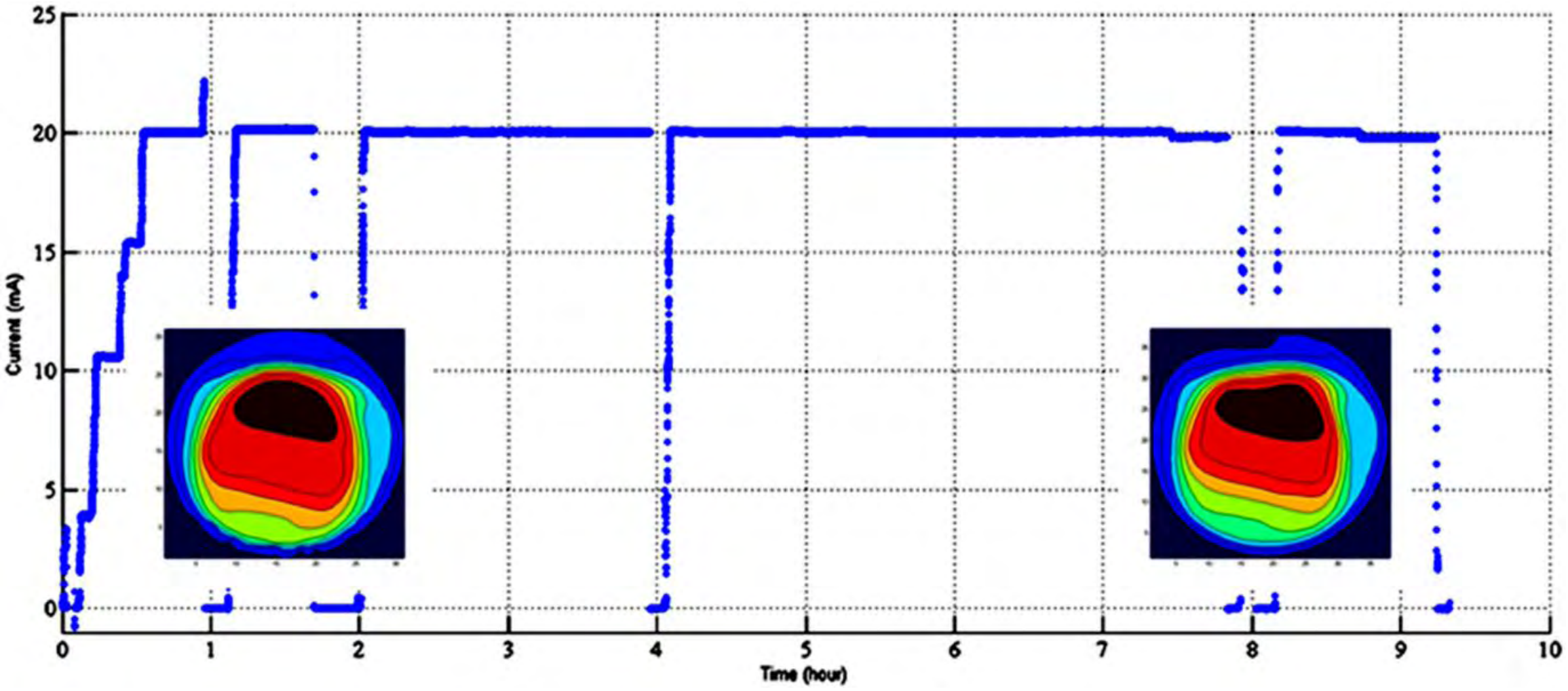

**Figure 7.14. The photocurrent extracted from a $K_2CsSb$ photocathode from the DC gun of the ERL prototype at Cornell University. A 20 mA average current was delivered for 8 hr. 2-D scans of the photocathode's QE show negligible degradation after the run (compare the QE maps before the run, on the left, and after the run, on the right). [Adapted figures with permission from [7.50]. Copyright 2011 by the American Physical Society]**

$Cs_2Te$ photocathodes are less sensitive than alkali-antimonide ones to poisoning by harmful residual gases [7.11]. Figure **7.15** plots the change in the QE of the former for typical residual gases found in a UHV system. $Cs_2Te$ photocathodes have demonstrated operational lifetimes for up to four months (Figure **7.16**) [7.50].

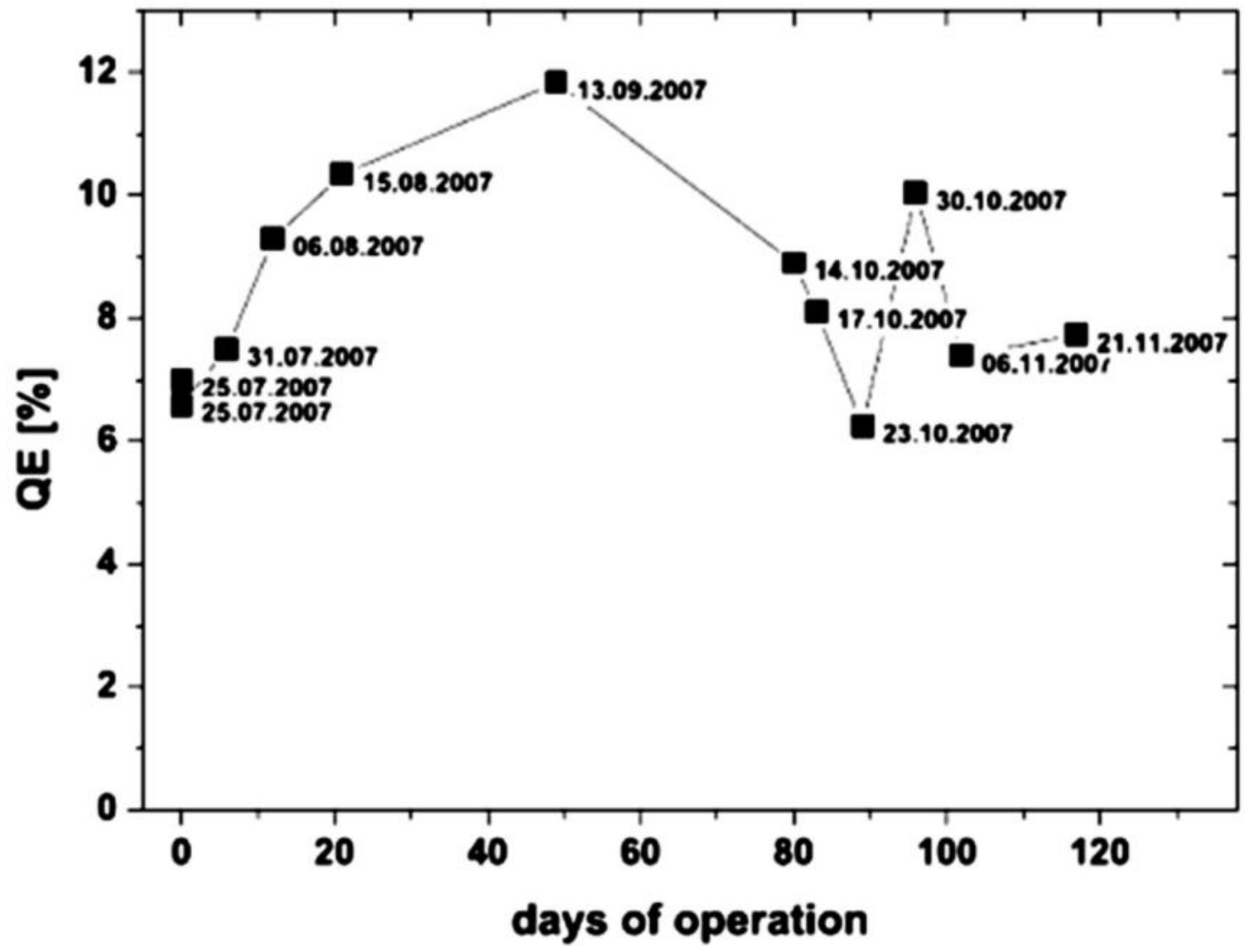

**Figure 7.15. QE measurement of a $Cs_2Te$ photocathode operated in a photoinjector showing an in-gun lifetime of about 4 months. [[7.51]; Available under Creative Common Attribution 3.0 License (www.creativecommons.org/licenses/by/3.0/us/) at www.JACoW.org.]**

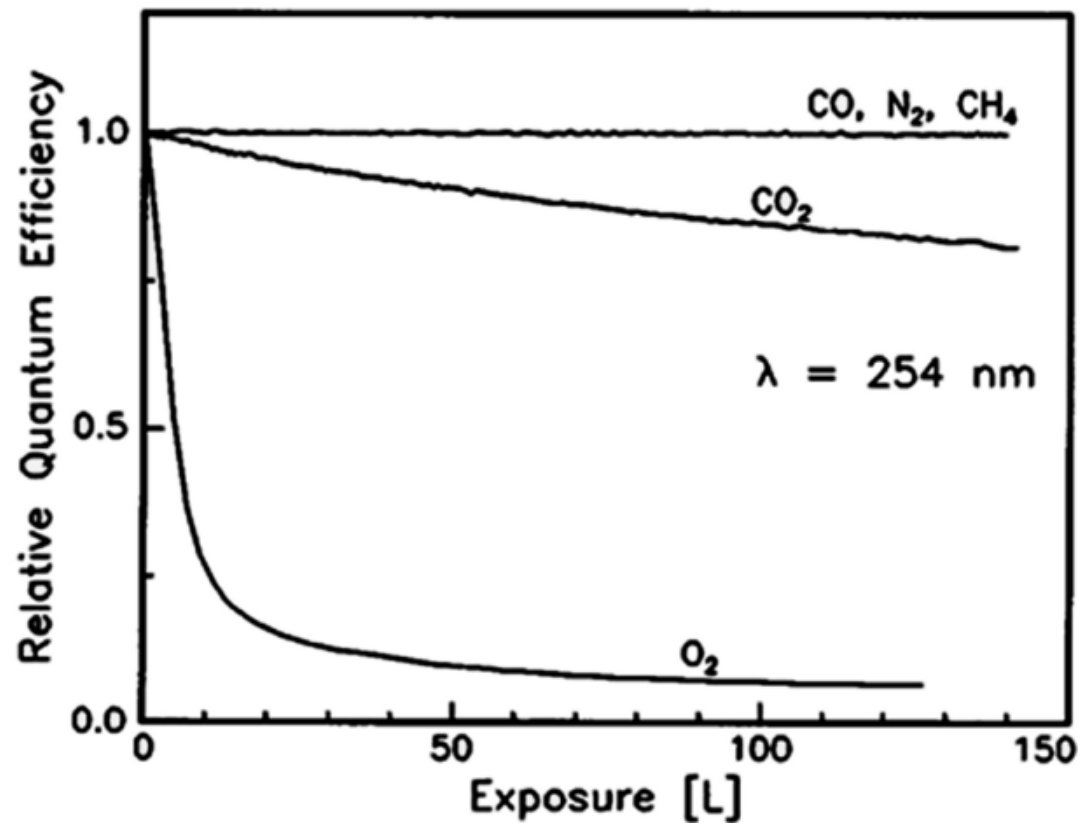

**Figure 7.16. Sensitivity of $Cs_2Te$ cathode to contaminants. [Reprinted with permission from [7.9]. Copyright 2011, American Institute of Physics]**

# 7.4 NEGATIVE ELECTRON AFFINITY CATHODES: III-V PHOTOCATHODES

## 7.4.1 Overview

III-V semiconductor materials (GaAs, in particular) have been operating in accelerators as a photocathode material in photoinjectors since about the last quarter of the 20$^{th}$ century. Initial interest in GaAs mainly was stimulated by polarized electron beam production (Chapter 8). To maximize the electrons' polarization, the laser's photon energy necessarily must be near the band gap, with modest QE (of ~0.1-1%). Apart from generating polarized beams, III-V semiconductors activated to negative electron affinity (NEA) (Figure **7.1**) possess several characteristics suitable for high average current photoinjectors. Among their advantages are very high quantum efficiencies (10-20% is typical, with reported values higher than 50%), prompt response time (< 1 ps) when excited with photon energies much larger than the band gap energy, very low thermal emittance, and mean transverse energy that is essential for low emittance photoinjectors. Furthermore, the quick activation with cesium on samples with atomically clean surfaces is another advantage; in many cases, degraded photocathodes readily are restored by depositing a monolayer of cesium either with or without high temperature treatment. The main drawback of these families of photocathodes lies in their high sensitivity to contaminants and stringent vacuum requirements ($10^{-9}$ Pa or better).

The material discussed herein covers high quantum efficiency, III-V semiconductor photocathodes suitable for high current applications. Much of the technology for the photocathode preparation is similar to that of the polarized sources (Chapter 8); therefore, the emphasis is on peculiarities specific to high QE photocathodes used in high current, low emittance photoinjectors. Chapter 4: DC/RF Injectors is a complementary chapter in this book covering technology aspects related to the performance, and preparation of GaAs photocathodes, and their transfer.

## 7.4.2 Properties and Preparation

III-V semiconductors, suitable as high efficiency photocathodes, are direct band gap materials that rely on their good photoemission properties *via* NEA, a condition established at their surface by the deposition of a single monolayer of very electropositive alkali metal (cesium), as shown in Figure **7.17**. The term direct band gap signifies that the valence band maximum (VBM) is aligned directly below the conduction band minimum (CBM) in momentum-energy band gap diagrams. The implication therein is that the absorption of a photon by an electron in the valence band and its subsequent promotion to the conduction band occurs as a single-step process. This transfer is depicted by a vertical line in diagrams of momentum-energy band gap because the photon carries a negligibly small momentum compared with much heavier electrons and holes in semiconductor. Those transitions, known as vertical- or direct-transitions, occur without needing an additional momentum transfer from the crystal lattice's vibrations, as is required for indirect band gap materials, where VBM and CBM have different momenta in the band-energy diagram. Consequently, the probability of optically promoted transitions across the band gap is much higher in direct band gap semiconductors, leading to their high quantum efficiency.

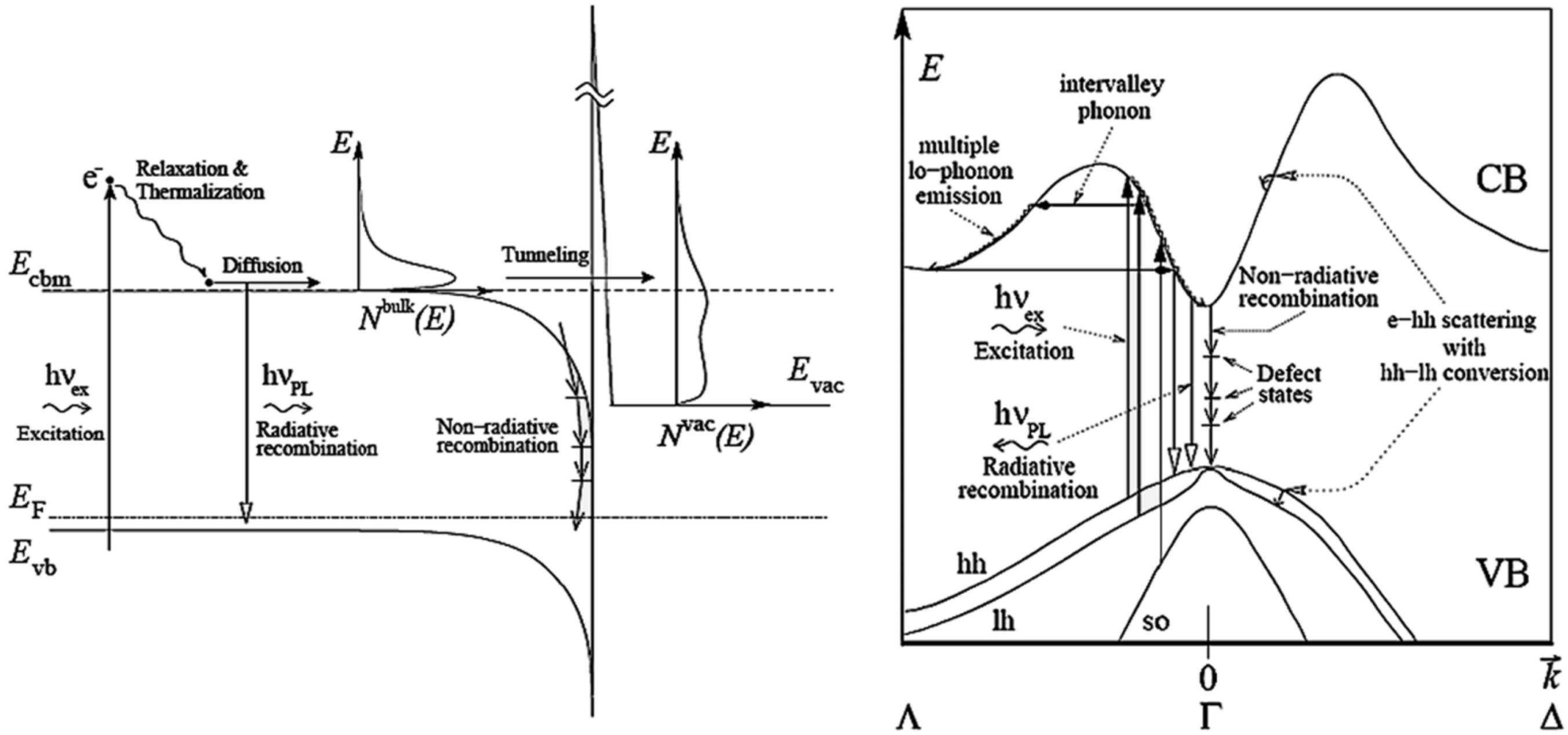

**Figure 7.17. Energy versus electron lattice momentum (left), and energy level structure near the surface (right).** ***CB*** **– conduction band; VB – valence band;** $E_F$ **– Fermi level;** ***hh, lh, so,*** **and** ***lo*** **represent heavy hole, light hole, split off energy bands, and longitudinal optical (phonons), respectively.** $\nu_{ex}$ **is the excitation frequency and** $\nu_{PL}$ **is the photoluminescence frequency. [7.52]**

There are excellent reviews of this class of materials, including a book by Bell [7.53], and a chapter by Escher [7.54]. There are several different materials in this category, with much interest traditionally devoted to IR sensitive devices (*e.g.*, InAs and related tertiary compounds). We discuss two materials as practical photocathodes for photoinjectors: GaAs and GaN. The first was used extensively in photoinjectors operating at high current, and has a conveniently large response in the visible range with the band gap of 1.42 eV at room temperature. On the other hand, GaN is a newer material for photoinjectors with a wide band gap of about 3.4 eV, and there is limited experience of its use. Finally, there is considerable flexibility in growing new high quality semiconductor structures with the desired properties for accelerators. We summarize some interesting developments employing tertiary structures (*e.g.*, AlGaAs) and other advanced ones for high QE unpolarized photocathodes in photoinjectors.

Achieving NEA (Figure **7.1** and Figure **7.14**) is central to a good QE of III-V photoemitters. Pure crystals with atomically clean surfaces are poor photoemitters with a work function of about 4 eV. Two main mechanisms establish the NEA: 1) A strong p-doping, inducing band bending near the surface; and, 2) through the donation of an electron toward the bulk by strongly electropositive alkali metal (cesium being the commonest although Li, Na, K, and Rb also have been used). Additional exposure to an oxidizing agent and cesium forms an even stronger dipole on GaAs, whereas GaN can be activated to NEA with only Cs. On the atomic level, NEA is understood as electron emission assisted by strong dipole fields formed preferentially so that a large negative electric field points towards the surface, accelerating electrons in the opposite direction. Table **7.2** shows some basic material properties for two of the III-V semiconductors detailed in this chapter.

| **Material** | **GaAs** | **GaN** |
|---|---|---|
| Structure | Zinc Blende | Zinc Blende or Wurtzite |
| Energy gap (@ 300 K) [eV] | 1.42 (direct) | 3.2 or 3.44 |
| Electron effective mass [Γ] | $0.067m_0$ | $0.13m_0$ or $0.20m_0$ |
| Melting point [K] | 1511 | 2773 |
| Thermal conductivity (@ 300 K) $\left[\frac{\text{W}}{\text{cm K}}\right]$ | 0.55 | 2.3 |
| Density $\left[\frac{\text{g}}{\text{cm}^3}\right]$ | 5.32 | 6.15 |

**Table 7.2. Some basic material properties of GaAs and GaN [7.55].**

#### *7.4.2.1 GaAs*

Preparing a bulk GaAs photocathode starts with obtaining a high quality crystal. Different atomic orientations can be used: (100) and (110) are typical, with (110) orientation cleaving easily. Manufacturers offer specific crystalline orientations (epitaxial-ready surface). A small angle to a specific atomic plane is attainable without any apparent detrimental effect on the photocathode's performance. Doping levels are important, preferably with high p-doping to ensure both the wafer's band bending and good conductivity. For example, common dopant densities (typically Zn) for high QE, unpolarized GaAs photocathodes used in high current applications is $0.6\text{-}2\times10^{19}$ cm$^{-3}$ with a low resistivity of 5-7 mΩ cm.

The surface of the semiconductor wafer is critical; it must be atomically clean before activation and, ideally, atomically flat. The preparation includes chemical cleaning ex-*situ* and heat- and hydrogen-cleaning in-*situ*. Table **7.3** gives typical steps used by two groups, one at Stanford University, and the other at Cornell University [7.56], [7.57]. No uniform procedure exists and the details of wafer preparation vary.

While the wet chemistry effectively removes carbon-containing impurities, oxygen and oxides forming on the surface of GaAs play a critical role both in achieving the high QE and assuring a smooth surface. A surface free from carbon contaminants, but having oxides does not lend itself to NEA activation. A layer of natural oxides forms readily in the atmosphere. To minimize exposure of the GaAs surface to oxygen, the Stanford group undertakes their wet chemistry in an Ar atmosphere. Some groups found that hydrogen cleaning effectively removes both oxygen and carbon once the semiconductor wafer is in vacuum. High temperature cleaning is an essential practice adopted by all groups employing GaAs; its main benefit is in breaking oxides. The exact value of the temperature is a critical parameter. Previous wet chemistry treatment in Ar atmosphere, effective at removing oxygen, allows a comparatively low temperature treatment (≤ 500 ˚C). First, the most volatile oxide AsO is desorbed at a temperature as low as 150 ˚C [7.58]. The more stable As-oxide, $As_2O_3$, breaks down at about 300 ˚C forming $Ga_2O_3$. Both $Ga_2O_3$ and $GaO_2$ are much more stable, requiring temperatures in excess of 500 ˚C to destabilize them. Gallium oxides are removed effectively at 580 ˚C; however, both Ga and As start to evaporate at that point. Also, 620 ˚C is known as the temperature of onset of non-congruent evaporation, *i.e.*, As starts leaving the surface preferentially, leaving a Ga-rich surface. Beyond that temperature, the structure of GaAs crystal is compromised entailing the deteriorated performance of the photocathode. Thus, high quantum efficiency photocathodes are routinely produced at Cornell University (15% at 532 nm) under heat treatment at 620 ˚C for 1-2 hours with no hydrogen cleaning. However, surface roughness increases substantially at these high temperatures [7.58]. Figure **7.18** shows the surface rms roughness as a function of heat cleaning temperature, along with depictions of a sample's surface [7.58]. Such a rough surface, while acceptable for photoinjectors that do not require the smallest emittances, are suboptimal for applications wherein emittance is a critical parameter [7.60].

| **Stanford [7.56]** | **Cornell [7.57]** |
|---|---|
| Wet chemistry:<br>$H_2SO_4$:$H_2O_2$:$H_2O$ (4:1:100) for 2 min<br>HCl:$H_2O$ (1:3) for 2 min<br>or $H_2SO_4$:$H_2O$ (1:1) for 30 sec | Wet chemistry:<br>Acetone and trichloroethylene cleaning<br>$H_2SO_4$:$H_2O_2$:$H_2O$ (20:1:1) for 1 min |
| Chemistry is done in Ar chamber connected to a load-lock system | Anodization and $NH_4OH$ treatment to limit QE active area; transported in air |
| Heated to 500 ˚C | Optional atomic hydrogen cleaning |
| Cs + $O_2$ co-deposition | Heat cleaning at ≤ 620 ˚C |
| | Cs + $NF_3$ "yo-yo" process |

Table 7.3. Cleaning and preparation steps for GaAs photocathodes [7.56], [7.57].

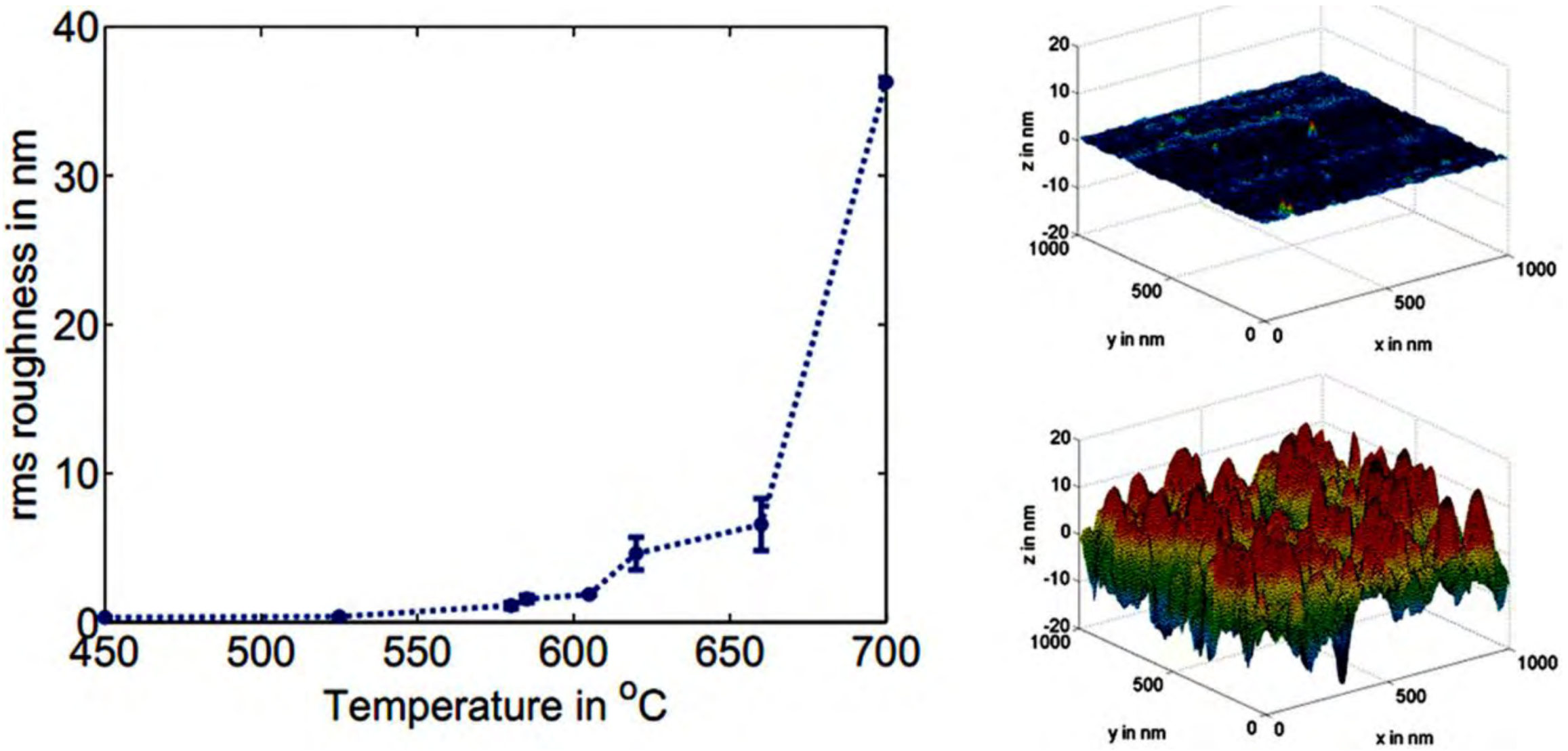

Figure 7.18. Surface roughness versus temperature for heat treatment of GaAs (left). Examples of AFM measurements on untreated GaAs (upper right) and after the heat treatment (lower right). [[1.1][2.1][4.1][5.1][6.1][7.1]; Available under Creative Common Attribution 3.0 License (www.creativecommons.org/licenses/by/3.0/us/) at www.JACoW.org.] [Reprinted with permission from [7.60]. Copyright 2011, American Institute of Physics]

Once a clean, oxide-free surface is established, the cathode is activated to NEA. QE is measured continuously during this activation process. Initially, about a monolayer of Cs is deposited resulting in an initial increase of QE (2-5% at 532 nm wavelength). As Cs is deposited, the QE rises to a peak and then starts to drop. Only the deposition of Cs continues until the QE reaches about half the value of the first peak. At that point, an oxidizing agent is used further to increase QE. $NF_3$ or $O_2$ are commonly used as the oxidizing agent, producing similar results. $NF_3$ reacts with Cs forming CsF at the surface, with nitrogen escaping the surface. However, detailed studies of the activation layer [7.61] revealed that the actual composition of the layer was complex, with N embedded in it. Two different techniques are used with good success: "Yo-yo" (Figure **7.19** left) and co-evaporation (Figure **7.19** Right). Using the "yo-yo" technique, the exposure of $NF_3/O_2$ and Cs is alternated, whereas for co-evaporation, it is held at a constant rate until the QE reaches its maximum. Studies indicate that electron affinity is close to 0 eV for only Cs and becomes -0.1 eV for a fully activated photocathode.

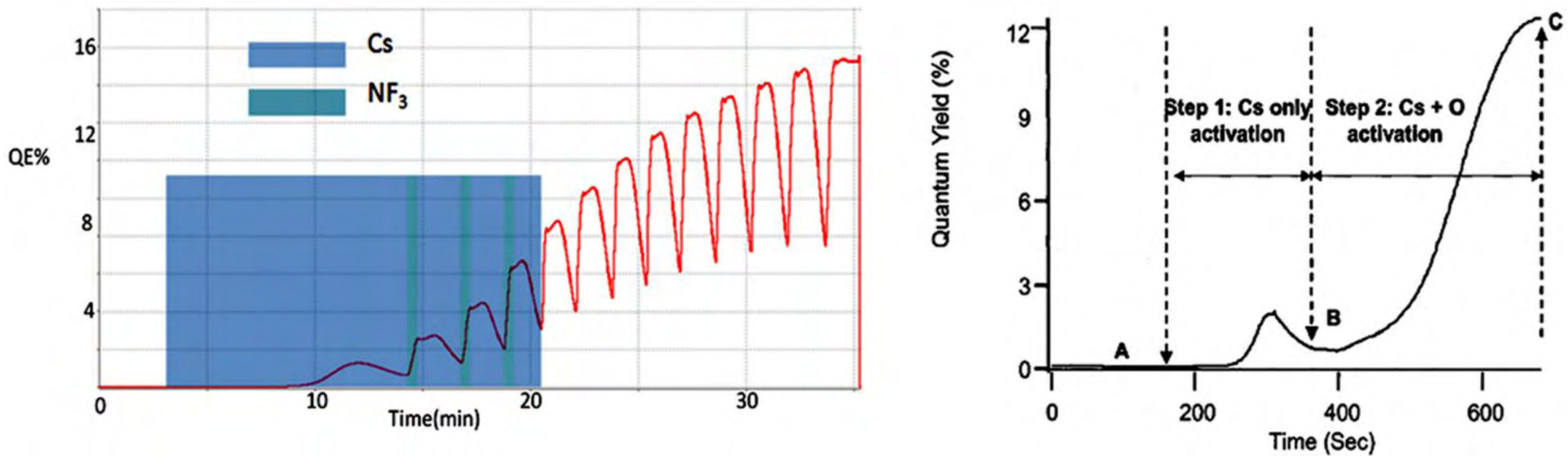

**Figure 7.19. "Yo-yo" activation with Cs and $NF_3$ (left, the activation pattern shown persists throughout the entire procedure), and co-evaporation of Cs and $O_2$ (right). [7.56]**

The typical dark lifetime of GaAs is on the order of days (even though CEBAF demonstrated the essentially infinite dark lifetime for their polarized photocathodes). A practical solution for rejuvenating the photocathode is reactivation with Cs that routinely restores the QE to 5% or higher (at 532 nm). High temperature cleaning is a reliable method to restore the QE of GaAs to its original performance, except after particularly heavy damages from ion back-bombardment and sputtering.

Because the vacuum requirements for GaAs are so stringent (vacuum of $\leq 10^{-9}$ Pa with RGA spectra free from water, oxygen, and carbon dioxide), transfer of this photocathode has not accomplished without affecting its sensitive activation layer. Presently, all guns operating with GaAs either have a preparation chamber attached to them (the preferred solution), or are activated inside the gun.

### *7.4.2.2 GaN*

Unlike GaAs, GaN is a new photocathode material that promises a more robust NEA material suitable for operating in less stringent vacuum conditions. As a wide band gap material, GaN requires UV illumination. Its large band gap leads to a stronger bonding of Cs to the surface, resulting in a NEA condition with only Cs. Therefore, the main appeal of GaN for accelerator applications is in its more stable Cs layer that may prove beneficial for high current operation, or when the photocathode must be at an elevated temperature. However, it is not expected that the activated layer noticeably would be more forgiving to chemical poisoning gases than GaAs, and therefore, the partial pressures of these species similarly must be small.

The preparation and activation procedures for GaN photocathodes resemble those for GaAs. Likewise, the material is typically strongly p-doped (~$1\times10^{19}$ $cm^{-3}$) to ensure high conductivity and band bending near the surface. However, extensive heat treatment is typically not required, as GaN seemingly is much more inert to oxide formation. For example, a simple vacuum-bakeout to 200 ˚C reportedly restored its QE to 50% of its original value after exposing the photocathode to a nitrogen atmosphere [7.62]. The activation is simpler, as mentioned earlier, with the use of cesium only generating excellent QEs of as high as 50% for simple bulk photocathodes, although requiring a UV spectral range more difficult to attain for the laser and the optics.

#### *7.4.2.3 Tertiary and Other Advanced Structures*

Considerable flexibility exists in researcher's ability to tune the band gap and other parameters for alloys such as $Al_xGa_{1-x}N$, $In_xGa_{1-x}N$, $GaAs_{1-x}P_x$, and $Al_xGa_{1-x}As$. For example, by varying the concentration of In in $In_xGa_{1-x}N$, the band gap can be matched to a laser operating in the visible range while retaining some benefits of GaN material. This can lead to both better QEs and improvements in other parameters relevant to accelerators, such as thermal emittance. Despite the promises of better performances compared to actual NEA photocathodes, those materials have not yet been tested in a real photoinjector. Particular attention must be paid when the semiconductor goes from a direct- to an indirect-band gap material. For example, despite its good QE, $GaAs_{0.55}P_{0.45}$ was found to have a noticeably higher thermal emittance than GaAs at the same laser wavelength, despite its larger band gap [7.63]. An explanation was offered, noting that GaP is an indirect band gap material, and the transition from direct (GaAs) to indirect occurs at about 45% concentration of phosphorus. This implies that additional momentum transfer occurs during photoemission to conserve crystal momentum, so engendering the increased transverse velocities of electrons leaving the surface.

Another common characteristic of more advanced structures is the ability to grow transmission-mode photocathodes on a transparent substrate. For example, the response time of thin films of III-V material can be tailored, even at wavelengths where the temporal tail can be significant for bulk crystal material [7.64]. Further benefits of a transmission-type photocathode may be in its very small thermal emittance, at the sub-thermal level, since the emitted electrons are forced to diffuse toward the surface and undergo scattering within the lattice, thermalizing in the process of doing so (discussed below).

Finally, graded photocathodes are another common approach. Here, the QE can be improved in two ways: First, by increasing the efficiency of light coupling into the material, *e.g.*, through matching the index of refraction of the photocathode and the substrate. For example, a normal bulk GaAs reflects about a third of the incident light, and using the index of refraction matching can better its efficiency by about a factor of three. The second approach is to reflect the electrons inside the material that are traveling away from the surface back to it *via* forming a potential barrier through varying the composition of the tertiary compound. For example, a graded $Al_xGa_{1-x}As$/GaAs photocathode recently demonstrated over 50% quantum efficiency in the green wavelengths [7.65], as depicted in Figure **7.20**.

### 7.4.3 Photocathode Performance

#### *7.4.3.1 Thermal Emittance*

Thermal emittance in NEA photocathodes has been studied in detail. These photocathodes remain the materials of choice when the smallest thermal emittance is required. Figure **7.21** summarizes a typical thermal emittance measurement (or MTE which is related to the thermal emittance) for GaAs obtained from a high voltage DC gun [7.63]. As expected, the thermal emittance is smallest when the photon energy closely matches the energy of the band gap, and increases for higher photon energies.

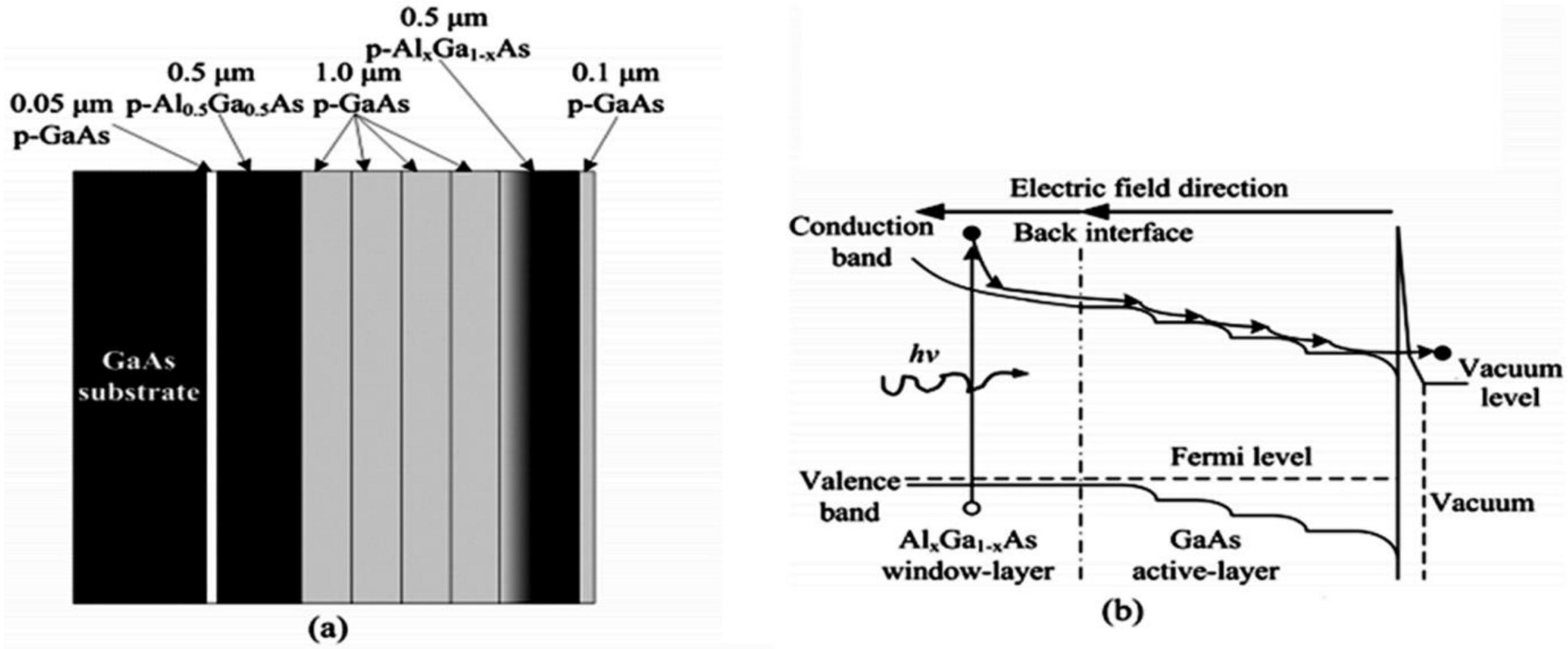

**Figure 7.20. An example of advanced transmission-mode graded photocathode that achieves over 30% QE in the transmission-mode and 60% in reflection-mode in the green spectral range. [Reprinted with permission from [7.65]. Copyright 2011, American Institute of Physics.]**

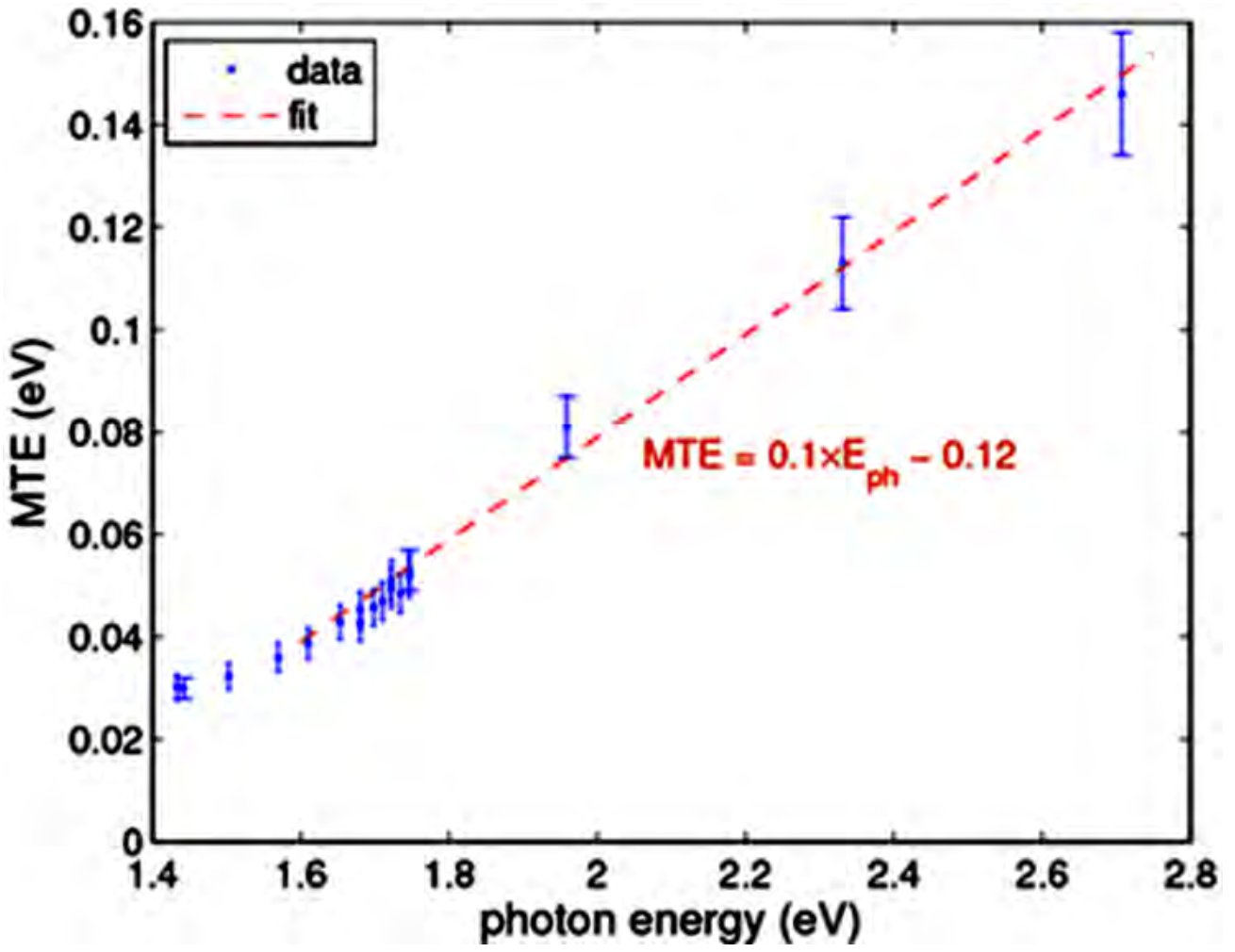

**Figure 7.21. Thermal emittance of GaAs photocathode as a function of photon energy. [Adapted with permission from [7.63]. Copyright 2008, American Institute of Physics.]**

This result is reproduced routinely in Cornell University's photoinjector operation, and shows remarkable insensitivity to the photocathode's QE. The situation is very different for $GaAs_{0.55}P_{0.45}$ that displayed about 50% variation in MTE when the QE dropped from 6% to 1% (0.200 eV to 0.135 eV, respectively, at 532 nm laser wavelength [7.63]. The explanation behind this, as argued earlier, is because this photocathode has a direct- and indirect-conduction band minimum valley, with emission from the latter being suppressed when the vacuum level is slightly raised (reducing the QE to 1%). We note that the MTE rises with photon energy at a rate slower than ⅓$E_{ph}$, as observed in the alkali-based photocathodes, and predicted by Equ. 7.7. This dependence is explained by the fact that a noticeable fraction of electrons escaping into vacuum is fully or partially thermalized down to the conduction band's minimum *via* electron-phonon collisions, or that the emission into the vacuum occurs in a preferential direction close to the surface normal.

So far, only one report of GaN emittance measurements [7.66] at a single laser wavelength (260 nm) has been produced. The result is a surprisingly large MTE = 0.9 eV. In this case, the excess energy is $E_{ph} - E_{gap}$ = 1.4 eV, and assuming a uniform emission into the vacuum hemisphere, the predicted MTE would have been about 1.4/3 = 0.46 eV, *i.e.*, a factor of two smaller. One explanation offered in [7.20] is that the band bending of GaN, which is larger than that in GaAs (1.2 V versus 0.2- to 0.5-V), increased the electrons' kinetic energy before their emission into vacuum. Additional measurements of this type of photocathode are required to arrive at definitive results.

Finally, it is important to point out that the thermal emittance measurements, demonstrated on a weekly basis in the operating photoinjector at Cornell University, as well as the measurements of thermal energy spread at the University of Heidelberg, Germany [7.67], do not reflect the theoretical findings that a very small reduced mass inside these semiconductors (*e.g.*, GaAs Γ-valley electron's effective mass is only $m^* = 0.067m_0$ of its vacuum mass ) should lead to a very small transverse velocity after the emission into vacuum, as required by transverse momentum conservation: $m^* v_{\perp,cryst} = m_0 v_{\perp,vac}$. This means that the theory predicts MTE, which can be as small as $kT(m^*/m_0)$, where $kT$ is thermal energy (1/40 eV at 300 K). Indeed, such results were reported by the Stanford group [7.68]. Possible explanations for why several well established groups cannot reproduce this result are the nano-scale roughness of the surface [7.60] or the scattering effect of the activation layer [7.69]. If such low thermal emittances can be reliably realized in photoinjectors, it will open a route toward generating very bright electron beams.

#### 7.4.3.2 Response Time

Response time is another critical parameter for photoinjectors, as the pulse duration needs to be considerably shorter than the wavelength of RF used for acceleration; and also, to allow for laser temporal shaping, which is effective for mitigation of emittance growth of space charge dominated beams. Response of 1 ps or shorter is typically desired. Early results from Max Planck Institute have shown that GaAs excited near the band gap has a long response time with a tail that extends to 100 ps. It was also shown in their work [7.70] that the result can be approximately explained by noting that the response time for diffusing electrons can be written as

$$\tau = \frac{1}{\alpha^2 D} \tag{7.10}$$

where $\alpha$ is the optical absorption constant and $D$ is the electron diffusion coefficient. Because the optical absorption constant varies substantially with the photon wavelength, it is therefore expected that the response time changes by many orders-of-magnitude with a shorter wavelength excitation. Indeed, the measurements performed at Cornell have shown that GaAs is a prompt photoemitter (< 1 ps) when excited with 520 nm laser wavelength. An example of data and comparison with Equ. 7.10 are shown in Figure **7.22**. Generally Equ. 7.10 overestimates the response time. Possible explanations are in the fact that additional recombination processes occurring at the surface are not accounted in this model.

We note that even for excitation with a near band gap photons, GaAs can be made into a fast photocathode in the transmission-mode albeit at the expense of its QE [7.64]. GaN's response time at 260 nm essentially approximated a delta function [7.66]. This is not very surprising given that the material's absorption length is very short at this photon energy.

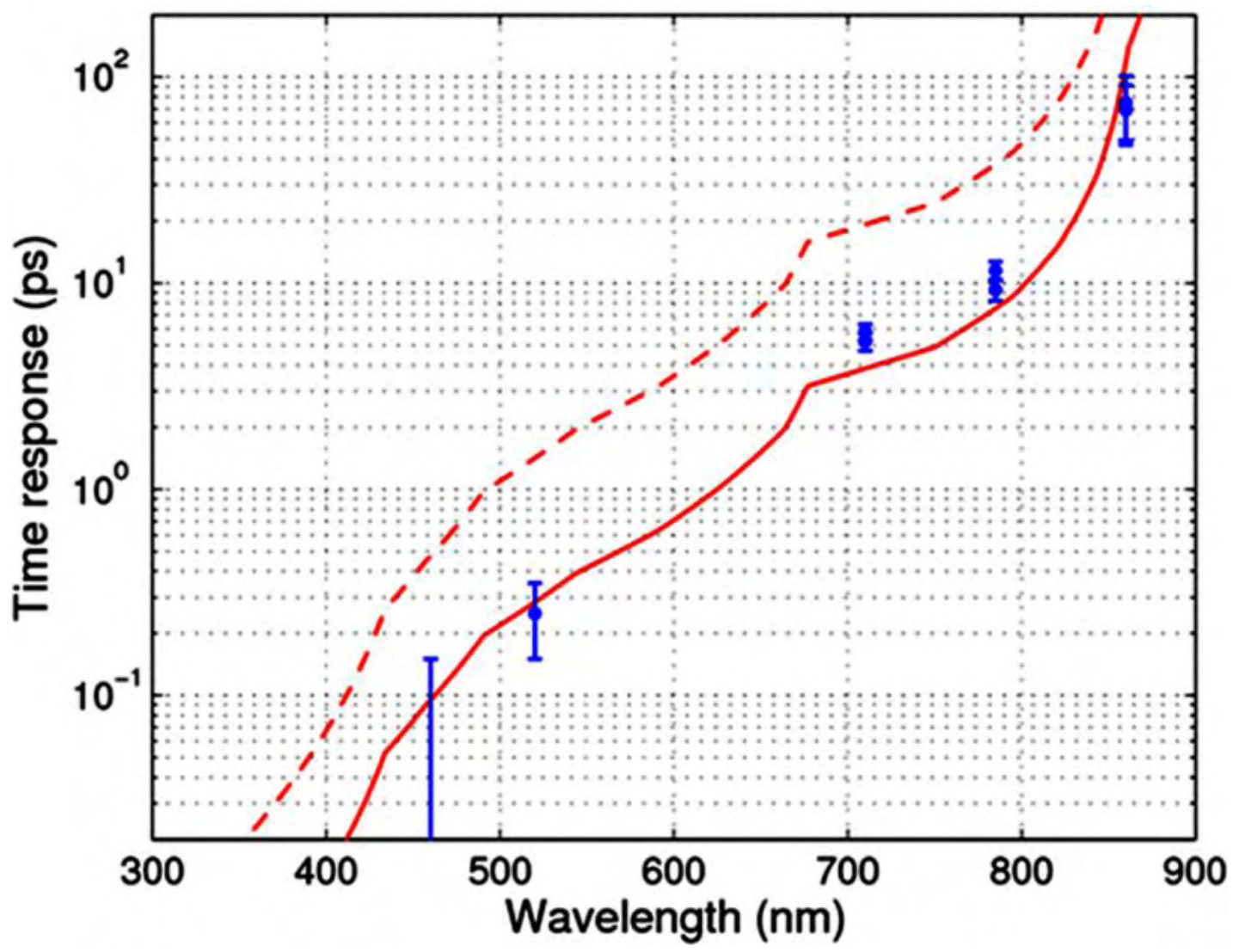

**Figure 7.22. Collection of the GaAs photoemission response time measurements [7.18], [7.26] vs. the laser wavelength along with the theoretical predictions (dashed line). Better agreement is obtained for $\tau = (\alpha^{-2} D^{-1})/5$ (solid line). The measurement at 460 nm was only able to place an upper estimate on the response time.**

### *7.4.3.3 High Current Operation*

High current operational experience with unpolarized high QE photocathodes mostly is limited to JLAB's experience of IR FEL, DC laser tests performed by the CEBAF injector group, and staff at Cornell University's Energy Recovery Linac photoinjector. The lifetime here depends on many factors, such as DC voltage versus the bunched structure of the beam, charge per bunch, average current, location of the laser spot on the photocathode, scattered laser light, beam losses, and the quality of the beamline's vacuum. For pulsed operation at JLAB FEL, a 1/*e* operational lifetime at 550 Coulombs at 5 mA was demonstrated with GaAs [7.71]. Extensive studies at CEBAF using the DC laser have resulted in 1/*e* lifetime of more than 1000 Coulombs for a 9.5 mA DC beam [7.72], *viz.*, the best for this photocathode operated at high average current. Cornell University's photoinjector can operates with GaAs at 20 mA average current with a 1/*e* lifetime of approximately an hour, with possible causes of this lifetime identified to be from the laser scattering due to rough in-vacuum metallic-polished mirrors, and the generation of beam halos. Figure **7.23** gives an example of beam current. Also, lifetime can be improved by simply moving the laser spot further off-axis since the location of the main damage is confined to the gun's electrostatic center. However, as expected, GaAs demonstrate a much reduced lifetime, as seen in Figure **7.14**, compared to alkali-based photocathodes, such as $K_2CsSb$. Other possible mechanisms for degrading lifetime are due to the somewhat elevated temperature of the photocathode when the beam is on, and to increased beam losses at higher bunch charges and currents. Additional studies on GaAs at elevated temperatures show a reduced dark lifetime [7.73] that can be resolved through designing efficient cooling of the photocathode in the presence of the incident laser beam.

Finally, we note the importance of the surface charge effect when operating high current, unpolarized NEA photocathodes, a phenomenon first observed when employing at high beam current at the SLAC polarized source using GaAs. Many references are available on this subject [7.74]–[7.76]. Essentially, the effect lies in the resistor-capacitor circuit-like behavior wherein an electron charge excited by the laser pulse is not extracted into vacuum and builds up a potential barrier at the photocathode's surface, causing a time dependent transient of the beam current, even though the laser's intensity stays constant. The effect is most strongly pronounced when the photon energy close to the semiconductor band gap is being used and not heavily p-doped crystals (*e.g.*, as typically required for polarized sources where generally the QE is very low).

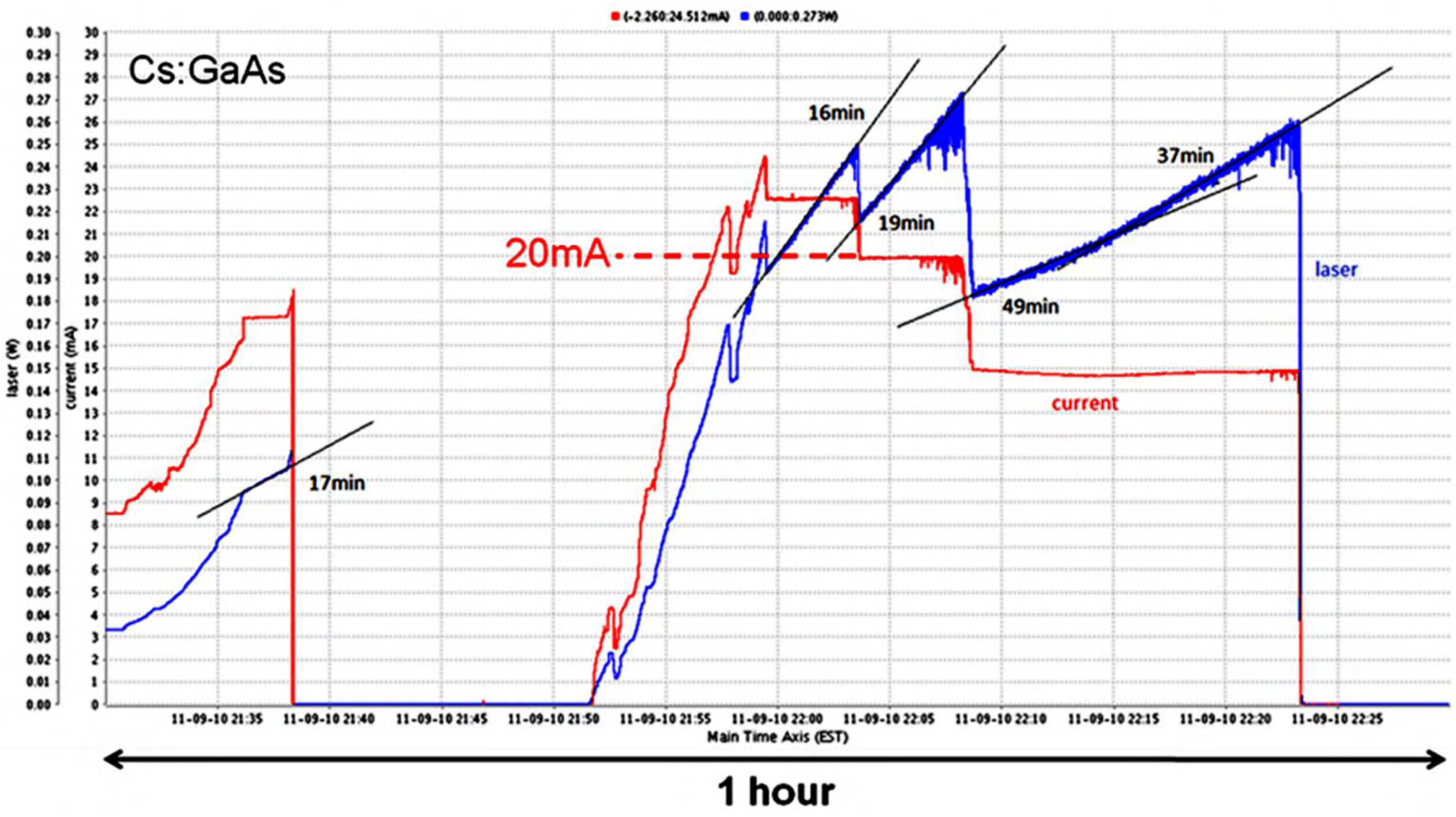

**Figure 7.23. An example of GaAs lifetime operated at high current at Cornell University's photoinjector. The laser's slow feedback ensures constant current (red line) by regulating its power (blue line) accordingly. 1/*e* lifetime is indicated for cases where the laser spot was close to the gun's electrostatic center, with the resulting short photocathode lifetime. By moving the spot further off-center, lifetime is increased substantially.**

This phenomenon also was observed at Cornell University's ERL photoinjector when operating at high currents, and using highly p-doped GaAs crystal. Employing fast beam current diagnostics with pulsed operation [7.77], revealed a small leading peak, lasting about 0.1 µs, in macropulses with a substantial average current, even though the laser pulse is constant over the macropulse's entire duration (Figure **7.24**). The effect became more pronounced as QE degraded, consistent with the SLAC observations. Overall, the effect played a small role for a 520 nm laser and a reasonable QE for heavily p-doped GaAs, though it can become important during a pulsed operation where the transient due to the leading peak of the surface charge effect may become undesirable.

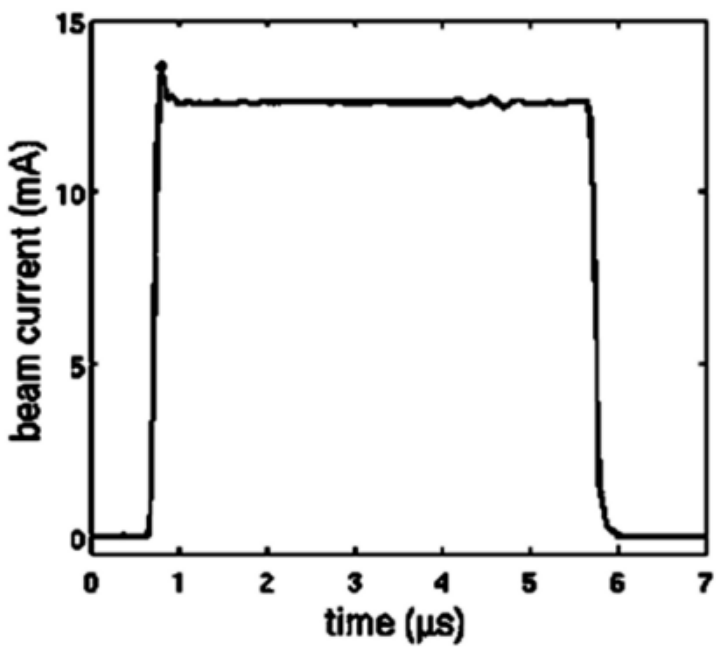

**Figure 7.24. An example of surface charge effect in GaAs operated at high average current [7.38]. The laser pulse is constant over the duration of the macropulse.**

## 7.5 CONCLUSION

Semiconductor photocathodes provide viable alternatives for generating average currents exceeding 30 mA with acceptable life time and intrinsic emittance for high average flux FEL and high current collider applications. Average current of > 30 mA, with life time in the order of days, intrinsic emittance of nearly the fundamental limit and response time in the picosecond time range has been generated with $K_2CsSb$ photocathodes. Similar performance with a slightly lower life time, but better intrinsic emittance has been produced with GaAs: Cs cathodes, as well. In the forthcoming years, with better understanding of physical properties controlling their performance, it may be possible to tailor the cathode to suit the application.

## 7.6 CONFLICT OF INTEREST AND ACKNOWLEDGEMENT

We confirm that this article content has no conflicts of interest and would like to acknowledge the support of the US National Science Foundation and Department of Energy under contract numbers NSF DMR-0807731, DOE DE-SC0003965 and DE-AC02-98CH10886.

***References***

[7.1] W. E. Spicer, "Photoemissive, photoconductive, and optical absorption studies of alkali-antimony compounds," *Phys. Rev.*, vol. 112, pp. 114-122, October 1958.

[7.2] I. V. Bazarov, B. M. Dunham and C. K. Sinclair, "Maximum achievable beam brightness from photoinjectors," *Phys. Rev. Lett.*, vol. 102, pp. 104801-1–104801-4, March 2009.

[7.3] C. K. Sinclair, "High voltage DC photoemission electron guns – current status and technical challenges," *ICFA Beam Dynamics Newslett.*, vol. 46, pp. 97-118, August 2008.

[7.4] D. H. Dowell I. Bazarov, B. Dunham *et al.*, "Cathode R&D for future light sources," *Nucl. Instrum. Meth. A*, vol. 622, pp. 685-697, October 2010.

[7.5] Z. Insepov, "Bialkali/multi-alkali response" [Online]. Available FTP: psec.uchicago.edu Directory: library/photocathodes File: zeke_Bialkali.png [Accessed on March 6, 2013].

[7.6] H. Sommer, *Photoemissive Materials: Preparation, Properties, and Uses*, New York: Wiley, 1968.

[7.7] D. H. Dowell, K. J. Davis, K. D. Friddell *et al.*, "First operation of a photocathode radio frequency gun injector at high duty factor," *Appl. Phys. Lett.*, vol. 63, pp. 2035-2037, October 1993.

[7.8] R. Nathan, C. H. B. Mee, "Photoelectric and related properties of the potassium-antimony-caesium photocathode," *Int. J. Electron.*, vol. 23, pp. 349-354, October 1967.

[7.9] A. di Bona, F. Sabary, S. Valeri *et al.*, "Auger and x-ray photoemission spectroscopy study on $Cs_2Te$ photocathodes," *J. Appl. Phys.*, vol. 80, pp. 3024-3030, September 1996.

[7.10] R. A. Powell, W. E. Spicer, G. B. Fisher *et al.*, "Photoemission studies of cesium telluride," *Phys. Rev. B*, vol. 8, pp. 3987-3995, October 1973.

[7.11] P. Michelato, A. Di Bona, C. Pagani *et al.*, "R&D activity on high QE alkali photocathodes for RF guns," in *Proc. 1995 Particle Accelerator Conf.*, 1995, pp. 1049-1051.

[7.12] P. Michelato, "Report of high quantum efficiency photocathode at Milano," in *Proc. AIP Conf.*, vol. 279, 1992, pp. 775-760.

[7.13] L. Cultrera, I. V. Bazarov, A. Bartnik *et al.*, "Thermal emittance and response time of a cesium antimonide photocathode," *Appl. Phys. Lett.*, vol. 99, pp. 152110-1–152110-3, October 2011.

[7.14] S. M. Johnson, "Optical absorption and photoemission in semitransparent and opaque $Cs_3Sb$ photocathodes," *Appl. Opt.*, vol. 32, pp. 2262, May 1993.

[7.15] M. Hagino and T. Takahashi, "Thickness of Cs-Sb films relative to the original Sb films," *J. Appl. Phys.*, vol. 37, pp. 3741-3743, September 1966.

[7.16] P. Michelato, P. Gallina *et al.*, "Alkali photocathode development for superconducting rf guns," *Nucl. Instrum. Methods A*, vol. 340, pp. 176-181, February 1994.

[7.17] Y. Nakazono, M. Kunihiro, A. Sakumi *et al.*, "Upgrade of cartridge-type exchangeable $Na_2KSb$ cathode gun," in *Proc. 2010 Int. Particle Accelerator Conf.*, 2010, pp. 4293-4295.

[7.18] R. W. Engstrom, *Photomultiplier Handbook*, Lancaster, PA: RCA/Burle, 1980.

[7.19] P. Dolizy, "Optical method for investigating alkali antimonide photocathodes," *Vacuum*, vol. 30, pp. 489-495, 1980.

[7.20] B. Erjavec, J. Šetina and L. Irmančnik-Belič, "Modelling of the alkali vapour-pressure dynamics during the epitaxial growth of semi-transparent photo-emissive layers," *Vacuum*, vol. 67, pp. 235-241, September 2002.

[7.21] A. Dubovoi, A. S. Chernikov, A. M. Prokhorov *et al.*, "Multialkali photocathodes grown by molecular beam epitaxy technique," in *Proc. SPIE*, 1991, vol. 1358, pp. 134-138.

[7.22] Natarajan, A. T. Kalghatgi, B. M. Bhat *et al.*, "Role of the cesium antimonide layer in the $Na_2KSb/Cs_3Sb$ photocathode," *J. Appl. Phys.*, vol. 90, pp. 6434-6439, December 2001.

[7.23] W. H. McCarroll, R. J. Paff and A. H. Sommer, "Role of Cs in the $(Cs)Na_2KSb$ (S-20) multialkali photocathode," *J. Appl. Phys.*, vol. 42, pp. 569-572, February 1971.

[7.24] D. H. Dowell, S. Z. Bethel and K. D. Friddell, "Results from the average power laser experiment photocathode injector test," *Nucl. Instrum. Meth. A*, vol. 356, pp. 167-176, March 1995.

[7.25] I. Bazarov, L. Cultrera, A. Bartnik *et al.*, "Thermal emittance measurements of a cesium potassium antimonide photocathode", *Appl. Phys. Lett.*, vol. 98, pp. 224101-1–224101-3, June 2011.

[7.26] T. Vecchione, I. Ben-Zvi, D. H. Dowell *et al.*, "A low emittance and high efficiency visible light photocathode for high brightness accelerator-based X-ray light sources," *Appl. Phys. Lett.*, vol. 99, pp. 034103-1–034103-3, July 2011.

[7.27] A. Braem, E. Chesi, W. Dulinski *et al.*, "Highly segmented large-area hybrid photodiodes with bialkali photocathodes and enclosed VLSI readout electronics," *Nucl. Instrum. Methods A*, vol. 442, pp. 128-135, March 2000.

[7.28] A. Burrill, I. Ben-Zvi, T. Rao *et al.*, "Multi-alkali photocathode development at Brookhaven National Lab for application in superconducting photoinjectors," in Proc. 2005 Particle Accelerator Conf., 2005, pp. 2672-2674.

[7.29] E. Shefer, A. Breskin, A. Buzulutskov *et al.*, "Laboratory production of efficient alkali-antimonide photocathodes," *Nucl. Instrum. Meth. A*, vol. 411, pp. 383-388, July 1998.

[7.30] K. Nakamura, Y. Hamana,Y. Ishigami *et al.*, "Latest bialkali photocathode with ultra high sensitivity," *Nucl. Instrum. Meth. A*, vol. 623, pp. 276-278, November 2010.

[7.31] D. Sertore, P. Michelato, L. Monaco *et al.*, "Review of the production process of TTF and PITZ photocathodes," in *Proc. 2005 Particle Accelerator Conf.*, 2005, pp. 671-673.

[7.32] L. Monaco, P. Michelato, D. Sertore *et al.*, "Multiwavelengths optical diagnostic during $Cs_2Te$ photocathodes deposition," in *Proc. 2010 Int. Particle Accelerator Conf.*, 2010, pp. 1719-1721.

[7.33] A. di Bona, F. Sabary, S. Valeri *et al.*, "Formation of the $Cs_2Te$ photocathode: auger and photoemission spectroscopy study," in *Proc. 1996 European Particle Accelerator Conf.*, 1996, pp. 1475-1477.

[7.34] F. Baumann, J. Kessler and W. Roessler, "Composition of Antimony Evaporating from Different Sources," *J. Appl. Phys.*, vol. 38, pp. 3398-3399, March 1967.

[7.35] H. Sugiyama, H. Kobayakawa, Y. Takeda *et al.*, "Effect of substrate on the quantum efficiency of cesium telluride thin-film photocathodes," *J. Japan Inst. Metals*, vol. 69, pp. 493-496, October 2005.

[7.36] L. Cultrera, I. V. Bazarov, J. V. Conway *et al.*, "Growth and characterization of bialkali photocathodes for Cornell ERL injector," in Proc. *2011 Particle Accelerator Conf.*, 2011, pp. 1942-1944.

[7.37] M. Krasilnikov, "Impact of the cathode roughness on the emittance of an electron beam," in *Proc. 2006 Free Electron Laser Conf.*, 2006, pp. 586-583.

[7.38] D. Sertore, S. Schreiber, K. Flöettmann *et al.*, "First operation of cesium telluride photocathodes in the TTF injector RF gun," *Nucl. Instrum. Meth. A*, vol. 445, pp. 422-426, May 2000.

[7.39] D. Sertore, "Photocathode for FLASH and PITZ: a summary," presented at 1st QE Photocathode RF Guns Workshop, Milan, Italy, October 5, 2006.

[7.40] *Photo-injectors for CTF3 and CLIC: Photocathodes* [Online]. Available: http://photoinjector.web.cern.ch/photoinjector/Photocathodes.htm [Accessed: January 31, 2012].

[7.41] S. Schreiber, P. Michelato, L. Monaco *et al.*, "On photocathodes used at the TTF photoinjector," in *Proc. Particle Accelerator Conf.*, 2003, pp. 2071-2073.

[7.42] D. Sertore, L. Monaco, P. Michelato *et al.*, "High QE photocathodes at FLASH," in *Proc. 2006 European Particle Accelerator Conf.*, 2006, pp. 2496-2498.

[7.43] D. Sertore, P. Michelato, L. Monaco *et al.*, "High quantum efficiency photocathode for RF guns," in *Proc. 2007Asian Particle Accelerator Conf.*, 2007, pp. 223-225.

[7.44] V. Miltchev, J. Bähr, H. J. Grabosch *et al.*, "Measurements of thermal emittance for cesium telluride photocathodes at PITZ," in *Proc. 2005 Free Electron Laser Conf.*, 2005, pp. 560-563.

[7.45] T. Rao, presented at 2011 Energy Recovery Linac, Tsukuba, Japan, October 2011.

[7.46] Femtosecond streak camera C6138 (FESCA-200), available by Hamamatsu Photonics K.K., Hamamatsu, Japan. Available at http://sales.hamamatsu.com/assets/pdf/hpspdf/e_c6138.pdf.

[7.47] T. Nordlund, "Streak Cameras for Time-Domain Fluorescence," in *Topics in Fluorescence Spectroscopy*, 1st Ed., Berlin: Springer-Verlag, 2002, Chap. 3, pp. 183-256.

[7.48] S. H. Kong, J. Kinross-Wright, D. C. Nguyen *et al.*, "Cesium telluride photocathodes," *J. Appl. Phys.*, vol. **77**, pp. 6031-6038, June 1995.

[7.49] K. Miyoshi, K. Kambe, A. Sakumi *et al.*, "Commissioning of $Na_3KSb$ photocathode rf gun in s-band linac at the university of Tokyo," in *Proc. 2009 Particle Accelerator Conf.*, 2009, pp. 587-589.

[7.50] L. Cultrera, J. Maxson, I. Bazarov *et al.*, "Photocathode behavior during high current running in the Cornell energy recovery linac photoinjector," *Phys. Rev. ST Accel. Beams*, vol. 14, pp. 120101-1–120101-12, December 2011.

[7.51] S. Lederer, S. Schreiber, P. Michelato *et al.*, "Photocathode studies at FLASH," in *Proc. 2008 European Particle Accelerator Conf.*, 2008, pp. 232-234.

[7.52] U. Weigel, "Cold intense electron beams from gallium arsenide photocathode," Ph.D. Thesis, Universität Heidelberg, Heidelberg, Germany, 2003.

[7.53] R. L. Bell, *Negative Electron Affinity Devices*, Oxford: Clarendon Press, 1973

[7.54] J. S. Escher, "NEA Semiconductor Photoemitters," in *Semiconductors and Semimetals* vol. 15, Oxford: Clarendon Press, 1981, Chap. 3, pp. 195-300.

[7.55] V. Siklitsky, *New Semiconductor Materials: Characteristics and Properties*, 2001. [Online] Available: http://www.ioffe.ru/SVA/NSM/ [Accessed: January 9, 2011].

[7.56] Z. Liu, "Surface characterization of semiconductor photocathode structures," Ph.D. Thesis, Stanford University, Stanford, California, 2005.

[7.57] X. Liu, unpublished.

[7.58] A. Guillén-Cervantes, Z. Rivera-Alvarez, M. López-Luna *et al.*, "GaAs surface oxide desorption by annealing in ultra high vacuum," *Thin Solid Films*, vol. 373, pp. 159-169, September 2000.

[7.59] S. Karkare, I. Bazarov, L. Cultrera *et al.*, “Effect of surface roughness on the emittance from GaAs photocathode,” in *Proc. 2011 Particle Accelerator Conf.*, 2011, pp. 2480-2482.
[7.60] S. Karkare and I.V. Bazarov, “Effect of nano-scale surface roughness on transverse energy spread from GaAs photocathodes,” *Appl. Phys. Lett.*, vol. 98, pp. 094104-1–094104-3, March 2011.
[7.61] Z. Liu, Y. Sun, S. Peterson *et al.*, “Photoemission study of Cs-$NF_3$ activated GaAs(100) negative electron affinity photocathodes,” *Appl. Phys. Lett.*, vol. 92, pp. 241107-4–241107-3, June 2008.
[7.62] O. Siegmund, J. Vallerga, J. McPhate *et al.*, “Development of GaN photocathodes for UV detectors,” *Nucl. Instrum. Meth. A*, vol. 567, pp. 89-92, November 2006.
[7.63] I. V. Bazarov, B. M. Dunham, Y. Li *et al.*, “Thermal emittance and response time measurements of negative electron affinity photocathodes,” *J. Appl. Phys.*, vol. 103, pp. 054901-1–054901-8, March 2008.
[7.64] K. Aulenbacher, J. Schuler, D. v. Harrach *et al.*, “Pulse response of thin III/V Semiconductor photocathodes,” *J. Appl. Phys.*, vol. 92, pp. 7536-7543, December 2002.
[7.65] Y. Zhang, B. Chang, J. Niu *et al.*, “High-efficiency graded band-gap $Al_xGa_{x-1}As$/GaAs photocathodes grown by metalorganic chemical vapor deposition,” *Appl. Phys. Lett.*, vol. 99, pp. 101104-1–101104-3, September 2011.
[7.66] I. V. Bazarov, B. M. Dunham, X. Liu *et al.*, “Thermal emittance and response time measurements of a GaN photocathode,” *J. Appl. Phys.*, vol. 105, pp. 083715-1–083715-4, April 2009.
[7.67] D. A. Orlov, M. Hoppe, U. Weigel *et al.*, “Energy distributions of electrons emitted from GaAs (Cs, O),” *Appl. Phys. Lett.*, vol. 78, pp. 2721-2723, April 2001.
[7.68] Z. Liu, Y. Sun, P. Pianetta and R. F. W. Pease, “Narrow cone emission from negative electron affinity photocathodes,” *J. Vacuum Sci. Technology B*, vol. 23, pp. 2758-2762, December 2005.
[7.69] D. C. Rodway and M. B. Allenson, “*In* situ surface study of the activating layer on GaAs (Cs, O) photocathodes,” *J. Physics D: Appl. Physics*, vol. 19, pp. 1353-1371, July 1986.
[7.70] P. Hartman, J. Bermuth, D. v. Harrach *et al.*, “A diffusion model for picosecond electron bunches from negative electron affinity GaAs photocathodes,” *J. Appl. Phys.*, vol. 86, pp. 2245-2249, August 1999.
[7.71] C. Hernandez-Garcia *et al.*, “Status of the Jefferson Lab ERL FEL DC photoemission gun,” in *Proc. 2009 Energy Recovery Linac*, 2009, pp. 37-39.
[7.72] J. Grames, R. Suleiman, P. A. Adderley *et al.*, “Charge and fluence lifetime measurements of a dc high voltage GaAs photogun at high average current,” *Phys. Rev. ST Accel. Beams*, vol. 14, pp. 043501-1–043501-12, April 2011.
[7.73] H. Iijima, C. Shonaka, M. Kuriki *et al.*, “A study of lifetime of NEA-GaAs photocathode at various temperatures,” in *Proc. 2010 Int. Particle Accelerator Conf.*, 2010, pp. 2323-2325.
[7.74] M. Woods, J. Clendenin, J. Frisch *et al.*, “Observation of a charge limit for semiconductor photocathodes,” *J. Appl. Phys.*, vol. 73, pp. 8531-8535, June 1993.
[7.75] G. A. Mulhollan, A. V. Subashiev *et al.*, “Photovoltage effects in photoemission from thin GaAs layers,” *Phys. Lett. A*, vol. 282, pp. 309-318, April 2001.
[7.76] H. Tang, R. K. Alley, H. Aoyagi *et al.*, “Study of nonlinear photoemission effects in III-V semiconductors,” in *Proc. 1993 Particle Accelerator Conf.*, 1993, pp. 3036-3038.
[7.77] F. Loehl, private communication.

# CHAPTER 8: CATHODES FOR POLARIZED ELECTRON BEAMS

MATTHEW POELKER

*Thomas Jefferson National Accelerator Facility*
*M/S Bldg. 5A, Room 500-17*
*12050 Jefferson Ave.*
*Newport News, VA 23606*

**Keywords**

Polarized Electrons, GaAs, Semiconductor, Superlattice, Strained Superlattice, Negative Electron Affinity, Vent-Bake Photogun, Load-Lock Photogun, Ion Bombardment, Cathode Degradation, Activation, Hydrogen Cleaning, Fiber Laser, Gain Switching, Spin Manipulation, Mott's Polarimeter

**Abstract**

Cathodes that can deliver spin polarized electrons with high quantum efficiency are unique and need special preparation and handling. In this chapter, we discuss the physics underpinning of GaAs as the source for polarized electrons and methods to increase the QE and the degree of polarization simultaneously. Since this cathode is very sensitive to contaminants, we also describe two methods of incorporating the cathode into the high voltage DC photogun. The back bombardment of ions on the cathode in this gun is the limiting factor for the life-time of such a cathode. We present strategies to mitigate this effect as well as step-by-step procedure for cathode preparation. Since the excitation of GaAs for spin polarized electrons is wavelength sensitive, we have devoted a section to this drive laser. We conclude with a brief section on polarimetry, spin manipulation and future directions of research that may benefit this field.

## 8.1 INTRODUCTION

Many accelerator-based nuclear- and particle-physics experiments require a spin-polarized electron beam [8.1], *viz* a beam wherein the spin axes of the electrons within each accelerated bunch are aligned in a preferential direction. Electron spin can be another "tool" in the physicist's tool bag; one that enhances studies of nuclear structure, the dynamics of strong interactions, electro-weak nuclear physics including parity-violation, physics beyond the Standard Model and more [8.2]. Electron beams at accelerator storage rings "self-polarize" *via* Sokolov-Ternov synchrotron spin-flip radiation; however, at other types of accelerators a direct source of polarized electrons is required.

The first polarized-electron source for an accelerator, based on the photo-ionization of state selected $^6$Li atoms, was developed at Yale University in the early '70s for the Stanford Linear Accelerator Center (SLAC) [8.3]. In 1977, a polarized electron source based on the Fano effect in rubidium was constructed for the Bonn synchrotron [8.4]. Other polarized sources were developed or proposed during the '70s, including an improved version of the $^6$Li photo-ionization source [8.5], a source based on the chemical ionization of metastable He atoms [8.6], and sources using the Fano effect in Cs [8.7]. Despite some technical demonstrations, none of these latter sources ever became operational in accelerators. After the 1974 demonstration of polarized photoemission from GaAs [8.8] at low voltage, researchers at SLAC constructed a high voltage source [8.9] to conduct the seminal parity violation experiment E122 [8.10] that verified Wienberg and Salam's predictions and thereby helped to establish the Standard Model of electro-weak physics. Since then, DC high voltage polarized electron sources based on GaAs photocathodes were built and have operated at several laboratories, including Nagoya University [8.11], the Mainz Microtron [8.12], [8.13], the MIT-Bates Laboratory [8.14], NIKHEF [8.15], Bonn University [8.16] and CEBAF/Jefferson Lab [8.17].

There are four basic requirements for constructing a DC high voltage spin-polarized electron source using GaAs photocathodes:

1. atomically clean GaAs photocathode material,
2. an appropriate high voltage cathode/anode accelerating structure free of field emission,
3. an ultra-high vacuum (UHV) chamber, and,
4. a suitable drive laser.

Paying proper attention to these requirements will enable researchers to build a good spin-polarized electron source, where ***good*** describes a source that delivers a highly polarized beam at the desired current for a long time.

I note that some of the topics above are described in other chapters, particularly Chapter 4 on DC photoinjectors and Chapter 9 on photoinjector drive lasers. To avoid redundancy, this chapter will focus on issues specifically related to generating spin-polarized electron beams.

## 8.2 GaAs: A SOURCE OF POLARIZED ELECTRONS

GaAs is a direct transition III-V semiconductor with zincblende crystal structure [8.18], [8.19]. It can absorb laser light across the broad visible spectrum, but only illumination with near-IR wavelengths provides polarized photoemission. This is understandable by looking at the energy level diagram of GaAs. Figure **8.1** shows detailed and simplified representations [8.20] of the band structure, as described in the figure caption. Electron spin-orbit coupling splits the $P_{1/2}$ and $P_{3/2}$ energy levels of the valence band into two states separated by 0.34 eV, *i.e.*, large enough to avoid optical excitation from the lower energy $P_{1/2}$ state. Polarized photoemission takes advantage of the quantum-mechanical selection rules, *viz*, for circularly polarized laser light, conservation of angular momentum requires that an electron's spin-angular momentum quantum number to change by one unit, $\Delta m_j = \pm 1$. Furthermore, some transitions are more favorable than others, as indicated by the transition probabilities in Figure **8.1(c)**. So, by using circularly polarized laser light with photon energy equal to near-band gap energy, the conduction band can be populated preferentially with a particular spin state.

Polarization is defined as

$$P = \frac{N\uparrow - N\downarrow}{N\uparrow + N\downarrow} \tag{8.1}$$

where N refers to the number of electrons in the conduction band of each spin state, "up" or "down". For bulk GaAs, the theoretical maximum polarization is 50%, corresponding to three electrons of the desired spin state and one electron with opposite spin. However, in practice, due to various depolarization mechanisms, such as the Bir-Aronov-Pikus process [8.21], the D'Yankonov-Perel process [8.22], the Elliot-Yafet process [8.23], [8.24], and radiation trapping [8.25], the maximum polarization from the bulk is typically 35%. A less detailed description simply attributes depolarization to imperfections within the photocathode material that reduce the diffusion length. This will prevent electrons from efficiently reaching the photocathode surface, as well as providing a greater probability for the electrons to depolarize.

The figure of merit of polarized-electron beam experiments is a measure of how quickly an experiment can be performed with a desired level of accuracy. The figure of merit scales with $P^2I$, where $P$ refers to polarization and $I$ refers to the beam current. Hence, there is a great incentive to increase the beam's polarization, particularly for experiments that cannot accommodate a high current, for example due to concerns over the failure of the target window or over cryogenic target boiling. Significant breakthroughs in the development of polarized electron sources occurred in the '90s, when groups at SLAC and Nagoya University independently developed a means to eliminate the heavy-hole/light-hole degeneracy at the valence-band maxima by introducing an axial strain within the GaAs crystal [8.26]–[8.28]. This was accomplished by growing GaAs atop GaAsP to introduce a strain resulting from the lattice mismatch between the GaAs and GaAsP crystal structures. Such a structure delivers polarization as high as ~75%, however the photocathode yield, or quantum efficiency (QE), typically is very low, just 0.1% (QE is discussed below). The GaAs surface layer is usually 50-100 nm thick. Thicker layers can provide higher QE, but then the strain relaxes and polarization is reduced.

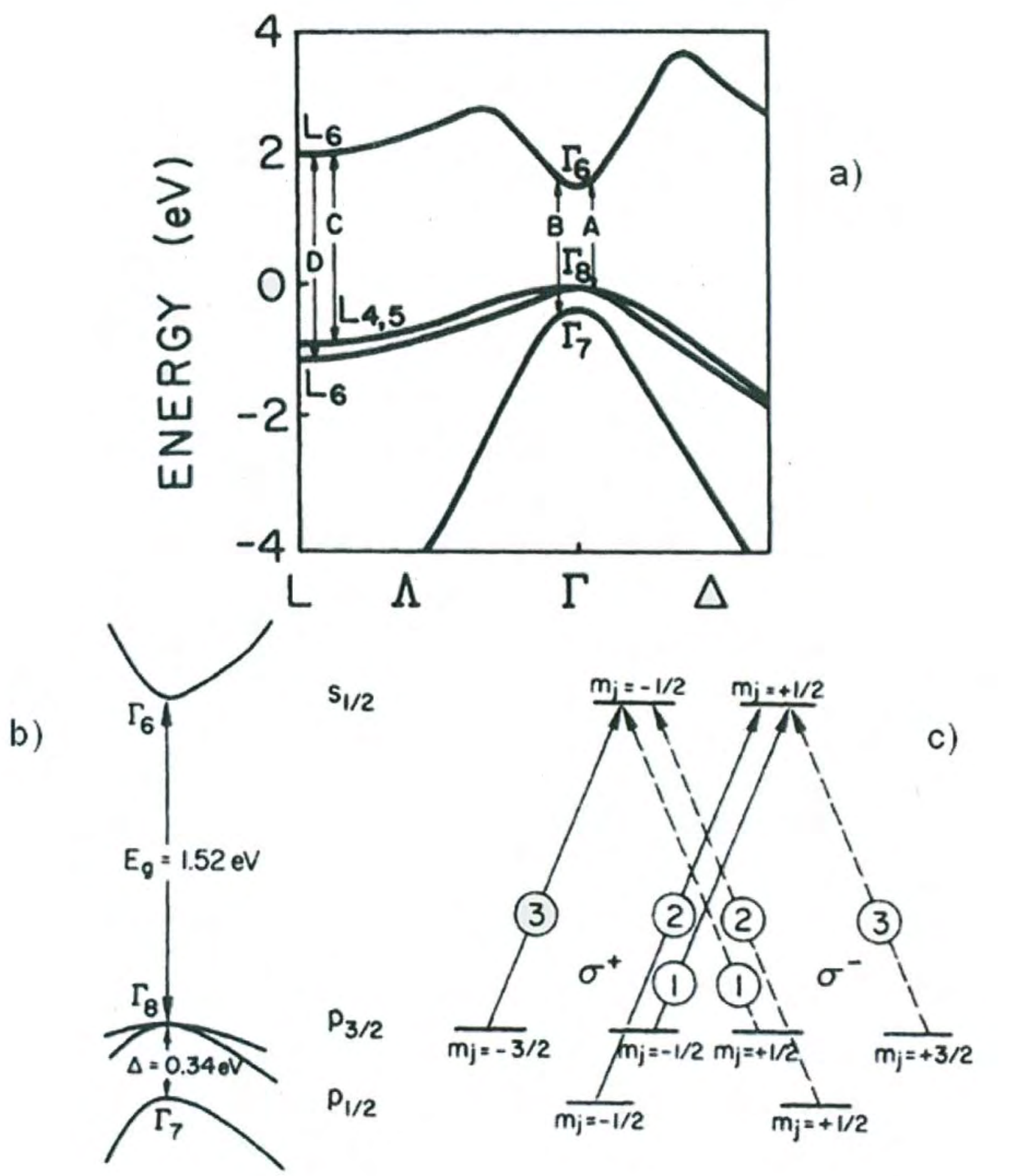

**Figure 8.1. Energy level diagrams of GaAs: (a) Detailed band structure of GaAs; (b) "close-up" view near valance band maxima/conduction band minima; and, (c) simplified view showing individual spin-angular momentum states and transition probabilities, circled. [Reprinted figures with permission from [8.20]. Copyright 1976 by the American Physical Society]**

Today's state of the art high polarization photocathode is the GaAs/GaAsP strained-superlattice structure [8.29], [8.30] consisting of a very thin GaAs surface layer (~5 nm) grown atop 10-20 pairs of thin, alternating layers of GaAsP and GaAs. These very thin GaAs layers maintain the strain, which improves polarization. Using many thin layers of GaAs/GaAsP raises the QE considerably higher than that obtained from a single (thicker) layer of strained GaAs since the electrons in sub-surface layers efficiently tunnel through the thin GaAsP layers. The net result is a polarization ~85% and a QE ~1%. Figure **8.2** represents each high polarization photocathode schematically with plots of polarization versus laser wavelength [8.31].

Both of the photocathode structures are commercially available [8.32], [8.33] thanks to collaborative R&D programs initiated by SLAC *via* the Department of Energy's (DOE) Small Business Innovative Research (SBIR) program. University groups in Japan and Russia have made similar photocathodes [8.34], [8.35] with different stoichiometric combinations of Ga, As and P, as well as In and Al to modify the band gap, and correspondingly the appropriate drive laser wavelength.

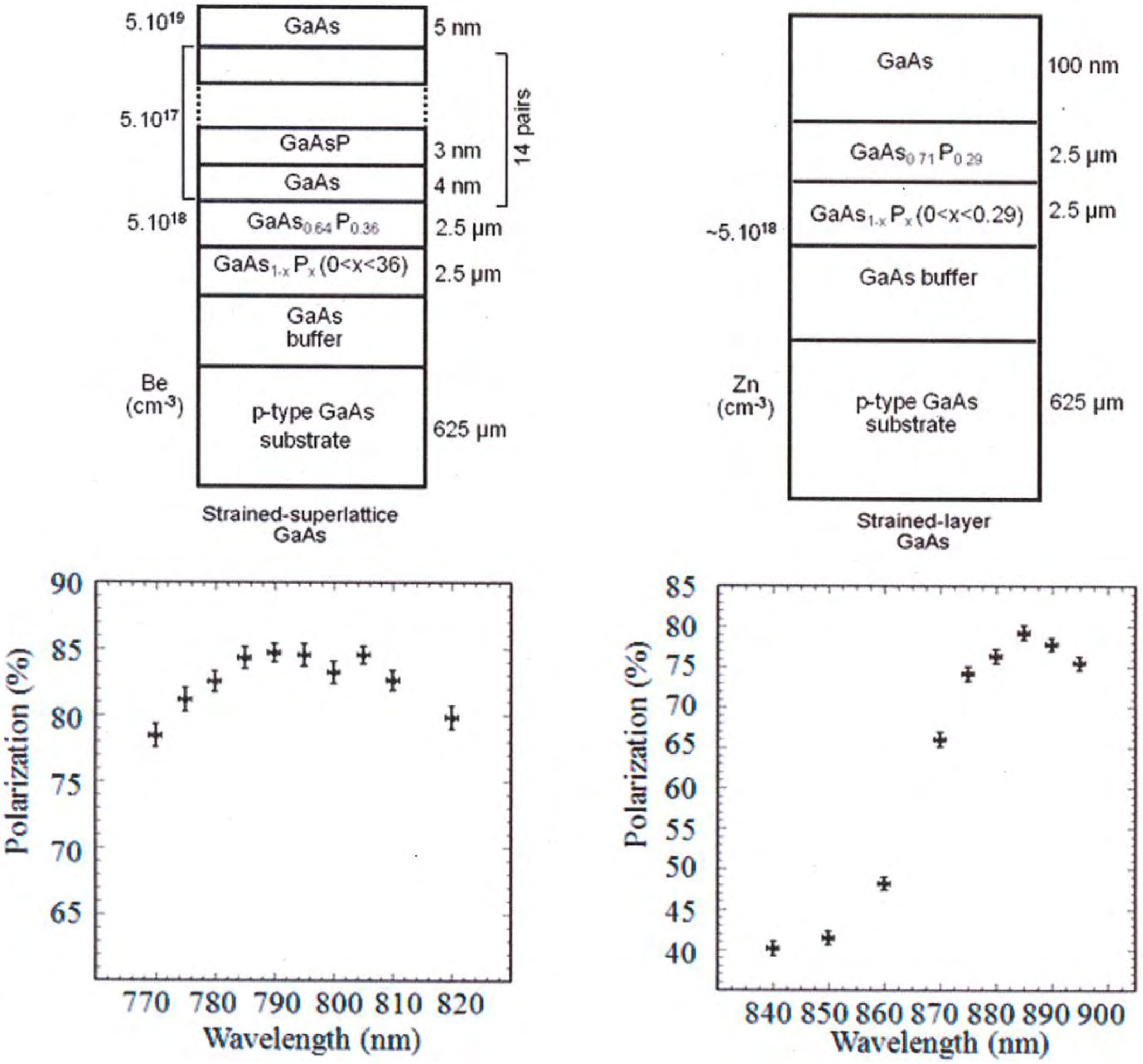

**Figure 8.2. Two types of GaAs photocathode structures that provide high polarization: Spectral dependence of polarization for (left) strained-superlattice GaAs and (right) strained-layer GaAs. [Reprinted figure with permission from [8.31]. Copyright 2005 by the American Physical Society]**

As detailed in Chapter 5, the emission of electrons from cathodes, such as GaAs, often is described by the three-step process [8.36] involving the absorption of light, the diffusion of electrons to the surface of the photocathode and the emission of the electrons into the gun vacuum chamber. GaAs is a strong absorber, absorbing most of the light within a few hundred nanometers. As described above, absorption of circularly polarized light with near-band gap energy preferentially populates the conduction band with spin-polarized electrons. These electrons diffuse in all directions and those that move toward the surface encounter a potential barrier known as the electron affinity (Figure **8.3(a)**). A requirement for efficient photoemission is that the GaAs is p-doped [8.8], which lowers the Fermi level throughout the material. The p-doping also lowers the conduction band at the surface of the photocathode, which, in turn decreases the electron affinity (Figure **8.3(b)**). Still, no significant photoemission is obtained until the potential barrier is reduced further; this is accomplished by adding a mono-layer of cesium and an oxidant to the photocathode (Figure **8.3(c)**), a process called “activation”. When the vacuum energy level is reduced below that of the conduction band of the intrinsic material, a negative electron affinity (NEA) condition is said to exist.

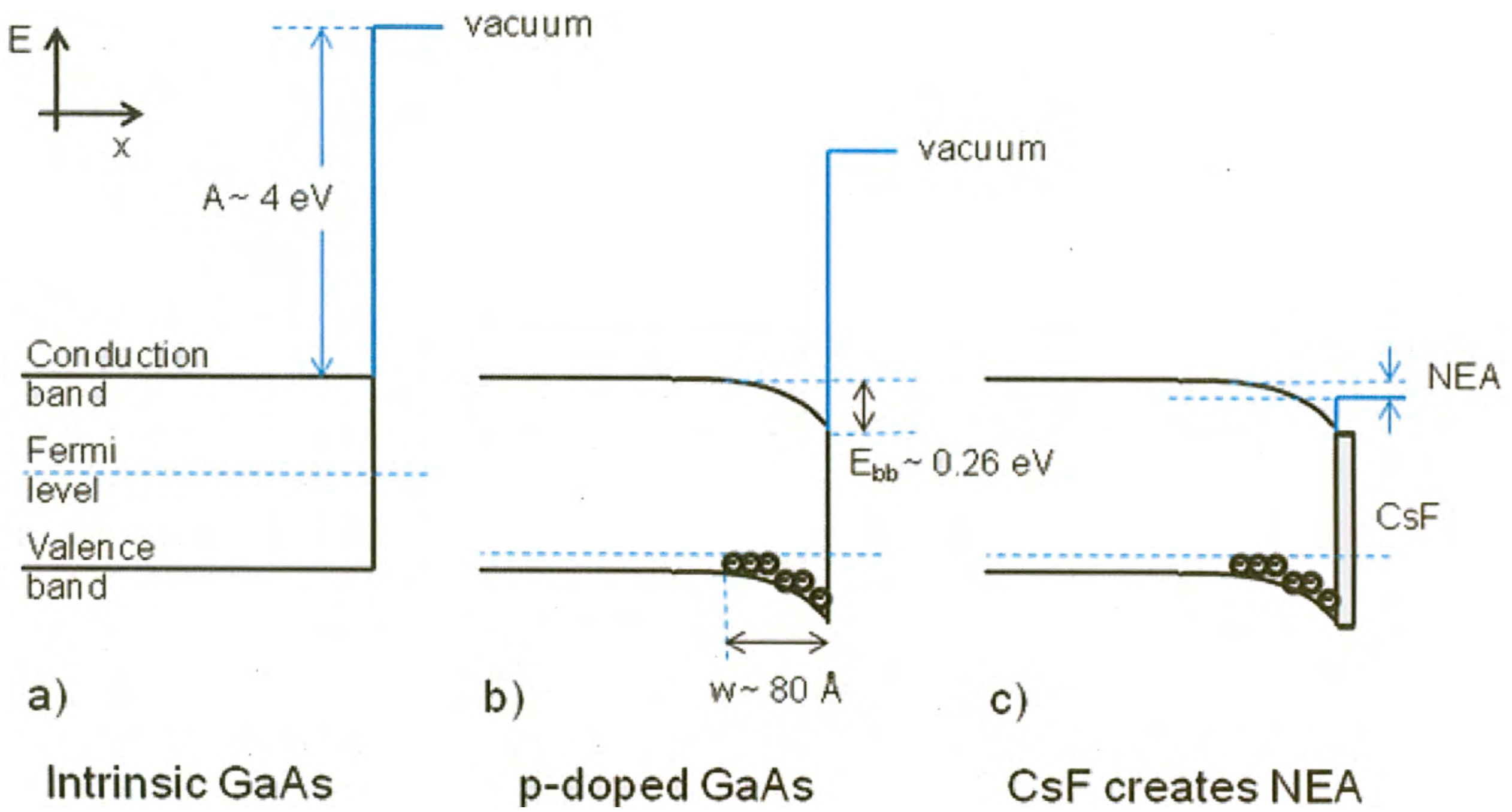

**Figure 8.3. Energy level diagram of GaAs at the vacuum interface. (a) Undoped GaAs, (b) p-doped GaAs, and (c) with Cs and oxidant applied to the surface. *A* is the electron affinity; $E_{bb}$, is the energy offset associated with band bending region**

While this sounds relatively simple, obtaining the expected amount of photoemission in practice can be difficult because the GaAs surface must be extremely clean and free of contamination at the atomic scale. Unfortunately, several steps are needed to insert a GaAs photocathode into a DC high voltage photogun, opening up many opportunities to contaminate the wafer. Once the photocathode is installed within the photogun, it must remain clean, meaning the photogun must function properly while delivering beam; the static vacuum inside the photogun must be extremely low and remain low while delivering the beam. The following sections describe typical photoguns and the steps required to install the clean photocathodes.

## 8.3 DESCRIPTION OF TYPICAL POLARIZED PHOTOGUN

DC high voltage, GaAs-based polarized photoguns are categorized as vent/bake or load-lock photoinjectors. In general, the former are considered easier to build, but require frequent maintenance, whereas the latter offer more accelerator up-time, once reliable sample manipulation has been demonstrated. Each type is described briefly below.

### 8.3.1 Vent/Bake Photoguns

Vent/bake photoguns must be vented to atmospheric pressure each time the photocathode is replaced and then baked for an extended time to recover the necessary vacuum level (see below). A typical vent/bake photogun shown in Figure **8.4** was successfully used at the CEBAF/Jefferson Laboratory for over ten years [8.17] and is described below. All the features needed to activate the photocathode to NEA, bias the photocathode at high voltage and generate high quality beam in an UHV environment are housed in a common vacuum chamber.

The photocathode is attached to the end of a long stalk extending into the gun vacuum chamber through the bore of the large cylindrical insulator, which is typical of the variety described in Chapter 4. Before activating the photocathode to NEA, the photocathode must be heated to ~500 ˚C to liberate loosely bound adsorbed gas. Higher temperatures can "boil off" some surface contamination (oxides, in particular), but not carbon. Temperatures > 630 ˚C must be avoided as this causes the GaAs to decompose due to preferential evaporation of arsenic. To heat the photocathode, the stalk is retracted ~50 mm to avoid heating other parts of the gun and a resistive heater is inserted into the atmospheric side of the stalk in close mechanical contact. After the heat treatment, the photocathode is left to cool to room temperature and moved back into position within the cathode electrode for activation and beam generation. The cathode electrode has a 25˚ focusing angle and the anode is ~6 cm away. This geometry optimizes transport for the CEBAF beam with a maximum field gradient of ~5 MV $m^{-1}$ when the cathode electrode is biased at -100 kV. Note that the cathode/anode geometry of each photogun depends heavily on the accelerator beam's specifications (*e.g.*, bunch charge) and typically is determined by computer simulations (*i.e.*, field mapping and particle tracking).

Non-evaporable getter (NEG) modules surround the cathode/anode gap, providing thousands of liters per second pumping for hydrogen. A small diode ion pump (not shown) pumps inert gasses, such as helium and methane that the NEGs do not pump efficiently. The photocathode is activated to NEA using cesium and flourine (or oxygen) sources located downstream of the anode. During activation, the drive laser can be directed on to the photocathode, or it can be illuminated from the side with a white light source, directed by a metallic mirror inside the vacuum chamber. The two chemicals, Cs and $NF_3$, are applied to the photocathode and metered while monitoring photocurrent that varies in a "yo-yo" manner (described below), although other groups follow different, but acceptable protocols. The chemical application is terminated once photocurrent ceases to increase appreciably, typically after ten yo-yos, with the net result corresponding to approximately one monolayer of chemical deposition. Cesium orginates from an alkali-metal dispenser from SAES Getters and is controlled by applying electrical current through a vacuum feedthough. The $NH_3$ is applied using a vacuum leak valve.

As mentioned above, the entire gun structure must be baked each time the photocathode is replaced. Bakeout temperature typically is ~250 ˚C and its duration is ~30 hr, although it can last longer if there is significant water vapor inside the vacuum chamber. High temperature bakeouts necessitate some precautions. For example, bare copper gaskets will oxidize during a bakeout. This is problematic because the oxide layer can "flake off" when flanges are disassembled, sometimes leading to a flange leak during a subsequent bakeout. To prevent this from happening, copper gaskets should be nickle-flashed and silver-plated to prevent oxidation. Silver-plated bolts are also recommended for the same reason. This ensures that nuts and bolts turn freely post-bakeout when the gun is disassembled. The NEG pumps can be electrically activated or passively activated to about 60% of their rated pump speed during the bakeout.

Besides the burden of vacuum-chamber bakeouts, which take days to complete, the most significant drawback of the vent/bake photogun design is the inadvertent application of cesium on the cathode electrode, which eventually leads to catastrophic field emission necessitating cathode electrode cleaning or replacement. The design in Figure **8.4** provides about seven full photocathode-activations before succumbing to field emission. Other gun designs at other laboratories fared better or worse; in hindsight, the results likely depended on the size of the anode hole and location of the cesium dispenser relative to the anode, which define the solid angle of cesium deposition at the photocathode and cathode electrode. Gun designs with small solid angle faired better than those that introduced more cesium on the cathode electrode.

### 8.3.2 Load-Lock Photoguns

Load-lock photoguns are comprised of multiple vacuum chambers separated by valves, with the vacuum improving from one chamber to the next, with the best vacuum obtained inside the gun's high voltage chamber. Reiterating, new photocathode samples can be installed without lengthy vacuum bake-outs of the entire gun; comprising one of the major benefits of the load-lock design. Another benefit is that cesium is not applied inadvertently to the cathode electrode since activation takes place inside another chamber. In this way, the cathode electrode remains pristine and it exhibits no field emission when biased at high voltage. Historically, most researchers have moved to a load-lock design to eliminate field emission.

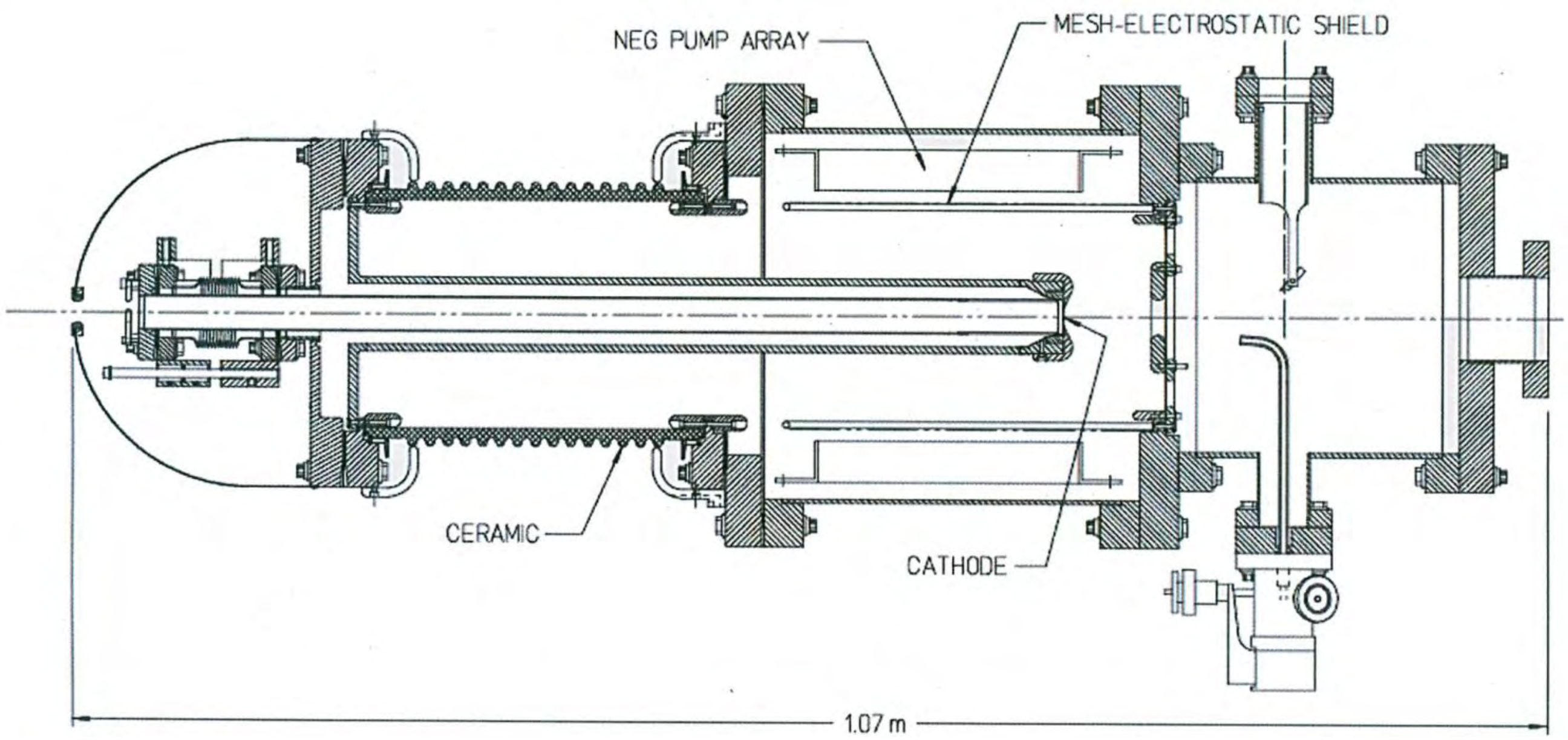

**Figure 8.4. The CEBAF/Jefferson Lab vent/bake 100 kV DC high voltage, spin polarized GaAs photogun. It rests in the horizontal plane with the drive laser light introduced through a vacuum window to the right (not shown). [Adapted from [8.37], with permission from Elsevier]**

The CEBAF/Jefferson Laboratory's load-lock gun is shown in Figure **8.5** [8.38]. It consists of four vacuum chambers: The high voltage chamber, the photocathode preparation chamber, a "suitcase" chamber used for replacing photocathode samples and an intermediary chamber that must be evacuated and baked each time the suitcase is attached. The suitcase normally is detached from the photogun and stored elsewhere. This approach helps to reduce the overall footprint of the photogun when in operation. Numerous alternative designs are used at laboratories worldwide [8.12], [8.13], [8.15], [8.16], [8.39], [8.40]. The ability to store multiple photocathode samples, reliable sample transport from one chamber to the next without dropping, and rapid heating and cooling of samples for fast turn-around at activation are some of the desirable features incorporated into most load-lock designs.

The high voltage chamber is similar to that of the vent/bake photogun described in Section 8.3.1, but lacks the components associated with photocathode activation. The NEG pump modules surround the cathode/anode gap and a small ion pump is used to pump inert gases.

Key features of the preparation chamber include storage for up to four pucks (each puck supports one photocathode), a mask for selective activation of a portion of the photocathode surface, puck heating to at least 600 ˚C, and good vacuum obtained using NEG and ion pumps. Photocathode activation takes place inside the preparation chamber using cesium and $NF_3$ similar to that described for the vent/bake photogun.

The preparation chamber has four magnetically coupled sample manipulators: one long manipulator with translation- and rotation-capability for moving pucks into or out of the gun's high voltage chamber cathode electrode; one short manipulator with translation- and rotation-capability for moving pucks from/on to the heater assembly and for transferring them to/from the long manipulator; and, two short manipulators with translation capability that hold pucks with additional photocathode samples.

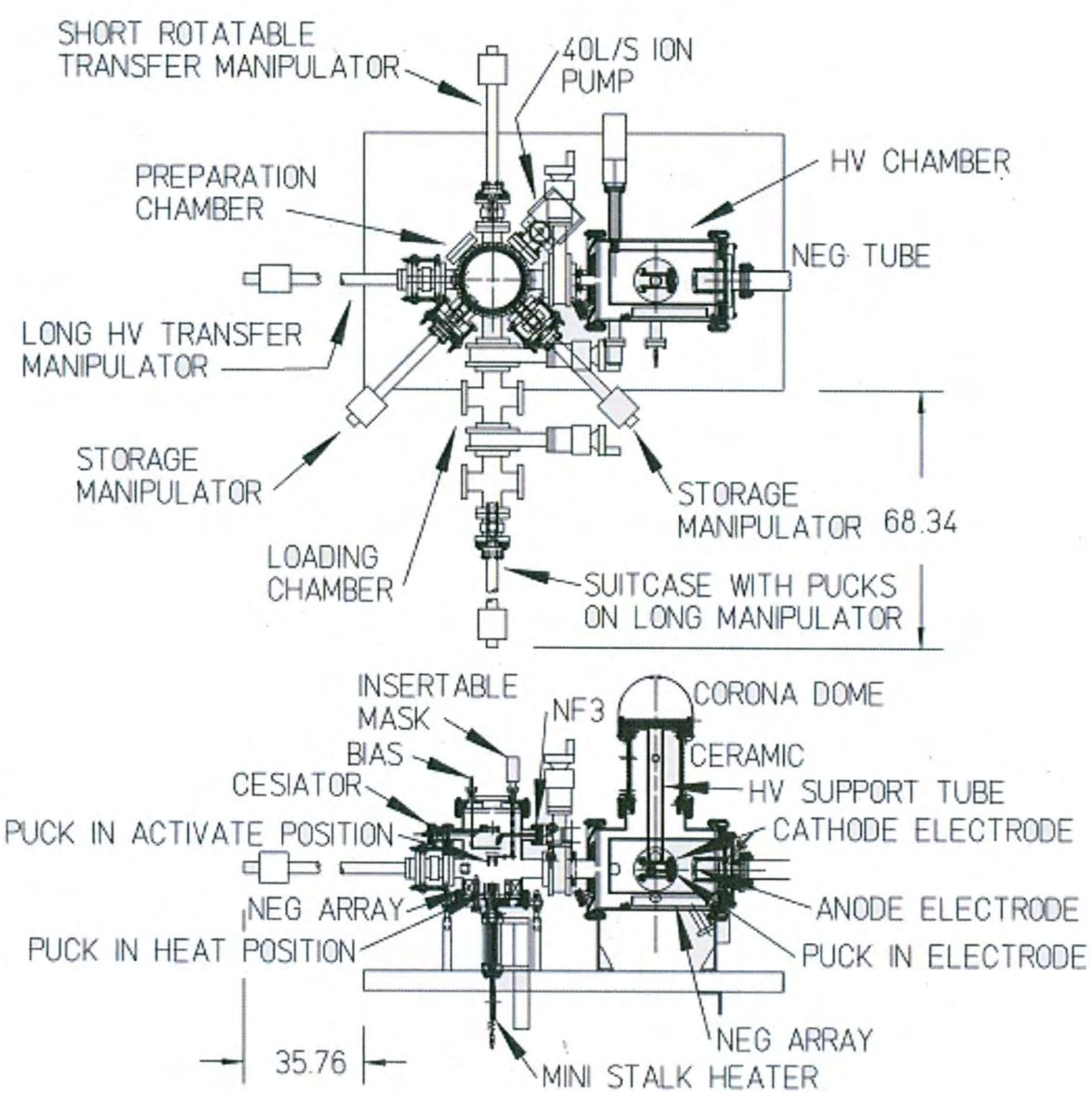

**Figure 8.5. (Top) The plan view shows the complete gun assembly with four vacuum chambers: Gun High Voltage Chamber (large bore ceramic insulator design), Preparation Chamber, Intermediary Chamber and "Suitcase". (Bottom) The side view shows some of the components inside the preparation chamber including a heater that also serves to move the puck toward a mask used to selectively activate only the center portion of the photocathode. [Reprinted figure with permission from [8.38]. Copyright 2011 by the American Physical Society]**

Care must be taken during the initial commissioning bake of the preparation chamber; the magnetic manipulators can develop excessive friction that limits functionality when heated above ~200 ˚C. Each magnetic manipulator is attached to a bellows assembly with adjustment screws for proper alignment to the electrode, heater and for other manipulators. Pumping inside the preparation chamber is provided by 40 L $s^{-1}$ ion pump and 1.5 WP-1250 NEG modules from SAES Getters with support rods removed and coiled into the bottom of the vacuum chamber. Pressure inside the preparation chamber is ~$1\times10^{-8}$ Pa, which is adequate for preparing a photocathode with high QE; however, improved vacuum would provide a longer dark lifetime, *i.e.*, the QE lifetime of the photocathode when it is not biased at high voltage or illuminated with laser light.

# 8.4 OPERATION OF A DC HIGH VOLTAGE, SPIN-POLARIZED GaAs PHOTOGUN

## 8.4.1 Ion Bombardment

The photocathode lifetime of modern DC high voltage GaAs photoguns is limited primarily by ion bombardment, or ion back-bombardment [8.41]. This is the mechanism wherein residual gas is ionized by the extracted electron beam and transported backward to the photocathode, where the ions adversely affect the photocathode's QE. Exactly how the ions degrade QE is the subject of much speculation. While it was determined that ions with sufficient kinetic energy penetrate the photocathode's surface [8.42], it is not known what they do to the photocathode. They might damage the GaAs crystal structure, act as trapped interstitial defects that reduce the electron diffusion length, or act as unwanted dopant species, adversely altering the photocathode's energy band structure. Impinging ions may also sputter away the chemicals used to reduce the work function at the surface of the photocathode. Predicting which ions are the most problematic (gas species and energy) awaits a detailed modeling study that considers many parameters. The study should include relevant ion species with appropriate ionization cross sections, accurate trajectories of both ions and electrons, sputtering yield of alkali (cesium) and oxidant (fluorine) used to create the NEA condition at the photocathode surface required for photoemission and the stopping depths of ions within the photocathode. Parameters such as optical absorption length, electron-diffusion length and active layer thickness are likely to be important factors as well.

The ions produced by the electron beam arrive at the photocathode in a manner determined by the electrostatic field of the cathode/anode structure. When the drive laser beam is positioned at the center of the photocathode, all of the ions are delivered to the same location. When it is moved radially outward, ions are produced at the location of the laser beam and along a "trench" connecting the point of origin to the electrostatic center of the photocathode. Furthermore, ions produced downstream from the anode can arrive at the photocathode and hit the electrostatic center. A typical "QE scan" of a GaAs photocathode is shown in Figure **8.6**, illustrating QE reduction due to ion bombardment.

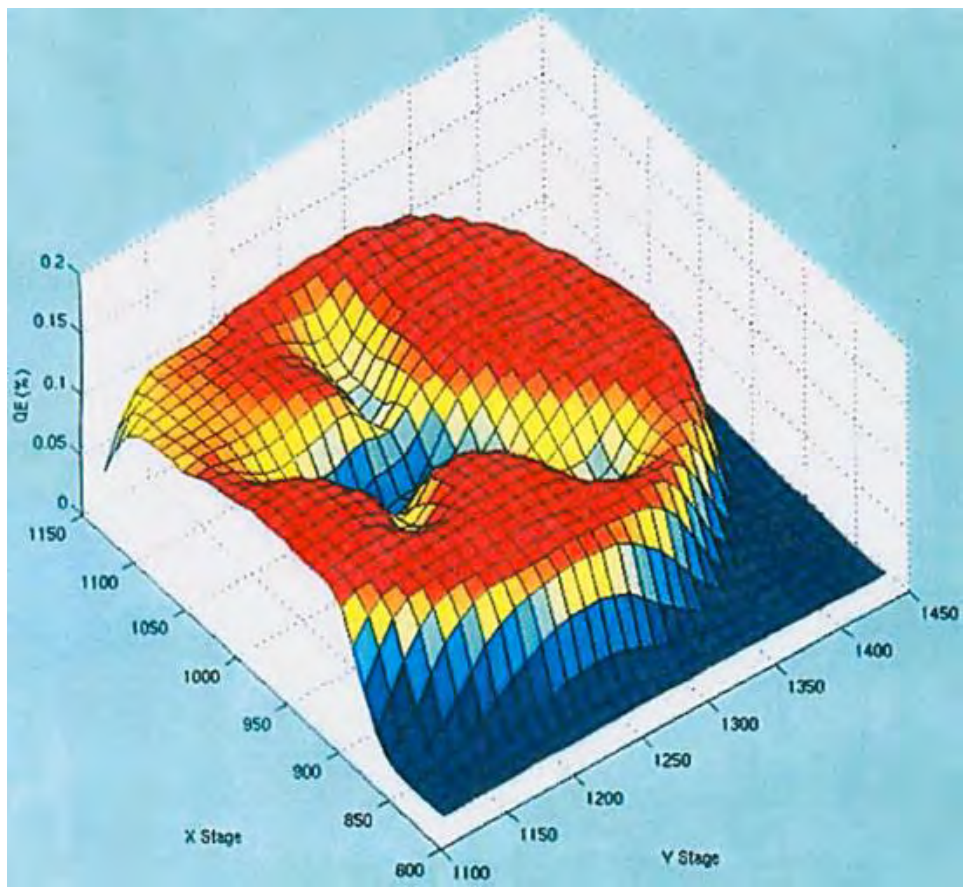

**Figure 8.6. False-color QE map across the surface of a photocathode that has been damaged by ions. The colors denote the QE value at each photocathode location. The electron beam was extracted from three different radial locations. Note QE "trenches" that terminate at a common "electrostatic center". [Adapted with permission from [8.43]. Copyright 2007, American Institute of Physics]**

The best strategy for minimizing QE decay associated with ion back-bombardment is to operate the photogun under an excellent vacuum. This includes static vacuum (no beam) and dynamic vacuum (while delivering the beam). A small cathode/anode gap is desirable to limit the number of ions created. However, small gaps produce a large gradient which in turn enhance the field emission from the cathode electrode, which can significantly degrade the gun's performance *via* chemical poisoning of the photocathode surface and enhanced ion bombardment. A recent study [8.38] showed that the best operating lifetime can be obtained by operating with a laser beam positioned away from the electrostatic center and with an active area that minimizes the creation of "halo" beam that might not be efficiently transported away from the photogun.

### 8.4.2 Vacuum

As discussed, the longest photocathode lifetime is attained by minimizing ion bombardment which recommends operating the photogun in an exceptionally good vacuum. Equ. 10.2 provides remarkably useful insight toward appreciating the vacuum aspects of the photogun:

$$P_{ult} = \frac{(Gas\ Load)}{(Pump\ Speed)} \tag{8.2}$$

where $P_{ult}$ is the ultimate pressure inside the gun. Obviously, it is beneficial to make the gas load inside the photogun small and the pump speed large.

To ensure a small gas load inside the photogun, several steps must be taken. First, proper UHV techniques must be practiced [8.44]. This includes constructing the photogun free of contamination. Manufactured parts typically are fabricated without oil or silicone lubricants and all components are cleaned in an ultrasonic bath of alkaline cleaner followed by acetone and hot de-ionized water. It also is very important to ensure that there are no virtual leaks inside the gun (*i.e.*, small spaces with trapped gas). For this reason, internal components are assembled with vented, silver-plated, stainless-steel screws.

After the photogun is constructed, it must be evacuated and baked to remove water vapor; typically, the gun is baked at 250 ˚C for 30 hr or longer. To assist this process, all CEBAF photoguns are built on tables with an insulating top large enough to accommodate oven panels that completely surround the gun. Heated air is directed into the enclosure using a 4 kW commercial heater system. The use of flowing hot air assures that the gun structure is heated uniformly without developing significant temperature differentials. As mentioned previously, nickel-flashed and silver-plated gaskets are recommended to avoid oxidation that can lead to flange leaks. The silver-plated, high-strength, stainless-steel bolts and stainless-steel nuts are easily disassembled post-bake. Belleville washers are used on flanges larger than 70 mm to assure reliable sealing during the expansion and contraction cycles of high temperature bakeouts.

When constructed properly, the gas load within the baked photogun originates from hydrogen out-gassing from the walls and internal components of the photogun. The typical out-gassing rate of 304 stainless steel is $1\times10^{-10}$ Pa L $s^{-1}$ $cm^{-2}$; with vacuum pumping from the NEGs and ion pump described below, it is not difficult to obtain pressure in the low $10^{-9}$ Pa range. Hotter bakeouts [8.45] provide lower out-gassing rates and proportionally lower pressure. High current applications benefit from extra effort to reduce the out-gassing rate of photogun components.

Ideally, when the valve to the beamline is opened, the gun vacuum should not degrade appreciably. This means that the beamline must be baked, and if space allows it is a good idea to incorporate a differential pump station near the gun to isolate the gun vacuum from the rest of the accelerator.
All modern DC high voltage, spin-polarized GaAs photoguns rely on NEG pumps and a small diode ion pump for removing inert gasses like He and methane which are not pumped by NEGs. NEG pumps provide thousands of liters per second pump speed for hydrogen gas, the dominant gas species inside a UHV chamber. NEG pumps are commercial items purchased from SAES Getters, and the pumps that rely on ST707 material can be activated at relatively low temperature (~400 ˚C). Typically, a photogun design incorporates many NEG modules connected in series and electrically isolated inside the gun. The pumps are activated (*i.e.*, heated) by passing current through them.

# 8.5 PHOTOCATHODE PREPARATION

As described, several steps must be taken to insert a GaAs photocathode into a DC high voltage photogun, so there are many opportunities to contaminate the wafer. These steps include:

1. Cutting a photocathode sample from a large wafer supplied by the vendor.
2. Anodizing the edge of the photocathode to eliminate unwanted photoemission from region not supported by a proper electrostatic field. This step can be eliminated if a mask is used at activation.
3. Mounting the photocathode sample to a support structure that eventually will be positioned within the cathode electrode.
4. Baking the photocathode and support structure to achieve the required vacuum level.
5. Heating the photocathode to ~500 ˚C to liberate loosely bound gas before activation to NEA.

The exact details of these steps vary somewhat depending on the specific design of the photogun; for example, whether it is vented and baked each time the photocathode is replaced, or installed *via* a load-lock vacuum apparatus wherein the photocathode is mounted on to a small support structure and moved between different vacuum chambers. The text below describes features common to both gun designs and highlights some of the relevant differences.

## 8.5.1 Cutting GaAs to the Appropriate Shape and Size

GaAs material typically is sold in large circular discs which are flat at one edge to indicate the direction of the cleave plane. The disc is usually ~600 μm thick and 50-75 mm in diameter. This large wafer must be cut into smaller samples for installing into photoguns. Originally at CEBAF/Jefferson Laboratory, samples were cut from large wafers using a circular-shaped cutting jig and diamond-paste slurry. The large wafer was sandwiched between glass slides using an acetone-soluble adhesive to protect the surface of the photocathode during cutting. This process was time consuming and invariably introduced a significant amount of contamination on the surface of the photocathode which needed to be removed using strong acids/bases or *via* hydrogen cleaning. This cutting technique was replaced with a far simpler cleaving technique. Now, a diamond-tipped scribe is used to cleave square samples from large wafers. Aside from the diamond-tip scribe, nothing touches the surface of the photocathode material during cleaving, and as a result, the photocathode surface is not contaminated.

## 8.5.2 Anodizing the Edge of the Photocathode to Limit QE

It is very important to eliminate unwanted and inadvertent photoemission from the edge of the photocathode – photoemission that is not properly tranported away from the gun results in the halo effect. Photoemission from the edge of the photocathode follows extreme trajectories, striking the vacuum chamber wall downstream of the gun and even hitting the anode plate. This degrades the vacuum in the gun hastening the photocathode’s QE decay. Anodizing the edge of the photocathode in an electrolytic bath is one way to eliminate photoemission from the edge.

A fixture was devised that holds the photocathode, sandwiched between two Viton O-rings. One O-ring prevents electrolytic fluid from contacting the center portion of the front face, while the other merely provides a surface to securely and carefully hold the wafer without breaking it. Clean distilled water with a few drops of phosphoric acid provides an adequate pH level for anodizing. In just a few seconds, a thick oxide layer forms on the photocathode's edge that assures no measureable photoemission; the oxide layer will not evaporate during bakeouts or photocathode heating.

For load-lock guns, an activation mask can be used to selectively activate only the center portion of the photocathode, eliminating the anodizing step and saving a considerable amount of time.

### 8.5.3 Mounting the Photocathode to Support Structure

Next, the small photocathode samples are indium-soldered to a molybdenum support (*i.e.*, the stalk or puck) at ~200 ˚C inside a nitrogen-filled glove box. Molydenum is a good material for supporting the photocathode sample because of its small coefficient of thermal expansion and UHV compatibility. Indium provides mechanical stability and good heat conduction (the GaAs must be heated to ~500 ˚C to remove the weakly bound gas before activation). A tantalum retaining ring then is placed over the GaAs wafer and crimped in place, thereby ensuring that the GaAs wafer is never dislodged inadvertently from the support structure.

### 8.5.4 Installing the Photocathode into the Photogun Vacuum Chamber: Bakeout, Heating and Activating the Photocathode

The GaAs wafer, mounted to its support structure and loaded into the gun vacuum chamber using a nitrogen-filled glove bag. The gun is pumped down with a clean, oil-free, rough pump. Once the pressure drops sufficiently low to energize the gun ion pump, the valve to the rough pump is closed. The entire photogun chamber is then baked, as described above. It is important to ensure that the GaAs photocathode remains clean during the bakeout. To do so, one must minimize the time spent venting and pumping down the vacuum chamber and vent the vacuum chamber with clean, dry nitrogen gas pressurized to assure minimal back-diffusion during the photocathode exchange. With these precautions, pump-down from atmospheric pressure is rapid, *i.e.* the pressure typically falls below $10^{-6}$ Pa within 20 min after starting pumping. It is good practice to not allow the pressure to rise above ~$1\times10^{-5}$ Pa during bakeout.

Once the bakeout is complete, the photocathode can be activated to NEA. First, the photocathode is heated to ~500 ˚C to liberate loosely bound adsorbed gas from its surface: two hours is sufficient at this temperature. Once the photocathode has cooled to ~30 ˚C, activation begins with the successive application first of cesium, then of $NF_3$ (or oxygen). During activation, the cathode is biased at ~-200 V and illuminated with light. On initial exposure to cesium, the photoemission current reaches a maximum and then decreases. A typical approach (called the "yo-yo" process) allows the photocurrent to decrease to about half of its maximum value before stopping the cesium exposure. On subsequent exposure to $NF_3$, the photocurrent rapidly increases to a new maximum, saturates and then slowly decreases. Further exposure to cesium quickly produces a rapid decrease in the photocurrent. Again, we allow the photocurrent to fall to about half and follow it with another exposure to nitrogen trifluoride. Typically, ten cycles of $Cs/NF_3$ are required to reach the final quantum efficiency.

To assess how well the photocathode was installed, it is customary to evaluate the QE. It can be written in terms of easily measured quanitities such that

$$\mathrm{QE} = \frac{(\#\ photoemitted\ electrons)}{(\#\ incident\ photons)} = 124\,\frac{I}{\lambda P_{inc}} \tag{8.3}$$

where $I$ is the photocurrent in microamperes, $\lambda$ is the laser wavelength in nanometers and $P_{inc}$ is incident laser power in milliwatts. Table **8.1** lists typical QE values from clean photocathode material illuminated with near-band gap light appropriate for high polarization.

| Material | Wavelength [nm] | QE [%] | Polarization [%] |
|---|---|---|---|
| "Bulk" GaAs | 780 | ~10 | ~35 |
| Strained layer GaAs/GaAsP | 850 | ~0.1 | ~75 |
| Strained-Superlattice GaAs/GaAsP | 780 | ~1 | ~85 |

**Table 8.1. Typical QE and polarization for common GaAs photocathodes.**

### 8.5.5 Hydrogen Cleaning

Edge-anodizing is a step that most often introduces contaminants onto the photocathode surface. Baking the photocathode within the gun at high pressure, *e.g.* due to excessive amounts of water in the gun, is another opportunity for contamination. There are many recipes for cleaning semiconductor surfaces with wet chemical solutions of strong acids and/or bases; however, experience at Jefferson Lab with wet chemical cleaning techniques was mixed. Moreover, these techniques remove the surface layer significantly, a situation that is unacceptable when using high polarization photocathodes. Therefore, we adopted an alternative cleaning procedure using atomic hydrogen, exposure to which has shown to remove surface contaminants such as carbon and oxygen from a wide variety of semiconductors [8.46]–[8.50]. Furthermore, as noted in [8.51], hydrogen atoms passivate the dangling bonds at the GaAs surface, leaving a relatively inert surface, ideal for bakeouts.

RF-dissociators and thermal crackers are common sources of atomic hydrogen. At the CEBAF/Jefferson Lab, the RF-dissociator approach is used, although there is some concern that this method roughens the photocathode surface. Molecular hydrogen from a small research-grade bottle is fed through a Pyrex cylinder at about 2 mPa (Figure **8.7**). A 12-turn coil surrounds the Pyrex tube and a plasma is formed when the applied RF (~50 W) is resonant with the circuit. Atomic hydrogen exits the chamber through a ~1 mm diameter hole and is guided *via* an aluminum tube (aluminum has a low recombination rate) to the photocathode sample about 15 cm away. The photocathode sample is held at 300 ˚C during hydrogen cleaning [8.48], [8.49]. A small turbo-molecular pump and an ion pump maintain pressure near the photocathode sample at ~$10^{-3}$ Pa during cleaning to provide a long mean-free path for the atoms and ensure the atoms hit the photocathode before recombining into molecules. Monte Carlo simulations predict that ~2.5% of the total atom flux reaches the photocathode. Under these conditions, the estimated atom flux at the cathode is ~$10^{17}$ atoms cm$^{-2}$ sec$^{-1}$, assuming 50% dissociation [8.51].

After hydrogen cleaning, the stalk and photocathode are installed within the photogun using a nitrogen-filled glove bag. Hydrogen cleaning has an added benefit of yielding a chemically inert surface that helps to keep the photocathode clean during the bake-out of the photogun. Load-lock gun systems also employ hydrogen cleaning for in-*situ* cleaning.

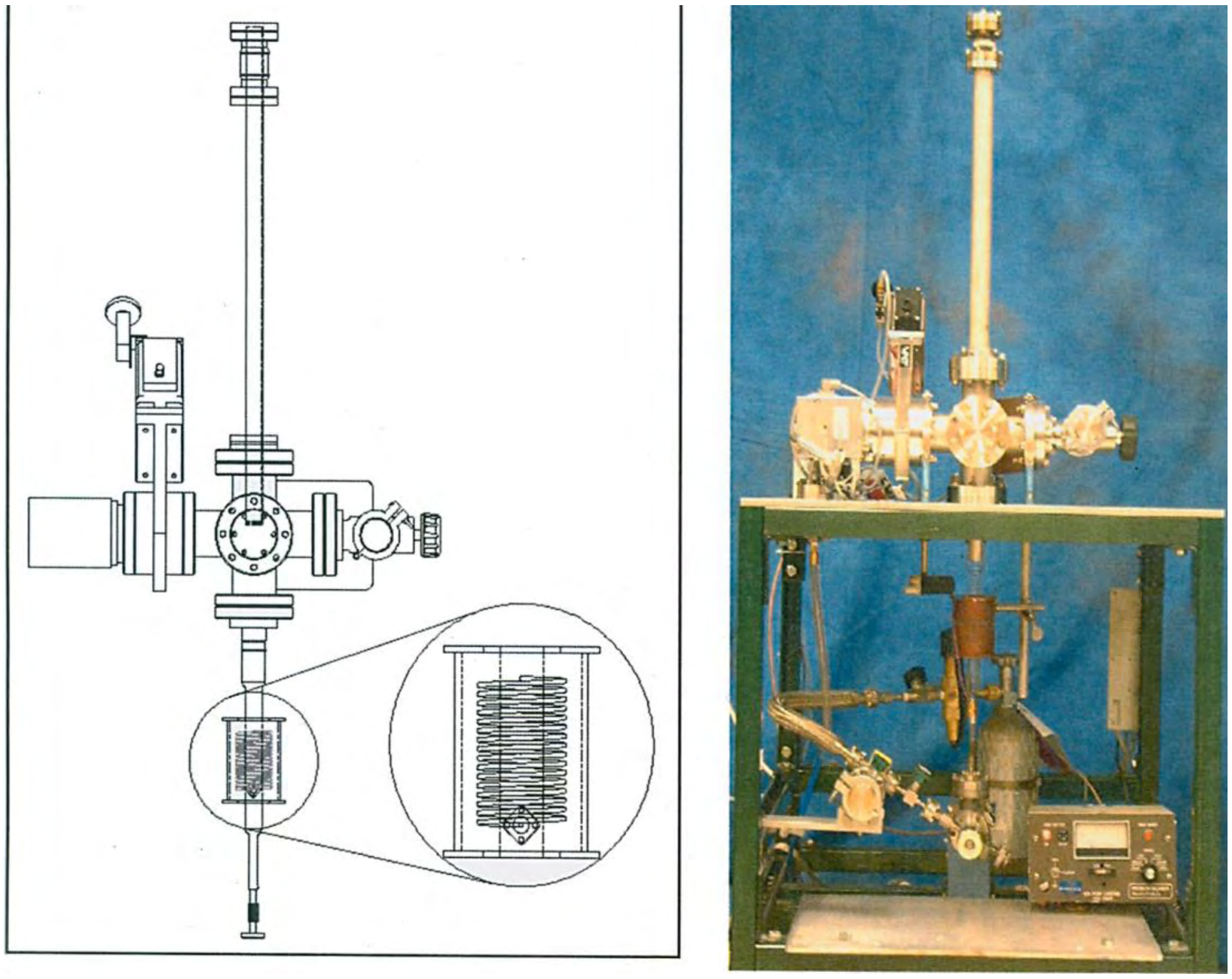

**Figure 8.7. The RF-dissociator, atomic hydrogen cleaning apparatus using at CEBAF/Jefferson Laboratory. [Reprinted/Adapted figures with permission from [8.31]. Copyright 2005 by the American Physical Society]**

## 8.6 DRIVE LASERS FOR A POLARIZED BEAM

A DC laser light source can be used to make an electron beam at an accelerator, but something must be done to create the appropriate RF time structure necessary for acceleration. Typically, this means using RF bunching cavities or RF choppers; however, bunching introduces energy spread and chopping is very inefficient, with a significant amount of the beam simply thrown away. At CEBAF, these ill effects were overcome by implementing synchronous photoinjection, a process whereby RF structure is created directly at the photocathode using an RF-pulsed drive laser. In the early 1990s, synchronous photoinjection with a GaAs photocathode had not been demonstrated. In fact, some thought it would not be possible, suggesting that GaAs would not respond quickly enough to the short-pulse light [8.36]. However, this concern proved unwarranted and synchronous photoinjection with GaAs now is widely used at many accelerators [8.53], [8.54], [8.55].

Mode-locked lasers often are used for synchronous photoinjection (see Chapter 9), but gain switching [8.56] is the preferred pulse-forming technique employed at CEBAF. Gain switching is a purely electrical technique that relies on diode lasers. By simply applying a ~1 W RF sine wave to the diode laser, ~30-50 ps optical pulses can be obtained at repetition rates between 100-3 000 MHz. This optical pulse-train can be easily locked to the accelerator RF frequency and a feedback loop is not required to maintain a fixed laser cavity length, as required when using mode-locked lasers. A gain-switched diode, however, can only produce a few milliwatts average power, and so, for most accelerators, a laser amplifier is required to boost power to an acceptable level. Diode lasers are available readily at wavelengths between 780 and 850 nm and single-pass traveling wave-tapered stripe diode amplifiers can generate ~100 mW. For higher power applications, fiber-based laser components from the telecom industry now are the best choice. Light at 1560 nm from a fiber-coupled, gain-switched seed laser can be sent to a fiber amplifier and then frequency doubled to produce Watts of useful light at 780 nm [8.57] (Figure **8.8**). Similar fiber-based systems are used to generate high power at 532 nm [8.55].

Standard optical components are used to deliver the laser light to the photocathode. The main distinction between polarized photoelectron guns and non-polarized photoguns is that the drive laser's light must be circularly polarized and have the appropriate wavelength. Thin mica wave-plates can generate circularly polarized light and the sign of the polarization can be flipped by moving wave-plates in/out of the laser beam's path. Faster spin-flipping can be achieved by using an electro-optic modulator called a Pockels cell, where two spin states are obtained by reversing the polarity of voltage applied to the Pockels cell. As an example, for typical experiments at CEBAF, the polarization direction flips at 30 Hz. Recently a technique was developed to flip polarization at rates up to 1000 Hz [8.58].

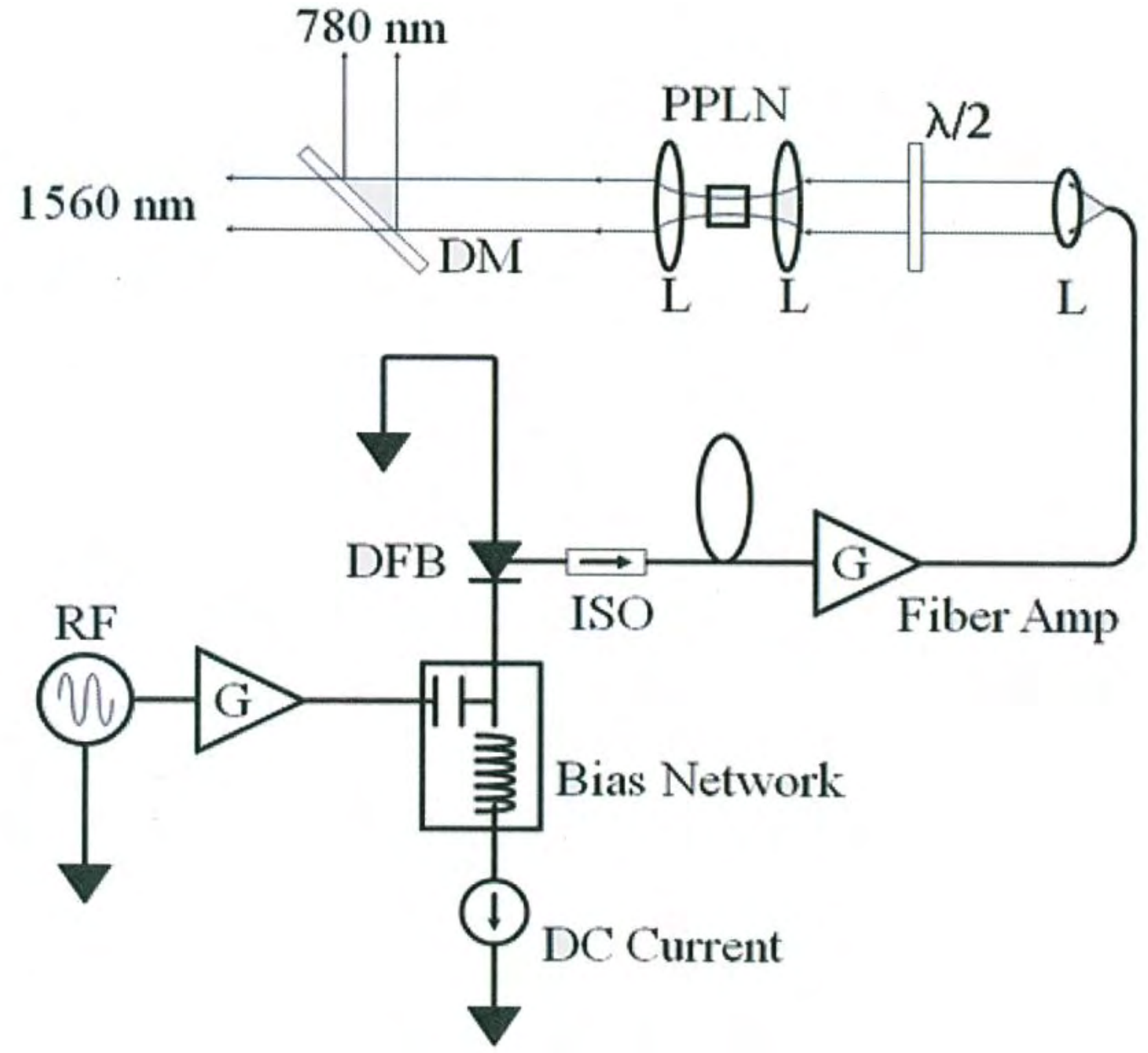

**Figure 8.8. Schematic of the fiber-based laser system with gain-switched diode master oscillator. The following list consists of the acronyms used in the figure: DFB, distributed feedback Bragg reflector diode laser; ISO, fiber isolator; G, amplifier gain; L, lens; PPLN, periodically poled lithium niobate frequency doubling crystal; and DM, dichroic mirror. [Reprinted figure with permission from [8.57]. Copyright 2006 by the American Physical Society]**

## 8.7 SPIN MANIPULATION

Polarized-beam experiments require a specific orientation of the direction of electron spin at the target, typically parallel to that of beam motion. Electrons leave the photocathode with their spin direction pointing parallel/antiparallel to the direction of beam motion, depending on the helicity of the laser's circular polarization (right or left circular) created by the Pockels cell. However, the spin direction precesses in the horizontal plane as the beam passes through the arcs and transport lines to the halls; this net spin precession must be "cancelled out" by orienting the spin direction at the injector by the opposite amount using a spin manipulator. At CEBAF, a Wien filter is used for spin manipulation [8.17]. It is a device with static electric- and magnetic-fields perpendicular to each other and to the velocity of charged particles passing through it (Figure **8.9**). Unit charged particles with a velocity of $\beta c = E\,B^{-1}$ remain undeflected in passing through the Wien filter, while the spin is rotated in the plane of the electric field. A window-frame dipole magnet provided the magnetic field. The magnet was terminated at each end with a nickel plate with a 20 mm diameter beam aperture. We carefully mapped the magnetic field of the full magnet, assembled on the Wien filter vacuum chamber, with a precision Hall probe. We then calculated the profile of the electric field plates, using the code POISSON [8.59], to produce an electric field whose profile closely matches that of the magnetic field. The Wien filter can rotate the spin by ±110° at 100 keV. The calibration and performance of this Wien filter is described in Grames *et al.* [8.60].

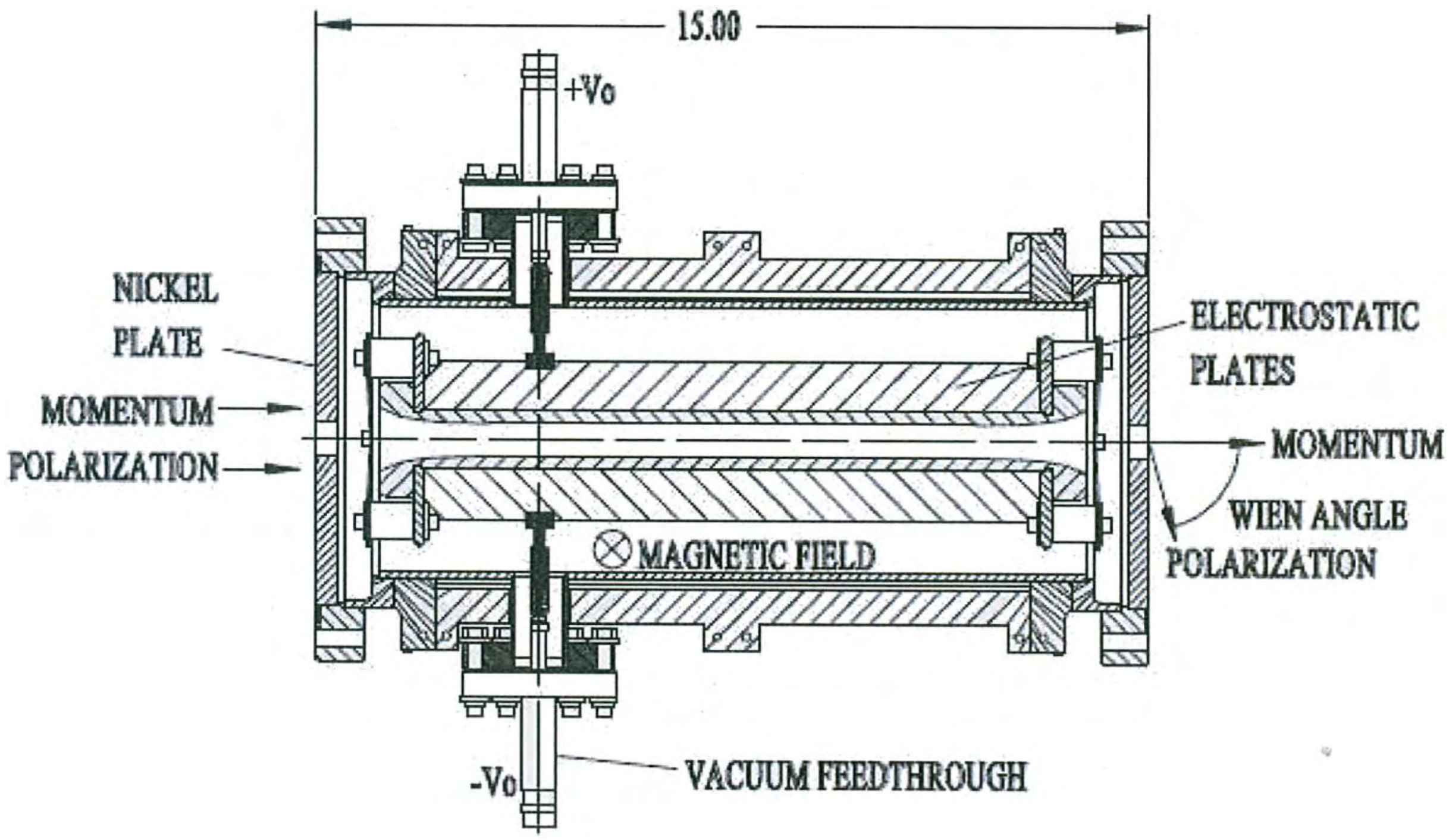

**Figure 8.9. The Wien filter spin manipulator used with CEBAF's second and third polarized electron sources. The magnet is not shown in this cutaway view. [[8.61]; Available under Creative Common Attribution 3.0 License (www.creativecommons.org/licenses/by/3.0/us/) at www.JACoW.org.]**

## 8.8 POLARIMETRY

After making a spin-polarized electron beam, the magnitude of the polarization must be measured. Typically, this is done using Mott polarimetry [8.62] that can accommodate electron beam energies between a few kiloelectron volts and a few megaelectron volts. A Mott polarimeter relies on the scattering asymmetry observed when spin-polarized electrons – with the spin vector oriented perpendicular to the scattering plane – strike the nuclei of an unpolarized target. To make a polarization measurement, a scattering asymmetry is measured using detectors that count the number of electrons that scatter to the right/left (or up/down) as the direction of the electron spin is flipped by changing the helicity of the circular polarization of the photogun drive laser's light (*via* the laser-table electro-optic Pockels cell, described above). The measured scattering asymmetry, $A$, is related to beam polarization, $P$, and the effective Sherman function, $S_{eff}$, as given below:

$$A = \frac{(N\rightarrow - N\leftarrow)}{(N\rightarrow + N\leftarrow)} = S_{eff} \quad (8.4)$$

The Sherman function, or analyzing power, is a term associated with the physics of the scattering process. The *effective* Sherman function describes the same process, but modified to account for real life experimental conditions, *e.g.* multiple scattering and detector acceptance. The most desirable characteristic of any polarimeter is a large and well known effective Sherman function; however, in practice, this value must be determined by computer simulation and/or detailed experimental measurements, *e.g.*, extrapolating target thickness, or retarding field scans.

The subject of Mott polarimetry is broad enough to be the focus of another book. Suffice to say, there are different types of Mott polarimeters that can be categorized loosely according to their electron beam energies: Low voltage retarding field Mott polarimeters, conventional gun-voltage Mott polarimeters and accelerator based MeV Mott polarimeters. Of these, the last are the most accurate (~1%) because experimental measurements can be predicted very accurately using a model that successfully accounts for multiple scatterings within the target foil [8.63]. For conventional Mott polarimetry at gun voltage (~100 kV), the effective Sherman function is determined empirically by performing a foil thickness extrapolation to deduce the asymmetry associated with single-scattering events [8.20]. For retarding field Mott polarimetry [8.64], a low voltage beam (~-200 V) is accelerated toward a thick target biased at ~20 kV. Electrons with a broad energy spectrum arrive at the detectors, but the single scattering events can be discerned by biasing the detectors at the photocathode voltage.

## 8.9 FUTURE R&D

Although a number of laboratories that provided spin-polarized electron beams have closed (NIKHEF, MIT-Bates and SLAC), there is still great interest within the scientific physics community to conduct experiments with polarized electron beams. Future R&D will focus on enhancing beam polarization and providing significantly higher current (milliamperes). These goals will spur exciting developments in the sub-fields of vacuum, drive lasers, high voltage and elimination of field emission. It seems reasonable to anticipate having a spin polarized beam from an RF gun, which is appealing for high-bunch charge accelerator applications that demand a small beam emittance.

## 8.10 CONFLICT OF INTEREST AND ACKNOWLEDGEMENT

We confirm that this article content has no conflicts of interest and would like to acknowledge the support of U.S. DOE Contract No. DE-AC05-84ER40150.

***References***

[8.1] J. Kessler, *Polarized Electrons*, Berlin: Springer-Verlag, 1985.

[8.2] E. Leader, *Spin in Particle Physics*, Cambridge: Cambridge University Press, 2001.

[8.3] M. J. Alguard, J. E. Clendenin, R. D. Ehrlich *et al.*, “A source of highly polarized electrons at the Stanford linear accelerator center,” *Nucl. Instrum. Meth.*, vol. 163, pp. 29-59, July 1979.

[8.4] W. von Drachenfels, U. T. Koch, Th. M. Müller *et al.*, “A pulsed source for polorized electrons with high repetition rate,” *Nucl. Instrum. Meth.*, vol. 140, pp. 47-55, January 1977.

[8.5] M. J. Alguard, J. E. Clendenin, P. S. Cooper *et al.*, “Depolorization effects in pulsed photoionization of state-selected lithium,” *Phys. Rev. A*, vol. 16, pp. 209-212, July 1977.

[8.6] L. A. Hodge, F. B. Dunning and G. K. Walters, “Intense source of spin-polarized electrons,” *Rev. Sci. Instrum.*, vol. 50, pp. 1-4, January 1979.

[8.7] P. F. Wainwright, M. J. Alguard, G. Baum *et al.*, “Applications of a DC Fano effect polarized election source to low-energy electron-atom scattering,” *Rev. Sci. Instrum.*, vol. 49, pp. 571-585, May 1978.

[8.8] D. T. Pierce, F. Meier and P. Zürcher, “Negative electron affinity GaAs: A new source of spin-polarized electrons,” *Appl. Physics Lett.*, vol. 26, pp. 670-672, June 1975.

[8.9] C. K. Sinclair, E. L. Garwin, R. H. Miller *et al.*, “A high intensity polarized electron source for the Stanford linear accelerator,” in *Proc.AIP Conf.*, vol. 35, 1976, pp. 424-431.

[8.10] C. Y. Prescott, W. B. Atwood *et al.*, “Parity non-conservation in inelastic electron scattering,” *Physics Lett. B*, vol. 77, pp. 347-352, August 1978.

[8.11] K. Wada, M. Yamamoto, T. Nakanishi *et al.*, “200 keV polarized electron source at Nagoya University,” in *Proc. 15$^{th}$ Int. Spin Physics Symp.*, 2003, 1063-1067.

[8.12] W. Hartmann, D. Conrath, W. Gasteyer *et al.*, “A source of polarized electrons based on photoemission of GaAsP,” *Nucl. Instrum. Meth. A*, vol. 286, pp. 1-8, January 1990.

[8.13] K. Aulenbacher, Ch. Nachtigall, H. G. Andresen *et al.*, “The MAMI source of polarized electrons,” *Nucl. Instrum. Meth. A*, vol. 391, pp. 498-506, June 1997.

[8.14] G. D. Cates, V. W. Hughes, R. Michaels *et al.*, “The Bates polarized electron source,” *Nucl. Instrum. Meth. A*, vol. 278, pp. 293-317, June 1989.

[8.15] M. J. J. van den Putte, C. W. De Jager, S. G. Konstantinov *et al.*, “The polarized electron source at NIKHEF,” in *Proc. AIP Conf.*, vol. 421, 1997, pp. 260-269.

[8.16] W. Hillert, M. Govin and B. Neff, “The 50 keV inverted source of polarized electrons at ELSA,” in *Proc. AIP Conf.*, vol. 570, 2000, pp. 961-964.

[8.17] C. K. Sinclair, P. A. Adderley, B. M. Dunham *et al.*, “Development of a high average current polarized electron source with long cathode operational lifetime,” *Phys. Rev. ST Accel. Beams*, vol. 10, pp. 023501-1–023501-21, February 2007.

[8.18] S. M. Sze, *Physics of Semiconductor Devices*, New York: John Wiley & Sons, 1981.

[8.19] J. S. Blakemore, “Semiconducting and other major properties of gallium arsenide,” *J. Appl. Physics*, vol. 53, pp. R123-R181, October 1982.

[8.20] D. T. Pierce and F. Meier, “Photoemission of spin-polarized electrons from GaAs,” *Phys. Rev. B*, vol. 13, pp. 5484-5500, June 1976.

[8.21] G. L. Bir, A. G. Aronov and G. E.Piku, “Spin relaxation of electrons due to scattering by holes,” *Soviet Physics J. Experimental Theoretical Physics*, vol. 42, 705, October 1975.

[8.22] M. I. D’Yakonov and V. I. Perel, “Spin orientation of electrons associated with the interband absorption of light in semiconductors,” *Soviet Physics J. Experimental Theoretical Physics*, vol. 33, pp. 1053-1059, November 1971.

[8.23] R. J. Elliott, “Theory of the effect of spin-orbit coupling on magnetic resonance in some semiconductors,” *Phys. Rev.*, vol. 96, pp. 266-279, October 1954.

[8.24] Y. Yafet, *Solid State Physics*, vol. 14, New York: Academic Press, 1963, Chap. 1, pp. 1.

[8.25] M. Zolotorev, “Effects of radiation trapping on polarization of photoelectrons from semiconductors,” in *Proc. Workshop Photocathodes Polarized Electron Sources Accelerator*, 1993, pp. 435-444.

[8.26] T. Maruyama, E. L. Garwin, R. Prepost *et al.*, “Observation of strain-enhanced electron-spin polarization in photoemission from InGaAs,” *Phys. Rev. Lett.*, vol. 66, pp. 2376-2379, May 1991.

[8.27] T. Nakanishi, H. Aoyagi, H. Horinaka *et al.*, “Large enhancement of spin polarization observed by photoelectrons from a strained GaAs layer,” *Physics Lett. A*, vol. 158, pp. 345-349, September 1991.

[8.28] T. Maruyama, E. L. Garwin, R. Prepost *et al.*, “Electron-spin polarization in photoemission from strained GaAs grown on $GaAs_{1-x}P_x$,” *Phys. Rev. B*, vol. 46, pp. 4261-4264, August 1992.

[8.29] T. Nakanishi, S. Okumi, K. Togawa *et al.*, “Highly polarized electrons from superlattice photocathodes,” in *Proc. AIP Conf.*, vol. 421, 1998, p. 300-310

[8.30] T. Maruyama, D.-A. Luh, A. Brachmann *et al.*, “Systematic study of polarized electron emission from GaAs/GaAsP superlattice photocathodes,” *Appl. Phys. Lett.*, vol. 85, pp. 2640-2642, September 2004.

[8.31] M. Baylac, P. Adderley, J. Brittian *et al.*, "Effects of atomic hydrogen and deuterium exposure on high polarization GaAs photocathodes," *Phys. Rev. ST Accel. Beams*, vol. 8, pp. 123501-1–123501-11, December 2005.

[8.32] SPIRE Semiconductor, L.L.C., 25 Sagamore Park Drive, Hudson, NH 03051, http://www.spirecorp.com/spire-bandwidth-semiconductor/index.php.

[8.33] SVT Associates, Inc., 7620 Executive Drive, Eden Prairie, MN 55344, http://www.svta.com.

[8.34] Yu. A. Mamaev, L. G. Gerchikov, Yu. P. Yashin *et al.*, "Optimized photocathode for spin-polarized electron sources," *Appl. Phys. Lett.*, vol. 93, pp. 81114-1–81114-3, August 2008.

[8.35] T. Nishitani, M. Tabuchi, Y. Takeda *et al.*, "Superlattice photocathode with high brightness and long NEA-surface lifetime," in *Proc. AIP Conf.*, vol. 1149, 2009, pp 1047-1051.

[8.36] W. E. Spicer and A. Herrera-Gómez, "Modern theory and applications of photocathodes," in *Int. 1993 Symp. Imaging Instrumuments*, 1993, pp. 18-33.

[8.37] M. L. Stutzman, P. Adderley, J. Brittian *et al.*, "Characterization of the CEBAF 100 kV dc GaAs photoelectron gun," *Nucl. Instrum. Meth. A*, vol. 574, pp. 213-220, May 2007.

[8.38] J. Grames, R. Suleiman, P. A. Adderley *et al.*, "Charge and fluence lifetime measurements of a DC high voltage GaAs photogun at high average current," *Phys. Rev. ST Accel. Beams*, vol. 14, pp. 043501-1–043501-12, April 2011.

[8.39] M. Breidenbach, M. Foss, J. Hodgson *et al.*, "An inverted-geometry, high voltage polarized electron gun with UHV load lock," *Nucl. Instrum. Meth. A*, vol. 350, pp. 1-7, October 1994.

[8.40] R. Alley, H. Aoyagi, J. Clendenin *et al.*, "The Stanford linear accelerator polarized electron source," *Nucl. Instrum. Meth. A*, vol. 365, pp. 1-27, November 1995.

[8.41] K. Aulenbacher, H. G. Andresen, T. Dombo *et al.*, "Operating experience with the MAMI polarized electron source," in *Proc. Workshop Photocathodes Polarized Electron Sources Accelerators*, 1994, pp. 1-12.

[8.42] M. L. Stutzman and J. Grames, "Superlattice photocathode damage analysis," in *Proc. AIP Conf.*, vol. 1149, 2008, pp. 1032-1037.

[8.43] J. Grames, P. Adderley, J. Brittian *et al.*, "A biased anode to suppress ion back-bombardment in a dc high voltage photoelectron gun," in *Proc. AIP Conf.*, vol. 980, 2007, pp. 110-117.

[8.44] P. A. Redhead, "Ultrahigh and Extreme High Vacuum," in *Foundations of Vacuum Science and Technology*, New York: Wiley, 1998, Chap. 11.

[8.45] C. D. Park, S. M. Chung, X. Liu *et al.*, "Reduction in hydrogen outgassing from stainless steels by a medium-temperature heat treatment," *J. Vac. Sci. Technology A*, vol. 26, pp. 1166-1171, August 2008.

[8.46] M. Yamada and Y. Ide, "Direct observation of species liberated from GaAs native oxides during atomic hydrogen cleaning," *Japanese J. Appl. Physics*, vol. 33, pp. L671-L674, March 1994.

[8.47] Y. Ide and M. Yamada, "Role of $Ga_2O$ in the removal of GaAs surface oxides induced by atomic hydrogen," *J. Vac. Sci. Technology A*, vol. 12, pp. 1858-1863, July 1994.

[8.48] E. Petit, F. Houzay and J. M. Moison, "Interaction of atomic hydrogen with native oxides on GaAs (100)," *J. Vac. Sci. Technology A*, vol. 10, pp. 2172-2177, July 1992.

[8.49] E. Petit and F. Houzay, "Optimal surface cleaning GaAs (001) with atomic hydrogen," *J. Vac. Sci. Technology B*, vol. 12, pp. 547-550, March 1994.

[8.50] S. Sugata, A. Takamori, N. Takado *et al.*, "GaAs cleaning with a hydrogen radical beam gun in an ultrahigh-vacuum system," *J. Vac. Sci. Technology B*, vol. 6, pp. 1087-1091, July 1988.

[8.51] Y. Okada and J. S. Harris, "Basic analysis of atomic-scale growth mechanics for molecular beam epitaxy of GaAs using atomic hydrogen as a surfactant," *J. Vac. Sci. Technology B*, vol. 14, pp. 1725-1728, May 1996.

[8.52] M. Poelker, K. P. Coulter, R. J. Holt *et al.*, "Laser-driven source of spin polarized atomic hydrogen and deuterium," *Nucl. Instrum. Meth. A*, vol. 364, pp. 58-69, September 1995.
[8.53] K. Aulenbacher, H. Euteneuer, D. von Harrach *et al.*, "High capture efficiency for the polarized beam at MAMI by R.F.-synchronized photoemission," in *Proc. 6th European Particle Accelerator Conf.*, 1998, pp. 1388-1390.
[8.54] S. Benson, G. Biallas, C. Bohn *et al.*, "First lasing of the Jefferson Lab IR demo FEL," *Nucl. Instrum. Meth. A*, vol. 429, pp. 27-32, June 1999.
[8.55] D. Ouzounov, H. Li, B. Dunham and F. Wise, "Fiber-based drive laser systems for the Cornell ERL electron photoinjector," in *Proc. SPIE Conf.*, vol. 7581, 2010, pp. 75810N-1–75810N-11.
[8.56] M. Poelker, "High power gain-switched diode laser master oscillator and amplifier," *Appl. Physics Lett.*, vol. 67, pp. 2762-2764, August 1995.
[8.57] J. Hansknecht and M. Poelker, "Synchronous photoinjection using a frequency-doubled gain-switched fiber-coupled seed laser and ErYb-doped fiber amplifier," *Phys. Rev. ST Accel. Beams*, vol. 9, pp. 063501-1–063501-5, June 2006.
[8.58] J. Hansknecht, "Pockels cell switching method," [Online]. Available: http://www.jlab.org/accel/inj_group/laser2001/pockels_files/pockels_switch_notebook.htm [Accessed: January 31, 2012]
[8.59] K. Halbach, "Program for inversion of system analysis and its applications to the design of magnets," Lawrence Berkley Radiation Laboratory, Technical Report No. UCRL-17436, January 1967.
[8.60] J. M. Grames, C. K. Sinclair, J. Mitchell *et al.*, "Unique electron polarimeter analyzing power comparison and precision spin-based energy measurment," *Phys. Rev. ST Accel. Beams*, vol. 7, pp. 042802-1–042802-18, April 2004.
[8.61] J. Grames, P. Adderley, J. Benesch *et al.*, "Two wien filter spin flipper," in *Proc. 2011 Particle Accelerator Conf.*, 2011, pp. 862-864.
[8.62] T. J. Gay and F.B. Dunning, "Mott electron polarimetry," *Rev. Sci. Instrum.*, vol. 63, pp. 1635-1651, October 1992.
[8.63] M. Steigerwald, "MeV Mott polarimetry at Jefferson Lab," in *Proc. AIP Conf.*, vol. 570, 2001, pp. 935-942.
[8.64] J. L. McCarter, M. L. Stutzman, K. W. Trantham *et al.*, "A low-voltage retarding-field Mott polarimeter for photocathode characterization," *Nucl. Instrum. Meth. A*, vol. 618, pp. 30-36, June 2010.

# CHAPTER 9: LASER SYSTEMS

TRIVENI RAO
*Brookhaven National Laboratory*
*Upton, NY 11973*
THOMAS TSANG
*Brookhaven National Laboratory*
*Upton, NY 11973*

**Keywords**
Diode Pumped Solid State Laser, Ti:Sapphire Laser, Fiber Laser, Mode-Locking, Phase Locking, RF-Laser Synchronization, Laser Shaping, Temporal Shaping, Spatial Shaping

**Abstract**
The laser system for the photoinjector is selected based on the cathode material and the electron beam parameters required for the application and hence can vary widely. This chapter is aimed not at describing one system in detail, but to give an overview of the options available. We encourage the readers to seek additional in-depth material by reviewing the references provided in this chapter. Since in majority of the systems, the general architecture typically is of Master Oscillator Power Amplifier (MOPA) configuration, we provide an overview of each of its components and their different operating modes. We then describe two most commonly used diode-pumped-solid-state (DPSS) laser systems: Nd: Vanadate and Ti: sapphire, as examples of the MOPA. Since the fiber lasers are gaining popularity, especially for high average current applications, their capabilities are also described. Irrespective of the exact configuration, all these photoinjector lasers have to meet a set of general conditions such as delivering the properly shaped beam at the correct time and energy with tight tolerance. We address this by discussing different beam shaping techniques, beam transport, synchronization and diagnostics.

## 9.1 INTRODUCTION

In previous chapters, we showed that the quantum efficiency (QE) and the photon energy required to overcome the work function for different cathode materials vary significantly.
Table 9.1 lists the photocathode materials used in various operating photoinjectors, the photon energy of the laser used, and the laser energy required to release a 1 nC charge.

| Cathode Material | QE/Photon Energy | Laser Energy for 1 nC |
|---|---|---|
| Copper | $1.4\times10^{-4}$/4.96 eV [9.1] | 35.4 μJ |
| Magnesium | $5\times10^{-4}$/4.66 eV [9.2] | 9.2 μJ |
| Lead | $2.7\times10^{-3}$/5.8 eV [9.3] | 2.2 μJ |
| $Cs_2Te$ | 0.092/4.66 eV [9.4] | 51 nJ |
| $K_2CsSb$ | 0.1/2.33 eV [9.5] | 23.3 nJ |
| GaAs:Cs, non-polarized | 0.1/2.33 eV [9.6] | 23.3 nJ |
| GaAs:Cs, polarized | 0.01/1.47 eV [9.6] | 147 nJ |

**Table 9.1. Photocathode materials used in existing photoinjectors, their respective QE, and the laser energy required to release 1 nC from the cathode.**

In this section, we highlight the architecture of the laser system capable of delivering the required specifications to the cathode at the appropriate wavelength. In order to assure a stable electron beam operation with the highest energy and minimum emittance, most RF injectors require that the repetition rate of the laser be phase-locked, either to the fundamental RF frequency of the injector that ranges from tens of megahertz to a few gigahertz or to a harmonic/sub-harmonic frequency of the injector. Furthermore, the laser may also need to run in the macro/micropulse mode, depending on the application.

In order to reduce the emittance growth related to RF fields, the pulse duration of the laser is typically a few degrees of the RF cycle. To reduce the energy fluctuation and maintain the longitudinal emittance of the electron beam the timing jitter should be of less than one degree of the RF cycle. The shortest pulse duration that the injector can accommodate is determined by the space charge limit imposed by the peak current and the transit time of the electron bunch in the cathode material. For a metal cathode with the electron velocity of ~3 500 m $s^{-1}$ and absorption length of ~1 nm, the transit time is ~285 fs. Hence, laser beams with a shorter pulse duration would not produce correspondingly shorter electron bunches. Furthermore, a nanocoulomb bunch charge with such a short pulse duration corresponds to a peak current of 3 500 A, which is well above the space charge limited regime, even in the presence of 100 MV $m^{-1}$ gradient field.

In a RF gun, the transverse spatial size of the electron beam is optimized to minimize the emittance growth due to RF and space charge effects. The space charge effects are proportional to the charge density, favoring a larger electron beam diameter for a given charge. However, the RF effects dictate that the radius of the electron beam should be well within the region wherein the accelerating field is normal to the cathode. In typical RF injectors, the electron beam's size, and hence the spatial laser beam size, are adjusted to < 1% of the RF wavelength in order to minimize the growth of transverse emittance; the laser pointing stability is usually kept within a few percent of this spot size.

Numerous simulations [9.7], [9.8] indicate that the emittance of the electron beam from the injector can be minimized by shaping the electron beam, and hence the laser beam, both transversely and longitudinally. Once again, the electron beam and the RF parameters dictate its exact dimensions.

## 9.2 LASER ARCHITECTURE

In general, the parameters of the laser systems driving the photoinjectors, such as the gain medium, energy, peak power, average power, repetition rate and pulse structure, are determined by the cathode and the electron beam characteristics for the intended application which can cover a wide range. The exact configuration of the laser system is application specific. In this section, we address only the general architecture and different options that are available. We evaluate their advantages and disadvantages. Only pulsed lasers are discussed since they are more relevant to the photoinjectors. We note that the laser field is evolving rapidly, making technical details obsolete rather quickly. Hence, different laser media are discussed only in terms of their fundamental properties that are relevant to the photoinjectors.

Before choosing a suitable laser system, the following characteristics must be specified based on the electron parameters and the cathode of choice: Laser wavelength, repetition rate, polarization, transverse- and longitudinal beam profile, pulse energy, pulse duration, pulse structure, pulse stability, pointing stability, timing accuracy and tolerance to the presence of pre/post pulse.

In most cases, the typical architecture is in the form of Maser Oscillator Power Amplifier (MOPA) arrangement. The weak light pulse generated from a laser master oscillator is increased in the subsequent amplifiers. The gain media in the master oscillator and the amplifier determine the laser wavelength and the shortest possible pulse duration. The upper state's lifetime and energy storable in this gain media, along with the damage threshold of the optics, dictate the maximum extractable energy and power from the system. Typically the master oscillator can meet most of the requirements of the photoinjector other than the pulse energy and the average power. Power amplifiers are used to increase the pulse energy and the average power without compromising overall performance. Additional optical elements are usually inserted in the system to tailor the light pulses for other needs, such as improvement of the signal-to-noise ratio, scribing micro/macropulse structure, and transforming the transverse- and longitudinal-beam profile.

### 9.2.1 Master Oscillator

The basic components of a laser oscillator are the gain medium, a pump source that excites the gain medium to the upper state of the lasing levels and a pair of high reflectivity mirrors on either side of the gain medium to form the laser cavity. Passage of the optical beam at the lasing wavelength through the gain medium increases its photon density, but is counteracted by the internal losses in the cavity, such as the scattering in the mirrors and absorption in the gain medium. The system oscillates when the gain in the cavity equals the total losses. The transverse parameters of the laser output are determined by the resonator geometry comprising of the mirrors, their placement and the overlap of the pump profile with the transverse electromagnetic fields supported by the laser resonating cavity. The cavity can be made to oscillate in a diffraction-limited single transverse mode, such as $TEM_{00}$ (Gaussian spatial distribution), by carefully choosing the geometry of the resonator and the pump beam. For a non-diffraction limited beam, the beam's quality factor $M^2$ (defines as $\theta/\theta_G$ where $\theta$ and $\theta_G$ are, respectively, the far-field divergence angles of the non-diffraction limited beam and the Gaussian beam) quantifies the deviation from a true Gaussian beam. The spectral distribution of the gain medium and the resonator define the spectrum of the laser output.

The temporal characteristics of the laser beam are dictated by the temporal behavior of the gain $g(t)$ and loss $\alpha(t)$ of the system; in steady-state, they are equal. However, by modulating $g(t)$ and $\alpha(t)$ a stable pulsed operation can be maintained. The gain can be adjusted by modulating the pump power. Some examples are flash lamp pumping and discharge pumping of the lasers, current modulation in a semiconductor laser [9.9]–[9.11], and synchronous pumping in a dye laser [9.12], [9.13]. Examples of the loss modulation include Q-switching, cavity dumping, and active or passive mode-locking. We discuss the general principles of Q-switching and mode-locking in the sub-section below. Current modulation that is specific to semiconductor lasers is described in Chapter 8. We will omit the discussion on older technologies of the synchronous pumping of dye laser and discharge pumping of excimer lasers, since they are not widely used in photoinjector applications. In most applications, flash lamp pumping was replaced by diode pumping, and hence, it will not be discussed either.

#### *9.2.1.1 Q-Switching*

The onset of laser oscillation occurs when the gain in the system equals the loss. Considering a laser cavity with a shutter, when the shutter is closed, the loss in the system is large and the Q of the cavity, *i.e.*, the ratio of the energy stored in the cavity to the energy loss per cycle, is small. When the gain medium is pumped to reach population inversion, the stored energy and the gain in the medium reach a high value without the onset of laser oscillation. If the shutter is quickly opened, the Q of the cavity suddenly switches to a high value so that the gain in the cavity far exceeds the loss and the stored energy can be released in an extremely short time. The time for which the energy can be stored in the gain medium is on the order of the upper state lifetime of the lasing levels. The pulse duration of the laser output is of the order of a few round-trip times of the laser cavity. Since this process involves fast switching of the Q of the cavity from low to high value, it is known as Q-switching [9.14]. Q-switching can be accomplished using acoustic-optic and electro-optic crystals [9.15]–[9.18] and saturable absorbers [9.14], [9.19]–[9.22] covering MHz to a few Hertz repetition rate.

When a laser is pumped continuously with repetitive Q-switching, many of the laser characteristics such as the peak and average power, pulse duration (longer at higher repetition rates) and the spatial beam quality are strongly influenced by the integrity of this repetitive switching. The pumping can be in the pulsed mode arriving shortly before the opening of the Q-switch, but strong enough to saturate the population inversion. Some drawbacks of this Q-switch technique are pre- and post-lasing due to the finite Q value, double-pulsing due to excessively high gain, and fast modulation of the Q-switched envelope due to multiple resonator modes supported in the cavity. However, employing the technique of injection seeding has resolved many of these issues, as detailed in [9.14]. The duration of the laser pulse is about a few round-trip times of the cavity typically in nanoseconds: the energy per pulse can be in the range of a few millijoules, making it an attractive, lower cost option. Shorter pulse durations can be obtained by Q-switching in pulse-transmission mode. In this scheme, the circulating power is allowed to build up in the cavity made up of 100% reflecting mirrors. At its peak, the reflectivity of one mirror is switched from 100% to 0% (transmission from 0% to 100%), thereby emitting the entire cavity's optical power. Depending on the length of the cavity, pulse durations ranging from 1-5 ns are feasible with such devices. Q-switched lasers delivering > 2 J of energy in a few nanoseconds duration with average powers of tens of watts are commercially available [9.23].

Cavity dumping, a mechanism similar to Q-switching, can be employed for obtaining shorter pulses at higher repetition rates. While in Q-switching, energy storage is primarily achieved *via* atomic population inversion; in cavity dumping, it is *via* the cavity's optical field. In this scheme, the intra-cavity power is built to a high value in a high Q cavity. A modulator inserted at the focus of the beam extracts the beam from the cavity by modulating the direction of its diffracted beam. Cavity-dumped Nd:YAG laser with repetition rates from 125 kHz to several megahertz and pulse duration to 25 ns have been built [9.24], [9.25].

#### 9.2.1.2 Mode-Locking

The shortest pulse obtainable from a laser cavity is the transform limit of the broadest bandwidth that the cavity can support. If we assume that all other optical elements in the cavity are broadband, an assumption justified for lasers operating with transform-limited pulse durations > 1 ps, then the inherent limitation stems from the bandwidth of the gain medium (the Ti:sapphire laser, with a bandwidth broad enough to support femtosecond pulse duration, is an exception to this condition). Since the laser cavities are much longer than the optical wavelengths, a typical laser cavity can simultaneously support hundreds of axial modes. In the frequency domain, the output contains a large number of discrete spectral lines spaced by the inverse of the round-trip time $t_R$ in the cavity. If the gain bandwidth of the cavity is $\Delta\nu_L$, the number of axial modes, $N$, within this bandwidth is $\Delta\nu_L * t_R$. These modes oscillate independently and their phases are distributed randomly. In a mode-locked laser, these modes are constrained to have fixed phase relationship to each other; hence, their amplitude in the frequency and time domains has a well defined distribution. This constraint is imposed by introducing a repetitive intra-cavity loss modulator with a period equal to an integral multiple of the cavity's round-trip time. The lasing then takes place around the minimum of the loss modulation and the duration of individual mode-locked pulses is ~$(t_R * N^{-1})$ [9.26]. This mode-locking is accomplished, in general, either passively with saturable absorbers [9.27]–[9.29] or actively [9.30], [9.31] with an amplitude or phase modulator. Commercial, mode-locked, diode-pumped solid-state (DPSS) lasers are readily available with average power in excess of 50 W, pulse duration of ~15 ps and repetition rates in the range of 50-100 MHz [9.23].

In a passively mode-locked laser, the amplitude of the strongest pulse from the noise is selectively emphasized, while the background noise is not. Since the timing of this pulse in the noise floor is random, if uncorrected, the laser output has a large timing jitter. In a drive laser for the photoinjector, the arrival time of the electron bunch, and hence the repetition rate of the laser, must then be synchronized accurately to an external RF clock to minimize timing jitter between laser and electron pulses. When saturable absorber such as SESAM mirror is used as a passive mode-locker and the cavity length of the laser oscillator must be actively controlled for phase-locking the RF frequency or its harmonics to the laser repetition rate. More in-depth discussion on this phase-locking technique appears later. With an active mode-locking, the synchronization is assured by matching the frequency of the modulator directly to the frequency of RF clock or its harmonic.

### 9.2.2 Power Amplifier

The low single-pulse energy from the oscillator is amplified with a series of power amplifiers. The number of amplifiers depends on the seed energy, the total energy required, the gain per amplifier, the acceptable signal-to-noise ratio (S/N) and the damage threshold of the components, especially the gain medium. Typical bulk oscillators run at ~100 MHz repetition rate with energy per pulse in the tens of nanojoules regime. As established in earlier chapters, to deliver nanocoulomb charge, this energy has to be increased to the microjoules regime in the UV for a metal photocathode of 0.1% QE, or to tens of nanojoules at 530 nm for GaAs or multialkali photocathodes. The S/N of the laser is not a significant concern in low average current injectors using metal photocathodes since the beam loss is not a big issue and the nonlinear conversion from IR to UV wavelength increases S/N significantly. The damage threshold is not a vital issue for delivering nanocoulomb charge at low repetition rates. However, for high average current, high repetition rate injectors, a high S/N is a key consideration in the laser design. Since the current generated from the noise has multiple deleterious effects, this current dilutes the beam quality; its trajectory does not follow that of the main beam, leading to unplanned beam loss and associated radiation and vacuum problems. These injectors require a S/N exceeding $10^6$ or better.

As in the oscillator, the amplifier's architecture also varies significantly from system to system. However, a few general considerations are valid for all of them. The primary interest in designing the amplifier is the gain and its efficiency in energy extraction. At low signal levels, the energy in the amplifier increases exponentially with the length of the gain medium. However, as the energy increases, its extraction affects the population in the upper laser level and eventually reaches a point where the driving signal depletes inversion density. At this saturation level, the gain is linear to the length of the gain medium. Operating the amplifier just above this saturation level minimizes the energy fluctuation. However, this could alter the spatial and temporal profile of the seed radiation. The design of the amplifier should take into account the distortions that can be introduced into the seed pulse during amplification. Some causes for spatial distortion are non-uniformity in pumping and the intrinsic and extrinsic anisotropy of the gain medium, gain saturation in the central part of the cylindrically shaped pump beam with exponential growth on the outer edge of the cylindrical region, diffraction due to the finite size of the gain medium, and lensing effects due to non-uniform temperature distribution of the gain medium. The temporal profile of the amplified pulse might deviate from the seed pulse due to time-dependent population inversion, finite bandwidth of the gain medium and nonlinear effects caused by the high intensity of the amplified beam. In the case of a chirped seed pulse, the frequency (time) dependent gain can also modulate the spectral distribution of the beam, thus its temporal distribution. Energy released outside the beam's spatial- and temporal-structure adds to the system's noise. Care should be taken to avoid feedbacks into the amplifiers to prevent damage, pre-lasing and spurious noise signals.

The amplifier system can be either a series of linear amplifiers, regenerative amplifiers, or a combination of both. In a regenerative amplifier, the seed-pulse is injected in a resonant cavity containing the excited gain medium. After several passes through the medium and the corresponding increase in energy, the pulse is ejected from the cavity typically by a Pockels cell [9.32]. The delay between the injection- and extraction-time determines the number of passes through the gain medium, as well as the highest repetition rate that the amplifier can support. Since the seed pulse is amplified by the gain medium several times, the overall gain is high. In this arrangement, the mode quality of the injected beam is matched to the cavity's mode; the transverse profile of the amplified beam is dictated by the regenerative amplifier which can be maintained at a high quality. However, parasitic lasing could lower the S/N and gain-narrowing could reduce the bandwidth of the amplified pulse, both effects must be addressed carefully. The pulse duration of the seed pulse must be considerably shorter than the cavity's round-trip time; accordingly, this scheme is more applicable to amplifying picosecond or shorter duration pulses, but not longer ones. The switching speed in the cavity and the Pockels cell limit the maximum repetition rate, typically to around 100 kHz. The resonant condition also limits the spot size to a single value, and its associated damage threshold dictates the maximum extractable energy. To mitigate damage in high peak power laser systems, the seed pulse is often stretched in time, amplified and then recompressed after extraction. Typically, the stretching process introduces up-chirp or down-chirp to the seed pulse, and hence, the technique is called chirped pulse amplification (CPA) [9.33]. Hundreds of kilohertz repetition rates [9.34] and Joules of energy [9.35] (although not necessarily simultaneously) are possible with CPA amplification. Ever higher repetition rate approaching MHz systems are recently introduced in the market. Pulse energies of millijoules in tens of kilohertz are available with moderately sized systems.

In a linear multi-pass amplifier, the seed beam passes through an excited gain medium one or more times, increasing energy at each pass. The stored energy in the medium, the layout of the optics and the damage threshold govern the number of passes. Since the gain medium is not a part of a resonant cavity, mode-matching is not essential and the amplified spatial mode is merely dictated by the seed pulse. However, good overlap with the pump beam is important for maximum efficiency and minimum ASE. The optics can be arranged such that the spot size of the seed radiation increases with the beam energy, minimizing saturation and damage problems. Also, with a higher seed pulse energy, a gain medium with larger aperture can be used. It is important that the profiles of the pump and the seed beams be properly matched to assure maximum energy extraction and minimum deviation of the amplified beam from the seed distribution. An average power of > 50 W was obtained at repetition rates up to 750 MHz [9.36] with a Nd:Vanadate slab as the gain medium in the amplifier. Although laser systems with peak powers of petawatts were built in several laboratories [9.37]–[9.39] and are commercially available as custom designs, their peak power is too high and repetition rates too low to be relevant for producing photoelectrons in conventional injectors.

We discuss part of the laser's architecture, specifically the pulse selection, timing synchronization, beam shaping, harmonic generation and beam transport in the subsequent sections.

## 9.3 BULK SOLID STATE LASER SYSTEMS: $\lambda > 1$ µm

One of the most common gain media used in the photoinjector applications is Nd based glass or crystal. Several vendors market DPSS Nd lasers that are in a mature developmental state. Some Yb based systems under development are expected to deliver > 100 W average power with pulse duration of ~10-50 ps and repetition rates ranging from a few hundred kilohertz to gigahertz. In the following sections, we describe their common features, capabilities and limitations.

Table **9.2** lists some of the bulk solid state (BSS) laser systems, their operating wavelengths, and pulse durations. The narrow gain bandwidth of the Nd in crystalline host materials limit the pulse duration to a few picoseconds. Shorter pulse duration, > 60 fs, is obtained with non-crystalline glass lasers.

| Host | Dopant | Wavelength [μm] | Band Width [nm] | Gain Cross Section [$\times 10^{-20}$ cm$^2$] | Upper State Lifetime [μs] |
|---|---|---|---|---|---|
| YAG | Nd | 1.064 | 0.6 | 33 | 230 |
| YLF | Nd | 1.047-1.0530 | 1.2 | 18 | 480 |
| Vanadate | Nd | | 0.8 | 300 | 100 |
| Phosphate Glass | Nd | 1.0535-1.054 | 24.3 | 4.5 | 323 |
| KGW | Yb | 1.026 | 25 | 2.8 | 250 |
| YAG | Yb | 1.030 | 6.3 | 2.0 | 950 |
| Phosphate Glass | Yb | 1.06-1.12 | 62 | 0.049 | 1 300 |
| Sapphire | Ti | 0.790 | 230 | 41 | 3.2 |

**Table 9.2. Selected list of BSS laser systems and their operating parameters [9.40].**

Figure **9.1** is a schematic of the laser system commercially developed for driving the ERL photoinjector at BNL. Although this system's parameters are specific for this application, several features are common to all drive lasers; hence, we give a detailed description of this system here.

The ERL laser system is designed to deliver up to 5 W at 355 nm, with pulse duration of ~10 ps, repetition rate of 9.38 MHz, synchronized to an external 703.5 MHz master clock with a sub-picosecond timing jitter. A pulse selection system is incorporated to allow the ramp up from a single pulse to a series of micropulses in a macropulse with variable repetition rate (up to 10 kHz), and ultimately, to a quasi-continuous 10 MHz to facilitate the recovery of electron energy in the LINAC.

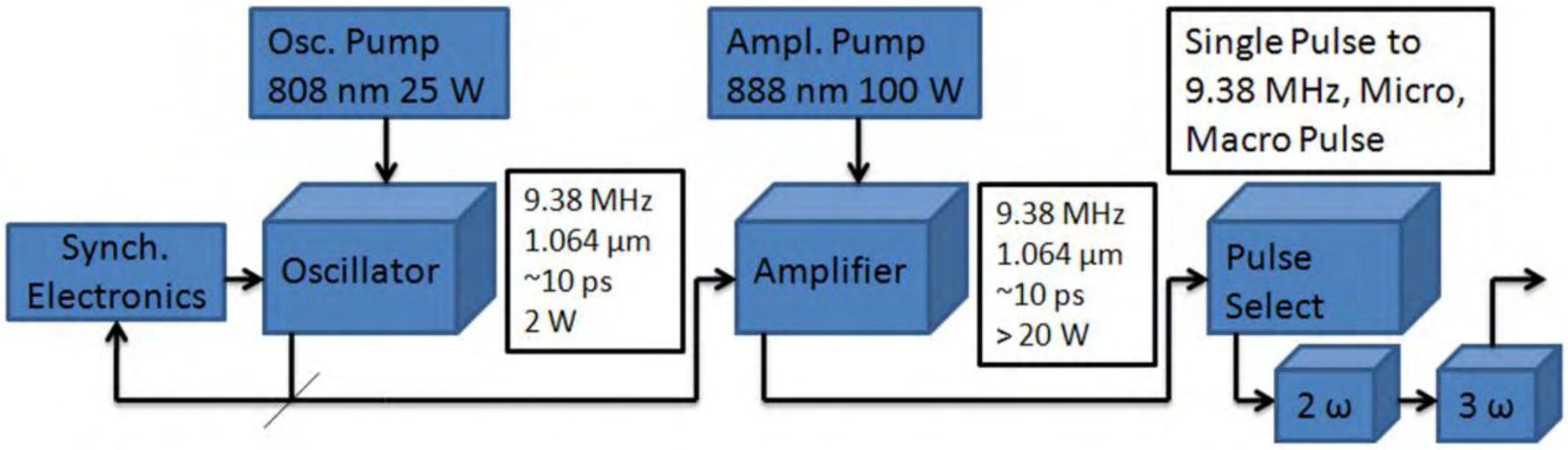

**Figure 9.1. Schematic of commercial laser system developed for the ERL photoinjector at BNL. Note that some parameters are given; however, parameters are application specific.**

### 9.3.1 Oscillator

The oscillator consists of a Nd:Vanadate crystal, a resonant cavity formed by a saturable absorber mirror (SAM) and an output coupler with 12% transmission. The 4×4×6 mm$^3$ laser crystal is end-pumped by 18 W fiber-coupled, diode laser, operating at 808 nm. The key feature of the oscillator is its repetition rate, too low for conventional CW mode-locking and too high for cavity dumping. To meet the challenge, a folded cavity with a resonator length of 16 m was custom-designed and built. Such a long cavity makes the laser very sensitive to misalignments compared to the conventional ~1 m long resonators. To isolate the oscillator from mechanical- and thermal-instabilities, the oscillator is built on a monolithic metal-block and is sealed off from the rest of the system. The SAM is mounted on a stepper motor-driven translational-stage with 25 mm travel range to accommodate slow drifts in the cavity length and to coarse tune the repetition rate of the laser. A small mirror that is a part of the resonant cavity is mounted on a piezo-driven stage with 9 µm travel range to compensate for fast changes in the cavity length and to preserve synchronization. The output of the oscillator is extracted through an AR-coated window. A camera mounted behind one of the cavity mirrors continually monitors the beam's profile and position.

Figure **9.2** is a schematic for synchronizing the oscillator to the master clock. The photodiode signal from the laser and the signal from master clock, down-converted to 9.38 MHz, are fed into a phase detector. The output signal is further amplified and used to drive the stepper motor and the piezo stages in the oscillator.

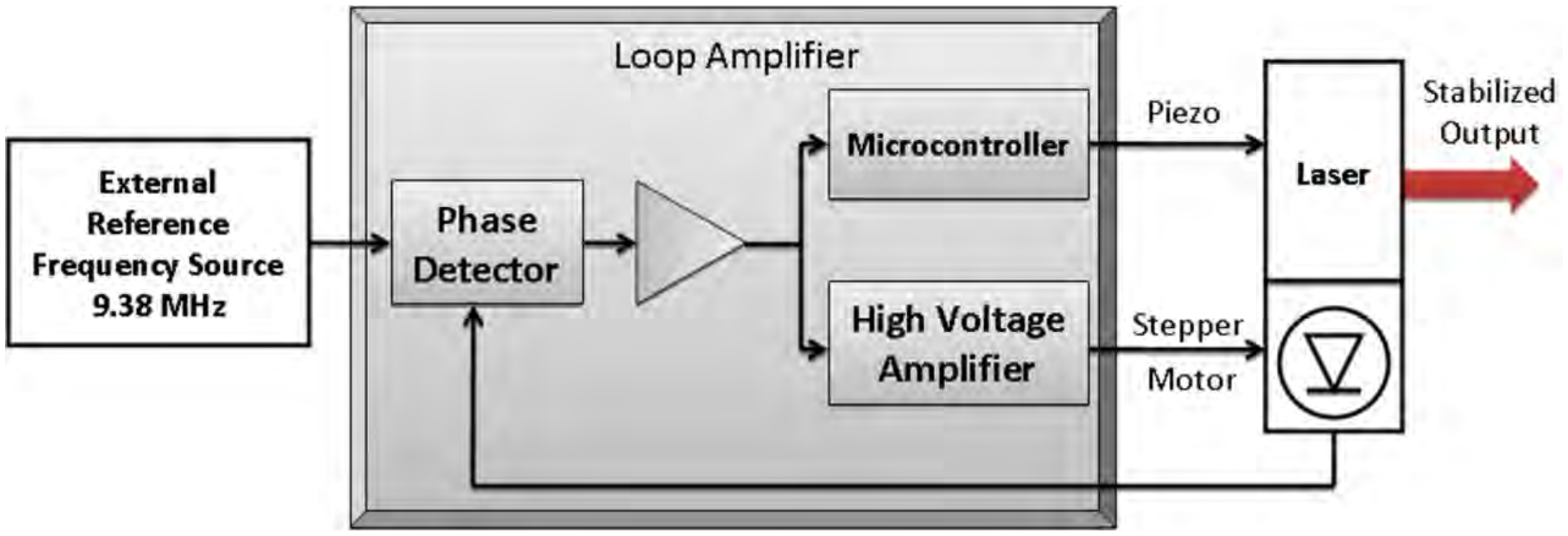

**Figure 9.2. Schematic of the lock to clock system used in BNL's ERL laser.**

### 9.3.2 Multipass Amplifier

The double-pass amplifier consists of a Vanadate crystal pumped by a 100 W diode laser operating at 888 nm. The weaker absorption at this wavelength reduces thermal problems, allowing a much higher pump power. The unabsorbed pump power in the first pass is reflected back into the crystal. A thin film polarizer that reflects vertical polarization and transmits horizontal polarization is inserted into the beam. A Faraday rotator changes the seed beam's polarization from vertical to horizontal after its second pass in the amplifier. The combination of thin film polarizer and the Faraday rotator allows the seed beam injection and amplified beam extraction out of the amplifier. The entire pulse train from the oscillator is amplified to eliminate the time-dependent changes in the thermal load induced by the seed- and amplified-pulses. The power and profiles of seed and amplified beams are monitored continually using the leakage signals at multiple locations. An optical isolator at the input end prevents feedback from the amplifier entering the oscillator. With 100 W pump power, > 20 W amplified power at 1.064 µm was delivered from the amplifier. Figure **9.3** Figure **9.4**, Figure **9.5** and Figure **9.6** display some performance characteristics of the amplified beam.

### 9.3.3 Pulse Selector

The pulse selector changes the repetition rate during the ramp-up process for the ERL and also alters its average current without changing the bunch charge. Since the S/N is an important consideration for the high current ERL, two significant changes were made to the conventional scheme: Since BBO can withstand a constant high voltage and can be oriented for the best contrast, it is used as the Pockels cell; and, the pulses are picked by the polarizing beam-splitter cube (PBS) when the voltage is applied to the Pockels cell, rather than when it is turned off. The high voltage is triggered externally to deliver pulses from single shot to a micro/macropulse configuration with variable number of micropulses within a macropulse and macropulse repetition rate variable up to 10 kHz. However, the entire 9.38 MHz pulse train also can be delivered to the cathode. A few selected pulse configurations are illustrated in Figure **9.7**. A variable attenuator, consisting of a half-wave plate and a PBS allows for the variation of single-pulse energy, and hence, bunch charge. A ground finished glass plate placed in front of a photodiode record the beam intensity and allow the adjustment of the variable attenuator to the desired level. The unpicked beam is rejected into a water-cooled beam dump.

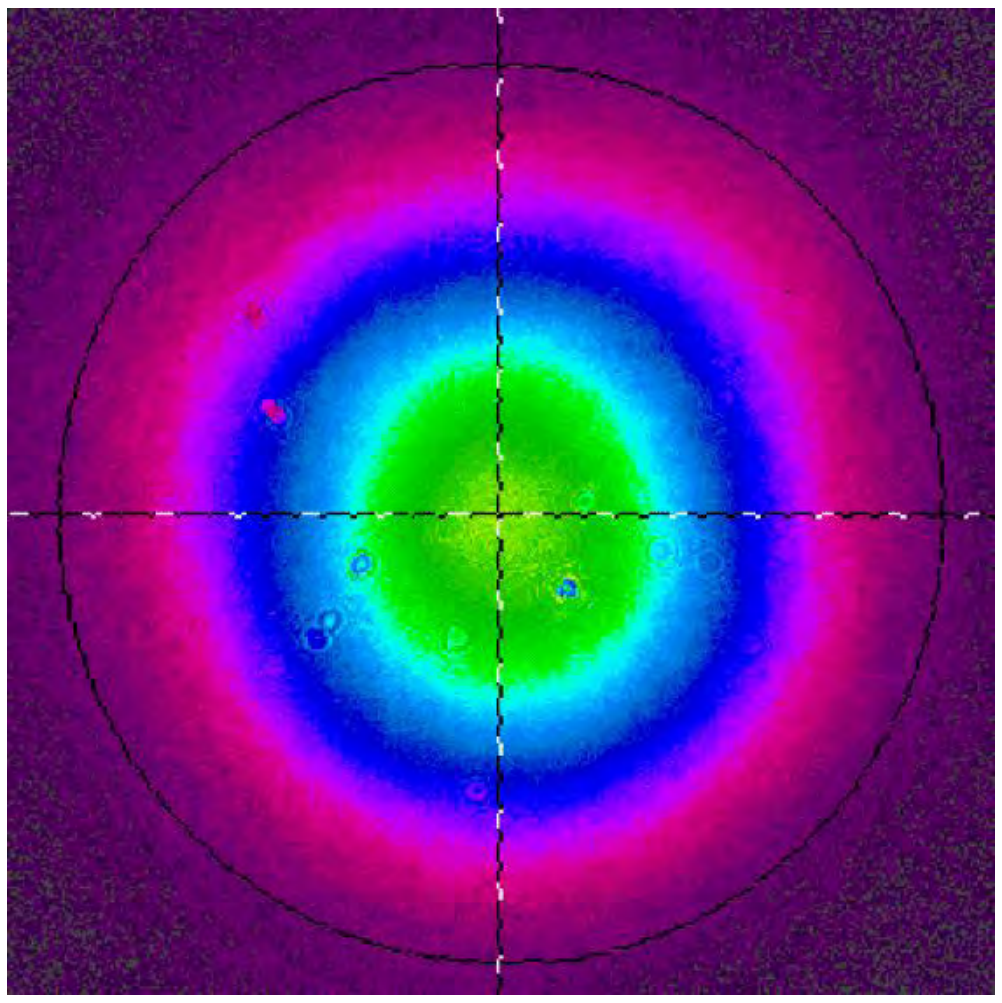

**Figure 9.3. Transverse profile of the amplified beam, core is nearly Gaussian. [[9.41]; Courtesy of Lumera Laser GmbH]**

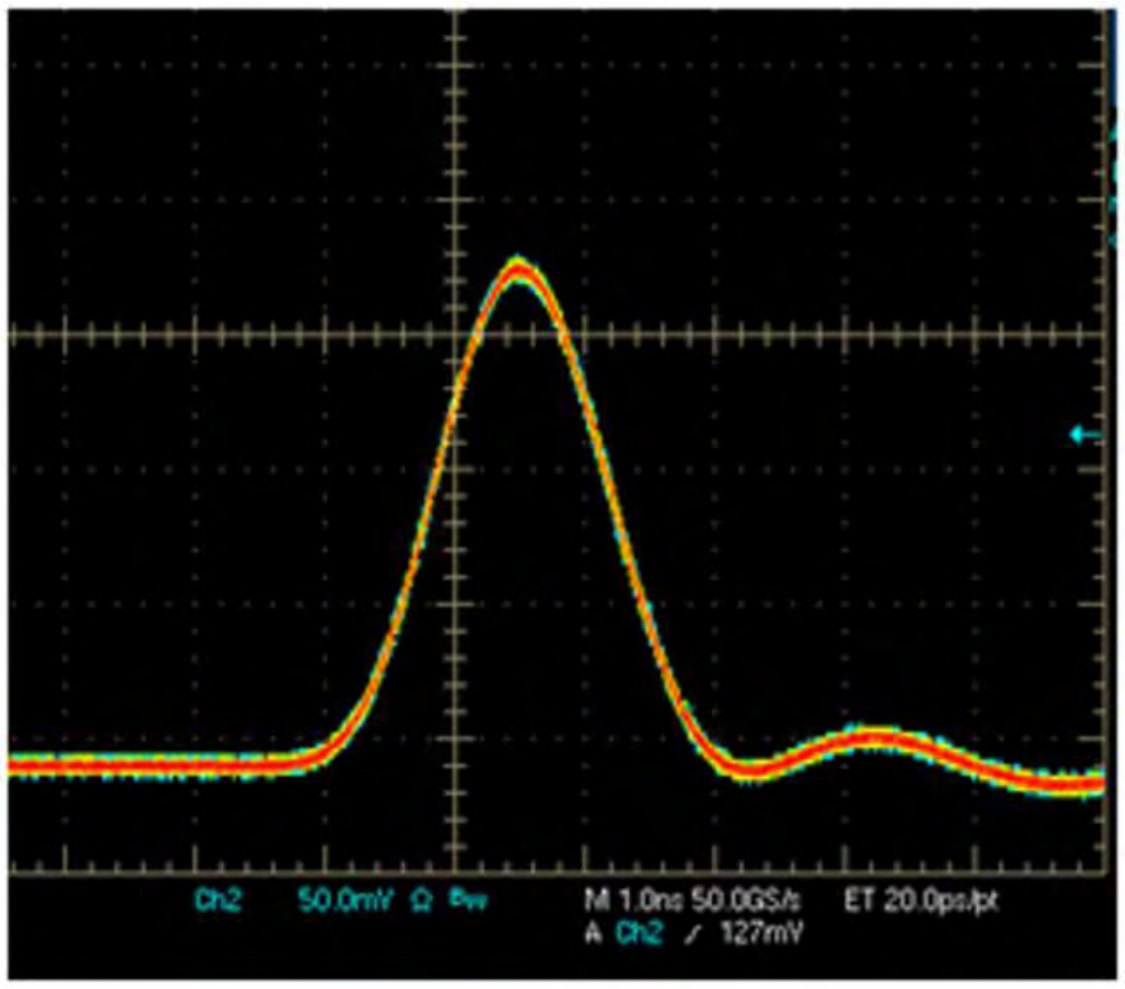

**Figure 9.4. IR energy stability. Contrast is better than 1/1800. [[9.41]; Courtesy of Lumera Laser GmbH]**

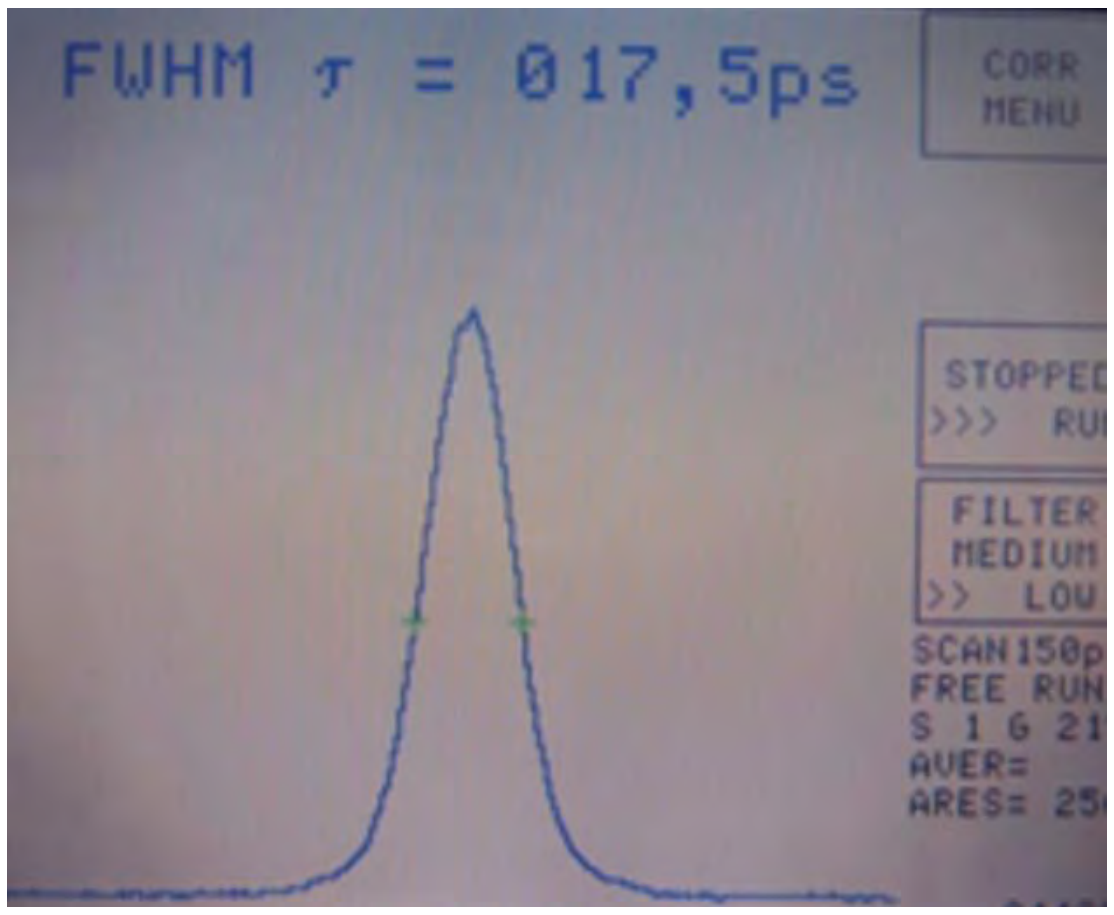

**Figure 9.5. Pulse duration of the IR beam, 12 ps FWHM. [[9.41]; Courtesy of Lumera Laser GmbH]**

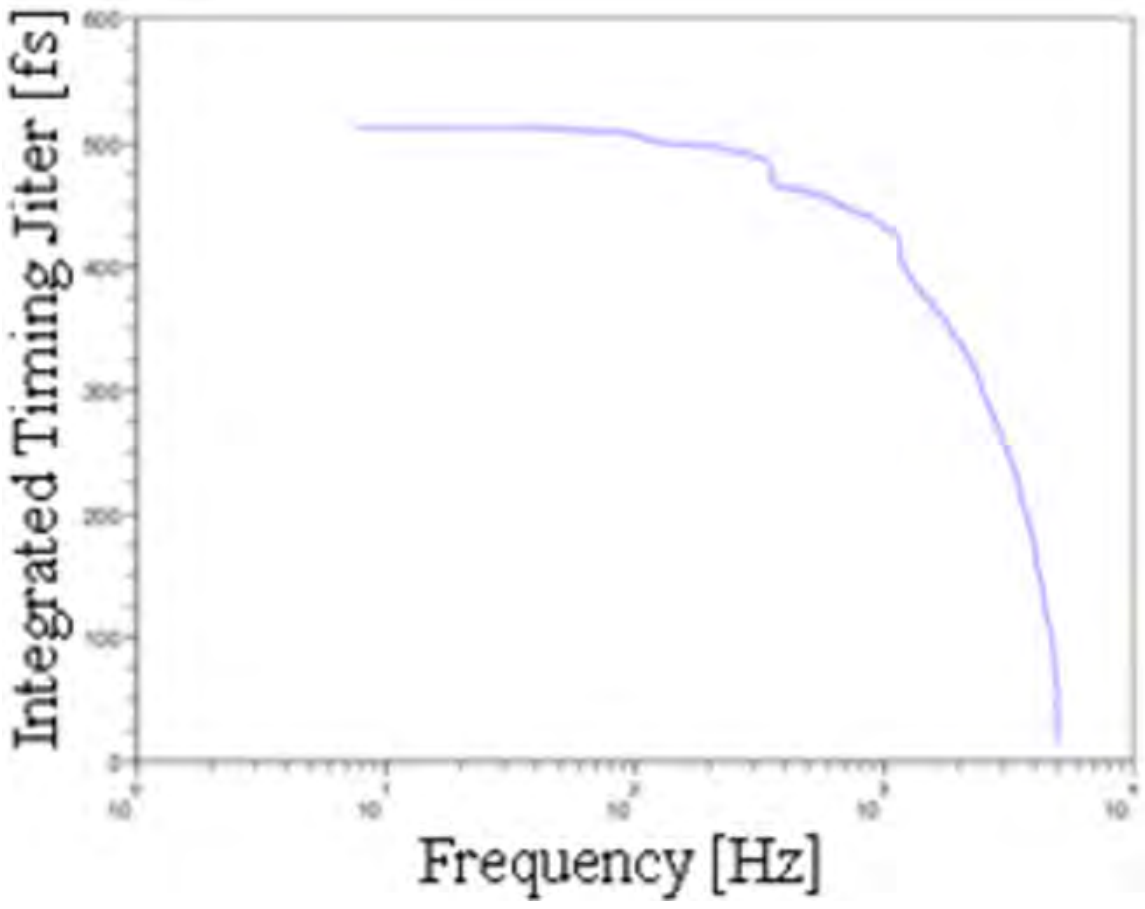

**Figure 9.6. Jitter of the IR beam with respect to the master clock. [[9.41]; Courtesy of Lumera Laser GmbH]**

### 9.3.4 Harmonic Crystals

The fundamental 1.064 µm radiation is first converted into 0.532 µm then 0.355 µm by harmonic crystals. The second harmonic crystal is a non-critically phase-matched LBO crystal maintained at 150 ˚C. A vertically polarized beam at 1.064 µm, focused to a beam waist of ~300 µm, is converted to a horizontally polarized 0.532 µm radiation with 50% efficiency. The transverse profile of this second harmonic beam is near Gaussian and the pulse duration is 8.2 ps. The residual 1.064 µm and 0.532 µm beams are re-collimated by an AR-coated achromatic lens for the subsequent third harmonic generation. If needed, the 0.532 µm beam can be separated from the fundamental by a dichroic mirror to exit the laser.

The third harmonic crystal is also a non-critically phase-matched LBO crystal maintained at 40 ˚C. The vertically polarized 1.064 μm and the horizontally polarized 0.532 μm radiation, are both focused down to a beam waist of 300 μm delivering vertically polarized 0.355 μm beam with an average power of ~5 W. The UV beam is separated by a series of dichroic mirrors and re-collimated to exit the laser. Leakage light from a Brewster window in the beam's path is used to monitor the UV power. Since the spot size in the crystal is small and both the average- and peak-powers are high, there is a very high probability of UV-induced surface degradation leading to damage to the harmonic crystals.

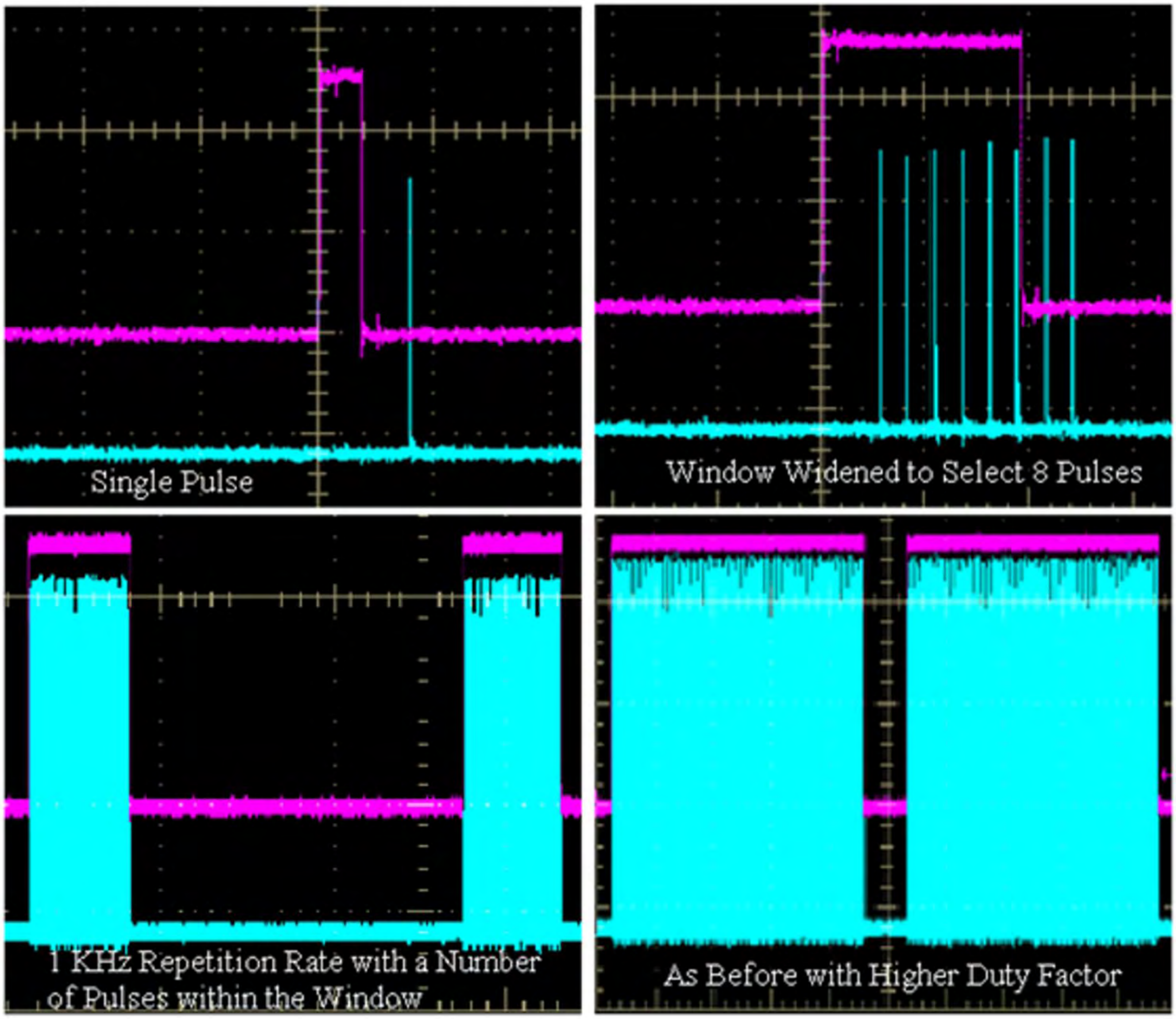

**Figure 9.7. Different configurations of pulses from the pulse picker: Single micropulse, multiple micropulses, 1 KHz macropulse with varying number of micropulses within the macropulse envelop. Magenta lines indicate the shape of the high voltage pulse. [[9.41]; Courtesy of Lumera Laser GmbH]**

The UV module is purged constantly with hydrocarbon-free air to increase the lifespan of the optical components and coatings. Furthermore, transversely oversized crystals mounted on an X-Y translation stage is used to move the crystals to a fresh location in case of damage. The pointing stability of the beam was measured to be ~3 microradians over 2 hr, well within the required range for the application. The beam walk off in the THG crystal causes the appearance of a halo in the transverse profile of the laser beam that must be filtered before it propagates to the photocathode.

Similar Nd based systems are used in a number of facilities including DESY [9.42], [9.43], the Accelerator Test Facility at BNL [9.44], JLab [9.45], SPring-8 [9.46] and Boeing [9.5] to drive different photocathodes.

## 9.4 BULK SOLID STATE LASER SYSTEMS: $\lambda < 1$ µm

### 9.4.1 Ti:sapphire Laser

Ever since Moulton's pioneering work on Ti:sapphire crystal ($Ti^{3+}:Al_2O_3$) [9.47] in the 1980s and the discovery of a robust, passive Kerr-lens mode-locking mechanism [9.48], [9.49] in the '90s, femtosecond Ti:sapphire lasers became the work-horse of nearly all ultrafast laser applications. The superb physical properties of Ti:sapphire and technological advances in optics led to its dominance in the ultrafast laser field. Table **9.2** lists some important properties of this gain medium. The high thermal conductivity of Ti:sapphire requires minimal to no cooling. The long fluorescence lifetime of 3.2 µs allows multiple passes assuring efficient photon amplification and energy extraction. The mature crystal-growth technology affords superior optical quality Ti:sapphire crystals up to tens of centimeters in diameter. Ti:sapphire has a low parasitic absorption at its lasing wavelengths, and a large absorption cross section of ~$4\times10^{-19}$ cm$^2$ at 450-550 nm wavelengths (peak at 490 nm), resulting in a figure-of-merit greater than 200. This is the ratio of the absorbance at the peak absorption wavelength of 490 nm to that of the lasing wavelength, ($\alpha_{490nm}/\alpha_{800nm}$). Its exceptionally broad emission bandwidth of 660-1200 nm can support optical-pulse duration of just a couple femtoseconds (Figure **9.8**). With the invention of all chirped and double-chirped dispersion-control mirrors, a few femtosecond-long optical pulse with octave spanning spectrum is readily achieved, with superb stability over long term continuous operation. The broad bandwidth associated with such short pulse require careful selection of optical components, to compensate dispersion effects pertaining not only to low order terms, but higher order (up to fifth) terms as well.

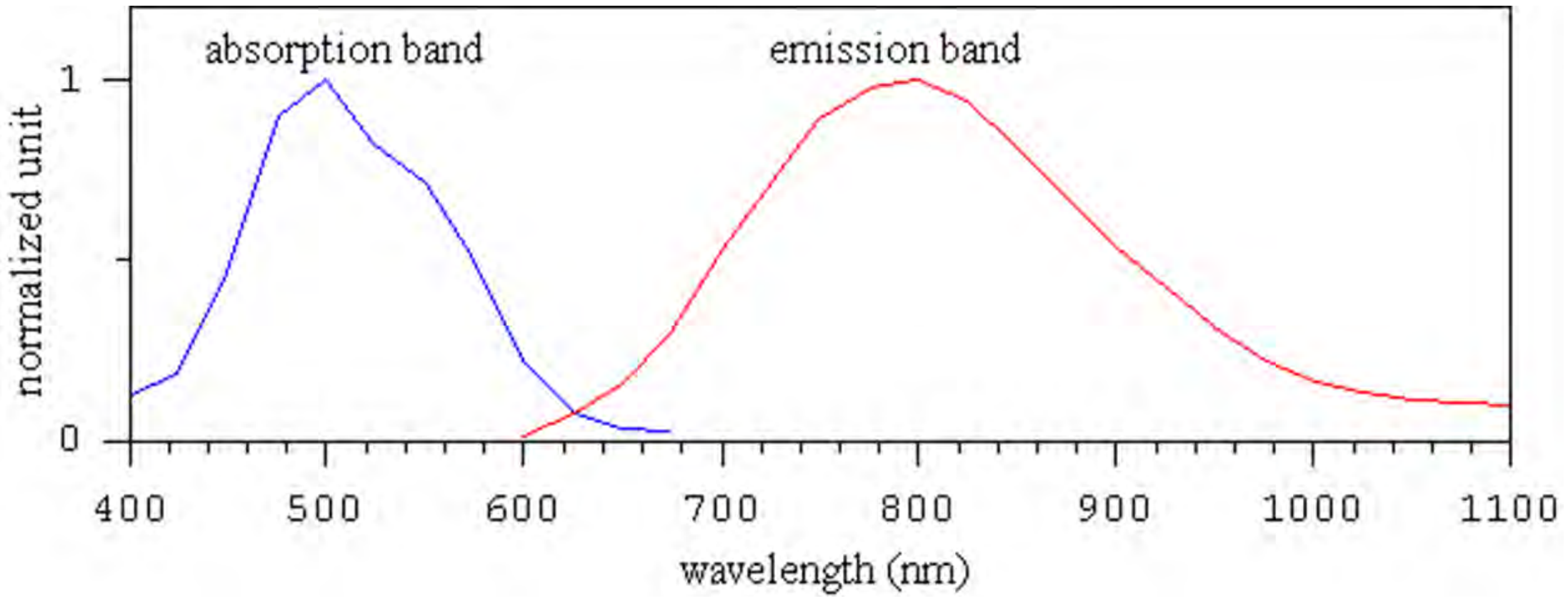

**Figure 9.8. The absorption and emission spectrum of a typical Ti:sapphire crystal.**

Such short laser pulses are advantageous in photoinjectors. As described in Chapter 1, electron beams with very small aspect ratio (ratio of longitudinal- to transverse-dimensions) emitted from the cathode will evolve under the RF field to have very small emittance. Such a beam requires a prompt photocathode and an ultrashort laser pulse, similar to that from the Ti:sapphire laser. Cathodes operating in multi-photon photoemission mode also would benefit from ultrashort, high intensity laser pulses on the cathode. Such an ultrashort pulse can provide sufficiently high peak power at energies well below the cathode's damage threshold. Furthermore, the photon energy of the doubled and tripled frequencies of Ti:sapphire laser is slightly above the work function of typical cathode materials used in photoinjectors. Thus, mode-locked

Ti:sapphire lasers are ideal to serve cathodes operating in both linear- and nonlinear-photoemission modes. Furthermore, using a single ultrafast Ti:sapphire laser to perform high resolution pump-probe experiments, such as electron pump and photon probe in ultrafast pulse radiolysis [9.50] and temporal electron bunch length characterization [9.51], photon pump and electron probe in ultrafast electron diffraction and electron microscopy [9.52], are very attractive because of the well-synchronized, short-pulse duration pump and probe.

### 9.4.2 Oscillator

In early days of femtosecond Ti:sapphire laser, oscillators were pumped by 5-10 W argon-ion lasers. Today, the 1-10 W diode-pumped, intracavity-frequency-doubled Nd:$YVO_4$ laser is the most preferred pump. As in Nd based systems, the cavity length of a femtosecond Ti:sapphire laser oscillator is adjusted to run at the $n^{th}$ sub-harmonic of a reference RF frequency relevant to the accelerator. Ti:sapphire lasers are mode-locked to deliver a train of ultrashort pulses. Mode-locking can be initiated in a prism-pair, dispersion-controlled Ti:sapphire laser oscillator without a saturable absorber (schematically shown in Figure **9.9**) by simply taping on one of the cavity mirrors. Such passive mode-locking techniques include using a hard aperture Kerr-lens and saturable absorber mirrors. Active mode-locking, using acoustic-optical (A-O) modulators, supports phase-locking to an external RF source. As in Nd laser systems, to preserve phase-locking, the repetition frequency is monitored and the cavity length adjusted accordingly.

As seen from Figure **9.9**, using a fused silica prism-pair readily assures the generation of 10-20 fs light pulses. In this design, a thin gain crystal with a high FOM is used to eliminate the residual third-order dispersion in the cavity and to minimize the fifth-order while preserving a high gain.

The invention of all chirped and double-chirped dispersion control mirrors has significantly changed the Ti:sapphire laser oscillator, see Figure **9.10**. Although the chirped mirror design lacks the wavelength tuning capability and relies on a mechanical device to initiate its mode-locking, octave spanning spectrum is readily achieved with good long term stability. However, they are generally not used in photoinjectors because of space charge limitations; rather an intentionally frequency chirped, longer light pulse is used to mitigate the space charge effects. An oscillator emits an average optical power of 0.3-2 W at the center wavelength at 800 nm with spectral bandwidth of 20-90 nm depending on configurations. However, it emits only 2-4 nJ of energy per pulse, *viz.*, insufficient for most photoinjectors. The transverse beam profile of femtosecond Ti:sapphire lasers often are slightly elliptical, not Gaussian. This can be corrected in the final amplified beam by implementing a well-tuned amplifier cavity, to result in a high quality $TEM_{00}$ Gaussian profile.

### 9.4.3 Pulse Stretcher and Power Amplifier

The output energy per pulse of a Ti:sapphire laser oscillator alone is generally not sufficient to deliver the nanocoulomb level charges needed for a photoinjector. Therefore, a laser amplifier or a series of power amplifiers normally are needed to boost the energy output of the oscillator.

The choices for amplifier configurations are regenerative- and multi-pass-amplifiers. In both schemes, to combat the self-focusing of a high peak power laser pulse in the gain medium that would lead to damage, the amplified pulse first is temporally stretched before amplification to reduce the peak power and recompressed afterwards, see Figure **9.11**, based on the chirp-pulse amplification technology developed by Strickland and Mourou [9.33].

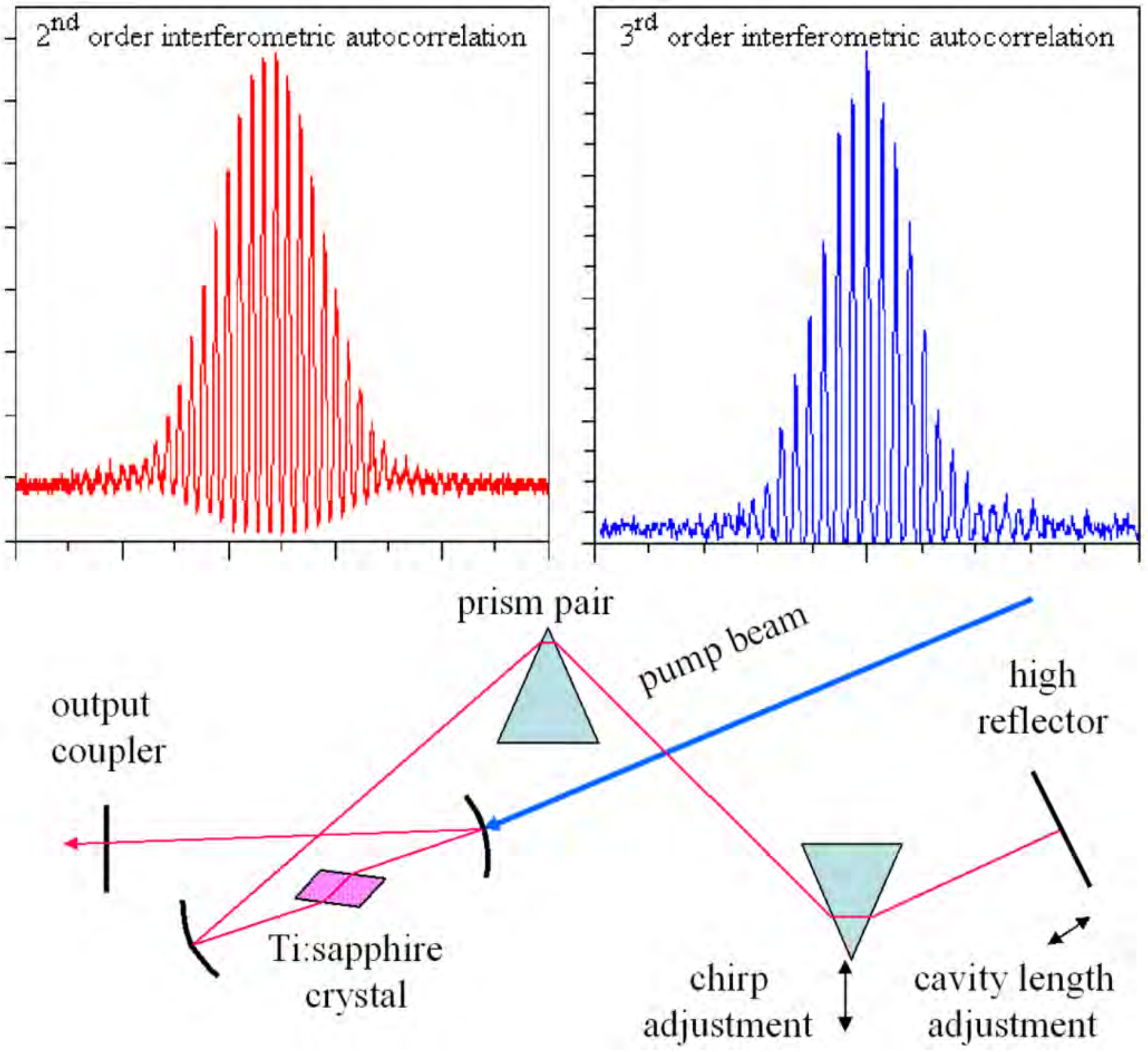

**Figure 9.9. Layout of a conventional prism dispersion-controlled, mode-locked femtosecond Ti:sapphire laser oscillator and a typical 2$^{nd}$ and 3$^{rd}$ order interferometric autocorrelation trace utilizing, respectively, the 2-photon photoconductivity of a GaP diode, and the third-harmonic generation at the air-glass interface of a piece of 16 µm thick cover glass, Δ*t* ~ 15 fs. The time axis of all interferometric signals is dictated by the fringe spacing, equals to an optical light field cycle; depicted is a 2.67 fs signal here at the carrier wavelength of 800 nm. [[9.53] (© 2005 Optical Society of America)]**

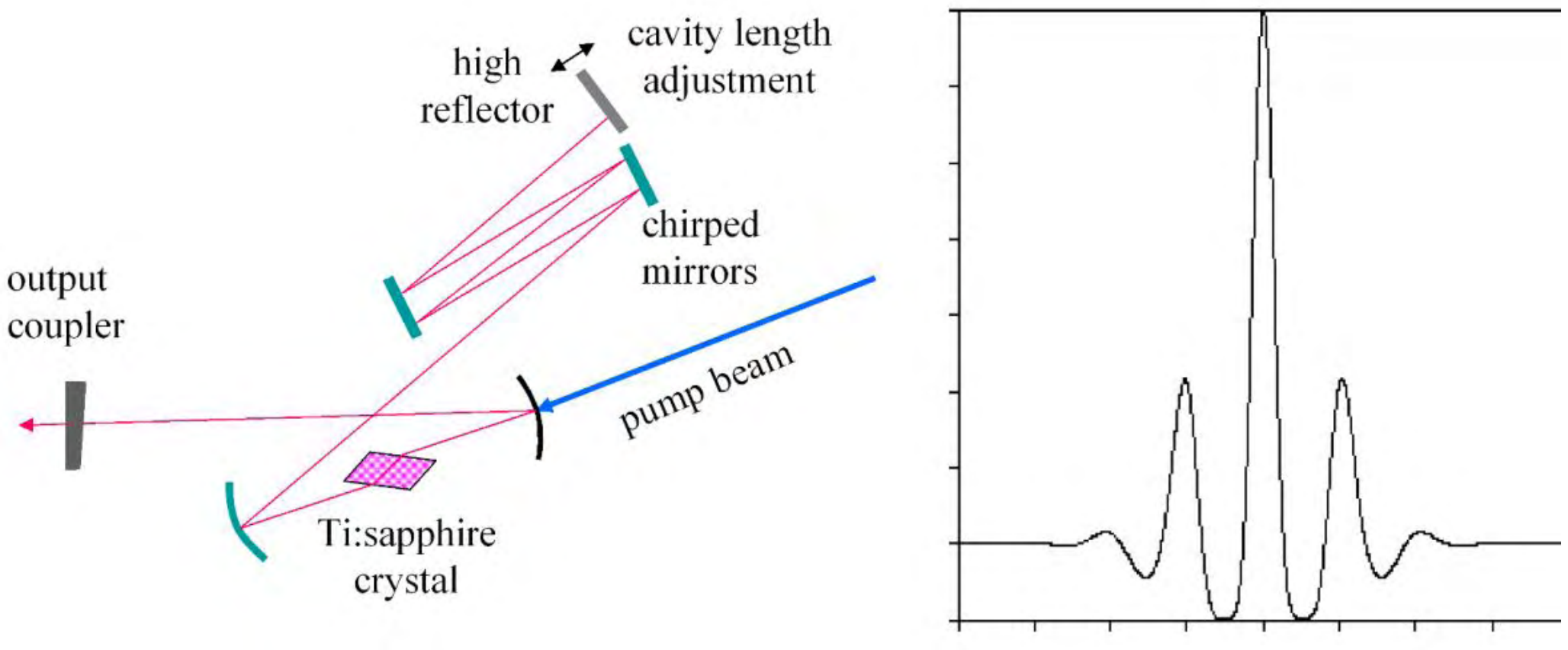

**Figure 9.10. Layout of all-chirped mirror, mode-locked femtosecond Ti:sapphire laser oscillator and a theoretical 5 fs light pulse.**

Although the simple Martinez [9.54] and Treacy [9.55] compression scheme was adequate to compensate picosecond light pulses, the ultrabroad bandwidth of femtosecond light pulse demands a simple, but quintic-phase-limited, aberration-free, chirped-pulse amplification system. However, the principle of chirp pulse amplification remains the same, where the ultrabroad bandwidth of a femtosecond optical pulse is first down-chirped using an optical stretcher and the chirped spectral components is then re-phased by a compressor. For temporal stretching, the all-reflective Offner stretcher design [9.56], [9.57] is preferred because of its compactness, simplicity and high performance in ultra-broadband light pulse operation. In temporal compression, grating pairs are still the only available optics that can compensate large amounts of group velocity delay dispersion with high transmission efficiency and wide tunability. Using a careful alignment procedure [9.58], a dual grating pair can reliably compress the ultrabroad bandwidth femtosecond light pulse to its Fourier transform-limited pulse width.

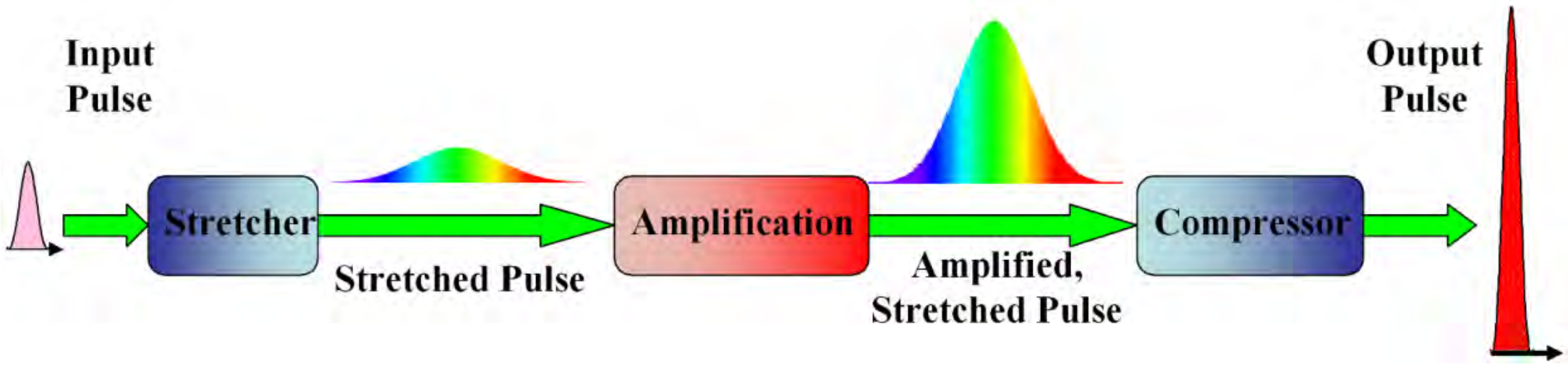

**Figure 9.11. Conceptual illustration of laser chirped pulse amplification.**

The first stage of the amplifier almost exclusively is a regenerative amplifier because of the maturity of the technology. Figure **9.12** is a generic layout of a regenerative amplifier. Although the output of a ~100 MHz repetition rate, femtosecond Ti:sapphire laser can exceed 1 W average power, usually 100-300 mW suffices for injection to a regenerative amplifier. In contrast, higher seed power is preferred in a multi-pass amplifier configuration.

Peak powers of terawatts at low repetition rates and average powers of hundreds of watts at ~100 MHz commonly are available in DPSS systems. New developments, such as cryogenically cooled DPSS systems are in the research and development stage and may be available within a decade.

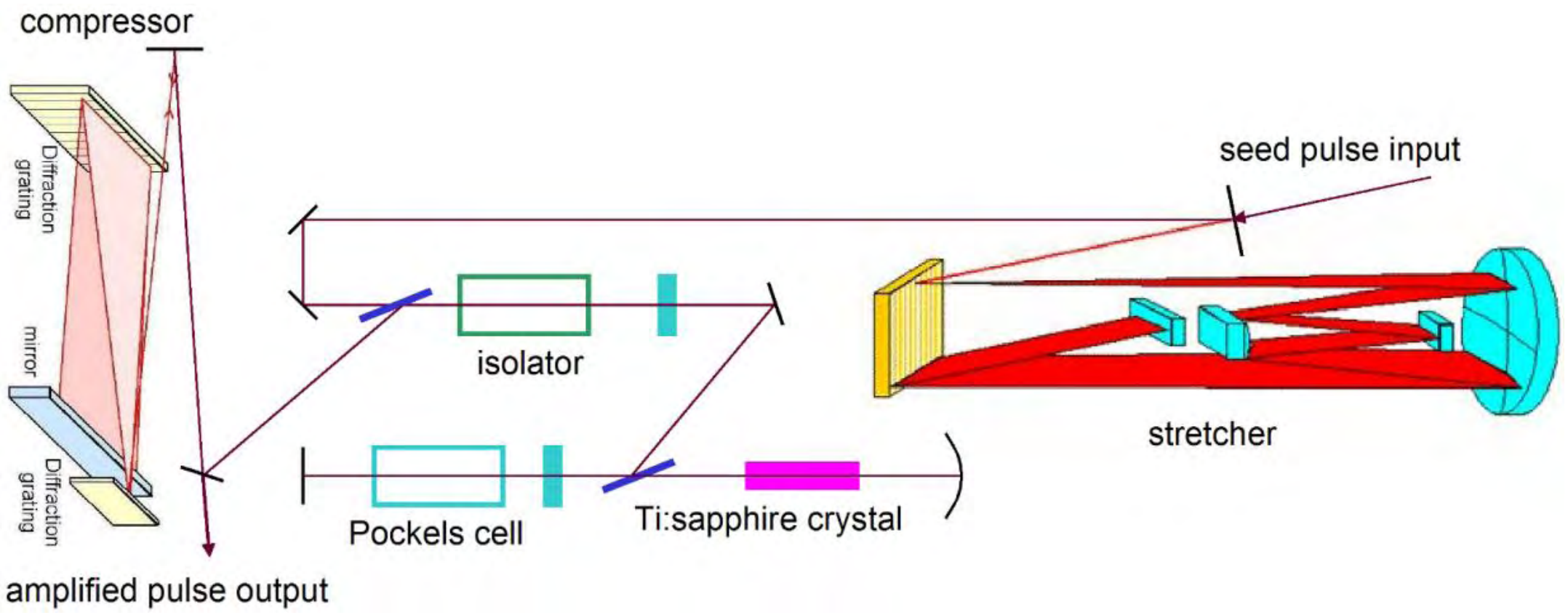

**Figure 9.12. Typical regenerative amplifier arrangement, employing an all-reflective Offner stretcher and a pair of grating compressor.**

The output power of the system can be increased by staging a series of amplifiers. The maximum power obtainable is limited by the components' damage threshold, the thermal effects, pump-power limitation and the required S/N ratio.

### 9.4.4 Harmonic Generation

Frequency up-conversion of the laser wavelength is accomplished by passing the high intensity laser beam through a nonlinear medium. The dipole moment induced by the laser field in the crystal contains the response of the crystal at both the fundamental and harmonic frequencies of the field. As each of these fields is described by their wave vectors, $k_{fundamental}$ and $k_{harmonic}$, constructive interference between the generated harmonic radiation will occur only if ($n_{fundamental}$ * $\omega_{fundamental}$) = ($n_{harmonic}$ * $\omega_{harmonic}$), where $n$ and $\omega$ are the refractive index and angular frequency, respectively. This condition is known as the phase matching condition. Since the refractive index in a birefringent crystal depends on the polarization of the light and the angle of incidence, one technique used in phase matching, known as angle tuning, is to select these two parameters appropriately for the fundamental and the harmonic frequency so that the phase matching condition is met over the entire length of the crystal. However, the broad bandwidth of the ultrashort laser pulses make it difficult to phase match the entire bandwidth over the full length of the crystal. Several techniques have been suggested for optimal phase matching over this large bandwidth [9.59]–[9.61].

The approach that has been used in some photoinjectors is using a thin crystal to minimize the walk-off, although this compromises the efficiency of the conversion process. LBO, BBO and KDP crystals are the most common materials for frequency doubling an ultrashort pulse of a Ti:sapphire laser because they can be phase matched over its entire tuning range and have low absorption both at the fundamental and second harmonic frequency. In selecting the crystal and the focusing geometry, care must be taken to evaluate the conversion efficiency, walk-off angle and damage threshold. Appropriate operating temperature and dimensions of the crystal, angle of incidence, divergence and the focal spot size of the laser must be determined. For high repetition rate lasers, the damage threshold due to the high average power should also be taken into account.

Figure **9.13** is a schematic of the Ti:sapphire laser system used at the LCLS; details and performance specifications are given in [9.62]. This well-characterized system was used to drive an X-ray FEL photoinjector. One major accomplishment is the high level of synchronization achievable between the laser, the electron beam and the user equipment. With a fiber-link system for distributing RF signals for the laser, a locking stability of 25 fs was attained between the reference and the laser [9.63]. Similar systems were used in SparC [9.64], SPring8 [9.65] and UCLA [9.66].

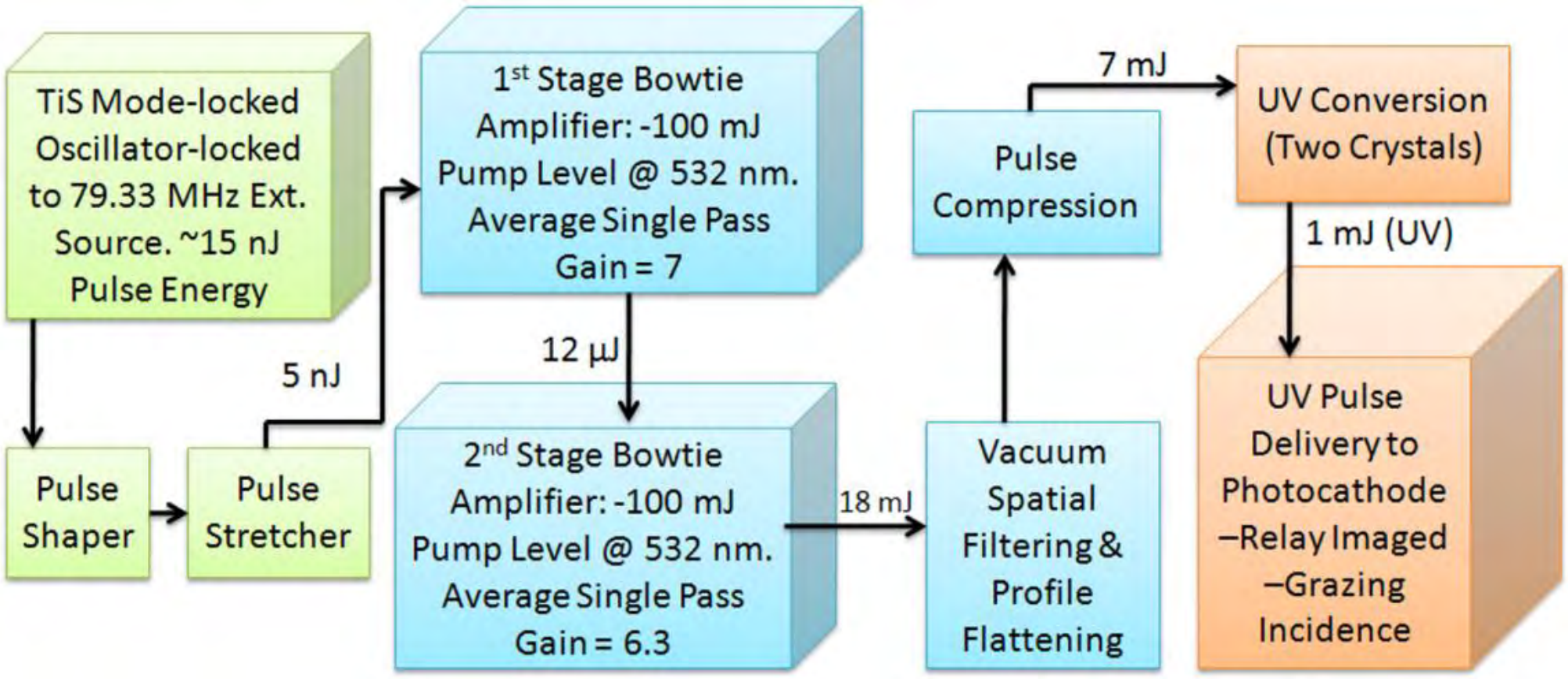

**Figure 9.13. Schematic of the LCLS Ti:sapphire laser system.**

## 9.5 FIBER LASER

The performance of fiber lasers has improved considerably in the past decade. This progress prompted their inclusion in this chapter, although their use in photoinjectors is not yet prevalent. As described earlier, laser beams with high peak power and high brightness were realized with diode-pumped, rare-earth doped, solid-state lasers. Typically, the gain medium is in the form of a rod. However, as pump power increases, the large energy difference between the absorbed and emitted photons elevated the heat load inside the gain medium. This power-dependent thermal distribution causes thermal lensing and heat-induced mechanical stress, leading to poor beam quality. Since fiber has a large surface area-to-volume ratio, heat dissipation in this medium is significantly easier than in BSS lasers. Furthermore, since the laser radiation is confined to the waveguide structure of the fiber, the lasing occurs naturally in the mode supported by the fiber. Hence, the beam quality depends primarily on the design of the fiber.

An optical fiber consists of a core surrounded by cladding. The refractive index of the cladding is chosen to be smaller than that of the core so that the total internal reflection guides the optical beam within the core. If the core supports only a single mode (the chosen mode will have very low loss and the others will experience high loss), then only this preferred mode will propagate. In a typical single-mode fiber, the dominant mode is $TEM_{00}$ and the beam's transverse profile is Gaussian. The most common gain media are different types of glass doped with rare-earth ions, Nd, Yb, or Er ions. Nd fiber lasers operate typically at 1.06 µm, while the tunability of a Yb fiber laser ranges from 1.03-1.1 µm [9.67]; that of an Er fiber ranges from 1.48-1.62 µm [9.68]. Fibers with dopants, such as praseodymium (Pr) and thulium (Tm), reach 2-3 µm wavelengths. Figure **9.14** is a schematic of the simplest form of a fiber laser [9.69].

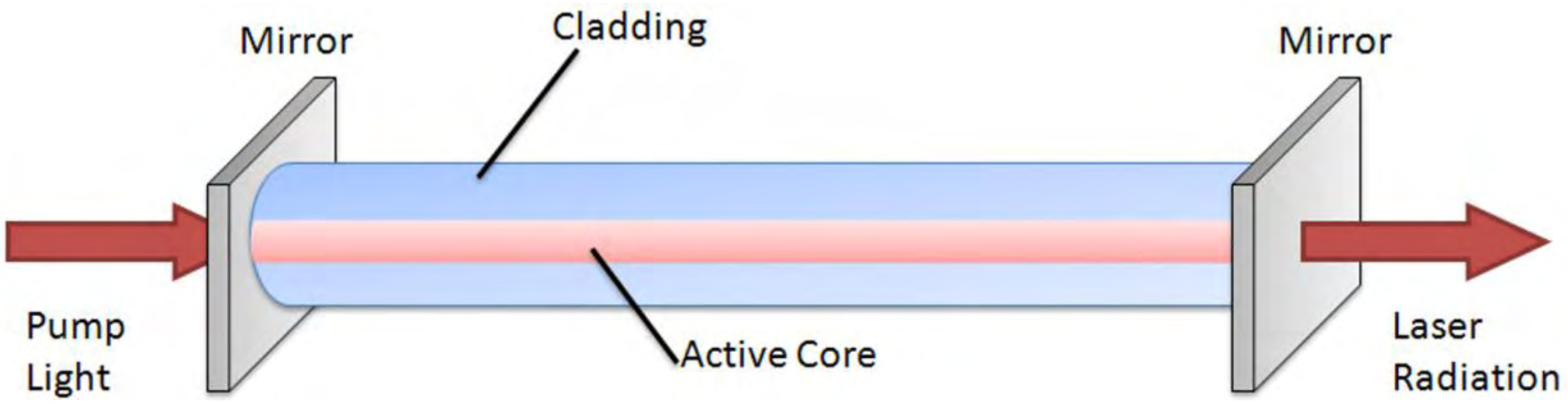

**Figure 9.14. Schematic of the simplest form of a fiber laser. Both the pump light and the laser radiation are guided through its core. [[9.69] (© 2010 Optical Society of America)]**

In a single-clad fiber, both the radiation from the pump and the laser are guided through the core. Since the core of a single mode fiber is typically in tens of micrometers, the output power is limited by the pump power that can be coupled into the fiber and the laser power that can be extracted without damaging the fiber. In this simple approach, the pump light must be of high mode quality for good coupling into the fiber. In pulse mode operation, the peak intensity of the laser radiation may be high enough to introduce nonlinear effects, such as self-phase modulation, four-wave mixing, self-focusing, and Raman and Brillouin scattering that should be avoided.

To increase the pump power and coupling efficiency, double-clad fibers with different inner cladding shapes were developed. Here, the core operates in single mode for laser radiation, while the inner cladding supports multimodes for launching the pump beam. Absorption efficiencies over 70% were achieved by optimizing the shape of the inner cladding and the position of the core within the inner cladding [9.70], [9.71].

For high-power operation, the core's mode volume must be increased without compromising the mode quality. The approach is to coil [9.72] a large mode area (LMA) fiber to increase the losses in the higher order modes, use of using photonic crystal-fibers (PCF), or use of fiber tapering, Figure **9.15** illustrates a cross section of a state-of-the-art PCF [9.69]. The 80 µm diameter, Yb/Al co-doped, active core is surrounded by three rings of small air holes with a pitch of 13.3 µm and diameter-to-spacing ratio of ~0.1 forming a 200 µm diameter inner cladding. The inner cladding guides the pump wavelength while the air holes confine the radiation in the doped region. The inner cladding is surrounded by an air clad consisting of ninety silica bridges, 400 µm thick and 10 µm long; this assures a ~0.6 numerical aperture for efficiently launching 976 nm multimode pump radiation into the fiber. A 1.5 mm fused-silica outer cladding surrounds this inner structure preserving its straightness and reducing bend-induced losses. With similar PCF fiber laser technology, but of different dimensions, 1.064 µm, 2 W, 10 ps, 80 MHz output from a laser oscillator was amplified to an average power of 48 W at a peak power of 60 kW in a near-diffraction-limited beam with 70% polarization extinction [9.73]. It was achieved by significantly reducing the nonlinearity induced in the fiber; this key impact paves the way for employing fiber lasers in high peak power applications.

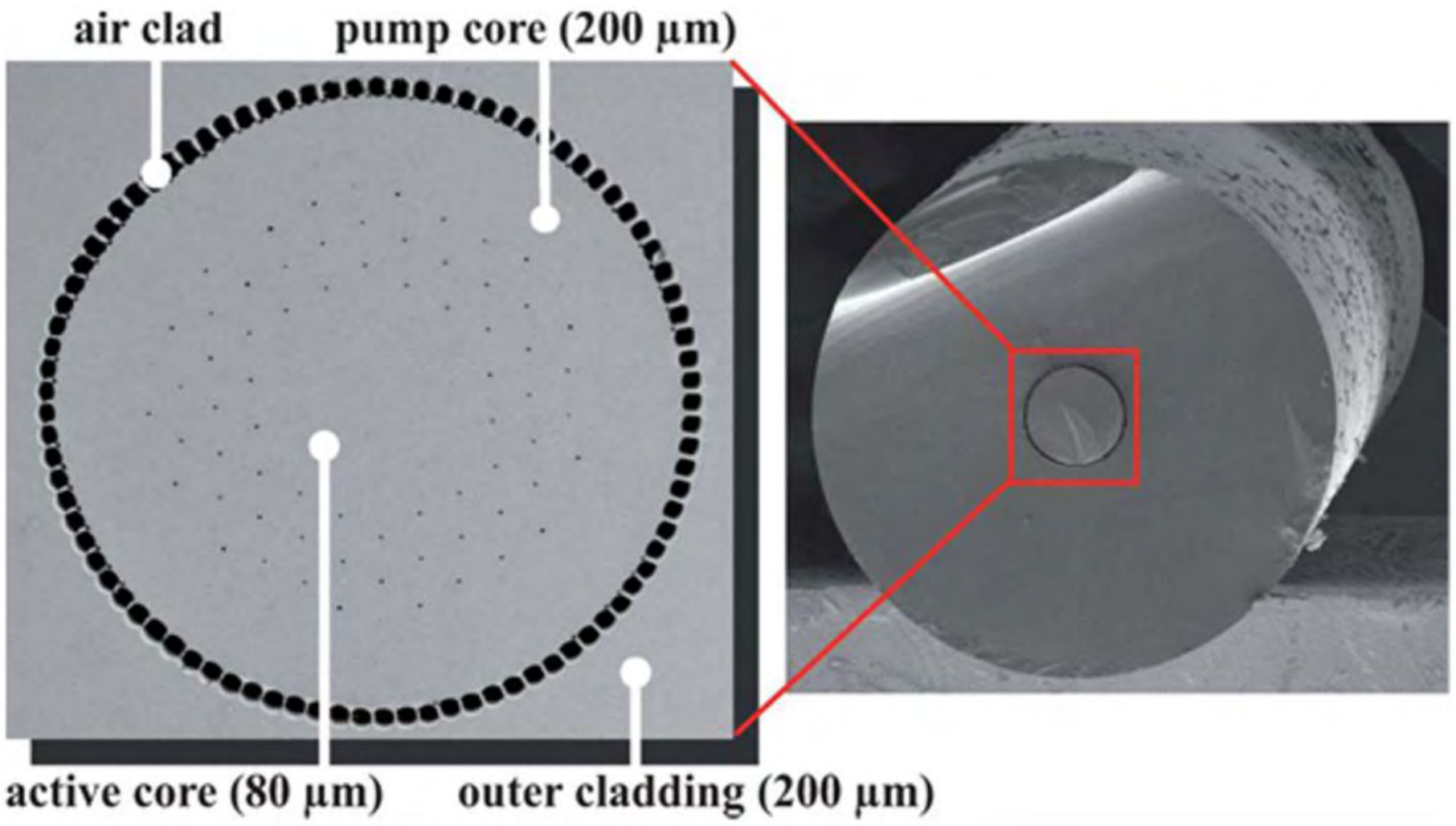

**Figure 9.15. Cross sectional and expanded view of an LMA rod type PCF. [[9.69] (© 2010 Optical Society of America)]**

However, at present only a few commercial fiber laser systems meet the specifications for the photoinjectors. However, there are several home-built [9.11], [9.74] and prototype [9.75] fiber lasers that meet the stringent photoinjector requirements. The design and operation of one of the prototypes is described below; this should be regarded only as a guideline because extensive research and development on high-power fiber lasers is ongoing.

### 9.5.1 Master Oscillator

A schematic of a fiber system is shown in Figure **9.16**. The compactness and the high efficiencies of fiber lasers make them attractive options for high repetition rate applications. The laser system can be an all fiber- or a hybrid-system. In the former, the oscillator also is a fiber laser, generating short laser pulses of low average- and peak-powers that are subsequently amplified. A number of techniques such as active mode-locking [9.76], passive soliton and saturable absorber mode-locking [9.77], [9.78] can produce ultrashort laser pulses from the fiber. Synchronization to the RF requires actively stabilizing the cavity length typically done by coiling the fiber in a cylindrical shaped expandable piezo drum.

Alternatively, in a hybrid system, electrically modulating the output from a semiconductor diode can supply the seed radiation for the fiber amplifier. This technique, as described in section 8.6 can be used to generate 30-50 ps laser pulses at 780 nm, with repetition rates ranging from 100 MHz to 3 GHz. Short pulse duration beam at desired pulse train mode can also be obtained by slicing the CW output of a diode laser with a fast electro-optic (E-O) amplitude modulator. A DC bias controller, superimposed with an impulse signal generated by a signal generator drives the modulator. By adjusting the amplitudes of the DC bias and the impulse signal, output pulses with duration of 50 ps can be generated. The repetition rate is varied by changing the pulse repetition rate of the signal generator. In the prototype system described in this section, the latter approach is adopted due to its simplicity.

In both of these approaches, the electronic control makes the synchronization to an external RF source relatively easy. However, the S/N of the output pulse, especially that of the E-O/A-O modulator, reaches only ~10[4]. This necessitates a smaller gain to be implemented on the subsequent amplifiers. This technique has been adapted to generate 1.06 µm radiation at a repetition rate of 700 MHz with a pulse duration of ~50 ps and single pulse energy of picojoules. Similar parameters at 1560 nm, which can be frequency up-converted after amplification for producing polarized electrons, can also be generated with this method.

### 9.5.2 Power Amplifier

The output power of fiber laser oscillator is typically insufficient for most photoinjectors; it is subsequently amplified in a series of fiber amplifiers. Power scaling in a fiber amplifier is limited by the onset of nonlinear effects and the maximum pump power that can be coupled into the fiber. The nonlinear effects include stimulated Raman scattering, stimulated Brillouin scattering and self-phase modulation. Although Watt to kilowatt high power pump diodes are commercially available, coupling such high power into micrometers to hundreds of micrometers diameter size optical fibers without encountering damage problems is still an issue, even for the large core fibers in millimeter diameter. Nonetheless, different techniques have been proposed to improve both the coupling efficiency and the out-coupled power [9.79], [9.80]. Using novel fiber coupling technique, multiple pump beams with a coupling efficiency of 67% into a final power amplifier, an amplified beam with up to 90 W at 1.06 µm, and 40 W at 0.532 µm, at ~50 ps pulse duration and 700 MHz repetition rate has been generated.

"Periodically poled nonlinear media" is the most preferred configuration for in-line harmonic generation where coupling from the amplifier to the nonlinear medium is straight forward and the conversion efficiency can typically reach > 70%. Even though the mode quality of the fundamental laser beam is $TEM_{00}$ in both the transverse directions, the resulting harmonic beam typically is elliptical because of residual walk-off effects. Hence, care might be taken to re-shape the beam prior to irradiating the cathode.

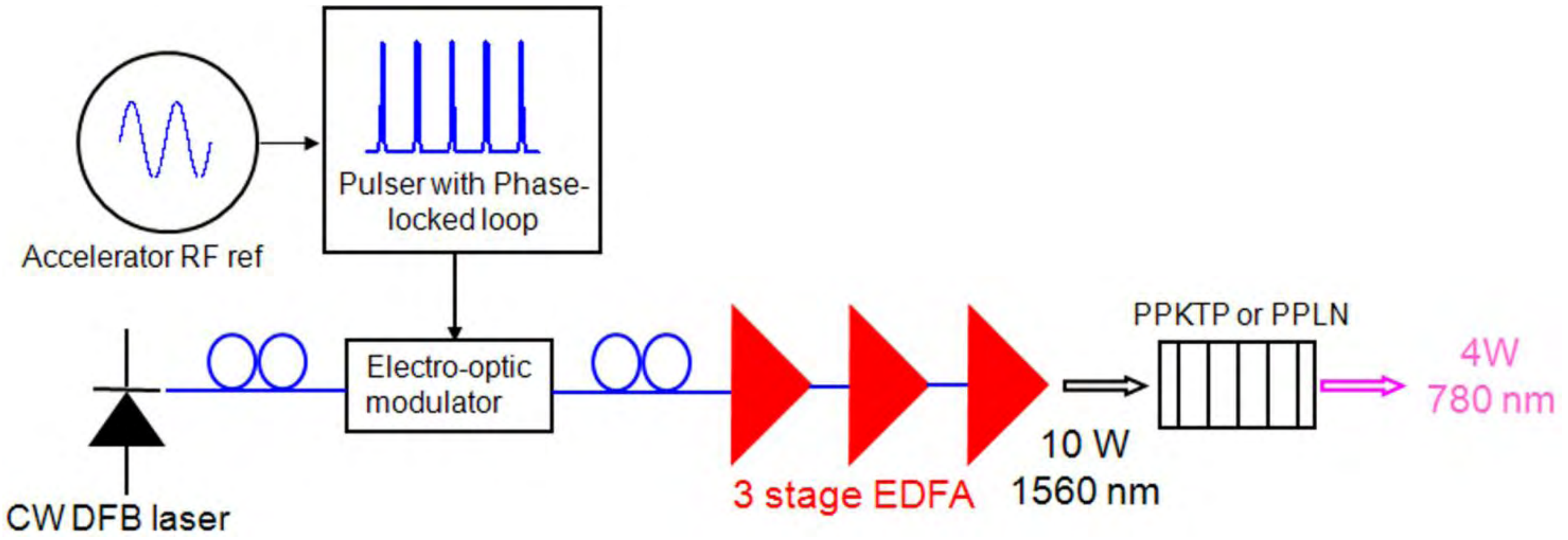

**Figure 9.16. Schematic of hybrid system.**

## 9.6 GENERAL CONSIDERATIONS

### 9.6.1 Beam Shaping (Transverse and Longitudinal)

As discussed in Chapter 1, various simulations [9.81], [9.82] suggest that modifying the transverse- and longitudinal-distribution of the electron beam from the nominal Gaussian to a beer can, an ellipsoidal, or a water-bag configuration (see Figure **9.17**), would lower the beam emittance [9.83]. As Luiten *et al.* demonstrated [9.84], when a 30 fs (pancake-shaped) electron beam of 100 pC charge from a 1 mm spot is launched into an RF accelerating field of 100 MV $m^{-1}$, the beam evolves into an ellipsoidal as it propagates through the RF injector. Their simulations revealed that, at a propagation distance of 200 mm, the emittance of the electron beam can be reduced from 3 µm for an initial Gaussian transverse-profile, to 1 µm for a flat-top profile, and to 0.4 μm for an ellipsoidal profile. Such pancake beams were produced and propagated by shining a femtosecond laser pulse on metal photocathode (prompt emitter) in RF injectors [9.85].

Since the spatial- and temporal-profile of a laser pulse can be transferred simply on to the photo-ejected electrons with minimum distortion, shaping the driving laser beam to the desired three-dimensional profile may offer an alternative route to obtaining an electron beam with ultralow emittance. Schemes to obtain spatiotemporal flat-top laser pulses were demonstrated at SPring8 [9.86], DESY [9.87], LCLS [9.88], [9.7] and EUROFEL [9.89].

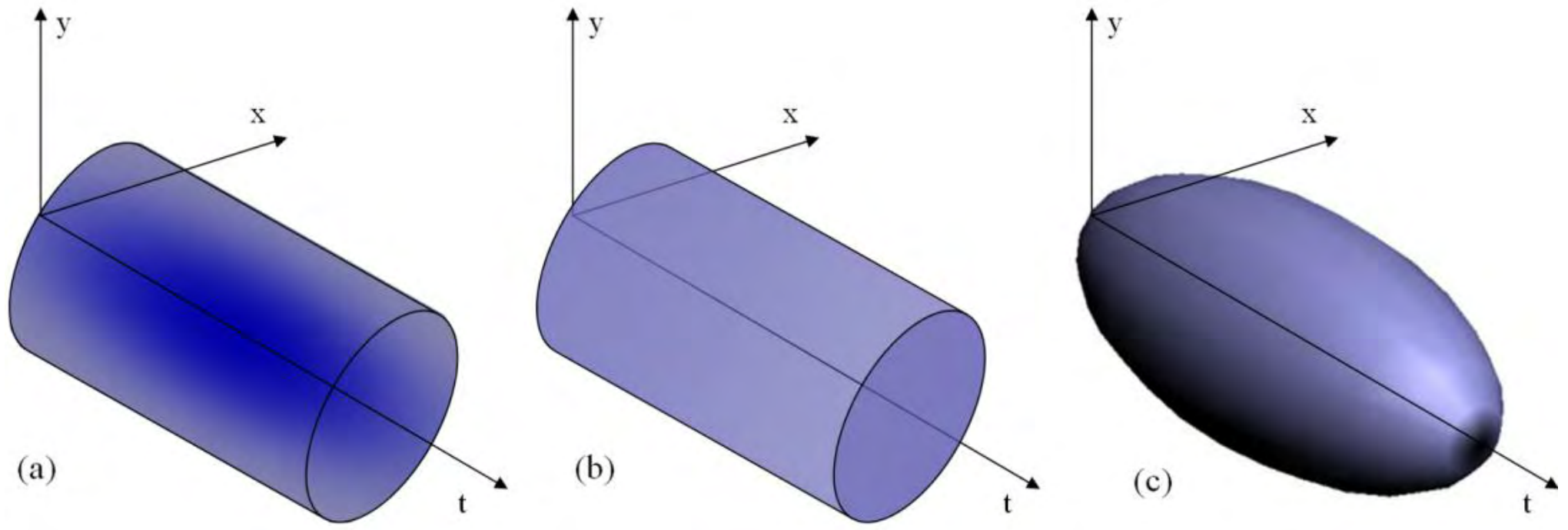

**Figure 9.17. Spatiotemporal profile of (a) conventional Gaussian intensity-distribution with a fixed cylindrical cross section, (b) beer-can wherein the light intensity is constant over the entire cylindrical volume, and (c) 3-D uniform ellipsoidal, where the intensity is constant, but with a time-dependent spatial; *x* and *y* are the spatial coordinates and *t* is the time axis.**

### 9.6.2 Temporal Shaping

The techniques used for temporal shaping depend strongly on the duration of the initial laser pulse. When laser pulses are longer than a nanosecond, they can be shaped directly by E-O modulators. These advanced modulators have tens of gigahertz speed [9.90]. Sub-picosecond and femtosecond light pulses intrinsically have large spectral bandwidth, and hence, they can be temporally shaped by modulating the laser pulse's spectral content. Weiner [9.91], who pioneered this approach, has extensively reviewed this technique.

Figure **9.18(a)** shows the basic principle of the process. A grating first disperses the frequency content of the laser pulse. The first Fourier lens transforms the angular dispersion from the grating to diffraction-limited spots at its focal plane, with well defined position to frequency correlation. Spatially patterned amplitude- and phase-masks manipulate the spectral content at this plane. A second lens-grating pair unfolds this modulated beam and recombines all frequency components into a single beam with a modulated temporal shape. For ideal shaping, the grating-lens combination without the mask must be dispersion free, that is, the output beams must be identical to the input when the phase mask is absent.

Using fixed amplitude- and phase-masks, 75 fs laser pulse at the wavelength of 0.62 μm was temporally shaped to ~2 ps pulse with rise and fall times of ~100 fs [9.92]. Employing a mechanical aperture of adjustable size, flat top UV pulses of ~8 ps FWHM were obtained from a 100 fs Ti:sapphire oscillator-amplifier system [9.93]. The major advantage of such static mask modulators is the ease of their fabrication and simplicity of the overall system. However, fabrication cannot be undertaken without a complete knowledge of the input pulse. Therefore, different masks must be designed, inserted, and aligned for different pulse shapes.

Temporal shape can be controlled actively by replacing the fixed mask with programmable liquid-crystal-spatial-light-modulator (LC-SLM), movable mirror, deformable mirror, or an A-O modulator. High quality square pulses were generated by modulating the spectrum of a 50 fs pulse from a Ti:sapphire laser with a computer-controllable SLM; a reduction in the beam emittance of a photoinjector was observed [9.94]. In that work, a LC-SLM with 128 pixels served as the phase mask and the period of the pixels was 100 µm with a 3 µm transparent gap between electrodes. The resolution of the phase shift on the LC-SLM was ~0.01π. Positioning this pulse shaper between the oscillator and the pulse stretcher lowered the likelihood of damaging the pulse shaping optics. The transmission of the laser power through the pulse shaper was about 60%. Using this device, the authors have compared the emittance of 0.6 nC beam with 9 ps FWHM Gaussian and square pulse shapes in the longitudinal dimension. Normalized emittance of 1.38 ± 0.06 π mm-mrad and 0.95 ± 0.03 π mm-mrad were obtained for the Gaussian and square shapes respectively, resulting in ~ 45% improvement in the emittance. The optimal bunch length for this charge was found to be in the range of 8-9 ps. Since the optimal bunch length depends strongly on the charge, current density, accelerating field and RF phase, this optimization needs to be executed for the parameters of interest.

A-O modulators with a programmable dispersive filter (commercially known as DAZZLER) have successfully shaped [9.95]–[9.97] picosecond- and femtosecond-pulses (see Figure **9.18(b)**). Employing a DAZZLER to shape the IR beam to a 15 ps long rectangular profile followed by frequency tripling to the UV, a 17-25% reduction in electron beam emittance was observed [9.97]. Although these active pulse-shaping techniques are well developed, powerful, and can deliver arbitrary temporal pulse shapes, they have limitations, *viz.*, their low optical-power handling capabilities make them applicable only to low repetition rates and the output may contain residual spatiotemporal distortions. Furthermore, very often only UV photons (the second-, third-, or fourth-harmonic of the light pulse) are sufficiently energetic to liberate electrons from a photocathode. Although it is highly desirable to shape the temporal pulse in the final UV pulses, the conventional SLMs do not function well at these short wavelengths. Recently, a UV version was introduced; its performance has not been thoroughly tested in these applications. If modulation is executed at the fundamental wavelength, then care must be taken so that no additional spatiotemporal distortions are introduced in the frequency-conversion process.

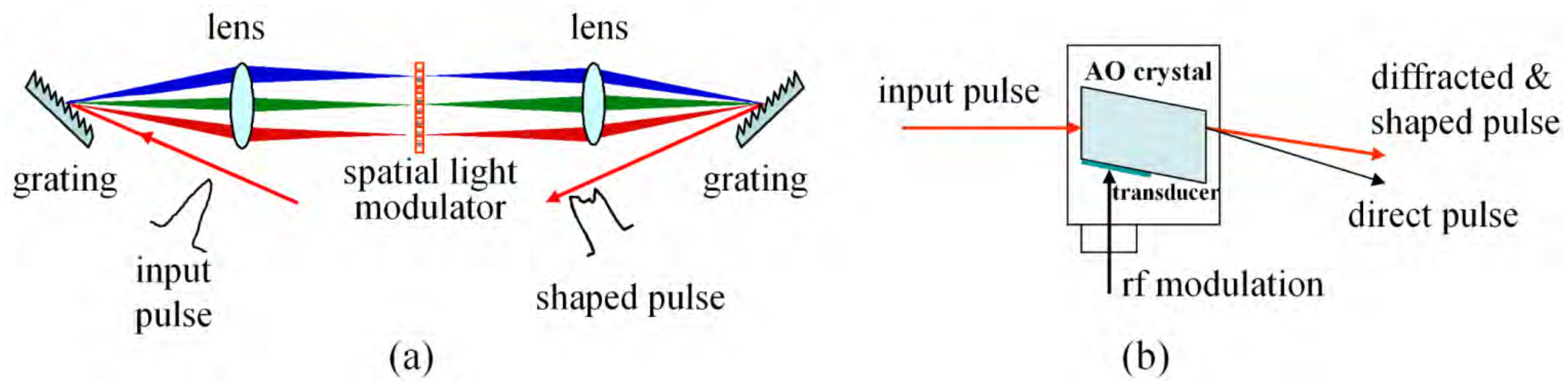

**Figure 9.18. Conventional temporal shaping using, (a) Spatial Light Modulator, and (b) DAZZLER.**

Alternatively, passive pulse stacking is commonly used to temporally shape energetic ultrashort laser pulses of hertz to megahertz repetition rate [9.98], [9.99]. This approach proved more successful with picosecond pulses wherein the spectral bandwidth is not large enough for accurate modulation in the spectral domain and the modulator's response time is not fast enough for modulation in the time domain. Such pulse stackers can be constructed either with a set of appropriately oriented birefringent crystals [9.100]–[9.102] or a conventional delay line [9.1], [9.103]. Both these passive techniques require precise interferometric alignment.

The conventional delay line consists of various optical elements to stack pulses in a cascading fashion. The number of optical elements required as well as the interferometric precision needed in pulse stacking greatly complicates the alignment and stabilization schemes to such a high precision.

With the birefringent crystal and a single polarized light pulse passes through a properly arranged birefringent crystal, the ordinary- and extraordinary-components of the light pulse propagate with different group velocities, resulting in a time-dependent change of polarization (Figure **9.19**). Coupling this output pulse with a polarizer splits the single input pulse into two output pulses, with the time delay between them dictated by the two refractive indices and the length of the crystal. The temporal pulse thus is stretched with a predetermined intensity modulation. Interferometric precision is achieved by accurately controlling the length of the crystal. This approach, which is relatively simple [9.101], results in stable pulse shape, but offers only limited control of the final pulse shape. Using a stack of three appropriately oriented $YVO_4$ birefringent crystals of thicknesses 24-, 12- and 6-mm respectively, a 10 ps laser beam at 532 nm was stretched to 60 ps [9.101]. The transmission efficiency of the arrangement is 62%, limited by the crystal's intrinsic absorption loss. Insertion loss can be lowered significantly by carefully selecting low absorption crystals.

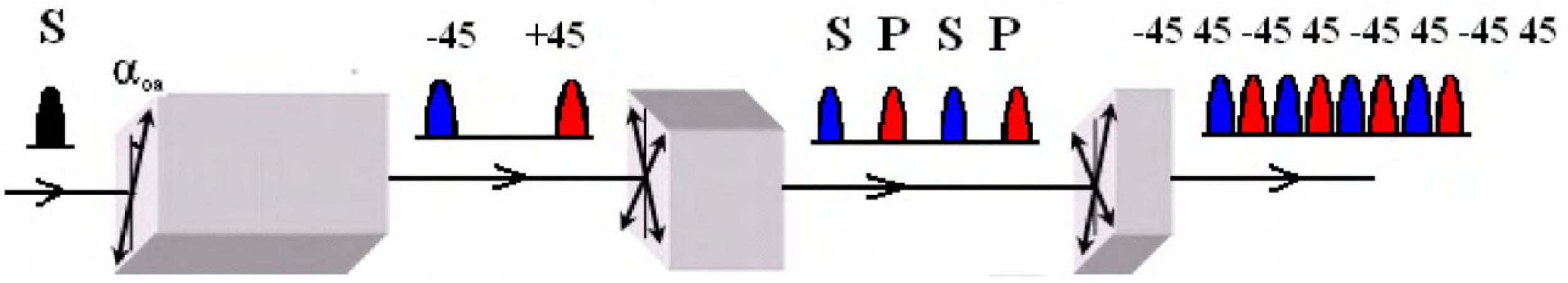

**Figure 9.19. Passive pulse-shaping by cascading several birefringent crystals. The optic axes are shown by thin lines on the surface of the each crystal. The polarization of the laser pulse is depicted by thick double arrows in each crystal. An input S-polarized pulse first is split into two replicas with various amplitude and temporal delays along the ±45° of the crystal axis. Subsequent crystals are arranged with the optics axis rotated 45° with respect to the preceding one. In this manner, a stack of n birefringent crystals will generate 2n replica pulses. It is possible to generate arbitrarily shaped pulse controlling their relative temporal delay and amplitude. [Adapted figure with permission from [9.101]. Copyright 2009 by the American Physical Society]**

### 9.6.3 Spatial Shaping

The conventional spatial beam shaping technique relies on imaging a predefined circular aperture using relay optics on to a photocathode. Although near flat-top profile can be achieved, the process introduces excessive optical loss with added risk of injecting airy fringe pattern onto the photocathode. Alternatively, the spatial profile may be shaped by controlling the optical transmission in the radial direction with either an inverse-Gaussian transmission filter [9.104], diffractive elements [9.105], [9.106], or by actively controlling the reflected/transmitted laser beams by controlling the mirror/lens in the beam path to produce the desired beam shape. Recently, commercial passive aspheric refractive/reflective optical shaping systems have become available [9.106]–[9.109]. They offer a robust yet simple design, with high optical transmission and high optical-power handling capability down to the UV wavelengths. Figure **9.20** depicts the ZEMAX simulated ray-diagram of the π-shaper (Newport Inc.), the simulated input- and output-beam profiles, and the experimental results [9.101]. With a Gaussian input-beam diameter of 4.7 mm, the π-shaper produces a high-quality flattop transverse profile with a diameter of 6.5 mm FWHM.

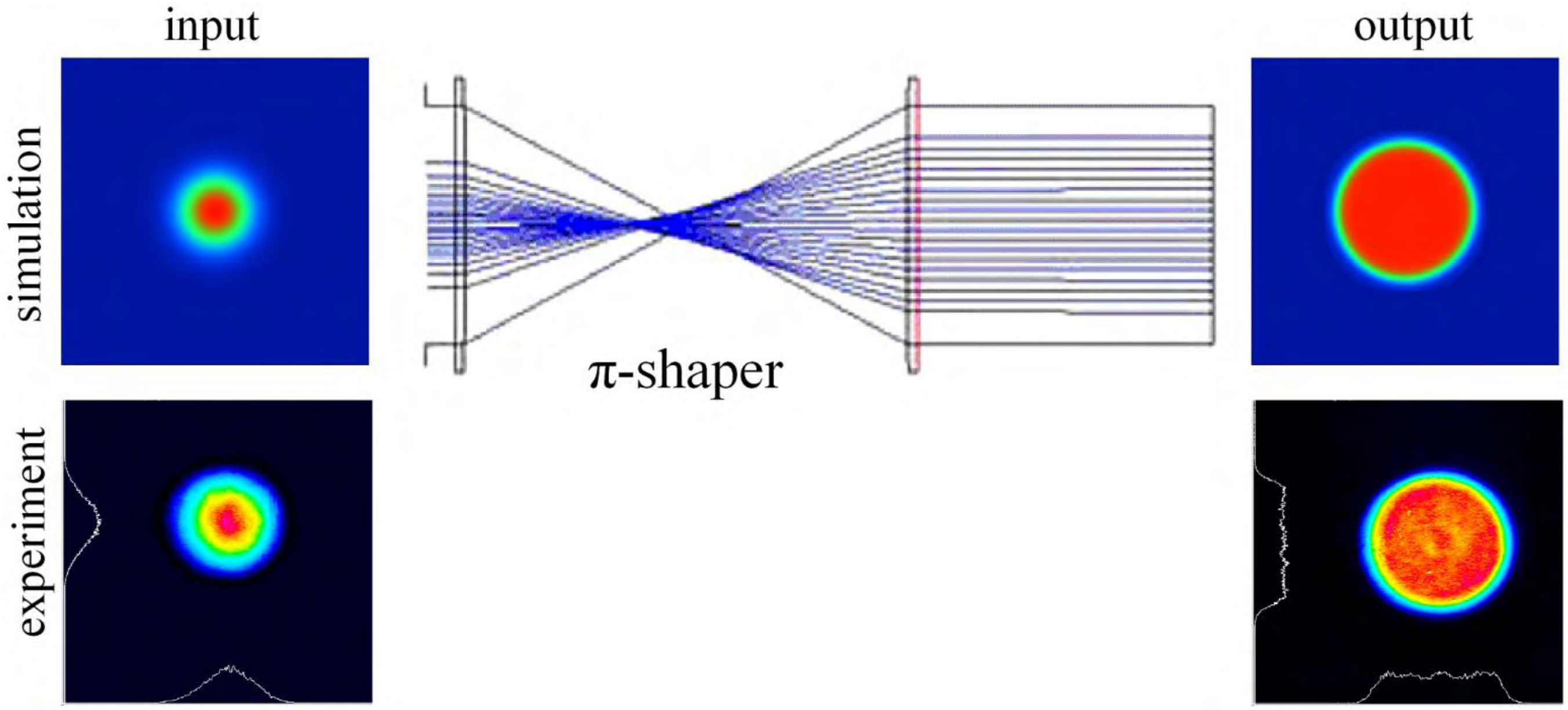

**Figure 9.20. Passive spatial π-shaper. Top: ZEMAX simulation of transverse beam distribution modified from a Gaussian at the input end to a uniform cylindrical output profile using a π-shaper. Bottom: Experimental profiles of input- and output-beam. [Adapted figure with permission from [9.101]. Copyright 2009 by the American Physical Society]**

The sensitivity of the π-shaper with respect to the angular misalignment (or tilt) and laser beam de-center (or beam lateral offset) was studied [9.101]. Using ZEMAX simulation it was shown and verified experimentally that the tolerance of this π-shaper is ±9 mrad in tilt angle and ±0.38 mm in de-center (Figure **9.21**). The experiment agrees well with the simulation. These tolerances also dictate the acceptable beam pointing error of an input laser beam.

The passive shapers are designed for specific beam-parameters, such as the size and divergence of the input and output spots; they work very well for Gaussian beams. Since both the position and angle of incidence could change due to pointing instability, the sensitivity of the beam shaper to misalignment is an important consideration in its design. Figure **9.21** illustrates the change in the beam profile due to shift in the location and angle of the centroid.

High-quality spatial profiles other than flat-top ones can be obtained with various optical-design modifications [9.110], [9.111]. A complete spatiotemporally shaped light pulse is generated by cascading a temporally shaped ultrafast light pulse followed by a spatial shaper.

Figure **9.22** illustrates the transverse- and longitudinal-beam profile of the beam [9.101] at different locations along the laser-beam transport that includes a pulse stacker and a π-shaper. The Gaussian transverse and temporal profile for the laser beam is depicted on the left side of the figure. Its flat-top profile obtained after passing through a beam stacker and a π-shaper near the laser is shown in the middle. The beam then is relay imaged onto the location of the photocathode ~9 m away from the laser. The flat-top image is a reproduced, with minor additional modulations, after the relay optics, shown on the right of Figure **9.22**. It was shown experimentally [9.100] that the electron beam profile mirrors that of the shaped laser beam.

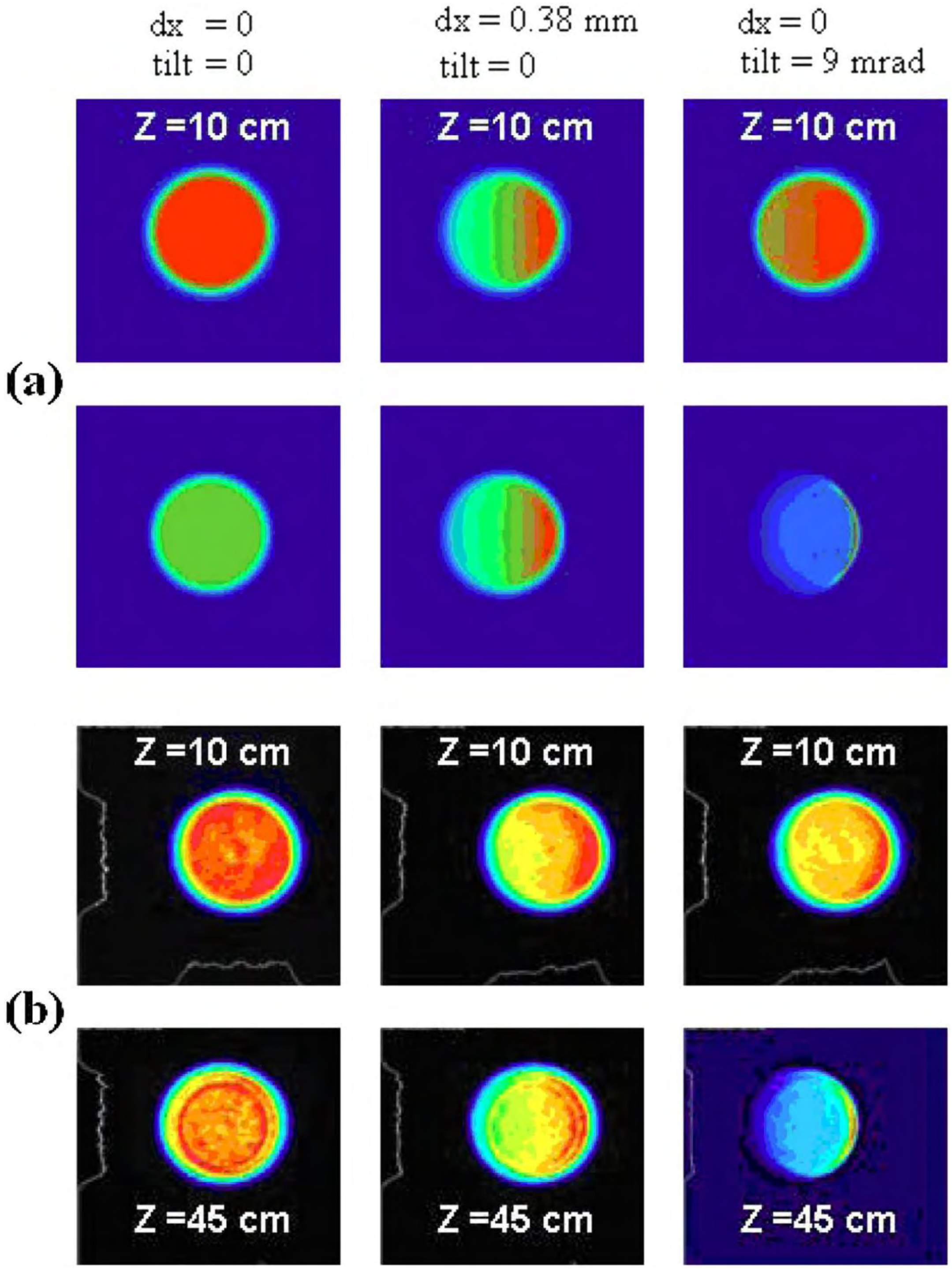

**Figure 9.21. Spatial beam profile at 10- and 45-cm away from the π-shaper under different alignment conditions, illustrating its sensitivity to beam shift and tilt with respect to the optical axis: (a) ZEMAX simulation; (b) corresponding experimental profiles. [Reprinted figure with permission from [9.101]. Copyright 2009 by the American Physical Society]**

The ellipsoidal profile is characterized with a constant energy density (thus, charge density) distribution at a given time slice, as shown in Figure **9.17(c)**. However, it displays a time-dependent change of beam size. This unique beam is characterized by spot size that is small at the beginning and the end of the pulse duration, but grows larger towards the middle. It is particularly challenging to generate such an ellipsoidal beam profile where the laser pulse must be shaped spatiotemporally in 3-D. Some suggested techniques use adaptive 2-D deformable mirrors or 2-D spatial light modulators in combination with a temporal shaper [9.85]. Recently, an elegant solution was proposed that uses the unique properties of an ultrashort light pulse. The basic principle is as follows: the modulation on the instantaneous frequency of a light pulse leads to time-dependent phase changes. The chromatic aberration of a subsequent focusing optics, converts this shift of the instantaneous frequency ($\delta\omega$) to a change in the focusing length of a lens ($\delta f$)

$$\delta f = \frac{-f_0}{n_0 - 1}\frac{\delta n}{\delta\omega}\delta\omega \tag{9.1}$$

where $f_0$ and $n_0$ are the nominal focus length and refractive index of the lens. In this manner, a time-dependent frequency change is mapped into a time-dependent focal length (Figure **9.23**). A DAZZLER, most suitably, realizes these modulations because therein the amplitude and phase of a light pulse can be controlled arbitrarily. Figure **9.23** illustrates the experimental arrangement and some results have been obtained using this method [9.112]. However, not only is generating this ellipsoidal profile complex, considering the ultrashort time scale on the change in spot size, but confirming this time-dependent behavior of beam size is equally difficult.

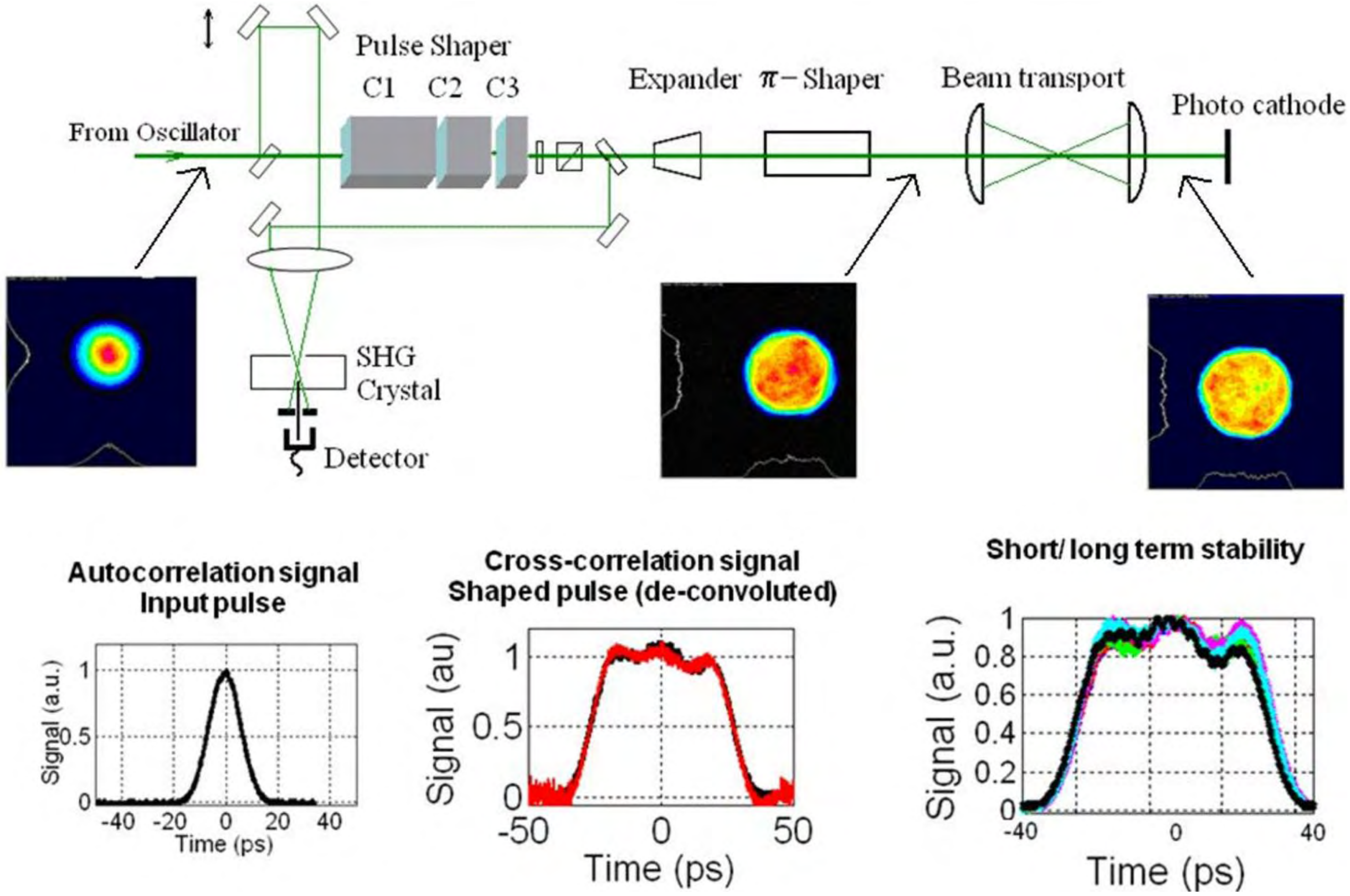

**Figure 9.22. Spatial- and temporal-profile of the laser beam as it propagates along the beam transport. Left: Output of the laser; Middle: Spatially shaped by beam shaper and temporally shaped by beam stacker near the laser; Right: Shaped beam after relay imaging to the cathode location, 9 m away from the laser. [Adapted figures with permission from [9.101]. Copyright 2009 by the American Physical Society]**

Despite the successful developments on spatiotemporal pulse-shaping techniques, they all have their advantages and disadvantages; and no one scheme suits all accelerator facilities. Indeed, spatiotemporal shaping is only the beginning in producing a low-emittance electron beam. Equal emphasis should be given to transporting the electron beam to preserve its quality. A well-designed spatiotemporal shaper with proven robustness, combined with a feedback system based on the electron beam characteristics and a self-learning algorithm eventually should lead to the next generation of ultrahigh brightness electron beams.

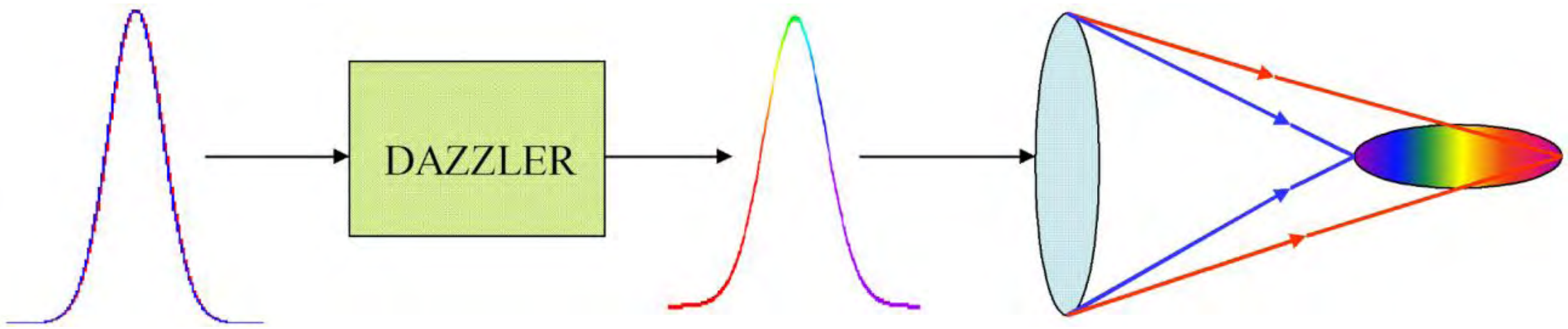

**Figure 9.23. Simplified schematic of a time-dependent frequency pulse mapped onto a time-dependent focal length that can produce an ellipsoidal spatiotemporal profile in 3-D.**

### 9.6.4 Beam Transport and Diagnostics

Preserving the beam profile as well as the timing synchronism is critical for generating reliable low emittance beams. It is a general practice to use low-dispersion vacuum windows, ultra-broadband silver- or gold-coated mirrors for low-power beams and ultra-broadband dielectric mirrors for high power beams. Typically the laser room is not in close proximity to the interaction region and the laser beam has to be transported over long distances with varying climate. Hence, careful design of the beam transport and monitoring the parameters are essential for reliable operation of the injector. Normally an overfilled iris or the image location of the spatial profiler is relay imaged on to the photocathode. It is beneficial to use a beam tracking code such as ZEMAX to trace the laser beam from start to finish and compare the results to the measurement for better understanding of the beam optics, especially when the beam is non-Gaussian. Enclosing the beam fully by evacuated pipes will minimize the climate related beam fluctuations. Use of remotely controlled mirrors, especially in high radiation area is necessary for online adjustment of the laser beam optics. Extensive use of irises and displays of beam position on these irises are recommended for correcting the beam path when needed. The laser energy, beam profile, location, pulse duration and timing should be monitored as much as possible and displayed in the laser room/control room to facilitate trouble shooting and correcting. Critical parameters should be displayed in the control room so that laser parameters can be correlated to electron beam parameters.

Other engineering aspects for the reliable and routine operation of all lasers, especially for photoinjectors are to hermetically enclose all oscillator components; temperature-stabilize the optical base plate containing all optics including the pump source; use low-profile optical mounts to minimize vibrations; minimize the distance between the pump and the gain crystal to reduce the effect of the beam pointing instability of the pump source; and, finally, regulate the temperature, cleanliness and humidity of laser room. All these parameters are paramount to a low (sub-100 fs) timing jitter needed for the photoinjector system.

### 9.6.5 Timing and Synchronization

Maintaining the frequency and phase relationship of the RF cavity in a photoinjector at all times is one of the most important parameters of an accelerator facility. An even more stringent requirement is that all the reference signals distributed throughout the facility should be synchronized accurately, since low-timing jitter leads to better experimental accuracy. Timing accuracy starts by examining the frequency stability of a high-precision, frequency synthesized RF clock, the repetition frequency of the laser oscillator and the transit time change for the reference signal. Assuming that the master clock determines the RF frequency and phase, we then look at other timing drifts. The cavity round-trip time, or the repetition rate, often drifts with time due to environmental disturbances, such as changes in temperature and acoustic noise. Likewise, changes in the cables' lengths could introduce timing drifts in trigger signals. In general, the laser is phase-locked to the master oscillator, or its sub-harmonic, to maintain its synchronism. A typical phase-lock [9.113] loop, shown in Figure **9.24**, is used to lock an RF source to the repetition rate of the laser to provide accurate timing.

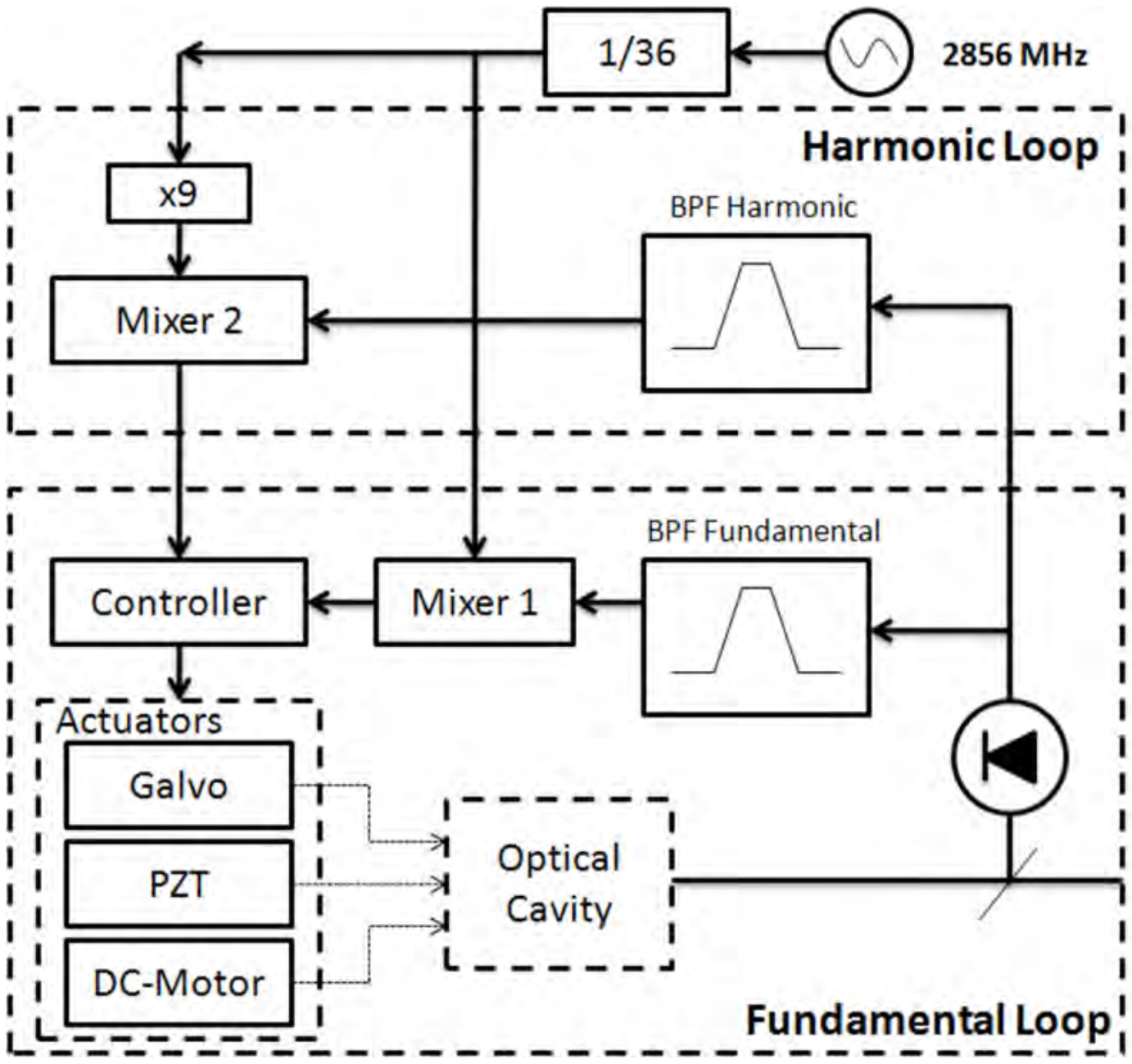

**Figure 9.24. Signal from the master oscillator is compared to the signal in the mixer from the laser, and the error signal from the mixer is used as a feedback to adjust the laser cavity length. [9.113]**

Phase-locking at the higher harmonics of an RF source would provide a better timing accuracy. As is evident from Figure **9.25**, the slope of the harmonics at zero crossing, depicted and highlighted, increases linearly with the harmonic number *n*, that is,

$$\left[\frac{\Delta V}{\Delta t}\right]_n > \left[\frac{\Delta V}{\Delta t}\right]_{n-1} \tag{9.2}$$

Therefore, a smaller $\Delta t$ can be registered by the same $\Delta V$ at the higher harmonic. This $\Delta V$ often is used as a feedback signal to control the length of the oscillator cavity. In practice, reaching the higher RF harmonics at the equivalent RF power as the fundamental requires much RF power. Consequently, analog phase-locking electronics often lock only to the first few RF harmonics, but this is sufficient for the oscillator's output to reach 50 fs timing jitter. As shown in the Nd system in Figure **9.24**, two feedback loops generally are employed; a slow feedback loop (100 ms to tens of seconds) corrects time drift due mostly to the thermal environmental effects and a fast feedback loop (< 100 ms) corrects all other electronic time-drifts.

Von der Linde developed an approach to measure the time jitter of a mode-locked femtosecond laser oscillator into the femtosecond range [9.114]. This quantitative timing jitter is obtained first by measuring the RF spectral intensity of the $n^{th}$-harmonic of the laser's repetition rate, then by integrating the single sideband phase-noise from 0.1 Hz to 100 kHz from the $n^{th}$-harmonic of the laser's repetition rate (Figure **9.26**). Using this approach, a timing jitter approaching 10 fs was measured.

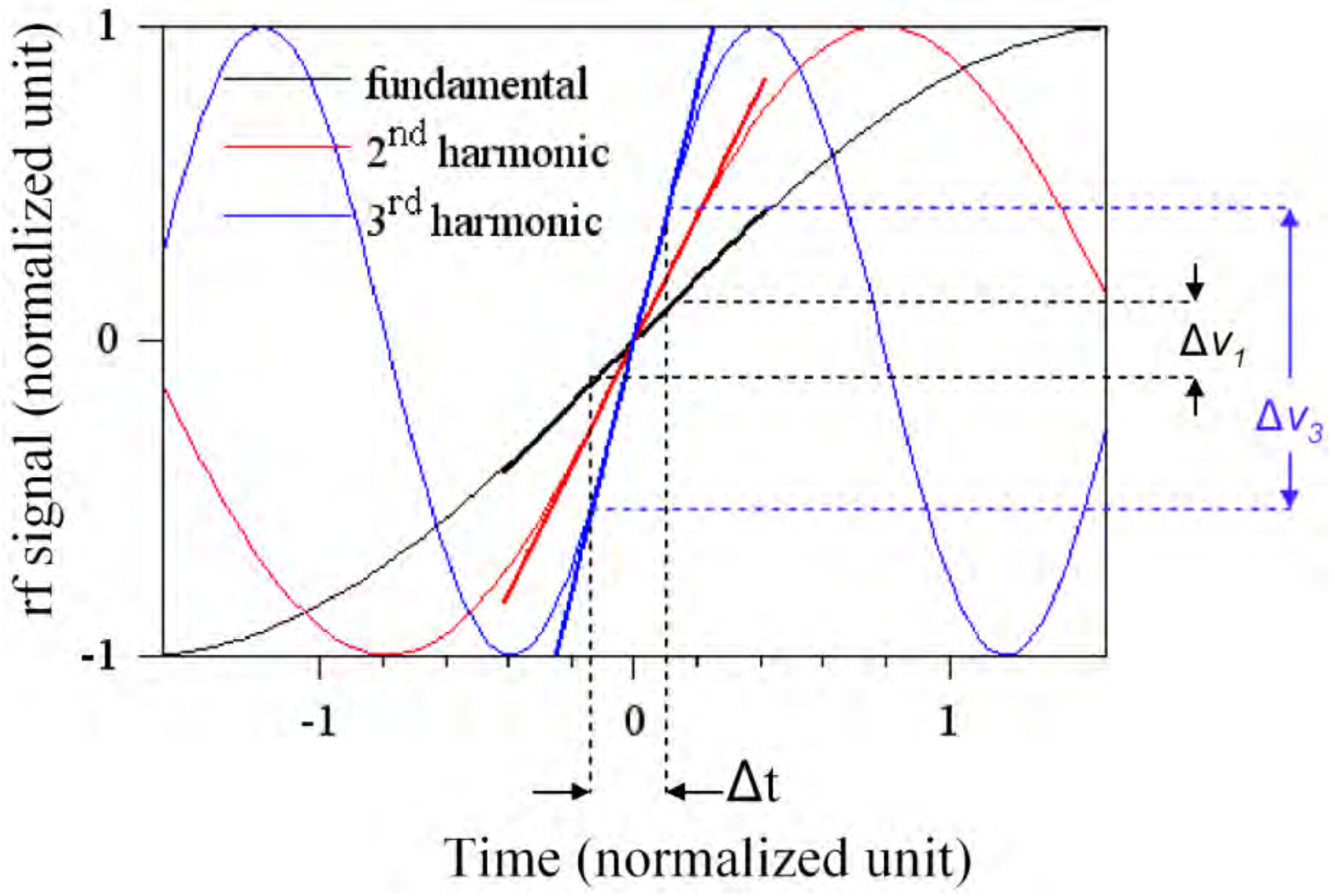

**Figure 9.25. Improvement of sensitivity with higher RF harmonic signal. The slopes of the 1st, 2nd and 3rd harmonics are highlighted about time $\Delta t$ = 0.**

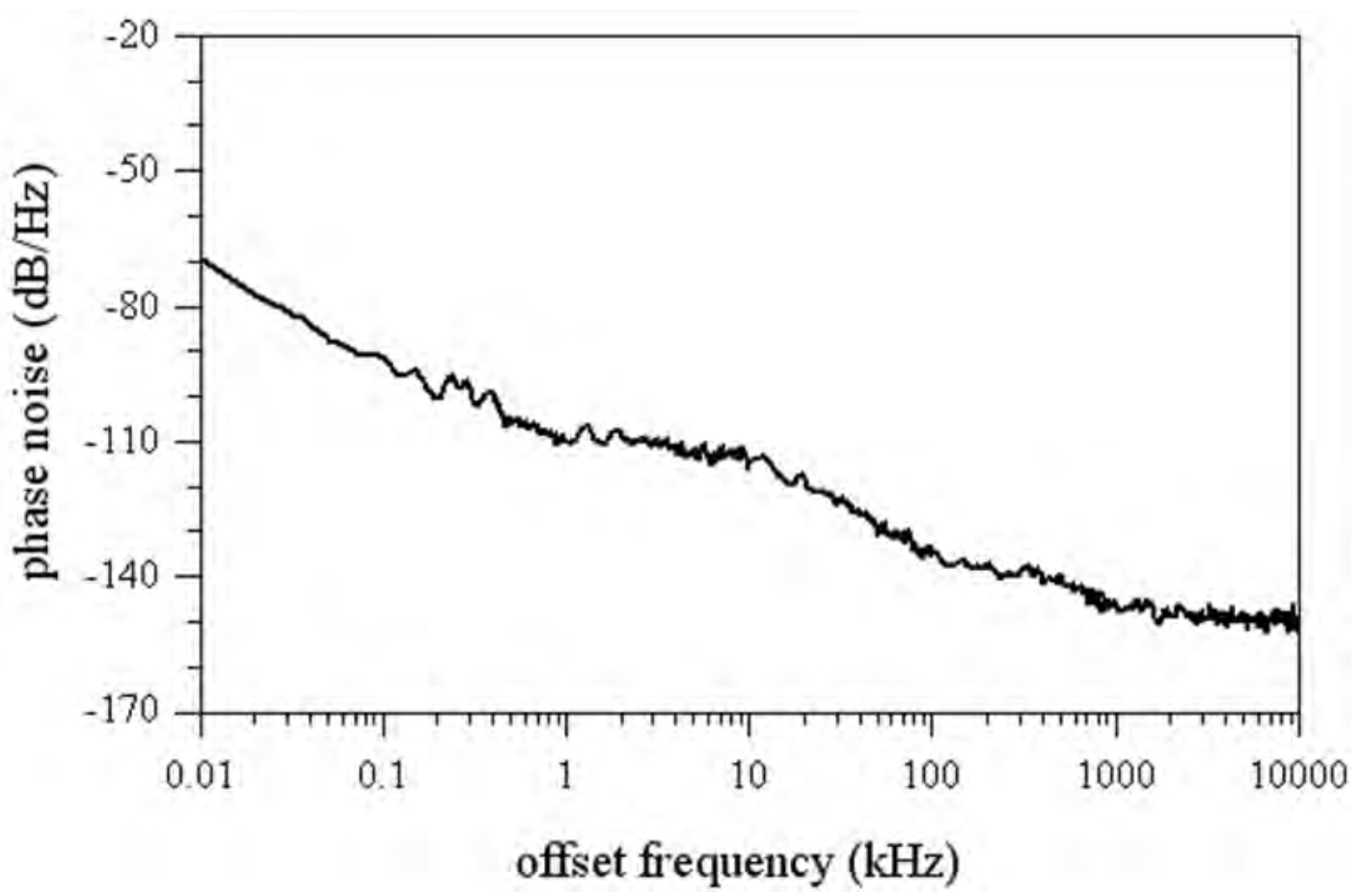

**Figure 9.26. Representative trace of the spectral intensity of the nth-harmonic of the laser repetition rate of an RF oscillator.**

Timing jitter does not degrade much when the seed pulse propagates through the chain of amplifiers operating in a stable environment. However, a larger timing jitter arises when the short laser pulse is delivered to different parts of a photoinjector facility that often are separated by 10-100 m and have large differences in temperature and environment. To deliver a femtosecond timing-link using a stabilized clock, Kärtner developed a bunch arrival time (BAT) fiber-link-timing synchronization unit [9.115]. It is a self-aligned, balanced, cross-correlator with no moving parts based on a type-II $KTiOPO_4$ optical-harmonic generation crystal, wherein an outgoing and a returned light pulse are cross-correlated through a long fiber link to generate a time-error signal and to close a feedback loop. Long-term timing stability reaching 10 fs precision has been established. These ultrastable timing levels of synchronization were achieved at LCLS after deploying these bunch arrival time (BAT) units.

## 9.7 CONFLICT OF INTEREST AND ACKNOWLEDGEMENT

We confirm that this article content has no conflicts of interest and would like to acknowledge the support of the U. S. Department of Energy under contract number DE-AC02-98CH10886.

***References***

[9.1] D. H. Dowell, F. K. King, R. E. Kirby *et al.*, "*In situ* cleaning of metal cathodes using a hydrogen ion beam," *Phys. Rev. ST Accel. Beams*, vol. 9, pp. 063502-1–063502-8, June, 2006.

[9.2] X. J. Wang, T. Srinivasan Rao, K. Batchelor *et al.*, "Measurements on photoelectrons from a magnesium cathode in a microwave electron gun," *Nucl. Instrum. Meth. A*, vol. 356, pp. 159-166, March 1995.

[9.3] J. Smedley, T. Rao and J. Sekutowicz, "Lead photocathodes," *Phys. Rev. ST Accel. Beams*, vol. 11, pp. 013502-1–0135022-9, January 2008.

[9.4] G. Suberlucq, "Technological challenges for high brightness photo-injectors," in *Proc. 2004 European Particle Accelerator Conf.*, 2004, pp. 64-68.

[9.5] D. H. Dowell, S. Z. Bethel and K. D. Friddell, "Results from the average power laser experiment photocathode injector test," *Nucl. Instrum. Meth. A*, vol. 356, pp. 167-176, March 1995.

[9.6] T. Rao, A. Burrill, X. Y. Chang *et al.*, "Photocathodes for energy recovery linacs," *Nucl. Instrum. Meth. A*, vol. 557, pp. 124-130, February 2006.

[9.7] C. Limborg-Deprey and P. R. Bolton, "Optimum electron distributions for space charge dominated beams in Photoinjectors," *Nucl. Instrum. Meth. A*, vol. 557, pp. 106-116, February 2006.

[9.8] O. J. Luiten, B. van der Geer, M. de Loos *et al.*, "Ideal waterbag electron bunches from an RF photogun," in *Proc. 2004 European Particle Accelerator Conf.*, 2004, pp. 725-727.

[9.9] K. Y. Lau, "Gain switching of semiconductor injection lasers," *Appl. Physics Lett.*, vol. 52, pp. 257-259, January 1988.

[9.10] M. Poelker, "High power gain-switched diode laser master oscillator and amplifier," *Appl. Physics Lett.*, vol. 67, pp. 2762-2764, November 1995.

[9.11] J. Hansknecht and M. Poelker, "Synchronous photoinjection using a frequency-doubled gain-switched fiber-coupled seed laser and ErYb-doped fiber amplifier," *Phys. Rev. ST Accel. Beams*, vol. 9, pp. 063501-1–063501-5, June 2006.

[9.12] J. M. Harris, R. W. Chrismlan and F. E. Lytle, "Pulse generation in a cw dye laser by mode-locked synchronous pumping," *Appl. Physics Lett.*, vol. 26, pp. 16-18, January 1975.

[9.13] D. L. MacFarlane and L. W. Casperson, "Pump pulse effects in synchronously pumped mode-locked dye lasers," *J. Optics Soc. America B*, vol. 6 pp. 292-299, March 1989.

[9.14] W. Koechner, *Solid-State Laser Engineering*, Berlin: Springer-Verlag, 1988, Chapter 8.

[9.15] F. J. McClung and R. W. Hellwarth, "Giant optical pulsations from ruby," *J. Appl. Physics*, vol. 33, pp. 828-829, March 1962.

[9.16] I. P. Kaminow and E. H. Turner, "Electrooptic light modulators," *Appl. Optics*, vol. 5, pp. 1612-1628, October 1966.

[9.17] C. L. Hu, "Linear electro-optic retardation schemes for the twenty classes of linear electro-optic crystals and their applications," *J. Appl. Physics*, vol. 38, pp. 3275-3284, July 1967.

[9.18] J. J. Zayhowski and C. Dill III, "Coupled-cavity electro-optically Q-switched Nd:$YVO_4$ microchip lasers," *Optics Lett.*, vol. 20, pp. 716-718, April 1995.

[9.19] J. Dong, A. Shirakawa, K. Takaichi *et al.*, "All-ceramic passively Q-switched Yb:YAG/$Cr^{4+}$:YAG microchip laser," *Electronics Lett.*, vol. 42, pp. 1154-1155, September 2006.

[9.20] G. J. Spuhler, L. Gallmann, R. Fluck *et al.*, "Passively modelocked diode-pumped erbium-ytterbium glass laser using a semiconductor saturable absorber mirror," *Electronics Lett.*, vol. 35, pp. 567-569, April 1999.

[9.21] B. H. Soffer, "Giant pulse laser operation by a passive, reversibly bleachable absorber," *J. Appl. Physics*, vol. 35, pp. 2551, 1964.

[9.22] P. Kafalas, J. I. Masters and E. M. E. Murray, "Photosensitive liquid used as a nondestructive passive Q-switch in a ruby laser," *J. Appl. Physics*, vol. 35, pp. 2349-2350, August 1964.

[9.23] PennWell 1421 S. Sheridan Road, Tulsa, OK 74112. PennWell Buyer's Guide [online]. Available: http://buyersguide.lfw.optoiq.com

[9.24] D. Mayden and R. B. Chesler, "Q-switching and cavity dumping of Nd:YAlG lasers," *J. Appl. Physics*, vol. 42, pp. 1031-1034, March 1971.

[9.25] H. Kruegle and L. Klein, "High peak power output, high PRF by cavity dumping a Nd:YAG laser," *Appl. Optics*, vol. 15, pp. 466-471.

[9.26] W. Koechner, *Solid-State Laser Engineering*, Berlin: Springer-Verlag, 1988, Chapter 9.

[9.27] A. J. De Maria, D. A. Stetser and H. Heynau, "Self mode-locking of lasers with saturable absorbers," *Appl. Physics Lett.*, vol. 8, pp. 174-176, April 1966.

[9.28] D. von der Linde, "Experimental study of single picosecond light pulses," *IEEE J. Quantum Electron.*, vol. QE-8, pp. 328-338, March 1972.

[9.29] D. Kopf, F. X. Kartner, U. Keller *et al.*, "Diode-pumped mode-locked Nd:glass lasers with an antiresonant Fabry–Perot saturable absorber," *Optics Lett.*, vol. 20, pp. 1169-1171, May 1995.

[9.30] D. J. Kuizenga and A. E. Seigman, "FM and AM mode locking of the homogeneous laser-part II: Experimental Results in a Nd:YAG laser with internal FM modulation," *IEEE J. Quantum Electron.*, vol. QE-6, pp. 709-715, November 1970.

[9.31] U. Keller, J. A. Valdmanis, M. C. Nuss *et al.*, "53 ps pulses at 1.32 μm from a harmonic mode-locked Nd:YAG laser," *IEEE J. Quantum Electron.*, vol. QE-24, pp. 427-430, February 1988.

[9.32] R. Paschotta. *Regenerative Amplifiers* [Online]. Available: http://www.rp-photonics.com/regenerative_amplifiers.html.

[9.33] D. Strickland and G. Mourou, "Compression of amplified chirped optical pulses," *Optics Commun.*, vol. 56, pp. 219-221, December 1985.

[9.34] C. Hoenninger, E. Mottay and M. Delaigue, "High Repetition Rate Ultrashort Pulse Picosecond Laser Amplifier," in *Proc. 2007 Int. Congr. Applications Lasers Electro-Optics*, 2007, Paper #M1207.

[9.35] J. L. Miller, L. A. Hackel, C. B. Dane *et al.*, "High power regenerative laser amplifier," U.S. Patent 5,285,310, February 8, 1994.

[9.36] A. Dergachev, M. A. Yakshin, P. F. Moulton *et al.*, *High-average-power picosecond drive source for photocathode injectors* [Online]. Available: http://www.qpeak.com/Meetings/CLEO-2005-PS-MOPA-d.pdf

[9.37] M. Perry. *The Amazing Power of the Petawatt* [Online]. Available: https://www.llnl.gov/str//MPerry.html

[9.38] *Vulcan Home Page* [Online]. Available: http://www.clf.rl.ac.uk/Facilities/Vulcan/12248.aspx

[9.39] X. Wei, Q. Zhu, X. Huang *et al.*, "Recent progress and future prospects of high-energy petawatt laser in LFRC, CAEP," in *Journal Physics: Conf. Series*, vol. 112, 2008, pp. 032011.

[9.40] E. Wolf, Ed., *Progress in Optics*, Vol. 34, Amsterdam: Elsevier Science, 2004.

[9.41] Lumera Laser GmbH, Opelstr. 10, 67661 Kaiserslautern, Germany.

[9.42] S. Schreiber, I. Will, D. Sertore *et al.*, "Running experience with the laser system for the RF gun based injector at the TESLA Test Facility linac," *Nucl. Instrum. Meth. A*, vol. 445, pp. 427-431, May 2000.

[9.43] I. Will, G. Kross and I. Templin, "The upgraded photocathode laser of the TESLA Test Facility," *Nucl. Instrum. Meth. A*, vol. 541, pp. 467-477, April 2005.

[9.44] V. Yakimenko. (2007, December 7). *Multi-Pulse Nd:YAG Laser Schematic Layout* [Online]. Available: http://www.bnl.gov/atf/core_capabilities/yaglayout.asp.

[9.45] M. Shinn. (2005, March 22). *Drive laser state-of-the-art: Performance, stability and programmable repetition rate* [Online]. Available: http://www.jlab.org/intralab/calendar/archive04/erl/talks/WG1/WG1_Shinn_Tue_0830.pdf

[9.46] A. Sakumi, "Synchronization between laser and electron beam at photocathode RF gun," in *Proc. 2005 Particle Accelerator Conf.*, 2005, pp. 3079-3081.
[9.47] P. F. Moulton, "Pulse-pumped operation of divalent transition-metal lasers," *IEEE J. Quantum Electron.*, vol. QE-18, pp. 1185-1188, August 1982.
[9.48] T. Brabec, Ch. Speilmann, P. F. Curley *et al.*, "Kerr lens mode locking," *Optics Lett.*, vol. 17, pp. 1292-1294, September 1992.
[9.49] L. Y. Liu, J. M. Huxley, E. P. Ippen *et al.*, "Self-starting additive-pulse mode locking of a Nd:YAG laser," *Optics Lett.*, vol. 15, pp. 553-555, May 1990.
[9.50] J. F. Wishart, A. R. Cook and J. R. Miller, "The LEAF picosecond pulse radialysis facility at Brookhaven National Laboratory," *Rev. Sci. Instrum.*, vol. 75, pp. 4359-4366, November 2004.
[9.51] I. Wilke, A. M. MacLeod, W. A. Gillespie *et al.*, "Single-shot electron-beam bunch length measurements," *Phys. Rev. Lett.*, vol. 88, pp. 124801-1–124801-4, March 2002.
[9.52] E. Fill, L. Veisz, A. Apolonski *et al.*, "Sub-fs electron pulses for ultrafast electron diffraction," *New J. Physics*, vol. 8, pp. 272-1–272-11, November 2006.
[9.53] T. Tsang, "Multiple-harmonic generation on ZnO nanocrystalline film," in *Tech. Dig, Lasers Electro-Optics/Quantum Electronics Laser Science and Photonic Applications Systems Technologies Conf.*, 2005, pp. 2103-2104.
[9.54] O. E. Martinez, "3000 times grating compressor with positive group velocity dispersion: Application to fiber compensation in 1.3-1.6 μm region," *IEEE J. Quantum Electron.*, vol. 23, pp. 59-64, January 1987.
[9.55] E. B. Treacy, "Optical pulse compression with diffraction gratings," *IEEE J. Quantum Electron.*, vol. 5, pp. 454-458, September 1969.
[9.56] A. Offner, "Unit power imaging catoptric anastigmat," U.S. Patent 3,748,015, June 21, 1971.
[9.57] G. Cheriaux, P. Rousseau, F. Salin *et al.*, "Aberration-free stretcher design for ultrashort-pulse amplification," *Optics Lett.*, vol. 21, pp. 414-416, March 1996.
[9.58] E. Miesak and R. Negres, "Alignment procedure for a dual grating pulse compressor," *Appl. Optics*, vol. 37, pp. 8146-8147, 1998.
[9.59] A. V. Tarasishin, S. A. Magnitskii and A. M. Zheltikov, "Matching phase and group velocities in second-harmonic generation in finite one-dimesnsional photonic band-gap structures," *Laser Physics*, vol. 11, pp. 31-38, 2001.
[9.60] B. A. Richman, S. E. Bisson, R. Trebino *et al.*, "Recent advances in achromatic phase matching for tunable and ultrashort second-harmonic generations," in *Tech. Dig. 1998 Conf. Lasers Electro-Optics*, 1998, pp. 105.
[9.61] P. Wasylczyk, I. A. Walmsley, W. Wasilewski *et al.*, "Broadband noncollinear optical parametric amplifier using a single crystal," *Optics Lett.*, vol. 30, pp. 1704-1706, July 2005.
[9.62] R. Akre, J. Castro, Y. Ding *et al.*, "Initial commissioning experience with the LCLS injector," in *Proc. 2007 Particle Accelerator Conf.*, 2007, pp. 1302-1304.
[9.63] J. M. Byrd, L. Doolittle, G. Huang *et al.*, "Femtosecond synchronization of laser system for the LCLS," in *Proc. 2010 Int. Particle Accelerator Conf.*, 2010, pp. 58-60.
[9.64] SPARC General Info [Online]. Available: http://www.lnf.infn.it/acceleratori/sparc/
[9.65] P. Musumeci. (2009, November). *Ultrashort laser pulses on the cathode: blow-out regime and multiphoton photoemission* [Online]. Available: http://pbpl.physics.ucla.edu/HBEB/PlenaryTalks/MUSUMECI.pdf
[9.66] P. Musumeci, L. Cultrera, M. Ferrario *et al.*, "Multiphoton photoemission from a copper cathode illuminated by ultrashort laser pulses in an rf photoinjector," *Phys. Rev. Lett.*, vol. 104, pp. 084801-1–084801-4, February 2010.

[9.67] D. A. Grukh, V. A. Bogatyrev, A. A. Sysolyatin *et al.*, “Broadband radiation source based on ytterbium-doped fibre with fibre-length-distributed pumping,” *Quantum Electron.*, vol. 34, pp. 247-248, 2004.

[9.68] A. Bellemare, M. Karásek, C. Riviere *et al.*, “A broadly tunable Erbium-doped fiber ring laser: Experimentation and modeling,” *IEEE J. Selected Topics Quantum Electron.*, vol. 7, pp. 22-29, January 2001.

[9.69] A. Tünnermann, T. Schreiber and J. Limpert, “Fiber lasers and amplifiers: An ultrafast performance evolution,” *Appl. Optics*, vol. 49, pp. F71-F78, September 2010.

[9.70] M. Nakazawa, Y. Kimura and K. Suzuki, “An ultra-efficient, erbium-doped fiber amplifier of 10.2 dB/mW at 0.98 μm pumping and 5.1 dB/mW at 1.48 μm pumping,” in *Proc. Topical Meeting Optical Amplifiers Applications*, 1990, pp. PDP1-1–PDP1-4.

[9.71] A. Liu and K. Ueda, “The absorption characteristics of circular, offset, and rectangular double-clad fibers,” *Optics Commun.*, vol. 132, pp. 511-518, December 1996.

[9.72] J. P. Koplow, D. D. A. V. Kliner *et al.*, “Single-mode operation of a coiled multimode amplifier,” *Optics Lett.*, vol. 25, pp. 442-444, April 2000.

[9.73] J. Limpert, A. Liem, M. Reich *et al.*, “Low-nonlinearity single-transverse-mode ytterbium-doped photonic crystal fiber amplifier,” *Optics Express*, vol. 12, pp. 1313-1319, April 2004.

[9.74] D. G. Ouzounov, I. V. Bazarov, B. M. Dunham *et al.*, “The laser system for the ERL electron source at Cornell University,” in *Proc. 2007 Particle Accelerator Conf.*, 2007, pp. 530-532.

[9.75] P. Madasamy, L. Coressel, D. R. Jander *et al.*, “Tunable pulse width, short pulse high power green laser,” presented at Conf. Lasers Electro-Optics, 2010.

[9.76] M. Nakazawa and E. Yoshida, “A 40-GHz 850-fs regeneratively FM mode-locked polarization-maintaining erbium doped fiber ring laser,” *IEEE Photonics Tech. Lett.*, vol. 12, pp. 1613-1615, December 2000.

[9.77] A. B. Rulkov, M. Y. Vyatkin, S. V. Popov *et al.*, “High brightness picosecond all-fiber generation in 525-1800 nm range with picosecond Yb pumping,” *Optics Express*, vol. 13, pp. 377-381, January 2005.

[9.78] M. H. Ober, M. Hofer, U. Keller *et al.*, “Self-starting diode-pumped femtosecond Nd fiber laser,” *Optics Lett.*, vol. 18, pp. 1532-1534, September 1993.

[9.79] H. Schlüter, C. Tillkorn, U. Bonna *et al.*, “Dense spatial multiplexing enables high brightness multi-kW diode laser systems,” in *Proc. SPIE – High-power Diode Laser Technology Applications IV*, 2006, vol. 6104, pp. 61040M-1–61040M-8.

[9.80] M. Ziad, M. Al-Haiari and W. Khalid. (2006). *Optical Caples* [Online]. Available: FTP: http://dar.ju.edu.jo Directory: mansour/optical/729projects/ File: fiber connections.doc

[9.81] C. Limborg-Deprey and H. Tomizawa, “Maximizing brightness in photoinjector,” in *Proc. 46$^{th}$ Physics Applications of High Brightness Electron Beams Workshop*, 2007, pp. 174-193.

[9.82] J. Luiten, B. van der Geer, M. de Loos *et al.*, “Ideal waterbag electron bunches from an RF photogun,” in *Proc 2004 European Particle Accelerator Conf.*, 2004, pp. 725-727.

[9.83] Y. Li and X. Chang, “Generation of ellipsoidal beam through 3D pulse shaping for a photoinjector drive laser,” in *Proc. 2006 Int. Linear Accelerator Conf.*, 2006, pp. 776-778.

[9.84] O. J. Luiten, S. B. van der Geer, M. J. de Loos *et al.*, “How to realize uniform three-dimensional ellipsoidal electron bunches,” *Phys. Rev. Lett.*, vol. 93, pp. 094802-1–094802-4, August 2004.

[9.85] P. Musumeci, J. T. Moody, R. J. England *et al.*, “Experimental generation and characterization of uniformly filled ellipsoidal electron-beam distributions,” *Phys. Rev. Lett.*, vol. 100, pp. 244801-1–244801-4, June 2008.

[9.86] H. Tomizawa, H. Dewa, T. Taniuchi *et al.*, “Adaptive shaping system for both spatial and temporal profiles of a highly stabilized UV laser light source for a photocathode RF gun,” *Nucl. Instrum. Meth A*, vol. 557, pp. 117-123, February 2006.

[9.87] I. Will and G. Klemz, “Generation of flat-top picosecond pulses by coherent pulse stacking in a multicrystal birefringent filter,” *Optics Express*, vol. 16, pp. 14922-14937, September 2008.

[9.88] C. Limborg-Deprey and H. Tomizawa, “Maximizing brightness in photoinjectors,” *Int. J. Modern Physics A*, vol. 22, pp. 3864-3881, 2007.

[9.89] D. Garzella, O. Gobert, Ph. Hollander *et al.*, “Temporal analysis and shape control of UV high energy laser pulses for photoinjectors,” in *Proc. 2006 Free Electron Laser*, 2006, p. 552-555.

[9.90] A. Yariv and P. Yeh, *Optical Waves in Crystals: Propagation and Control of Laser Radiation*, New York: John Wiley & Sons, 1984.

[9.91] A. M. Weiner, “Femtosecond pulse shaping using spatial light modulators,” *Rev. Sci. Instrum.*, vol. 71, pp. 1929-1960, May 2000.

[9.92] A. M. Weiner, J. P. Heritage and E. M. Kirschner, “High-resolution femtosecond pulse shaping,” *J. Optics Soc. America B*, vol. 5, pp. 1563-1572, August 1988.

[9.93] S. Cialdi, C. Vicario, M. Petrarca *et al.*, “Simple scheme for ultraviolet time-pulse shaping,” *Appl. Optics*, vol. 46, pp. 4959-4962, August 2007.

[9.94] J. Yang, F. Sakai, T. Yanagida *et al.*, “Low-emittance electron-beam generation with laser pulse shaping in photocathode radio-frequency gun,” *J. Appl. Physics*, vol. 92, pp. 1608-1612, August 2002.

[9.95] R. Akre, D. Dowell, P. Emma *et al.*, “Commissioning the Linac Coherent Light Source injector,” *Phys. Rev. ST Accel. Beams*, vol. 11, pp. 30703-1–30703-20, March 2008.

[9.96] C. Vicario, A. Ghigo, M. Petrarca *et al.*, “Laser temporal pulse shaping experiment for SPARC photoinjector,” in *Proc. 2004 European Particle Accelerator Conf.*, 2004, pp. 1300-1302.

[9.97] H. Loos, M. Boscolo, D. Dowell *et al.*, “Temporal E-beam shaping in an S-band accelerator,” in *Proc. 2005 Particle Accelerator Conf.*, 2005, pp. 642-644.

[9.98] C. W. Siders, J. L. W. Siders, A. J. Taylor *et al.*, “Efficient high-energy pulse-train generation using a $2^n$-pulse Michelson interferometer,” *Appl. Optics*, vol. 37, pp. 5302-5305, August 1998.

[9.99] S. Zhou, D. Ouzounov, H. Li *et al.*, “Efficient temporal shaping of ultrashort pulses with birefringent crystals,” *Appl. Optics*, vol. 46, pp. 8488-8492, December 2007.

[9.100] I. V. Bazarov, D. G. Ouzounov, B. M. Dunham *et al.*, “Efficient temporal shaping of electron distributions for high-brightness photoemission electron guns,” *Phys. Rev. ST Accel. Beams*, vol. 11, pp. 40702-1–040702-6, April 2008.

[9.101] A. K. Sharma, T. Tsang and T. Rao, “Theoretical and experimental study of passive spatiotemporal shaping of picosecond laser pulses,” *Phys. Rev. ST Accel. Beams*, vol. 12, pp. 33501-1–33501-9, March 2009.

[9.102] I. Will, “Generation of flat-top picosecond pulses by means of a two-stage birefringent filter,” *Nucl. Instrum. Meth. A*, vol., pp. 119-125, July 2008.

[9.103] Y. Park, M. H. Asghari, T. J. Ahn *et al.*, “Transform-limited picosecond pulse shaping based on temporal coherence synthesization,” *Optical Express*, vol. 15, pp. 9584-9599, July 2007.

[9.104] S. P. Chang, J. Kuo, Y. Lee *et al.*, “Transformation of Gaussian to coherent uniform beams by inverse-Gaussian transmittive filters,” *Appl. Optics*, vol. 37, pp. 747-752, February 1998.

[9.105] M. Yun, M. Wang, Q. Wang *et al.*, “Laser beam shaping system with a radial birefringent filter,” *J. Modern Optics*, vol. 54, pp. 129-136, January 2007.

[9.106] X. Tan, B. Gu, G. Yang *et al.*, “Diffractive phase elements for beam shaping: A new design method,” *Appl. Optics*, vol. 34, pp. 1314-1320, March 1995.

[9.107] P. W. Rhodes and D. L. Shealy, "Refractive optical systems for irradiance redistribution of collimated radiation: Their design and analysis," *Appl. Optics*, vol. 19, pp. 3545-3553, October 1980.

[9.108] J. A. Hoffnagle and C. M. Jefferson, "Design and performance of a refractive optical system that converts a Gaussian to a flattop beam," *Appl. Optics*, vol. 39, pp. 5488-5499, October 2000.

[9.109] T. Takaoka, N. Kawano, Y. Awatsuji *et al.*, "Design of a reflective aspherical surface of a compact beam-shaping device," *Optical Rev.*, vol. 13, pp. 77-86, March 2006.

[9.110] Y. Li and J. W. Lewellen, "Generating a quasiellipsoidal electron beam by 3D laser-pulse shaping," *Phys. Rev. Lett.*, vol. 100, pp. 074801-1–074801-4, February 2008.

[9.111] S. Zhang, S. Benson, J. Gubeli *et al.*, "Investigation and evaluation on pulse stackers for temporal shaping of laser pulses," in *Proc. 32nd Int. Free Electron Linac Conf.*, 2010, pp. 394-397.

[9.112] Y. Li, S. Chemerisov and J. Lewellen, "Laser pulse shaping for generating uniform three-dimensional ellipsoidal electron beams," *Phys. Rev. ST Accel. Beams*, vol. 12, pp. 020702-1–020702-11, February 2009.

[9.113] M. Bellaveglia, A. Gallo and C. Vicario, "SPARC photo-injector synchronization system and time jitter measurement," LNF of INFN, Roma, Italy, EUROFEL-Report-2006-DS3-027, 2006.

[9.114] D. von der Linde, "Characterization of the noise in continuously operating mode-locked lasers," *Appl. Physics B*, vol. 39, pp. 201-217, April 1986.

[9.115] J. Kim, J. Chen, Z. Zhang *et al.*, "Long-term femtosecond timing link stabilization using a single-crystal balanced cross correlator," *Optics Lett.*, vol. 32, pp. 1044-1046, May 2007.

# CHAPTER 10: RF SYSTEMS

WOLFGANG ANDERS

*Helmholtz-Zentrum Berlin for Materials and Energy*

*Albert-Einstein-Str. 15, D-12489*

*Berlin, Germany*

AXEL NEUMANN

*SRF Science and Technology/BESSY II*

*Helmholtz-Zentrum Berlin for Materials and Energy*

*Albert-Einstein-Str. 15, D-12489*

*Berlin, Germany*

**Keywords**

RF Transmitters, Tetrode, Klystron, Solid State RF Amplifier, Modulators, Transmission Line, Circulator, Low-level RF, RF Control System, Fundamental Power Coupler, Waveguide Coupler, Coaxial Coupler, Higher-Order Mode Coupler

**Abstract**

The RF system is responsible for feeding adequate power to the injector with required phase and amplitude stability. The low level RF system provides the input for the high power amplifier and controls the phase and amplitude *via* feedback and feedforward loops. The high power RF amplifies the milliwatt level input to kilowatt/megawatt level. The first step in developing the high power RF system is to establish the power level and the tolerances required. In this chapter, we discuss the methods to determine these values followed by a description of the essential components, namely, the preamplifier, high power amplifier, transmission line, fundamental power couplers and higher order mode couplers. We describe major errors sources for field stability of the RF cavities such as frequency shift due to temperature fluctuation, Lorentz detuning, micro-phonics. We provide a sequence of steps to design and operate the LLRF a generic system with examples for controlling the LLRF.

This chapter is subdivided into six sections: After an introduction, we discuss the essential aspects of the high power RF transmitters and transmission lines, the low-level RF (LLRF) system and the issues related to the fundamental- and higher order mode-couplers.

## 10.1 OVERVIEW OF APPLICATIONS AND REQUIREMENTS

The high-power RF system of a photoinjector is responsible for feeding adequate power with the required stability into the accelerating RF cavity. It amplifies a signal in the milliwatt power range from the master oscillator to a typical power level, from kilowatts to a few megawatts, which is fed to the fundamental power coupler of the accelerating cavity. Its main components are the preamplifier, the tube or solid-state-based high power amplifier, with the corresponding power supplies and the transmission line, including a circulator. Depending upon its application, the RF system is designed for continuous wave (CW) or in pulsed operation mode.

The LLRF system is responsible for controlling the amplitude and phase *via* feedback loops and provides the input signal for the high power RF system. The feedback loops are realized either in analog technology or in digital technology based on field programmable gate-array processors.

An overview of fundamental- and higher order mode- (HOM) couplers for the RF cavities will complete this chapter. Fundamental power couplers (FPC) transmit high power levels from the transmission line into the cavity cell that is under vacuum. HOM couplers also damp unwanted frequency modes excited by the beam in the cavity cell. Fundamental mode and HOM coupler technologies differ significantly for normal conducting- (NC) and superconducting- (SC) cavities.

**Power Requirements**

The application of an RF photoinjector determines the main design parameters of the corresponding RF cavity. The decision between normal or superconducting and between the pulsed or CW mode of operation determines the field-control strategy and also the coupling strength of the supplied forward power wave to the cavity. The coupling strength, in the absence of the beam, given by the ratio between external and intrinsic quality factor, $Q_{ext}$ and $Q_0$ respectively, finally gives the half-bandwidth $f_{½}$ of the system expressed

$$f_{½} = \frac{f_0}{2Q_L} \tag{10.1}$$

with

$$Q_L = \left(\frac{1}{Q_{ext}} + \frac{1}{Q_0}\right)^{-1} \tag{10.2}$$

and,

$$Q_{ext} = \frac{\omega_0 U}{P_{ext}} \tag{10.3}$$

where $Q_L$ is the loaded quality factor, $f_0$ is the cavity's resonance frequency, and $\omega_0$ the cavity's angular resonance frequency. $Q_{ext}$ is the ratio of stored energy in a cavity, $U$, to the power $P_{ext}$ leaking through the coupler in every RF period, $\omega_0$. Three factors drive the choice of $Q_{ext}$: The amount of beam current to be accelerated; the expected cavity detuning; and, the available power. The required power to establish a given accelerating field, $V_{cav}$, at a given $Q_L$, the geometric shunt impedance ($R/Q$), the beam current $I_b$ and accelerating phase $\Phi_b$ is given by

$$P_f = \frac{V_{cav}^2}{4(R/Q)Q_L} * \left[\left(1 + \frac{(R/Q)Q_L I_b}{V_{cav}}\cos(\Phi_b)\right)^2 + \left(\frac{\Delta f}{f_{½}} + \frac{(R/Q)Q_L I_b}{V_{cav}}\sin(\Phi_b)\right)^2\right] \tag{10.4}$$

Besides the beam loading, power is determined by the cavity's detuning, $\Delta f$, over the cavity's half-bandwidth, $f_{½}$. This relationship is of major concern for narrow-band superconducting cavities wherein the amount of detuning, especially for pulsed operation, easily exceeds the cavity's bandwidth. On the other hand, the selection of the coupling impacts the field's stability.

The field's phase stability, $\sigma_\Phi$, is given by the cavity's frequency detuning, $\sigma_f$, and $f_{½}$

$$\sigma_\Phi = \arctan\left(\frac{\sigma_f}{f_{½}}\right) \tag{10.5}$$

Figure **10.1** shows the required RF power due to microphonic noise. We depict the RF power versus the cavity's bandwidth for different microphonic noise levels for a SC cavity used in a low current application with 2 kW beam loading. The RF power required for stable operation is a multiple of the beam loading. Piezo actuator-based fast cavity tuning loops can compensate for part of the microphonic detuning in attempting to minimize the RF power of the installation, but it is challenging to design these loops [10.1].

When specifying the overhead transmitter power, we also must take into account the losses in the transmission line and in the circulator, in addition to losses in the cavity's walls, beam loading and microphonic needs.

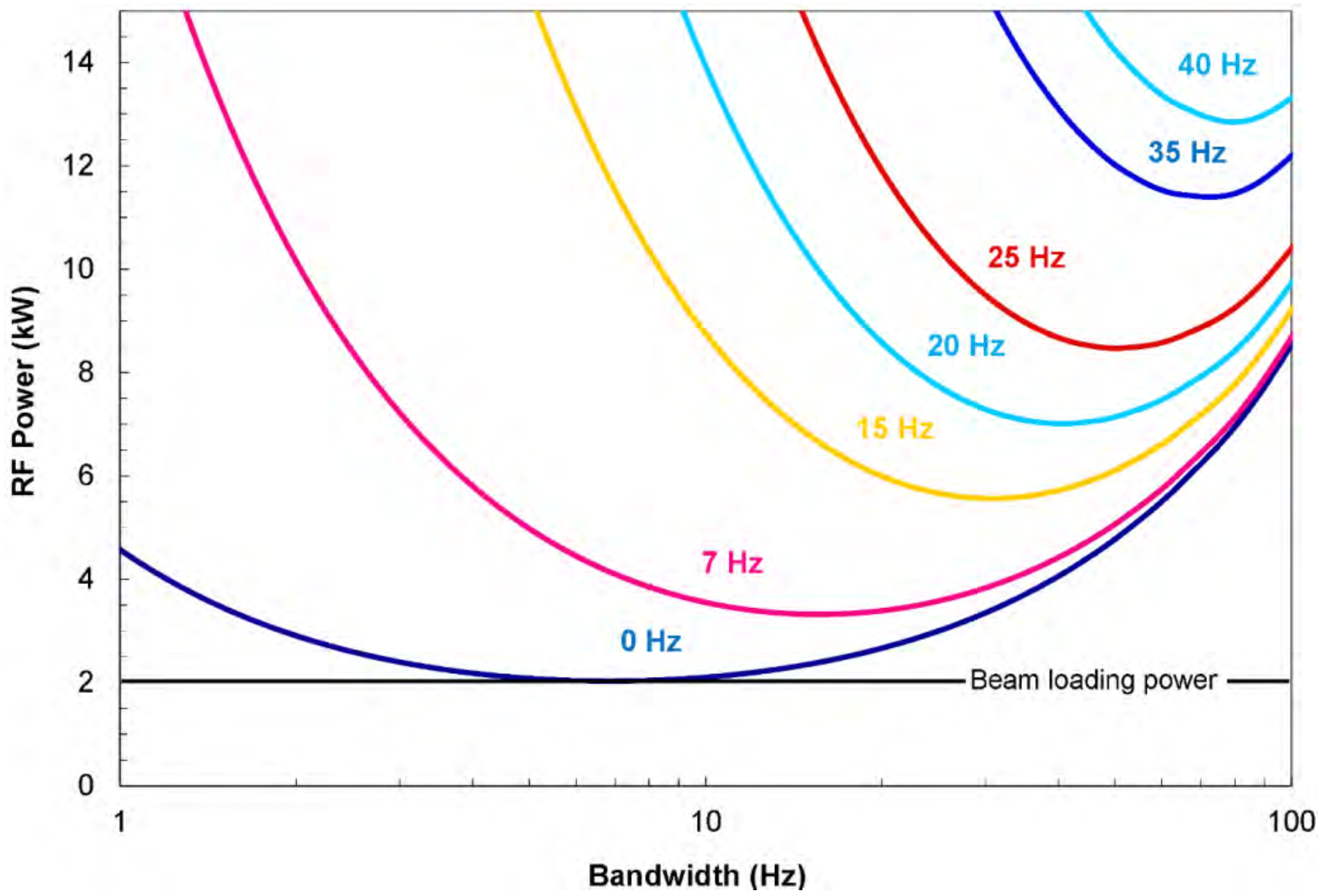

**Figure 10.1. Required RF power to compensate for microphonics for a TESLA cavity operating at 20 MV m$^{-1}$. Depicted is the RF power versus the cavity bandwidth for different microphonic noise levels a system with 2 kW beam loading. The RF power required for stable operation is significantly more than that for the beam loading. [Courtesy of J. Knobloch]**

Three different categories define the power level required in photoinjector application:

- CW or pulsed operation
- normal conducting or superconducting cavity
- beam currents below or above 1 mA

CW operation is used for the low energy, superconducting free electron laser (FEL) [10.2] and the Energy Recovery Linac (ERL) [10.3], while pulsed operation is chosen for applications with normal conducting linac structures [10.4], [10.5]. Pulsed high energy FEL accelerators with a superconducting linac [10.6] nowadays are equipped with a NC pulsed gun cavity [10.7]. Because gun cavities are operated with high accelerating gradients to preserve the emittance, normal conducting gun cavities require RF power at megawatt levels, as achieved in pulsed operation. Superconducting gun cavities generally are operated in the CW mode. The wall losses in superconducting cavities amount to a mere few watts. Beam loading and microphonic detuning define the power level of the transmitter. Special injectors for high current ERLs with beam currents of 100 mA to 1 A need transmitters with power levels over 100 kW [10.3], [10.8], [10.9]. This breaks down the transmitters into three categories:

- CW transmitter < 50 kW for SC injector cavities with low beam loading
- CW transmitter > 50 kW for SC injector cavities with high beam loading
- pulsed transmitters at megawatt power levels for NC injector cavities

For CW transmitters operated below 50 kW, solid-state technology and inductive output tubes (IOTs) are good choices. For CW operated transmitters above 50 kW, klystron and IOT tubes are competing concepts, while pulsed megawatt power-level transmitters are realized with klystrons. However, we note that progress in technology will increase the power level for economic applications of solid-state transmitters.

## 10.2 TRANSMITTERS

The transmitter is the device that amplifies the milliwatt level RF signal from the low-level RF system to the kilowatt- or megawatt-power level. The following are the major components of a transmitter:

1. preamplifier
2. power stage
3. set of power supplies
4. interlock systems
5. control system
6. cooling system

This section describes the most relevant devices for the several power stages. We show, using the example of a CW operated IOT transmitter, how to specify the stability of the power supplies from the general specifications of the transmitter and the data of the power stage. For pulsed transmitters, we detail the different circuit principles for the power modulators. We briefly mention the preamplifier and other auxiliary systems.

### 10.2.1 Power Stage

The power stage is the dominant part of a RF system. In the past century, high-power RF was generated only by electron tubes, such as tetrodes, klystrons and IOTs. Solid-state technology became much more relevant with the increase in the power levels and efficiency of single RF transistors that, in turn, resulted in an increasing range of economical RF-transmitter solutions.

The efficiency and harmonic content of the output signal of a power amplifier depend on which one of the three classes of operation is employed, *i.e.*, A, B, or C: In class-A operations, the active element conducts throughout the whole RF cycle. Its harmonic content is low and its linear amplification range is high; the drawback is a low efficiency, especially during partial load operation. Even without any input signal, power is wasted. Class-A operation is used with klystrons.

In class-B operations, the amplifier is conducting only during the positive half of the RF cycle. Here, the efficiency is higher than in class-A, but so is the harmonic content. In class-C operations, less than half of the positive RF cycle is used, yielding a very high harmonic content and the highest efficiency; however, there is no longer a linear dependence of the output signal to the input.

Tetrodes, IOTs and solid-state amplifiers usually are employed in a mix of class-A and class-B, compromising between the harmonic content, the linear behavior of the amplification and high efficiency.

#### *10.2.1.1 Tetrode*

Tetrodes are mostly used at frequencies below 200 MHz. Figure **10.2** is a schematic of a tetrode. The anode and screen grid are at a positive potential, and the control grid at a negative potential with respect to the cathode. The control grid determines the intensity of the beam in the tetrode, while the screen grid prevents capacitive feedback from the anode to the control grid.

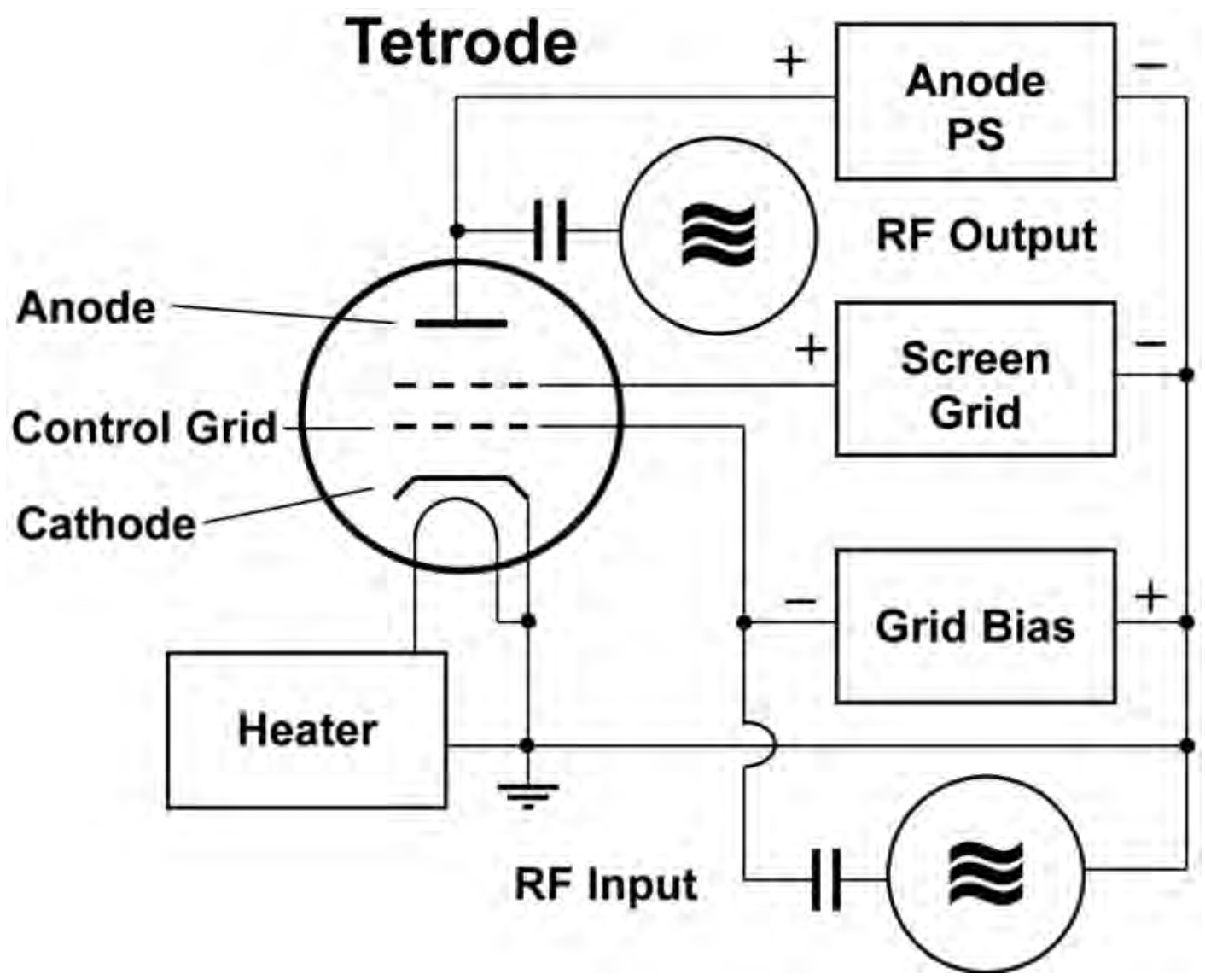

**Figure 10.2. Schematic diagram of a tetrode. The RF input controls the current from the cathode to the anode *via* the grid voltage.**

The RF tetrode has a coaxial design, with the cathode inside and the anode outside, allowing for effective cooling of the anode. Output power is limited by the maximum current density available from the cathode and the maximum power that can be dissipated from the anode. The dimensions and spacing of the electrodes are limited to avoid

- Variation in signal levels – the length of the anode must be less than the free space wavelength of the signal to be amplified.
- Higher order modes between the anode and the screen grid – the perimeter of the anode must be less than the free-space wavelength.
- Transient time problems – spacing between the anode and cathode must be less than the RF period. Increasing the anode potential to overcome this problem may cause discharge between the electrodes.

The coaxial design of the tetrode allows the construction of coaxial input and output circuits. The most common arrangement is a grounded control grid. The input- and output-circuits are well isolated because both grids are at RF ground. However, as the anode current flows in the input circuit, both the input impedance and the gain are lower than for a grounded cathode.

#### *10.2.1.2 Klystron*

The klystron is a vacuum electron tube with a linear output-input characteristic. Figure **10.3** is a schematic diagram of a four-cavity klystron; the main parts of which are:

1. an electron gun to produce the high current electron beam
2. a set of cavities to modulate the velocity of the beam
3. an RF input coupler at the first cavity

4. an RF output coupler at the last cavity
5. an axial magnetic field to confine the electron beam, usually generated by a solenoid electromagnet
6. the collector for the spent electrons

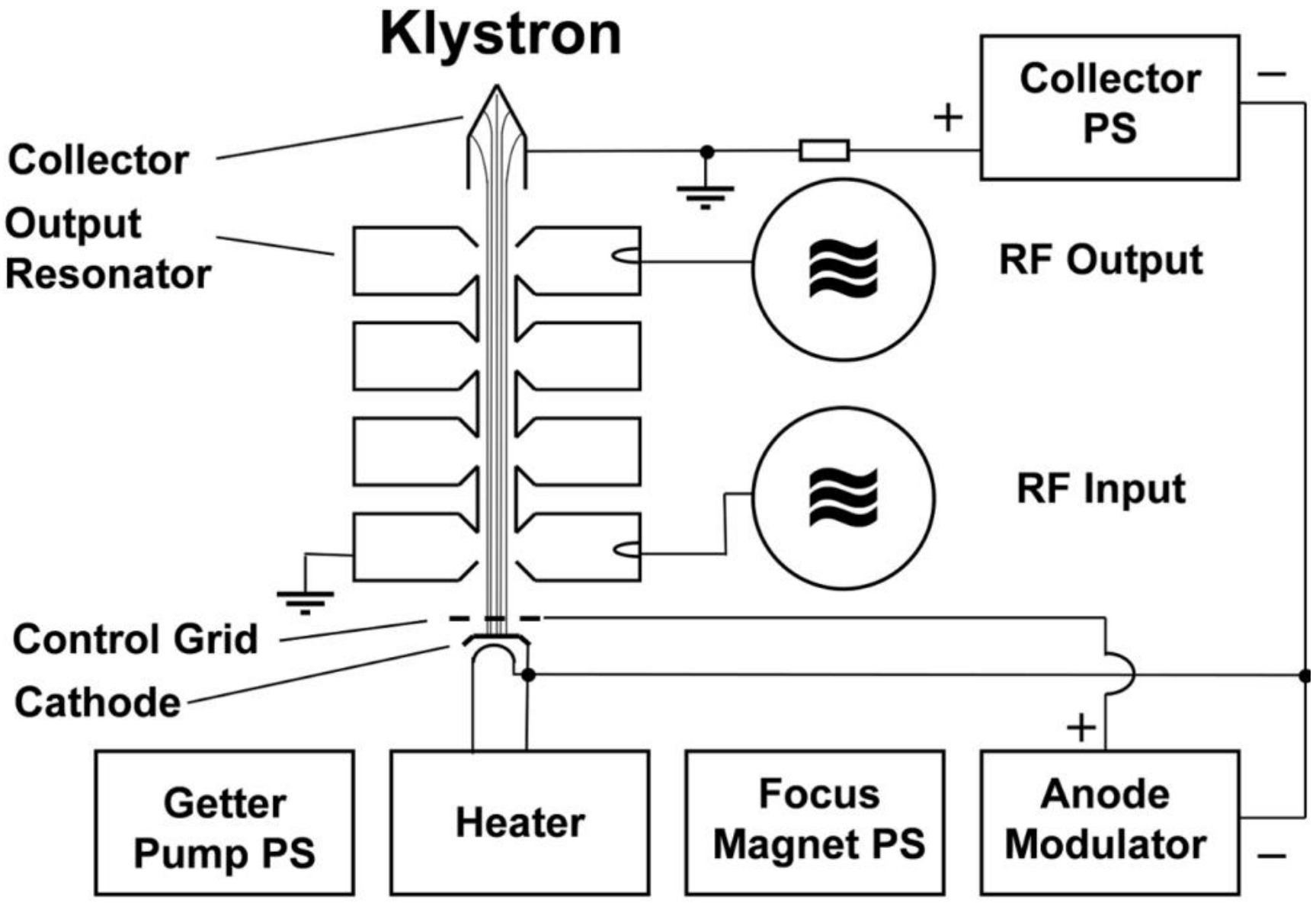

**Figure 10.3. Schematic diagram of a four-cavity klystron showing the electron gun area with the cathode, the set of cavities including the input cavity with an RF coupler, and the output cavity with the RF output. The solenoid magnet (not shown) along the klystron focuses the beam and the collector gathers up the spent beam.**

The DC electron beam, produced in a thermionic electron gun, is accelerated by the voltage between the cathode and the collector. Typical voltages range from tens of kilovolts for small klystrons, up to hundreds of kilovolts for klystrons at the megawatt power-level. The electron beam passes through a cavity resonator excited in the $TM_{010}$ mode, establishing an axial electric field across the gap in the drift tube. Depending on its phase, the RF field in the cavity accelerates or delays the electrons as they cross it. On leaving the cavity, the electron beam is velocity-modulated. After a particular length of drift tube, the "slow" electrons are caught up by the faster ones, causing current modulation. If a cavity is placed at this point, the electron beam induces an RF field in it. By using several cavities (typically four to seven) and by tuning them in a special way to frequencies slightly above and below the RF frequency, the current modulation can be optimized. Such tuning also determines the bandwidth of the klystron. The output cavity is placed at the point of maximum current modulation, thus inducing a high electric RF field. The output coupler extracts the power from the cavity. The typical efficiency of a klystron is 40-65% with respect to the DC beam power.

The beam is collimated in the tube by solenoid magnets to minimize losses into the klystron walls. This is essential because the power stored in the beam, if misdirected, could easily melt the walls of the tube. The spent beam is collected in the high power collector.

In a linear (class-A) operation mode, the output power is nearly linear with respect to the input power at constant collector voltage, but the gain of the tube falls when the drive power is increased beyond the level at which maximum output power is reached. Therefore, it is necessary to limit the drive power to avoid oscillations of the amplitude loop. During operations with too high input power, the electrons of the beam might be accelerated back and could damage the cathode.

Megawatt range (especially in pulsed operation) klystrons mostly are operated in saturated operation mode to run the tube at its maximum efficiency. Here, the input power of the klystron is set on a level in which the klystron is operated in saturation. A variation in the drive power will not affect the output. In saturated operation mode, the output power level of the klystron is adjusted by changing the collector voltage.

Klystrons have undergone considerable development and can be designed for a high frequency and output power range. The gain of a klystron typically is 40-55 dB; hence, the driver amplifier has to deliver only a few watts of power.

#### *10.2.1.3 Inductive Output Tube*

The inductive output tube (IOT) was invented in the late 1930s by A. Haeff [10.10] and L. Nergaard [10.11] a few years before the principles of the klystron were known. The IOT is a gridded tube, like a triode with an additional output resonator. The technology needed to build a high power IOT was not ready then, so velocity-modulated tubes, like the klystron, became popular as power RF amplifiers. In the late 1970s, progress in technology brought the IOT back into focus, and at the turn of the millennium, it became the most dominant TV transmitter tube. Figure **10.4** is a schematic of an IOT amplifier.

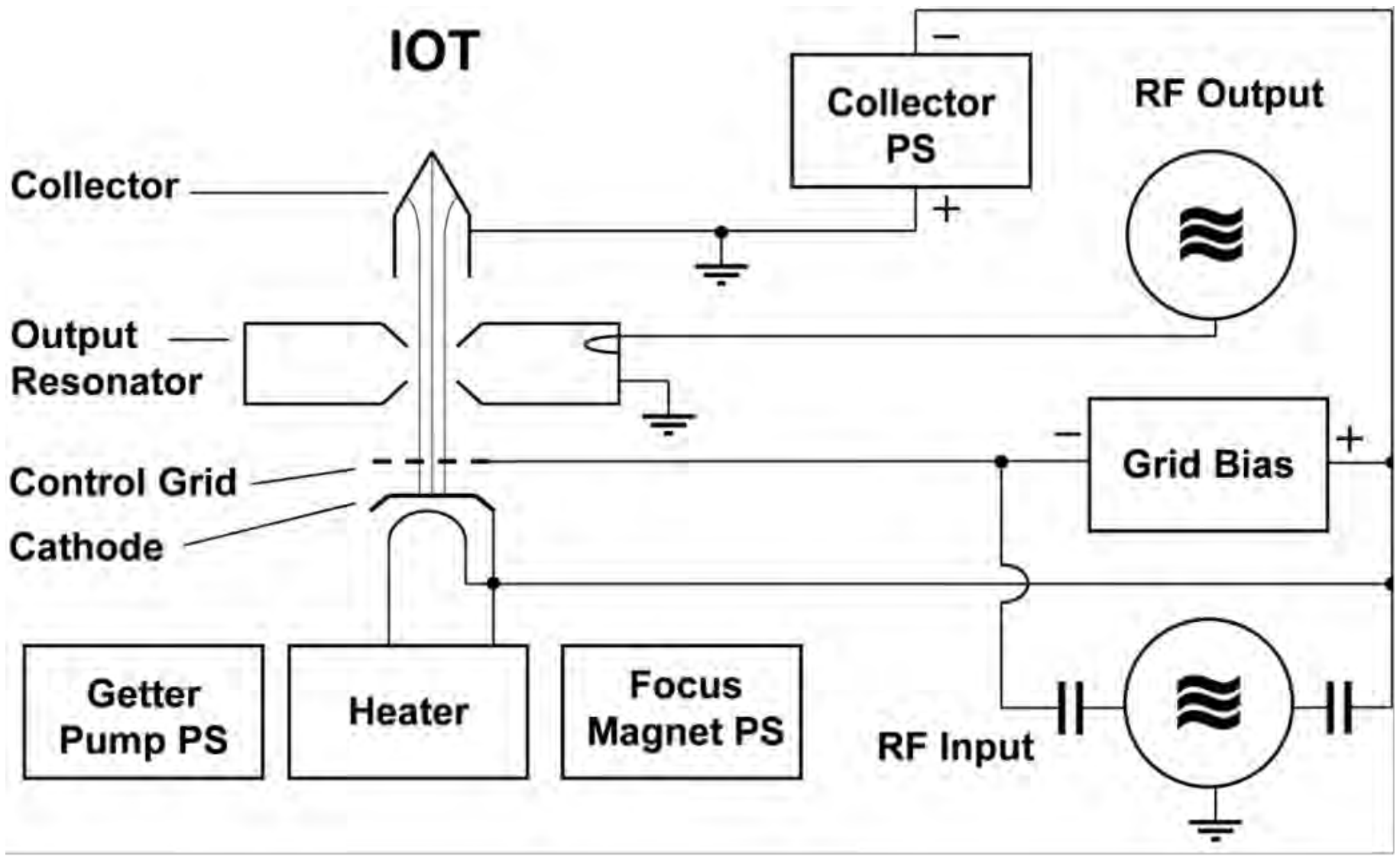

**Figure 10.4. Schematic of an IOT. The DC electron beam, produced in a gun and modulated by a grid, induces an RF field in the cavity. The power of this field is coupled by an output coupler to the transmission line. The solenoid magnet (not shown) along the IOT focuses the beam and the collector intercepts the spent beam.**

The IOT has five main parts:

1. an electron gun that generates the high current electron beam
2. a grid to modulate the density of the beam
3. an RF output coupler at the output cavity
4. an axial magnetic field, usually from an electromagnet, to confine the electron beam
5. the collector for the spent electrons

The DC electron beam, produced in a thermionic electron gun, is accelerated by the voltage of the collector's high voltage power supply. The electron beam passes through a grid biased such that the beam current flows only during the positive half of the input RF cycle. The bunched beam induces an RF field in the output resonator loaded by the output coupler. A solenoid magnet focuses the electrons in the tube. The high power collector gathers up the spent electrons. The input operation is similar to that of a tetrode, while the output operation resembles that of a klystron. Hence, some people refer to this device as a "klystrode."

Advantageously, the design of the IOT is simpler than that of a klystron, and it is smaller, easier to handle and weighs less. It is operated in class-AB mode, assuring good efficiency over a large dynamic range even though its typical maximum efficiency of ~65% is achieved only at the maximum power level. The class-AB operation makes the IOT a good choice (in terms of overall efficiency) in all systems with amplitude modulated applications, such as microphonic dominated operation of superconducting cavities. The major drawback of an IOT is its low gain, typically 23 dB, thus requiring a high power preamplifier, which raises the cost of an IOT-based transmitter.

#### *10.2.1.4 Solid-State Amplifiers*

Solid-state technology for radio frequency power applications still is under development [10.12]. The critical issue is that the heat produced in the small crystal must be conducted properly to the cooling structures. Figure **10.5** depicts the maximum achievable RF power by one transistor, and its efficiency versus frequency.

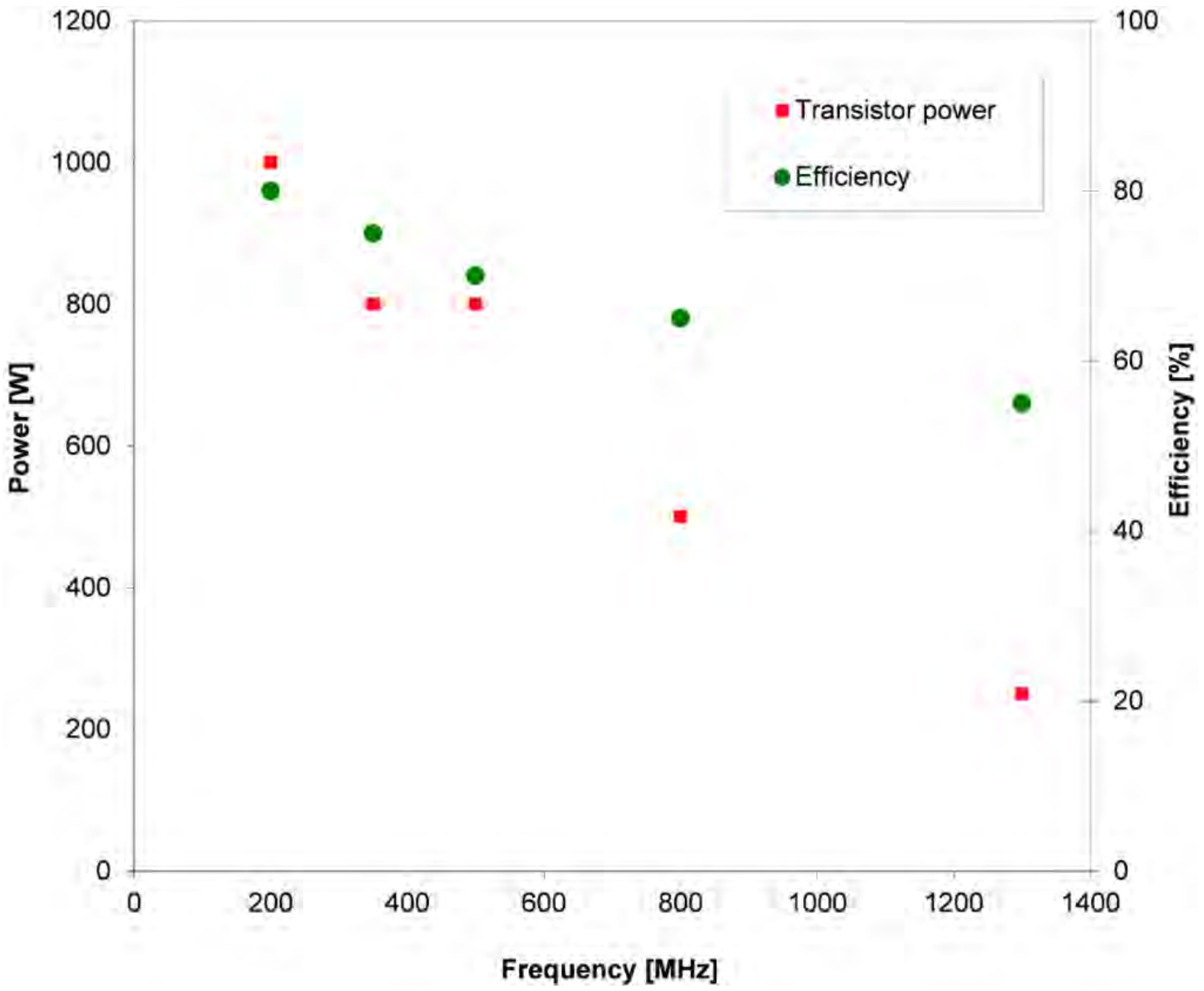

**Figure 10.5. Maximum achievable RF power of a single transistor, and its efficiency versus frequency.**

Obviously, in constructing a solid-state, high power RF amplifier, the power of tens or hundreds of RF transistors must be combined, as illustrated in Figure **10.6**. Consequently, multiple power combiners are needed. New ideas utilize cavity combiners to add the power of up to a hundred modules in one device [10.13]: to date, this has been demonstrated only for a combiner with two inputs [10.14].

Adding up the power of a large number of transistors is the big advantage of solid-state technology, creating a modular system with high redundancy, and therefore high reliability. A tube failure in a klystron or IOT transmitter, where the generation of RF power is localized in one tube, shuts down the whole transmitter. With an appropriate modular design, only one part of the transmitter is shut down in such a failure and the overall transmitter continues operating on a reduced power level. The defective power supply or RF module can be changed during the following maintenance cycle. Depending on the construction, it may even be possible to change modules while the system is operating, leading to a high reliability of the transmitter.

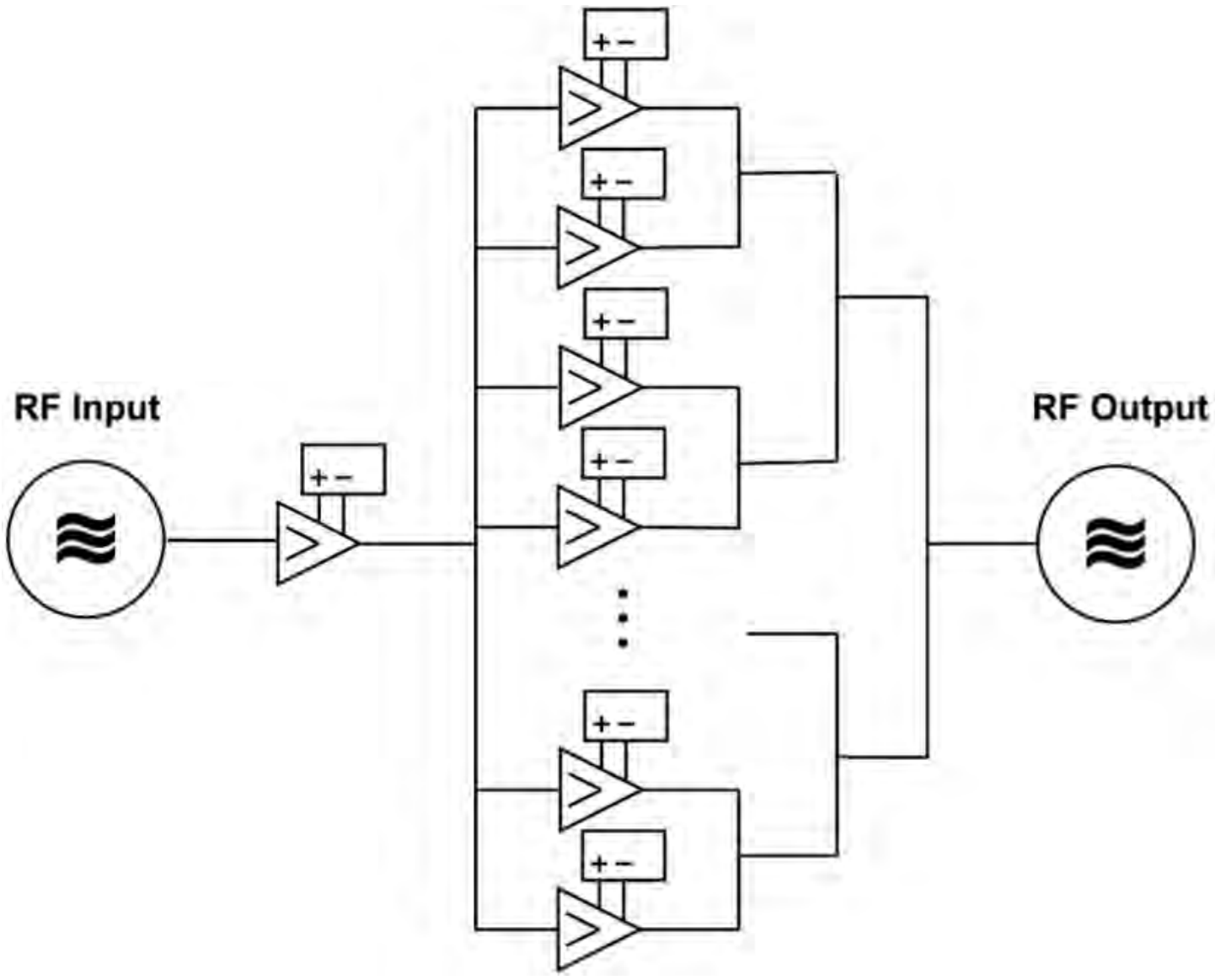

**Figure 10.6. Schematic diagram of a solid-state amplifier. Up to a few hundred medium power amplifier modules are combined in a cascade. Combiners have a high number of inputs that reduces the depth of the cascade. Each amplifier module has its own power supply, assuring a high degree of redundancy.**

The SOLEIL storage ring (Paris, France) contains a 354 MHz, 180 kW transmitter built from 315 W modules each powered by their own 600 W switched power supply [10.12]. It has operated for over 20 000 hr without a single RF power trip.

In contrast to electron tubes that need a set of several high voltage power supplies, solid-state RF modules need merely one low voltage power supply. RF transistors built in or before 2008 have been operated typically with a 28 V supply voltage. The new generation transistors operating at 50 V and thereby reducing the currents make solid-state technology easier to handle at high power levels.

Most RF modules are equipped with their own circulator, including a load to protect the transistor against reflections. The most significant drawback of solid-state technology is its restricted overall efficiency, including the combiner network, which still falls short of the values of electron tubes. Furthermore, given the complexity of the combiners, the price of a transmitter is only competitive to electron tubes at lower power levels and low frequencies. The answer to this problem is developing transistors with more power per single device.

### 10.2.2 Preamplifiers

Solid-state preamplifiers most often are those commercially available. Since the typical gain of a klystron is ~40-55 dB, the preamplifier power level ranges from 5 W for a 50 kW for CW klystron transmitter, to 500 W for a pulsed 20 MW system. The preamplifier for an IOT transmitter must be more powerful because the IOT’s gain is only 20-23 dB. In a solid-state transmitter, the preamplifier is integrated. It is essential to specify and measure the noise of the preamplifier as carefully as the main power stage! Many amplifiers on the market, commonly used by the telecommunication industry, meet the required power level; however, most do not fulfill the stability requirements. By exchanging the integrated power supply for a more stable one, these amplifiers serve well in the RF systems of accelerators.

### 10.2.3 Power Supplies for CW Transmitters

The typical application of photoinjectors is in accelerators with short electron bunches that need very stable RF conditions. Typical specifications for the stability of the RF voltage in the cavity are 0.1$^\circ$ in phase and 0.1% in amplitude, or better. To reduce the load on the amplitude and phase loops of the LLRF system, the design of the transmitter should ensure that the noise of the RF at its output meets these requirements without using a loop. With the recent substantial progress in the development of switched power supplies, it is possible to realize this by carefully specifying the stability of the power supplies.

In this section, we discuss the transmitter's specifications using the example of an IOT amplifier [10.15]. The specifications for a klystron or tetrode transmitter are similar because the set of power supplies is alike, as evident by juxtaposing the power supplies in Figure **10.2**, Figure **10.3** and Figure **10.4**. We do not discuss here the specification of solid-state transmitter power supplies. However, stability issues, particularly dynamic stability, can be applied in the same way.

#### *10.2.3.1 Transfer Characteristics of the Electron Tube*

To translate the given specification in phase and amplitude stability into specifications of voltage stability for the power supplies, we must know the transfer characteristics of the IOT in amplitude and phase in relation to the different power supplies. These parameters must be provided by the manufacturer of the tube, or must be measured on a prototype.

Figure **10.7** illustrates the dependencies of phase and gain for voltage changes of the collector and the grid power supply for a 1.3 GHz, 16 kW IOT (Type E2V IOT116LS), as an example. The phase gain versus the collector voltage amounts to 7˚ per kilovolt at the most. Hence, for a phase stability specification of 0.1˚ maximum, the voltage stability should not exceed 15 V or $6\times10^{-4}$ at 23 kV.

Looking at the requirements for voltage stability in relation to the amplitude stability of the transmitter, the gain variation versus the voltage of the collector's power supply is 0.5 dB $kV^{-1}$. To achieve a stability of 0.1% in amplitude, the voltage variation of the collector's power supply should not exceed a value of 8 V or $3\times10^{-4}$ at 23 kV. The requirements for the amplitude stability are the more stringent; hence, the specification for the stability of the collector power supply should be better than $3\times10^{-4}$.

The same logic applies to the grid power supply. The gain stability gives the stronger specification, resulting in requirement for the grid bias of $7\times10^{-4}$. We note that the grid power supply is on a high voltage level that entails a greater effort in installation. However, if the grid power supply is applied to ground level, the noise of the collector power supply must be added to it, leading to much stronger specifications for the collector's power supply!

#### *10.2.3.2 Static- and Dynamic-Stability of the Collector Power Supply*

The collector's power supply is a high voltage power supply operated on ground level in the voltage regulation mode. The class-AB operation mode of the IOT causes variable beam currents depending on the power level of the RF output, necessitating a dynamic specification of the voltage stability. In Figure **10.8**, we depict the collector voltage while pulsing the RF [10.15]. The details of the ripple are evident when the magnification is increased (zooming in), as are the overshoots. By varying the slope of the pulse, information is gained that is dependent on the bandwidth. With an exact specification, the manufacturer of the power supply can design it to meet the desired requirements.

It is advisable to limit the output current and voltage in the power supply. To reduce the thermal stress on the tube when switching on, the voltage should be ramped up in a few seconds, thereby extending the lifespan of the tube. Stored energy at the output of the power supply and discharge time should be specified according to the manufacturer's standards. A "wire-test" is usually specified to demonstrate the fulfillment of the specifications. In this test, a thin wire (~0.2 mm) should not be destroyed when it shorts the output of the power supply.

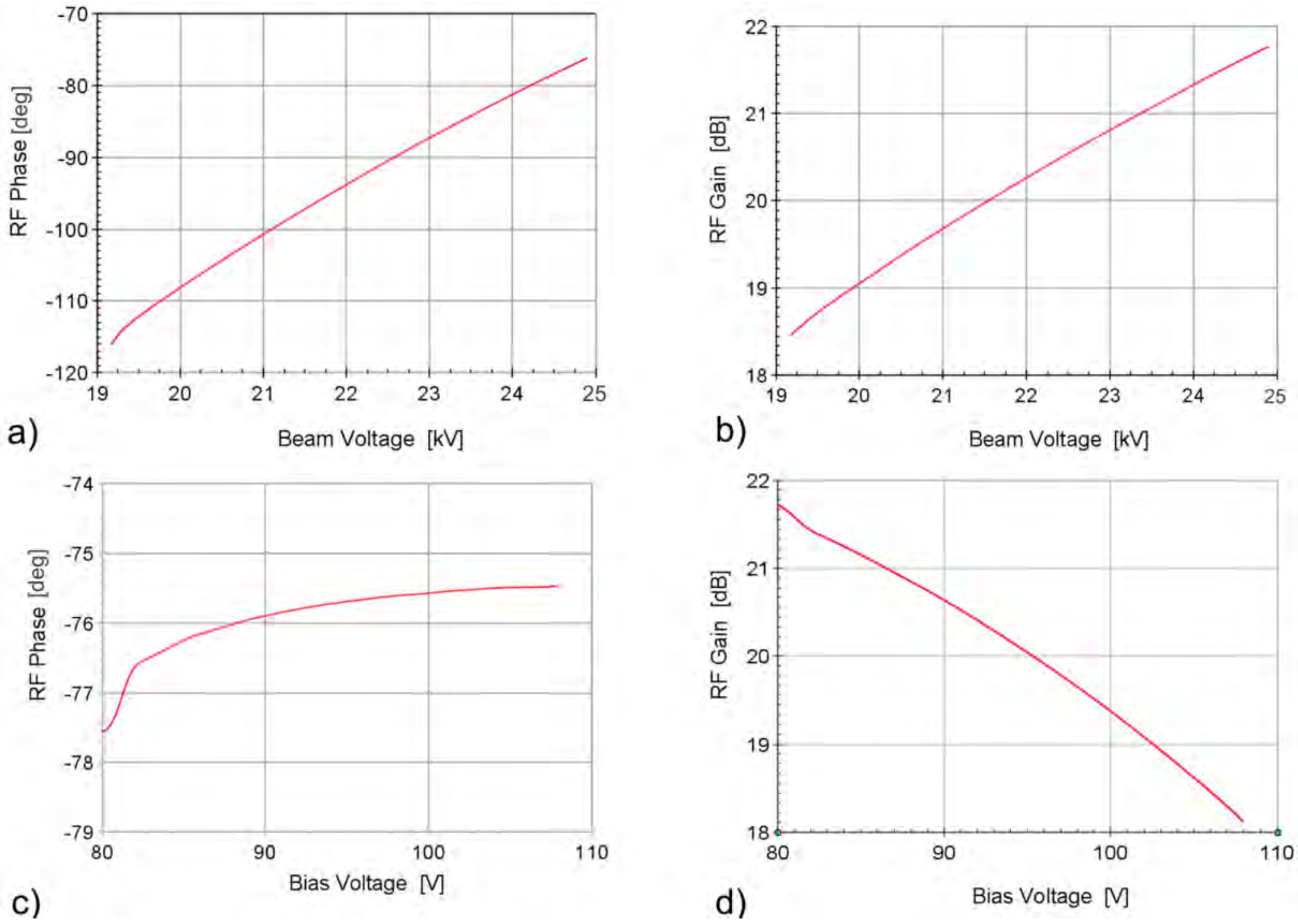

**Figure 10.7. Transfer characteristics of E2V IOT116LS IOT: a) Depicts the phase shift; b) the amplitude gain versus collector voltage; c) depicts the phase shift; and d) the amplitude gain versus grid bias voltage.**

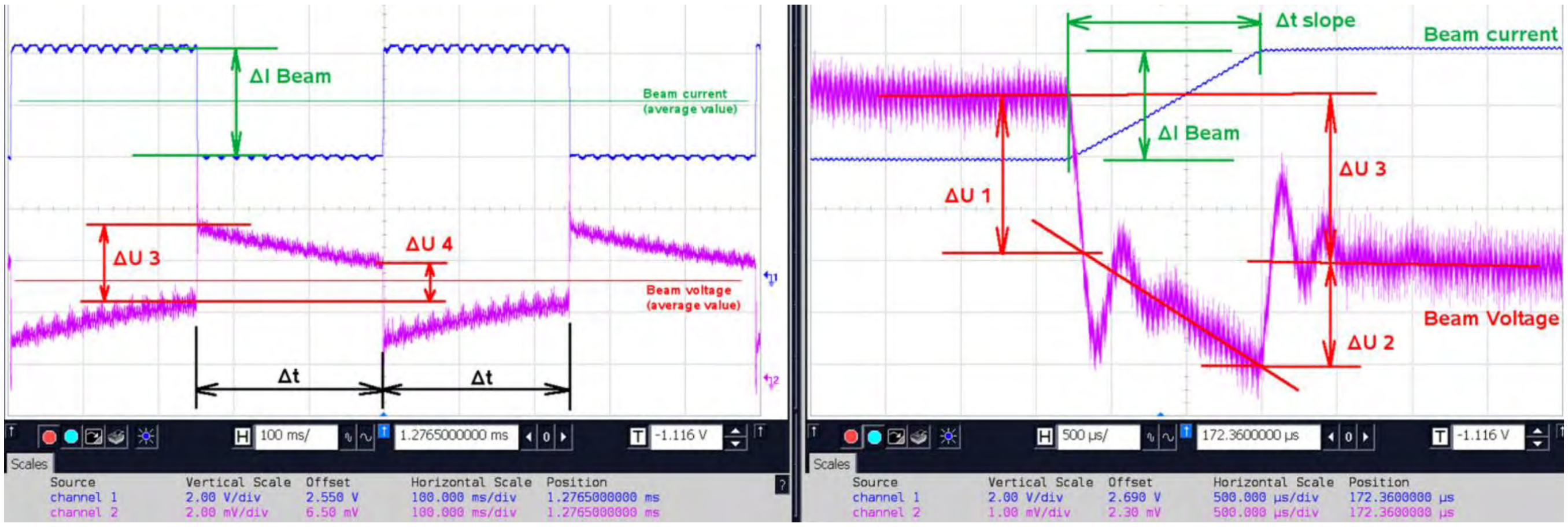

**Figure 10.8. Voltage of the collector power supply of an IOT transmitter shown on an oscilloscope. Blue represents the beam current in the IOT and purple is the AC component of the collector voltage. When the beam current is abruptly changed ($\Delta I_{Beam}$) by changing the RF drive power, the voltage regulation of the power supply is not able to follow instantly, resulting in a voltage step $\Delta U$. The right picture is a zoomed view of a rising edge of the RF. By varying the rise-time of the current, characteristics of the power supply can be determined. [[10.15]; Adapted under Creative Common Attribution 3.0 License (www.creativecommons.org/licenses/by/3.0/us/) at www.JACoW.org.]**

### *10.2.3.3 Stability of the Grid-Bias Power Supply*

The grid-bias power supply is operated at the high voltage potential of the collector power supply in the voltage-regulated mode. It should have a low impedance, otherwise instabilities can occur in the tube: a capacitor at the grid will help to maintain stability. Installing a current limiter, acting on the drive power of the tube, is recommended. In typical operation, the power supply is operated as a current drain, but for ion- or grid-emissions, it also must serve as a current source, hence, the grid bias power supply should be bipolar. The dynamic requirements and the discharge characteristics are the same as the collector power supply.

#### *10.2.3.4 Heater, Focusing and Getter Pump*

The other power supplies do not need the same strict specifications as the collector and the grid bias power-supplies. Nevertheless, some special issues must be considered.

The heater's power supply should be either voltage- or current-regulated, operating on ground potential; the latter is gentler on the heater when switching on the power supply, and the internal resistance of the heater is very low. The heater should always be powered by a DC source. Even though thermionic emission is a slow process, a 50-60 Hz powered heater definitely influences the stability of the RF output. The heater is integrated in a system operated with a high voltage. Therefore, precautions should be taken to ensure that the power supply is protected from arcing.

The power supply for the focusing element is current-regulated and responsible for powering the magnet to focus the beam inside the tube. If focusing is switched off, the beam hitting the walls of the IOT can destroy the tube. Therefore, a fast interlock must be activated if the current in this power supply is out of a defined current window.

The power supply to the getter pump does not require special voltage stabilization. It must be protected against arcs in the tube and should generate an interlock signal if the current exceeds a defined threshold, indicating that the vacuum pressure in the tube is too high.

An example of two IOT based transmitters is shown in Figure **10.9** [10.15] operated at the HoBiCaT facility at the Helmholtz-Zentrum Berlin (HZB). Left is the 30 kW, 1.3 GHz system operated at the HoBiCaT facility. The right picture shows the 80 kW, 500 MHz transmitter operated at the Metrology Light Source (MLS) [10.16]. As the voltage of the collector power supply is 41 kV, the high voltage part of this power supply is built in an isolating oil tank. The concept, specifications and controls of both transmitters are the same.

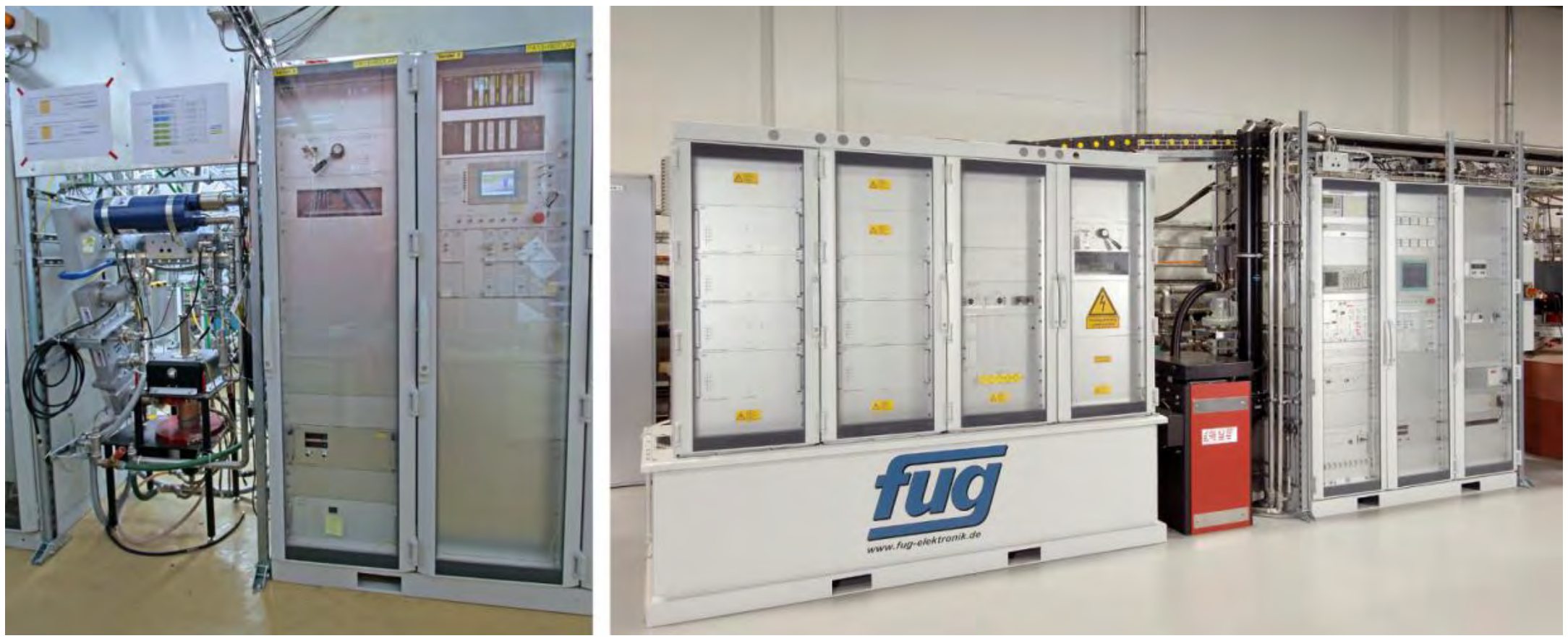

**Figure 10.9. Example of two IOT based transmitters. On the left is a 30 kW, 1.3 GHz system operated at the HoBiCaT facility at the Helmholtz-Zentrum Berlin. On the right is an 80 kW, 500 MHz transmitter operated at the MLS. [[10.15]; Adapted under Creative Common Attribution 3.0 License ([www.creativecommons.org/licenses/by/3.0/us/](www.creativecommons.org/licenses/by/3.0/us/)) at [www.JACoW.org](www.JACoW.org).]**

### 10.2.4 Modulators for Pulsed Transmitters

Pulsed RF systems are used in high power systems with klystrons, most of which run on megawatt peak RF power levels in the pulse. An overview on different techniques is given in [10.17]. Typical parameters for a modulator are [10.18]

| Parameter [units] | Value |
|---|---|
| Pulse Voltage [kV]: | 100-500 |
| Pulse Current [A]: | 50-500 |
| Pulse Length: | |
| - Short Pulse [μs]: | 1-50 |
| - Long Pulse [μs]: | 50 to some 1000 |
| Repetition Rate [Hz]: | 1-1000 |

**Table 10.1. Typical parameters of a klystron modulator.**

#### *10.2.4.1 Design principles*

A typical pulse modulator consists of

- capacitor bank for storing energy
- charging unit for the capacitor bank
- switching unit
- voltage pulse transforming unit
- pulse-forming network (PFN)
- general control and interlock systems
- klystron as load

The energy storage usually is a capacitor-based system. The requirements on the capacitors should include the ability to handle the extremely high currents and high frequency transients for the short rise-and-fall times of the pulses. Low inductance, and sometimes even cooling, is needed to handle the energy flow characteristic of high repetition rate systems. The voltage drop on the capacitor within the pulse is a main contribution to the instability of RF power in the cavity. The required capacitance can be calculated directly from the maximum allowed voltage drop. The charging units are power supplies providing the average power needed to recharge the energy storage system between two pulses. The switching unit dominates the design of a modulator. There are two basic choices for it:

- systems that switch only ON
- systems that switch ON and OFF

An ON-switch device suffices for storage systems such as a PFN, or if the load defines the pulse length. In the past 50 years, gas switches, like thyratrons, generally were used, but recently high power semiconductors became increasingly common. Solutions were found by combining a series of GTOs (gate turn-off thyristor) and GCTs (gate-commutated thyristor) operated up to 4.5 kV.

In other cases, an ON/OFF-switch device, like MOSFETs, IGBTs, or IGCTs must be used; sometimes saturable magnet cores are employed as an additional switch element.

The limits of all these switch devices lie in their voltage, current and repetition rate. Nowadays, semiconductors are limited to 6.5 kV in voltage and 3 kA in pulse current, so that switch devices must be combined for the input power of a klystron. Figure **10.10** shows three different design principles and circuits.

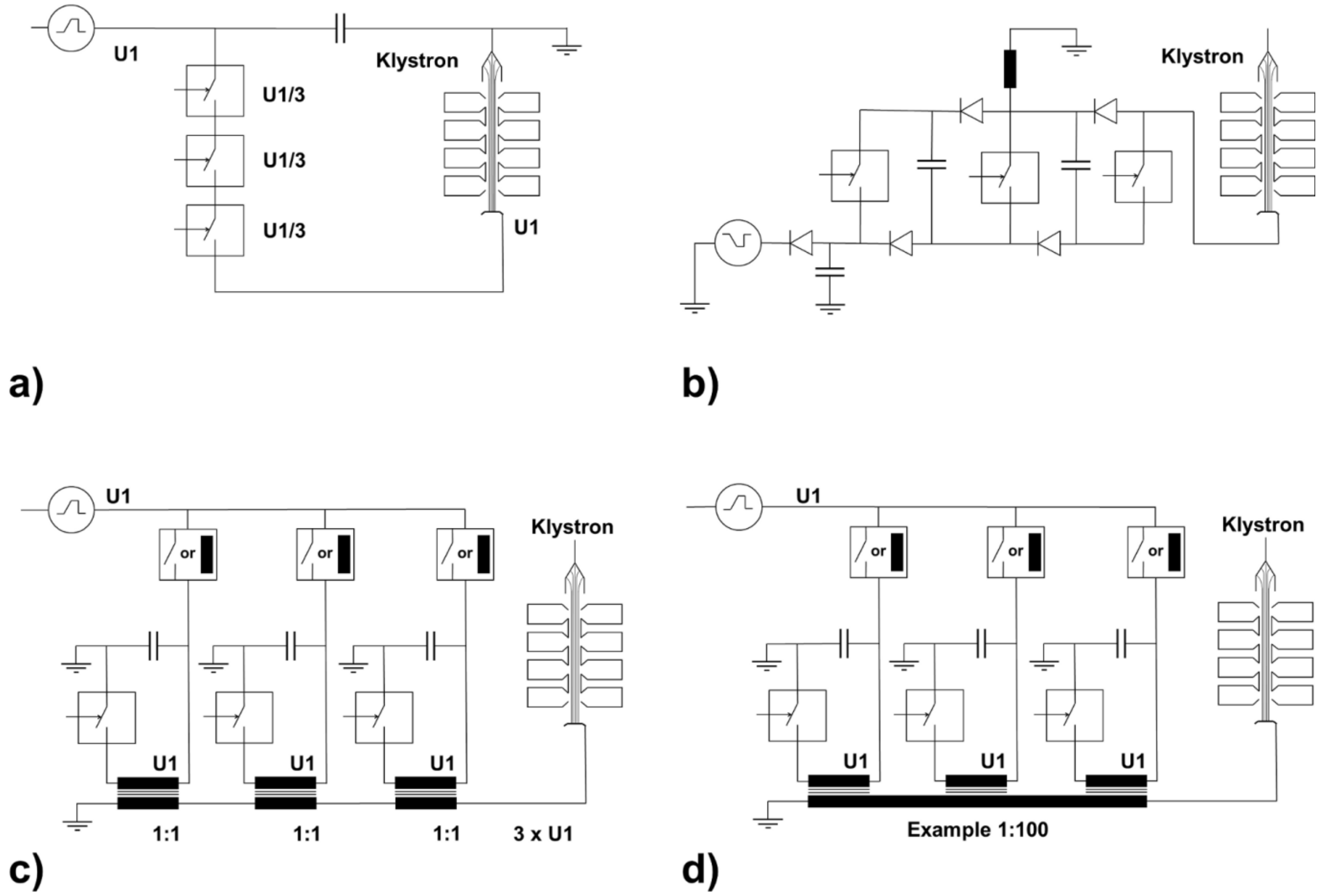

**Figure 10.10. Different design principles for klystron modulators [10.18]: (a) Direct switch; (b) Marx Generator; and, (c) and (d) two inductive adders of different designs. [Courtesy of J. Knobloch]**

### *10.2.4.2 Direct Switch*

In principle, it is possible to switch (Figure **10.10(a)**) the whole voltage with a serial combination of switches to the load. The big challenge is in synchronizing the trigger circuits for each switch. This assembly requires a DC power supply with full-load voltage [10.19].

### *10.2.4.3 Marx Generators*

The Marx generator (Figure **10.10(b)**) also is a serial combination of switching cells [10.20]; their main components are the switch and the capacitor. All devices work at the load voltage divided by the number of cells. Consequently, the power supply also is designed in the low kilovolt region. The disadvantage of a Marx generator is that it only allows a negative pulse voltage.

### *10.2.4.4 Inductive Voltage and Inductive Current Adder*

The voltage adder has a secondary coil to each primary coil. The output voltage in Figure **10.10(c)** is the sum of all group windings transformed from the primary side [10.21]. The output current is the same as that of the primary coil. Typically, a 1 to 1 ratio is used to achieve a low inductance. The input voltage equals the output voltage divided by the steps, and therefore can be reduced down to kilovolts.

The current adder (Figure **10.10(d)**) has one secondary coil to all primary winding coils. The ratio could be higher than 1:100, so the voltage is transformed *via* the winding relation. The power supply has the same specifications as the voltage adder type.

#### *10.2.4.5 Pulse Forming and Bouncers*

The switch technology provides flexibility in the pulse length dynamic. However, the transformers from the core design are optimized to one specific pulse length and rise time. This means that we do not have a pulse-transformer concept for pulse length covering the range from microseconds to some milliseconds. Hence, there is a separation between short and long pulse modulators, especially with respect to the transformer design. Concepts for pulse shape and pulse qualities also differ:

- Passive pulse forming: PFNs are a combination of capacitors and tunable coils.
- Active pulse forming: Control circuit or with a secondary bounce-circuit against the droop.

Figure **10.11** [10.19] shows the modulator built by Puls-Plasmatechnik, Dortmund (PPT) [10.22] for feeding the 10 MW (Thales TH1801) multi-beam klystron. The modulator is constructed of a switched capacitor bank using a series stack of IGCTs with bouncer droop compensation and pulse transformer matching of 1:12. It can switch a 12 kV pulse voltage and 2 kA pulse current. The pulse width of the modulator can be varied from 50 μs to 2 ms. For pulse lengths less than 1 ms, a smaller pulse transformer will yield better rise and fall times.

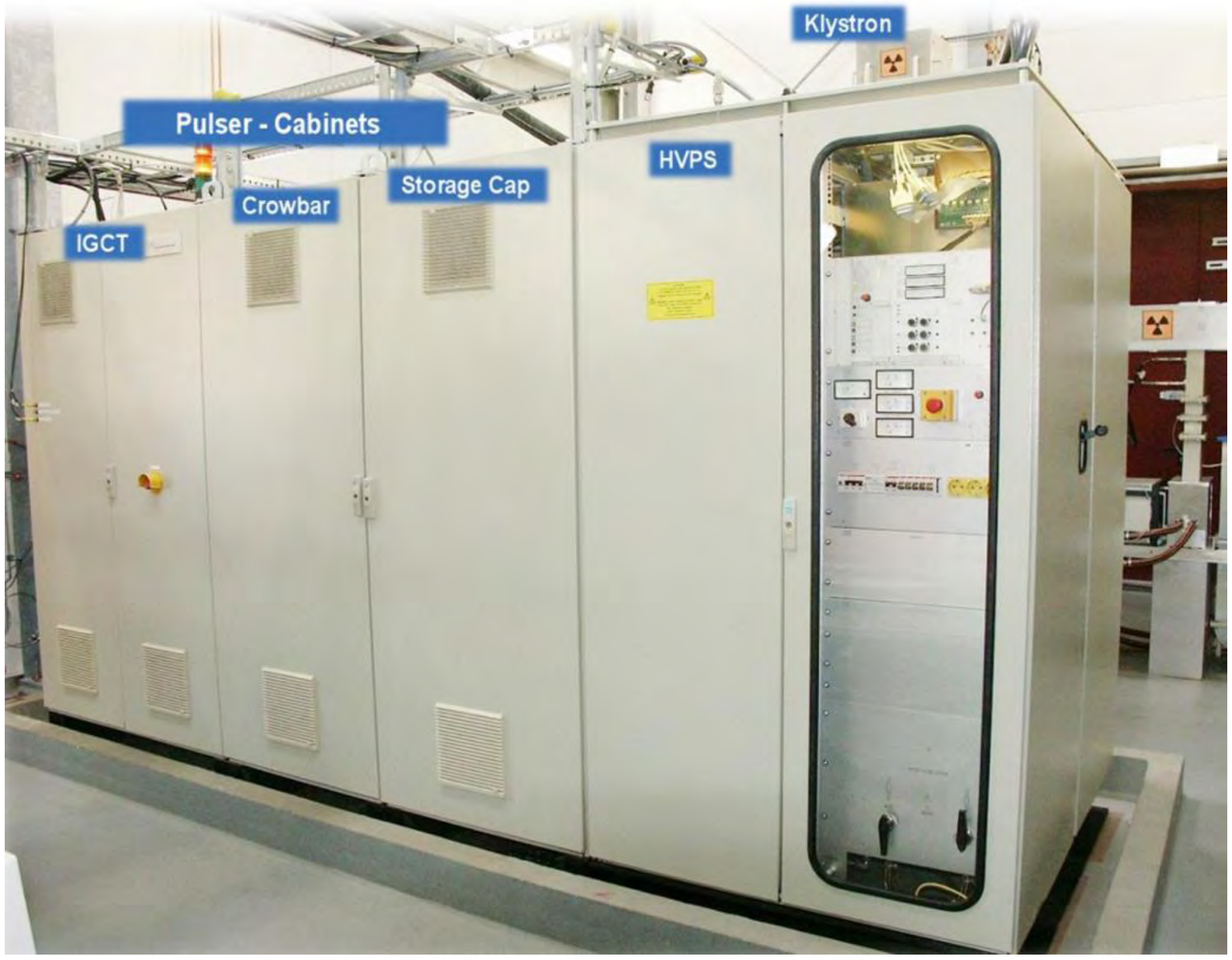

**Figure 10.11. Klystron modulator built by PPT to operate the Thales TH1801, a 10 MW multi-beam klystron. The modulator consists of a capacitor bank using a series stack of IGCTs with bouncer droop compensation and pulse transformer matching of 1:12. It can switch a 12 kV pulsed voltage and 2 kA pulsed current. The pulse width can be varied from 50 μs to 2 ms. [[10.19]; Adapted under Creative Common Attribution 3.0 License (www.creativecommons.org/licenses/by/3.0/us/) at www.JACoW.org.]**

### 10.2.5 Other Parameters to be Specified for an RF Transmitter

The following list contains the main parameters for the general infrastructure of each transmitter, which must be defined.

- fast interlock, acting in the 100 μs range
- connection to radiation-protection interlock
- connection to cavity signals
- interface to control system, type of PLC (programmable logic controller)

- accelerator control system interface
- metering instruments
- analog signals
- (water) cooling
- stability of mains
- temperature and humidity
- acceptable level of acoustic vibration

We emphasize a few of these points.

There are some expensive devices involved in a transmitter, so there is a need for installing fast-acting protection circuits because the reacting time of the PLC is insufficient. A fast, hard-wired interlock therein switches off the RF power *via* a PIN switch within microseconds. Other such emergencies include the following: Too high a reflection at the tube's output; voltage on the focusing magnet out of range; too high a drive power; arcs in the circulator, a transmission line, or the coupler of the cavity; too high a voltage in the cavity; or, bad vacuum in the cavity or in the radiation-safety interlock.

A second point is defining the internal control system, the analog measuring sockets, the metering instruments, and (often debatable) the interface to the accelerator control system. It always is worthwhile to define all signals which are required to be transmitted to the control system.

We highlight a final point, *viz.*, specifications of water quality and the maximum speed of water flow. As long as most accelerator laboratories use demineralized water, only a few materials must be considered. Materials suitable for these cooling circuits are stainless steel, copper, tin-free cast bronze, lead-free solder, ceramics and some plastics. This has to be specified!

## 10.3 TRANSMISSION LINE AND CIRCULATOR

There are only two practical choices of the transmission line for transmitting high power; either a coaxial line or a rectangular waveguide.

A waveguide transmission line acts as a high-pass filter and can handle hundreds of kilowatts: The lower the frequency cutoff, the larger the waveguide. All power losses are on the outside wall of the waveguide because there is no inner conductor. When the transmission line is used in megawatt applications, it is filled with pressurized insulation gas. The advantage of a waveguide is its power capacity, but the disadvantages are its large size and low flexibility, making it difficult to mount, especially through the labyrinth of the accelerator's radiation shielding.

Coaxial transmission lines can consist of fixed aluminum- or copper-tubes, or as a flexible line made from corrugated tubes. A coaxial line can be used from DC to an upper frequency limit, where waveguide modes can propagate. In Equ. 10.6, we calculated the cutoff frequency, $f_c$ (in Hertz), using the diameter of the inner conductor (in mm) of the coaxial line, $d$, the inner diameter of the outer conductor, $D$, the velocity of light, $c = 3\times10^{11}$ mm s$^{-1}$, and the relative dielectric constant of the material located between the two conductors, $\varepsilon_r$.

$$f_c = \frac{2c}{\pi\sqrt{\varepsilon_r}} * \frac{1}{D+d} \tag{10.6}$$

This limit differs slightly for straight lines and other devices, like bending elbows. Detailed data should be obtained from the manufacturer. About ⅔ of the power dissipation is in the inner conductor and ⅓ in the outer conductor. This causes stronger heating of the inner conductor than the outer one, which additionally is cooled by the surrounding air. Therefore, at high power levels, they have different temperatures, which lead to two different thermal expansions for the two conductors. For long straight lines, length-compensation is needed. In flexible coaxial lines, this can cause reflections in bends wherein the heated inner conductor with a larger bending radius is pressed out of the center of the line, disturbing the impedance of the transmission line. Hence, there is a dimension and power limit of the coaxial line at about 80 $kW_{CW}$ at 500 MHz, for instance, with a 6 and ⅛ inch line, or 9 $kW_{CW}$ at 1.3 GHz with a 3 inch line. The power limit can be shifted by letting air flow between the inner and outer conductor, resulting in convection cooling of the inner conductor. The main advantage of coaxial lines is that they are easy to handle, especially flexible ones. Furthermore, since they are smaller than waveguides, they can be installed easily even if the radiation-protection labyrinths are tight.

Cavities in accelerator applications are operated under high and varying reflection conditions. This will cause problems due to standing waves in the output resonator of the electron tube or in the transistors in a solid-state transmitter. Three-port circulators can protect the RF power source from reflections of the cavity. Using ferromagnetic resonance, they divert the RF power into a matched load. Power fed into port 1 always comes out of port 2; power fed back into port 2 is directed to the matched load on port 3.

## 10.4 RF FUNDAMENTAL POWER-COUPLERS

A fundamental input power-coupler (FPC) is the device that transfers the RF power from the waveguide into the cavity. It matches the generator's output to the impedance of the cavity-beam system that varies from matched impedance to full reflection, depending on the beam's intensity. There are numerous technical requirements. It also must provide a vacuum barrier between the cavity's vacuum and the feeding RF transmission line. In addition, it must meet the cleanliness conditions required for superconducting injector cavities. On the mechanical side, the coupler must cover the thermal shrinkage during cool-down in SC systems and maintain a low static heat-load and minimize mechanical stress on the ceramic windows. Some couplers also have to allow some mechanical flexibility for variable coupling strengths to control the external $Q$ value.

Since the energy of the beam in injector cavities is quite low and photoinjectors are used to generate low emittance beams, special attention has to be paid to the wakefield kicks produced by the FPC.

When choosing the design of the FPC, the main parameters to consider are the following:

- power level, peak/average
- pulsed/CW operation mode
- RF frequency
- adjustability of coupling strength, if needed
- mechanical boundary conditions
- cooling
- heat leakage (for SC injector cavities)
- sensitivity to multipacting
- coupler kicks on the beam

Several reviews on couplers have been published [10.23]–[10.26]. Three different design principles are of interest for injector cavities. For each approach, we discuss their advantages and difficulties, and give examples. For adapting an injector cavity, detailed calculations have to be undertaken to predict its electromagnetic-, mechanical-, thermal- and multipacting-properties. Before becoming operational, all couplers must be carefully conditioned.

### 10.4.1 Waveguide Couplers

A simple concept underlies waveguide couplers: A rectangular waveguide with a ceramic window is connected to the cavity cell in a normal conducting cavity, or to the beam pipe close to the cavity cell in superconducting cavities [10.27]. Usually, coupling strength is fixed and depends on the position of the waveguide with respect to the coupling cavity cell and the size and shape of the coupling iris. Coupling can be adjusted over ~1 order-of-magnitude, using an external 3-stub waveguide tuner in the waveguide transmission line [10.28]. A waveguide coupler needs more space than a coaxial one, but cooling the former is easier. Cooling is from the outside only, since there is no inner conductor. There is considerable heat leakage of waveguide couplers in superconducting cavities due to the waveguide's size; good thermal anchoring is essential.

Special attention must be paid to the ceramic window that serves as a vacuum barrier between the cavity that is under vacuum and the waveguide that is under air or insulation gas. The soldered joint between the waveguide and the ceramics of the window must be designed very carefully because of the different thermal expansion coefficients of ceramics and the waveguide's metal. Cooling is achieved by attaching copper braids or cooling channels to the wall of the waveguide, or by blowing air onto the ceramics in case of a warm window. Figure **10.12** shows a waveguide coupler tested up to 800 $kW_{CW}$ at a frequency of 700 MHz [10.29]. Most such windows are made of $Al_2O_3$ ceramic that exhibits a high secondary electron emission coefficient. To reduce multipacting, coating it with a thin layer (a few nanometers) of Ti or TiN is essential. Designing the system's geometry without a direct line of sight to the accelerated beam also prevents multipacting [10.30].

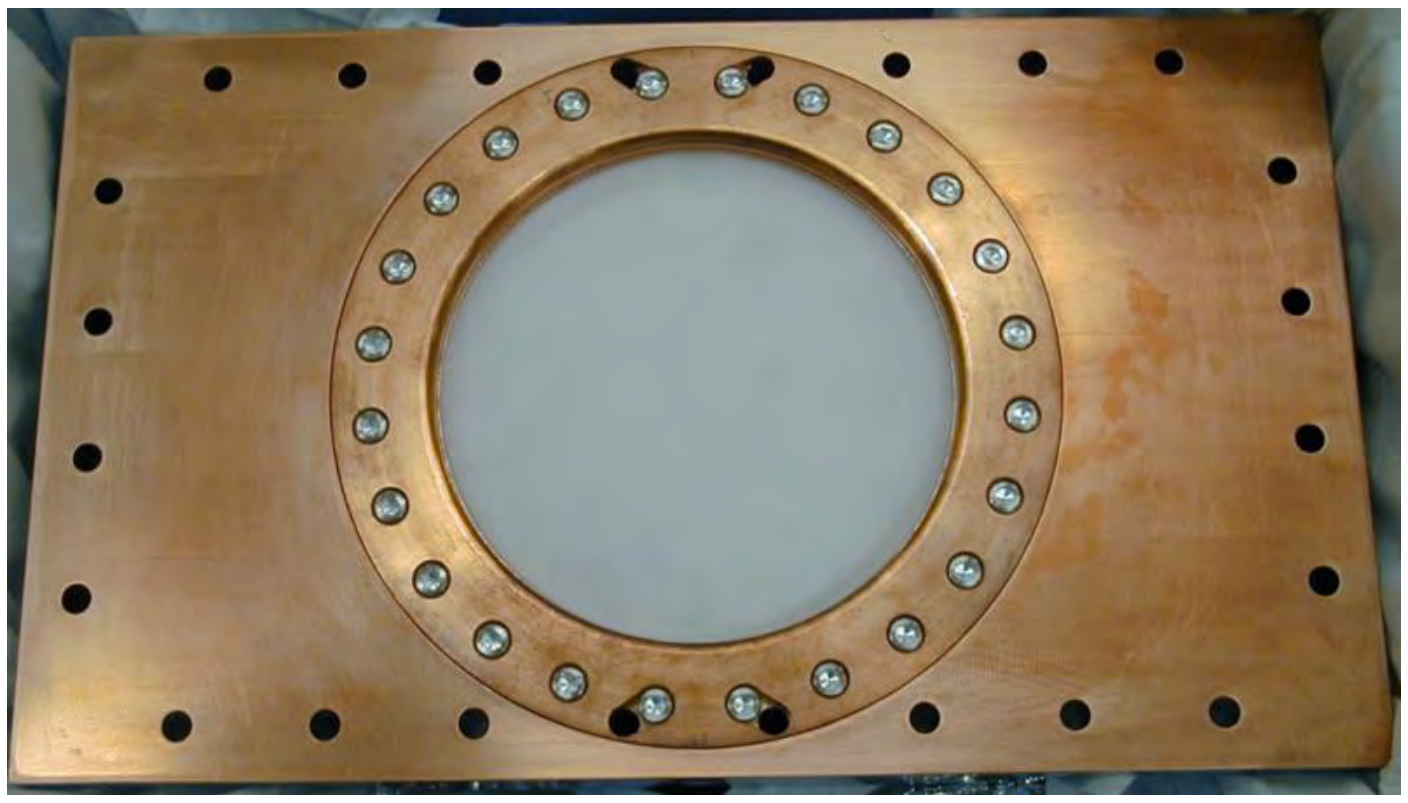

**Figure 10.12. Waveguide window tested up to 800 $kW_{CW}$ at 700 MHz. [[10.29]; Available under Creative Common Attribution 3.0 License (www.creativecommons.org/licenses/by/3.0/us/) at www.JACoW.org.]**

### 10.4.2 Coaxial Couplers

Coaxial couplers are more compact than waveguide couplers. In NC cavities, inductive coupling with a coupling loop is popular. SC systems use capacitive coupling *via* the inner conductor, which protrudes into the beam pipe, close to the cavity. These techniques support adjustable coupling factors by turning the loop or changing the penetration depth of the coupler finger. Cooling is more complicated because about ⅔ of the losses are dissipated in the inner conductor. Water- or air-cooling can be applied for warm inner conductors.

Figure **10.13(a)** is a cross section of the variable coupling, coaxial coupler, developed by Cornell, for use at 1.3 GHz SC cavities; 61 $kW_{CW}$ capability was demonstrated [10.31]. It is based on the TTF3 coupler [10.32] developed by Deutsches Elektronen Synchrotron (DESY) (2-3 $kW_{CW}$) and the BESSY coupler (10 $kW_{CW}$) [10.33] developed by HZB. The difference between them is the strength of cooling. In pulsed operation at FLASH, the average power of the TTF3 coupler is low. To use them in CW operated systems, heating of the bellows is the power limiting issue. HZB modified the coupler by introducing a gas-cooling system for the inner conductor. We note that the heat conductivity of the window ceramics has a maximum close to the temperature of liquid nitrogen (and is higher than the temperature coefficient of copper), so the cold inner conductor is cooled *via* the ceramics at the liquid nitrogen intercept. To reduce the coupler kick, the tip of the antenna has the same contour as the beam pipe.

Power capacity can be increased by using fixed coupling without avoiding bellows at the inner conductor of the coupler. Figure **10.13(b)** shows a fixed coupling FPC at 1.3 GHz developed for the Koh Ene Ken (KEK) ERL project (Japanese National Laboratory for High-Energy Physics). In the first tests, it demonstrated 100 $kW_{CW}$ and 130 kW in pulsed operation, limited by the test environment [10.34]. A higher power capability is expected. It is a scaled version of the KEKB coupler, which has demonstrated 370 $kW_{CW}$ at 508 MHz [10.35]. KEKB is the B-meson factory at KEK.

### 10.4.3 Coaxial Waveguide Couplers

Figure **10.14** illustrates the combination of both coupler principles [10.7]. A waveguide to coax transition is realized by using the beam pipe following the gun cavity as the outer conductor of the coaxial line. The electron beam and the laser light pass through the coupler inside the inner conductor of the coaxial line. By this mode of construction, the coupling of the RF power to the cavity is highly symmetrical and the coupler kicks to the beam are small. The disadvantage of this principle is the longitudinal length and the difficulty in implementing HOM dampers.

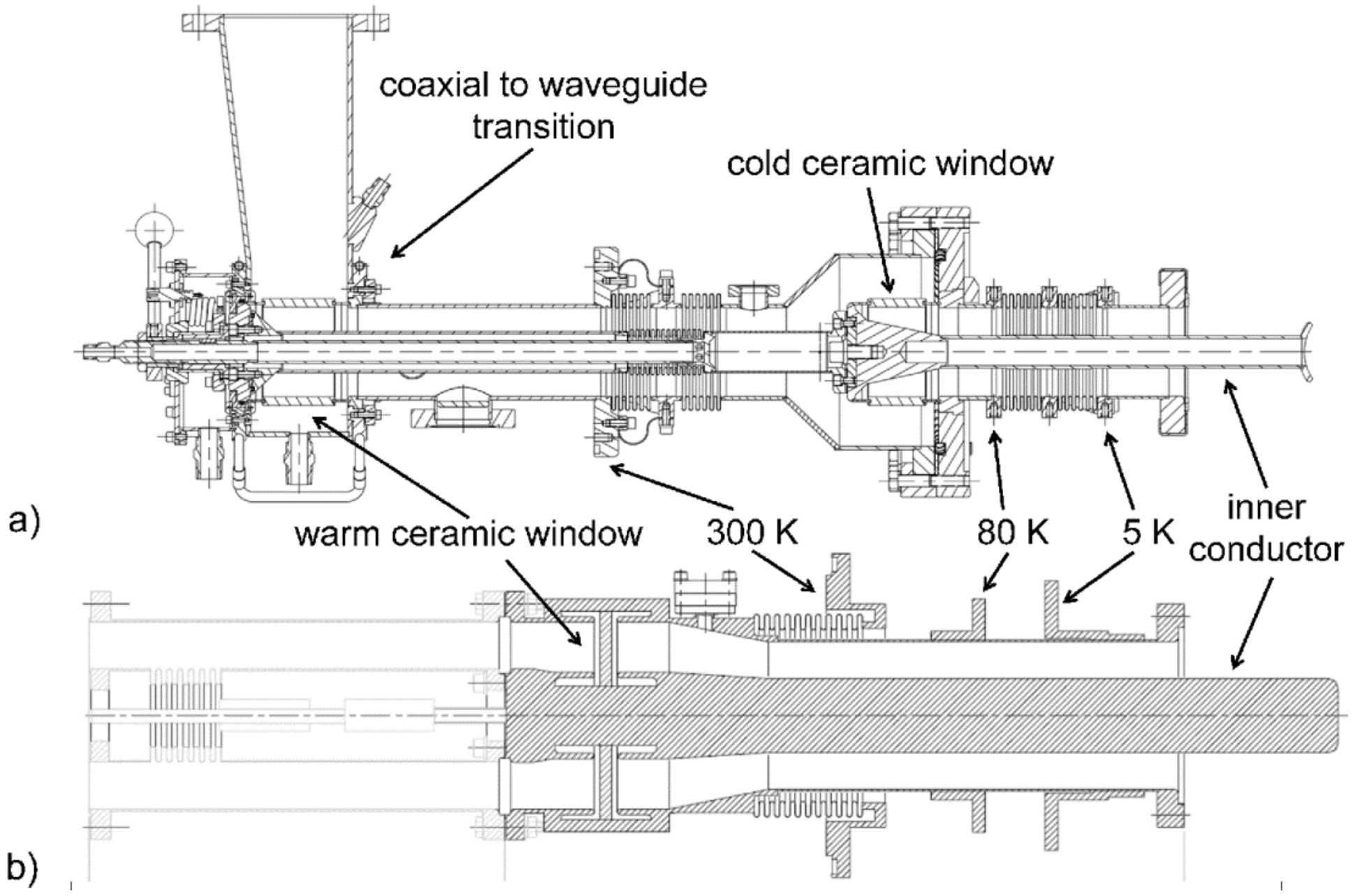

**Figure 10.13. Coaxial fundamental power couplers with thermal intercepts at 5 K, 80 K and 300 K. (a) Variable coupler, developed by Cornell, with a demonstrated power capability of 61 $kW_{CW}$ at 1.3 GHz using cylindrical windows. (b) Fixed coupling FPC by KEK using a ceramic disc as window. First tests demonstrated 100 $kW_{CW}$ at 1.3 GHz but higher values are expected. [[10.31]; Adapted under Creative Common Attribution 3.0 License (www.creativecommons.org/licenses/by/3.0/us/) at www.JACoW.org.]**

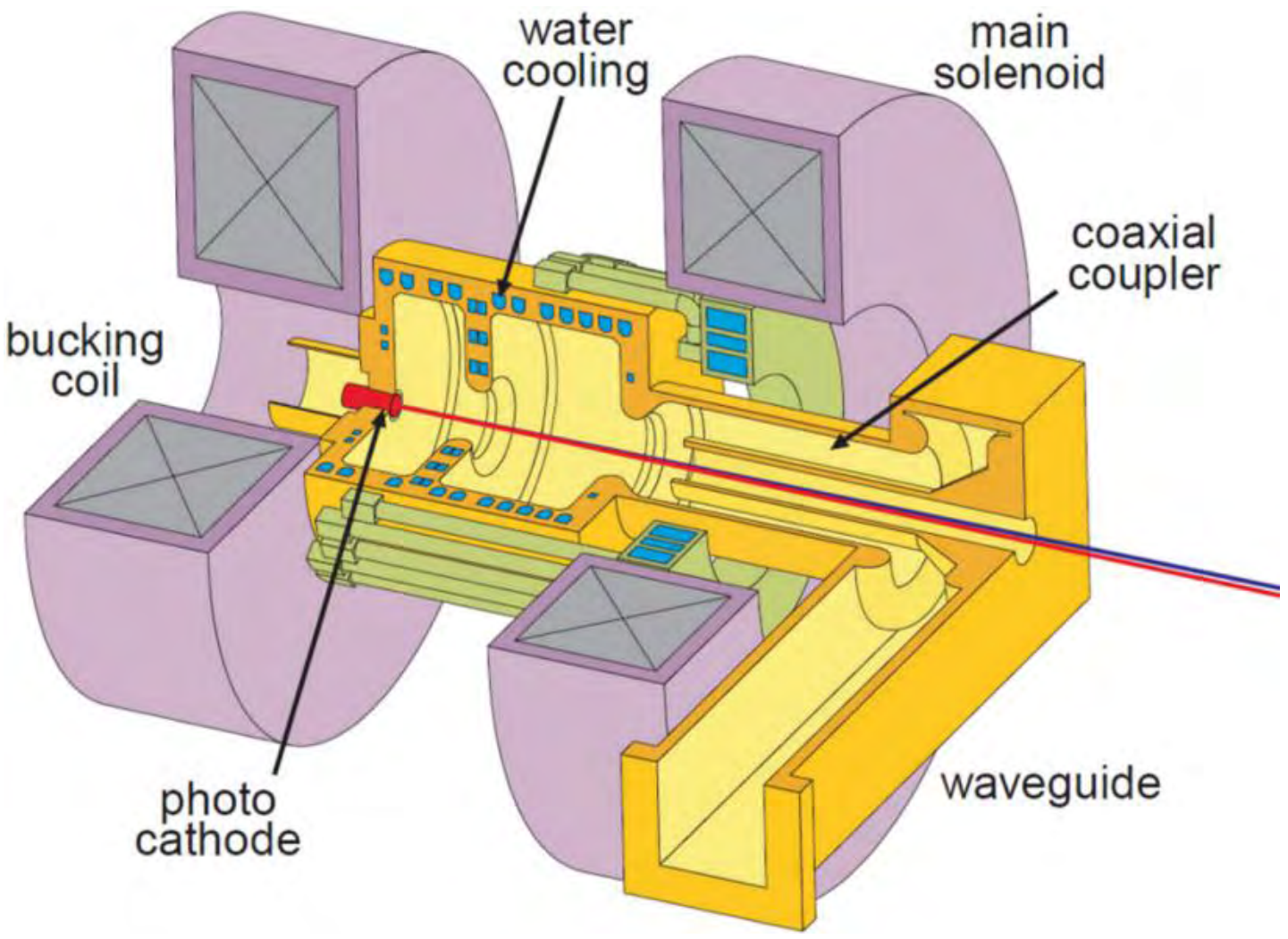

**Figure 10.14. Coaxial waveguide coupler as used at PITZ (Photoinjector Test Facility Zeuthen), and at TTF (TESLA Test Facility with TESLA, the Tera Electron Volt Energy Superconducting Linear Accelerator) at DESY. A waveguide to coaxial transition is realized by using the beam pipe as the outer conductor of the coaxial line. The electron beam and the laser light pass through inside of the inner conductor of the coaxial line. This coupler has a high degree of symmetry and the coupler kicks to the beam are small. [Courtesy of K. Flöttmann]**

# 10.5 HIGHER ORDER MODE COUPLERS

An accelerating cavity is designed to resonate in the fundamental $TM_{010}$ mode, but a wide spectrum of other modes also can be excited. The frequencies of these modes are higher than that of the fundamental mode, so they are called higher order modes (HOMs). When the beam passes through the cavity and a frequency component of the beam spectrum has a frequency close to that of the HOM mode, it can excite the HOM. Depending on the $Q$-value of the HOM, energy stored in this mode can revert back to the beam, causing longitudinal- and transverse-instabilities that degrade its emittance and broaden its energy. The strength of the HOM is given by its shunt impedance, $R$, and its quality factor, $Q$; both values must be kept small.

Monopole modes can cause longitudinal instabilities in the beam and increase its energy spread; dipole-, quadrupole- and sextupole-modes can lead to transverse instabilities. Monopole and dipole modes inflict the highest impedance. Thus, when designing the damper, the modes' different polarizations have to be account for.

We mention that care must be taken so only the HOMs are damped and the fundamental accelerating mode left unaffected; hence, rejection filters or other filtering techniques are essential.

An overview on HOM damping techniques is given in [10.26], [10.36].There are three design principles for injector cavities of interest: Waveguide couplers; coaxial couplers; and, beam pipe couplers. We discuss and illustrate an example of each of these types.

## 10.5.1 Waveguide HOM Couplers

A waveguide is a frequency high-pass device, simplifying rejection of the fundamental mode. The cut-off frequency is set such that the lowest HOM can propagate in the guide, but the fundamental mode is rejected. More than one damper is required because dipole modes have two polarizations. Detailed calculations are necessary to ensure that all HOM modes can propagate to the dampers and no modes are trapped in the cavity. For NC cavities, the HOM dampers are attached directly at the cavity cell: for SC cavities, they are placed at the beam pipe close to the cavity. Figure **10.15** shows a cavity using six waveguide HOM couplers on a five-cell SC cavity [10.37]. The terminating loads are at room temperature. The waveguide dampers are transversal space consuming devices, but their longitudinal space requirement is low.

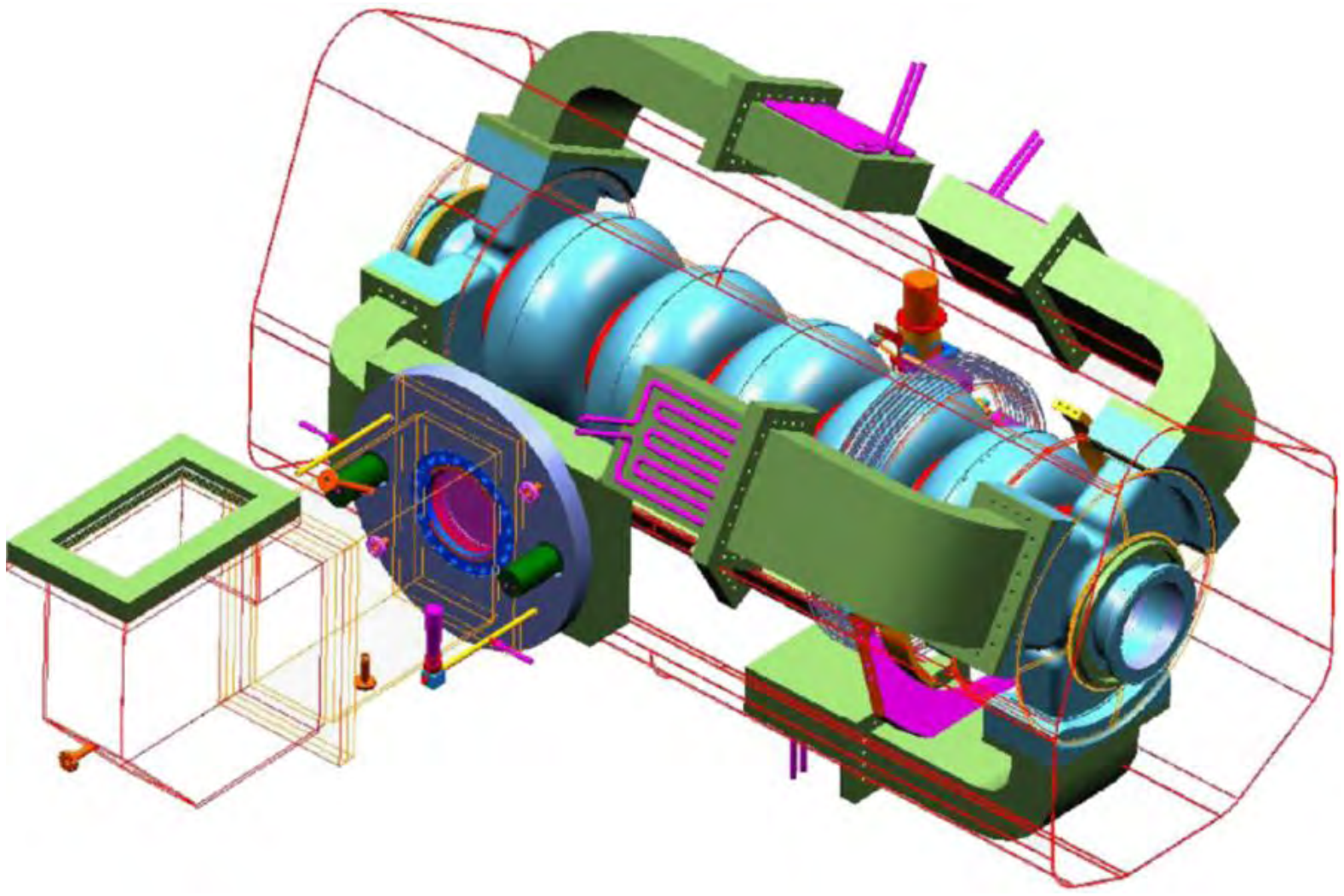

**Figure 10.15. Six waveguide HOM couplers are attached to a five-cell SC cavity developed by JLab (Jefferson Laboratory) to ensure heavy damping of all polarizations of modes. The terminating loads are at room temperature. [[10.37]; Available under Creative Common Attribution 3.0 License (www.creativecommons.org/licenses/by/3.0/us/) at www.JACoW.org.]**

Above the cut-off of the waveguide, all coupled modes from the cavity can propagate to the load. The terminating loads at the end of the waveguide can be used at room temperature even in SC cavities. Ferrite loads may be applied in vacuum when using special materials with low outgassing rates. In separating the cavity's vacuum and the load with a window, we need to remember that most ceramics have limited transmission characteristics at the very high frequencies such as those contained in the HOM spectrum of an injector for short electron bunches.

The absorbing waveguide load is separated from the cavity and can be maintained at room temperature even in SC cavities making the waveguide couplers a good choice in applications with very high HOM power loads. Flanges and windows should be placed far enough from the cavity to ensure that the power level of the fundamental mode of the cavity is low at both.

### 10.5.2 Coaxial HOM Couplers

A coaxial HOM coupler consists of an inductive coupled pick-up loop, or a capacitive coupled antenna to transfer the energy of the HOMs from the cavity. These antennas might be resonant fingers for damping single modes, or broadband matched couplers, feeding the HOM power to an external load *via* cable. In all cases, it is essential to reject the fundamental frequency using a filtering unit, which makes their design more difficult. This rejection filter must be carefully tuned, typically to rejection values of -70 dB or better. The HOM dampers are attached directly at the cavity cell in NC cavities, or at the beam pipe close to the cavity for SC cavities.

Figure **10.16** illustrates the HOM coupler with double fundamental-mode rejection filter used at KEK [10.38]; it is a development of the TESLA-type HOM coupler [10.39] having a single rejection filter. An antenna in the beam tube close to the cavity picks up the HOM energy. The geometry of the pickup loop results in a rejection filter for the cavity's fundamental frequency. The filter frequency is tuned by changing (deforming) the distance from the far end stub to the HOM filter housing. The HOM energy is coupled with a feedthrough into a cable. The TESLA HOM coupler is cooled solely by the heat conduction of the niobium beam pipe. For higher CW applications, this type of cooling is insufficient so that the inner conductor warms up and eventually quenches, causing Q-switches in the cavity. CW operation is made possible by using sapphire as an insulating material [10.40]; its high thermal conductivity leads to better cooling of the inner conductor.

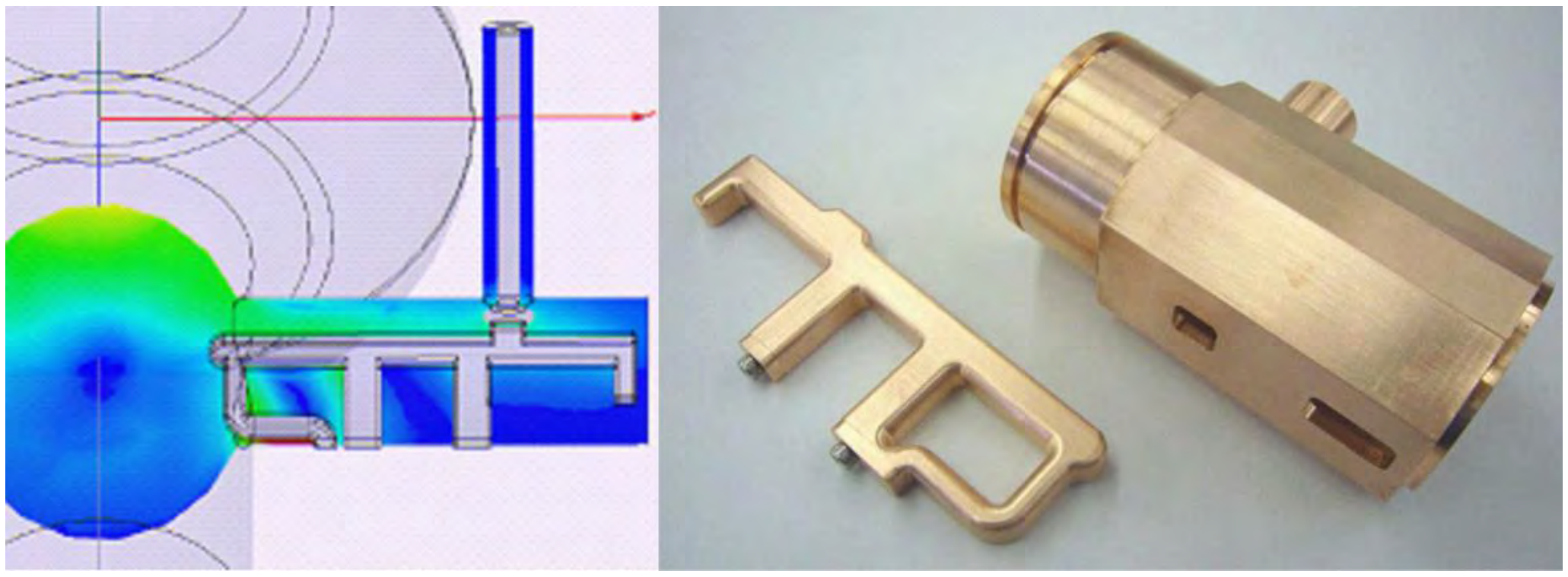

**Figure 10.16. Coaxial HOM couplers developed by KEK that employ a double notch filter to increase the rejection of the accelerating mode of the cavity. The colors in the left picture symbolize the magnetic field strength. [[10.38]; Available under Creative Common Attribution 3.0 License (www.creativecommons.org/licenses/by/3.0/us/) at www.JACoW.org.]**

Coaxial HOM couplers need to match the required frequency characteristics to fulfill multipacting restrictions and to account for boundary conditions by power limitations. This means that extensive electromagnetic- and thermal-calculations are a prerequisite. We note that the ceramics or sapphires at the feed through have restricted transmission characteristics at very high frequencies. The frequency limitation of sapphire is at about 6 GHz! The space consumption of a coaxial HOM coupler is low. Its power capability mostly is limited by the feed through, attached cable and the cooling applied.

### 10.5.3 Beam Pipe HOM Couplers

Beam pipe HOM couplers afford another option for superconducting cavities. The beam pipe opening at a SC cavity is large enough to act as a waveguide above cutoff for HOM frequencies while rejecting the fundamental mode. The beam pipe houses high power ferrite loads at liquid nitrogen- or room-temperature to damp the propagation. Due to the fact that the beam pipe is acting as a waveguide to the HOM, the cutoff of the beam pipe has to be low such that all HOM modes can propagate to the absorbers. It is important to ensure that the dipole modes at lower frequencies can propagate into the beam pipe. Because the beam pipe opening of NC cavities is small and thereby the cutoff is high, beam pipe HOM absorbers are not applicable for NC cavities.

Using different ferrite materials achieves smooth damping characteristics over a wide frequency range. Figure **10.17(a)** [10.41] shows a beam pipe ferrite absorber developed at Cornell University. Three different ferrite materials are used to give overall good damping characteristics up to 40 GHz.

The power dissipation capability of beam pipe absorbers depends highly upon the mass of ferrite used. The drawback of beam pipe absorbers is their longitudinal space consumption, *viz.* lowering the filling factor of modules. A second problem occurs when cooling the ferrites to the temperature of liquid nitrogen to reduce static losses in SC systems; the conductivity of ferrites drops, resulting in electrostatic charging by the passing beam, discharges and strong kicks to the beam. So far, there is no solution, although research is underway. Figure **10.17(b)** and Figure **10.17(c)** show a ferrite ring operated in the FLASH accelerator [10.42]. The designed power capability is 100 W and frequency ranges up to tens of gigahertz.

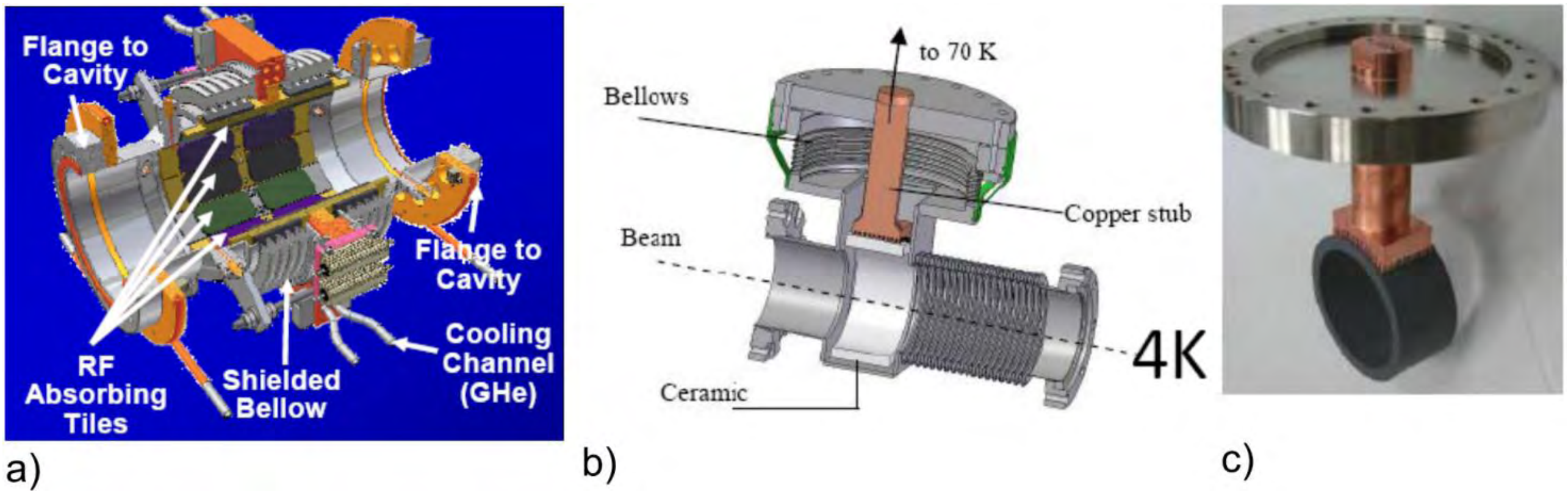

**Figure 10.17. Beam pipe HOM coupler using ferrites and ceramics. (a): Three different materials are used to cover a wide frequency range. (b) and (c): A ceramic ring adapted to the beam pipe diameter. [[10.41]; Adapted under Creative Common Attribution 3.0 License (www.creativecommons.org/licenses/by/3.0/us/) at www.JACoW.org.] [[10.42]; Available under Creative Common Attribution 3.0 License (www.creativecommons.org/licenses/by/3.0/us/) at www.JACoW.org.] [[10.42]; Adapted under Creative Common Attribution 3.0 License (www.creativecommons.org/licenses/by/3.0/us/) at www.JACoW.org.]**

## 10.6 LOW-LEVEL RF CONTROL FOR PHOTOINJECTOR RF CAVITIES

In very general terms, the major task of the LLRF is to maintain a constant energy gain of the beam in the accelerating cavities. This is achieved by measuring the field level of the accelerating mode in the cavity at a given phase of the bunched beam with respect to the field and determine the deviation of that value to a given reference value. Finally, we must assure the compensation of that deviation by altering the low power signal that drives the RF power amplifier.

There are four major types of application:

1. Normal conducting cavities, *e.g.* made from copper, for high power RF operation at low duty factor, pulsed mode to accelerate a beam organized in bunch trains, as in the X-FEL project [10.6]
2. Normal conducting cavities at low- to intermediate-field levels at high duty factor for high current application [10.43]
3. Superconducting cavities for low beam current CW operation, *e.g.* the ELBE FEL [10.43]
4. Superconducting cavities for high beam current CW operation being developed for future ERL projects [10.44]

Two major questions drive the design of the LLRF system: First, what is the needed amplitude and phase stability of the field compared with the reference? This query leads to the issue of the required accuracy with which the field vector is to be sampled and processed. Second, what is the bandwidth of the system to be controlled? Is it more a broadband NC cavity driven in pulsed mode, where the pulse’s ramping time and flat-top properties have to be controlled properly; or, does it pertain to handling a bandwidth in the Hertz regime, such as the ones in low beam current, CW driven, superconducting cavity?

Often, it is useful to start designing the LLRF system by analyzing the possible sources of errors that detune the cavity and entail deviations of the field vector from the reference. This is accompanied by properly evaluating the RF reference system that determines the noise of the LLRF and the jitter and drift between the major components driving the photoinjector, *viz.*, the laser system and the controlled accelerating field. A helpful tool for setting the boundary conditions for stable operation within requirements is to mathematically simulate the RF cavity and the feedback system, including the known or modeled error sources [10.45].

Keeping the power requirements (Equ. 10.4) in mind, we summarize the following error sources for field stability of RF cavities:

**NC Cavity:**

- In comparison to a SC cavity, NC cavities have low intrinsic quality factor of $Q_0 \approx 10^4$ due to their high surface resistance. For an L-band cavity, this corresponds to a natural bandwidth of several kilohertz. Nevertheless, due to the high ohmic-wall losses, the cavity must be cooled and any shift in temperature detunes it. Values of 23 kHz per degree Celsius were measured for a NC one-and-a-half-cell photoinjector cavity [10.46].
- In high power pulsed operation, overshoots and oscillations may occur that must be compensated by fine tuning the feedback parameters.

**SC Cavity, Low Beam Loading, High Power Pulsed Mode:**

- The components of the electric- and magnetic-fields of the standing RF wave interact with the currents in the cavity wall, exerting a net force on the cavity wall, the so-called Lorentz force detuning [10.47]. Usually, detuning is about 1-3 Hz $m^2$ $MV^{-2}$, that detunes a cavity operating at 35 MV $m^{-1}$ by more than a kilohertz. In the pulsed mode, this is comparable to a step response of the mechanical cavity system. Thus, over the length of the pulse, decaying mechanical eigenmodes alter the field level; an effect which can only partly be compensated by tuning systems [10.48].

**SC Cavity, Low Beam Loading, CW Operation:**

- The external quality factors of cavities operated at CW, with low beam loading, are optimized to cope with the external mechanical cavity-detuning. In CW operation, microphonic detuning is the major source of errors in field stability. Mechanical oscillations from the vacuum pump and the liquid helium supply are coupled *via* beam pipes or the helium vessel to the cavity's mechanical eigenmodes. Even a nanometer deformation shifts the frequency by 0.3 Hz [10.1].
- The thin-walled niobium cavities are very susceptible to variations in helium pressure. Detunings of 1.46 Hz $Pa^{-1}$ due to pressure dependencies were measured [10.49]. Fast piezo-based tuning schemes and passive stiffening systems are needed to minimize this source of error.
- Even in the CW mode, the Lorentz force detuning may create instabilities. Residual field fluctuations from strong microphonics cause further detuning due to the Lorentz forces. If there is insufficient power in the control loop, the cavity field may trip, the so-called ponderomotive instability that appears in generator driven loops [10.50], [10.51].

**Beam Related Effects:**

- An electron bunch passing through the cavity extracts energy from the field proportional to its charge as a function of the RF phase. This can be measured as a transient in the field's amplitude. Due to shot-to-shot variations in laser intensity, charge fluctuates, leading to a linear increase of the beam's energy spread with that fluctuation. This effect depends on the field decay time constant, the gain of the RF feedback loop and the bunch's repetition rate.
- Time jitter of the laser with respect to the reference and phase errors of the controlled RF field might produce an increase of the beam's energy spread after the injector's cavity. Even for electron beams that are not fully relativistic, RF phase dependent velocity bunching occurs that may increase the timing jitter of the beam in the drift sections after the injector.

**RF Control Related Effects:**

- The measurement of the cavity field at the pick-up probe, or *via* the reflected power signal, may be falsified by measurement noise, drifts, noise in the RF cabling and nonlinearities in the RF mixers.
- The total noise of a system limits its performance. During high gain operation, amplified noise creates a gain dependent increase in field errors; hence, noise limits field control through the gain of the system. Thus, in addition to using low noise hardware throughout the entire reference system, including all oscillators and the clock generation for the LLRF, must be optimized for low phase noise.
- In RF amplifiers, the gain and phase response of the output wave alter as a function of the driving input signal. This effect is predictable, and thus can be compensated by using feedforward tables, or an additional feedback loop [10.52].
- Fast fluctuations of the power supply, *e.g.* the klystron cathode, causes a ripple in the output power due to velocity effects of the beam in the klystron's cavity and drift sections. Effectively, this acts as phase noise modulation on the forward power. Compensation is achieved by measuring the klystron ripple and implementing an additional feedback loop [10.53].
- Noise in the RF feedback loop may excite unwanted modes of the cavity pass-band, as was reported for TESLA cavities [10.54] during vertical acceptance tests, and also during LLRF operation at a horizontal test stand at HZB. This issue depends on the phase of the other pass-band modes with reference to the π-mode in the cell where the field is measured with a pick-up antenna. Furthermore, the total time delay of the feedback loop, together with that phase dependence, sets the boundary for stable operation [10.55].

Mathematically, all these effects are summarized in the following formulas. $E_{kin}$, the energy gained by the particle bunch, is a function of the experienced accelerating field $E_{acc}$ given by its current longitudinal position, $z$, and the elapsed time $t$ since its extraction from the photocathode,

$$E_{acc}(t,z) = [E_z(z)\,(1 + \sigma_A)]\cos\left(\Phi_b + \sqrt{\sigma_{\Phi_{inj}}^2 + \sigma_\Phi} + \omega t\right) \tag{10.7}$$

$$E_{kin}(t,z) = e\int E_{acc}(t,z)\,dz \tag{10.8}$$

with $\Phi_b$ the injection phase with respect to the RF field with frequency $\omega$, the phase error due to the laser jitter, $\sigma_{\Phi_{inj}}$, and the phase- and relative-amplitude errors of the RF field, $\sigma_\Phi$ and $\sigma_A$, respectively. $e$ is the particle's charge. The LLRF system needs to resolve these field errors and compensate for them by altering the forward wave coupled to the cavity. Time jitter of the laser system at least can be minimized by assuring proper synchronization between the RF and laser system.

### 10.6.1 RF Control Strategies: Field and Tuning Control

Modern field control systems implement a huge variety of functions to operate, control and survey a cavity. To allow for flexible implementation of filters, feedback- and feedforward- algorithms, and rather complex codes to cope with the complex mechanical cavity transfer function; nowadays, most LLRF systems are built on digital hardware. Analog systems still are in operation and often special fast loops, *e.g.* for klystron control, are sometimes implemented as an analog circuit. However, with the availability of fast ADC converters with a more than 14-bit resolution, most modern systems rely on digital implementation. It is easier to incorporate the LLRF system in the accelerator's control system, supporting well correlated measurements between the beam's properties and the stability of the LLRF cavity.

Furthermore, digital hardware is easy to configure with several modes of operation, remotely controllable and eases changing of the system's parameters without needing to exchange hardware components.

Nowadays, there are a huge variety of digital LLRF systems operating worldwide [10.52], [10.53], [10.56]–[10.58]. Mostly, they were developed for operation of main linac- or storage ring-cavities, even allowing the vector sum control of up to eight cavities [10.59]; they were not specifically designed to operate photoinjector cavities. Nevertheless, due to their intrinsic flexibility, it is easy to adapt them to controlling the injector. Depending on the personnel available at a laboratory, the choice is between commercial systems based on Digital Signal Processing (DSP) boards [10.56], [10.60], or custom design, based on their own hardware solutions (*e.g.* [10.53]).

The LLRF feedback loop itself mostly is implemented as a standard Proportional-Integral (PI) controller, as described in any control theory textbook (*e.g.* [10.61]) and is a rather simple algorithm. For good performance, it must be fast, with a high resolution better than 14 bits sampling ADCs. This is essential to achieve a small group delay, below a microsecond or even several hundred nanoseconds. A high sampling rate is mandatory for a small phase lag of the loop to allow a high gain margin. High-bit resolution is required to resolve small field deviations, so supporting control in the < 1/1000 regime. Also, the system should enable a fast change of control parameter during operation and a deterministic handling of the processed signals, the prerequisite for a numerical stable time-discrete system.

A Field Programmable Gate Array (FPGA) chip can meet all these features [10.62]. The core feedback functions of nearly all modern LLRF systems are performed by an FPGA-based digital system. It directly accesses the sampled signals of the fast ADCs and often features some SRAM memory blocks to load different algorithms to the FPGA or supports the transfer of parameters and data. For general operability, the FPGA is embedded into some general DSP system that houses control loops operating on a slower time scale (as cavity tuning), the communication to the accelerator control system (*e.g.* EPICS [10.63]), a state machine or similar to monitor the cavity state.

In the following subsection, the set-up and functioning of a digital LLRF system is explained in the context of a generic system.

### 10.6.2 A Generic System

An LLRF field control system has two major components: the analog RF hardware consisting of RF mixers, vector modulators, low-pass and band-pass filters, and small amplifiers. The digital hardware including the fast ADCs and DACs, onboard memory, slow ADCs/DACs, and FPGA, plus DSP chips for the signal processing and controller implementation. Furthermore, there may be auxiliary hardware to record fast interlock events, amplifiers for piezo tuners and stepping motor control for driving mechanical cavity tuner or plunger tuner.

The analog RF hardware has the major task of providing the control loop with low power signals of all measurable power signals describing the state of a cavity. These are the forward power provided by the high power amplifier to the cavity *via* the fundamental power coupler, the power reflected at the cavity due to (wanted) impedance mismatch, and finally, the transmitted power to measure the field vector. As L-band cavities operate at 1.3 GHz, several gigahertz ADCs with 14-bit resolution are needed to measure the field. The analog RF hardware thus is often referred to as the down-converter. It transforms the cavity frequency by mixing it with a reference to a measurable intermediate frequency (IF) that still contains the amplitude and phase information of the accelerating mode. This is a valid procedure, as all error sources changing the cavity field happen on a slower timescale than the cavity frequency; the LLRF system, therefore, measures the envelope of the field. Lower frequency structures (some hundred megahertz) are controlled by directly sampling the field vector.

#### *10.6.2.1 Principle of Field Detection*

The amplitude and phase of the field often is measured by sampling the real- and imaginary-components of the RF signal by the intermediate step of down-conversion to a lower frequency. This method often is termed IQ demodulation, quadrature demodulation, or complex demodulation. I and Q represent the "In-Phase" and "In-Quadrature," respectively, and are the complex representation of the RF signal. To reconstruct the signals' content in accordance with the Nyquist criterion, it should be sampled at twice the rate of the highest frequency component. Therefore, the cavity pick-up signal is down-converted to an IF, still containing the side-band information of the cavity's field variations.

The demodulation contains three steps:

- down-conversion or down-mixing by an RF mixer circuit
- low-pass filtering to remove the higher frequency content of the mixing process resulting in the IF
- IQ sampling of the IF, *e.g.* by oversampling to resolve the field's I and Q components

Figure **10.18** shows the principle of the IQ detection by oversampling of the RF signal four times. The sinusoidal IF signal is sampled every $\pi/2$ in the unity circle representation. A measurement of the real amplitude at that time gives, as a consecutive data stream, the components I, Q, -I, -Q, I, and so on. As the sampling starts randomly at some phase, the feedback loop itself is a phase delay and all IQ components have to be shifted to the first quadrant of the unity circle; also, the IQ components are rotated in phase by a two-by-two rotation matrix to supply the digital control algorithm with the real- and imaginary-field components.

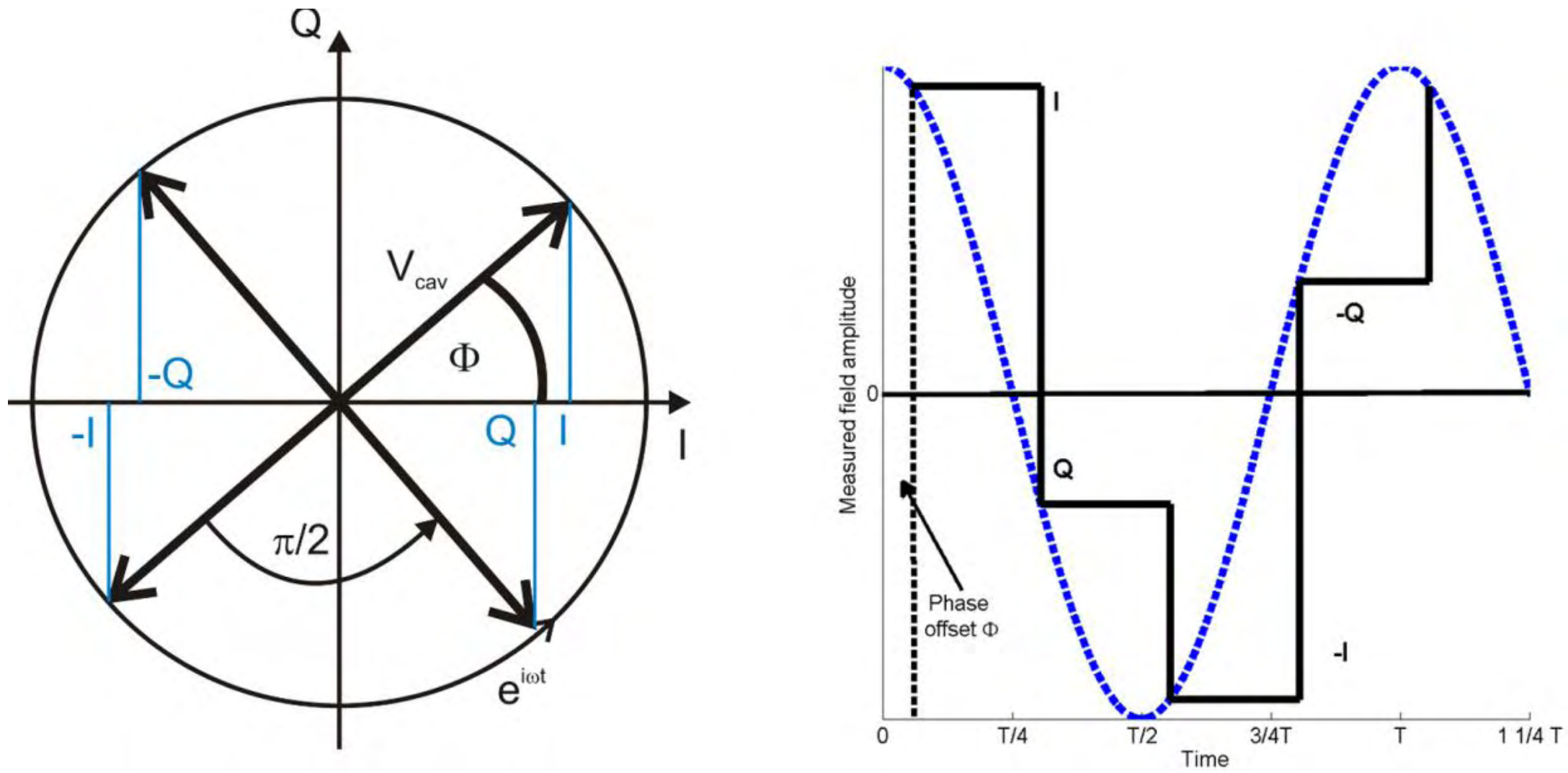

**Figure 10.18. Principle of detecting the field vector component (IQ detection) by four times oversampling of the IF signal following down-conversion. Each consecutive sample is shifted by $\pi/2$ in phase, thus representing the In-Phase and In-Quadrature component of the field. Every two samples thus deliver information to reconstruct the field vector. The measurement starts to sample field components at an arbitrary phase offset $\Phi_0$, which has to be corrected for by determining the loop's phase shift.**

After the signal is processed by the feedback in the FPGA, there are two ways to reconstruct the phase and amplitude corrected output signal to the cavity: the IQ components are fed *via* two individual DACs to a vector modulator varying the forward power-wave; or, the IQ components internally modulate a digital up-converter, similar to a vector modulator, and the output IF after the DAC is mixed with the shifted LO signal to the 1.3 GHz forward-power signal.

### *10.6.2.2 An Example*

Figure **10.19** shows a generic field control system for a photoinjector RF cavity. The RF reference and clock system supplies all analog signals to operate the cavity and LLRF system: these are the RF cavity's fundamental mode; a frequency, shifted by the IF frequency, for mixing the three power signals of the cavity down to the IF; the sampling clock frequency of the ADCs (often four times the IF); and, the reference frequency to lock the laser to the cavity RF. For example, assuming a cavity operating at 1.3 GHz, the reference system would provide a 1.3 GHz signal twice, at least three 1.28 GHz signals for an IF of 20 MHz, and a clock signal of 80 MHz.

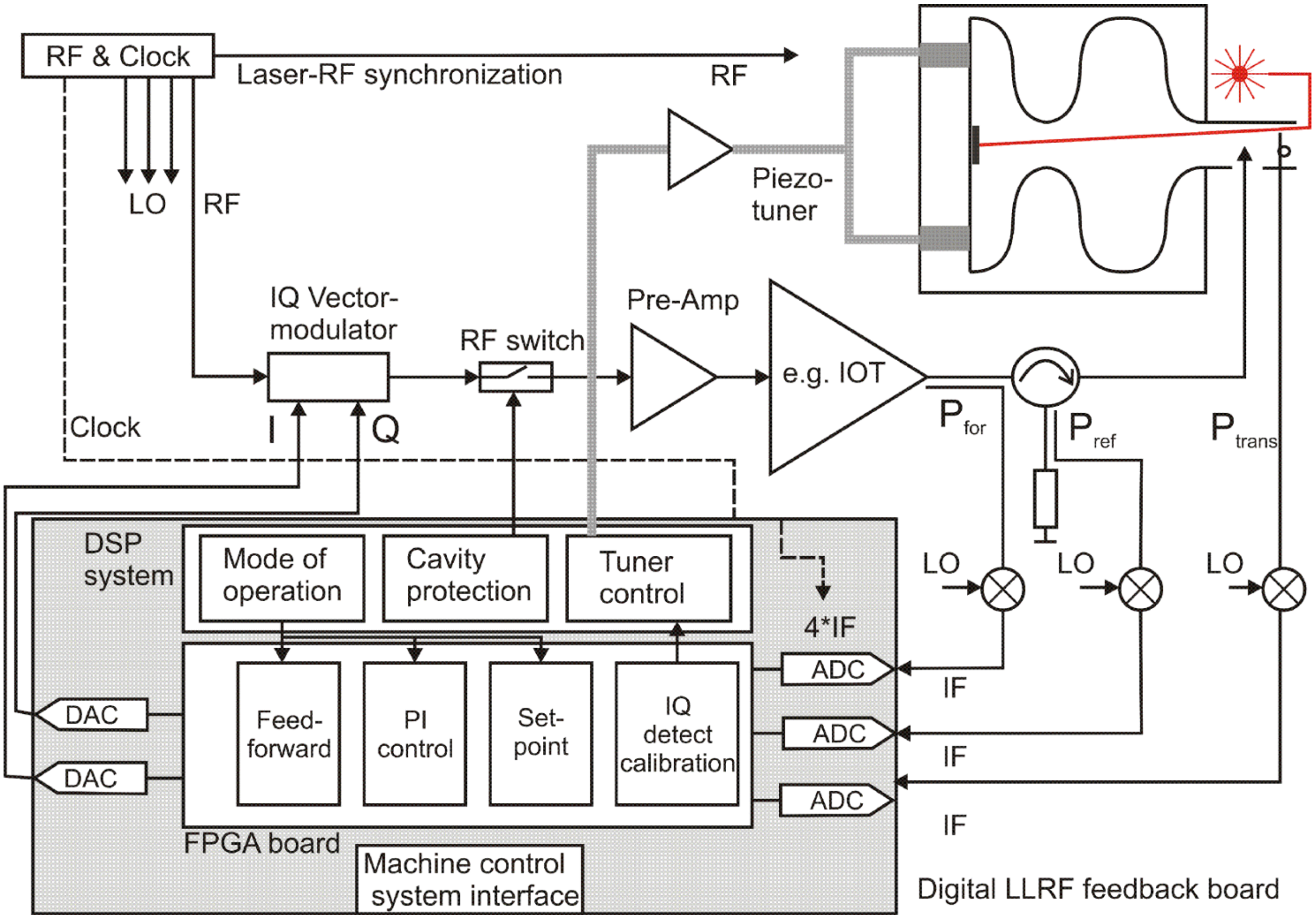

**Figure 10.19. Layout of a generic digital single-cavity RF control system in a generator-driven set-up. It incorporates a fast field programmable-gate array, a digital signal processing unit to detect cavity RF interlock events, fast- and slow- tuning control algorithms and general procedures for different modes of operation. Further, there is an interface to the machine-control system. The reference for the field-control system is the RF and clock generation system that also synchronizes the photoinjector's laser system.**

The 1.3 GHz forward signal from the reference system drives the cavity. With a vector modulator, it can be shifted in amplitude and phase by the LLRF system, providing the In-Phase and In-Quadrature components for vector modulation to correct the measured field level. In addition, there is some RF switch, *e.g.*, a pin diode, to switch the forward power during an interlock event. This signal then is amplified *via* a pre-amplifier by the main high power unit, to supply the cavity with the needed forward power. As reflections would destroy the IOT or klystron, a circulator directs the reflected power to a water-cooled load.

The power signals usually are obtained by weakly coupling to the transmission line with a directional coupler. The level of the transmitted power obtained by the pick-up antenna intrinsically is low; only weak coupling is needed to measure the field. These signals are down-converted at RF mixers to the IF. The In-Phase- and In-Quadrature-field components are detected by four times oversampling, thus providing the LLRF system every two sampling steps with the full field vector, or amplitude and phase. In this example, the IQ pair of values would be updated at a rate of a 40 MHz. Depending on the bunch's repetition rate, a much lower data rate often is sufficient for control, such that an initial filtration can be achieved by averaging over 40 samples, for example. To reject the excitation of other modes of the pass-band, the transmitted power IQ signal can be processed by special notch filters at the mode's frequency and, in general, low-pass filtered. The amplitude and phase of the filtered signal is compared to those of the desired set point and the resulting error signal is amplified by the PI control gain settings.

In pulsed operation or repetitive errors, often a feedforward value is added to the amplified control signal. This IQ pair finally is connected to the vector modulator, *via* the DACs, to steer the forward power, thereby to close the feedback loop.

In parallel to the FPGA for LLRF feedback, there may be further DSP systems that contain feedback and feedforward loops for tuning control and receiving the tuning angle of the cavity from the LLRF loop. Furthermore, all kinds of algorithms may be programmed on that DSP system to change automatically loop parameters on the FPGA, or to implement different modes of operation, such as ramping the field in CW-operated cavities by using the tuner.

### 10.6.3 Steps towards an LLRF Control

As this part of the book only can give a short introduction into LLRF control design and operation, the following list offers an overview of how to prepare a LLRF feedback system for operating the cavity.

1. Evaluate the required performance of the LLRF control system by calculating the needed phase and amplitude stability of the cavity's accelerating mode. This is obtained by tolerance studies of particle tracking codes.
2. Consider the frequency of the cavity's fundamental mode and further modes in the pass-band. How close is the next pass-band mode to the accelerating π-mode?
3. Determine the cavity's bandwidth and simulate the error sources that could detune the cavity mode (for CW or pulsed operation) to gain insights to the physics that define boundary conditions for a stable operation.
4. Plan tuning schemes operating parallel to the LLRF feedback loop.
5. Assess the needed loop bandwidth to compensate for detuning effects and beam loading effects; consider the beam's repetition rate.
6. Perform a stability analysis using methods of control theory to determine the attainable margins in gain and phase; especially determine a goal for the loop's group delay, and thus the needed sampling time constant of the time-discrete system.
7. Sketch a field control design including LLRF control, tuner control and necessary interlock events to protect the cavity and modes of operation, which will be useful for photoinjector operation.
8. Appraise the available hardware, whether to buy a fully integrated commercial system or to choose a full in-house development. Especially consider ADC/DAC sampling rates, bit resolution and the number of NAND gates of the FPGA chip.
9. Choose a programming code; *i.e.*, closer to FPGAs, such as VHDL or Verilog, or some higher programming language with the needed translation tools, like MATLAB™.

10. Prepare for the interface between the local LLRF system and the accelerator's control system.
11. Design and build a low-noise plus low-drift RF clock and reference system.
12. Program and test the feedback code supported by LLRF simulations at a low power level using a known dummy cavity. A test using signal sources is useful for first tests.
13. Once these steps are completed, start with low power tests at low cavity field amplitudes to test the full system to identify any further inconsistencies.

## 10.7 CONFLICT OF INTEREST AND ACKNOWLEDGEMENT

We confirm that this article content has no conflicts of interest and would like to acknowledge the support of *Deutsches Bundesministerium für Bildung und Forschung* and *Land Berlin*. We would also like to thank H. Bohlen from CPI, G. Blokesch from PPT, E. Weihreter, H.G. Hoberg, and Jens Knobloch from HZB and Ti Ruan from Soleil for fruitful discussions.

***References***

[10.1] A. Neumann, W. Anders, O. Kugeler *et al.*, "Analysis and active compensation of microphonics in continuous wave narrow-bandwidth superconducting cavities," *Phys. Rev. ST Accel. Beams*, vol. 13, pp. 082001-1–082001-15, August 2010.

[10.2] E. Buechner, F. Gabriel, E. Grosse *et al.*, "The ELBE-Project at Dresden-Rossendorf," in *Proc. 2000 European Particle Accelerator Conf.*, 2000, pp. 732-734.

[10.3] M. Abo-Bakr, W. Anders, T. Kamps *et al.*, "BERLinPro - a prototype ERL for future synchrotron light sources," in *Proc. 2009 Superconducting RF Conf.*, 2009, pp. 223-227.

[10.4] P. Emma, R. Akre, J. Arthur *et al.*, "First lasing and operation of an ångstrom-wavelength free-electron laser," *Nature Photonics*, vol. 4, pp. 641-647, August 2010.

[10.5] A. Fabris, P. Craievich, P. Delgiusto *et al.*, "Progress of the S-Band RF systems for the FERMI@ELETTRA Linac," in *Proc. 2009 Particle Accelerator Conf.*, 2009, pp. 2039-2041.

[10.6] M. Altarelli, R. Brinkmann, M. Chergui *et al.*, Eds., *The European X-Ray Free-Electron Laser*, Technical design report, Hamburg: DESY XFEL Project Group, 2006.

[10.7] B. Dwersteg, K. Flöttmann, J. Sekutowicz *et al.*, "RF gun design for the TESLA VUV free electron laser," *Nucl. Instrum. Meth. A*, vol. 393, pp. 93-95, July 1997.

[10.8] S. Belomestnykh, Z. Conway, J. Dobbins *et al.*, "CW RF systems of the Cornell ERL injector," in *Proc. 2008 Linear Accelerator Conf.*, 2008, pp. 857-859.

[10.9] I. Ben-Zvi *et al.*, "The ERL high-energy cooler for RHIC," in *Proc. 2006 European Particle Accelerator Conf.*, 2006, pp. 940-944.

[10.10] A. V. Haeff, "Electron Discharge Device," U.S. Patent 2,284,733, June 2, 1942.

[10.11] L. S. Nergaard, "Electron Discharge Device," U.S. Patent 2,329,778, September 21, 1943.

[10.12] M. Di Giacomo, "Solid state RF amplifiers for accelerator applications," in *Proc. 2009 Particle Accelerator Conf.*, 2009, pp. 757-761.

[10.13] M. Langlois, J. Jacob and J. M. Mercierl, "Cavity combiners for transistor amplifiers," in *6$^{th}$ Workshop CW High Average Power RF*, 2010.

[10.14] P. Pérez, B.Baricevic, P. Sánchez *et al.*, "High power cavity combiner for RF amplifiers," in *Proc. 2006 European Particle Accelerator Conf.*, 2006, pp. 3215-3217.

[10.15] H.-G. Hoberg, W. Anders, A. Heugel *et al.*, "Highly stable IOT based 1.3 GHz/16 kW RF power source for CW operated SC Linacs," in *Proc. 14$^{th}$ Int. Conf. RF Superconductivity*, 2009, pp. 679-682.

[10.16] K. Buerkmann, M. Abo-Bakr, W. Anders *et al.*, "Status of the Metrology Light Source," in *Proc. 2006 European Particle Accelerator Conf.*, 2006, pp. 3299-3301.

[10.17] International Study Group, "International study group progress report on linear collider development," SLAC-Report-559, 2000, Chapter 4.
[10.18] G. Blokesch, private communication, 2010.
[10.19] W. Kaesler, "A long-pulse modulator for the TESLA test facility (TTF)," in *Proc. 2004 Int. Linear Accelerator Conf.*, 2004, pp. 459-461.
[10.20] G. E. Dale, H. C. Kirbie, W. B. Haynes *et al.*, "Design, and application of a diode-directed solid-state marx-modulator," in *Proc. 15$^{th}$ Pulsed Power Conf.*, 2005, pp. 1211-1214.
[10.21] D. Bortis, J. Biela and J. W. Kolar, "Active gate control for current balancing in parallel connected IGBT modules in solid state modulators," in *2007 Pulsed Power Conf.*, 2007, pp. 1323-1326.
[10.22] Puls-Plasmatechnik GmbH, Feldstraβe 56, 44141, Dortmund, Germany.
[10.23] B. Rusnak, "RF Power and HOM Coupler Tutorial," in *11$^{th}$ Int. Workshop RF Superconductivity*, 2003.
[10.24] S. Belomestnykh, "Overview of input power coupler developments, pulsed and CW," in *Proc. 13$^{th}$ Int. Workshop RF Superconductivity*, 2007.
[10.25] M. Champion, "RF input couplers and windows: Performances, limitations, and recent developments," in *Proc. 1995 Workshop RF Superconductivity*, 1995, pp. 195-221.
[10.26] H. Padamsee, *RF Superconductivity*, Weinheim: Wiley-VCH Verlag, 2009.
[10.27] V. Nguyen, H. L. Phillips and J. Preble, "Development of a 50 kW CW L-band rectangular window for Jefferson Lab FEL cryomodule," in *Proc. 1999 Particle Accelerator Conf.*, 1999, pp. 1459-1461.
[10.28] V. Veshcherevich and S. Belomestnykh, "Correction of the coupling of CESR RF cavities to klystrons using three-post waveguide transformers," LEPP Cornell University, Ithaca, NY, Report SRF020220-02, 2002.
[10.29] R. Rimmer, E. F. Daly, W. R. Hicks *et al.*, "The JLab ampere-class cryomodule," in *Proc. 2001 Workshop RF Superconductivity*, 2001, pp. 567-570.
[10.30] J. Preble, "CEBAF energy upgrade program including re-work of CEBAF cavities," in *Proc. 2007 Workshop RF Superconductivity*, 2007, pp. 756-760.
[10.31] V. Veshcherevich, S. Belomestnykh, P. Quigley *et al.*, "High power tests of input couplers for Cornell ERL injector," in *Proc. 2007 RF Superconductivity*, 2007, pp. 517-519.
[10.32] W.-D. Möeller for the TESLA Collaboration, "High power coupler for the TESLA test facility," in *Proc. 1999 Workshop RF Superconductivity*, 1999, pp. 577-581.
[10.33] W. Anders, J. Knobloch, O. Kugeler *et al.*, "CW operation of superconducting TESLA cavities," in *Proc. 2007 Workshop RF Superconductivity*, 2007, pp. 222-226.
[10.34] S. Noguchi, E. Kako, M. Sato *et al.*, "Present status of superconducting cavity system for CERL injector linac at KEK," in *Proc. 2010 Int. Particle Accelerator Conf.*, 2010, pp. 2944-2946.
[10.35] H. Nakai, K. Akai, E. Ezura *et al.*, "Status report of superconducting RF activities in KEK," in *Proc. 2001 Workshop RF Superconductivity*, 2001, pp. 76-80.
[10.36] J. Sekutovicz, "HOM damping and power extraction from superconducting cavities," in *Proc. 2006 Linear Accelerator Conf.*, 2006, pp. 506-510.
[10.37] R. A. Rimmer, G. Ciovati, E. F. Daly *et al.*, "The JLAB ampere-class cryomodule conceptual design," in *Proc. 2006 European Particle Accelerator Conf.*, 2006, pp. 490-492.
[10.38] K. Watanabe, H. Hayano, S. Noguchi *et al.*, "New HOM coupler design for ERL injector at KEK," in *Proc. 2007 Workshop RF Superconductivity*, 2007, pp. 530-535.
[10.39] J. Sekutovicz, "Higher order mode coupler for TESLA," in *Proc. 6$^{th}$ Workshop RF Superconductivity*, 1993, pp. 426-439.
[10.40] C. E. Reece, E. F. Daly, T. Elliott *et al.*, "High thermal conductivity cryogenic RF feedthroughs for higher order mode couplers," in *Proc. 2005 Particle Accelerator Conf.*, 2005, pp. 4108-4110.

[10.41] V. Shemelin, P. Barnes, B. Gillett *et al.*, "Status of HOM load for the Cornell ERL injector," in *Proc. 2006 European Particle Accelerator Conf.*, 2006, pp. 478-480.
[10.42] J. Sekutowicz, A. Gössel, N. Mildner *et al.*, "Beam tests of HOM absorber at FLASH," in *Proc. 2010 Int. Particle Accelerator Conf.*, 2010, pp. 4092-4094.
[10.43] A. Todd, "State-of-the-art electron guns and injector designs for energy recovery linacs (ERL)," *Nucl. Instrum. Meth. A,* vol. 557, pp. 36-44, November 2006.
[10.44] A. Arnold, H. Büttig, D. Janssen *et al.*, "Development of a superconducting radio frequency photoelectron injector," *Nucl. Instrum. Meth. A*, vol. 577, pp. 440-454, July 2007.
[10.45] A. Neumann and J. Knobloch, "Cavity and linac RF and detuning control simulations," in *Proc. 2007 Workshop RF Superconductivity*, 2007, pp. 627-631.
[10.46] F. Marhauser, "High power tests of a high duty cycle, high repetition rate RF photo-injector gun for the BESSY FEL," in *Proc. 2006 European Particle Accelerator Conf.*, 2006, pp. 68-70.
[10.47] J. C. Slater, "Microwave electronics," *Rev. Modern Physics*, vol. 18, pp. 441-512, October 1946.
[10.48] M. Liepe, W.-D.Möeller and S. N. Simrock, "Dynamic lorentz force compensation with a fast piezoelectric tuner," in *Proc. 2001 Particle Accelerator Conf.*, 2001, pp. 1074-1076.
[10.49] A. Neumann, W. Anders, M. Dirsat *et al.*, "CW superconducting RF photoinjector development for energy recovery linacs," in *Proc. 2010 Linear Accelerator Conf.*, 2010, pp. 998-1000.
[10.50] D. Schulze, "Ponderomotorische stabilität von hochfrequenzresonatoren und resonatorregelungssystemen," Ph.D. Thesis, Kernforschungszentrum Karlsruhe, Karlsruhe, Germany, 1972.
[10.51] A. Mosnier, "RF feedback systems for SC cavities," in *Proc. 1998 European Particle Accelerator Conf.*, 1998, pp. 174-178.
[10.52] P. Baudrenghien, G. Hagmann, J. C. Molendijk *et al.*, "The LHC low level RF," in *Proc. 2006 European Particle Accelerator Conf.*, 2006, pp. 1471-1473.
[10.53] M. Liepe, S. Belomestnykh, J. Dobbins *et al.*, "Experience with the new digital RF control system at the CESR storage ring," in *Proc. 2005 Particle Accelerator Conf.*, 2005, pp. 2592-2594.
[10.54] G. Kreps, A. Gössel, D. Proch *et al.*, "Excitation of parasitic modes in CW cold tests of 1.3 GHz TESLA-type cavities," in *Proc. 2009 Workshop RF Superconductivity*, 2009, pp. 289-291.
[10.55] E. Vogel, "High gain proportional rf control stability at TESLA cavities," *Phys. Rev. ST Accel. Beams*, vol. 10, pp. 052001-1–052001-12, May 2007.
[10.56] A. Salom and F. Perez, "Digital LLRF for ALBA storage ring," in *Proc. 2008 European Particle Accelerator Conf.*, 2008, pp. 1419-1421.
[10.57] S. Michizono, H. Katagiri, T. Matsumoto *et al.*, "Performance of digital LLRF system for STF in KEK," in *Proc. 2008 Linear Accelerator Conf.*, 2008, pp. 1048-1050.
[10.58] T. Allison, K. Davis, H. Dong *et al.*, "CEBAF new digital LLRF System extended functionality," in *Proc. 2007 Particle Accelerator Conf.*, 2007, pp. 2490-2492.
[10.59] V. Ayvazyan, K. Czuba, Z. Geng *et al.*, "LLRF control system upgrade at FLASH," in *Proc. 2010 Personal Computers Particle Accelerator Conf.*, 2010, pp. 150-152.
[10.60] H.-S. Kim, H.-J. Kwon, K.-T. Seol *et al.*, "LLRF control using a commercial board," in *Proc. 2008 Linear Accelerator Conf.*, 2008, pp. 1057-1059.
[10.61] W. L. Brogan, *Modern Control Theory*, 3rd ed., Upper Saddle River: Prentice Hall, 1991.
[10.62] U. Meyer-Baese, *Digital Signal Processing with Field Programmable Gate Arrays*, 2nd Ed., Berlin: Springer-Verlag, 2004.
[10.63] L. R. Dalesio, M. R. Kraimer and A. J. Kozubal, "EPICS Architecture," in *Proc. Int. Conf. Accelerator Large Experimental Physics Control Systems*, 1991, pp. 278-282.

# CHAPTER 11: DIAGNOSTICS

SIEGFRIED SCHREIBER

*Deutsches Elektronen-Synchrotron*
*Notkestraße 85, D-22603*
*Hamburg, Germany*

**Keywords**
Charge Measurement, Faraday Cup, Toroid, Transverse Shape, Optical Transition Radiation, Wire Scanner, Bunch Length Measurement, Longitudinal Bunch Shaping, Streak Camera, Deflecting Cavity, Electro-Optic Sampling, Beam Position Monitor, Energy Measurement, Energy Spread Measurement, Emittance, Slit Mask Technique, Quadrupole Scanning, Tomographic Reconstruction

**Abstract**
Understanding the physics of photoinjectors requires accurate and high resolution measurements of its beam properties. In photoinjectors, the beam is bunched. Critical parameters to be measured are the charge of the bunch, its transverse- and longitudinal-size and shape, its emittance, the energy and energy spread, and its position in the accelerator. This chapter gives an overview of the basic techniques in measuring and monitoring the beam properties.

## 11.1 INTRODUCTION

Photoinjectors produce bunches of electrons that are quickly accelerated to relativistic energies. The injectors often aim for state-of-the art beam properties; for instance, X-ray, free electron lasers (FEL) require a very small emittance and bunches with peak currents at the limit of today's technology. Under these circumstances, measuring the properties of the electron beam is of the utmost importance. The measured data is compared to simulations, and thus provides an essential tool for understanding the underlying physics. In an operating facility, measurements of the beam properties are used to set-up, tune and match the beam optics for further acceleration.

The beam's basic properties are the bunch charge, its transverse- and longitudinal-size and shape, its emittance, energy and energy spread, and its position in the accelerator. In the following chapter, I discuss the most common techniques to measure and monitor these basic parameters. I start with the charge, followed by the transverse- and longitudinal-size and shape, the beam position, and the energy and energy spread. Finally, I discuss measuring the emittance and Twiss parameters. For further reading, I recommend Strehl's comprehensive overview on beam instrumentation and diagnostics [11.1].

## 11.2 BUNCH CHARGE

Electron beams generated by photoinjectors are bunched. The charge of a single bunch typically ranges from a few picocoulombs to a few nanocoulombs, and the number of electrons roughly $10^9$. However, this may vary by orders-of-magnitude depending on the specific application.

RF guns emit dark current. Field emission from surfaces of the gun body or the cathode leads to a spurious current, partly accelerated and detectable on many beam diagnostic devices. Dark current is a background to measurements of electron bunch properties, but also a source of activation of beamline components. In a photoinjector, one would like to know the amount of dark current emitted and accelerated by the RF gun.
Novel, non-destructive dark current monitors are in development. An example is a monitor laid out as a resonant cavity operating with the RF frequency of the accelerator. Dark current is bunched with the RF frequency and thus excites a field in the cavity which is measured with suitable pick-up antennas. A resolution of 40 nC has been achieved. [11.2]. A nice feature of this monitor is that it also measures bunch charges with femtocoulomb resolution. Analyzing the high-order mode signals, an estimate of the bunch length with picosecond resolution is achievable as well.

### 11.2.1 Faraday Cups

A standard way, and also the simplest one, to measure the charge of electron bunches is with a Faraday cup [11.3]. Faraday cups span a wide dynamic range and are easy to calibrate. They are often used in applications with small bunch charges and are suitable for measuring dark currents, as well.

Figure **11.1** shows an experimental set-up with a Faraday cup to measure charge or current. The cup itself is a metallic structure absorbing the electron beam. It is important that it absorbs all electrons of the bunch, that the electron shower is well contained and that all scattered electrons from the cup (secondaries) are captured. Electrons or ions from showers created elsewhere must not hit the cup. Also, care must be taken to avoid leakage current to the ground.

To contain the shower, the cup has a certain thickness and a geometrical form – typically of a cup – to capture all secondaries. Some designs first use low Z material to avoid backscatter, followed by high Z material, like tungsten or lead, to contain the shower. The Faraday cup is electrically isolated from the beamline and all other components and the electrical connection is made to the measurement device only. In the simplest configuration, the cup is connected *via* a standard 50 Ω cable (for instance, RG58) to an oscilloscope of appropriate resolution, set to an input impedance of 50 Ω. Preferably, the cup should be inside the vacuum system to avoid the passage of electrons through a window generating ions in air.

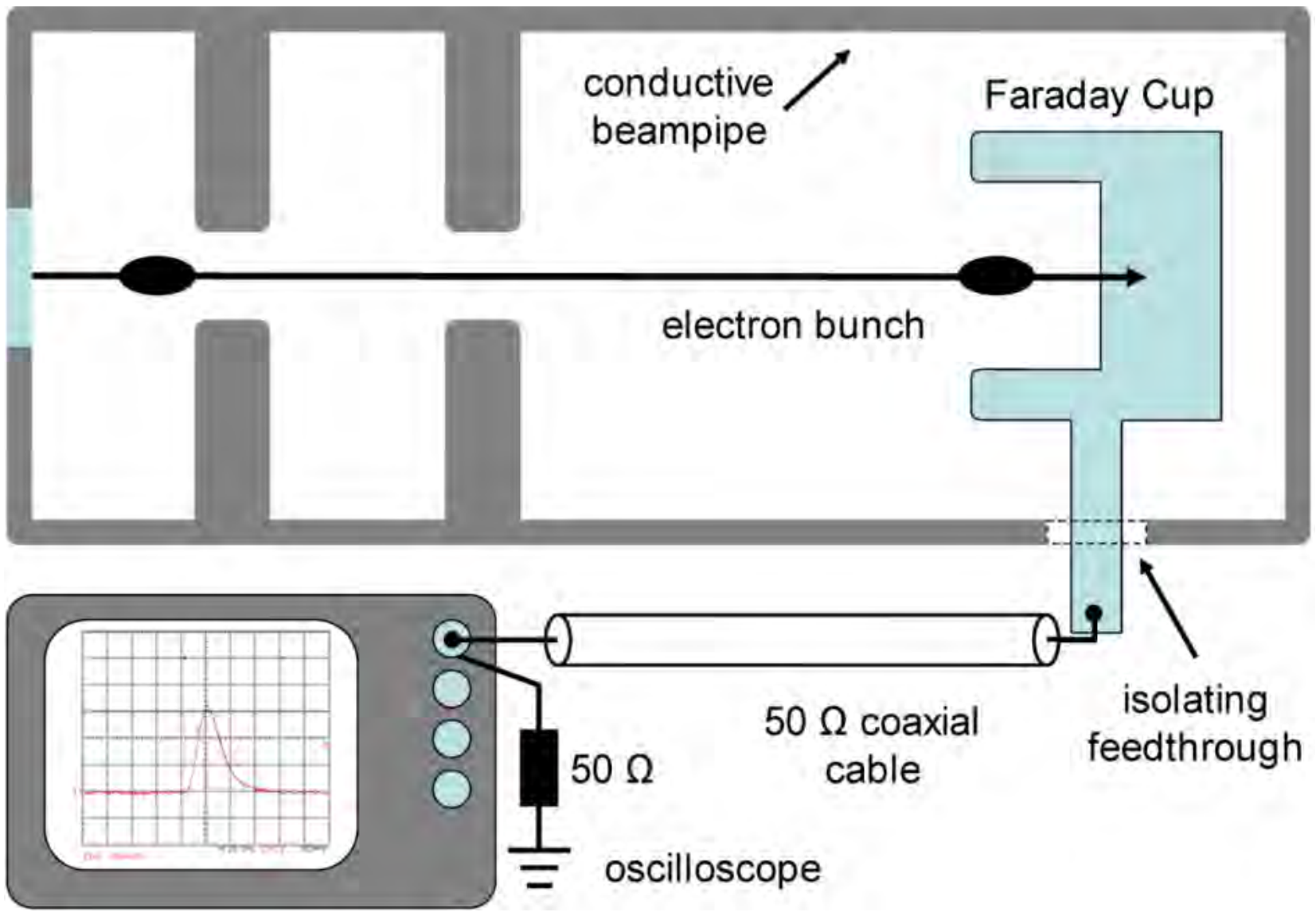

**Figure 11.1. Basic set-up to measure the charge of an electron bunch with a Faraday cup. The bunch emerges from the source and passes through some accelerating section before it hits the cup. The cup is electrically isolated from the beam pipe. The current pulse is measured with an oscilloscope terminated to 50 Ω.**

The time dependent current of the pulse, $I(t)$, is simply given by

$$I(t) = \frac{U(t)}{R} \tag{11.1}$$

with the voltage $U(t)$ measured by the oscilloscope with an input impedance of $R$ = 50 Ω. The charge is obtained by integrating over the current

$$Q = \int I(t)dt = \frac{1}{R}\int U(t)dt \tag{11.2}$$

The integration is over the duration of the beam pulse and requires careful subtraction of the background. The integration measurement feature of standard oscilloscopes can be used. The background is evaluated by using the same integration measurement set-up, but with the beam switched off.

This procedure gives an absolute measurement of the current or charge with a precision in the 10% range. High precision measurements necessitate more effort: very accurate measurements require special modeling of the cup, especially to avoid backscatter and to contain the shower.

To have a clean pulse and avoid reflections and ringing, the Faraday cup arrangement should also have an impedance of about 50 Ω.

The cup should be inside the beam vacuum, either in a fixed position acting as a beam dump, for instance in a diagnostic section, or mounted on an actuator allowing it to be removed from the beamline when required. For large average currents, the cup may need to be water-cooled. For low energy beams, the weight of the cup still is reasonable and can be handled easily.

The oscilloscope can be replaced by a data acquisition system based on fast analog-digital converters (ADCs) to digitize the output of a sample-and-hold integrator.

The electron bunches in an RF gun are short, typically in the 1-10 ps range, a parameter that must be taken into account in designing the electronics.

### 11.2.2 Toroids

Faraday cups are destructive devices since they absorb the beam. An elegant way to measure the beam current or bunch charge is by inducing a current in a coil or toroid placed around the beam pipe. The coil acts as a current transformer [11.4]–[11.9] and can be built large enough to fit over the diameter of standard beam pipes. The toroidal design assures that the current- or charge-measurement is independent of the beam position.

Figure **11.2** is a sketch of the type of toroids used at FLASH. In this specific design, two-halves of a torus are clamped together forming a ring. The material is Vitrocvac 6025 [11.10], an amorphous alloy with a high permeability of $\mu_r \approx 10^5$ (for low frequencies of $f$ < 100 kHz) and a high electrical resistance for a fast decay of Eddy currents [11.11]. At higher frequencies, the permeability decreases like $\mu_r \propto 1/f$.

The clamped design offers the nice advantage of allowing mounting and removing of the toroid without opening the beam vacuum.

A small ceramic ring mounted into the beam pipe breaks the electrical contact, allowing the field of the electron bunch to interact with the toroid. The wall current, which normally flows along the pipe, is now guided around the toroid's housing with a low impedance bypass. Figure **11.3** is a 3-D model of a toroid used at FLASH and at the European XFEL.

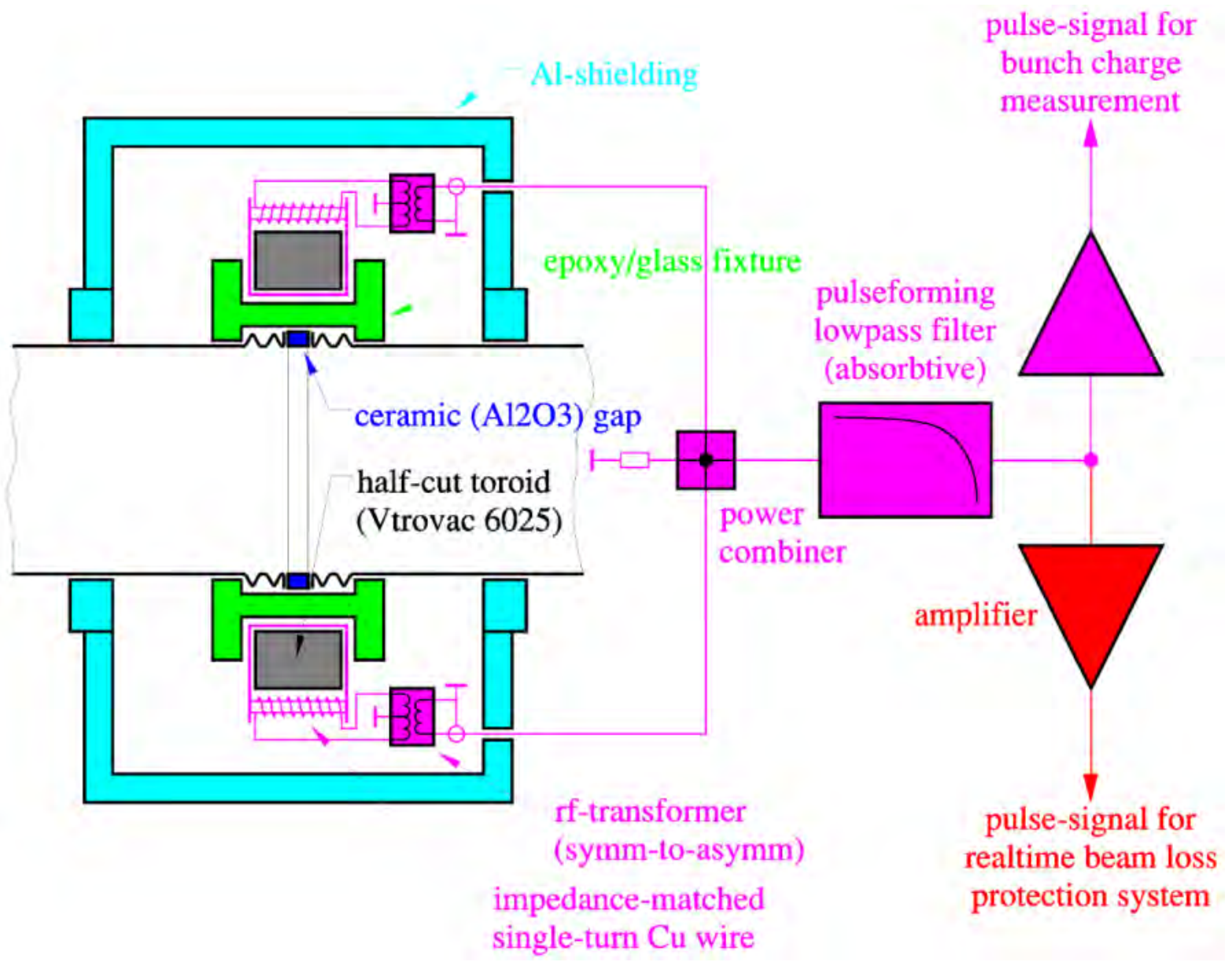

**Figure 11.2. Sketch of a toroid system with read-out. The transformer coil (gray) is placed around the ceramic gap (dark blue). In this design, two quarter torus pieces of Vitrovac 6025 [11.10] are clamped together, each having its own readout. The signals are combined and distributed to the read-out system. One branch is used for charge measurements and one to measure beam losses between adjacent toroids. [Courtesy of M. Wendt, FNAL]**

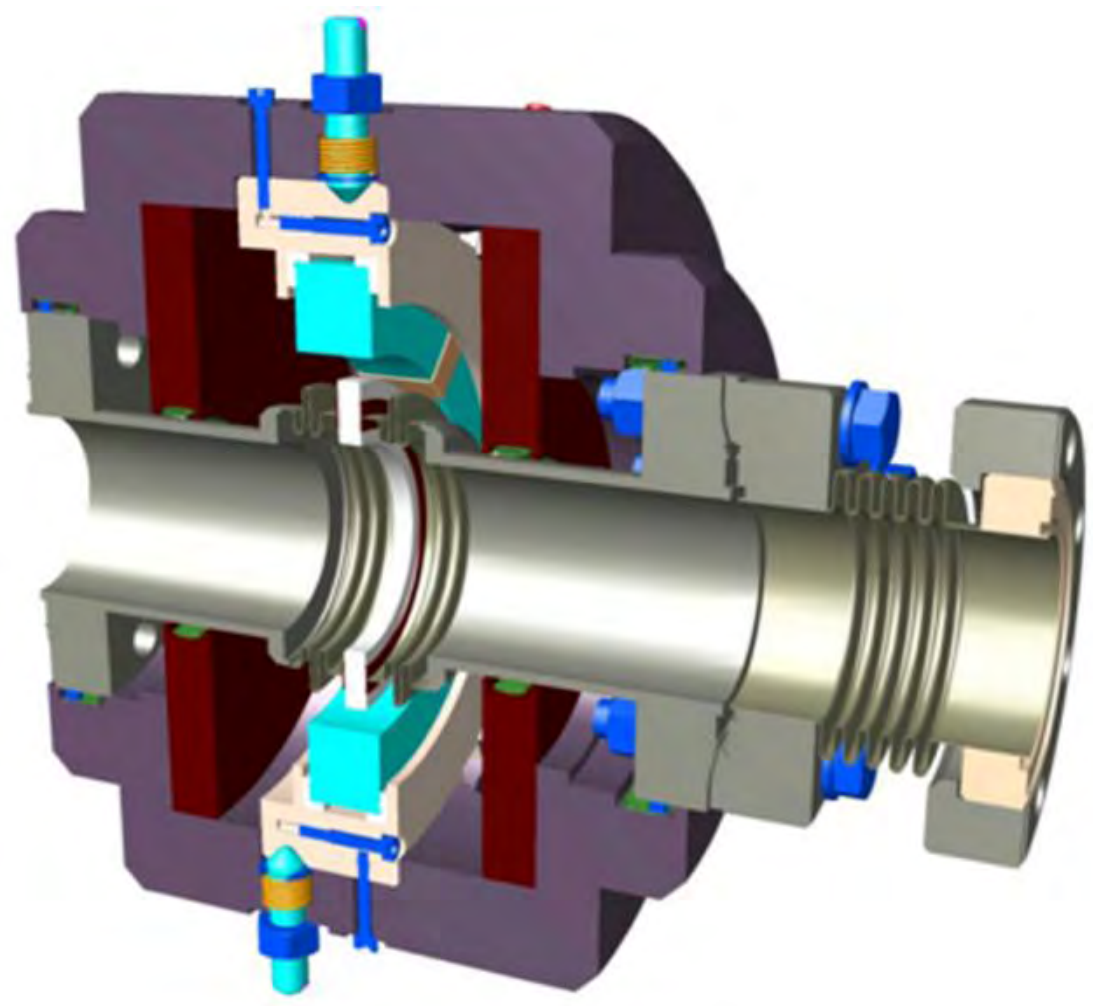

**Figure 11.3. Toroid assembly at FLASH and the European XFEL. The drawing shows a cut out of the toroid's housing and the vacuum chamber. The transformer coil (cyan) is visible over the ceramic gap (white) that is welded into the vacuum chamber. The housing (violet) serves as a shield and a bypass for the wall current. Read-out connectors also are shown. [Courtesy of M. Siemens, DESY]**

The ceramic gap may be metalized to avoid charging up. This especially is advisable for RF guns with a high level of dark current. In this case, the impedance of the bypass must be much smaller than the impedance of the metallization.

For low charges, amplifiers must be mounted close to the toroid. Typically, the analog signal is transported with a 50 Ω cable to the data acquisition system.

In contrast to Faraday cups, toroids must be calibrated because the transformer ratio and the amplifier gain may not be known exactly. In fact, calculating the transformer response is rather complex. Here, the reader is referred to Strehl's book [11.1] and the JUAS lectures given by Forck [11.11].

Following Strehl's evaluation of a passive transformer – without an additional amplifier – the voltage response $U(t)$ is given approximately by

$$U(t) = -I_p \frac{R}{N_w} e^{-t/\tau_d} \tag{11.3}$$

The beam current, $I(t)$, is idealized to a step function with a peak current of $I_p$. This simplified model yields us some basic properties of a toroid transformer. The output voltage is proportional to the inverse number of windings $U \propto N_w^{-1}$. In contrast, the transformer inductance $L$ is proportional to $L \propto N_w^2$. The sensitivity $S$, defined as the ratio of the output voltage to the beam current, is given by

$$S = \frac{U(0)}{I_p} = \frac{R}{N_w} \propto \frac{1}{N_w} \tag{11.4}$$

and the droop time constant $\tau_d$ , of the exponential voltage droop is given by

$$\tau_d = \frac{L}{R_L + R} \approx \frac{L}{R} \propto N_w^2 \tag{11.5}$$

where $R$ is the load resistance of the system, and $R_L$ is the resistance of the cables in the secondary circuit, which is usually much smaller than $R$.

For a good low frequency response and a large droop time constant, the number of windings $N_w$ should be high. On the other hand, for good sensitivity or voltage response, the number of windings should be low. However, using operational amplifiers, the large load resistance is overcome so that the droop time due to the very small $R_L$ is considerably increased with $\tau_d \approx L\, R_L^{-1}$ (Equ. 11.5) without needing to increase $N_w$, thus allowing one to choose a rather small number of windings to attain good sensitivity.

Evaluating the rise time, $\tau_r$, of the output signal for high frequencies involves taking into account the stray capacitance, $C_s$, and the stray inductance, $L_s$. The former is caused by the capacitance between the windings, the windings and the toroid, as well as the cables. The inductance of the toroid decreases due to the decreasing permeability with $1/f$. For frequencies ~100 MHz, the stray inductance becomes the dominant contribution [11.11]. $\tau_r$ is given by

$$\tau_r = \sqrt{L_s\, C_s} \tag{11.6}$$

The rise time is defined the same way as is the droop time in Equ. 11.3,

$$U(t) \propto (1 - e^{-t/\tau_r}) \tag{11.7}$$

With a typical stray capacitance and inductance, the rise time is on the order of nanoseconds: this is independent of the pulse duration of the electron bunch, which is in the picosecond range. The left of Figure **11.4** shows an example of a voltage output signal of an electron bunch. Here, the length of the voltage pulse is 20 ns, the length of the electron bunch is only 10 ps, and the bunch charge is 1.2 nC.

Calibrating the toroids is straightforward. A pulse of a known charge is fed through a loop around the coil and the coil response is measured. For instance, a temporally rectangular pulse generated by a standard pulse generator with a length of 50 ns and a voltage of 1 V terminated with 50 Ω has a charge of 1 nC. The calibration constant is obtained by integrating the resulting voltage pulse, very similar to in the case of the Faraday cup (Equ. 11.2 in Section 11.2.1). The beam charge is then measured by integrating the toroid current pulse and applying the calibration constant. A standard 500 MHz scope terminated to 50 Ω can be employed, or better, a suitable data acquisition system with a sample-and-hold integration. As discussed above, the shape of the output pulse is independent of the electron pulse. For simplicity, it is possible to measure the peak voltage only, omitting the integration.

Many photoinjectors produce pulse trains. FLASH, for instance, has trains of hundreds of individual pulses with 1 μs spacing between them. The right of Figure **11.4** shows an example of a pulse train of 30 bunches. For pulse train durations comparable to the droop time constant $\tau_d$ (Equ. 11.5), the voltage droop from pulse to pulse must be accounted for. A typical droop is about a few percent per microsecond.

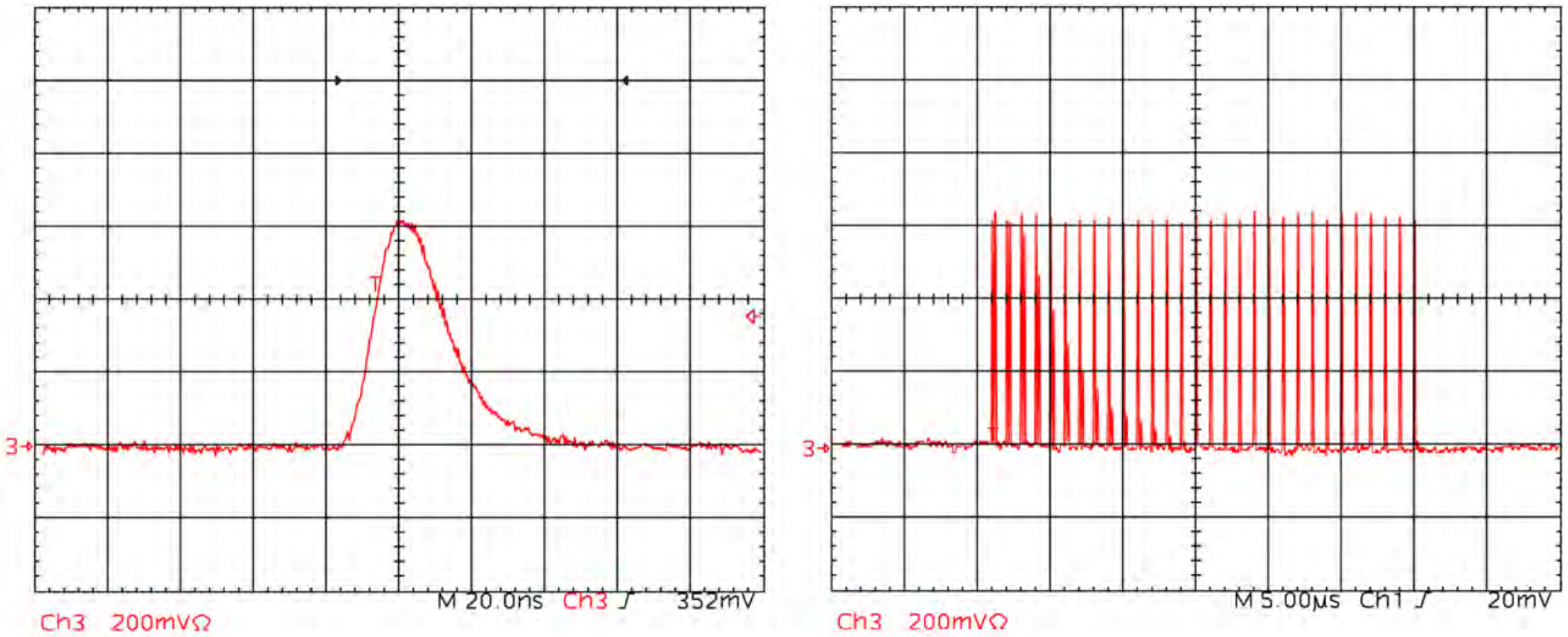

**Figure 11.4. Examples of toroid signals measured with an oscilloscope. The voltage response of a single electron bunch (left), and of a train of 30 bunches with a spacing of 1 μs (right). Although the duration of the bunch is 10 ps, the width of the voltage response of a single bunch is 20 ns. A small voltage droop along the bunch train is visible. The bunch charge in this example is 1.2 nC (vertical scale 200 mV per division, horizontal scale 20 ns (left) and 5 μs (right) per division and a termination of 50 Ω).**

To correct for the droop, the signal height is measured as the difference between the peak voltage and the baseline. Each pulse in the train is sampled at its peak and at the baseline, close to the pulse (a few nanoseconds earlier).

Care must be taken with external magnetic fields that may saturate the toroid. Vitrovac 6925, for instance, saturates at 0.58 T. [11.10] RF gun designs use large, strong solenoids to compensate for the space charge induced emittance growth. The toroid should be mounted at an appropriate distance from the solenoid.

For good toroid designs, the linearity of the charge measurement is better than $10^{-4}$ with a resolution on the order of 1 pC in a wide dynamic range.

## 11.3 TRANSVERSE SHAPE

A good knowledge of the transverse bunch shape at various locations along the injector beamline is mandatory for tuning the injector for a small transverse beam emittance and for matching the beam optics to the accelerator's optical design. This section details the measurement of the transverse shape: measuring the emittance is covered in Section 11.7.

A transverse image of the beam usually is obtained by letting the electron beam hit a screen that in turn emits photons in the visible wavelength spectrum. The emitted light is guided out of the beam pipe through a window and imaged by an appropriate imaging system on a high resolution detector. Nowadays, commercial CCD cameras with suitable optics (objectives or lenses) provide the required magnification and resolution.

Another method to determine the transverse shape is based on using thin metallic wires hit by the electron beam, for example, a fine grid of several wires or a so-called wire scanner, which is a single wire moving through the beam. Either the electrons scattered by the wire or the current induced by it is measured as a function of the wire's position. The wire scanner provides a 1-D, averaged projected profile of the beam. It has the advantage of being almost non-destructive and is a fast measurement of the beam size. In contrast, although wire grids are destructive, they yield a single shot 2-D charge distribution. However, they are rarely used since it is easier to realize a screen system with an excellent resolution of 10 μm.

### 11.3.1 Screens

The simplest method to measure the transverse bunch shape and size is to use a screen in combination with a suitable imaging system, which could be a simple commercial objective, or a more dedicated set-up of several lenses and other optical components. In most cases, the images are recorded by a commercial CCD camera.

Most screen monitors use flat screens made of scintillating, fluorescent material or having a polished metallic surface (in the case of optical transition radiation), inserted at an angle of 45˚ with respect to the beam. The backward radiation emitted from the screen is guided out of the beam pipe through a window and transported into the imaging system. The vacuum window should be radiation hard to remain transparent in the range of optical wavelengths for a long time; windows made out of the best quality fused silica are an example. To reduce γ-radiation hitting the camera, an additional 45˚ mirror often is used in the optical system. Figure **11.5** is a sketch of a typical set-up.

The simple arrangement described works for most applications. For high resolution measurements, care must be taken to polish the optical transition radiation (OTR) screens to optical quality. For fluorescent or scintillating screens, other factors may influence the resolution, such as grain size, crystal thickness, orientation angle with respect to the beam and the camera, and the choice of optics.

The depth-of-focus effects need to be considered only when a very high resolution for large spot sizes is required. For screens mounted at 45˚ with respect to the beam, this effect may limit the resolution since radiation created on the screen from different parts of the beam have different distances to the CCD of the camera, and thus are imaged differently. The beam image itself is not distorted. Since the electrons are moving almost at the speed of light, the screen acts like a normal mirror. To retain the best resolution, it is also important that the beam hits the point of the screen for which the imaging system was adjusted, usually the center.

When the screen is placed normal to the electron beam, the path length of the radiation is equal over the entire bunch. However, the set-up then is more complicated: a mirror is required to guide the emitted photons out of the beam pipe. Figure **11.5** illustrates an arrangement using forward radiation from an optically transparent crystal. Also, a screen coated on the back side with a material emitting the radiation can be employed.

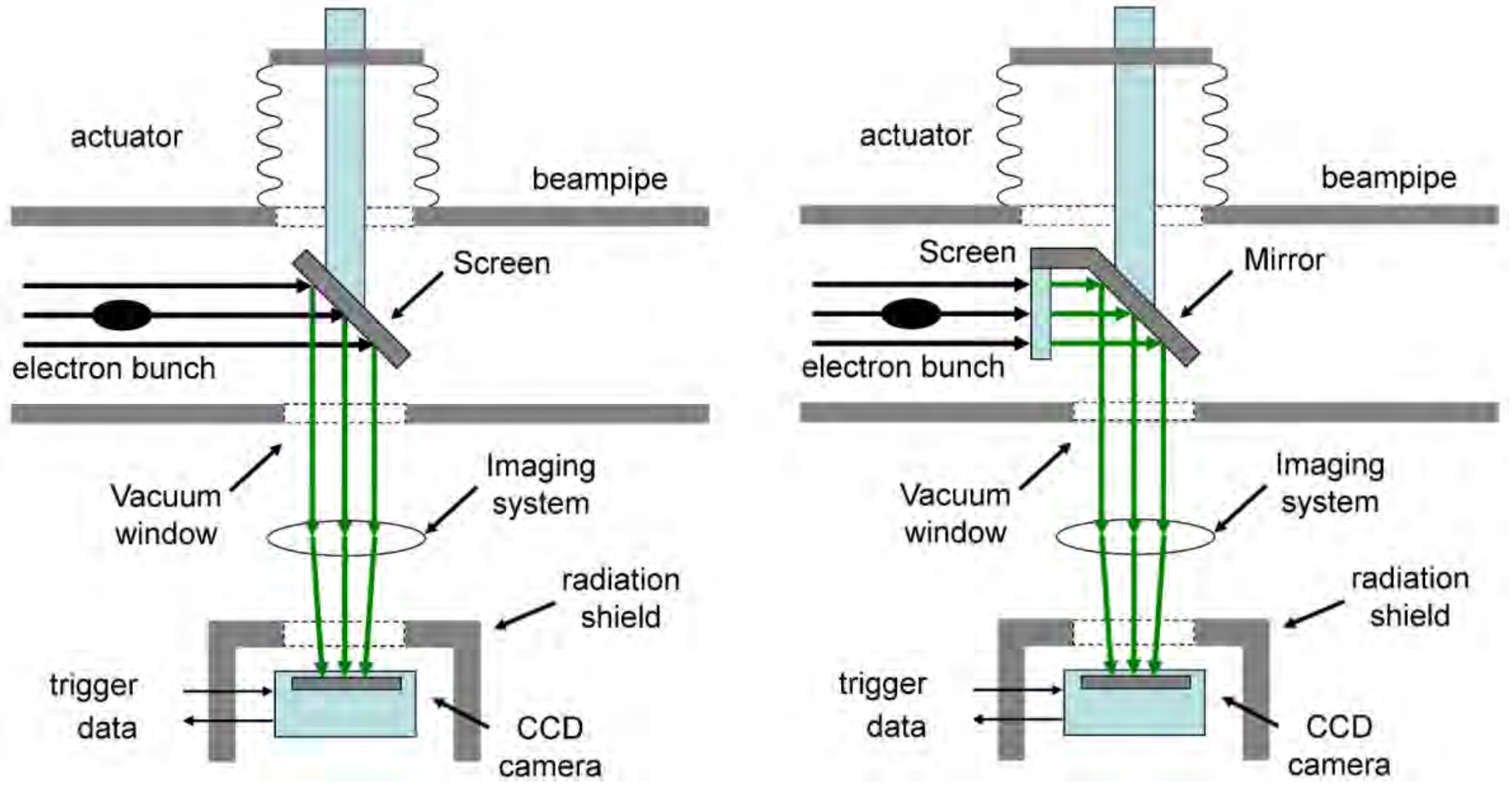

**Figure 11.5. Typical set-up of a screen system to measure the transverse shape of an electron beam. A standard system uses a flat screen with fluorescent material or a polished metallic surface for optical transition radiation at an angle of 45° with respect to the beam (left). An optical system images the screen surface onto a CCD camera. A system avoiding the depth-of-focus issue places the screen normal to the electron beam direction (right). In this arrangement, the screen is either optically transparent or the radiation is created at the back surface to allow imaging over an additional mirror (here shown with an angle of 45°). In both versions, the screen and mirror are retractable from the beamline.**

For this type of set-up, a complication occurs when coherent optical transition radiation has to be considered. Such radiation is also emitted in the forward direction by the radiator and the backwards direction by the mirror. Interference between the two sources will be visible, complicating the analysis of the image. The interference pattern depends also on the beam energy, since the slippage between the two radiation sources is given by the difference between the speed of light and that of the electrons. Special experiments were undertaken to measure this effect [11.12], [11.13]. In practice, where coherent optical transition radiation is not expected, the number of photons from high efficiency radiators is considerably larger than the number of photons emitted by optical transition radiation, so this effect can be neglected.

The screen typically has a size of approximately a centimeter and can be moved in and out of the beam pipe, either by pneumatic actuators or motorized drives. Motorized actuators are used when several different screens mounted on a single actuator must be inserted. The actuator has end switches and intermediate position sensitive switches to insert the different screens properly. In machines with a large number of high charge bunches per second, the switches are also used by the machine protection system to prevent damaging the screens. The number of bunches allowed is limited when the screen is inserted.

Transverse bunch sizes in a typical injector are in the millimeter scale; for higher beam energies, they are on the order of 100 μm. Even smaller sizes, some tens of micrometers, are obtained, for instance, during solenoid scans. Measuring the beam size as a function of the RF gun solenoid field is a typical measurement to validate simulation codes.
The applied optical system must take into account the expected size of the beam. Typically, a screen monitor system should be able to measure bunch sizes down to 50 μm, or even less. A good system should aim for a resolution on the order of 10 μm. For some cases, a simple system with a commercial objective will suffice. Others may require a more complicated and flexible system. The optical elements should always be of high quality: achromatic lenses are used, and suitable apertures may be needed to reduce chromatic aberration. Retractable neutral density filters adjust the light intensity to avoid saturating the CCD camera. Wavelength filters to select a specific spectral range are sometimes included.

The pixel size of a standard CCD camera with a ½-inch sensor is about 10 µm × 10 µm. The rms resolution, $\sigma_p$, of a pixel sensor of size, $l_{pix}$, is limited to

$$\sigma_p = \frac{l_{pix}}{\sqrt{12}} \tag{11.8}$$

Here we assume, that always several pixels are hit by the beam, so that the rms beam size $\sigma_x$ in one transverse direction[1] can be determined using the second moment of the distribution

$$\langle x^2 \rangle = \frac{\sum_{i=1}^{n} W_i \,(x_i - \langle x \rangle)^2}{\sum_{i=1}^{n} W_i} \tag{11.9}$$

with $\langle x \rangle$ being the first moment defined by

$$\langle x \rangle = \frac{\sum_{i=1}^{n} W_i \, x_i}{\sum_{i=1}^{n} W_i} \tag{11.10}$$

The sum is taken over all pixels in a certain row or column indicated by $i = 1\ldots n$, $x_i$ is the transverse coordinate, and $W_i$ the number of counts of pixel $i$. A beam with a size less than one pixel appears as one pixel only and cannot be resolved with a resolution better than the pixel size.

The transverse coordinate is calculated from the pixel number by applying a calibration constant such that $x_i = ki$. The calibration constant $k$ is given by the pixel size $l_{pix}$ and the magnification $M$ of the optical system where $k = l_{pix}\, M^{-1}$. Since neither the pixel size nor the magnification is known exactly, the calibration constant is obtained using special marks on the screen with a known distance or with special calibration screens. A remotely controlled light source to illuminate the calibration marks eases the calibration procedure and should be included in the design of the screen station.

The counts in each CCD pixel are proportional to the number of photons hitting this pixel, which, in turn, is proportional to the charge of this specific part of the beam. To obtain a projected size, the pixel counts in one row or column are summed up first.

Using the second moment, the rms beam size is given by

$$\sigma_x = \sqrt{\langle x^2 \rangle} \tag{11.11}$$

Care has to be taken to correctly subtract the background. The second moment is sensitive to spurious pixel counts far from the beam center. High pixel counts created, for instant, by broken pixels or γ-rays need to be eliminated. A background image is usually taken under the same condition as with beam, but the beam is switched off. This is best done by blocking the drive laser of the RF gun while keeping all other parameters the same.

[1]The coordinate system used is left-handed Cartesian, with the horizontal coordinate $x$, positive to the right in beam direction, the vertical coordinate $y$, positive up, and $z$ pointing in the direction of the beam. In order to simplify the notation, we use $x$, even though a similar relation holds for $y$, as well.

Figure **11.6** is an example of the profile of a transverse beam measured using optical transition radiation with a CCD camera.

Usually, the shape of the beam is not Gaussian. However, if it is close to a Gaussian distribution, a fit to a Gaussian function can be used to determine the beam size. In this case, the rms beam size is equal to the standard deviation $\sigma$ of the Gaussian function.

A prerequisite to achieving a resolution below 10 µm with a standard CCD camera is that the magnification of the optical system is at least $M$ = 1. However, not only does the pixel size limit the resolution of the system, but also the properties of the whole optical system.

With a magnification of 1, it is often impossible to image the whole screen onto the CCD sensor. Let us take, as an example a CCD sensor having 658 × 494 pixels with a pixel size of 10 µm × 10 µm (a typical ½-inch sensor). With a magnification of 1, an area of 6.6 mm × 4.9 mm is imaged, the size of the CCD chip. Assuming a screen with dimensions of 20 mm × 20 mm, a magnification of 0.3 would be required to image the whole screen. Therefore, many screen systems have arrangements with multiple magnifications. Figure **11.7** shows a system with three magnifications [11.14]. Three magnifications, 1, 0.39 and 0.25, are realized with three different lenses, which are remotely movable into and out of the optical pass.

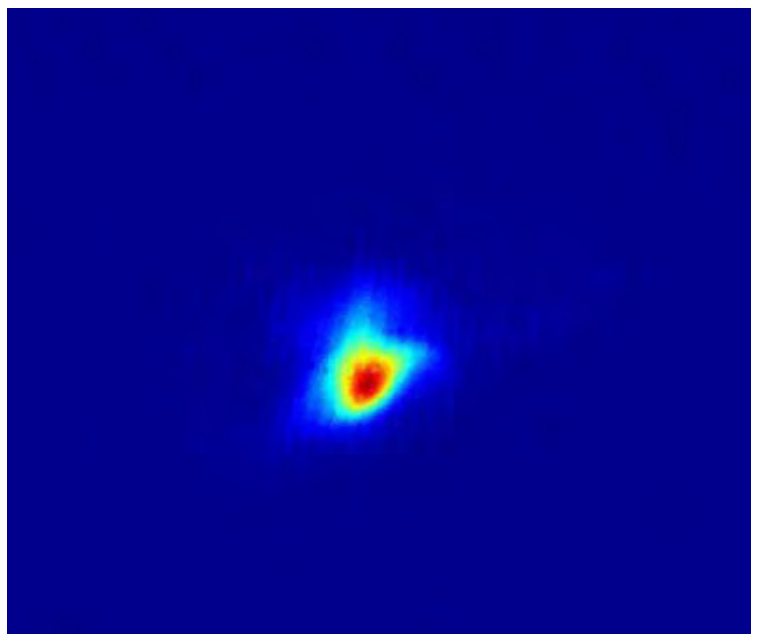

**Figure 11.6. Example of an image of an electron beam taken with an OTR screen at FLASH. The beam size in this example is 200 µm rms. The false colors encode the pixel count: Blue is very low, green to yellow is increasing and red represents high pixel counts. The pixel count is proportional to the numbers of photons hitting a specific pixel.**

Commercial digital CCD cameras are used nowadays. The camera is connected to a computer where the read-out software is running, for instance, with an IEEE 1394 interface, an Ethernet connection, or by other means. A beam signal triggers the camera. The control system provides an online image of the electron beam. Beam profiles and sizes are determined from the recorded images by dedicated image analysis software.

The exposure time of standard digital camera is several microseconds. Therefore, obtaining single shot images of the electron bunch is possible only for machines with low repetition rate or with specially designed fast cameras.

To obtain single shot images of high repetition rate bunch trains, gated intensified cameras are used. Compared to standard cameras, they are expensive and use image-intensifiers based on a photocathode together with a micro-channel plate to amplify the electron signal. The electrons then are back-converted to light using a phosphor screen and subsequently imaged by a CCD camera. This set-up allows nanosecond scale gating of the electron signal, opening the possibility for measuring a single bunch in bursts of bunches, where individual ones are only a few nanoseconds apart. I note that the number of bunches with a given charge and size allowed to hit a screen is limited by the damage threshold of the coating on the screen or the screen material itself.

The properties of the screen monitor considered above do not depend on the choice of the screen type: the set-up is similar for both, fluorescence screens and for screens producing OTR. These two types of screens have special characteristics which need to be considered when choosing a screen suitable for a specific application.

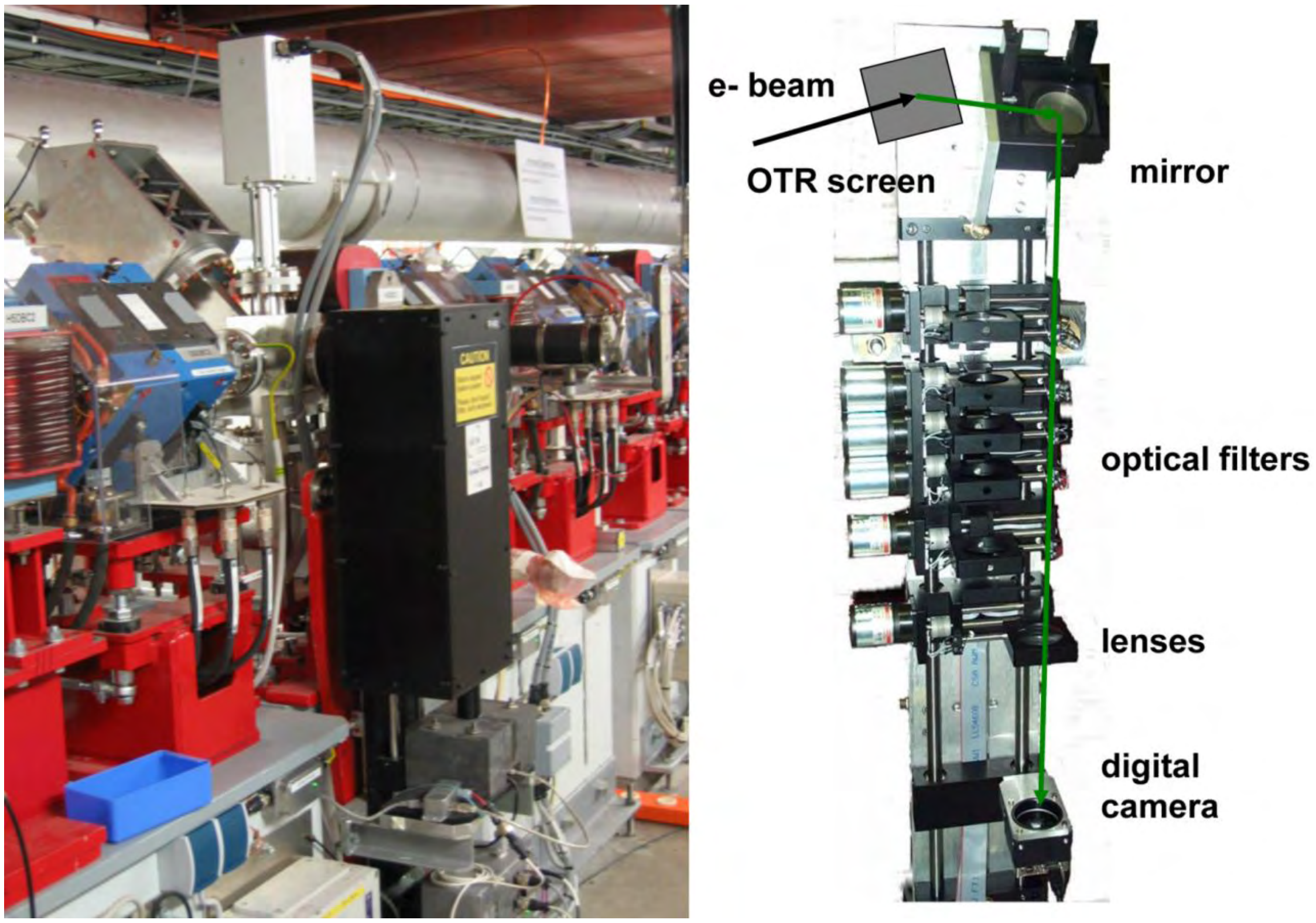

**Figure 11.7. OTR screen system of FLASH. The picture on the left shows the system as installed at FLASH; on the right, the cover was removed to reveal the optical system with optical filters (attenuators), lenses and the CCD camera. On top of the beamline, the actuator that moves screens into the beam is visible. Two different screens and a calibration screen can be placed in the beam path. Three remotely controlled lenses allow one to choose between three different magnifications, optical filters and attenuators. A commercial digital CCD camera [11.15] with a pixel size of 9 µm × 9 µm is used. The resolution of the measurement of the beam size is 11 µm.**

### 11.3.2 Fluorescence Screens

Scintillators or fluorescent screens[2] are an appropriate choice for measuring beam size at energies << 50 MeV, or when a high photon yield is required.

Earlier, scintillators of zinc sulfide or activated plastic were the standard screen material. However, they frequently exhibited damage due to radiation. A radiation-hard material was developed in the late ‘60s, *i.e.*, chromium activated aluminum oxide scintillators. [11.16] A subsequent step forward was the introduction of alumina ceramic fluorescent screens at CERN (Chromox-6). [11.17] Chromox is a commercial material, but was developed with collaboration with CERN. [11.18] For low charge bunches and where a strong photon signal is required, this type of screens is a suitable choice. It is a robust aluminum oxide, ceramic-doped with chromium oxide for ultra-high vacuum applications (typically 99.4% $Al_2O_3$ and 0.5% $Cr_2O_3$). A typical plate is 1 mm thick, with a favorable grain size of 10-15 μm. The fluorescence lifetime at room temperature is 3 ms. [11.17] The light yield depends on the thickness of the plate; a 1 mm thick plate gives enough light for a beam with about $10^9$ electrons per square-centimeter. Chromox may saturate and therefore is not suitable for high precision measurement of bunches with a high charge density; and due to the long fluorescence lifetime, they are not used for high repetition rate beams either.

---

[2]I do not explicitly distinguish between scintillation and fluorescence. Scintillators emit light when particles transverses the material; fluorescence occurs when photons excite the material’s atoms, which decay back to the ground level. Particles in an electromagnetic shower like electrons, photons, and positrons emit light through many processes not detailed herein.

Ce:YAG, Yttrium aluminum oxide crystals doped with, for instance, Cerium (1%) has been proven to be linear in a wide charge density range [11.19] with a fast decay time and therefore is typically applied at RF gun photoinjectors. Commercial Ce:YAG crystals are available at a size of about 1 cm × 1 cm and a couple hundred of micrometers thick. The crystal is transparent and is usable in both configurations shown in Figure **11.5**. Table **11.1** lists basic properties of Ce:YAG.

| **Property [units]** | **Value** |
|---|---|
| Index of Refraction | 1.82 |
| Wavelength of Peak Emission [nm] | 550 |
| Density [g/cm$^3$] | 4.55 |
| Radiation Length [cm] | 3.6 |
| Photon Yield [photons per MeV] | $8\times10^3$ |
| Cerium Concentration [with respect to Yttrium] | 0.18% |
| Decay Time [ns] | 70 |

**Table 11.1. Basic properties of Ce:YAG. [11.20], [11.21]**

Saturation at high charge densities is an issue for Ce:YAG, as well. Comparison with data from OTR and wire scanners show that Ce:YAG saturates at 0.01 nC μm$^{-2}$. [11.20] Besides the depth of focus problem described above, the optical resolution is limited by the crystal's thickness and light reflection from its back surface.

### 11.3.3 Optical Transition Radiation

Transition radiation is produced whenever a charged particle passes through a surface of any material. The transition between two media with different dielectric properties forces the electromagnetic field of the electron to adapt to the properties of the new media. The difference in the fields is emitted as transition radiation in the forward and backward directions with respect to the boundary surface. The forward radiation is emitted in the direction of the electron beam trajectory and the backward radiation in the direction of the specular reflection (Figure **11.8**).

The angular distribution of transition radiation is energy dependent: it peaks at an angle of $\theta_{peak}$

$$\theta_{peak} = \frac{1}{\gamma} = \frac{E}{m_e c^2} \qquad (11.12)$$

where $\gamma$ is the relativistic factor, $E$ is the electron energy, $m_e$ the electron mass and $c$ the speed of light.

At ultra relativistic energies and assuming a perfectly reflecting metallic surface, the photon intensity per solid angle and per unit frequency emitted by a single electron with charge $e$ is approximated by a simple expression

$$I(\theta,\omega) = \frac{e^2}{(4\pi\varepsilon_0)\,\pi^2 c}\,\frac{\theta^2}{(\gamma^{-2} + \theta^2)^2} \qquad (11.13)$$

where $e$ is the electron charge magnitude and $\varepsilon_0$ the permittivity of free space. The angle $\theta$ is defined with respect to the direction of specular reflection for the backward radiation, and with respect to the electron beam trajectory for the forward radiation (Figure **11.8**). Equ. 11.13 is valid for $\gamma >> 1$ and $\theta << 1$.

Figure **11.8** illustrates the angular distribution of transition radiation emitted by an electron with an energy of 1 GeV. Although the distribution peaks at an angle of $\theta_{peak} = \gamma^{-1}$, due to the long tails, a large part of the photons are emitted at angles significantly larger than $\theta_{peak}$.

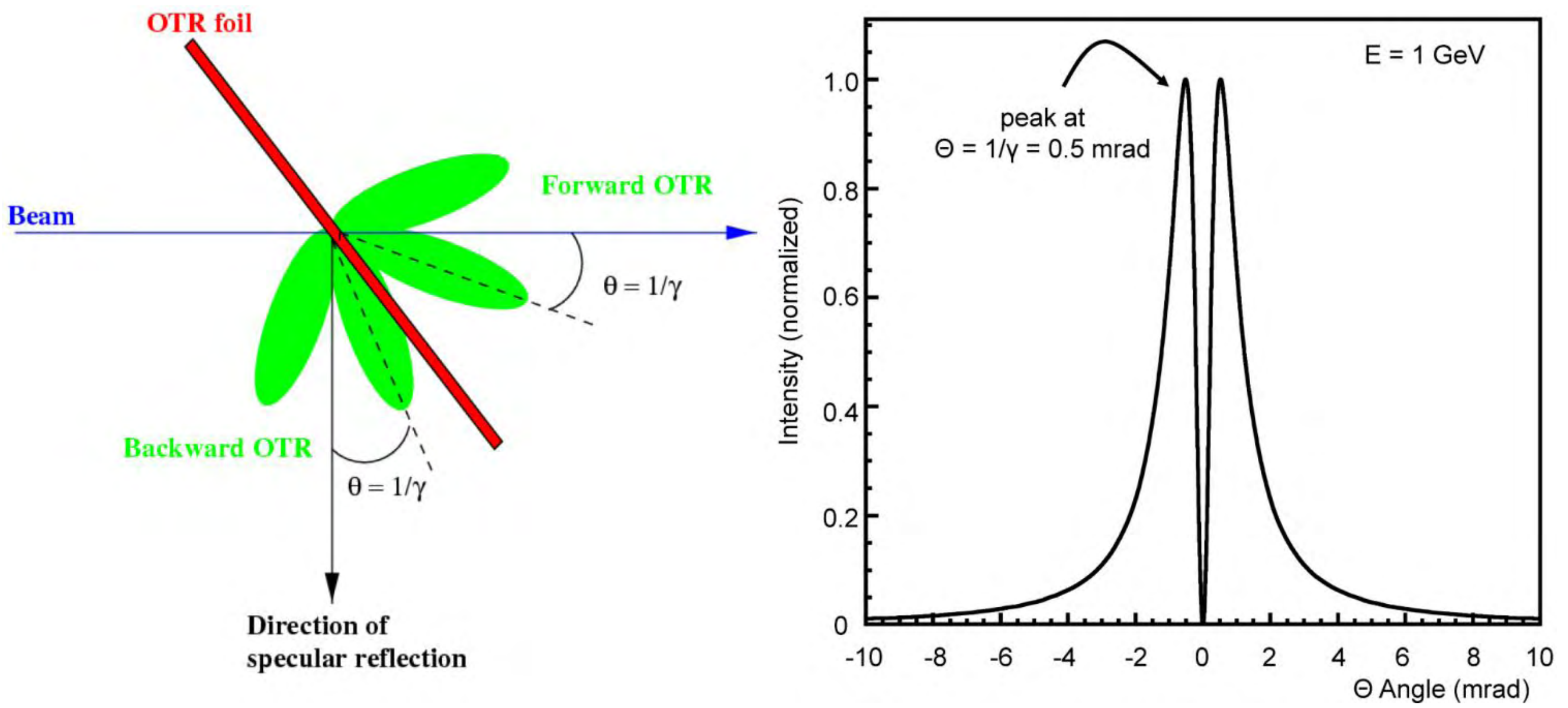

**Figure 11.8. Optical transition radiation (OTR) produced when an electron beam traverses a medium; here, a plane metallic foil is shown (left). The radiation is emitted in the forward direction (the direction of the beam trajectory) and in the direction of the specular reflection (backward). The angular distribution of OTR reveals a ring-like structure peaking at $\theta_{peak} = 1/\gamma$ (right). The distribution shown was calculated for a beam energy of 1 GeV. [Courtesy of K. Honkavaara, DESY] [Adapted from [11.22], with permission from Elsevier]**

For non-relativistic energies, $\gamma \approx 1$, as is the case for RF guns, a more complicated formula for the angular distribution applies [11.23]. In this case the distribution is asymmetric. Figure **11.9** shows the angular distribution for $E$ = 1 MeV and $E$ = 10 MeV. In both cases, the screen is tilted by 45˚ with respect to the beam direction.

Transition radiation has a wide spectrum, but typically the visible wavelengths (OTR) are used when measuring the transverse size and shape of an electron beam. In the optical wavelength range, standard commercial high resolution CCD cameras are available.

The advantages of OTR are its linear response to the electron beam charge and its fast response time, supporting single shot measurements. Also, the measurement set-up is quite simple: typically, a metallic, thin, flat, mirror-like screen with optical quality is used as a radiator target. For example, a 300 μm thin silicon wafer coated with aluminum is a suitable choice. Figure **11.6** is an example of an image of an electron beam using an OTR system at FLASH.

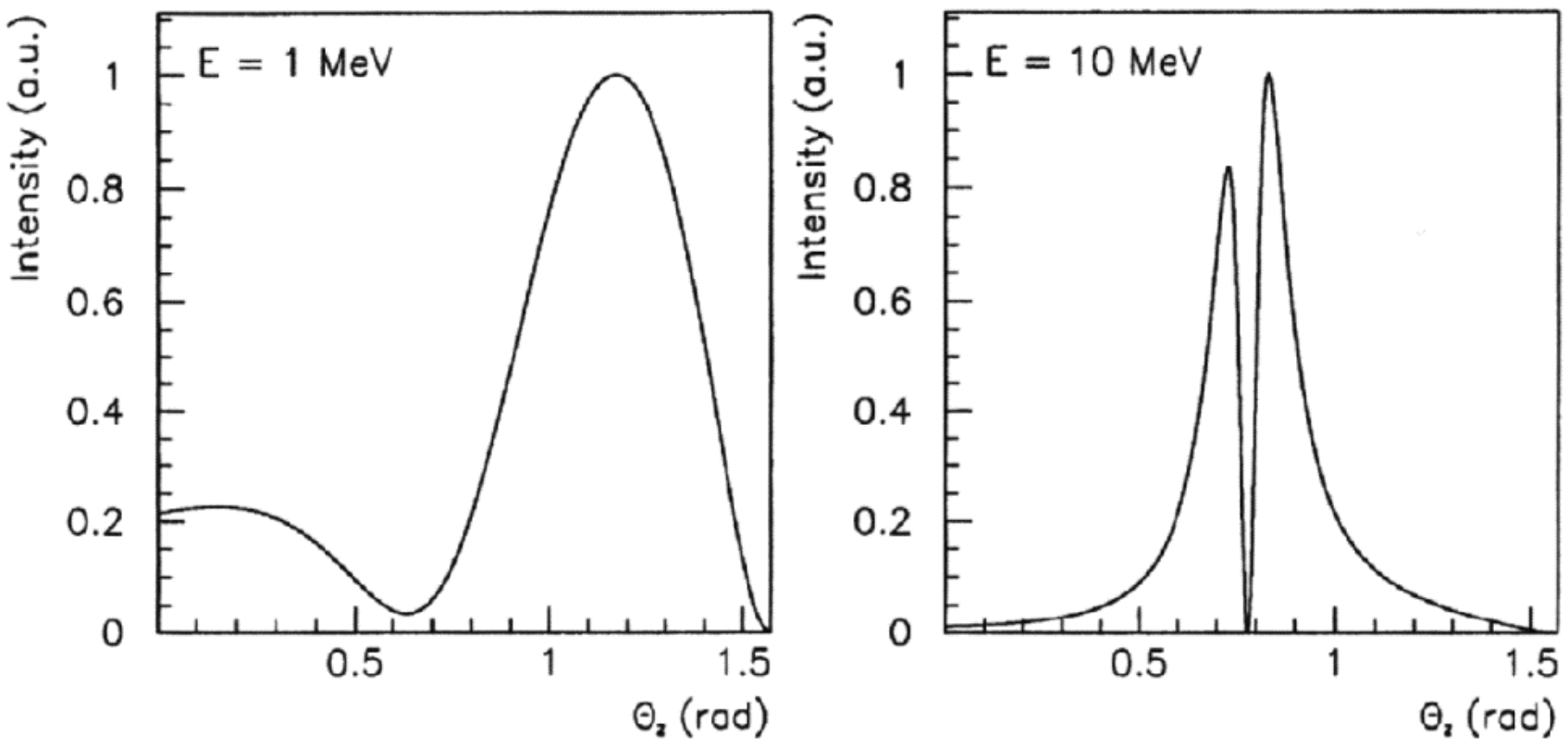

**Figure 11.9. Angular distributions of optical transition radiation for low energy electrons, 1 MeV (left) and 10 MeV (right). The screen is tilted by 45° with respect to the beam direction. In contrast to high energies, the angular distribution is asymmetric. [Courtesy of K. Honkavaara, DESY]**

OTR is a standard method used for beam imaging at high electron beam energies. However, for low beam energies (below ~50 MeV), the number of photons emitted may be insufficient for high resolution imaging of the electron beam's transverse shape with standard optics.

The number of photons emitted per electron, $N_\gamma$, in the frequency band [$\omega_1$, $\omega_2$] is on the order of the fine structure constant, $\alpha = e^2 (4\pi\varepsilon_0\hbar c)^{-1} \approx 1/137$, and increases logarithmically as a function of the electron energy

$$N_\gamma = \frac{2\alpha}{\pi}\left(\ln(2\gamma) - \frac{1}{2}\right)\ln\left(\frac{\omega_2}{\omega_1}\right) \tag{11.14}$$

At an electron beam energy of 10 MeV, about 100 electrons are needed to emit a single photon in the optical wavelength range ($\lambda$ = 350-750 nm). As considered above, the OTR angular distribution depends on the electron energy: the smaller the beam energy, the wider is the distribution. The angular acceptance of a typical screen monitor is about a couple hundred milliradians. Therefore, at low energies, it often is practically impossible to transport enough photons to a standard CCD camera.

Then, fluorescent screens, as discussed in the previous section Section 11.3.2, should be chosen.

### 11.3.4 Wire Scanners

Wire scanners are expensive and technically demanding devices. Their disadvantage compared to screens is that the wire is scanned through the beam giving an average projected profile, not a single shot 2-D image. For this reason, wire scanners are applied only if the technology is at hand, or where they are indispensable. There are a number of examples where wire scanners are successfully used for high repetition rate accelerators, like the SLAC linac SLC, LEP at CERN and others [11.24]–[11.27]. Wire scanners are also applicable where the coherent effects of OTR hamper the proper analysis of screen images, such as LCLS and FLASH.

A wire scanner consists of a thin wire, typically tungsten or carbon that is moved through the electron beam. Usually the radiation scattered off the wire is measured with a suitably placed scintillator and read-out with a photo multiplier. The thickness of the wire should be less than the expected beam size. A typical thickness is 10 or 30 µm. The response of the read-out system to a given charge hitting the wire has to be accounted for. The wire is mounted on a fork, which is moved with a motorized system, as illustrated in Figure **11.10**.

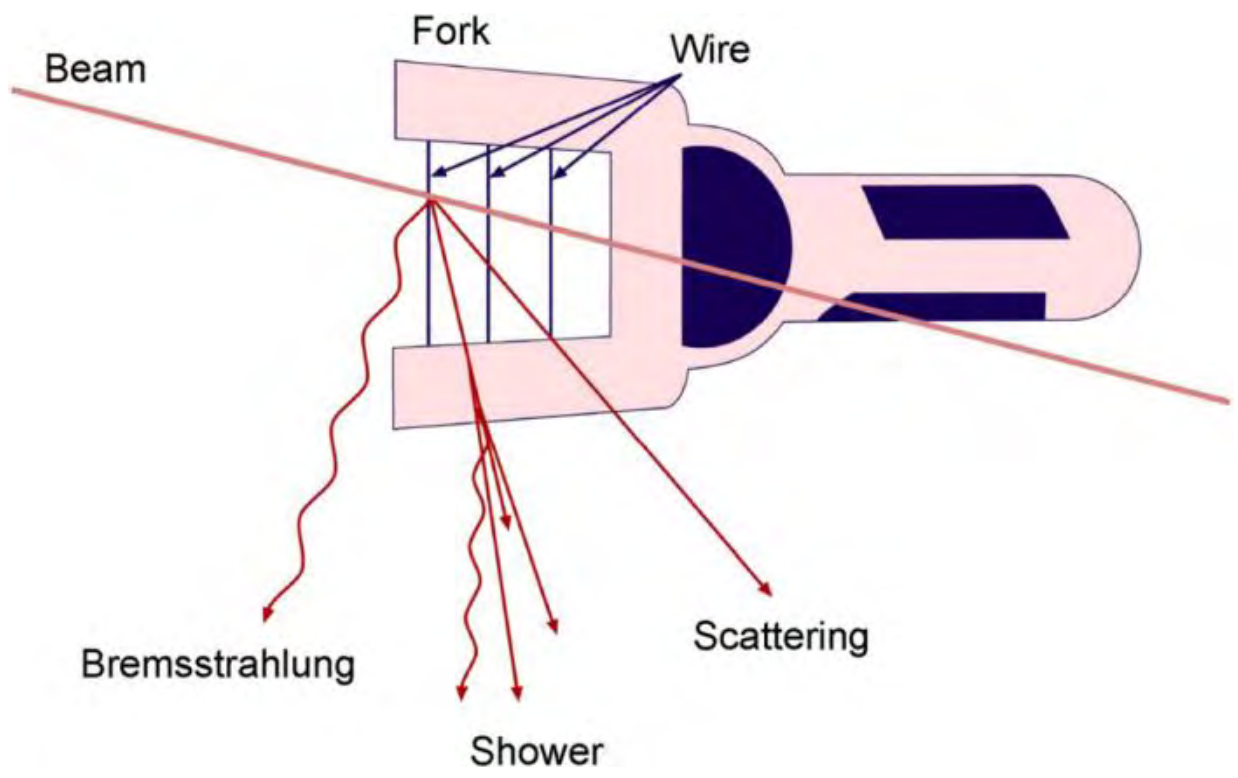

**Figure 11.10. Sketch of wires mounted on a fork. Scattered radiation is measured while the wire moves through the beam. [Reprinted from [11.27] with permission from Elsevier.]**

The wire's position is measured with precise encoders. The step size or the encoder's resolution should be a small fraction of the beam size. Typically, a resolution of 10 µm is obtained. In some systems, the fork has wires mounted horizontally, vertically and also at an angle of 45˚. Such systems can measure the horizontal- and vertical-profile together with a possible correlation at the same scan. Figure **11.11** has examples of wire scans in the FLASH undulator.

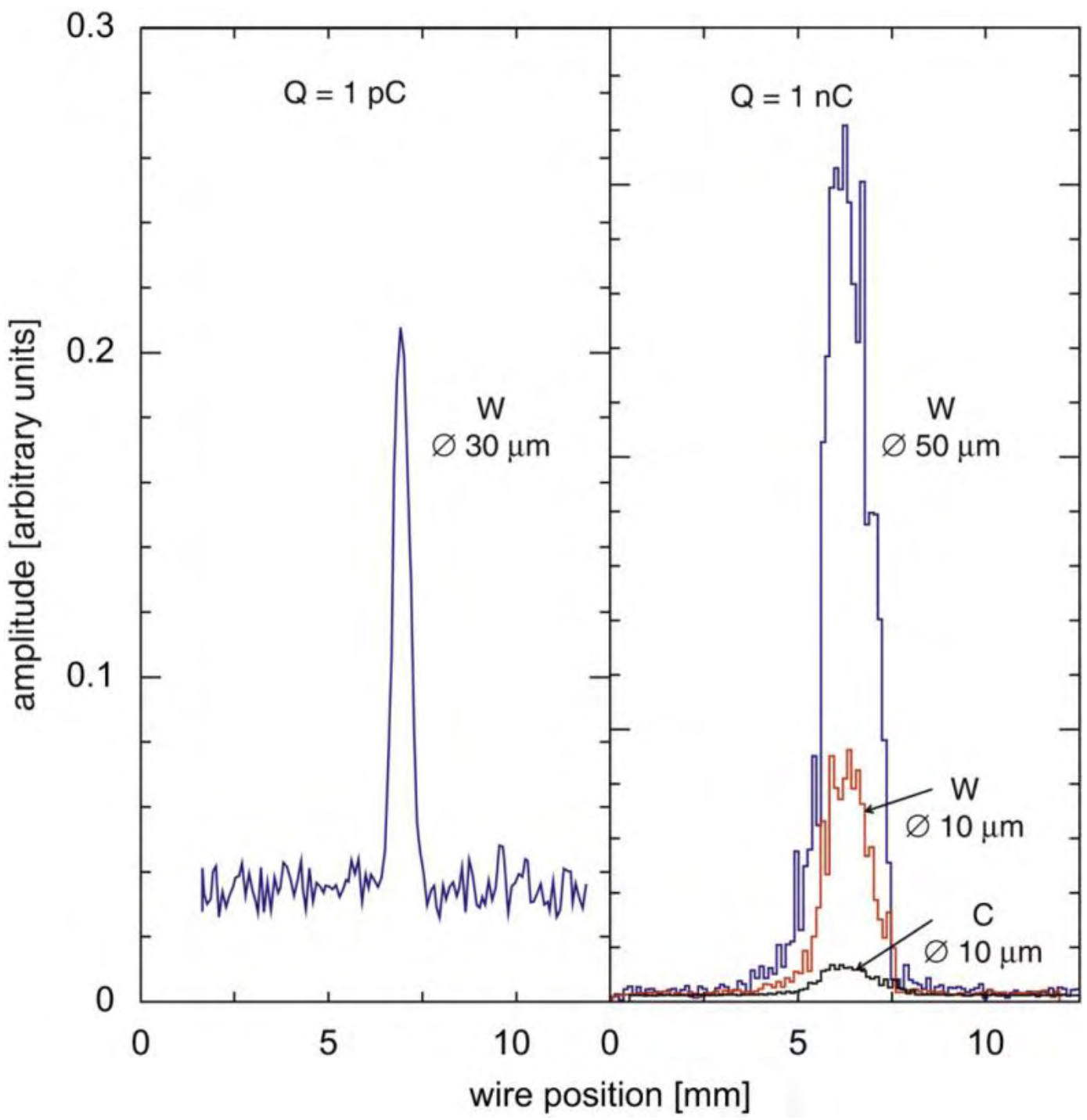

**Figure 11.11. Example of several wire scans of the electron beam at FLASH in the undulator section. While the wire is scanned through the beam, signals from the scintillator panels are recorded. Scans are shown for tungsten wires with varying diameters for a very small (1 pC) and normal bunch charge (1 nC). A scan with a 10 µm carbon wire is also shown. [Reprinted from [11.27] with permission from Elsevier.]**

For all wire scanner systems, care must to be taken to avoid breaking the wire, for example due to heating by the electron beam [11.28] or mechanical stress. Also, vibrations of wires during fast scans must be considered and damped. [11.29]

## 11.4 BUNCH LENGTH MEASUREMENTS AND LONGITUDINAL BUNCH SHAPE

Another important electron beam parameter is the bunch length and its shape. A typical bunch length at the RF gun is on the order of millimeters, corresponding to durations of a couple of picoseconds. In some applications, shorter bunches are realized down to the 100 fs scale. In FELs, the bunch is compressed to tens of micrometers to achieve the required peak current in the kiloamperes range. Compression usually is obtained *via* magnetic chicane bunch compressors.

Determining longitudinal bunch properties is more complicated than measuring transverse ones. A direct time resolved measurement requires ultrafast detectors and devices that are not readily available. Often the longitudinal phase space is transformed into the transverse phase space allowing the use of standard methods like screens viewed with standard cameras.

Most methods use incoherent and coherent radiation produced by the electron beam, for example, streak cameras and interferometers. This radiation can be synchrotron radiation that is emitted when the electron bunch moves along a curved trajectory, *e.g.*, through a bending dipole magnet. At beam energies above 100 MeV, synchrotron radiation is emitted in the optical wavelength range required for streak cameras. Another approach is to use transition or diffraction radiation; this is useful for beam energies above 25 MeV. The former is emitted when the beam passes through a screen and the latter as the beam travels through an aperture, for instance a slit. OTR is discussed previously in Section 11.3.3. For beam energies below 25 MeV, Cherenkov radiation can be used, which is generated when a particle travels inside a medium wherein the speed of the particle exceeds the velocity of light. Aerogel with a very low refractive index, or Sapphire with its very high refractive index, are used as Cherenkov radiators.

An example of directly measuring the electron bunch length in the time domain is the transverse deflecting RF structure. Here, the beam is streaked by an RF field. The longitudinal charge distribution in the bunch is mapped onto a transverse screen and imaged with standard imaging methods. A deflecting structure can serve for all beam energies.[1]

### 11.4.1 Streak Camera

The streak camera provides a convenient way to measure electron bunch lengths in the millimeter and sub-millimeter range. The method is based on the prompt emission of light in the optical wavelength range, for instance OTR (see Section 11.3.3), synchrotron radiation [11.30], or Cherenkov radiation [11.31]. An important feature is that the light pulse is a “replica” of the longitudinal shape of the electron pulse. Suitable optics guide the light pulse to the streak camera. Pulse broadening due to diffraction effects must be avoided. In the streak camera, the incident light is passed through a narrow slit and is then converted into photoelectrons by a high efficiency photocathode, like S-20. The electrons are accelerated and deflected transversally by a streak tube applying a fast linear deflecting field. Typically, a streak speed of 2-5 ps $mm^{-1}$ is obtained. The deflection is proportional to the arrival time of the electrons, and thus, to the time structure of the incident light pulse. A phosphor screen (P-20) including an image intensifier converts the streaked electrons back to photons that are finally imaged to a CCD chip of a high resolution, low noise camera. Figure **11.12** shows the principal features of the layout of a streak camera.

[1]At the high energy end of several gigaelectron volts, a deflecting structure may be inappropriate.

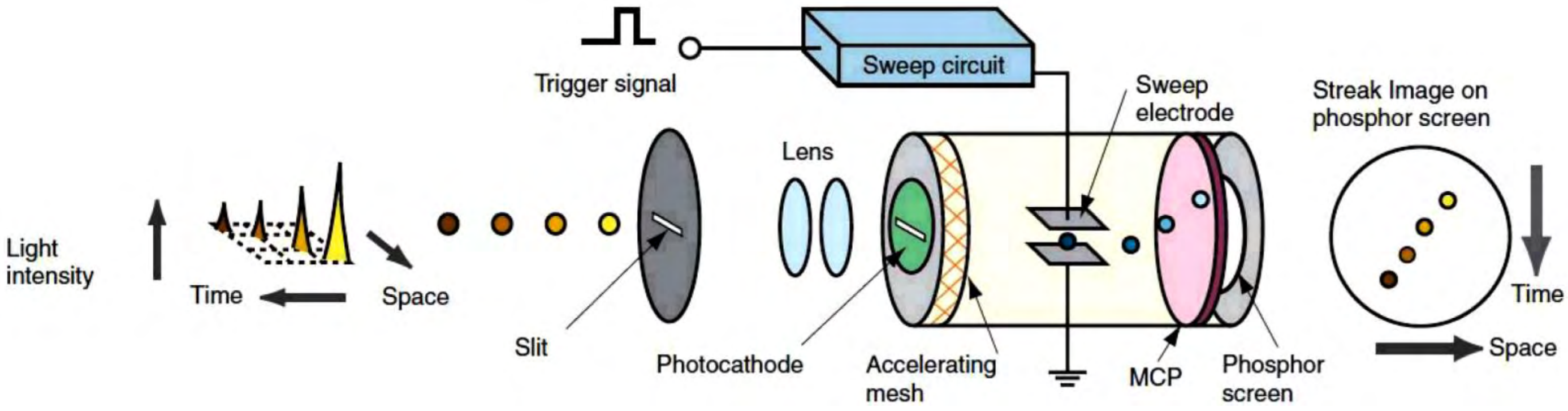

**Figure 11.12. Principal layout of a streak camera. The light passes through a narrow slit and is then converted into photoelectrons by a high efficiency photocathode. The electrons are accelerated and deflected transversally by a streak tube. The streaked electrons are converted back to photons by a phosphor screen and amplified by an image intensifier. Finally, the image is recorded by a CCD camera. [Figure from [11.32]; Courtesy of Hamamatsu Phontoics]**

The pixel size of the CCD chip is about 10 μm, small enough to not limit the resolution. With a streak speed of 2 ps mm$^{-1}$, this size would be good for a resolution of 20 fs. The slit size selected should be on the order of the pixel's size. The resolution of a streak camera mainly is limited by space charge effects of the electron beam emitted by the cathode, dispersion effects in the optics and by the quality and speed of the streak tube. A resolution of 200 fs rms is attained by a state-of-the-art camera. [11.32], [11.33]

For low charge beams, photon yield may be an issue. Even though streak cameras can detect single photons, a substantial amount of light could be lost by optical filters used to reduce dispersion effects. In addition, OTR has a low photon yield at low energies. An increased yield is obtained by Cherenkov radiators (Aerogel [11.34] or Sapphire). Synchrotron radiation is used for beam energies above 100 MeV, where the photon yield in the range of optical wavelengths is reasonable.
Figure **11.13** is an image of an electron bunch measured with a FESCA-200 [11.32]. The figure shows the time domain of the beam, where the streak direction is vertical and the horizontal direction is the transverse profile of the beam. The resolution of a streak camera is good enough for the typical bunch lengths obtained at RF guns. Streak cameras can also be used to measure the pulse length of the photoinjector laser. For compressed bunches with femtosecond scale duration, other methods need to be applied (detailed in the next sections).

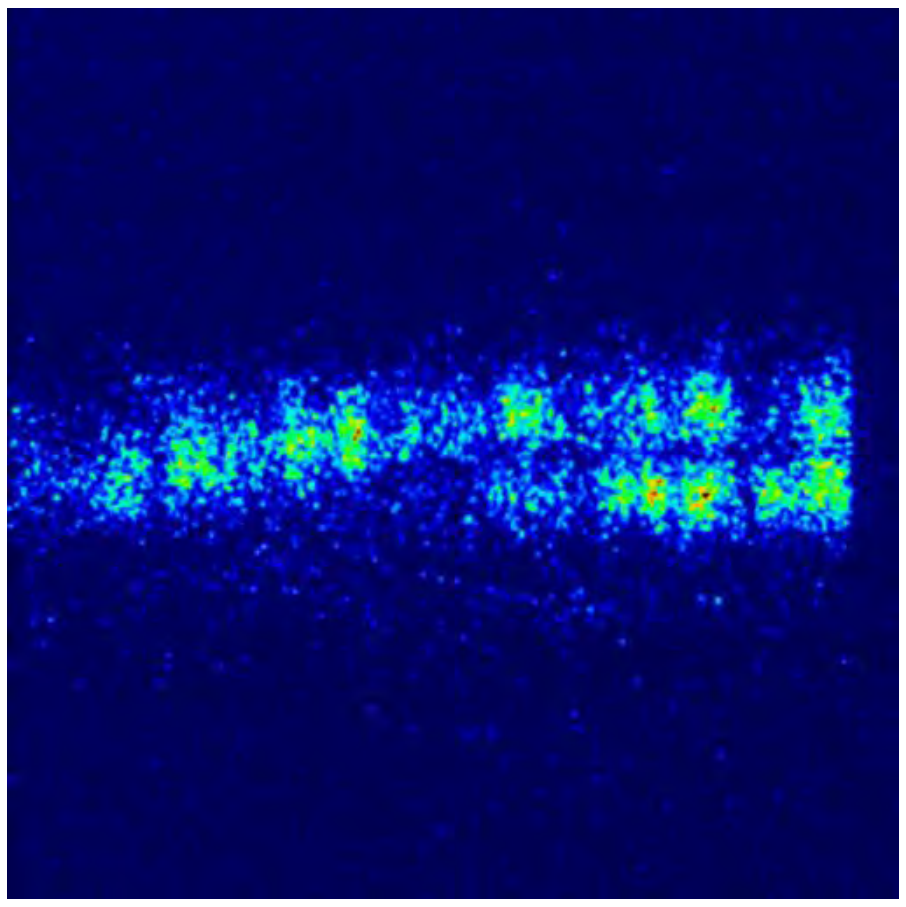

**Figure 11.13. Example of an image obtained by a high resolution streak camera. The streak direction is vertical. The full range of the vertical scale is 100 ps with a resolution of 1 ps. A sub-structure is visible in the time domain with a picosecond scale, which is a function of horizontal position. This demonstrates the instrument's power of detection.**

### 11.4.2 Transverse Deflecting RF Structure

The transverse deflecting RF structure (TDS) provides a direct method to measure the shape and length of the electron bunch. [11.35]–[11.38] Figure **11.14** displays a schematic drawing of a SLAC S-band TDS. The operating principle is similar to that of a streak camera (Figure **11.15**). The RF field of the structure deflects the electron beam, either vertically or horizontally. [11.39]–[11.41] Usually, the bunch is set to the zero crossing of the field, where the time dependence of the RF field is linear and has the steepest slope.

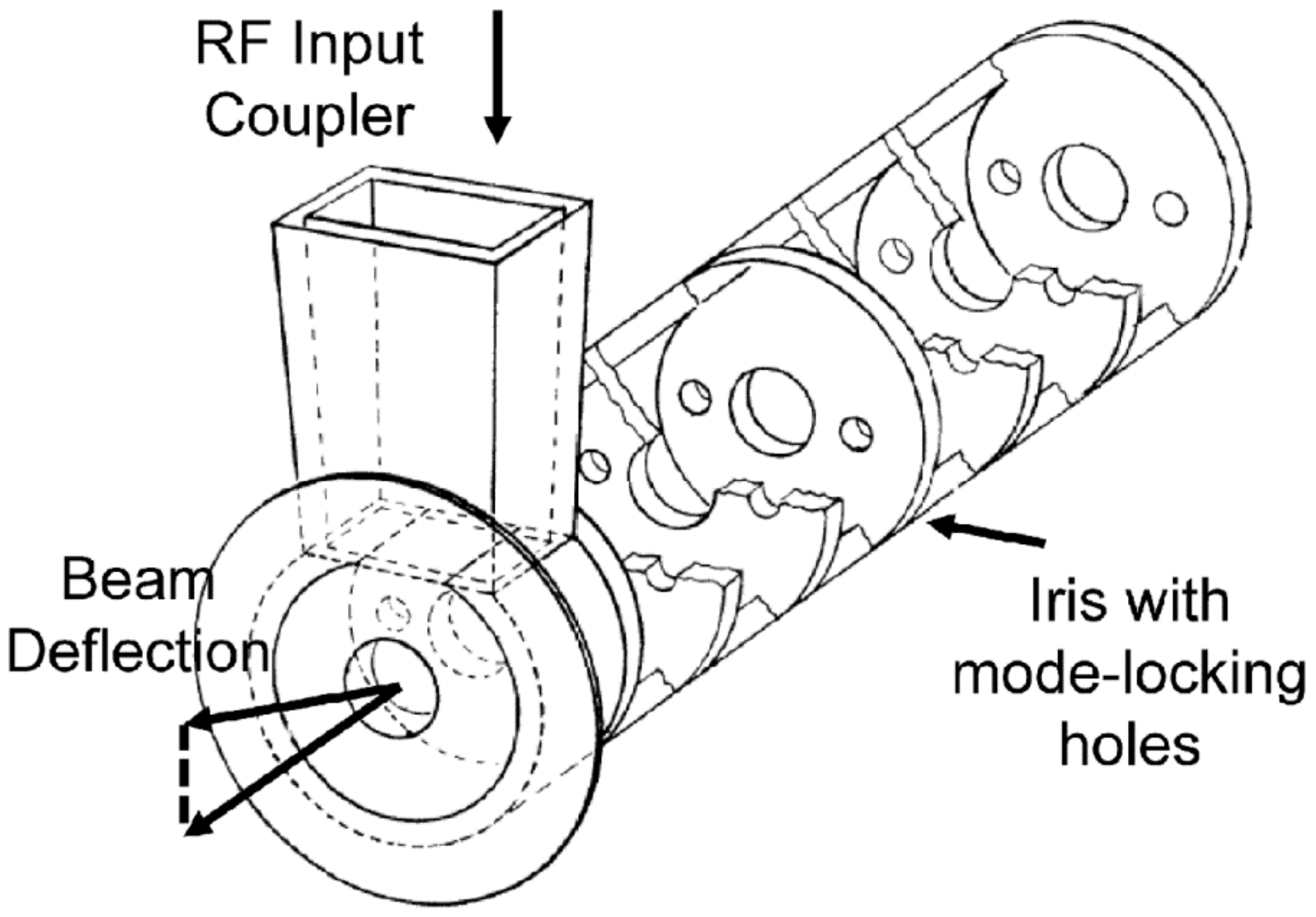

**Figure 11.14. Schematic drawing of a SLAC S-band TDS. The kick is vertical in this drawing. [Reprinted with permission from [11.40]. Copyright 1964, American Institute of Physics.]**

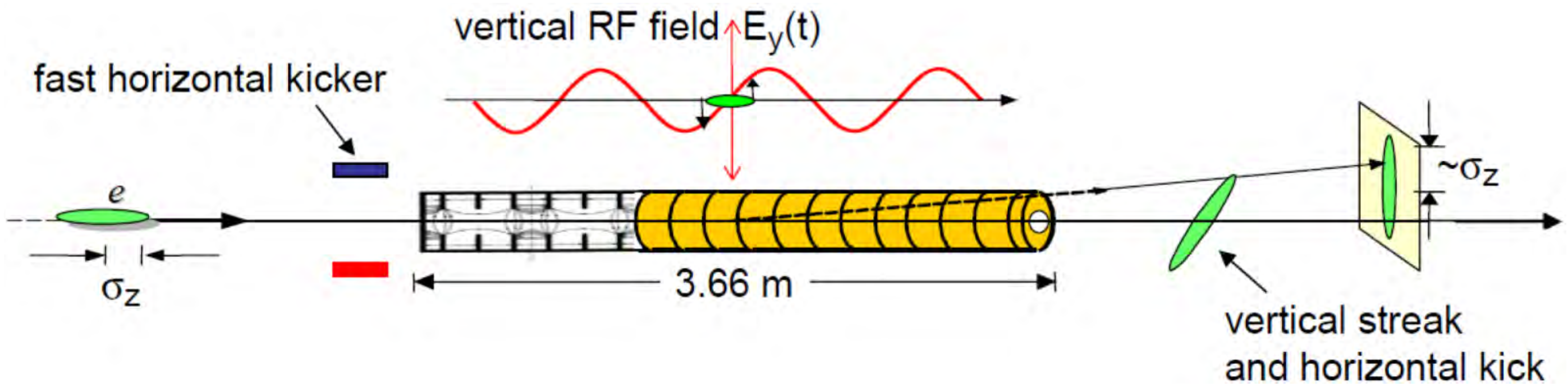

**Figure 11.15. Schematic drawing of the operating principle of a transverse deflecting structure (adapted from [11.42]). The beam enters from the left and is "streaked" by the RF cavity. For bunch trains, a fast kicker may be used to pick one bunch and move it to an off-axis screen letting all other bunches safely pass the screen. [[11.42]; Courtesy of C. Behrens, DESY]**

The deflection is proportional to the field, and thus, due to the time dependence of the field, is proportional to the arrival time of the electrons, or in other words, to the longitudinal position $z$ of the electron in the bunch. As explained in [11.37], the small kick angle $\Delta x'$ as a function of longitudinal position along the bunch $z$ is given approximately by

$$\Delta x'(z) = \frac{eV_0}{pc}\sin(kz+\phi) \approx \frac{eV_0}{p_z c}\left[\frac{2\pi}{\lambda}z\cos(\phi)+\sin(\phi)\right] \tag{11.15}$$

where $V_0$ is the peak voltage, $p$ is the beam momentum, $\lambda$ is the RF wavelength of the structure, $k$ is the wave number such that $k = 2\pi\,\lambda^{-1}$, and $\phi$ is the RF phase. At zero crossing, $\phi = 0$. The approximation is made that $|z| << \lambda/(2\pi)$, and only the linear term in $z$ is retained.

The streaked electron bunch is imaged on a screen placed downstream of the TDS. Including the transport of the beam from the TDS to the screen with an angular to spatial element defined as

$$R_{12} = \sqrt{\beta_d\beta_s}\,\sin(\Delta\Psi) \tag{11.16}$$

the position $\Delta x$ of the deflected electrons on the screen is obtained by [11.37]

$$\Delta x(z) = \frac{eV_0}{pc}\,R_{12}\left[\frac{2\pi}{\lambda}\,z\cos(\phi) + \sin(\phi)\right] \tag{11.17}$$

with the beta-functions $\beta_d$ at the deflector and $\beta_s$ at the screen position, and $\Delta\Psi$ is the betatron phase advance from the deflector to the screen.

From this we can calculate the rms size $\sigma_x$ of the streaked beam on the screen

$$\sigma_x^2 = \sigma_{x,s}^2 + \sigma_z^2 R_{12}^2\left[\frac{eV_0}{pc}\,\frac{2\pi}{\lambda}\,\cos(\phi)\right]^2 \tag{11.18}$$

The size of the streaked beam depends on the rms length of the bunch, $\sigma_z$, and also on the rms size of the unstreaked transverse bunch at the screen location, $\sigma_{x,s}$, *i.e.*, the beam size when no deflecting voltage is applied. The bunch length will dominate the width of the streaked bunch if the chosen voltage is high enough, such that

$$V_0 > \frac{\lambda}{\pi\sigma_z|\cos(\phi)|}\,\frac{\sigma_x\gamma E_e}{R_{12}} \tag{11.19}$$

with the electron rest energy of $E_e$ = 511 keV. The unit of the voltage $V_0$ in Equ. 11.19 is in volts if $E_e$ is given in electron volts.

As an example, for LCLS ($p$ = 5 GeV $c^{-1}$, $\beta \approx 50$ m, $\sigma_z$ = 24 µm, $\sigma_{x,s} \approx 70$ µm), a voltage of more than 10 MV should be applied for an S-band structure ($\lambda$ = 10.5 cm). [11.37]

The bunch length is calculated from Equ. 11.18 with the following expression

$$\sigma_z = \sqrt{\sigma_x^2 - \sigma_{x,s}^2}\,\frac{\lambda}{2\pi}\,\frac{1}{eV_0}\,\frac{pc}{|\sin(\Psi)\cos(\phi)|}\,\frac{1}{\sqrt{\beta_d\beta_s}} \tag{11.20}$$

An advantageous phase advance is near $\pi/2$ and an RF phase near zero-crossing, $\cos(\phi) = \sin(\Psi) = 1$. If the beam has a momentum chirp along the bunch, which for example is the case for beams compressed by chicane bunch compressors, a correction term has to be applied. As shown in [11.37], the effect of a linear momentum correlation along the bunch can be eliminated by choosing $\phi = 0$, or by taking the average bunch length measured at the two phases $\phi = \phi_0$ and $\phi = \pi - \phi_0$, respectively. The rms time resolution obtained with a TDS can be as small as 10 fs.

The streaked beam can be passed through a dispersive section in the plane that is not streaked. Then, the image directly shows the longitudinal phase space: Time on one coordinate and energy on the other. Figure **11.16** is an example of the measurement of the longitudinal phase space at FLASH. A single shot measurement of the longitudinal phase space is a powerful tool to measure and tune the performance of an FEL, for example.

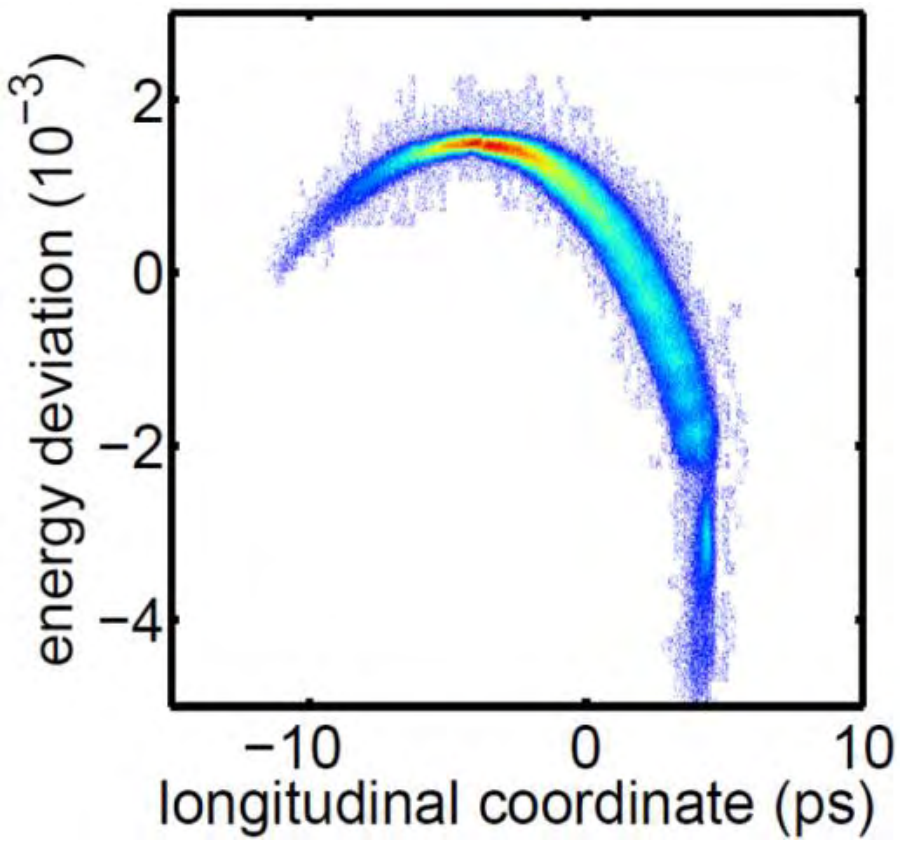

**Figure 11.16. Example of a single shot image showing the longitudinal phase space distribution of a single electron bunch measured with the deflecting cavity LOLA at FLASH. The colors indicate the charge density: blue is a low density and red is a high density. [Courtesty of C. Behrens]**

Burst mode machines with narrowly spaced bunches use a fast kicker system to select one bunch from the bunch train. The bunch to be analyzed is kicked on an off-axis screen, while all other bunches proceed on their usual path to the dump. In this case, the RF pulse of the TDS must be shorter than the bunch distance.

### 11.4.3 Frequency Domain Measurements

Frequency domain measurements use coherent radiation emitted by the electron bunch (transition, diffraction, or synchrotron radiation). A bunch emits coherently in wavelengths longer than the bunch length or the longitudinal microstructures of the bunch.

The radiation power spectrum of the coherent radiation, $P_{coh}(\omega)$, emitted by a bunch of $N$ particles can be expressed by

$$P_{coh}(\omega) = P_s(\omega)\, N^2\, |F(\omega)|^2 \qquad (11.21)$$

where $P_s(\omega)$ is the radiation power spectrum of a single particle, $F(\omega)$ is the longitudinal form factor of the bunch and $\omega$ is the frequency of the emitted radiation. $P_s(\omega)$ can be calculated analytically, and thus the measurement of the coherent radiation power spectrum directly yields $|F(\omega)|$. The longitudinal charge distribution, *i.e.*, bunch shape, is the inverse Fourier transform of the form factor.

A Martin-Puplett interferometer is an example of a device used for measuring longitudinal bunch shape. [11.43], [11.44]

It is based on the measurement of the autocorrelation function of the incoming radiation pulse in the terahertz wavelength range. Electron bunches with a length in the 50 μm to millimeter scale radiate coherently in the terahertz spectral range. As a radiator, diffraction radiation from off-axis screens or from a slit screen is often used. The coherent radiation pulse is guided out of the electron beam pipe through a suitable window. Diamond is the best choice, and single crystal quartz is also used. It is important that the window is transparent to the wavelength of interest and the beam path is evacuated to avoid frequency dependent absorption of the radiation, for example, by water. Wire grids are used to polarize and split the pulse into two beams. A beam splitter, which transmits one polarization and reflects the other, is used to divide and recombine the two pulses after flipping their polarization by roof mirrors. The recombined radiation is elliptically polarized; the degree of polarization depends on the path difference between the two arms. By varying the path length of one arm, the autocorrelation function is measured. The horizontal- and vertical-polarization components are measured by two broadband detectors, such as pyroelectric detectors or Golay cells. A Fourier transform of the autocorrelation function gives the magnitude of the form factor. However, this method does not include phase information, which is also required to determine the longitudinal charge distribution. The phase must be calculated analytically, for example by using a Kramers-Kronig relation. [11.45] In practice, the data analysis for this method is quite complicated and several frequency dependent corrections need to be considered.

### 11.4.4 Electro-Optical Sampling Method

Electro-optical (E-O) sampling is a technique to measure the shape and length of short bunches. It is based on detecting the electric field of the electron bunch as it passes closely to a non-linear optical crystal, for example, ZnTe. The electric field inside the crystal is influenced by the electron bunch and can be probed by an initially linearly polarized femtosecond laser pulse. The laser pulse becomes elliptically polarized and the measurement of its polarization yields information on the longitudinal distribution of the electron bunch. Electro-optical sampling methods are detailed in [11.42]. Figure **11.17** is an example of a set-up for temporally encoded electro-optic detection.

Using a Ti:Sa probe laser pulse of 30 fs (FWHM) duration (energy of > 100 μJ) and a 300 μm thick $\beta$-barium borate (BBO) crystal, a resolution of better than 100 fs can be reached. [11.42] Temporally resolved, electro-optic detection was first demonstrated at FELIX. [11.46]

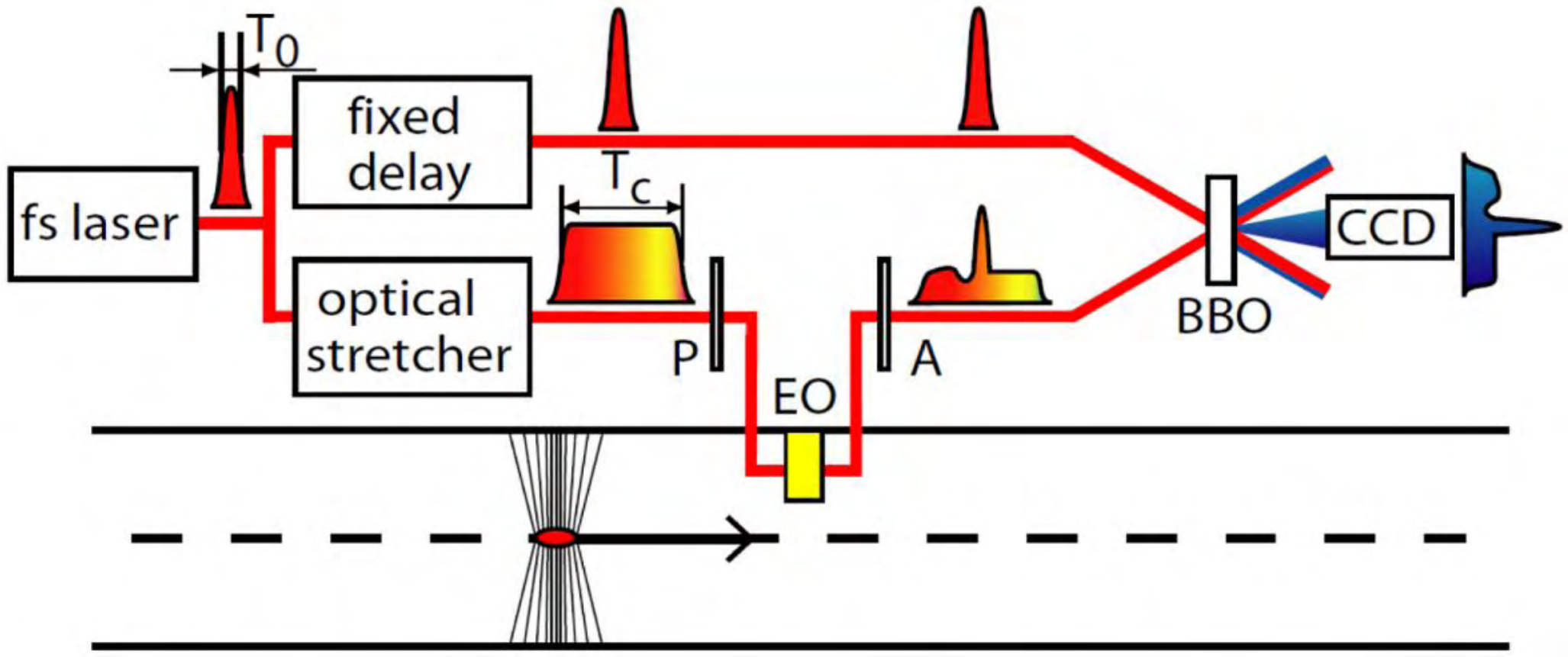

**Figure 11.17. Schematic of a temporally encoded electro-optic detection set-up. The short laser pulse with a length of $T_0$ is split into two parts: One part is stretched to several picoseconds ($T_C$) and sent through the polarizer (P) and the E-O crystal (EO) in parallel with the Coulomb field of the electron bunch, while the second part remains unstretched. In the E-O crystal, the stretched laser pulse acquires an elliptic polarization with an ellipticity proportional to the electric field of the electron bunch, and with the same temporal structure. The analyzer (A) turns the elliptical polarization into an intensity modulation, which is then sampled by the short pulse in a single-shot cross-correlator. [Adapted from [11.47], with permission from Elsevier]**

## 11.5 BEAM POSITION

Beam position monitors (BPMs) are used to measure the transverse beam position in the beamline. BPMs are usually placed along the entire beamline at appropriate locations to determine the orbit, to understand the optics response, to determine the beam energy in spectrometers and to evaluate beam jitter. In linacs, a good single shot resolution is desirable to be able to observe shot-to-shot effects. For accelerators operating in burst modes, individual bunches in a train should be resolved clearly.

There are several different types of BPMs: The basic types are broadband pickups and resonant cavity monitors. In a broadband pickup, the beam induces a signal on four electrodes which are arranged symmetrically around the beam pipe. Examples of this kind of monitor are button- and stripline-BPMs. In resonant cavity monitors, a dipole mode is excited in a cylindrical cavity by an off-axis beam. For both types of monitors, the signal depends on the transverse beam position and the bunch charge. Therefore, four symmetrically arranged electrodes (for button- and stripline-monitors) or slots (for cavity monitors) are needed to achieve a linear, charge independent measurement of the beam position. Reentrant cavity monitors are special type of cavity monitors, which are operated below the resonant frequency with a large bandwidth.

The typical single shot resolution of button monitors is ~50 µm rms, *viz.*, sufficient for most applications, especially for injectors where the beam size is still large. In general, the resolution should be less than 10% of the expected beam size. This is because most applications require a beam position jitter of less than 10% of the beam size – to be resolved by the BPMs. Examples of a very demanding requirement are undulators for FELs. Here, the single shot resolution should be in the 1 µm range. This is achieved with cavity type monitors.

The signals from the pick-ups are amplified. The signal's amplitude or its arrival time is measured with suitable sample-and-hold detectors together with good resolution ADCs. One method to determine the beam position is to measure the difference between two signals from two opposite pick-ups, normalized by their sum. For example, the horizontal beam position $x$ is determined by

$$x = k\,\frac{h_l - h_r}{h_l + h_r} \tag{11.22}$$

where $h_l$ and $h_r$ are the signal amplitudes of pick-ups mounted left and right of the vacuum chamber. For instance, the calibration constant $k$ is determined by measuring the known response of corrector magnets on the beam position. The absolute zero position (center of the vacuum chamber) is determined by carefully equalizing the amplifier gain of both arms.

## 11.6 ENERGY AND ENERGY SPREAD

A dispersive section of the electron beamline is used to determine the energy (or momentum) and energy spread of the beam. Such a section consists typically of a deflecting dipole magnet with a known homogeneous field and a screen located in the dispersive section. Often, beam position monitors are included.

The momentum, $p$, of an electron is given by

$$p = eBr \tag{11.23}$$

with the electron charge *e*, the magnetic field *B*, and the radius of the track inside the magnet, *r*. The beam energy, *E*, is obtained *via* the usual relation $E^2 = p^2c^2 + m_e^2c^4$. In practical units, Equ. 11.23 can be written as

$$p\left[\frac{\text{GeV}}{\text{c}}\right] = 0.3\,B\,[\text{T}]\,r\,[\text{m}] \tag{11.24}$$

For accurate energy measurements, the beam must be placed on the nominal pass through the dipole magnet where the field and the radius of curvature are known. A set of BPMs is used to determine the orbit through the dipole field. The dipole current is adjusted such that the beam is centered in the beamline downstream, the dispersive arm. Centering is assured using BPMs or calibrated screens. The beam momentum is then given by Equ. 11.23. Hysteresis effects have to be taken into account. A careful measurement of the field as a function of the current is mandatory.

Beam centering is not required if its position is measured accurately upstream and downstream of the dipole, and these positions must be considered when determining the relation between the beam energy and the dipole current.
The energies of the individual electrons in the bunch differ somewhat from each other. Electrons with different energies follow different paths through the dipole, and therefore end up at different locations on the downstream screen.

The momentum, *p*, of an electron at position $x_D$, with respect to the nominal center position occupied by an electron with momentum $p_0$, is

$$p = p_0\left(1 + \frac{x_D}{D}\right) \tag{11.25}$$

The quantity *D* is called the dispersion and has the dimension of length. Thus, with a dispersion of 10 mm and a screen- or position-resolution of 10 μm, a momentum resolution of $10^{-3}$ is obtained.

The energy spread can be determined from the transverse beam spot size. If the dipole deflects in the horizontal plane, the horizontal dimension gives the energy spread. The transverse size of the bunch needs to be taken into account. The spot width caused by the energy spread should be significantly larger than the beam's natural size. Therefore, for accurately measuring energy spread, the beam should be well focused using quadrupoles or a solenoid magnet upstream of the dipole.

A screen in a dispersive section is also often used to determine the on-crest phase of an accelerating module. With the phase on crest, energy is maximized and the correlated energy spread minimized. For adjusting the right compression for FELs, determining the on-crest phase within a degree of RF-phase is mandatory.

The correlated energy spread is induced by the sinusoidal shape of the RF field. The uncorrelated energy spread is an intrinsic property of the electron source, for example, RF gun-based sources have a very small energy spread of a few kiloelectron volts. Its smallness may induce unwanted collective effects during bunch compression due to space charge or coherent synchrotron radiation; these need to be mitigated by the so-called laser heaters, for example. [11.48]

## 11.7 EMITTANCE

Beam emittance characterizes the beam dynamics and is an important measure of its quality.

The emittance of an electron bunch, $\varepsilon$, usually is defined in terms of measurable quantities that are also applicable in simulation codes [11.49]

$$\varepsilon = \frac{1}{\bar{p}_z}\sqrt{\langle x^2\rangle\langle p_x^2\rangle - \langle xp_x\rangle^2} \tag{11.26}$$

with $x$ being the Cartesian transverse horizontal coordinate of the electron and $p_x$ the momentum component in horizontal direction. An equivalent equation holds for the vertical component, $y$. For simplicity, whenever the notation $x$ is used, the expression holds for both Cartesian coordinates. The brackets $\langle\ \rangle$ define the second central moment of the distributions, or in other words, the mean square value. For a definition, see [11.49]. Note that $\bar{p}_z$ is the average longitudinal momentum of the electrons. The coordinate $z$ is in direction of acceleration. The emittance defined in this way is an rms emittance.

In experiments, the terms with transverse momenta $p_x$ are often estimated by the measured beam divergence, such as

$$\sigma_x' = \left\langle\sqrt{\frac{p_x}{p_z}}\right\rangle^2 = \langle\sqrt{x'}\rangle^2 \tag{11.27}$$

the rms beam size is

$$\sigma_x = \sqrt{\langle x^2\rangle} \tag{11.28}$$

Using the beam divergence, we write the rms emittance in the following form, often called trace space[2] emittance

$$\varepsilon_{\text{rms}} = \sqrt{\langle x^2\rangle\langle x'^2\rangle - \langle xx'\rangle^2} \tag{11.29}$$

Equ. 11.26 shows that the emittance is adiabatically damped with the average longitudinal beam momentum, $\bar{p}_z$. Therefore, it is convenient to define a normalized emittance, $\varepsilon_n$, to be

$$\varepsilon_n = \frac{\bar{p}_z}{m_e c}\varepsilon_{geo} \approx \gamma\varepsilon_{geo} \tag{11.30}$$

where $\varepsilon_{qeo}$ is sometimes called the geometrical emittance. Here, $\gamma$ is the relativistic factor as defined in Equ. 11.12. Although the geometrical emittance decreases as a function of increasing beam energy, the normalized emittance stays constant. Therefore, normalized emittance commonly is used to characterize the beam.

The emittance, as defined above, is equivalent to the Courant-Snyder invariant, $W = \varepsilon$, which is given by [11.50]

---

[2]Usually, we talk about *phase space* when we use the space coordinates $x$ together with the momentum $p_x$ of the particles; *trace space* is used to express, that the angular coordinate $x'$ is used rather than $p_x$.

$$W = \gamma x^2 + 2\alpha x x' + \beta x'^2 \tag{11.31}$$

$\alpha$, $\beta$ and $\gamma$ are the Twiss-parameters, and $x' = dx/dz$; $z$ is the direction of motion. These parameters should not be confused with the relativistic $\beta$ and $\gamma$ parameters. The relation between emittance and beam size in terms of the Courant-Snyder beta-function, $\beta(z)$, is

$$\sigma_x^2(z) = \varepsilon \beta_x(z) \tag{11.32}$$

As an example, a beam size of $\sigma_x = 100$ μm with a beta-function $\beta_x(z)$ of 10 m is achieved with an emittance of $\varepsilon = 1$ nm.
To determine the emittance, the second central moments $\langle x^2 \rangle$, $\langle x'^2 \rangle$ and $\langle xx' \rangle$ must be measured. This can be done by different methods: A slit mask, a quadrupole scan, or a multi-monitor method.

A typical emittance of a 1 nC electron bunch produced by a photoinjector RF gun is between 1-2 mm mrad. Often the emittance is expressed in micrometers, omitting the angular unit.

A detailed description of how to measure emittance, including a brief theoretical background and a discussion on systematic measurement errors, is given in [11.51].

### 11.7.1 Slit Mask Technique

The slit technique typically is used at low beam energies, where the space charge effects of the drifting beam would not allow a measurement of the true beam divergence. If we neglect space charge effects, a drifting beam along $z$ from $z = 0$ to $z = L$ would change its transverse size, $\sigma_x(z)$, according to its emittance by

$$\sigma_x(L) = \sigma_x(0) + \sigma_x' L \tag{11.33}$$

A beam with this behavior is called an *emittance* dominated beam. $\sigma_x'$ is the beam divergence as defined in Equ. 11.27.

In a space charge dominated beam, we can cut out small portions of the beam such that the resulting beamlets are now emittance dominated. After a certain drift, the beamlets are imaged with an OTR or fluorescence screen, where their size is measured.

To do this, a mask with one or several slits cuts the beam into these beamlets. The size of the slits is small enough so that the effect of space charge can be neglected. As an example, for a typical beam with a charge of 1 nC, a beam size of 1 mm, a nominal emittance of 1-10 μm, and a slit width of 50 μm are usually taken. The thickness of the mask is large enough to provide a good stopping power of the electrons cut. A stainless steel mask for a beam with a few megaelectron volts typically is 5 mm thick. The separation between the slits is larger than the slit's width (1 mm) in order to keep the beamlets well separated after the drift. The drift distance to the screen is about 50 cm. The values above are an example taken from [11.52]. A mask with several slits is difficult to manufacture, but has the advantage of a simultaneous measurement of the whole beam. In contrast, the single slit needs to be scanned through the beam.
A detailed description of the slit mask technique is given in [11.52], [11.53], where space charge effects are also discussed.

Figure **11.18** shows a schematic layout of a multi-slit-based emittance measurement. The space charge dominated beam passes through a mask with multiple slits yielding emittance dominated beamlets.

The parameters needed to determine the emittance as defined in Equ. 11.29 are derived from the beam image: the overall envelope formed by the beamlets gives $\langle x \rangle^2$, the distance between the beamlet centroids is proportional to $\langle xx' \rangle$ and the width of the beamlets gives $\langle x'^2 \rangle$. There are several techniques for calculating these parameters from the measured slit images. [11.53], [11.54]

Instead of using slits, a mask consisting of many small holes, the so-called “pepper-pot,” is used for single shot, 2-D emittance measurements. [11.55]

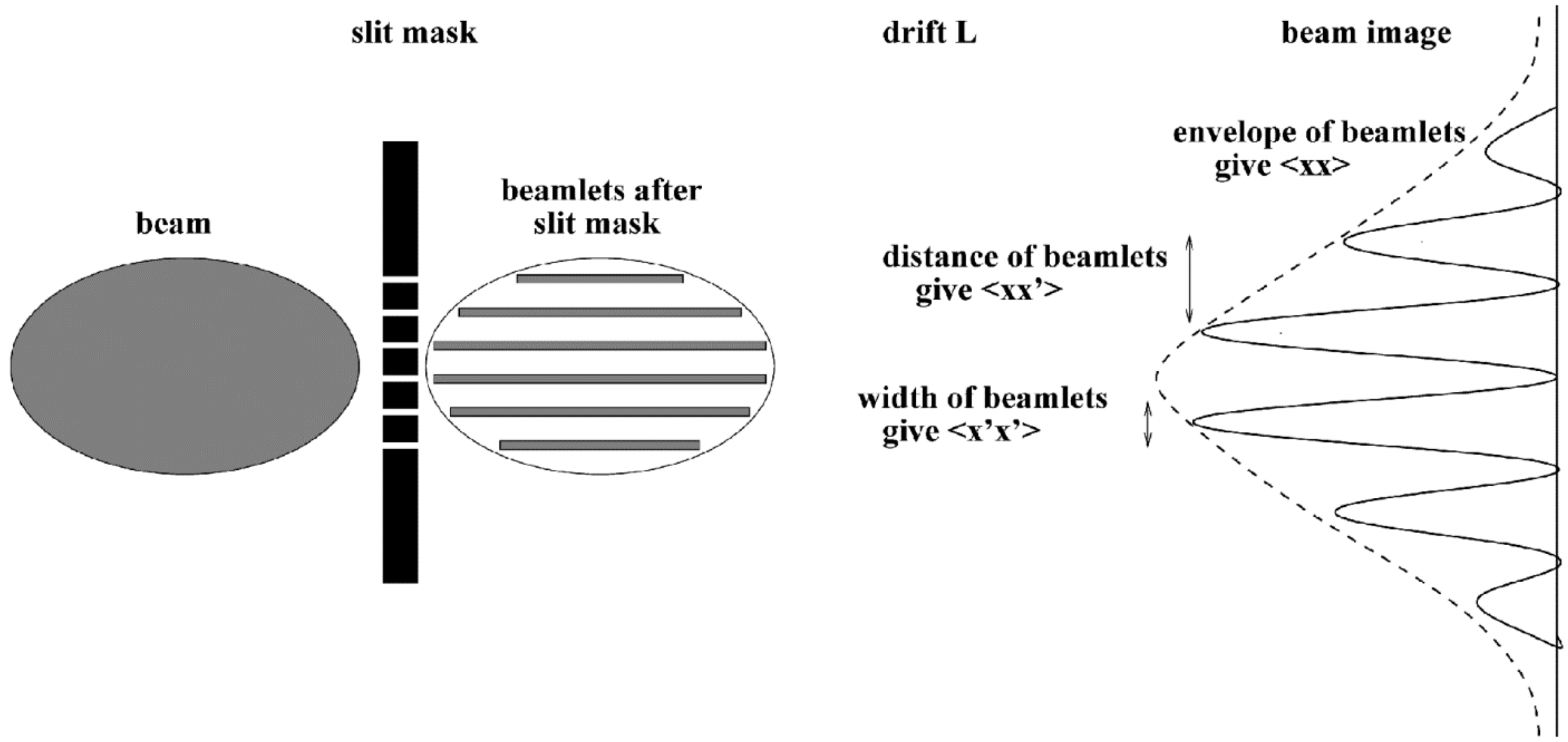

**Figure 11.18. Schematic scheme of an emittance measurement based on multi-slits. The beam is cut into small beamlets which drift the distance *L* to the observation screen. [Courtesy of K. Honkavaara, DESY]**

Slits are used at the exit of RF guns, for high beam energies after considerable acceleration, to say 100 MeV or more; thereafter, slits become inappropriate and other methods are more convenient. They are discussed in the next sections.

### 11.7.2 Quadrupole Scan Method

In the quadrupole scan method, the beam size is measured for different settings of one or more quadrupole magnets placed upstream of the beam size monitor (a screen or a wirescanner). The beam optics (transfer matrices) for different quadrupole settings is calculated and a parabolic fit of the beam sizes measured for the different settings gives the needed beam parameters to determine the emittance. The principle of this method is equivalent to measuring $M^2$ for laser beams, where $M^2$ is the ratio of the beam parameter product, waist size times divergence, of the given laser beam to that of an ideal Gaussian beam. The “focusability” of the beam is a measure of its emittance.

The disadvantage of this method is that the beam optics must be changed for each measurement point and thus it is not suitable if beam losses induced by the scan cannot be transported safely to the dump.
In injectors, the beam power usually is small so that the beam can be dumped somewhere in the beam pipe. Another issue is that very small spot sizes may be obtained at the focal point of the quadrupole. The space charge density at the focus can be so large that its effect is not negligible and needs to be considered.

There may be a problem with resolution; small beam sizes, say less than 50 μm, are difficult to measure with standard methods. On the other hand, good resolution comes with a small field view, which precludes measuring larger spot sizes. To gain a good dynamic range, several magnifications of the optical viewing system are required.
Let us use matrix formalism to describe the beam optics. The transverse trace space of the beam is described by the beam matrix *M*, which is a function of the direction of motion *z* (as an example, refer to [11.51], [11.56])[3]

$$M(z) = \begin{pmatrix} M_{11} & M_{12} \\ M_{21} & M_{22} \end{pmatrix} \tag{11.34}$$

The matrix elements are defined as the second moments of the transverse coordinate *x*, also expressed in terms of the Twiss-parameters and the emittance

$$M_{11} = \langle xx \rangle = \beta\varepsilon \text{ ; } M_{12} = M_{21} = \langle xx' \rangle = -\alpha\varepsilon \text{ ; } M_{22} = \langle x'x' \rangle = \gamma\varepsilon \tag{11.35}$$

The motion of the beam, from a position $z_0$ to a position *z* further downstream along the beamline, can be described by a transfer matrix *R*. The beam matrix at position *z* is then calculated by

$$M(z) = R\, M(z_0)\, R^T \tag{11.36}$$

The emittance can now be determined by measuring the transverse beam size at a location *z* as a function of the current of a quadrupole magnet. The transfer matrix then becomes a function of the current or the magnet field gradient, *g*, of the quadrupole. The square of the beam size $\langle x^2 \rangle = M_{11}$ is given by

$$M_{11}(z) = M_{11}(z_0)\, R_{11}(g)^2 + 2M_{12}(z_0)\, R_{11}(g)\, R_{12}(g) + M_{22}(z_0)\, R_{12}(g)^2 \tag{11.37}$$

The transport matrix includes drifts and the quadrupole. Figure **11.19** shows a simulated quadrupole scan.

With a fit to the measured data, the relevant matrix elements are evaluated to determine the emittance according to Equ. 11.29, using Equ. 11.37 and Equ. 11.35.

To overcome potential systematic errors due to space charge effects, using two or more quadrupoles will keep the beam spot size large enough. Furthermore, this method has the advantage that the beam size can be kept sufficiently small so that the entire beam fits on to the screen.

In the next section, I describe the multi-monitor method that avoids the space charge effect and does not require a change of the beam optics.

[3]Often, the Greek letter $\sigma$ is used for the beam matrix. We use *M* to avoid confusion with the beam size $\sigma$.

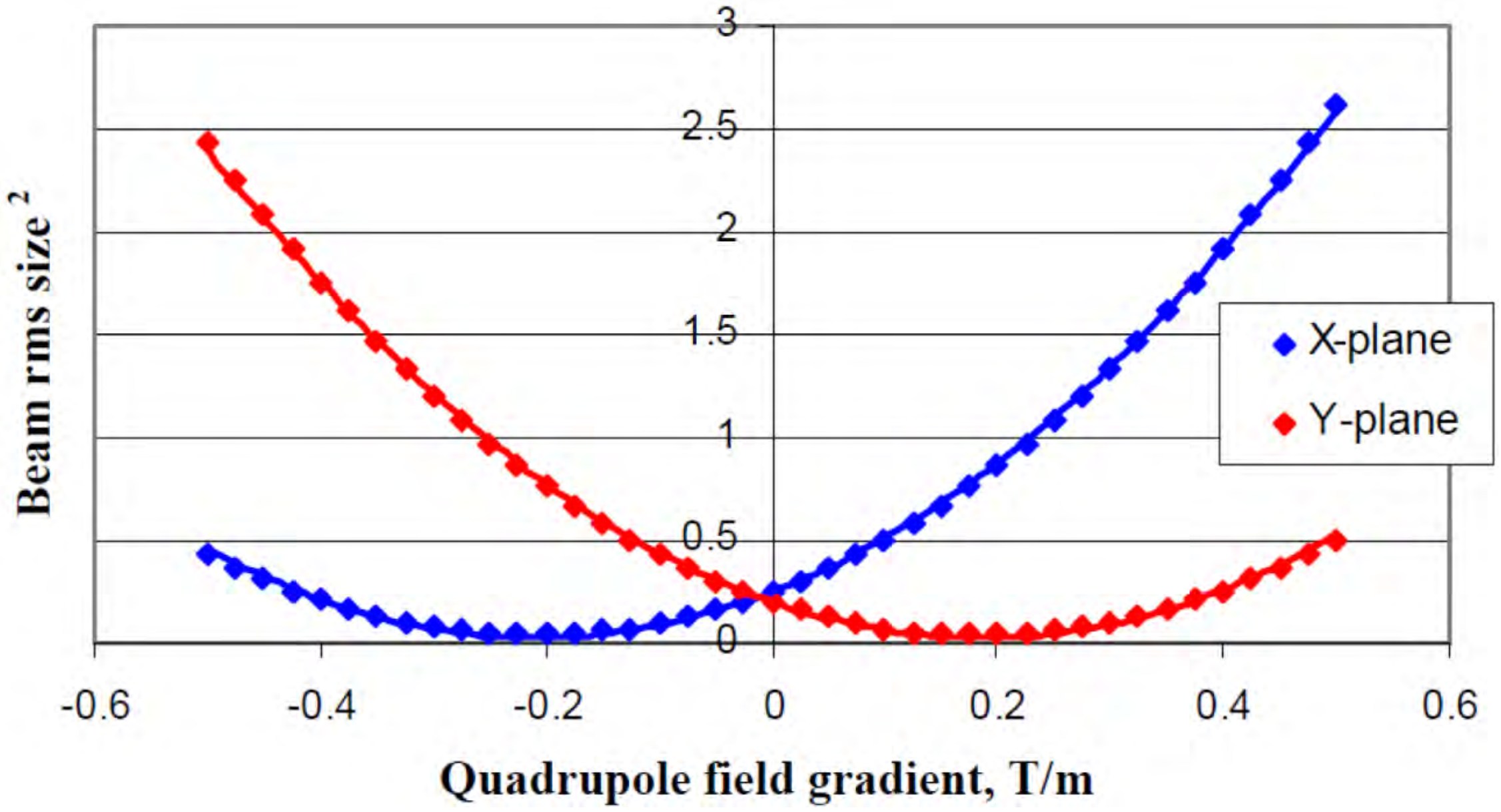

**Figure 11.19. Simulated quadrupole scan. The quadrupole gradient is plotted as a function of the squared-beam size at a certain position *z* downstream from the quadrupole. In this example, the focal point for the horizontal- and vertical-coordinate appears at different gradients. The beam size is given in millimeters. [[11.56]; Courtesy of S. Skelton]**

### 11.7.3 Multi-Monitor Method

In a multi-monitor method, the beam size is measured at three or more locations without changing the beam optics [11.57]. A typical set-up consists of a period FODO magnet lattice of focusing- and defocusing-quadrupoles, together with four screens or wirescanners. A FODO lattice is a periodic structure. One period consists of a focusing element (F), usually a quadrupole, a drift (O), a defocusing element (D) and an additional drift (O).

The parameters required for assessing emittance are determined from the measured sizes of the beam by fitting the transport function or by using tomographic phase space reconstruction methods. The advantage of the multi-monitor method is that the beam optics is fixed and the beam size on the screens can be chosen such that a good resolution is obtained. Fixed beam optics avoid a scan of the magnetic field of quadrupoles. Hysteresis effects of the quadrupole magnets, beam losses during scans are avoided and set-up time is reduced. However, a set-up with a FODO lattice, including a matching section, requires considerable space for beamline elements; several quadrupoles and screen stations have to be realized. The matching section is useful to adapt the optics to be optimal for the emittance measurement. Once such a section is established in an accelerator, it can also serve to match the optics emerging from the injector to the accelerator. The measurement yields all Twiss-parameters from Equ. 11.31 ($\alpha$, $\beta$, $\gamma$) required to match to the optics of the accelerator.

As an example, Figure **11.20** shows the matching and FODO section in the FLASH injector that is 10 m long, with 12 quadrupoles and 4 OTR/wirescanner stations.

Loehl *et al.* give a detailed description of emittance measurements at FLASH using the multi-monitor method [11.14], [11.51].

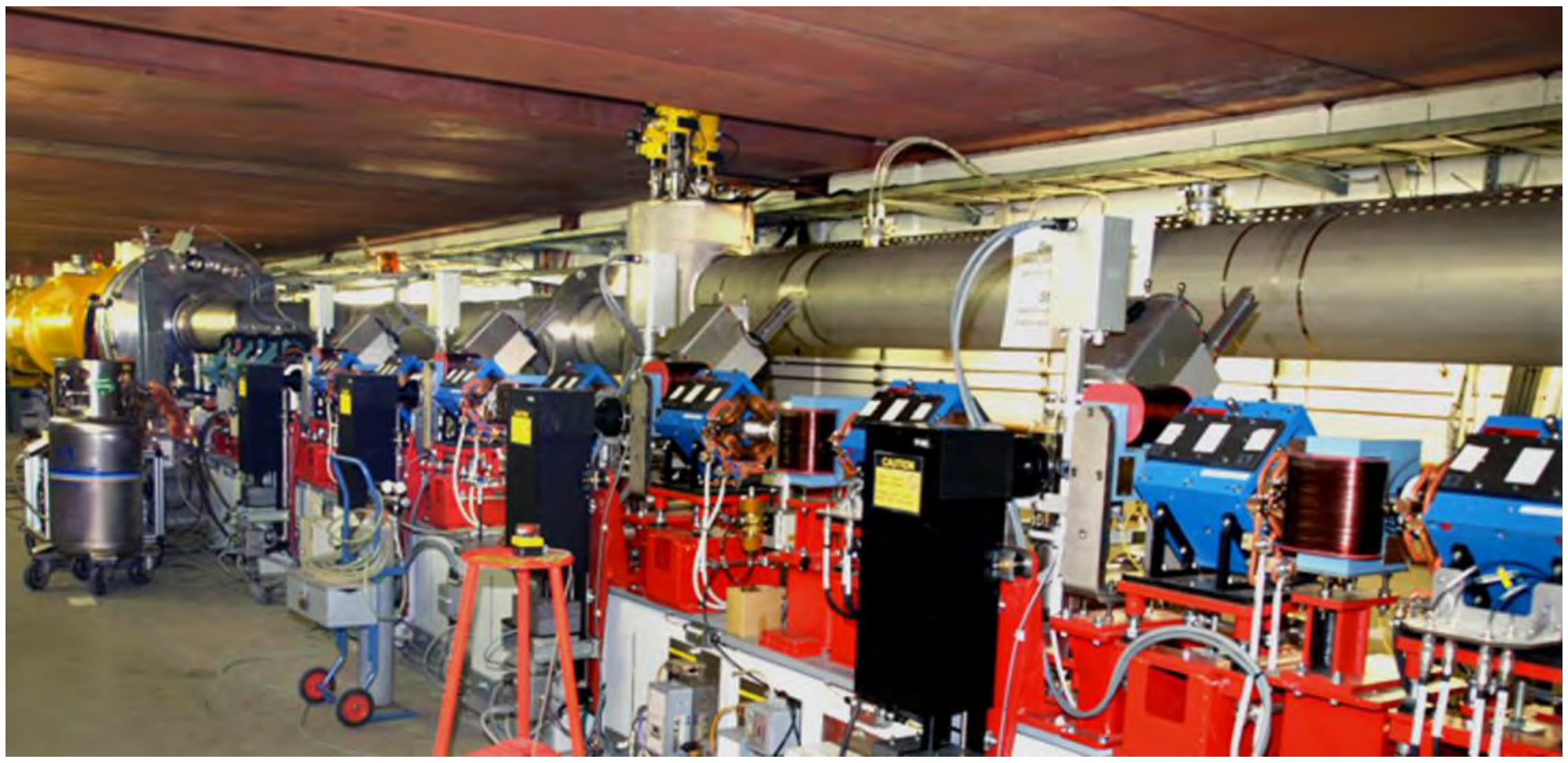

**Figure 11.20. Picture of the diagnostic section of the FLASH injector. Four screen stations equipped with wire scanners in a FODO lattice with several quadrupole magnets are employed in measuring Twiss-parameters of the beam.**

Using Equ. 11.36 and Equ. 11.35, we can write a similar expression as for the quadrupole scan method (Equ. 11.37)

$$M_{11}(z_i) = M_{11}(z_0)\, R_{11}(z_i)^2 + 2M_{12}(z_0)\, R_{11}(z_i)\, R_{12}(z_i) + M_{22}(z_0)\, R_{12}(z_i)^2 \tag{11.38}$$

Here, the matrix elements are evaluated at a position $z = z_i$. When we now measure the squared-beam size $\langle x_i^2 \rangle = M_{11}(z_i)$ at three different locations $z_i$, where $i = 1, 2, 3$, we obtain the three relevant beam matrix elements from

$$\begin{pmatrix} \langle x_0^2 \rangle \\ \langle x_0 x_0' \rangle \\ \langle x_0'^2 \rangle \end{pmatrix} = \Re^{-1} \begin{pmatrix} \langle x_1^2 \rangle \\ \langle x_2^2 \rangle \\ \langle x_3^2 \rangle \end{pmatrix} \tag{11.39}$$

with the matrix $\Re$ given by

$$\Re = \begin{pmatrix} R_{11}^2(z_1) & R_{11}(z_1)R_{12}(z_1) & R_{12}^2(z_1) \\ R_{11}^2(z_2) & R_{11}(z_2)R_{12}(z_2) & R_{12}^2(z_2) \\ R_{11}^2(z_3) & R_{11}(z_3)R_{12}(z_3) & R_{12}^2(z_3) \end{pmatrix} \tag{11.40}$$

The emittance is then calculated with Equ. 11.29 where the Twiss parameters at $z = z_0$ are determined by Equ. 11.35.

The information about the beam lattice is contained in the matrix $\Re$ and needs to be known accurately. A discussion of the systematic errors involved appears in [11.51].

Correctly choosing the phase advance between the screens is a key in optimizing the system's resolution power. A phase advance of 180° between screens would mean that the same information is always measured. The optimal phase advance would be one wherein screens are equally placed in the 180° phase advance space, taking into account that 0˚ and 180˚ measure the same beam. Therefore, we obtain 60˚ for three screens, 45˚ for four screens, and so on.

In the discussion above, I assumed a vanishing dispersion. In the presence of a non-zero dispersion, this method can be extended to encompass the dispersion function and the angular dispersion function. Here, at least six measurements of the beam size at appropriate places are needed.

It is also possible to measure the coupling between the horizontal and the vertical plane; at least four measurements are required for this.

Even though three measurements would suffice to give the Twiss parameters, more screens allow using fitting methods, thereby increasing the measurement's precision.

### 11.7.4 Tomographic Reconstruction of the Phase Space

The phase space description with Twiss parameters assumes a homogeneous distribution of the electrons in the phase space. Since this is often not the case, tomographic methods generate more information about the distribution of electrons in the phase space. These methods also access information about possible tails in the distribution. With the pure statistical approach using an emittance defined by the second-order moments of the distribution, tails have a large weight, and thus greatly influence the rms emittance.

The tomographic reconstruction uses measurements of the phase space from different projections. Many methods require a large number of projections, which is difficult to realize experimentally. [11.58] The maximum entropy algorithm (MENT) provides a reconstruction from a few projections only. A comprehensive description of the MENT algorithm is given in [11.59]–[11.62].

Temporally resolved slice emittance measurements using the quadrupole scan technique together with a transverse deflecting cavity is a good method of reconstructing the transverse phase space as a function of longitudinal position within the electron bunch using tomographic methods as detailed in [11.63] and [11.64].

## 11.8 CONFLICT OF INTEREST AND ACKNOWLEDGEMENT

I confirm that this article content has no conflicts of interest. I would like to thank Katja Honkavaara (DESY) for her help in preparing this manuscript, her continuous support and fruitful discussions on many scientific details.

***References***

[11.1] P. Strehl, *Beam Instrumentation and Diagnostics*, Berlin: Springer, 2005.
[11.2] D. Lipka, W. Kleen, J. Lund-Nielsen *et al.*, "Dark current monitor for the European XFEL," in *Proc. 10th European Workshop Beam Diagnostics Instrumentation Particle Accelerators*, 2011, May 2011.
[11.3] K. L. Brown and G. W. Tautfest, "Faraday-cup monitors for high-energy electron beams," *Rev. Sci. Instrum.*, vol. 27, pp. 696-702, September 1956.
[11.4] L. Bess and A. O. Hanson, "Measurement of the electron current in a 22-mev betatron," *Rev. Sci. Instrum.*, vol. 19, pp. 108-110, February 1948.
[11.5] L. Bess, J. Ovadia and J. Valassis, "External beam current monitor for linear accelerators," *Rev. Sci. Instrum.*, vol. 30, pp. 985-988, November 1959.
[11.6] R. Yamada, "New magnetic pickup probe for charged particle beams," *Japanese J. Appl. Phys.*, vol. 1, pp. 92-100, 1962.

[11.7] S. N. Gardiner, J. L. Matthews and R. O. Owens, "An accurate non-intercepting beam current integrator for pulsed accelerator beams," *Nucl. Instrum. Meth.*, vol. 87, pp. 285-290, October 1970.
[11.8] K. B. Unser, "Design and preliminary tests of a beam intensity monitor for LEP," in *Proc. 1989 Particle Accelerator Conf.*, 1989, pp. 71-73.
[11.9] J. Bergoz, "Current monitors for particle beams," *Nucl. Physics A*, vol. 525, pp. 595-600, April 1991.
[11.10] Vacuumschmelze GmbH & Co. KG, Gruener Weg 37, 63450 Hanau, Germany, http://www.vacuumschmelze.de/.
[11.11] P. Forck, "Lecture notes on beam instrumentation and diagnostics," Archamps: Joint University Accelerator School, January 3-March 11 2011, http://www-bd.gsi.de/conf/juas/juas_script.pdf.
[11.12] L. Wartski, S. Roland, J. Lasalle *et al.*, "Interference phenomenon in optical transition radiation and its application to particle beam diagnostics and multiple-scattering measurements," *J. Appl. Phys.*, vol. 46, 3644-3653, August 1975.
[11.13] L. Wartski, "Study of optical transition radiation produced by 30 to 70 MeV energy electrons: Applications to the diagnostics of beams with charged particles," Ph.D. Thesis, Paris-Sud University, Orsay, France, 1976.
[11.14] F. Löhl, S. Schreiber, M. Castellano *et al.*, "Measurements of the transverse emittance at the FLASH injector at DESY," *Phys. Rev. ST Accel. Beams*, vol. 9, pp. 092802-1–092802-6, September 2006.
[11.15] Basler AG, An der Strusbek 60-62, 22926 Ahrensburg, Germany, http://www.baslerweb.com/index.html.
[11.16] R. W. Allison Jr., R. W. Brokloff, R. L. McLaughlin *et al.*, "A radiation-resistant chromium-activated aluminum oxide scintillator," Lawrence Berkley Radiation Laboratory, Technical Report No. UCRL-19270, July 1969.
[11.17] C. D. Johnson, "The development and use of alumina ceramic fluorescent screens," European Laboratory for Particle Physics, Technical Report No. CERN/PS/90-42(AR), October 1990.
[11.18] MAC-A994R, available by Morgan Technical Ceramics, 55-57 High Street, Windsor, Berkshire, U.K., SL4 1LP. Available at http://www.morgantechnicalceramics.com/download.php?5151434153695931675250727139694b714e33346d696d4675694a784e6150714d6c54785842453058532f2b.
[11.19] W. S. Graves, E. D. Johnson and P. G. O'Shea, "A high resolution electron beam profile monitor," in *Proc. 1997 Particle Accelerator Conf.*, 1997, pp. 1993-1995.
[11.20] A. Murokh, J. Rosenzweig, V. Yakimenko *et al.*, "Limitations on the resolution of YAG:Ce beam profile monitor for high brightness electron beam," in *Proc. 2$^{nd}$ Int. Committee Future Accelerators Advanced Accelerator Workshop Physics High Brightness Beams*, 2000, pp. 564-580.
[11.21] Saint-Gobain Crystals, 18900 Great Lakes PKWY, Hiram, OH 44234-9681, USA, "YAG(Ce) Yttrium Aluminum Garnet Scintillation Material," http://www.detectors.saint-gobain.com/uploadedFiles/SGdetectors/Documents/Product_Data_Sheets/YAG-Data-Sheet.pdf.
[11.22] X. Artru, R. Chehab, K. Honkavaara *et al.*, "Resolution power of optical transition radiation: Theoretical considerations," *Nucl. Instrum. Meth. B*, vol. 145, pp. 160-168, October 1998.
[11.23] K. Honkavaara, Optical Transition Radiation in High Energy Electron Beam Diagnostics, Helsinki: Helsinki Institute of Physics, 1999.
[11.24] R. Jung and R. J. Colchester, "Development of beam profile and fast position monitors for the LEP injector linacs," *IEEE Trans. Nucl. Sci.*, vol. 32, pp. 1917-1919, October 1985.
[11.25] R. Fulton, J. Haggerty, R. Jared *et al.*, "A high resolution wire scanner for micron-size profile measurements at the SLC," *Nucl. Instrum. Meth. A*, vol. 274, pp. 37-44, January 1989.
[11.26] M. C. Ross, "Wire scanner systems for beam size and emittance measurements at SLC," in *AIP Conf. Proc.*, vol. 229, 1991, pp. 88-106.

[11.27] U. Hahn, N. V. Bargen, P. Castro *et al.*, “Wire scanner system for FLASH at DESY,” *Nucl. Instrum. Meth. A*, vol. 592, pp. 189-196, July 2008.

[11.28] J. Camas, C. Fischer, J. J. Gras *et al.*, “Observation of thermal effects on the LEP wire scanners,” in *Proc. 1995 Particle Accelerator Conf.*, vol. 95, 1995, pp. 2649-2651.

[11.29] C. Yan, P. Adderley, D. Barker *et al.*, “Superharp – a wire scanner with absolute position readout for beam energy measurement at CEBAF,” *Nucl. Instrum. Meth. A*, vol. 365, pp. 261-267, November 1995.

[11.30] Ch. Gerth, J. Feldhaus, K. Honkavaara *et al.*, “Bunch length and phase stability measurements at the TESLA test facility,” *Nucl. Instrum. Meth. A*, vol. 507, pp. 335-339, July 2003.

[11.31] M. Uesaka, T. Ueda, T. Kozawa *et al.*, “Precise measurement of a subpicosecond electron single bunch by the femtosecond streak camera,” *Nucl. Instrum. Meth. A*, vol. 406, pp. 371-379, April 1998.

[11.32] Femtosecond streak camera C6138 (FESCA-200), available by Hamamatsu Photonics K.K., Hamamatsu, Japan. Available at http://jp.hamamatsu.com/resources/products/sys/pdf/eng/e_c6138.pdf.

[11.33] T. Watanabe, M. Uesaka, J. Sugahara *et al.*, “Subpicosecond electron single-beam diagnostics by a coherent transition radiation interferometer and a streak camera,” *Nucl. Instrum. Meth. A*, vol. 437, pp. 1-11, November 1999.

[11.34] J. Roensch, J. Rossbach, G. Asova *et al.*, “Investigations of the longitudinal beam properties at the photoinjector test facility in Zeuthen,” in. *Proc. 2006 Free Electron Laser Conf.*, 2006, pp. 597-600.

[11.35] R. H. Miller, R. F. Koontz and D. D. Tsang, “The SLAC injector,” *IEEE Trans. Nucl. Sci.*, vol. 12, pp. 804-808, June 1965.

[11.36] X.-J. Wang, “Producing and measuring small electron bunches,” in *Proc. 1999 Particle Accelerator Conf.*, 1999, pp. 229-233.

[11.37] P. Emma, J. Frisch and P. Krejcik, “A Transverse RF deflecting structure for bunch length and phase space diagnostics,” Stanford Linear Accelerator Laboratory, Technical Report No. LCLS-TN-00-12, August 2000.

[11.38] R. Akre, L. Bentson, P. Emma *et al.*, “A transverse RF deflecting structure for bunch length and phase space diagnostics,” Stanford Linear Accelerator Center, Technical Report SLAC-PUB-8864, June 2001.

[11.39] O. H. Altenmueller, R. R. Larsen and G. A. Loew, “Investigations of traveling wave separators for the Stanford two-mile linear accelerator,” Stanford Linear Accelerator Center, Technical Report No. SLAC-R-0017, August 1963.

[11.40] O. H. Altenmueller, R. R. Larsen and G. A. Loew, “Investigation of traveling wave separators for the Stanford two-mile linear accelerator,” *Rev. Sci. Instrum.*, vol. 35, pp. 438-442, April 1964.

[11.41] G. A. Loew and O. H. Altenmueller, “Design and applications of RF separator structures at SLAC,” Stanford Linear Accelerator Center, Technical Report No. SLAC-PUB-0135, August 1965.

[11.42] B. Steffen, “Electro-optic methods for longitudinal bunch diagnostics at FLASH,” Ph.D. Thesis, Universität Hamburg, Hamburg, Germany, 2007.

[11.43] L. Frölich and O. Grimm, “Bunch length measurements using a Martin-Puplett interferometer at the VUV-FEL,” Ph.D. Thesis, Universität Hamburg, Hamburg, Germany, 2005.

[11.44] R. Lai and A. J. Sievers, “On using the coherent far IR radiation produced by a charged-particle bunch to determine its shape: I Analysis,” *Nucl. Instrum. Meth. A*, vol. 397, pp. 221-231, October 1997.

[11.45] R. Lai and A. J. Sievers, “Determination of a charged-particle-bunch shape from the coherent far infrared spectrum,” *Phys. Rev. E*, vol. 50, pp. R3342-3344, November 1994.

[11.46] G. Berden, S. P. Jamison, A. M. MacLeod *et al.*, “Electro-optic technique with improved time resolution for real-time, nondestructive, single-shot measurements of femtosecond electron bunch profiles,” *Phys. Rev. Lett.*, vol. 93 , 114802-1–114802-4, September 2004.

[11.47] S. P. Jamison, G. Berden, A. M. MacLeod *et al.*, "Electro-optic techniques for temporal profile characterisation of relativistic coulomb fields and coherent synchrotron radiation," *Nucl. Instrum. Meth A*, vol. 557, pp. 305-308, February 2006.

[11.48] E. L. Saldin, E. A. Schneidmiller and M. V. Yurkov, "Longitudinal space charge-driven microbunching instability in the TESLA Test Facility linac," *Nucl. Instrum. Meth. A*, vol. 528, pp. 355-359, August 2004.

[11.49] K. Floettmann, "Some basic features of the beam emittance," *Phys. Rev. ST Accel. Beams*, vol. 6, pp 034202-1–034202-7, March 2003.

[11.50] E. D. Courant and H. S. Snyder, "Theory of the alternating-gradient synchrotron," *Ann. Physics*, vol. 3, pp. 1-48, January 1958.

[11.51] F. Löhl, "Measurements of the transverse emittance at the VUV-FEL," Ph.D. Thesis, Universität Hamburg, Hamburg, Germany, 2005.

[11.52] S. G. Anderson, J. B. Rosenzweig, G. P. LeSage *et al.*, "Space-charge effects in high brightness electron beam emittance measurements," *Phys. Rev. ST Accel. Beams*, vol. 5, 014201-1–014201-12, January 2002.

[11.53] S. G. Anderson, "Creation, manipulation, and diagnosis of intense, relativistic picosecond photo-electron beams," Ph.D. Thesis, University of California Las Angeles, Las Angeles, Californaia, 2002.

[11.54] L. Staykov, "Characterization of the transverse phase space at the photo-injector test facility in DESY, Zeuthen site," Ph.D. Thesis, Universität Hamburg, Hamburg, Germany, 2008.

[11.55] S. C. Hartmann, N. Barov, C. Pellegrini *et al.*, "Initial measurements of the UCLA RF photoinjector," *Nucl. Instrum. Meth. A*, vol. 340, pp. 219-230, February 1994.

[11.56] S. Skelton. (2007). *Multi-quadrupole scan for emittance determination at PITZ* [Online]. Available FTP: www-zeuthen.desy.de Directory: students/2007/doc File: skelton.pdf

[11.57] The lattice of the FLASH injector beam diagnostics section to measure the emittance and the Twiss parameters was designed by P. Piot, Fermilab: the code to calculate the emittance using the multi-screen method is based on a similar code provided to us by P. Emma, SLAC.

[11.58] M. Geitz, "Investigation of the transverse and longitudinal beam parameters at the TESLA test facility linac," Ph.D. Thesis, Universität Hamburg, Hamburg, Germany, 1999.

[11.59] J. J. Scheins, "Tomographic reconstruction of transverse and longitudinal phase space distributions using the maximum entropy algorithm," TESLA Technology Collaboration, Technical Report No. 2004-08, May 2004.

[11.60] G. Minerbo, "MENT: A maximum entropy algorithm for reconstructing a source from projection data," *Computer. Graph. Image Process.*, vol. 10, pp. 48-68, May 1979.

[11.61] G. N. Minerbo, O. R. Sander and R. A. Jameson, "Four-dimensional beam tomography," *IEEE Trans. Nucl. Sci.*, vol. 28, pp. 2231-2233, June 1981.

[11.62] C. T. Mottershead, "Maximum entropy beam diagnostic tomography," *IEEE Trans. Nucl. Sci.*, vol. 32, pp. 1970-1972, October 1985.

[11.63] M. Röhrs, C. Gerth, H. Schlarb *et al.*, "Time-resolved electron beam phase space tomography at a soft x-ray free-electron laser," *Phys. Rev. ST Accel. Beams*, vol. 12, pp. 050704-1–050704-13, May 2009.

[11.64] M. Röhrs, C. Gerth, H. Schlarb *et al.*, "Time-resolved electron beam phase space tomography at a soft x-ray free-electron laser," *Phys. Rev. ST Accel. Beams*, vol. 12, pp. 050704-1–050704-13, May 2009.